THESE

pour l'obtention du grade de Docteur, délivré par

l'ECOLE NORMALE SUPERIEURE DE LYON

**Ecole Doctorale** N°512
InfoMaths - Informatique et Mathématiques de Lyon

**Discipline :** INFORMATIQUE (Informatique)

Soutenue publiquement le 17 septembre 2025, par :

**Giulia GIUSTI**

---

# Exploiter le Potentiel de la Linéarité dans la Différenciation Automatique et la Cryptographie Computationnelle

*Exploiting the Potential of Linearity in Automatic Differentiation and Computational Cryptography*

---

Après avis de :

| | |
|---|---|
| Bruce KAPRON, Professeur, Université de Victoria | Rapporteur |
| Patrick BAILLOT, Directeur de recherche, Université de Lille / CNRS | Rapporteur |

Devant le jury composé de :

| | |
|---|---|
| Bruce KAPRON, Professeur, Université de Victoria | Rapporteur |
| Patrick BAILLOT, Directeur de recherche, Université de Lille / CNRS | Rapporteur |
| Claudia FAGGIAN, Chargée de recherche, Université Paris Cité / CNRS | Examinatrice |
| Elaine PIMENTEL, Professeure, University College de Londres | Examinatrice |
| Michele PAGANI, Professeur des universités, École Normale Supérieure de Lyon | Directeur de thèse |
| Ugo DAL LAGO, Professeur, Université de Bologne | Co-directeur de thèse |

## Abstract

The concept of linearity plays a central role in both mathematics and computer science, albeit with different yet complementary meanings and applications. In mathematics, linearity is a fundamental structural property of functions and vector spaces, forming the foundation of linear algebra and functional analysis. In computer science, linearity is predominantly associated with resource-sensitive computation. In particular, linear logic expresses the use of an assumption exactly once, providing a natural framework to represent computational systems where the consumption of resources (such as time, memory, or data access) is explicit and trackable. This dual significance makes linearity a key concept in the development of programming languages, type systems, and formal models that express both quantitative properties, such as computational complexity, and functional properties, such as composability of transformations. The synthesis of mathematical and computational interpretations of linearity thus enables novel methodologies for the analysis and verification of complex systems, fostering the integration of theoretical rigour with practical applicability in the formal study of computation.

The general goal of this work is to exploit linear logic for modelling programming paradigms built atop the notion of linearity. More precisely, the thesis is structured in two parts: *ADLL* (Automatic Differentiation, Linear Logic and Lambda Calculus) and *CryptoBLL* (Computational Cryptography Bounded Linear Logic and Lambda Calculus). In the former, we apply linear logic to Automatic Differentiation in order to model the notion of linear function over real numbers and the related transposition operation, whereas in the latter linear logic is used to model constraints related to the computational complexity of attackers in the context of Computational Cryptography.

In the field of Automatic Differentiation, there are mainly two approaches that explicitly use a linear type system: the purely theoretical approach rooted in proof theory and type systems and the implementation-oriented approach defined in JAX, a Python library designed by Google for machine learning research. In contrast, other mainstream frameworks such as PyTorch and TensorFlow, despite their extensive support for automatic differentiation, do not integrate linear type systems as a foundational aspect of their design. The ADLL part of the thesis focuses on finding a connection between JAX's linear type system and linear logic, aiming to bridge the gap between theoretical foundations and practical implementations.

In the literature of modern cryptography there are several attempts to define a formal calculus aimed at modelling cryptographic constructions and proofs according to the computational model. In such models, a tension is evident between the need to be expressive, so as to capture proofs by reduction, and the need to keep the model simple enough, masking the details of probability and complexity. The CryptoBLL part aims at defining a framework that can be used for the automatic analysis of protocols in the context of computational cryptography, trying to alleviate the tension described above.

# Contents

# Introduction

The concept of linearity plays a central role in both mathematics and computer science, albeit with different yet complementary meanings and applications. In mathematics, linearity is a structural principle that governs a wide range of disciplines. At its core, it captures the idea that operations should preserve addition and scalar multiplication, properties that underpin the theory of *vector spaces*, *linear transformations*, and systems of linear equations. These ideas form the backbone of linear algebra, functional analysis, and countless applied domains — from physics and engineering to statistics and machine learning. In computer science, linearity is primarily associated with resource-sensitive computation. Unlike theoretical models, real-world computations are constrained by limited resources. It is within this landscape that *linear logic*, introduced by Jean-Yves Girard in 1987 [51], emerged as a revolutionary formalism. In linear logic, the use of assumptions is restricted so that each must be used *exactly once*, unless explicitly stated otherwise. This directly incorporates resource sensitivity into the logical framework and provides a solid foundation for systems that require careful tracking and management of resource consumption.

The synthesis of mathematical and computational interpretations of linearity has become a central focus in Programming Languages (PL) Theory, aimed at providing rigorous methodologies for analyzing and verifying complex systems. This integration of theoretical rigour with practical computation fosters a deeper understanding of the principles governing programming constructs, enabling the development of tools that are both theoretically sound and practically applicable.

A particularly prominent setting for this synthesis is the $\lambda$-calculus, a formal system introduced by Alonzo Church [30], which serves as the theoretical foundation of functional programming and offers a minimal yet powerful model of computation. In its classical form, it permits unrestricted reuse of variables and functions, a feature that stands in contrast to the resource constraints inherent in many practical computing systems. Evaluation strategies play a central role in determining the behaviour of programs written in $\lambda$-calculus. Two foundational strategies are Call-by-Value (CBV) and Call-by-Name (CBN) [96, 92]. In CBV, arguments to functions are evaluated before the function is applied, ensuring that each expression is reduced to a value exactly once. This strategy is common in strict functional languages and aligns naturally with many models of resource-aware computation. In contrast, CBN delays evaluation of function arguments until their values are actually needed, which can avoid unnecessary computations but may duplicate work if the same expression is evaluated multiple times. A more general and unifying framework that encompasses both CBV and CBN is Call-by-Push-Value (CBPV), introduced by Paul Blain Levy [78]. CBPV decomposes computation into two orthogonal categories: values, which are static data, and computations, which may involve effects or delayed execution. By distinguishing these two roles explicitly, CBPV can represent CBV and CBN as particular mode choices within a single operational structure. This makes it a powerful tool for analyzing evaluation order, program effects, and resource usage in a uniform and fine-grained way.

The integration of linearity into $\lambda$-calculus results in what is known as *linear $\lambda$-calculus* and marks a significant evolution in PL Theory and formal systems. By enforcing that functions consume their arguments exactly once, linear $\lambda$-calculus offers a model in which resource consumption is encoded at the level of the language's type system. This innovation enables formal reasoning about programs that interact within constrained environments. A key element driving this development is the *Curry-Howard correspondence* [64], a deep and powerful connection between logic and computation. At its core, this correspondence establishes a structural analogy between formal proofs and typed programs: logical propositions correspond to types, and proofs correspond to programs. Under this interpretation, writing a program is akin to constructing a proof, while type-checking ensures logical soundness. In particular, if we consider a linear $\lambda$-calculus whose type system has a Curry-Howard correspondence with linear logic, variables and resources are treated as assumptions that must be used exactly once, directly mirroring the logic's constraints. This embeds resource management directly into the structure of the program, allowing formal guarantees about resource consumption.

Within this theoretical landscape, logical relations have emerged as a foundational semantic technique for reasoning about program behaviour [97]. Originally developed to prove properties like strong normalization for the simply typed $\lambda$-calculus, logical relations have since evolved into a versatile and general method for establishing a wide range of semantic properties, including contextual equivalence, type soundness, and parametricity. A logical relation is typically defined as an inductively structured family of relations indexed by the types of a language, designed to relate terms that exhibit extensionally equivalent behaviour at each type. This methodology facilitates abstraction from syntactic details, enabling the establishment of properties of interest in a type-directed, compositional, and semantically grounded way. Consequently, logical relations are particularly well-suited for analyzing complex systems, especially in contexts involving resource constraints, higher-order functions, polymorphism, and effectful computation.

The ongoing convergence of these ideas has significant implications for the formal verification of programs and systems. Linearity allows us to express that a program is efficient, safe, and well-behaved with respect to its resource consumption. This thesis builds upon this rich theoretical landscape by exploring how linear logic and linear $\lambda$-calculus can be leveraged to model programming paradigms based on the notion of linearity. More precisely, the thesis is structured in the following two parts:

- **ADLL: Automatic Differentiation, Linear Logic and Lambda Calculus**
  Automatic Differentiation (AD) provides efficient methods for computing the derivative (or, more generally, the gradient or Jacobian) of a function specified by a computer program. Since computing derivatives is a key ingredient in the resolution of all sorts of optimization problems, it is not surprising that AD grew into a large field with applications to a host of scientific domains, most notably machine learning, where AD is used in many learning processes, like in the implementation of the gradient descent algorithm [19].

  Traditionally, AD focused on first-order imperative programs and its scope was often limited to straight-line programs (also known as *computational graphs* [18]), which were enough for most practical purposes, such as expressing neural networks. However, the advancements in deep learning, along with the rapid development of automatic differentiation frameworks (e.g.[3, 90]), have significantly broadened its applicability. At the same time, AD has attracted considerable attention from PL research, particularly in efforts to develop a formal understanding of its principles and establish its theoretical foundations, e.g. [47, 112, 102, 2, 65, 26, 83].

From the viewpoint of PL theory, AD methods boil down to *program transformations*: writing $\mathbb{R}$ for the set of real numbers, and given a program $M$ computing a function $[\![M]\!] : \mathbb{R}^n \to \mathbb{R}$ whose gradient $\nabla[\![M]\!]$ exists, the goal is to output another program $grad(M)$ that computes $\nabla[\![M]\!]$. The PL viewpoint addresses a series of questions which can be tackled with theoretical tools, such as:

(i) *soundness*: does $grad(M)(\vec{r})$ evaluate to $\nabla[\![M]\!](\vec{r})$, for all $\vec{r} \in \mathbb{R}^n$?

(ii) *efficiency*: how does it cost to evaluate $grad(M)$ with respect to the evaluation of $M$?

(iii) *modularity*: if $P$ is a subprogram of $M$ then is $grad(P)$ a subprogram of $grad(M)$? or, if this is not literally the case, may the computation of the former be reused in computing the latter?

Linearity is essential in AD as it ensures efficient resource management by restricting data to be used exactly once, preventing redundancy and reducing memory overhead. This restriction not only enhances efficiency by minimizing redundant computations but also guarantees the correctness of gradient calculations, offering formal assurances about the accuracy of differentiation. By managing memory and computational resources more effectively, linearity enables optimizations that improve performance, particularly in large-scale or resource-constrained applications. This formal control over resource usage ensures that AD processes remain both precise and scalable. The general purpose of ADLL is to show how linear logic can help in understanding AD program transformations and in giving answers to the questions (i)-(iii) listed above.

In the literature, there are mainly two prominent approaches that explicitly use a linear type system: a theoretical approach exemplified by the work of Mazza and Pagani [26, 83], and an implementation-oriented approach defined in JAX [93].

The work of Mazza and Pagani is part of a broader research effort exploring linear type systems in the context of automatic differentiation. A pioneering paper is [2], which formalizes backward-mode automatic differentiation in a simple while-language and proves soundness via a denotational model. The concept of linearity plays a crucial role in more categorical and type-theoretic approaches to automatic differentiation. Notable examples include CHAD [108, 81], which leverages a categorical and compositional perspective grounded in linear logic, and the Dialectica system [71], which introduces a differentiation operator via a refined linear type system. Another significant contribution is the work by Smeding and Vákár [105] that discusses an implementation of dual-numbers reverse-mode automatic differentiation in Haskell, building upon the foundational approach of [26]. Their work introduces several optimizations aimed at improving both the efficiency and scalability of the differentiation process.

JAX is a Python library developed by Google that combines NumPy-like APIs with a powerful system of composable function transformations designed for high-performance numerical computing and machine learning research. In JAX, the automatic differentiation system is specifically referred to as *Autodiff*. Linear A, introduced in [93], is a model language that formalizes the core principles of Autodiff and offers a minimal, type-theoretic foundation for reasoning about differentiation, while also establishing the soundness of the transformations in Autodiff. The typing discipline of Linear A integrates a form of linearity, requiring that variables associated with tangent data be managed through explicit constructors for duplication and erasure. In contrast, other mainstream frameworks such as PyTorch [90] and TensorFlow [3], despite offering extensive support for automatic differ-entiation and being widely adopted in the machine learning community, do not incorporate linear type systems as a core element of their design.

The goal of this part of the thesis is to investigate whether the type system of Linear A has a logical interpretation, particularly with respect to the linearity constraints. Specifically, we aim to explore whether these constraints can be understood within a substructural logic that controls the contraction or weakening of hypotheses — concepts that are logically analogous to the duplication or erasure of data.

In order to investigate the logical foundations of Linear A, we define an encoding into a linear $\lambda$-calculus, called $\lambda$LL, which establishes a Curry-Howard correspondence with a fragment of linear logic [51]. The $\lambda$LL calculus extends the linear logic $\lambda$-calculus to the ground type of the real numbers $\mathbb{R}$ with a set of functional symbols which are associated with differentiable functions. We then formally define AD transformations in $\lambda$LL and we prove their soundness: *qualitatively*, by proving that they compute the same as their counterparts in Autodiff, and *quantitatively*, by showing that they achieve comparable efficiency. Moreover, leveraging the correspondence with linear logic, we show that our framework allow us to skip a fundamental transformation in Autodiff, known as unzipping, while preserving efficiency. By skipping unzipping, our approach enhances the modularity of the AD system, particularly in the presence of programs with independent subroutines, and offers potential improvements in compositional reasoning and parallelism.

- **CryptoBLL: Cryptography, Bounded Linear Logic and Lambda Calculus**
  Cryptographic protocols are designed to ensure secure communications over insecure channels. The growing size of networks and their dependence on cryptographic protocols require a high level of security in protocols.

  In modern cryptography, security properties are rigorously defined and the verification of protocols with respect to these properties is carried out using mathematical proofs. A successful approach in this line of research consists in modelling messages using symbolic expressions subject to an equational theory, which represents the capabilities of the attacker. Originally proposed by Dolev and Yao [42], this idea has been refined over the years, giving rise to multiple models known as *formal* or *symbolic*. The latter capture large classes of attackers and allow an automated verification of the protocols that is reflected in the design of several concrete tools. However, tools based on the symbolic model struggle to support primitives with rich algebraic properties. For this reason, the security property in the symbolic model is often weaker than the notion of security based on a second model, called the *computational model* [54, 70]. In this model, the adversaries are represented by Turing machines that work in probabilistic polynomial time, the cryptographic operations are seen as functions on bitstrings and the properties of security are defined in terms of the probability and computational complexity of the attackers. A central concept in this framework is *computational indistinguishability*, which formalizes the idea that two distributions (typically, the outputs of cryptographic constructions) are considered secure if no efficient adversary can distinguish them with more than negligible advantage. Security proofs in the computational model are thus based on proving that an adversary cannot distinguish between the real protocol and an idealized version, according to this notion of indistinguishability. However, the computational model presents several limitations regarding the excessive complication of probabilistic reasoning, which in turn leads to a low scalability of the model as the number of parties involved in the analyzed protocol increases.

  In computational cryptography, linearity facilitates accurate modelling of adversarial constraints by limiting resource usage. This approach closely mirrors linear logic, which formalizes resource management by ensuring that each resource is consumed exactly once unless explicitly duplicated. By modelling the attacker's capabilities in this manner, linearity aligns with complexity-theoretic assumptions about bounded attackers, guaranteeing that resource usage is both explicit and verifiable.

In the literature there are several attempts to define a formal calculus aimed at modelling cryptographic constructions and proofs according to the computational model. The so called Universal Composability (UC) model, introduced by Canetti more than twenty years ago [28, 27], provides a rigorous framework for defining and analyzing the security of cryptographic protocols under composition. The UC model ensures that a protocol remains secure when composed with others by comparing real-world execution with an idealized execution in a secure environment. This composability is a key strength, enabling reliable integration into larger systems. However, UC's complexity makes proofs intricate and difficult to verify, with computational assumptions sometimes obscuring protocol intuition. Additionally, UC struggles with higher-order cryptographic settings, where protocols manipulate or generate other protocols. UC assumes a fixed, first-order interaction structure, where parties follow prescribed roles and defined interfaces. However, higher-order constructions, like passing oracles or instantiating sub-protocols at runtime, challenge these assumptions, making it difficult to ensure security. As a result, even if individual components are UC-secure, their composition in a higher-order context may not preserve security. These limitations have driven efforts aimed at determining if it is possible to either simplify it or to capture it by way of a calculus or process algebra (e.g. [29, 79, 73, 16]). In all the aforementioned works, a tension is evident between the need to be expressive, so as to capture UC proofs, and the need to keep the model simple enough, masking the details of probability and complexity as much as possible.

The second part of this thesis aims at defining a framework that can be used for the automatic analysis of protocols in the context of computational cryptography, trying to alleviate the tension described above. In particular, we define a language based on $\lambda$-calculus, called $\lambda$BLL, which is flexible enough to model cryptographic experiments in the sense of computational cryptography. The $\lambda$BLL calculus extends the pure $\lambda$-calculus with probabilistic polynomial-time function symbols and global references, and its linear type system has a correspondence with Bounded Linear Logic [52, 63, 38] to capture complexity constraints of adversarial computations.

The security proofs based on computational indistinguishability [70, 55] are expressed through equations over terms in the $\lambda$BLL language. However, equational reasoning at the syntactic level alone is not sufficient to capture more subtle aspects of such cryptographic proofs. In order to overcome these limitations, we develop an approximate logical relation that is sound with respect to computational indistinguishability in a higher-order $\lambda$-calculus with probabilistic effects and references. The resulting framework enables a structured and compositional proof methodology, where cryptographic proofs are formulated as a series of equational steps, each step justified via the logical relation. If two terms are related by the logical relation, their observable behaviours are computationally indistinguishable to any probabilistic polynomial-time adversary. This approach abstracts away from low-level simulation-based arguments typically used in cryptographic proofs, offering a higher-level and compositional reasoning methodology.

Both parts of the thesis are unified by the underlying insight that linearity is not just a mathematical convenience or a syntactic constraint, but a powerful conceptual framework. It enables precise control over how information and resources flow through a computation and provides a foundation for reasoning about correctness, efficiency, and security in a compositional way. By drawing connections between differentiation, cryptographic reasoning, and the logical structure of computation, this thesis aims to contribute to a broader understanding of how linearity can serve as a guiding principle in the design of languages, logics, and systems. It highlights the versatility of linear logic not only as a theoretical tool but also as a practical framework for the modelling, analysis, and verification of complex computational paradigms. The connection between the two parts of the thesis and the unifying role of linearity are graphically summarized in Figure 1.

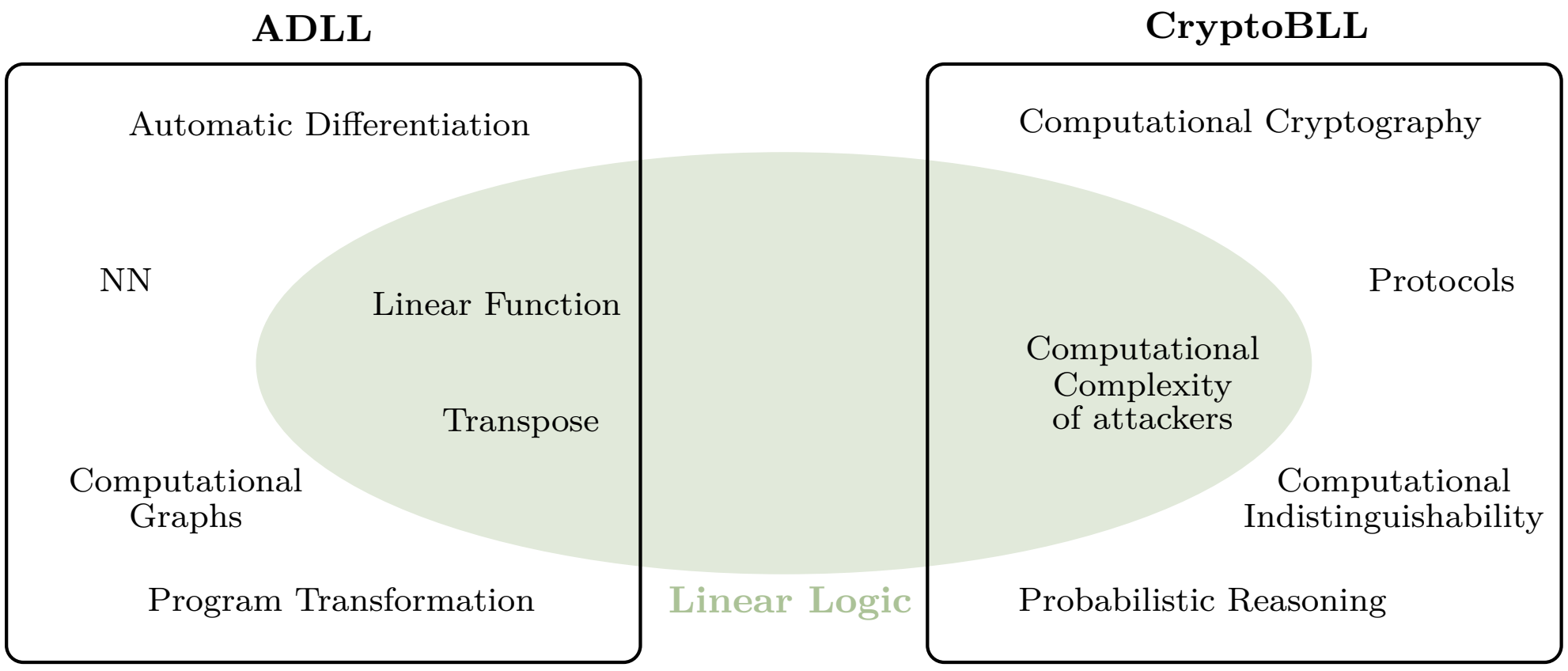

Figure 1: Conceptual summary of the thesis structure. The left side illustrates the first part of the thesis (ADLL), where linear logic is applied to Automatic Differentiation in order to model the notion of linear function over real numbers and the related transposition operation. The right side represents the second part (CryptoBLL), where linear logic is used to model constraints related to the computational complexity of attackers in the context of Computational Cryptography. The central role of linear logic unifies both contributions as a foundational framework.

# Part I

# ADLL

**Automatic Differentiation, Linear Logic and Lambda Calculus**

Chapter 1

# Introduction to Automatic Differentiation

Consider a program $P$ that computes a real function $[\![P]\!]$ from $\mathbb{R}^n$ to $\mathbb{R}$. Automatic differentiation (AD) refers to a family of algorithms for evaluating derivatives and gradients of numerical functions, such as $[\![P]\!]$, by leveraging the source code of programs like $P$. This approach contrasts with other methods, such as numerical differentiation, which computes small differences in $[\![P]\!]$, and symbolic differentiation, which manipulates closed forms of $[\![P]\!]$. AD can be presented as a program transformation, similar to a compilation procedure or a domain-specific interpretation applied to $P$.

The directional derivative $\mathrm{D}_{\vec{u}}[\![P]\!](x)$ (assuming it exists) intuitively indicates how much a small perturbation at point $x$ along the direction given by the tangent vector $\vec{u} \in \mathbb{R}^n$ affects the output of $[\![P]\!]$. The gradient $\nabla[\![P]\!](x)$, on the other hand, is a vector that points in the direction of the steepest ascent of $[\![P]\!]$ at $x$. AD primarily operates in two modes: the *forward mode*, which efficiently computes the derivative $\mathrm{D}[\![P]\!]$, and the *backward or reverse mode*, which generates a program that evaluates $\nabla[\![P]\!]$.

The literature on AD dates back to the 60s (e.g. [113]), and we can acknowledge three distinct periods or trends. Initially, AD focused on low-level programs with very simple programming primitives. Only narrow fragments of programming languages like FORTRAN or C were considered, encompassing floating-point variables, arrays, branching, goto statements, and while-loops. This approach was a natural choice to ensure the efficiency of the computation while maintaining enough structure to share intermediate results between different subroutines of a program.

Since computing derivatives is a key ingredient in the resolution of all sorts of optimization problems, it is not surprising that AD grew into a large field with applications to a host of scientific domains, most notably *machine learning*, where AD is used in many learning processes, like in the implementation of the gradient descent algorithm [20]. Hence, a second period or generation of AD has advanced towards comprehensive AD systems for large general-purpose programming languages such as C++ or Python. This approach has surged in the last decade with the development of industrial deep learning libraries like TensorFlow [3], PyTorch [90], and Jax [10]. These libraries apply AD to complex programs which define numerical functions (e.g., neural networks) dynamically and incorporate increasingly complex programming features such as procedure calls, recursive functions, user-defined types, classes, and more.

Returning to a more academic line of research, a third phase or trend is characterised by efforts to formalise these techniques within an idealised framework[1]. The aim is to develop a formal system that models the core principles of modern AD implementations, abstracting from specific programming language details and other features such as parallel computation and floating-point arithmetic. The goals of this theoretical approach are manifold: to establish soundness proofs, which become less straightforward as program complexity increases; to elucidate the assumptions underlying such proofs, such as program termination, smoothness, and data persistence; to analyse asymptotic complexity (as opposed to performance evaluation in practical testing); and to decompose AD while drawing connections with other concepts in the theory of programming languages.

[1] The term "formalisation" may be misunderstood as providing a mechanised proof in a proof assistant. This is too narrow in scope here: by formalisation, we refer to a general theoretical analysis of an algorithm or program transformation — providing definitions and precise statements that can be proven or refuted by counterexamples.

In this chapter we first introduce the two modes of AD using a standard framework for the evaluation of functions over real numbers and then we describe how this work fits into the last period of research in Automatic Differentiation.

## 1.1 The Two Modes of Automatic Differentiation

As mentioned above, there are two main modes of AD: forward mode and reverse mode. The terminology refers to the execution flow of the computation: the forward mode propagates tangent vectors from the inputs of the program $P$ to its output, while the backward mode traces back from the output to the inputs.

In this section, we first introduce a standard framework for evaluating functions that we will use to describe how forward and reverse AD transform $P$ to compute $\mathrm{D}[\![P]\!]$ and $\nabla[\![P]\!]$, respectively.

### 1.1.1 Framework for Evaluating Functions

Automatic Differentiation (AD) can be viewed as a transformation of a program $P$ computing a complex function $[\![P]\!]$ from $\mathbb{R}^n$ to $\mathbb{R}$, which is composed of multiple primitive operations. These primitive operations encompass basic arithmetic operations (addition, subtraction, multiplication, and division), unary operations like sign switching, as well as transcendental functions, including the exponential, logarithmic, and trigonometric functions. In the transformations performed by AD the standard computation of $P$ is enhanced by the simultaneous calculation of the derivatives of the primitive operations. This approach is grounded in the principle that all numerical computations can ultimately be reduced to compositions of a finite set of primitive operations, for which derivatives are already well-known. By applying the chain rule, the derivatives of these primitive operations are combined to obtain the derivative of the overall computation.

In the context of AD, the transformation of a program $P$ is typically represented using a *computational graph* and its *evaluation trace*, that visually and structurally represents how the mathematical function $[\![P]\!]$ is computed. The evaluation trace is the sequence of primitive operations executed during the function's evaluation, capturing both their execution order and data dependencies. This trace is essential for tracking intermediate values and enabling the efficient propagation of derivatives through the computational graph during differentiation.

In the following we adopt a framework for evaluating functions similar to the one introduced by Griewank and Walther in [61, Chapter 2], where a function from $\mathbb{R}^n$ to $\mathbb{R}$ such that $(x_1, \ldots, x_n) \mapsto [\![P]\!](x_1, \ldots, x_n)$ is represented by variables $v_i$ such that: variables $v_{x_i} = x_i$ for $i \in \{1, ..., n\}$ are the input variables, $v_i$ for $i \in \{1, ..., l\}$ are the intermediate variables and $z$ is the output variable. Moreover, we graphically represent the function $[\![P]\!]$ by using computational

This stands in contrast to more "experimental methods" based on testing and real-world runtime evaluations, which offer a different yet complementary approach to program analysis. We do not delve into mechanised proofs of AD using proof assistants in this paper; this remains an ultimate goal and such theoretical modelling is a preliminary step.

graphs, introduced by Bauer [18] and its evaluation trace. Formally, a computational graph is a Directed Acyclic Graph (DAG) where the vertices are variables and the edges represents the dependence relations between these variables.

Let us illustrate this framework with an example that will be referenced throughout the subsequent chapters of this part of the thesis.

**Illustrative Example 1.** Consider the function $g : \mathbb{R}^2 \to \mathbb{R}$ computing $z = (sin(x)*y)+cos(x)$. In Figure 1.1a we present a program $P$ which computes $g$, using a python-like syntax. The computational graph in Figure 1.1b graphically describes the dependencies of intermediate values of $g$. More precisely, $v_x$ and $v_y$ are the input variables, $v_i$ for $1 \leq i \leq 4$ are the intermediate variables and $z$ is the output variable. Finally, the evaluation trace in Figure 1.1c lists the sequence of operations executed during the evaluation of $g$.

```
def P(x,y):
    v1 = sin(x)
    v2 = v1*y
    v3 = cos(x)
    v4 = v2+v3
    return v4
```

(a) Program $P$ computing $z = (sin(x) * y) + cos(x)$.

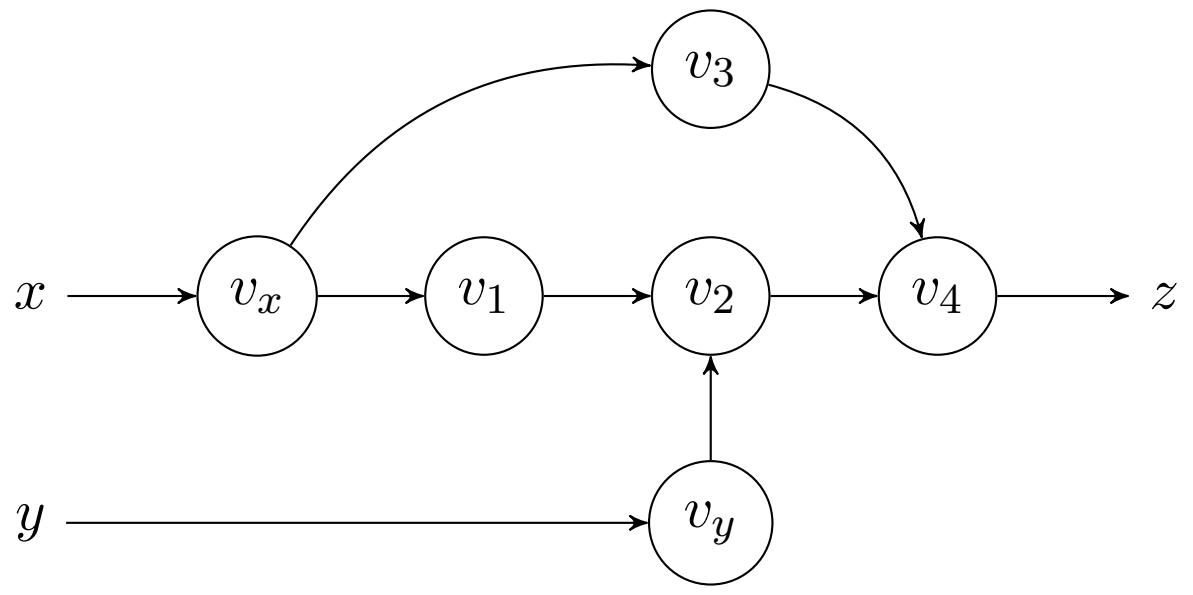

(b) Computational Graph for $P$.

$$
\begin{aligned}
v_x &= x \\
v_y &= y \\
v_1 &= sin\ v_x \\
v_2 &= v_1 * v_y \\
v_3 &= cos\ v_x \\
v_4 &= v_2 + v_3 \\
z &= v_4
\end{aligned}
$$

(c) Evaluation Trace.

Figure 1.1: Framework evaluating function $g$ computing $z = (sin(x) * y) + cos(x)$.

### 1.1.2 Forward Mode

The conceptually simplest AD transformation is known as forward mode differentiation. At the core of forward mode AD lies the *chain rule*, which describes how to compute the derivative of a composition of functions. Given a function $y = f(g(x))$ composed of two derivable functions $f$ and $g$, then by applying the chain rule its derivative is given by:

$$\frac{\partial y}{\partial x} = \frac{\partial f}{\partial g} \cdot \frac{\partial g}{\partial x}$$

where: $\frac{\partial f}{\partial g}$ is the derivative of $f$ with respect to $g(x)$, indicating how changes in $g(x)$ affect $f(g(x))$, and $\frac{\partial g}{\partial x}$ is the derivative of $g$ with respect to $x$, capturing how changes in $x$ affect $g(x)$. This principle extends to more deeply nested or complex compositions. In general, the chain rule describes how small changes in the inputs affect the overall computation, making it crucial in computing derivatives of complex functions.

Forward mode AD applies the chain rule by propagating derivatives together with function values throughout a computation. This is accomplished using *dual numbers*, an extension of the real numbers that allows both quantities to be tracked simultaneously in a single evaluation. Specifically, dual numbers enable the systematic application of the chain rule through nested compositions of functions, efficiently maintaining derivative information at each step. Consider again the composition $y = f(g(x))$, where $x$ is replaced by a dual number $(x, \epsilon)$ representing a perturbation $x + \epsilon$ on $x$. In this dual number the primal part $x$ represents the real value of the input and the tangent part $\epsilon$ is a symbolic infinitesimal that tracks the derivative information. When evaluating the inner function $g(x)$ at this dual number, the result is $(g(x), \epsilon \cdot \frac{\partial g}{\partial x})$, where the primal part gives the value of $g(x)$ and the tangent part contains the derivative $\frac{\partial g}{\partial x}$. Then, when applying the outer function $f$, the dual number is used to propagate the derivative as follows:

$$f(g(x), \epsilon \cdot \frac{\partial g}{\partial x}) = (f(g(x)), \epsilon \cdot \frac{\partial f}{\partial g} \cdot \frac{\partial g}{\partial x})$$

Thus, the value of $y$ is $f(g(x))$, and the derivative $\frac{\partial y}{\partial x}$ is efficiently computed as $\frac{\partial f}{\partial g} \cdot \frac{\partial g}{\partial x}$, with the derivative information being propagated through the entire composition.

In our setting, forward mode AD can be seen as evaluating the function computed by a program $P$ using the concept of dual numbers: each numeric variable $x$ in $P$ is paired with a sibling variable $\dot{x}$, where $x$ is called the *primal* and $\dot{x}$ the *tangent*. Therefore, a tangent variable stores the differential information related to the primal sibling, arising from small perturbations in the inputs of $P$. Suppose that we are given a program $P$ such that $[\![P]\!] : \mathbb{R}^n \to \mathbb{R}$ whose computation is represented by a primal evaluation trace as described in Subsection 1.1.1. The forward mode transformation start by associating to each input variable $v_{x_i}$ a tangent input variable $\dot{v}_{x_i}$. It then proceeds by applying the chain rule to each primitive operation in the computation of the intermediate variables $v_i$, obtaining tangent variables $\dot{v}_i$ for $i \in \{1, \ldots, l\}$ computing the derivatives. After iterating over the intermediate variables we obtain the required derivative in $\dot{v}_n$, the forward transformation terminates by setting the tangent output $\dot{z}$ to $\dot{v}_n$. The result of this transformation is a program $\mathcal{F}(P)$ computing the function and its derivative $\mathrm{D}[\![P]\!]$. It is worth noting that, to compute a single partial derivative $\frac{\partial [\![P]\!]}{\partial x_i}$, forward mode AD sets the tangent of $x_i$, denoted by $v_{x_i}$, to 1 and all other tangents input variable to 0. Consequently, the full gradient $\nabla[\![P]\!]$ can be obtained by executing $\mathcal{F}(P)$ $n$ times, once for each input variable.

**Illustrative Example 2** (Forward Mode AD)**.** Consider the program $P$ in Example 1 computing $g(x, y) = (sin(x) * y) + cos(x)$. Mathematically, we have that the directional derivative of $g$ is

$$\mathrm{D}(g)(x, \dot{dx}, y, \dot{dy}) = (cos(x) * y) * \dot{dx} + sin(x) * \dot{dy} - sin(x) * \dot{dx}$$

By applying the forward mode as described above we obtain the program $\mathcal{F}(P)$ in Figure 1.2a and its computational graph in Figure 1.2b where the parts in blue are those added by the forward transformation. We can easily observe that the program $\mathcal{F}(P)$ in Figure 1.2a is computing the function $g(x, y)$ and its derivative $\mathrm{D}(g)(x, \dot{dx}, y, \dot{dy})$. Moreover, the primal evaluation trace in Figure 1.2c is augmented by the tangent operations in Figure 1.2d.

It is worth noting that the computational complexity of $\mathcal{F}(P)$ is asymptotically linear in the complexity of the original program $P$, assuming that the basic numerical operations in the program $P$ have constant complexity.

$$g(x,y) = (sin(x) * y) + cos(x)$$
$$\mathrm{D}(g)(x, \dot{dx}, y, \dot{dy}) = (cos(x) * y) * \dot{dx} + sin(x) * \dot{dy} - sin(x) * \dot{dx}$$

```
def F(P)(x,dẋ,y,dẏ):
    v1 = sin(x)
    v̇1 = cos(x)*dẋ
    v2 = v1*y
    v̇2 = v̇1*y+dẏ*v1
    v3 = cos(x)
    v̇3 = (-sin(x))*dẋ
    v4 = v2+v3
    v̇4 = v̇2+v̇3
    return (v4,v̇4)
```

(a) Program obtained by applying forward mode AD to $P$.

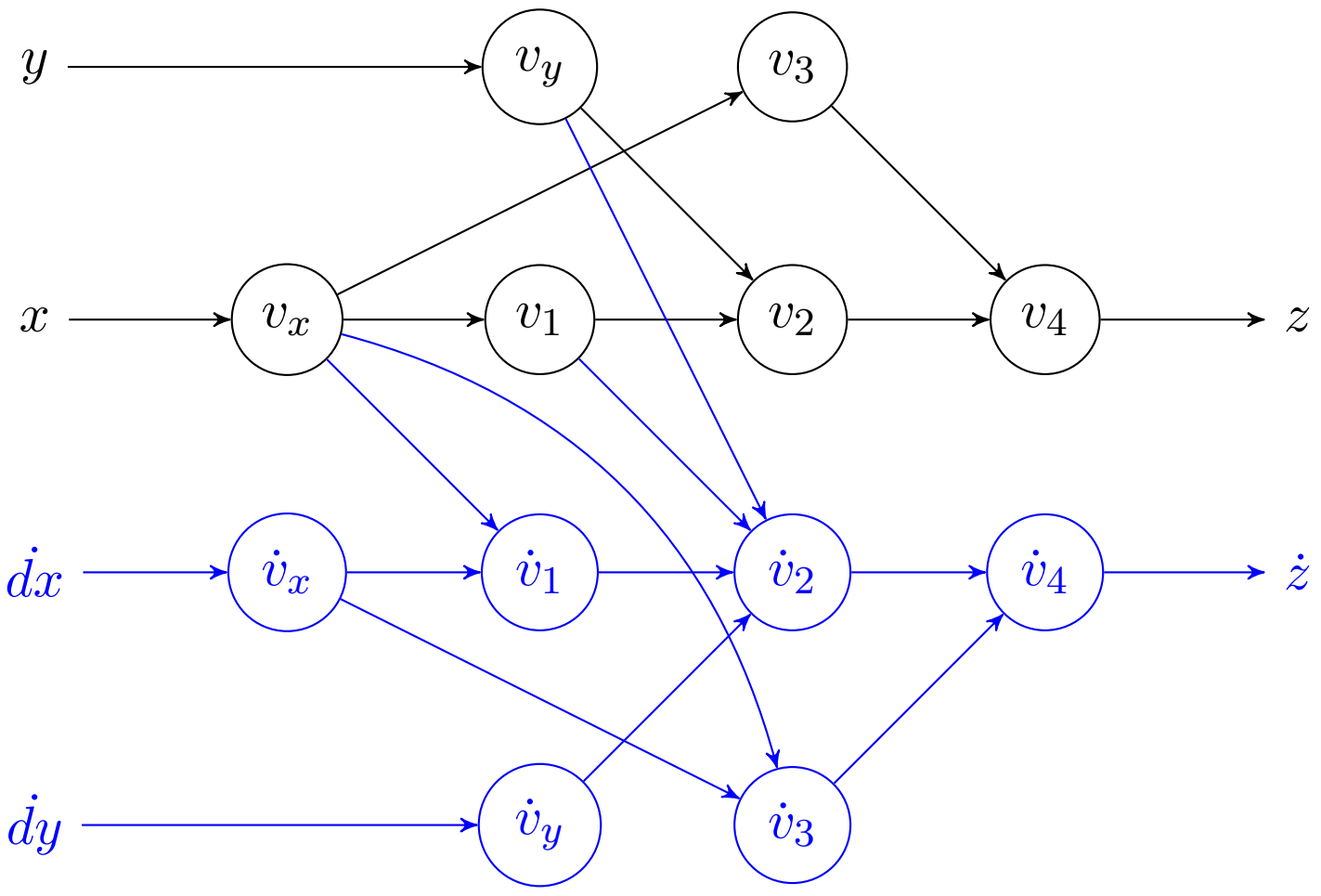

(b) Computational Graph for $\mathcal{F}(P)$.

$$v_x = x \qquad \dot{v}_x = \dot{dx}$$
$$v_y = y \qquad \dot{v}_y = \dot{dy}$$
$$v_1 = sin\ v_x \qquad \dot{v}_1 = cos\ v_x * \dot{v}_x$$
$$v_2 = v_1 * v_y \qquad \dot{v}_2 = \dot{v}_1 * v_y + \dot{v}_y * v_1$$
$$v_3 = cos\ v_x \qquad \dot{v}_3 = -sin\ v_x * \dot{v}_x$$
$$v_4 = v_2 + v_3 \qquad \dot{v}_4 = \dot{v}_2 + \dot{v}_3$$
$$z = v_4 \qquad \dot{z} = \dot{v}_4$$

(c) Primal Evaluation Trace.

(d) Tangent Evaluation Trace.

Figure 1.2: Forward Mode applied to the program $P$ computing the function $g$.

### 1.1.3 Reverse Mode or Backward Mode

Reverse mode AD is another way of computing both the value of a function and its derivative, but it takes a different approach compared to forward mode AD. Instead of computing derivatives as the computation progresses, reverse mode computes the derivatives backward starting from the output and propagating toward the inputs. This approach is particularly efficient when there are more input variables than output variables. In reverse mode, we still use dual numbers: each numeric variable $x$ in $P$ is paired with a sibling variable $\overline{x}$, where $x$ is called the *primal* and $\overline{x}$ the *adjoint*.

Suppose that we are given a program $P$ such that $[\![P]\!] : \mathbb{R}^n \rightarrow \mathbb{R}$ whose computation is represented by a primal evaluation trace as described in Subsection 1.1.1. In reverse mode AD, derivatives are computed during the second phase of a two-phase process. In the first phase, the function is evaluated from the input to the output. More precisely, it initializes the primal input variables $v_{x_i}$ to the corresponding primal value $x_i$ and proceed by computing the intermediate variables $v_i$, recording the dependencies in the computational graph through a bookkeeping procedure. In the second phase, reverse mode AD proceeds backward, from the output to the inputs. It starts by initializing the last adjoint intermediate variable $\overline{v_n}$ to the adjoint input value $\overline{z}$. It then propagates the adjoints $\overline{v_i}$ in reverse (from $\overline{v_{n-1}}$ to $\overline{v_1}$) by applying the chain rule to compute the derivatives of each intermediate variable. After iterating over the intermediate variables, it propagates the adjoints through the adjoint input variables, obtaining the required derivatives in $\overline{v_{x_i}}$ for $n \leq i \leq 1$. Finally, the reverse transformation terminates by setting the adjoint outputs $\overline{x_i}$ to $\overline{v_{x_i}}$ for $1 \leq i \leq n$. The result of this transformation is a program $\mathcal{R}(P)$ computing the original function and its gradient $\nabla[\![P]\!]$ backward.

**Illustrative Example 3** (Reverse Mode AD)**.** Consider again the program $P$ computing the function $g$ such that $z = (sin(x) * y) + cos(x)$. Mathematically, we have that the backward gradient of $g$ is

$$\nabla g(x, y, \overline{z}) = (\overline{z} * (y * cos(x) - sin(x)), sin(x) * \overline{z})$$

By applying the reverse mode as described above we obtain the program $\mathcal{R}(P)$ in Figure 1.3a and its computational graph in Figure 1.3b where the parts in red are those added by the reverse transformation.

Since reverse mode is tricky to understand, let's proceed step by step. In the first phase, reverse AD applied to $P$ evaluates the function from the input to the output, generating the primal evaluation trace in Figure 1.3c and the black part of the computational graph in Figure 1.3b. In the second phase, reverse mode AD applied to this example starts by initializing $\overline{v_4}$ to $\overline{z}$. It then proceeds by propagating the adjoints in reverse, applying the chain rule. Intuitively, we are computing the contribution $\overline{v_i} = \frac{\partial z}{\partial v_i}$ for $1 \leq i \leq 3$ of the change in each variable $v_i$ to the change in the output $z$ as follows

$$\overline{v_3} = \overline{v_4} * \frac{\partial v_4}{\partial v_3} = \overline{v_4} * \frac{\partial(v_2 + v_3)}{\partial v_3} = \overline{v_4} * 1$$
$$\overline{v_2} = \overline{v_4} * \frac{\partial v_4}{\partial v_2} = \overline{v_4} * \frac{\partial(v_2 + v_3)}{\partial v_2} = \overline{v_4} * 1$$
$$\overline{v_1} = \overline{v_2} * \frac{\partial v_2}{\partial v_1} = \overline{v_2} * \frac{\partial(v_1 * v_y)}{\partial v_1} = \overline{v_2} * v_y$$

Afterward, it propagates the adjoints through the adjoint input variables as follows

$$\overline{v_y} = \overline{v_2} * \frac{\partial v_2}{\partial v_y} = \overline{v_2} * \frac{\partial(v_1 * v_y)}{\partial v_y} = \overline{v_2} * v_1$$
$$\overline{v_x} = \overline{v_3} * \frac{\partial v_3}{\partial v_x} + \overline{v_1} * \frac{\partial v_1}{\partial v_x} = \overline{v_3} * \frac{\partial(cos\ v_x)}{\partial v_x} + \overline{v_1} * \frac{\partial(sin\ v_x)}{\partial v_x} = \overline{v_3} * (-sin\ v_x) + \overline{v_1} * cos\ v_x$$

Take the variable $\overline{v_x}$ as an example, we can observe in Figure 1.3b that the only way it can effect $z$ is through effecting $v_1$ and $v_3$ (arcs exiting from vertex $v_x$ in the computational graph). Therefore, the contribution of $v_x$ to the changes in $z$ is given by $\frac{\partial z}{\partial v_x} = \frac{\partial z}{\partial v_3}\frac{\partial v_3}{\partial v_x} + \frac{\partial z}{\partial v_1}\frac{\partial v_1}{\partial v_x}$ which is equivalent to $\overline{v_x} = \overline{v_3} * \frac{\partial v_3}{\partial v_x} + \overline{v_1} * \frac{\partial v_1}{\partial v_x}$. Finally, the reverse transformation terminates by setting the adjoint outputs to the corresponding output adjoint variable as follows

$$\overline{dx} = \overline{v_x}$$
$$\overline{dy} = \overline{v_y}$$

The result of the procedure just described is summarized in Figure 1.3d (reading it from bottom to top as the adjoint variables are listed in the same order used in the primal evaluation trace to explicit the correspondence).

## 1.2 Goals of ADLL

Consider the three lines of research outlined at the beginning of the chapter. Our contribution fits into the third line, which is devoted to the formalization of AD techniques within an idealized framework. This work starts from the paper [93], which formalises how AD, in particular its reverse mode, is implemented in libraries like JAX and Dex. In the context of JAX, the AD mechanism is specifically referred to as *Autodiff*. Recall the notation introduced earlier: a program $P$ computes a real function $[\![P]\!]$ from $\mathbb{R}^n$ to $\mathbb{R}$, and there are forward and reverse modes for computing the directional derivative $\mathrm{D}[\![P]\!]$ and the gradient $\nabla[\![P]\!]$, respectively. It is well-known that these two notions are dual to each other, in the sense that $\mathrm{D}_{\vec{u}}[\![P]\!](x) = \nabla[\![P]\!](x) \cdot \vec{u}$ for any point $x$ and tangent vector $\vec{u} \in \mathbb{R}^n$. The peculiarity of JAX is to start from this fact and to implement the reverse mode as a composition of three intermediate program transformations: the forward mode, denoted here as $\mathcal{F}$, the unzipping $\mathcal{U}$, and the linear transpose $\mathcal{T}$:

$$\nabla[\![P]\!] \approx [\![\mathcal{T}(\mathcal{U}(\mathcal{F}(P)))]\!] \tag{1.1}$$

The implementation of $\mathcal{F}$ implements the forward mode by using the concept of dual numbers, as described in the previous section (see Subsection 1.1.2).

The unzipping transformation $\mathcal{U}$ divides the program $\mathcal{F}(P)$ into two subroutines: $\mathcal{F}(P)^p$, which computes all primal outputs of $\mathcal{F}(P)$, and $\mathcal{F}(P)^t$, which computes all tangent outputs. Specifically, the primal computation is independent of the tangent values, whereas the tangent computation generally depends on the primal values. Therefore, $\mathcal{F}(P)^t$ is defined as a program that takes as input the tangent variables associated with the inputs of $P$, along with a sequence of primal variables that store the values computed by $\mathcal{F}(P)$ affecting certain tangent variables. This sequence of variables corresponds to the *tape* in some AD literature [91, 100].

The program $\mathcal{F}(P)^t$ indeed performs only linear algebraic operations, such as vector addition and scalar multiplication. Formally, the transpose of a linear map $X \mapsto Y$ is a linear map $Y^* \mapsto X^*$ where $X^*$ (resp. $Y^*$) is the algebraic dual of $X$ (resp. $Y$). Hence, the final transformation $\mathcal{T}$ transposes $\mathcal{F}(P)^t$, resulting in a program giving the adjoint of $\mathrm{D}[\![P]\!]$, i.e. the gradient of $[\![P]\!]$.

The paper [93] formalises these three transformations using a simply typed calculus called Linear A (Figure 2.1). The typing discipline integrates a form of linearity: tangent variables are subject to specific constructors for copying or erasure, denoted `dup` and `drop`, respectively. Some typing rules are intricate, particularly the one governing primal/tangent compositions, such as `let` $(x, \dot{y}) = e_1$ `in` $e_2$. Additionally, rules manipulating purely primal or purely tangent tuples are constrained to tuples of variables.

$$g(x, y) = (sin(x) * y) + cos(x)$$
$$\nabla g(x, y, \overline{z}) = (\overline{z} * (y * cos(x) - sin(x)), sin(x) * \overline{z})$$

```
def R(P)(x,y,z̄):
    v1 = sin(x)
    v̄1 = z̄*y
    dx̄ = z̄ *(-sin(x))+ v̄1*cos(x)
    v2 = v1*y
    v3 = cos(x)
    v4 = v2+v3
    dȳ = z̄*v1
    return (v4,(dx̄,dȳ))
```

(a) Program obtained by applying forward mode AD to $P$.

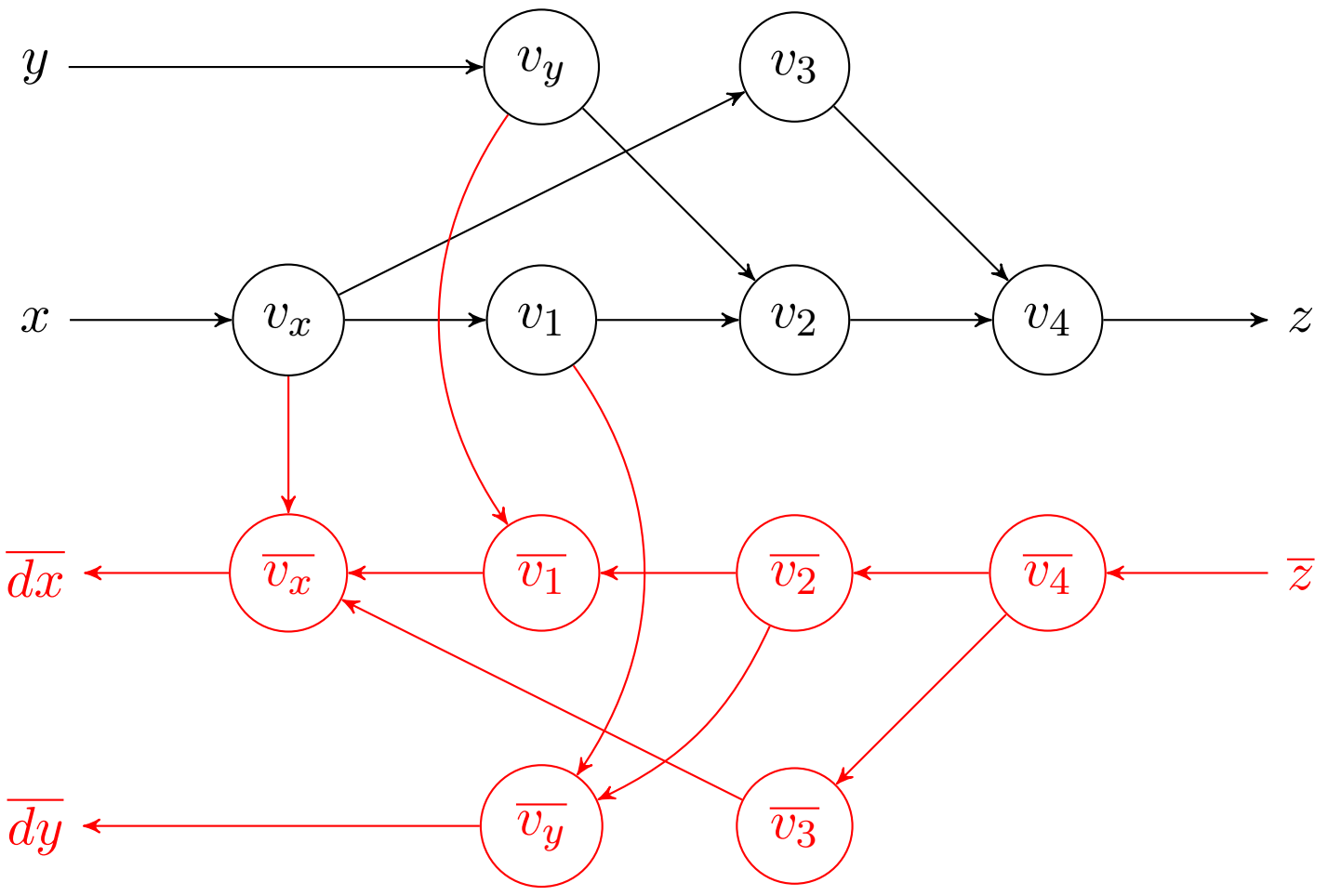

(b) Computational Graph for $\mathcal{R}(P)$.

$$\begin{aligned} v_x &= x \\ v_y &= y \\ v_1 &= sin\ v_x \\ v_2 &= v_1 * v_y \\ v_3 &= cos\ v_x \\ v_4 &= v_2 + v_3 \\ z &= v_4 \end{aligned}$$

(c) Primal Evaluation Trace.

$$\begin{aligned} \overline{dx} &= \overline{v_x} \\ \overline{dy} &= \overline{v_y} \\ \overline{v_x} &= \overline{v_3} * (-sin\ v_x) + \overline{v_1} * cos\ v_x \\ \overline{v_y} &= \overline{v_2} * v_1 \\ \overline{v_1} &= \overline{v_2} * v_y \\ \overline{v_2} &= \overline{v_4} * 1 \\ \overline{v_3} &= \overline{v_4} * 1 \\ \overline{v_4} &= \overline{z} \end{aligned}$$

(d) Adjoint Evaluation Trace.

Figure 1.3: Reverse Mode applied to the program $P$ computing the function $g$.

Linear A is a "domain-specific" calculus designed precisely for formalising JAX AD, and in this regard, it successfully achieves its objective of proving soundness, which essentially corresponds to Equation (1.1). However, it remains unclear whether the grammar expression can be embedded into a more general calculus that has independent interests. Additionally, it is uncertain whether the typing system has a logical interpretation, particularly whether the linearity constraints correspond to a substructural logic that controls the contraction or weakening of hypotheses (which are the logical equivalents to data copying or erasure).

We bridge this gap by encoding Linear A into a linear $\lambda$-calculus, denoted $\lambda$LL, which establishes a Curry-Howard correspondence with a fragment of linear logic (LL) [51]: types correspond with formulas, programs with proofs and the operational semantics is associated with cut-elimination, a crucial procedure in proof-theory proving consistency. Linear logic is a substructural logic featuring two families of conjunctions: the multiplicative conjunction $\otimes$ and the additive conjunction $\&$. It also includes an exponential modality that relates these two families through the exponential isomorphism $!A \otimes !B = !(A\&B)$, and governs the structural rules of weakening and contraction: only hypotheses of type $!A$ can be used multiple times in a proof. The dependency of the conclusion on the hypotheses in a proof (or equivalently, of the output on the inputs in a $\lambda$LL term) is governed by the LL implication $A \multimap B$. In our encoding, primal tuples from Linear A are associated with a multiplicative conjunction of exponentiated types $!A \otimes !B$, while tangent tuples correspond to an additive conjunction $A\&B$ (see Chapter 4). This setup allows the exponential isomorphism to establish a precise link between these two data types (Remark 6).

In this context, `dup` represents the additive diagonal ($A \multimap A\&A$), and `drop` applied to a primal-tangent pair is the weakening rule on the primal part ($!A \multimap \mathtt{1}$, where $\mathtt{1}$ is the neutral element of $\otimes$) and the terminal rule on the tangent part ($A \multimap \top$, where $\top$ is the neutral element of $\&$).

### 1.2.1 Motivations

Embedding a domain-specific calculus like Linear A into a $\lambda$-calculus such as $\lambda$LL, which enjoys a Curry-Howard correspondence with LL, offers several advantages. It allows us to import the cut-elimination rewriting from LL, giving a well-behaving $\beta$-reduction (see Figure 3.4 and Corollary 2 and 5). This step provides an operational semantics for $\lambda$LL, which was lacking for Linear A, and facilitates the proof of various term equivalences. In fact, we introduce a logical relation equivalence (see Section 3.3) which guarantees the soundness of our encoding of Linear A (Theorem 6) and verify different translations. Furthermore, we can restate the cost-preservation of Linear A transformations by referring to the number of floating-point operations (flops in the following) evaluated along a specific rewriting strategy, essentially implementing a call-by-value evaluation strategy (see Section 3.5).

Furthermore, by encoding Linear A into a linear $\lambda$-calculus as $\lambda$LL we have placed it within the same theoretical framework as many other formalizations, such as [26, 108], enabling a formal comparison of them, which we plan as a future work.

Another benefit of $\lambda$LL lies in the modularity of our AD system. This benefits is tied to $\lambda$LL's ability to define the transpose transformation directly within the image set of the $\mathcal{F}$ transformation, without requiring the unzipping transformation (Section 5.4). This is a notable byproduct, as the unzipping process imposes an order on the different phases of the back-propagation algorithm. This order typically involves a forward phase (computing $\mathcal{F}(P)^p$ following the input-output execution flow of $P$) followed by a backward phase (computing $\mathcal{T}(\mathcal{F}(P)^t)$, reversing the execution flow of $P$). In $\lambda$LL, various mixings of these two phases can be represented: applying the $\mathcal{U}$ transformation, introduced in Equation 1.1, results in a complete separation of the forward and backward phases, similar to Linear A. However, if $\mathcal{U}$ is not applied (or applied selectively to some sub-terms), terms can represent intermediate computations that blend aspects of both the forward and backward passes. This improves the modularity of the reverse mode and might be particularly beneficial for programs with independent subroutines (see Section 5.4).

### 1.2.2 Outline of ADLL

Chapter 2 introduces Linear A, recalling the main definitions of Autodiff and adapting the notations from [93]. This serves as a compendium to ensure that this part of the thesis is self-contained, although for a more comprehensive understanding, we refer the reader to the original paper.

Chapter 3 introduces the linear $\lambda$-calculus $\lambda$LL, its typing system (Section 3.1) and $\beta$-reduction (Section 3.2). Moreover, our calculus enjoys important properties such as: subject reduction (Theorem 4), strong normalisation (Corollary 2), and confluence (Theorem 5). Section 3.3 defines the logical relation $\sim$ comparing $\lambda$LL terms with respect to their extensional behaviour at ground types. Moreover, in Section 3.5 we give an adaptation of the work cost notion presented in [93, Section 4.3], where we establish bounds on the number of flops evaluated in a $\beta$-reduction sequence for a specific (yet comprehensive) family of reduction strategies dubbed *safe-reduction*. This strategies are applicable to a set of terms, denoted as *safe*, which includes those representing Linear A.

Our main original contributions begin in Chapter 4, where we define a translation $\delta$ from Linear A to $\lambda$LL (Figure 4.1). We prove the soundness of this translation with respect to the extensional semantics of Linear A (Proposition 6). Chapter 5 focuses on the AD System of $\lambda$LL. Sections 5.1, 5.2, and 5.3 respectively define forward, unzipping, and transpose transformations on $\lambda$LL terms and establish their commutativity with the $\delta$ translation modulo $\sim$-equivalence (see Theorem 7, Theorem 9, and Theorem 12). Section 5.4 demonstrates also how to skip the unzipping transformation and shows the beneficial effects in modularity and parallelism in our AD system. Chapter 6 presents a quantitative type system for $\lambda$LL that embed an extended version of the cost model introduced in Section 3.5 allowing us to refine the notion of safe term. Section 6.2 show that a quantitative version of subject reduction and discuss the modifications in the cost model of $\lambda$LL within the quantitative setting. Section 6.3 provides quantitative soundness for AD System of $\lambda$LL.

Finally, the first part of the thesis concludes by summarising the main results in Figure 6.2 and giving some perspectives.

The illustrative example introduced earlier in Section 1.1 will be progressively extended across the various chapters of this part of the thesis and will serve to highlight the differences between AD system of $\lambda$LL and Autodiff.

### 1.2.3 Related Works

Let us mention [20] as a smooth and modern introduction to AD. The literature is abounding in this last decade even if we restrict to the third period described above. Apart from the already cited [93], let us mention some other approaches.

A pioneering paper is [2] formalising the backward mode in a simple while-language and proving soundness via a denotational model. Soundness of forward AD has been achieved in variants of simply typed $\lambda$-calculus, by e.g. [17], using an open logical relation, and [65], using diffeologies. The paper [26] formalises both forward and backward AD in a linear $\lambda$-calculus and proves a qualitative and a form of quantitative soundness. We leave for future work an in-depth comparison with [26], but let us point out that our work focalises on Autodiff, defining the backward mode out of the forward mode (recall (1.1)). On the contrast, [26] defines the two modes separately. Let us mention [105], discussing an implementation of [26] in Haskell improving it with a number of optimisations.

The difference between primal and tangent products (which we model by using, respectively, the LL connectives $\otimes$ and $\&$) is also at the core of more categorical accounts, like CHAD [108, 81] or the Dialectica system [71]. Comparing these quite different approaches with the one discussed in this part of the thesis is not trivial and we leave it to future work. Still from a categorical side, we highlight a series of papers that provide an axiomatization of reverse derivatives [31, 32, 34, 33], offering an abstract framework for expressing the equation $\mathrm{D}_{\vec{u}}[P](x) = \nabla[P](x) \cdot \vec{u}$, or more general gradient-based optimisations, e.g. [35]. However, our approach differs from this literature in key ways. While these works address derivatives in more general spaces, we focus on a practical library limited to functions over tuples of real numbers. However, unlike this literature, which emphasises semantic soundness, our primary concern is efficiency. Our transformations are designed to maintain the flops workload of the original AD algorithms, enabling gradient computation with a numeric cost comparable to that of partial derivatives. This perspective significantly shifts the focus, as discussed at the beginning of Section 5.3.

All these papers present AD as program transformations (in the same line as we do here), however an alternative approach is also quite popular, consisting in looking at AD as a kind of domain specific interpreter. The two approaches go under the names of *define-then-run* and *define-by-run*. Let us mention [109] as a pedagogical and modern account to this latter, providing a (machine-checked) proof of soundness by means of a separation logic.

# Chapter 2

# JAX

JAX is a Python library designed by Google for machine learning research. Its API for numerical functions is based on NumPy, a collection of functions used in scientific computing. In addition to its NumPy API, JAX includes an extensible system of composable function transformations for machine learning research, we have focused on the transformation related to differentiation (Autodiff) which supports both forward and reverse mode automatic differentiation. This framework represents a good starting point for our research because it is very well documented and maintained both from a theoretical [93] and an implementation point of view. Moreover, its popularity is rising in the deep-learning industry because Python and NumPy are widely used and familiar, making JAX simple, flexible, and easy to adopt.

In this chapter, we revisit the fundamental definitions of Autodiff through the introduction of Linear A, adapting the notation established in [93]. This exposition is intended to serve as a self-contained reference; nonetheless, for a more thorough and detailed description, we refer to the original work.

## 2.1 Syntax and Semantics of Linear A

Linear A is a model language formalising the core of Autodiff – an implementation of AD in projects like JAX and Dex. The main feature is that the syntax marks which variables store primal values and which variables carry tangent values.

In this section we revisit the core of Linear A as presented in [93], with some minor notational variations. In the original paper, Linear A is a first order language, as it supports top-level function definitions. However, we omit this feature here, as it is not relevant to our purposes.

### 2.1.1 Grammars of Types and Expressions

*JAX types* are nested tuples of the ground type of reals:

$$\tau, \sigma ::= \ \mathbb{R} \mid \mathtt{1} \mid \tau \otimes \sigma \qquad \text{(JAX Types)}$$

JAX considers two disjoint copies of this set: $\mathit{Type} \uplus \{\cdot\} \times \mathit{Type}$. The elements from the first copy are called *primal types* and the ones from the second copy are called *tangent types*.

We adopt Church-style typing: the type of each variable is fixed, once and for all, by a function $\mathtt{typeof} : \mathit{Var} \to \mathit{Type} \uplus \{\cdot\} \times \mathit{Type}$ which associates each variable with a primal or tangent type. Variables then inherit the primal/tangent terminology and we denote by $\dot{x}$ a tangent variable, i.e. a variable supposed to have a tangent type $(\cdot; \tau)$. This latter notation allows for omitting the tag $\cdot$ on the tangent types, so simply writing $\dot{x} : \tau$ instead of $\dot{x} : (\cdot; \tau)$.

$$\frac{}{x : \tau; \dot{y} : \sigma \vdash^{\text{Jax}} (x; \dot{y}) : (\tau; \sigma)} \qquad \frac{\Gamma_1; \dot{\Gamma}_1 \vdash^{\text{Jax}} e_1 : (\tau_1; \sigma_1) \quad \Gamma_2, x : \tau_1; \dot{\Gamma}_2, \dot{y} : \sigma_1 \vdash^{\text{Jax}} e_2 : (\tau; \sigma)}{\Gamma_1 \cup \Gamma_2; \dot{\Gamma}_1, \dot{\Gamma}_2 \vdash^{\text{Jax}} \texttt{let } (x; \dot{y}) = e_1 \texttt{ in } e_2 : (\tau; \sigma)}$$

$$\frac{}{; \vdash^{\text{Jax}} \otimes() : (\mathtt{1}; \mathtt{1})} \qquad \frac{\Gamma, x_1 : \tau_1, x_2 : \tau_2; \dot{\Gamma} \vdash^{\text{Jax}} e : (\tau; \sigma)}{\Gamma, z : \tau_1 \otimes \tau_2; \dot{\Gamma} \vdash^{\text{Jax}} \texttt{let } \otimes (x_1, x_2) = z \texttt{ in } e : (\tau; \sigma)}$$

$$\frac{}{x_1 : \tau_1, x_2 : \tau_2; \vdash^{\text{Jax}} \otimes(x_1, x_2) : (\tau_1 \otimes \tau_2; \mathtt{1})} \qquad \frac{\Gamma; \dot{\Gamma} \vdash^{\text{Jax}} e : (\tau; \sigma)}{\Gamma, z : \mathtt{1}; \dot{\Gamma} \vdash^{\text{Jax}} \texttt{let } \otimes () = z \texttt{ in } e : (\tau; \sigma)}$$

$$\frac{}{; \vdash^{\text{Jax}} \dot{\otimes}() : (\mathtt{1}; \mathtt{1})} \qquad \frac{\Gamma; \dot{\Gamma} \vdash^{\text{Jax}} e : (\tau; \sigma)}{\Gamma; \dot{\Gamma}, \dot{z} : \mathtt{1} \vdash^{\text{Jax}} \texttt{let } \dot{\otimes}() = \dot{z} \texttt{ in } e : (\tau; \sigma)}$$

$$\frac{}{; \dot{x}_1 : \tau_1, \dot{x}_2 : \tau_2 \vdash^{\text{Jax}} \dot{\otimes}(\dot{x}_1, \dot{x}_2) : (\mathtt{1}; \tau_1 \otimes \tau_2)}$$

$$\frac{\Gamma; \dot{\Gamma}, \dot{x}_1 : \tau_1, \dot{x}_2 : \tau_2 \vdash^{\text{Jax}} e : (\tau; \sigma)}{\Gamma; \dot{\Gamma}, \dot{z} : \tau_1 \otimes \tau_2 \vdash^{\text{Jax}} \texttt{let } \dot{\otimes}(\dot{x}_1, \dot{x}_2) = \dot{z} \texttt{ in } e : (\tau; \sigma)}$$

$$\frac{}{; \vdash^{\text{Jax}} \dot{0}_\tau : (\mathtt{1}; \tau)} \qquad \frac{}{; \vdash^{\text{Jax}} \underline{r} : (\mathbb{R}; \mathtt{1})} \qquad \frac{}{x_1 : \mathbb{R}, ..., x_n : \mathbb{R}; \vdash^{\text{Jax}} \underline{f}(x_1, ..., x_n) : (\mathbb{R}; \mathtt{1})}$$

$$\frac{}{; \dot{x} : \tau, \dot{y} : \tau \vdash^{\text{Jax}} \dot{x} \dot{+} \dot{y} : (\mathtt{1}; \tau)} \qquad \frac{}{x : \mathbb{R}; \dot{y} : \tau \vdash^{\text{Jax}} x \dot{*} \dot{y} : (\mathtt{1}; \tau)}$$

$$\frac{}{; \dot{x} : \tau \vdash^{\text{Jax}} \text{dup}(\dot{x}) : (\mathtt{1}; \tau \otimes \tau)} \qquad \frac{\Gamma; \dot{\Gamma} \vdash^{\text{Jax}} e : (\tau; \sigma)}{\Gamma; \dot{\Gamma} \vdash^{\text{Jax}} \text{drop}(e) : (\mathtt{1}; \mathtt{1})}$$

Figure 2.1: Linear A Typing Rules

Figure 2.1 shows the typing rule for Linear A expressions. A *judgment* is defined as $\Gamma; \dot{\Gamma} \vdash^{\text{Jax}} e : (\tau; \sigma)$, where $e$ is the typed expression, called the *subject of the judgement*, $\Gamma = \{x_1 : \tau_1, \ldots, x_n : \tau_n\}$ is a set of primal variables, $\dot{\Gamma} = \{\dot{y}_1 : \sigma_1, \ldots, \dot{y}_m : \sigma_m\}$ is a set of tangent variables[1], and the type $(\tau; \sigma)$ of the expression $e$ is a pair, giving respectively the type of the primal and the tangent result of $e$.

We write $\Gamma_1 \cup \Gamma_2$ for the union of two primal contexts. We use commas to denote disjoint unions, so when we write $\dot{\Gamma}_1, \dot{\Gamma}_2$ we suppose that $\dot{\Gamma}_1$ and $\dot{\Gamma}_2$ have no variable in common, otherwise the rule does not hold. Similarly for $\Gamma, x : \tau$.

The grammar of Linear A expressions can be conveniently described by the following grammar.

$$\begin{aligned} e ::= \;& (x; \dot{y}) \mid \texttt{let}\ (x; \dot{y}) = e_1\ \texttt{in}\ e_2 \\ & \mid \otimes() \mid \otimes(x_1, x_2) \mid \texttt{let}\ \otimes() = z\ \texttt{in}\ e \mid \texttt{let}\ \otimes(x_1, x_2) = z\ \texttt{in}\ e \\ & \mid \dot{\otimes}() \mid \dot{\otimes}(\dot{x_1}, \dot{x_2}) \mid \texttt{let}\ \dot{\otimes}() = \dot{z}\ \texttt{in}\ e \mid \texttt{let}\ \dot{\otimes}(\dot{x_1}, \dot{x_2}) = \dot{z}\ \texttt{in}\ e \\ & \mid \underline{r} \mid \underline{f}(x_1, \ldots, x_n) \mid \dot{0}_\tau \mid \dot{x} \dot{+} \dot{y} \mid x \dot{*} \dot{y} \mid \text{dup}(\dot{x}) \mid \text{drop}(e) \end{aligned} \tag{Linear A}$$

Variables are introduced by pairs $(x; \dot{y})$ of a primal and a tangent variable (notice the semicolon separator). In accordance, expressions compose by a primal/tangent `let` which is the most peculiar operator of Linear A. The original paper [93] considers $n$-ary introduction and elimination rules for both primal and tangent tuples. We consider only binary tuples in their primal and tangent versions as it is enough for the Autodiff transformations. The extension $n$-ary primal and tangents tuple is simple, but notational more heavy and not essential for our results.

Finally, we suppose numeric constants $\underline{r}$ and $\underline{f}$ for, respectively, real numbers and $n$-ary numeric functions, e.g. $\underline{f} \in \{\underline{exp}, \underline{*}, \underline{+}, \ldots\}$. We suppose also a bound $b$ to the possible arity $n$ of the numeric functions. In fact, for short, we consider here only unary and binary $\underline{f}$ (e.g. $n = 1$ and $n = 2$), the general case being trivial. We suppose that all functions are differentiable and come together with their partial derivatives $\underline{\partial_i f}$.

Numeric functions act over primal variables. We have in addition the sum $\dot{+}$ over tangent variables and the product $\dot{*}$ between a primal variable and a tangent one. Note that primal variables can be duplicated or erased in the environments but they cannot depend on tangent variables. On the contrast, tangent variables can be modified only by linear operators, but may depend on primal variables through scaling $\dot{*}$. Finally, Linear A has an explicit copying operator dup over tangents and a drop operator erasing both primal and tangent results.

The set of primal (resp. tangent) free variables of an expression $FV(e)$ (resp. $FV^t(e)$) is defined as usual by induction on $e$, with the `let` operators as binders.

### 2.1.2 Semantics and Workload

The *semantics of an expression* $\Gamma; \dot{\Gamma} \vdash^{\text{Jax}} e : (\tau; \sigma)$ is defined as a pair of two functions $[\![e]\!]^{\mathsf{p}}$ and $[\![e]\!]^{\mathsf{t}}$: the former maps real vectors $\underline{\vec{r}}$ associated with $\Gamma$ to a real vector $[\![e]\!]^{\mathsf{p}}_{\underline{\vec{r}}}$ for $\tau$ giving the primal result of $e$; the second map $[\![e]\!]^{\mathsf{t}}$ takes in input both a real vector $\underline{\vec{r}}$ for $\Gamma$ and a real vector $\underline{\vec{s}}$ for $\dot{\Gamma}$ and returns a real vector $[\![e]\!]^{\mathsf{t}}_{\underline{\vec{r}};\underline{\vec{s}}}$ for $\sigma$, giving the tangent result of $e$.

Formally, a vector for $\mathbb{R}$ is a real number $\underline{r}$, a vector for the empty sequence `1` is the zero vector of dimension zero, a vector for $\tau \otimes \sigma$ is a pair $(\underline{\vec{r}}, \underline{\vec{s}})$ of a vector for $\tau$ and a vector of $\sigma$. A vector for a typing environment $\Gamma$ (or $\dot{\Gamma}$) is a map $\underline{\vec{r}}$ associating each $x : \tau \in \Gamma$ with a vector $\underline{\vec{r}}(x)$ for $\tau$. Given a subset $\mathcal{X} \subseteq \Gamma$, we write by $\underline{\vec{r}}|\mathcal{X}$ the restriction of $\underline{\vec{r}}$ to the variables in $\mathcal{X}$. The semantics $[\![e]\!]^{\mathsf{p}}$ and $[\![e]\!]^{\mathsf{t}}$ are then defined inductively on $e$ in the obvious way. For example, the definition of $[\![\texttt{let } (x;\dot{y}) = e_1 \texttt{ in } e_2]\!]^{\mathsf{t}}_{\vec{\underline{r}};\vec{\underline{s}}}$ first computes both $[\![e_1]\!]^{\mathsf{p}}$ and $[\![e_1]\!]^{\mathsf{t}}$ by taking into account the values of the primals and tangents free in $e_1$ and then computes $[\![e_2]\!]^{\mathsf{t}}$ by affecting the values $[\![e_1]\!]^{\mathsf{p}}$ and $[\![e_1]\!]^{\mathsf{t}}$ to the variables $x$ and $\dot{y}$ bounded by the `let`. More formally:

$$[\![\texttt{let } (x;\dot{y}) = e_1 \texttt{ in } e_2]\!]^{\mathsf{t}}_{\vec{\underline{r}};\vec{\underline{s}}} = [\![e_2]\!]^{\mathsf{t}}_{\vec{\underline{r}}|FV(e_2),x\mapsto[\![e_1]\!]^{\mathsf{p}}_{\vec{\underline{r}}|FV(e_1)};\vec{\underline{s}}|FV^t(e_2),\dot{y}\mapsto[\![e_1]\!]^{\mathsf{t}}_{\vec{\underline{s}}|FV^t(e_1)}}$$

where $[\![\texttt{let } (x;\dot{y}) = e_1 \texttt{ in } e_2]\!]^{\mathsf{p}}_{\vec{\underline{r}}} = [\![e_2]\!]^{\mathsf{p}}_{\vec{\underline{r}}|FV(e_2),x\mapsto[\![e_1]\!]^{\mathsf{p}}_{\vec{\underline{r}}|FV(e_1)}}$.

[1] In fact, Church-style typing does not require explicit contexts in a typing judgement, as this can be inferred by the set of free variables and the typing function `typeof`. We however decided to keep it explicit, in order to underline in the typing rules the different roles of the two kinds variables: primal and tangents.

A notion of *workload* $\mathcal{W}^{\texttt{Jax}}(e)$ is introduced in [93, Section 4.3], which basically estimates the number of flops performed in the computation of $[\![e]\!]^{\mathsf{t}}$. In particular, every non-linear primitive costs 1, linear addition $\dot{+}$ and linear multiplication $\dot{*}$ cost 1 per scalar $\mathbb{R}$ type presents in the result, and drop$(e)$ costs $\mathcal{W}^{\texttt{Jax}}(e)$ plus 1 for every scalar type $\mathbb{R}$ in the output of $e$. Let us also define the workload $\mathcal{W}^{\texttt{Jax}}(\tau)$ of a type $\tau$ as the number of occurrences of the type $\mathbb{R}$ in it. Given a finite set of variables $\mathcal{V}$, we will write $\mathcal{W}^{\texttt{Jax}}(\mathcal{V})$ for the sum $\sum_{x:\tau\in\mathcal{V}}\mathcal{W}^{\texttt{Jax}}(\tau)$.

### 2.1.3 Notational Conventions and Linear B

The syntax of Linear A is very restrictive and some syntactic sugar is necessary for manipulating the "purely primal" or "purely tangent" parts of an expression. We use the following syntactic sugars for primal/tangent expressions and purely primal/tangent `let` constructs

$$\begin{aligned}
x &\approx \texttt{let } \dot{y} = \dot{\otimes}() \texttt{ in } (x;\dot{y})\\
\dot{y} &\approx \texttt{let } x = \otimes() \texttt{ in } (x;\dot{y})\\
\texttt{let } x = e_1 \texttt{ in } e_2 &\approx \texttt{let } (x;\dot{y}) = e_1 \texttt{ in let } \dot{\otimes}() = \dot{y} \texttt{ in } e_2\\
\texttt{let } \dot{y} = e_1 \texttt{ in } e_2 &\approx \texttt{let } (x;\dot{y}) = e_1 \texttt{ in let } \otimes() = x \texttt{ in } e_2
\end{aligned}$$

Moreover, we consider pairs of expressions of different kind and tensors of expressions of same kind:

$$\begin{aligned}
(e_1;e_2) &\approx \text{let } x = e_1 \text{ in let } \dot{y} = e_2 \text{ in } (x;\dot{y})\\
\dot{\otimes}(e_1,e_2) &\approx \text{let } \dot{x} = e_1 \text{ in let } \dot{y} = e_2 \text{ in } \dot{\otimes}(\dot{x},\dot{y})\\
\otimes(e_1,e_2) &\approx \text{let } x = e_1 \text{ in let } y = e_2 \text{ in } \otimes(x,y)
\end{aligned}$$

The typing rules for these notations can be found in Figure 2.2.

We denote $n$-fold tangent tuples $\dot{\otimes}(e_1,\dot{\otimes}(e_2,\ldots,e_n),\ldots)$ as an $n$-ary tangent tuple $\dot{\otimes}(e_1,\ldots,e_n)$. We can use shortcut like $\dot{\otimes}(e_i)_{i=1}^n$, or even $\dot{\otimes}(e_i)_i$ if 1 and $n$ are clear from the context or irrelevant. We adopt similar writings for types: $\dot{\otimes}(\tau_i)_{i=1}^n$ or $\dot{\otimes}(\tau_i)_i$.

Given $\theta = (\dot{x}_1,\ldots,\dot{x}_n)$ where $\dot{x}_i : \tau_i$, we define the syntactic sugar $\dot{\otimes}\theta$ by induction on $\theta$ as follows

$$\dot{\otimes}\theta \approx \begin{cases}\dot{\otimes}() & \text{if } \theta = (\,)\\ \dot{x} & \text{if } \theta = (\dot{x})\\ \dot{\otimes}(\dot{x},\dot{\otimes}\theta') & \text{otherwise we can suppose } \theta = \dot{x},\theta'\end{cases}$$

Similarly, we can define the type $\otimes\theta$. Moreover, we define the syntactic sugar $\texttt{let } \dot{\otimes}\theta = \dot{z} \texttt{ in } e$ by induction on $\theta$ as follows

$$\texttt{let } \dot{\otimes}\theta = \dot{z} \texttt{ in } e \approx \begin{cases}\texttt{let } \dot{\otimes}() = \dot{z} \texttt{ in } e & \text{if } \theta = (\,)\\ \texttt{let } \dot{x} = \dot{z} \texttt{ in } e & \text{if } \theta = (\dot{x})\\ \texttt{let } \dot{\otimes}(\dot{x},\dot{y}) = \dot{z} \texttt{ in let } \dot{\otimes}\theta' = \dot{y} \texttt{ in } e & \text{otherwise we can suppose } \theta = \dot{x},\theta'\end{cases}$$

$$\frac{}{x:\tau;\vdash^{\text{Jax}} x:(\tau;\mathtt{1})} \qquad \frac{\Gamma_1;\dot\Gamma_1 \vdash^{\text{Jax}} e_1:(\tau_1;\mathtt{1}) \qquad \Gamma_2, x:\tau_1;\dot\Gamma_2 \vdash^{\text{Jax}} e_2:(\tau;\sigma)}{\Gamma_1\cup\Gamma_2;\dot\Gamma_1,\dot\Gamma_2 \vdash^{\text{Jax}} \mathtt{let}\ x = e_1\ \mathtt{in}\ e_2:(\tau;\sigma)}$$

$$\frac{}{;\dot x:\tau \vdash^{\text{Jax}} \dot x:(\mathtt{1};\tau)} \qquad \frac{\Gamma_1;\dot\Gamma_1 \vdash^{\text{Jax}} e_1:(\mathtt{1};\sigma_1) \qquad \Gamma_2;\dot\Gamma_2,\dot y:\sigma_1 \vdash^{\text{Jax}} e_2:(\tau;\sigma)}{\Gamma_1\cup\Gamma_2;\dot\Gamma_1,\dot\Gamma_2 \vdash^{\text{Jax}} \mathtt{let}\ \dot y = e_1\ \mathtt{in}\ e_2:(\tau;\sigma)}$$

$$\frac{\Gamma_1;\dot\Gamma_1 \vdash^{\text{Jax}} e_1:(\sigma_1;\mathtt{1}) \qquad \Gamma_2;\dot\Gamma_2 \vdash^{\text{Jax}} e_2:(\mathtt{1};\sigma_2)}{\Gamma_1\cup\Gamma_2;\dot\Gamma_1,\dot\Gamma_2 \vdash^{\text{Jax}} \otimes(e_1;e_2):(\sigma_1\otimes\sigma_2;\mathtt{1})}$$

$$\frac{\forall i\in\{1,2\},\quad \Gamma_i;\dot\Gamma_i \vdash^{\text{Jax}} e_i:(\sigma_i;\mathtt{1})}{\Gamma_1\cup\Gamma_2;\dot\Gamma_1,\dot\Gamma_2 \vdash^{\text{Jax}} \otimes(e_1,e_2):(\sigma_1\otimes\sigma_2;\mathtt{1})} \qquad \frac{\forall i\in\{1,2\},\quad \Gamma_i;\dot\Gamma_i \vdash^{\text{Jax}} e_i:(\mathtt{1};\tau_i)}{\Gamma_1\cup\Gamma_2;\dot\Gamma_1,\dot\Gamma_2 \vdash^{\text{Jax}} \dot\otimes(e_1,e_2):(\mathtt{1};\tau_1\otimes\tau_2)}$$

Figure 2.2: Derived JAX Typing Rules for Syntactic Sugar.

Moreover, given $\theta = (\dot x_1,\dots,\dot x_n)$ and two partitions $\theta_1$ and $\theta_2$ of $\theta$, let $\dot y_1 : \otimes\theta_1$ and $\dot y_2 : \otimes\theta_2$, we define the *fusion expression* as

$$\overline{\sigma}^{\text{Jax}}_{\dot y_1,\dot y_2;\theta} = \begin{array}{l}\mathtt{let}\ \dot\otimes\theta_1 = \dot y_1\ \mathtt{in}\\ \mathtt{let}\ \dot\otimes\theta_2 = \dot y_2\ \mathtt{in}\ \dot\otimes\theta\end{array} \tag{2.1}$$

Observe that $\overline{\sigma}^{\text{Jax}}_{\dot y_1,\dot y_2;\theta}$ is well-typed as $;\dot y_1 : \otimes\theta_1, \dot y_2 : \otimes\theta_2 \vdash^{\text{Jax}} \overline{\sigma}^{\text{Jax}}_{\dot y_1,\dot y_2;}\theta:(\mathtt{1};\otimes\theta)$.

One crucial step of Autodiff is to split the primal part from the tangent part of an expression before performing the transpose transformation. The following fragment of Linear A, called *Linear B* in [93], uses the conventions introduced above in order to define a 3-sorted grammar, giving purely primal ($e^p$) and purely tangent ($\dot e$) expressions and pairs ($d$) of each of them possibly prefixed by a stack of primal let-definitions:

$$\begin{aligned} d ::= {} & (e^p;\dot e) \mid \mathtt{let}\ x = e^p\ \mathtt{in}\ d \\ & \mid \mathtt{let}\ \otimes() = z\ \mathtt{in}\ d \mid \mathtt{let}\ \otimes(x_1,x_2) = z\ \mathtt{in}\ d \end{aligned} \tag{Linear B}$$

$$\begin{aligned} e^p ::= {} & x \mid \mathtt{let}\ x = e^p_1\ \mathtt{in}\ e^p_2 \mid \underline{r} \mid \underline{f}(x_1,x_2) \mid \mathrm{drop}(e^p) \\ & \mid \otimes() \mid \otimes(e^p_1,e^p_2) \mid \mathtt{let}\ \otimes() = z\ \mathtt{in}\ e^p \mid \mathtt{let}\ \otimes(x_1,x_2) = z\ \mathtt{in}\ e^p \end{aligned} \tag{Primal}$$

$$\begin{aligned} \dot e ::= {} & \dot x \mid \mathtt{let}\ \dot x = \dot e_1\ \mathtt{in}\ \dot e_2 \mid \mathrm{dup}(\dot x) \mid \dot 0_\tau \mid \dot x \dot + \dot y \mid x \dot * \dot y \mid \mathrm{drop}(\dot e) \\ & \mid \dot\otimes() \mid \dot\otimes(\dot e_1,\dot e_2) \mid \mathtt{let}\ \dot\otimes() = \dot z\ \mathtt{in}\ \dot e \mid \mathtt{let}\ \otimes(\dot x_1,\dot x_2) = \dot z\ \mathtt{in}\ \dot e \end{aligned} \tag{Tangent}$$

Notice that a primal $e^p$ (resp. tangent $\dot e$) in Linear B is typed as $\Gamma;\vdash^{\text{Jax}} e^p:(\tau;\mathtt{1})$ (resp. $\Gamma;\dot\Gamma \vdash^{\text{Jax}} \dot e:(\mathtt{1};\tau)$).

**Illustrative Example 4** (Linear B Expression)**.** Consider again the function $g$ computing $z = (sin(x) * y) + cos(x)$. The purely primal expression in Linear B (see grammar Primal) is the following

$$\begin{aligned} e^p = {} & \mathtt{let}\ v_1 = \underline{sin}\ x\ \mathtt{in} \\ & \mathtt{let}\ v_2 = v_1\ \underline{*}\ y\ \mathtt{in} \\ & \mathtt{let}\ v_3 = \underline{cos}\ x\ \mathtt{in} \\ & \mathtt{let}\ v_4 = v_2\ \underline{+}\ v_3\ \mathtt{in} \\ & v_4 \end{aligned} \tag{2.2}$$

which is well-typed as $x : \mathbb{R}, y : \mathbb{R}; \vdash^{\texttt{Jax}} e^p : (\mathbb{R}; \mathtt{1})$. It is important to note that $\underline{*}$ and $\underline{+}$ represent the primal operations, not the tangent operations, which are indicated using dot notation.

## 2.2 Autodiff Transformations

The peculiarity of Autodiff, the AD system of JAX, is to separate the proper differentiation from direction reversal. This separation is achieved by decomposing reverse mode AD, as introduced in Chapter 1, into three distinct transformations: forward, unzipping, and transpose. In this section, we delve into the details of these three components, examining how they contribute individually and collectively to the overall reverse mode process in JAX.

### 2.2.1 Forward

The transformation $\mathcal{F}^{\texttt{Jax}}$ takes a purely primal expression $e^p$ in Linear B and a mapping $\phi = \{x_i \to \dot{y}_i\}_{i=1}^n$ which associates each primal $x_i : \tau$ free in $e^p$ with a corresponding tangent variable $\dot{y}_i : \tau$ and returns an expression in Linear A, which is a pair of a primal (computing the same value as $e^p$) and a tangent.

The definition is by induction on the grammar of $e^p$ above (see grammar Primal) and can be found in Figure 2.3. The main base case is the transformation of a numeric functional constant $\underline{f}$ (which we detail for $\underline{f}$ binary), implementing the chain rule $(f \circ g)' = (f' \circ g) \cdot g'$, where the primals $x_1, x_2$ are the outputs of $g$ and the tangents $\dot{y}_1, \dot{y}_2$ give the derivative $g'$ (under the form of a vector of partial derivatives). Then $\mathcal{F}^{\texttt{Jax}}(\underline{f}(x_1, x_2))$ returns a Linear A expression having in the primal position the image of the inputs along $f$, and in the tangent position the product of the derivative of $f$ at $(x_1, x_2)$ (under the form of a bunch of two primal variables $w_1, w_2$) with the tangent variables $\dot{y}_1, \dot{y}_2$.

The core case is the definition of $\mathcal{F}^{\texttt{Jax}}(\texttt{let}\ x = e_1^p\ \texttt{in}\ e_2^p)$, composing $\mathcal{F}^{\texttt{Jax}}(e_1^p)$ with $\mathcal{F}^{\texttt{Jax}}(e_2^p)$ by using the primal/tangent `let` $(x; \dot{y}) = \ldots$ `in` ... of Linear A. The typing requires disjoint sets of tangent variables in the environments of the `let` and the `in` expressions. Then the composition is post-processed by a bunch of `dup` terms in order unify these environments. Anticipating Chapter 4, our encoding will show this construction is a simple instance of the contraction of the environment of the linear logic additive conjunction &.

Finally, the following theorem shows that the forward transformation $\mathcal{F}^{\texttt{Jax}}$ places the derivative computation within the tangent part of an expression in Linear A, preserves the workload up to a constant factor and it is sound.

**Theorem 1** (Type, Work-Preservation and Soundness $\mathcal{F}^{\texttt{Jax}}$: Adaptation of Theorem 5.1 in [93])**.** Given an expression in Primal $\Gamma; \vdash^{\texttt{Jax}} e : (\tau; \mathtt{1})$ and a renaming $\phi$ of $\Gamma$ into tangent variables, then

1. $\Gamma; \phi(\Gamma) \vdash^{\texttt{Jax}} \mathcal{F}_\phi^{\texttt{Jax}}(e^p) : (\tau; \tau)$ is a well-typed expression of Linear A.

2. there exists a constant $c$ such that $\mathcal{W}^{\texttt{Jax}}(\mathcal{F}_\phi^{\texttt{Jax}}(e^p)) \leq c \cdot \mathcal{W}^{\texttt{Jax}}(e)$

3. for every numeral sequences $\underline{\vec{r}}$ for $\Gamma$ and $\underline{\vec{s}}$ for $\phi(\Gamma)$, we have that: $[\![\mathcal{F}_\phi^{\texttt{Jax}}(e^p)]\!]_{\underline{\vec{r}}}^{\mathtt{p}} = [\![e^p]\!]_{\underline{\vec{r}}}^{\mathtt{p}}$ and the tangent results $[\![\mathcal{F}_\phi^{\texttt{Jax}}(e^p)]\!]_{\underline{\vec{r}},\underline{\vec{s}}}^{\mathtt{t}}$ are equal to the matrix-vector product of the Jacobian of $e^p$ at $\underline{\vec{r}}$ with the vector $\underline{\vec{s}}$, provided this is true for the primitive derivatives.

*Sketch Proof.* By structural induction on $e$. The statement of this theorem is an adaptation to our notation of the Theorem 5.1 in [93], we refer to it for the detailed proof. □

$$\mathcal{F}^{\texttt{Jax}}_{x\to\dot{y}}(x) \overset{\text{def}}{=} (x;\dot{y})$$

$$\begin{aligned}
\mathcal{F}^{\texttt{Jax}}_{\phi_1,\phi_2,\{z_i\to\dot{u}_i\}_{i=1}^k}(\texttt{let } x = e_1^p \texttt{ in } e_2^p) \overset{\text{def}}{=} & \texttt{let } \dot{a}_1 = \text{dup}(\dot{u}_1) \texttt{ in} \\
& \texttt{let } \dot{\otimes}(\dot{w}_1,\dot{v}_1) = \dot{a}_1 \texttt{ in} \\
& \dots \\
& \texttt{let } \dot{a}_k = \text{dup}(\dot{u}_k) \texttt{ in} \\
& \texttt{let } \dot{\otimes}(\dot{w}_k,\dot{v}_k) = \dot{a}_k \texttt{ in} \\
& \texttt{let } (x;\dot{y}) = \mathcal{F}^{\texttt{Jax}}_{\phi_1,\{z_i\to\dot{w}_i\}_{i=1}^k}(e_1) \texttt{ in} \\
& \mathcal{F}^{\texttt{Jax}}_{\phi_2,\{z_i\to\dot{v}_i\}_{i=1}^k}(e_2)
\end{aligned}$$

where $\text{Dom}(\phi_i) = FV(e_i^p) \setminus (FV(e_1^p) \cap FV(e_2^p))$ and $\{z_i\}_{i=1}^k = FV(e_1^p) \cap FV(e_2^p)$

$$\mathcal{F}^{\texttt{Jax}}_{\emptyset}(\otimes()) \overset{\text{def}}{=} (\otimes();\dot{\otimes}())$$

$$\mathcal{F}^{\texttt{Jax}}_{\{x_i\to\dot{y}_i\}_{i=1}^2}(\otimes(x_1,x_2)) \overset{\text{def}}{=} (\otimes(x_1,x_2);\dot{\otimes}(\dot{y}_1,\dot{y}_2))$$

$$\mathcal{F}^{\texttt{Jax}}_{\phi,z\to\dot{w}}(\texttt{let } \otimes() = z \texttt{ in } e^p) \overset{\text{def}}{=} \mathcal{F}^{\texttt{Jax}}_{\phi}(e^p)$$

$$\begin{aligned}
\mathcal{F}^{\texttt{Jax}}_{\phi,z\to\dot{w}}(\texttt{let } \otimes(x_1,x_2) = z \texttt{ in } e^p) \overset{\text{def}}{=} & \texttt{let } \otimes(x_1,x_2) = z \texttt{ in} \\
& \texttt{let } \dot{\otimes}(\dot{y}_1,\dot{y}_2) = \dot{w} \texttt{ in} \\
& \mathcal{F}^{\texttt{Jax}}_{\phi,\{x_i\to\dot{y}_i\}_{i=1}^2}(e^p)
\end{aligned}$$

$$\mathcal{F}^{\texttt{Jax}}_{\emptyset}(\underline{r}) \overset{\text{def}}{=} (\underline{r};\dot{0}_{\mathbb{R}})$$

$$\begin{aligned}
\mathcal{F}^{\texttt{Jax}}_{\phi,\{x_i\to\dot{y}_i\}_{i=1}^n}(\underline{f}(x_1,\dots,x_n)) \overset{\text{def}}{=} & \texttt{let } w_1 = \underline{\partial_1 f}(x_1,\dots,x_n) \texttt{ in} \\
& \dots \\
& \texttt{let } w_n = \underline{\partial_n f}(x_1,\dots,x_n) \texttt{ in} \\
& \texttt{let } \dot{z}_1 = w_1 \dot{*}\, \dot{y}_1 \texttt{ in} \\
& \dots \\
& \texttt{let } \dot{z}_n = w_n \dot{*}\, \dot{y}_n \texttt{ in} \\
& (\underline{f}(x_1,\dots,x_n);\dot{z}_1 \dot{+} \dots \dot{+} \dot{z}_n)
\end{aligned}$$

$$\mathcal{F}^{\texttt{Jax}}_{\phi}(\text{drop}(e^p)) \overset{\text{def}}{=} \text{drop}(\mathcal{F}^{\texttt{Jax}}_{\phi}(e^p))$$

Figure 2.3: Forward Transformation in JAX

**Remark 1** (Constant Overhead in Work Preservation Forward)**.** It is worth observing that the forward mode transformation preserves the workload up to a constant overhead $c$. Indeed, for a function symbol of arity $n$, the forward transformation involves computing partial derivatives (with cost $n$), evaluating the function itself (cost 1), performing $n-1$ binary additions, and evaluating $n$ scalar multiplications, resulting in a total cost bounded by $3n$. Since the arity of primitive numeric functions in the language is bounded by a fixed constant $b$, we can choose a constant $c > 3b$ such that the cost of the forward transformation remains within a constant factor of the original computation. This ensures that forward mode differentiation does not introduce asymptotic overhead.

Let us illustrate how the forward transformation operates through our toy example.

**Illustrative Example 5** (Forward Transformation JAX)**.** Consider the expression $e^p$ in Equation 2.2 and the renaming $\phi = \{x \to \dot{dx}, y \to \dot{dy}\}$, we proceed by applying the forward transformation defined in Figure 2.3 obtaining exactly the following Linear A expression in Figure 2.4 which is well-typed as $x : \mathbb{R}, y : \mathbb{R}; \dot{dx} : \mathbb{R}, \dot{dy} : \mathbb{R} \vdash^{\mathtt{Jax}} \mathcal{F}^{\mathtt{Jax}}_{\{x \to \dot{dx}, y \to \dot{dy}\}}(e^p) : (\mathbb{R}; \mathbb{R})$.

Let us apply some simplifications for readability and get the following expression

$$
\begin{aligned}
\mathcal{F}^{\mathtt{Jax}}_{\{x \to \dot{dx}, y \to \dot{dy}\}}(e^p) \approx\ & \mathtt{let}\ \dot{a} = \mathrm{dup}(\dot{dx})\ \mathtt{in} \\
& \mathtt{let}\ \dot{\otimes}(\dot{dx}_1, \dot{dx}_2) = \dot{a}\ \mathtt{in} \\
& \mathtt{let}\ (v_1; \dot{v_1}) = (\underline{sin}\ x; \underline{cos}\ x \mathbin{\dot{*}} \dot{dx}_1)\ \mathtt{in} \\
& \mathtt{let}\ (v_2; \dot{v_2}) = (v_1 \mathbin{\underline{*}} y; (y \mathbin{\dot{*}} \dot{v_1}) \mathbin{\dot{+}} (v_1 \mathbin{\dot{*}} \dot{dy}))\ \mathtt{in} \\
& \mathtt{let}\ (v_3; \dot{v_3}) = (\underline{cos}\ x; (-\underline{sin}\ x) \mathbin{\dot{*}} \dot{dx}_2)\ \mathtt{in} \\
& \mathtt{let}\ (v_4; \dot{v_4}) = (v_2 \mathbin{\underline{+}} v_3; \dot{v_2} \mathbin{\dot{+}} \dot{v_3})\ \mathtt{in} \\
& (v_4; \dot{v_4})
\end{aligned}
\tag{2.3}
$$

where the part in blue is clearly computing the directional derivative of $g$ which is $\mathrm{D}(g)(x, \dot{dx}, y, \dot{dy}) = (cos(x) * y) * \dot{dx} + sin(x) * \dot{dy} - sin(x) * \dot{dx}$, in accordance with soundness property of forward transformation (Claim 3 of Theorem 1).

We can observe the double occurrence of the tangent variable $\dot{dx}$ in the body of the program in Figure 1.2a is managed by the a duplication and the unpacking of a tangent tuple in Equation 2.3 (first two lines).

### 2.2.2 Unzipping

The transformation $\mathcal{U}^{\mathtt{Jax}}$ disentangles primal and tangent values mapping Linear A into Linear B. The definition is by induction on the grammar of Linear A and consists in splitting each primal/tangent `let` into a purely primal `let` and a purely tangent `let`, moving the former towards the root of the syntactic tree and the latter towards the tangent leaves. The final result $\mathcal{U}^{\mathtt{Jax}}(e)$ will be a stack of purely primal `let`'s in the form $\mathtt{let}\ x_1 = e_1^p\ \mathtt{in}\ \ldots \mathtt{let}\ x_n = e_n^p\ \mathtt{in}\ \ldots$ followed by a pair of a primal value and a purely tangent expression. Figure 2.5 gives the definition of $\mathcal{U}^{\mathtt{Jax}}$, denoting by a metavariable $E$ such a stack of purely primal `let`'s.

An easy induction gives that for any Linear A expression $e$, $\mathcal{U}^{\mathtt{Jax}}(e)$ is a well-typed expression of Linear B of the same type and workload as $e$ and the same semantics.

$$
\begin{aligned}
\mathcal{F}^{\texttt{Jax}}_{\{x\to \dot{dx}, y \to \dot{dy}\}}(e^p) = {} & \texttt{let } \dot{a} = \mathrm{dup}(\dot{dx}) \texttt{ in} \\
& \texttt{let } \dot{\otimes}(\dot{dx}_1, \dot{dx}_2) = \dot{a} \texttt{ in} \\
& \texttt{let } (v_1; \dot{v_1}) = \begin{pmatrix} \texttt{let } w_1 = \underline{cos}\ x \texttt{ in} \\ \texttt{let } \dot{z_1} = w_1 \mathbin{\dot{*}} \dot{dx}_1 \texttt{ in} \\ (\underline{sin}\ x; \dot{z_1}) \end{pmatrix} \texttt{ in} \\
& \texttt{let } (v_2; \dot{v_2}) = \begin{pmatrix} \texttt{let } w_2 = y \texttt{ in let } w_3 = v_1 \texttt{ in} \\ \texttt{let } \dot{z_2} = w_2 \mathbin{\dot{*}} \dot{v_1} \texttt{ in let } \dot{z_2} = w_3 \mathbin{\dot{*}} \dot{dy} \texttt{ in} \\ (v_1 \mathbin{\underline{*}} y; \dot{z_2} \mathbin{\dot{+}} \dot{z_3}) \end{pmatrix} \texttt{ in} \\
& \texttt{let } (v_3; \dot{v_3}) = \begin{pmatrix} \texttt{let } w_4 = -\underline{sin}\ x \texttt{ in} \\ \texttt{let } \dot{z_4} = w_4 \mathbin{\dot{*}} \dot{dx}_2 \texttt{ in} \\ (\underline{cos}\ x; \dot{z_4}) \end{pmatrix} \texttt{ in} \\
& \texttt{let } (v_4; \dot{v_4}) = \begin{pmatrix} \texttt{let } w_5 = 1 \texttt{ in let } w_6 = 1 \texttt{ in} \\ \texttt{let } \dot{z_5} = w_5 \mathbin{\dot{*}} \dot{v_2} \texttt{ in let } \dot{z_6} = w_6 \mathbin{\dot{*}} \dot{v_3} \texttt{ in} \\ (v_2 \mathbin{\underline{+}} v_3; \dot{z_5} \mathbin{\dot{+}} \dot{z_6}) \end{pmatrix} \texttt{ in} \\
& (v_4; \dot{v_4})
\end{aligned}
$$

Figure 2.4: Forward Transformation of JAX applied to the expression $e^p$ defined in Equation 2.2.

**Theorem 2** (Type, Work-Preservation and Soundness $\mathcal{U}^{\texttt{Jax}}$: Adaptation of Theorem 6.1 in [93])**.** Given an expression in Linear A $\Gamma; \dot{\Gamma} \vdash^{\texttt{Jax}} e : (\tau, \sigma)$, then

1. $\Gamma; \dot{\Gamma} \vdash^{\texttt{Jax}} \mathcal{U}^{\texttt{Jax}}(e) : (\tau, \sigma)$ is a well-typed expression of Linear B.
2. $\mathcal{W}^{\texttt{Jax}}(\mathcal{U}^{\texttt{Jax}}(e)) = \mathcal{W}^{\texttt{Jax}}(e)$
3. for every numeral sequences $\underline{\vec{r}}$ for $\Gamma$ and $\underline{\vec{s}}$ for $\phi(\Gamma)$, we have that: $[\![e]\!]^{\mathsf{p}}_{\underline{\vec{r}}} = [\![\mathcal{U}^{\texttt{Jax}}(e)]\!]^{\mathsf{p}}_{\underline{\vec{r}}}$ and $[\![e]\!]^{\mathsf{t}}_{\underline{\vec{r}},\underline{\vec{s}}} = [\![\mathcal{U}^{\texttt{Jax}}(e)]\!]^{\mathsf{t}}_{\underline{\vec{r}},\underline{\vec{s}}}$.

*Sketch Proof.* By structural induction on $e$. The statement of this theorem is an adaptation to our notation of the Theorem 6.1 in [93], we refer to it for the detailed proof. □

$$\frac{}{\mathcal{U}^{\mathtt{Jax}}((x;\dot{y})) \stackrel{\text{def}}{=} (x;\dot{y})}$$

$$\frac{\mathcal{U}^{\mathtt{Jax}}(e_1) \stackrel{\text{def}}{=} E_1 \ \mathtt{in}\ (e_1^p;\dot{e_1}) \qquad \mathcal{U}^{\mathtt{Jax}}(e_2) \stackrel{\text{def}}{=} E_2 \ \mathtt{in}\ (e_2^p;\dot{e_2})}{\mathcal{U}^{\mathtt{Jax}}(\mathtt{let}\ (x;\dot{y}) = e_1 \ \mathtt{in}\ e_2) \stackrel{\text{def}}{=} E_1 \ \mathtt{in}\ \mathtt{let}\ x = e_1^p \ \mathtt{in}\ E_2 \ \mathtt{in}\ (e_2^p; \mathtt{let}\ \dot{y} = \dot{e_1} \ \mathtt{in}\ \dot{e_2})}$$

$$\frac{}{\mathcal{U}^{\mathtt{Jax}}(\otimes()) \stackrel{\text{def}}{=} (\otimes();\dot{\otimes}())} \qquad \frac{}{\mathcal{U}^{\mathtt{Jax}}(\otimes(x_1,x_2)) \stackrel{\text{def}}{=} (\otimes(x_1,x_2);\dot{\otimes}())}$$

$$\frac{\mathcal{U}^{\mathtt{Jax}}(e_1) \stackrel{\text{def}}{=} E \ \mathtt{in}\ (e_1^p;\dot{e_1})}{\mathcal{U}^{\mathtt{Jax}}(\mathtt{let}\ \otimes() = z \ \mathtt{in}\ e_1) \stackrel{\text{def}}{=} \mathtt{let}\ \otimes() = z \ \mathtt{in}\ E \ \mathtt{in}\ (e_1^p;\dot{e_1})}$$

$$\frac{\mathcal{U}^{\mathtt{Jax}}(e_1) \stackrel{\text{def}}{=} E \ \mathtt{in}\ (e_1^p;\dot{e_1})}{\mathcal{U}^{\mathtt{Jax}}(\mathtt{let}\ \otimes(x_1,x_2) = z \ \mathtt{in}\ e_1) \stackrel{\text{def}}{=} \mathtt{let}\ \otimes(x_1,x_2) = z \ \mathtt{in}\ E \ \mathtt{in}\ (e_1^p;\dot{e_1})}$$

$$\frac{}{\mathcal{U}^{\mathtt{Jax}}(\dot{\otimes}()) \stackrel{\text{def}}{=} (\otimes();\dot{\otimes}())} \qquad \frac{}{\mathcal{U}^{\mathtt{Jax}}(\dot{\otimes}(\dot{x_1},\dot{x_2})) \stackrel{\text{def}}{=} (\otimes();\dot{\otimes}(\dot{x_1},\dot{x_2}))}$$

$$\frac{\mathcal{U}^{\mathtt{Jax}}(e_1) \stackrel{\text{def}}{=} E \ \mathtt{in}\ (e_1^p;\dot{e_1})}{\mathcal{U}^{\mathtt{Jax}}(\mathtt{let}\ \dot{\otimes}() = \dot{z} \ \mathtt{in}\ e_1) \stackrel{\text{def}}{=} E \ \mathtt{in}\ (e_1^p; \mathtt{let}\ \dot{\otimes}() = \dot{z} \ \mathtt{in}\ \dot{e_1})}$$

$$\frac{\mathcal{U}^{\mathtt{Jax}}(e_1) \stackrel{\text{def}}{=} E \ \mathtt{in}\ (e_1^p;\dot{e_1})}{\mathcal{U}^{\mathtt{Jax}}(\mathtt{let}\ \dot{\otimes}(\dot{x_1},\dot{x_2}) = \dot{z} \ \mathtt{in}\ e_1) \stackrel{\text{def}}{=} E \ \mathtt{in}\ (e_1^p; \mathtt{let}\ \dot{\otimes}(\dot{x_1},\dot{x_2}) = \dot{z} \ \mathtt{in}\ \dot{e_1})}$$

$$\frac{}{\mathcal{U}^{\mathtt{Jax}}(\underline{r}) \stackrel{\text{def}}{=} (\underline{r};\dot{\otimes}())} \qquad \frac{}{\mathcal{U}^{\mathtt{Jax}}(\underline{f}(x_1,\ldots,x_n)) \stackrel{\text{def}}{=} (\underline{f}(x_1,\ldots,x_n);\dot{\otimes}())}$$

$$\frac{}{\mathcal{U}^{\mathtt{Jax}}(\dot{0}_\tau) \stackrel{\text{def}}{=} (\otimes();\dot{0}_\tau)} \qquad \frac{}{\mathcal{U}^{\mathtt{Jax}}(\dot{x} \dot{+} \dot{y}) \stackrel{\text{def}}{=} (\otimes();\dot{x} \dot{+} \dot{y})}$$

$$\frac{}{\mathcal{U}^{\mathtt{Jax}}(x \dot{*} \dot{y}) \stackrel{\text{def}}{=} (\otimes();x \dot{*} \dot{y})} \qquad \frac{}{\mathcal{U}^{\mathtt{Jax}}(\mathrm{dup}(\dot{x})) \stackrel{\text{def}}{=} (\otimes();\mathrm{dup}(\dot{x}))}$$

$$\frac{\mathcal{U}^{\mathtt{Jax}}(e_1) \stackrel{\text{def}}{=} E \ \mathtt{in}\ (e_1^p;\dot{e_1})}{\mathcal{U}^{\mathtt{Jax}}(\mathrm{drop}(e_1)) \stackrel{\text{def}}{=} E \ \mathtt{in}\ (\mathrm{drop}(e_1^p);\mathrm{drop}(\dot{e_1}))}$$

Figure 2.5: Unzipping Transformation in JAX

**Illustrative Example 6** (Unzipping Transformation JAX)**.** Consider the expression $\mathcal{F}^{\texttt{Jax}}_{\{x\to\dot{dx},y\to\dot{dy}\}}(e^p)$ in Figure 2.4, we proceed by applying the unzipping transformation defined in Figure 2.5 obtaining exactly the Linear B expression in Figure 2.6 which is well-typed as $x:\mathbb{R}, y:\mathbb{R}; \dot{dx}:\mathbb{R}, \dot{dy}:\mathbb{R} \vdash^{\texttt{Jax}} \mathcal{U}^{\texttt{Jax}}(\mathcal{F}^{\texttt{Jax}}_{\{x\to\dot{dx},y\to\dot{dy}\}}(e^p)):(\mathbb{R};\mathbb{R})$.

Let us apply some simplifications for readability and get the following expression

$$\mathcal{U}^{\texttt{Jax}}(\mathcal{F}^{\texttt{Jax}}_{\{x\to\dot{dx},y\to\dot{dy}\}}(e^p)) \approx \begin{array}{l} \texttt{let } v_1 = \underline{sin}\ x \texttt{ in} \\ \texttt{let } v_2 = v_1 \mathbin{\underline{*}} y \texttt{ in} \\ \texttt{let } v_3 = \underline{cos}\ x \texttt{ in} \\ \texttt{let } v_4 = v_2 \mathbin{\underline{+}} v_3 \texttt{ in} \\ \texttt{let } w_1 = \underline{cos}\ x \texttt{ in} \\ \texttt{let } w_4 = -\underline{sin}\ x \texttt{ in} \\ \left( v_4;\ \begin{array}{l} \texttt{let } \dot{a} = \text{dup}(\dot{dx}) \texttt{ in} \\ \texttt{let } \dot{\otimes}(\dot{dx}_1, \dot{dx}_2) = \dot{a} \texttt{ in} \\ \texttt{let } \dot{v}_1 = w_1 \mathbin{\dot{*}} \dot{dx}_1 \texttt{ in} \\ \texttt{let } \dot{v}_2 = (y \mathbin{\dot{*}} \dot{v}_1) \mathbin{\dot{+}} (v_1 \mathbin{\dot{*}} \dot{dy}) \texttt{ in} \\ \texttt{let } \dot{v}_3 = w_4 \mathbin{\dot{*}} \dot{dx}_2 \texttt{ in} \\ \texttt{let } \dot{v}_4 = \dot{v}_2 \mathbin{\dot{+}} \dot{v}_3 \texttt{ in} \\ \dot{v}_4 \end{array} \right) \end{array} \tag{2.4}$$

where the purely tangent computation (part in blue) is entirely moved to the end, the transpose operation will act on this part. This expression in Equation 2.4 computes exactly the same as the expression obtained after the forward mode in Equation 2.3, in accordance with soundness property of unzipping (Claim 3 of Theorem 2).

### 2.2.3 Transpose

The transformation $\mathcal{T}^{\texttt{Jax}}$ is an endotransformation of Linear B transposing the tangent part of an expression and keeping invariant the primal part.

The core of the definition of $\mathcal{T}^{\texttt{Jax}}$ is on purely tangent expressions $\Gamma;\dot{\Gamma} \vdash^{\texttt{Jax}} \dot{e}:(\texttt{1};\tau)$ giving a $\Gamma;\dot{u}:\tau \vdash^{\texttt{Jax}} \mathcal{T}^{\texttt{Jax}}_{\theta;\dot{u}:\tau}(\dot{e}):(\texttt{1};\otimes\theta)$ depending on an enumeration $\theta$ of $\dot{\Gamma}$ and a free tangent variable $\dot{u}$ associated with output of $\dot{e}$. The output type of $\mathcal{T}^{\texttt{Jax}}_{\theta;\dot{u}:\tau}(\dot{e})$ is $(\texttt{1},\otimes\theta)$, where $\otimes\theta$ represents the nested product $\tau_1 \otimes (\tau_2 \otimes \dots (\tau_{n-1} \otimes \tau_n) \dots)$ for $\theta = (\dot{y}_1:\tau_1,\dots,\dot{y}_n:\tau_n)$. Figure 2.7 gives this definition.

The definition of $\mathcal{T}^{\texttt{Jax}}(\texttt{let } \dot{x} = \dot{e}_1 \texttt{ in } \dot{e}_2)$ reverses the order of the composition and the dependence between the two sub-expressions $\dot{e}_1$ and $\dot{e}_2$. The transpose first computes $\mathcal{T}^{\texttt{Jax}}(\dot{e}_2)$ storing its result in a pair $(\dot{u}_1,\dot{y})$ and then performs $\mathcal{T}^{\texttt{Jax}}(\dot{e}_1)$ by using the result $\dot{u}_1$ which is associated to the $\dot{x}$ dependence of $\dot{e}_2$ from $\dot{e}_1$. Notice also the duality between dup and $\dot{+}$ and between drop and $\dot{0}$. In our encoding into λLL, dup will be the diagonal in the additive conjunction & and drop the introduction of the neutral element of this conjunction ⊤.

The transpose transformation is then lifted to the primal constructs of Linear B by a simple commutation: $\mathcal{T}^{\texttt{Jax}}_{\theta;\dot{u}:\tau}((e^p;\dot{e})) \stackrel{\text{def}}{=} (e^p;\mathcal{T}^{\texttt{Jax}}_{\theta;\dot{u}:\tau}(\dot{e}))$ and $\mathcal{T}^{\texttt{Jax}}_{\theta;\dot{u}:\tau}(\texttt{let } x = e^p \texttt{ in } d) \stackrel{\text{def}}{=} \texttt{let } x = e^p \texttt{ in } \mathcal{T}^{\texttt{Jax}}_{\theta;\dot{u}:\tau}(d)$, and similarly for the other constructs (see Figure 2.8).

Finally, the following theorem shows that the transpose transformation $\mathcal{T}^{\texttt{Jax}}$ returns a well-typed expression in Linear B, preserves an amortized workload criterion for transposition and it is sound.

$$
\begin{aligned}
\mathcal{U}^{\texttt{Jax}}(\mathcal{F}^{\texttt{Jax}}_{\{x\to\dot{dx},y\to\dot{dy}\}}(e^p)) = {} & \texttt{let } v_1 = \underline{sin}\ x \texttt{ in} \\
& \texttt{let } v_2 = v_1 \mathbin{\underline{*}} y \texttt{ in} \\
& \texttt{let } v_3 = \underline{cos}\ x \texttt{ in} \\
& \texttt{let } v_4 = v_2 \mathbin{\underline{+}} v_3 \texttt{ in} \\
& \texttt{let } w_1 = \underline{cos}\ x \texttt{ in} \\
& \texttt{let } w_2 = y \texttt{ in let } w_3 = v_1 \texttt{ in} \\
& \texttt{let } w_4 = -\underline{sin}\ x \texttt{ in} \\
& \texttt{let } w_5 = 1 \texttt{ in let } w_6 = 1 \texttt{ in} \\
& \left(\begin{array}{ll}
 & \texttt{let } \dot{a} = \mathrm{dup}(\dot{dx}) \texttt{ in} \\
 & \texttt{let } \dot{\otimes}(\dot{dx}_1, \dot{dx}_2) = \dot{a} \texttt{ in} \\
 & \texttt{let } \dot{z}_1 = w_1 \mathbin{\dot{*}} \dot{dx}_1 \texttt{ in} \\
 & \texttt{let } \dot{v}_1 = \dot{z}_1 \texttt{ in} \\
 & \texttt{let } \dot{z}_2 = w_2 \mathbin{\dot{*}} \dot{v}_1 \texttt{ in let } \dot{z}_2 = w_3 \mathbin{\dot{*}} \dot{dy} \texttt{ in} \\
v_4; & \texttt{let } \dot{v}_2 = \dot{z}_2 \mathbin{\dot{+}} \dot{z}_3 \texttt{ in} \\
 & \texttt{let } \dot{z}_4 = w_4 \mathbin{\dot{*}} \dot{dx}_2 \texttt{ in} \\
 & \texttt{let } \dot{v}_3 = \dot{z}_4 \texttt{ in} \\
 & \texttt{let } \dot{z}_5 = w_5 \mathbin{\dot{*}} \dot{v}_2 \texttt{ in let } \dot{z}_6 = w_6 \mathbin{\dot{*}} \dot{v}_3 \texttt{ in} \\
 & \texttt{let } \dot{v}_4 = \dot{z}_5 \mathbin{\dot{+}} \dot{z}_6 \texttt{ in} \\
 & \dot{v}_4
\end{array}\right)
\end{aligned}
$$

Figure 2.6: Unzipping Transformation of JAX applied to the expression $\mathcal{F}^{\texttt{Jax}}_{\{x\to\dot{dx},y\to\dot{dy}\}}(e^p)$ defined in Figure 2.4.

**Theorem 3** (Type, Work-Preservation and Soundness $\mathcal{T}^{\texttt{Jax}}$: Adaptation of Theorem 7.1 in [93])**.** Given an expression in Linear B $\Gamma;\dot{\Gamma} \vdash^{\texttt{Jax}} d : (\tau,\sigma)$ and an enumeration $\theta$ of $\dot{\Gamma}$ , then

1. $\Gamma;\dot{u}:\sigma \vdash^{\texttt{Jax}} \mathcal{T}^{\texttt{Jax}}_{\theta;\dot{u}:\sigma}(d) : (\tau, \otimes\theta)$ is a well-typed expression of Linear B, where $\otimes\theta$ represents the nesting product of the types in $\theta$.

2. $\mathcal{W}^{\texttt{Jax}}(\mathcal{T}^{\texttt{Jax}}_{\theta;\dot{u}:\sigma}(d)) + \mathcal{W}^{\texttt{Jax}}(\dot{\Gamma}) \leq \mathcal{W}^{\texttt{Jax}}(d) + \mathcal{W}^{\texttt{Jax}}(\sigma)$

3. The function $[\mathcal{T}^{\texttt{Jax}}_{\theta;\dot{u}:\sigma}(d)]^{\mathsf{t}}$ is equal to the "transpose" of the function computed by the purely tangent computation of $d$. Formally, for every numeral sequences $\underline{\vec{r}}$ for $\Gamma$, $\underline{\vec{s}}$ for $\dot{\Gamma}$ and $\underline{w}$ for $\{\dot{u}:\sigma\}$, we have that:

$$[d]^{\mathsf{p}}_{\underline{\vec{r}}} = [\mathcal{T}^{\texttt{Jax}}_{\theta;\dot{u}:\sigma}(d)]^{\mathsf{p}}_{\underline{\vec{r}}} \quad \text{and} \quad \langle [d]^{\mathsf{t}}_{\underline{\vec{r}},\underline{\vec{s}}}, \underline{w}\rangle = \langle \underline{\vec{s}}, [\mathcal{T}^{\texttt{Jax}}_{\theta;\dot{u}:\sigma}(d)]^{\mathsf{t}}_{\underline{\vec{r}},\underline{w}}\rangle$$

   where $\langle\cdot,\cdot\rangle$ is the dot product.

*Sketch Proof.* By structural induction on $d$. The statement of this theorem is an adaptation to our notation of the Theorem 7.1 in [93], we refer to it for the detailed proof. □

In order to better understand why work-preservation for transposition requires amortization, as described in Claim 2 of the theorem above, consider the following example. Let $d$ be an

$$
\begin{aligned}
\mathcal{T}^{\mathtt{Jax}}_{\dot{x}:\tau;\dot{u}:\tau}(\dot{x}) &\stackrel{\text{def}}{=} \dot{u} \\
\mathcal{T}^{\mathtt{Jax}}_{\theta;\dot{u}:\tau}(\mathtt{let}\ \dot{x} = \dot{e_1}\ \mathtt{in}\ \dot{e_2}) &\stackrel{\text{def}}{=} \mathtt{let}\ \dot{\otimes}(\dot{x},\dot{u_2}) = \mathcal{T}^{\mathtt{Jax}}_{\dot{x}:\sigma,\theta\cap FV^t(\dot{e_2});\dot{u}:\tau}(\dot{e_2})\ \mathtt{in} \\
&\qquad \mathtt{let}\ \dot{u_1} = \mathcal{T}^{\mathtt{Jax}}_{\theta\cap FV^t(\dot{e_1});\dot{x}:\sigma}(\dot{e_1})\ \mathtt{in}\ \overline{\sigma}^{\mathrm{Jax}}_{\dot{u}_1,\dot{u}_2;\theta} \\
\mathcal{T}^{\mathtt{Jax}}_{\emptyset;\dot{u}:\mathbf{1}}(\dot{\otimes}()) &\stackrel{\text{def}}{=} \mathtt{let}\ \dot{\otimes}() = \dot{u}\ \mathtt{in}\ \dot{\otimes}() \\
\mathcal{T}^{\mathtt{Jax}}_{\theta;\dot{u}:\tau\otimes\sigma}(\dot{\otimes}(\dot{e_1},\dot{e_2})) &\stackrel{\text{def}}{=} \mathtt{let}\ \dot{\otimes}(\dot{u_1},\dot{u_2}) = \dot{u}\ \mathtt{in}\ \dot{\otimes}(\mathcal{T}^{\mathtt{Jax}}_{\theta\cap FV^t(\dot{e_1});\dot{u_1}:\tau}(\dot{e_1}),\mathcal{T}^{\mathtt{Jax}}_{\theta\cap FV^t(\dot{e_2});\dot{u_2}:\sigma}(\dot{e_2})) \\
\mathcal{T}^{\mathtt{Jax}}_{\theta;\dot{u}:\tau}(\mathtt{let}\ \dot{\otimes}() = \dot{z}\ \mathtt{in}\ \dot{e}) &\stackrel{\text{def}}{=} \dot{\otimes}(\dot{\otimes}(),\mathcal{T}^{\mathtt{Jax}}_{\theta\cap FV^t(\dot{e});\dot{u}:\tau}(\dot{e})) \\
\mathcal{T}^{\mathtt{Jax}}_{\theta;\dot{u}:\tau}(\mathtt{let}\ \otimes(\dot{x_1},\dot{x_2}) = \dot{z}\ \mathtt{in}\ \dot{e}) &\stackrel{\text{def}}{=} \mathtt{let}\ \dot{w} = \mathcal{T}^{\mathtt{Jax}}_{\dot{x_1}:\tau_1,\dot{x_2}:\tau_2,\theta\cap FV^t(\dot{e});\dot{u}:\tau}(\dot{e})\ \mathtt{in} \\
&\qquad \mathtt{let}\ \dot{\otimes}(\dot{y_1},\dot{z}) = \dot{w}\ \mathtt{in} \\
&\qquad \mathtt{let}\ \dot{\otimes}(\dot{y_2},\dot{y_3}) = \dot{z}\ \mathtt{in}\ \dot{\otimes}(\dot{\otimes}(\dot{y_1},\dot{y_2}),\dot{y_3}) \\
\mathcal{T}^{\mathtt{Jax}}_{\dot{x}:\tau;\dot{u}:\tau\otimes\tau}(\mathrm{dup}(\dot{x})) &\stackrel{\text{def}}{=} \mathtt{let}\ \dot{\otimes}(\dot{z},\dot{w}) = \dot{u}\ \mathtt{in}\ \dot{z}\dot{+}\dot{w} \\
\mathcal{T}^{\mathtt{Jax}}_{\{\dot{x}:\tau,\dot{y}:\tau\};\dot{u}:\tau}(\dot{x}\dot{+}\dot{y}) &\stackrel{\text{def}}{=} \mathrm{dup}(\dot{u}) \\
\mathcal{T}^{\mathtt{Jax}}_{\emptyset;\dot{u}:\tau}(\dot{0}_\tau) &\stackrel{\text{def}}{=} \mathrm{drop}(\dot{u}) \\
\mathcal{T}^{\mathtt{Jax}}_{\theta;\dot{u}:\mathbf{1}}(\mathrm{drop}(\dot{e})) &\stackrel{\text{def}}{=} \dot{0}_{\otimes\theta} \\
\mathcal{T}^{\mathtt{Jax}}_{\dot{y}:\tau;\dot{u}:\tau}(x\dot{*}\dot{y}) &\stackrel{\text{def}}{=} x\dot{*}\dot{u}
\end{aligned}
$$

Figure 2.7: Transpose on purely tangent expressions of JAX

$$
\begin{aligned}
\mathcal{T}^{\mathtt{Jax}}_{\theta;\dot{u}:\tau}((e^p;\dot{e})) &\stackrel{\text{def}}{=} (e^p;\mathcal{T}^{\mathtt{Jax}}_{\theta;\dot{u}:\tau}(\dot{e})) \\
\mathcal{T}^{\mathtt{Jax}}_{\theta;\dot{u}:\tau}(\mathtt{let}\ x = e^p\ \mathtt{in}\ d) &\stackrel{\text{def}}{=} \mathtt{let}\ x = e^p\ \mathtt{in}\ \mathcal{T}^{\mathtt{Jax}}_{\theta;\dot{u}:\tau}(d) \\
\mathcal{T}^{\mathtt{Jax}}_{\theta;\dot{u}:\tau}(\mathtt{let}\ \otimes(x_1,x_2) = z\ \mathtt{in}\ d) &\stackrel{\text{def}}{=} \mathtt{let}\ \otimes(x_1,x_2) = z\ \mathtt{in}\ \mathcal{T}^{\mathtt{Jax}}_{\theta;\dot{u}:\tau}(d)
\end{aligned}
$$

Figure 2.8: Transpose Transformation in JAX

expression in Linear B that computes a function, which takes a tangent variable as input and uses the `dup` construct to duplicate it $n$ times. Then the transposed expression of $d$, let us denote it by $\mathcal{T}^{\mathtt{Jax}}(d)$, will then perform $n$ addition operations (as detailed in the case for `dup` in the Figure 2.7). Therefore in this example we have $\mathcal{W}^{\mathtt{Jax}}(d) = 1$, since the `dup` operation incurs no cost. However, after transposition, $\mathcal{W}^{\mathtt{Jax}}(\mathcal{T}^{\mathtt{Jax}}(d)) = n$, which accounts for the $n$ addition operations required for the sums. The discrepancy in the cost arises because the transposed expression must manage the $n$ additional inputs created by the duplication. This additional work is amortized by crediting an extra $(n + 1) - 1$, where the $n$ extra inputs correspond to the cost of the sums in the transposed expression. Thus, the overall cost is balanced by the accumulation of these additional inputs during transposition.

**Illustrative Example 7** (Transpose Transformation JAX)**.** Consider the expression $\mathcal{U}^{\mathtt{Jax}}(\mathcal{F}^{\mathtt{Jax}}_{\{x\to\dot{dx},y\to\dot{dy}\}}(e^p))$ in Figure 2.6 and an enumeration $\theta$ of $\{\dot{dx} : \mathbb{R}, \dot{dy} : \mathbb{R}\}$, we then proceed by applying the transpose transformation defined in Figure 2.8 and Figure 2.7. We obtain an expression in Linear B typed as $x : \mathbb{R}, y : \mathbb{R}; \dot{\overline{z}} : \mathbb{R} \vdash^{\mathrm{Jax}} \mathcal{T}^{\mathtt{Jax}}_{\theta;\dot{\overline{z}}}(\mathcal{U}^{\mathtt{Jax}}(\mathcal{F}^{\mathtt{Jax}}_{\{x\to\dot{dx},y\to\dot{dy}\}}(e^p)))$ :

$(\mathbb{R}; \otimes(\mathbb{R},\mathbb{R}))$ and after some simplifications for readability we get the following expression

$$
\begin{aligned}
\mathcal{T}^{\mathsf{Jax}}_{\theta;\dot{\overline{z}}}(\mathcal{U}^{\mathsf{Jax}}(\mathcal{F}^{\mathsf{Jax}}_{\{x\to \dot{dx}, y\to \dot{dy}\}}(e^p))) \approx\ & \texttt{let } v_1 = \underline{sin}\ x \texttt{ in} \\
& \texttt{let } v_2 = v_1 \mathbin{\underline{*}} y \texttt{ in} \\
& \texttt{let } v_3 = \underline{cos}\ x \texttt{ in} \\
& \texttt{let } v_4 = v_2 \mathbin{\underline{+}} v_3 \texttt{ in} \\
& \texttt{let } w_1 = \underline{cos}\ x \texttt{ in} \\
& \texttt{let } w_4 = -\underline{sin}\ x \texttt{ in} \\
& \left( v_4; \begin{array}{l}
\texttt{let } \dot{\overline{b_1}} = \mathrm{dup}(\dot{\overline{z}}) \texttt{ in let } \dot{\otimes}(\dot{\overline{v_2}}, \dot{\overline{v_3}}) = \dot{\overline{b_1}} \texttt{ in} \\
\texttt{let } \dot{\overline{b_2}} = \mathrm{dup}(\dot{\overline{v_2}}) \texttt{ in let } \dot{\otimes}(\dot{\overline{v_{21}}}, \dot{\overline{v_{22}}}) = \dot{\overline{b_2}} \texttt{ in} \\
\texttt{let } \dot{\overline{v_1}} = y \mathbin{\dot{*}} \dot{\overline{v_{21}}} \texttt{ in} \\
\texttt{let } \dot{\overline{dx}} = (w_4 \mathbin{\dot{*}} \dot{\overline{v_3}}) \mathbin{\dot{+}} (w_1 \mathbin{\dot{*}} \dot{\overline{v_1}}) \texttt{ in} \\
\texttt{let } \dot{\overline{dy}} = v_1 \mathbin{\dot{*}} \dot{\overline{v_{22}}} \texttt{ in} \\
\dot{\otimes}(\dot{\overline{dx}}, \dot{\overline{dy}})
\end{array} \right)
\end{aligned}
\tag{2.5}
$$

where the part in red is clearly computing the gradient backward of $g$ which is $\nabla g(x, y, \overline{z}) = (\overline{z} * (y * cos(x) - sin(x)), sin(x) * \overline{z})$. More precisely, the red part of Equation 2.5 computes the transpose of the function calculated by the blue part of Equation 2.4, according to the soundness property of transpose transformation (Claim 3 of Theorem 3).

We can observe the triple occurrence of the adjoint variable $\dot{\overline{z}}$ in the body of the program in Figure 1.3a is managed by two duplication and the unpacking of a tangent tuples in Equation 2.5 (first two red lines).

# Chapter 3

# $\lambda$LL

We introduce $\lambda$LL as an extension of the linear logic $\lambda$-calculus (see e.g. [4, 66, 82, 110, 15, 44, 115]) to the ground type of the real numbers $\mathbb{R}$ with a set of functional symbols which are associated with differentiable functions. Section 3.1 gives the grammar of $\lambda$LL terms and their typing rules (Figure 3.1), Section 3.2 defines its operational semantics as a rewriting relation, $\beta$-reduction (Figure 3.4), satisfying the properties of subject reduction, progress, strong normalisation and confluence (Subsections 3.2.1, 3.2.2, 3.2.3). Section 3.3 extends $\beta$-equivalence to a logical equivalence $\sim$ which will be used to compare $\lambda$LL terms. Finally, Subsection 3.4 discusses some examples, notably dealing with the specialised operators over real numbers, and Subsection 3.5 updates the notion $\mathcal{W}$ of workload given in Subsection 2.1.2 and identifies a class of $\lambda$LL terms for which the evaluation to a $\beta$-normal form can be bounded by $\mathcal{W}$. This class includes all terms that will be used in the sequel for encoding Linear A expressions and its transformations.

The presentation of $\lambda$LL follow a standard pattern and the acquainted reader might will jump to Sections 3.3, 3.4 and 3.5 to have an immediate preview of the special features that will be used to study Autodiff.

## 3.1 Syntax and Type System

The grammar of types is defined as follows:

$$A, B, C ::= \mathbb{R} \mid A \multimap B \mid \mathtt{1} \mid A \otimes B \mid \top \mid A \& B \mid !A \qquad \text{(Types)}$$

Linear types distinguish between a resource of a given type $A$ that is called exactly once from a resource of *exponential modality* $!A$ which can be called at will (zero, one, or many times).

Linear A data-types are nested tuples, representing multidimensional numeric arrays. We express them with two different families of types, mirroring the distinction between primal and tangent data.

$$D, E ::= \mathbb{R} \mid \mathtt{1} \mid !D \otimes !E \qquad (\otimes\text{-sequence types})$$
$$L, H ::= \mathbb{R} \mid \top \mid H \,\&\, L \qquad (\&\text{-sequence types})$$

In Remark 6, we will observe that any $\otimes$-sequence type can be seen as a retraction of an exponential of a &-sequence type. The generation rules for the syntax of well-typed terms of $\lambda$LL are given in Figure 3.1, together with the typing rules. As for Linear A, we adopt a Church style typing: each variable has its type fixed once and for all. It is convenient to handle destructors

$$\frac{}{x : A \vdash x : A}\, v \qquad \frac{}{!x : !A \vdash x : A}\, !_e \qquad \frac{!\Gamma \vdash M : A}{!\Gamma \vdash !M : !A}\, !_i \qquad \frac{\Delta \vdash M : B}{!x : !A, \Delta \vdash M : B}\, !_w$$

$$\frac{p : A, \Delta \vdash M : B}{\Delta \vdash \lambda p.M : A \multimap B}\, \multimap_i \qquad \frac{!\Gamma_1, \Delta_1 \vdash M : A \multimap B \qquad !\Gamma_2, \Delta_2 \vdash N : A}{!\Gamma_1 \cup !\Gamma_2, \Delta_1, \Delta_2 \vdash MN : B}\, \multimap_e$$

$$\frac{}{\vdash () : \mathtt{1}}\, 1_i \qquad \frac{\Delta \vdash M : B}{() : \mathtt{1}, \Delta \vdash M : B}\, 1_e$$

$$\frac{!\Gamma_1, \Delta_1 \vdash M : A \qquad !\Gamma_2, \Delta_2 \vdash N : B}{!\Gamma_1 \cup !\Gamma_2, \Delta_1, \Delta_2 \vdash (M, N) : A \otimes B}\, \otimes_i \qquad \frac{p : A, q : B, \Delta \vdash M : C}{(p, q) : A \otimes B, \Delta \vdash M : C}\, \otimes_e$$

$$\frac{}{\Delta \vdash \langle\,\rangle : \top}\, \top$$

$$\frac{\Delta \vdash M_1 : A_1 \qquad \Delta \vdash M_2 : A_2}{\Delta \vdash \langle M_1, M_2 \rangle : A_1 \,\&\, A_2}\, \&_i \qquad \frac{p_i : A_i, \Delta \vdash M : B}{\langle p_1, p_2 \rangle : A_1 \,\&\, A_2, \Delta \vdash M : B}\, \&_{ei},\ i \in \{1, 2\}$$

$$\frac{r \text{ real number}}{\vdash \underline{r} : \mathbb{R}}\, R \qquad \frac{f \text{ unary map}}{\vdash \underline{f} : !\mathbb{R} \multimap !\mathbb{R}}\, F_1 \qquad \frac{f \text{ binary map}}{\vdash \underline{f} : !\mathbb{R} \otimes !\mathbb{R} \multimap !\mathbb{R}}\, F_2 \qquad \frac{}{\Delta \vdash \underline{0} : \mathbb{R}}\, Z$$

$$\frac{}{\vdash \dot{+} : \mathbb{R} \,\&\, \mathbb{R} \multimap \mathbb{R}}\, S \qquad \frac{}{\vdash \dot{*} : \mathbb{R} \multimap \mathbb{R} \multimap \mathbb{R}}\, M$$

Figure 3.1: λLL Typing Rules. We examine the cases where $\underline{f}$ is unary and binary, the general $n$-ary case being immediate.

as *patterns* binders, these latter being constructors of pairwise distinct variables:

$$p, q ::= \ x \mid !x \mid () \mid (p, q) \mid \langle p, q \rangle \qquad \text{(Patterns)}$$

where we suppose $FV(p) \cap FV(q) = \emptyset$, so a variable occurs at most once in a pattern. We say that a pattern is *exponential* whenever it is of the form $!x$. We use meta-variables $p^{\otimes}, q^{\otimes}$ (resp. $p^{\&}, q^{\&}$) for denoting patterns of ⊗-sequence types (resp. &-sequence types).

A typing environment $\Gamma$ is a finite set of patterns. We write $!\Gamma$ whenever all patterns in the environment are exponential.

It is however convenient to give the raw terms in the following grammar:

$$M, N ::= x \mid \lambda p^A.M \mid MN \mid () \mid (M, N) \mid !M \mid \quad \mid \langle\,\rangle \mid \langle M_1, M_2 \rangle \mid \underline{r} \mid \underline{f} \mid \dot{+} \mid \dot{*} \qquad (\lambda\text{LL})$$

where, as for Linear A, $\underline{r}$ and $\underline{f}$ are meta-variables varying over, respectively, numerals for real numbers, and $n$-ary numeric functions (for $n \geq 1$). We suppose that all numerical functions are differentiable and are equipped with their partial derivatives $\underline{\partial_i f}$. We suppose also a bound $b$ to the maximal arity $n$ of a numeric function primitive in the language (henceforth we will suppose $b = 2$). We have dedicated symbols for the specialised sum $\dot{+}$ and product $\dot{*}$ with scalars which have a different typing with respect to the typing of their sibling numerical functions.

**Typing rules.** The rules generating the set of the well-typed λLL terms are given in Figure 3.1. These rules use *judgments* of the form $\Delta \vdash M : A$, for a *typing environment* $\Delta = \{p_1 : A_1, \ldots, p_n : A_n\}$, a raw term $M$ and a type $A$.

We adapt the same conventions as for Linear A, in particular commas stand for “disjoint unions”. A difference with Linear A is that now typing environments are sets of patterns, not

$$\frac{\Delta, p : A \vdash M : B}{\Delta, \S^u p : \S A \vdash M : B}\ \S_\ell \qquad \frac{\Delta \vdash M : B}{\Delta, \S^u p : \S A \vdash M : B}\ \S_w \qquad \frac{!\Delta, \S\Sigma \vdash M : B}{!\Delta, \S\Sigma \vdash \S M : \S B}\ \S_r$$

Figure 3.2: Derived $\lambda$LL typing rules for the syntactic sugar of the affine modality $\S A \stackrel{\text{def}}{=} \mathtt{1} \,\&\, A$.

simply variables: so $\Delta$ is disjoint from $\Delta'$ means that no variable appears in both a pattern of $\Delta$ and a pattern of $\Delta'$. Namely, the rule $\multimap_e$ in Figure 3.1 is asking that the free variables in common between $M$ and $N$ belong to an exponential pattern in the environment. Notice that a variable $x : !A$ of exponential type is not an exponential pattern and so cannot be duplicated or erased. This restriction is known to be necessary to guarantee the subject reduction in a linear type system[1].

All rules except those in the last two lines are standard in linear logic [51]. Notice that $\dot{+}$ takes an additive pair $\mathbb{R} \,\&\, \mathbb{R}$ and returns $\mathbb{R}$, while $\dot{*}$ morally takes a multiplicative pair $\mathbb{R} \otimes \mathbb{R}$ and returns $\mathbb{R}$, reflecting the difference in linear algebra between addition, which is a linear operation, and scalar multiplication, which is a bilinear operation.

The set of free variables $FV(M)$ of a term $M$ is defined as usual, in particular $FV(p)$ is the set of variables appearing in a pattern $p$ and $FV(\lambda p.M) = FV(M) \setminus FV(p)$. Giving a typing environment $\Gamma = p_1 : A_1, \ldots, p_n : A_n$, we define $FV(\Gamma) \stackrel{\text{def}}{=} \bigcup_i FV(p_i)$. We may write $\lambda p^A.M$ if we wish to explicit the type of $p$.

**Notational conventions.** We adopt some syntactic sugar. First of all, we will often use the let notation for an application of an abstraction to a term: $\mathtt{let}\ p = N\ \mathtt{in}\ M$ is syntactic sugar for $(\lambda p.M)N$.

It is known that the formula $\mathtt{1} \,\&\, A$ expresses in linear logic the affine resource of type $A$: a value of type $A$ that can be used at most once. This modality will be used in our encoding of Autodiff, so we introduce the following notation:

$$\S A \stackrel{\text{def}}{=} \mathtt{1} \,\&\, A, \qquad \S M \stackrel{\text{def}}{=} \langle (\,), M \rangle, \qquad \S p \stackrel{\text{def}}{=} \langle (\,), p \rangle.$$

In particular, $\S p$ is a pattern of type $\S A$, whenever $p$ is a pattern of type $A$. We have then the derivation rules in Figure 3.2. For example, $\S_w$ is obtained by one $!_w$-rule introducing $u : \mathtt{1}$ in the environment and then by one $\&_{e1}$-rule. The $\S_r$-rule requires that all pattern in the environment are either exponential variables or pattern of the form $\S p$.

Some additional notation will be useful for the additive tuples. First, we may denote the $n$-fold additive product $\langle M_1, \langle M_2, \ldots M_n \rangle \ldots \rangle$ as an $n$-ary tuple $\langle M_1, M_2, \ldots, M_n \rangle$. We can use shortcut like $\langle M_i \rangle_{i=1}^n$, or even $\langle M_i \rangle_i$ if 1 and $n$ are clear from the context or irrelevant. We adopt similarly writing for the types: $\&_{i=1}^n A_i$ or $\&_i A_i$.

It will be convenient to index $n$-ary additive tuples with sequences of variables, so let us introduce the following notation. Let $\theta = (x_1, \ldots, x_n)$ be an enumeration of a finite set of variables (typically the free variables in a term). Let us write $x \in \theta$ (resp. $X \subseteq \theta$) for meaning that $x$ (resp. $X$) varies over all elements (resp. sets of elements) in $\theta$. For any $x \in \theta$, $\theta(x)$ denotes the position of $x$ in $\theta$. We write by $\theta \setminus X$ (resp. $\theta \cap X$) the enumeration obtained from $\theta$ by removing (resp. retaining) the variables in $X$. For example, if $\theta = (x_1, x_2, x_3, x_4)$, then $\theta \setminus \{x_2, x_3\} = (x_1, x_4)$ and $\theta \cap \{x_2, x_3\} = (x_2, x_3)$. We may write $\theta \setminus x$ if it is clear we are meaning $\theta \setminus \{x\}$. Finally, $x, \theta$ is the sequence obtained by adding $x$ to the beginning of $\theta$

[1] In fact, $f : A \multimap !B, x : A \vdash (\lambda y^{!B}.(y, y))(fx) : !B \otimes !B$ would be derivable if it were possible to copy variables of exponential type, but $f : A \multimap !B, x : A \vdash (fx, fx) : !B \otimes !B$ would not. Similar examples are known in the literature, see e.g. [111]. The solution adopted here is in the spirit of Barber's dual intuitionistic linear logic [15], based on dual environments.

(supposing that $x \notin \theta$), e.g. if $\theta = (x_1, x_2, x_3)$, then $x, \theta = (x, x_1, x_2, x_3)$. Moreover, $p, \theta$ for a pattern $p = \langle q_1, q_2 \rangle$ is the sequence obtained by adding the elements of the pattern $q_1, q_2$ at the beginning of $\theta$ respecting the order, e.g. if $\theta = (x_1, x_2, x_3)$ and $p = \langle y_1, \langle y_2, y_3 \rangle \rangle$ then $p, \theta = y_1, y_2, y_3, \theta = (y_1, y_2, y_3, x_1, x_2, x_3)$.

Given $\theta = (x_1, \dots, x_n)$, where $x_i : A_i$, and a family of terms $\{M_x\}_{x \in \theta}$, we write $\&\theta \approx \&_i A_i$, $\langle M_x \rangle_{x \in \theta} \approx \langle M_{x_1}, \dots, M_{x_n} \rangle$. Notice in particular that, if $n = 0$ (i.e. $\theta$ is empty), then $\&\theta \approx \top$ and $\langle M_x \rangle_{x \in \theta} = \langle \rangle$. If $n = 1$ (i.e. $\theta$ is a singleton $(x_1)$), $\&\theta \approx A_1$ and $\langle \theta \rangle = M_{x_1}$.

Given $\&_{i=1}^n A_i$ and a set $\mathcal{I}$, we define the *splitting* and *fusion* terms:

$$\sigma_{\mathcal{I}}^{\&_{i=1}^n A_i} \stackrel{\text{def}}{=} \lambda \langle x_i \rangle_{i=1}^n . \langle \langle x_i \rangle_{i \in \mathcal{I}}, \langle x_i \rangle_{i \notin \mathcal{I}} \rangle , \qquad \overline{\sigma}_{\mathcal{I}}^{\&_{i=1}^n A_i} \stackrel{\text{def}}{=} \lambda \langle \langle x_i \rangle_{i \in \mathcal{I}}, \langle x_i \rangle_{i \notin \mathcal{I}} \rangle . \langle x_i \rangle_{i=1}^n . \tag{3.1}$$

Note that these terms have the following types:

$$\sigma_{\mathcal{I}}^{\&_{i=1}^n A_i} : \&_{i=1}^n A_i \multimap (\&_{i \in \mathcal{I}}^n A_i) \,\&\, (\&_{i \notin \mathcal{I}}^n A_i) , \qquad \overline{\sigma}_{\mathcal{I}}^{\&_{i=1}^n A_i} : (\&_{i \in \mathcal{I}}^n A_i) \,\&\, (\&_{i \notin \mathcal{I}}^n A_i) \multimap \&_{i=1}^n A_i .$$

In fact, $\sigma_{\emptyset}^{A} : A \multimap \top \,\&\, A$ is (one direction of) the neutrality law, $\sigma_{\{2\}}^{A_1 \& A_2} : A_1 \,\&\, A_2 \multimap A_2 \,\&\, A_1$ is symmetry, and $\sigma_{\{1,2\}}^{A_1 \& (A_2 \& A_3)} : A_1 \,\&\, (A_2 \,\&\, A_3) \multimap (A_1 \,\&\, A_2) \,\&\, A_3$ is associativity.

Linear sum uses prefix notation, but we allow infix notation if $M$ is a pair: $N_1 \dot{+} N_2 \approx \dot{+} \langle N_1, N_2 \rangle$. We also extend the specialised operators $\underline{0}$, $\dot{+}$ and $\dot{*}$ to any &-sequence type:

$$\begin{aligned} 0_{\&_i H_i} &\stackrel{\text{def}}{=} \langle 0_{H_i} \rangle_i , \\ \dot{+}_{\&_i H_i} &\stackrel{\text{def}}{=} \lambda \langle \langle h_i \rangle_i, \langle h'_i \rangle_i \rangle . \langle \dot{+}_{H_i} \langle h_i, h'_i \rangle \rangle_i , \\ \dot{*}_{\&_i H_i} &\stackrel{\text{def}}{=} \lambda x . \lambda \langle h_i \rangle_i . \langle \dot{*}_{H_i} (x, h_i) \rangle_i . \end{aligned} \tag{3.2}$$

Notice that these are closed terms of type: $0_H : H$, $\dot{+}_H : H \,\&\, H \multimap H$, $\dot{*}_H : \mathbb{R} \multimap H \multimap H$.

## 3.2 Reduction and Rewriting Properties

The $\beta$-reduction relation $\to$ of λLL is defined as the contextual closure, with respect to the evaluation contexts given in Figure 3.3, of the $\beta$-reduction rules shown in Figure 3.4.

$$\gamma[\,] ::= [\,] \mid \lambda p^A . \gamma[\,] \mid \gamma[\,] N \mid M \gamma[\,] \mid (\gamma[\,], N) \mid (M, \gamma[\,]) \mid !\gamma[\,] \mid \langle \gamma[\,], M \rangle \mid \langle M, \gamma[\,] \rangle$$

Figure 3.3: Grammar of one-hole contexts. Moreover, we call *exponential safe context* a context generated by the sub-grammar not having $!\gamma[\,]$.

The rule $\beta_\lambda$ replaces a pattern $p$ by a term $V$, supposing this latter has a "structure compatible with $p$". This is formalised by the notion of *a value $V$ for a pattern $p$.* A value for a variable $x$ of type $A$ is any term of type $A$, a value for $(\,)$ is $(\,)$, a value for $!x$ is $!M$ for $M$ any term, a

$$\begin{aligned} &\beta_\lambda : (\lambda p . M) V \to M\{V/p\} && \beta_F : \underline{f}(!\underline{r}_1, !\underline{r}_2) \to !\underline{f(r_1, r_2)} \\ &\beta_{\dot{+}} : \dot{+} \langle \underline{r}_1, \underline{r}_2 \rangle \to \underline{(r_1 + r_2)} && \beta_{\dot{*}} : \dot{*} \underline{r}_1 \underline{r}_2 \to \underline{(r_1 \cdot r_2)} \end{aligned}$$

Figure 3.4: $\beta$-reduction. $\beta_\lambda$ supposes $V$ to be a value for the pattern $p$. Steps $\beta_F$, $\beta_{\dot{+}}$ and $\beta_{\dot{*}}$ are called *numeric steps* or *flops*.

value for $(p_1, p_2)$ (resp. $\langle p_1, p_2 \rangle$) is $(V_1, V_2)$ (resp. $\langle V_1, V_2 \rangle$) where $V_i$ is a value for $p_i$. We then generalise the standard variable substitution $M\{N/x\}$ to the *substitution* $M\{V/p\}$ *of a pattern* $p$ *for a value* $V$ *in a term* $M$, by dispatching all components in $V$ to the free occurrences in $M$ of $FV(p)$, i.e.:

$$M\{()/()\} \stackrel{\text{def}}{=} M$$
$$M\{!N/!x\} \stackrel{\text{def}}{=} M\{N/x\}$$
$$M\{(V_1, V_2)/(p_1, p_2)\} \stackrel{\text{def}}{=} M\{\langle V_1, V_2 \rangle / \langle p_1, p_2 \rangle\} \stackrel{\text{def}}{=} M\{V_1/p_1\}\{V_2/p_2\}$$

We denote by $\rightarrow^*$ and $=_\beta$ respectively the reflexive-transitive and the equivalence closure of $\rightarrow$. A *$\beta$-normal form*, $\beta$-nf for short, is a term $M$ s.t. there is no $N$ s.t. $M \rightarrow N$.

All reductions are closed by one-hole contexts. A *one-hole context* $\gamma[\,]$ is a term with a sole occurrence of an hole $[\,]$, this latter being a place holder which will be replaced for a term $M$ (with possible capture of free variables) generating a new term denoted by $\gamma[M]$. The grammar of the one-hole contexts of $\lambda$LL is given in Figure 3.3.

The $\beta$-reduction is designed on the top of the LL cut-elimination. This yields a well behaving rewriting system, satisfying crucial properties such as subject reduction (Subsection 3.2.1), progress (Subsection 3.2.2), strong normalisation and confluence (Subsection 3.2.3).

### 3.2.1 Subject Reduction

Subject reduction guarantees that types are preserved along reduction. This property holds in $\lambda$LL and it is formalized in Theorem 4. We prove it by means of a pattern substitution lemma (Lemma 3) which uses the hypothesis that an exponential pattern must have the basic form $!x$ for a variable $x$ (see Remark 2).

Given a type derivation $\Pi$, we define the size of such derivation $s(\Pi)$ as the number of derivation rules of $\Pi$ and we proceed as follows to prove the pattern substitution lemma.

**Lemma 1.** If $\Gamma \vdash M : A$, then $FV(M) \subseteq FV(\Gamma)$.

*Sketch Proof.* By induction on a derivation of $\Gamma \vdash M : A$. ☐

**Lemma 2.** If $FV(p) \cap FV(M) = \emptyset$, then $M\{V/p\} = M$.

*Sketch Proof.* By induction on $M$. ☐

**Lemma 3** (Pattern Substitution)**.** Given two derivable judgements $!\Gamma_1, \Delta_1, p : A \vdash M : B$ and $!\Gamma_2, \Delta_2 \vdash V : A$ such that

1. $V$ is a value for $p$,
2. $FV(!\Gamma_i, \Delta_i) \cap FV(\Delta_{3-i}) = \emptyset$ for $i \in \{1, 2\}$,

then we have that the judgment $!\Gamma_1 \cup !\Gamma_2, \Delta_1, \Delta_2 \vdash M\{V/p\} : B$ is derivable.

*Proof.* Taking the judgements as in the hypotheses of the lemma, for any derivation $\Pi_1$ of $!\Gamma_1, \Delta_1, p : A \vdash M : B$ and $\Pi_2$ of $!\Gamma_2, \Delta_2 \vdash V : A$, we give a derivation of $!\Gamma_1 \cup !\Gamma_2, \Delta_1, \Delta_2 \vdash M\{V/p\} : B$ by induction on the lexicographically ordered pair $(s(\Pi_2), s(\Pi_1))$. Notice that the condition 1 in the lemma hypothesis is necessary to assure that the substitution $M\{V/p\}$ is well-defined, while condition 2 assure that the typing environment $!\Gamma_1 \cup !\Gamma_2, \Delta_1, \Delta_2$ is a set of patterns of pairwise distinct variables.

We split depending on the last derivation rule in $\Pi_1$ or $\Pi_2$.

- If the last rule of $\Pi_1$ is a rule $r$ among $\{!_w, \&_{ei}, \otimes_e, 1_e\}$ acting on a pattern in $!\Gamma_1, \Delta_1$, then the immediate subderivation of $\Pi_1$ is $\Pi_1'$ of $!\Gamma_1', \Delta_1', p : A \vdash M : B$ for suitable $!\Gamma_1', \Delta_1'$, such that $FV(\Delta_1') \subseteq FV(\Delta_1)$ and $FV(!\Gamma_1', \Delta_1') \subseteq FV(!\Gamma_1, \Delta_1)$.[2]

  We then have: $FV(!\Gamma_1', \Delta_1') \cap FV(\Delta_2) \subseteq FV(!\Gamma_1, \Delta_1) \cap FV(\Delta_2) = \emptyset$, as well as $FV(!\Gamma_2, \Delta_2) \cap FV(\Delta_1') \subseteq FV(!\Gamma_2, \Delta_2) \cap FV(\Delta_1) = \emptyset$. We can then apply the induction hypothesis on $\Pi_1'$ and $\Pi_2$, getting a derivation of $!\Gamma_1' \cup !\Gamma_2, \Delta_1', \Delta_2 \vdash M\{V/p\} : B$. We can then conclude by possibly applying the rule $r$ for getting $!\Gamma_1 \cup !\Gamma_2, \Delta_1, \Delta_2 \vdash M\{V/p\} : B$.

- The case where the last rule of $\Pi_2$ is a rule $r$ among $\{!_w, \&_{ei}, \otimes_e, 1_e\}$ acting on a pattern in $!\Gamma_2, \Delta_2$ is analogous to the previous case.

- For the other cases, we can then suppose that the last rules of $\Pi_1$ and $\Pi_2$ are not rules acting on $!\Gamma_i, \Delta_i$. We then split in further sub-cases depending if the last rules of $\Pi_1$ acts on the pattern $p : A$ or acts on the term $M$.

  Let us consider first the cases related to the last rule $r$ of $\Pi_1$ acting on the pattern $p : A$.

  - If $r$ is of type $!_w$, then $p = !x$ and $V = !N$ for some term $N$. By cases inspection, one can infer that the last rule of $\Pi_2$ is a $!_i$, as it cannot be a rule acting on $!\Gamma_2, \Delta_2$ by the above hypothesis. We conclude in particular that $\Delta_2$ is empty.

    Notice that the subderivation $\Pi_1'$ above $r$ has conclusion $!\Gamma_1, \Delta_1 \vdash M : B$. By a suitable sequences of $!_w$ we then get $!\Gamma_1 \cup !\Gamma_2, \Delta_1, \vdash M : B$. We then conclude as by Lemma 1 $x \notin FV(M)$ so by Lemma 2, $M\{!V/!x\} = M$.
  - If $r$ is of type $\&_{ei}$, then $p = \langle p_1, p_2 \rangle : A_1 \& A_2$ and $V = \langle V_1, V_2 \rangle$ for some values $V_i$ for $p_i$.

    Notice that the subderivation $\Pi_1'$ above $r$ has conclusion $!\Gamma_1, \Delta_1, p_i : A_i \vdash M : B$.

    By cases inspection, one can infer that the last rule of $\Pi_2$ is a $\&_i$, as it cannot be a rule acting on $!\Gamma_2, \Delta_2$ by the above hypothesis. Therefore we have a subderivation $\Pi_2'$ above such a rule of the judgement $!\Gamma_2, \Delta_2 \vdash V_i : A_i$.

    We can then apply the induction hypothesis to $\Pi_1'$ and $\Pi_2'$ and getting a derivation of $!\Gamma_1 \cup !\Gamma_2, \Delta_1, \Delta_2 \vdash M\{V_i/p_i\} : B$. We then conclude as by Lemma 1 $FV(p_{3-i}) \notin FV(M)$ so by Lemma 2, $M\{V_i/p_i\}\{V_{3-i}/p_{3-i}\} = M\{V_i/p_i\}$.
  - If $r$ is of type $\otimes_e$, then $p = (p_1, p_2) : A_1 \otimes A_2$ and $V = (V_1, V_2)$ for some values $V_i$ of $p_i$.

    Notice that the subderivation $\Pi_1'$ above $r$ has conclusion $!\Gamma_1, \Delta_1, p_1 : A_1, p_2 : A_2 \vdash M : B$.

    Also, by cases inspection, one can infer that the last rule of $\Pi_2$ is a $\otimes_i$, as it cannot be a rule acting on $!\Gamma_2, \Delta_2$ by the above hypothesis. Therefore we have two subderivations $\Pi_{2,1}$ and $\Pi_{2,2}$ above such a rule of the judgements, resp., $!\Gamma_{2,1}, \Delta_{2,1} \vdash V_1 : A_1$ and $!\Gamma_{2,2}, \Delta_{2,2} \vdash V_2 : A_2$ such that $!\Gamma_2 = !\Gamma_{2,1} \cup !\Gamma_{2,2}$ and $\Delta_2 = \Delta_{2,1}, \Delta_{2,2}$.

    By applying the induction hypothesis on $\Pi_1'$ and $\Pi_{2,1}$ we then get a derivation of $!\Gamma_1 \cup !\Gamma_{2,1}, \Delta_1, \Delta_{2,1} \vdash M\{V_1/p_1\} : B$. We then apply the induction hypothesis on this latter derivation and $\Pi_{2,2}$ (notice that $s(\Pi_{2,2}) < s(\Pi_2)$ and so we have $(s(\Pi_{2,2}), k) < (s(\Pi_2), s(\Pi_1))$ for any $k$), so to get $!\Gamma_1 \cup !\Gamma_{2,1} \cup !\Gamma_{2,2}, \Delta_1, \Delta_{2,1}, \Delta_{2,2} \vdash M\{V_1/p_1\}\{V_2/p_2\} : B$ and we can conclude.
  - If $r$ is of type $1_e$, then $p = () : \mathtt{1}$ and $V = ()$.

    Notice that the subderivation $\Pi_1'$ above $r$ has conclusion $!\Gamma_1, \Delta_1 \vdash M : B$.

[2] Notice on the contrary that we can have $FV(!\Gamma_1') \not\subseteq FV(!\Gamma_1)$ whenever $r = \&_{ei}$ acting on a pattern $!x \in \Gamma_1'$ and so transferring the result into $\Delta_1$.

Also, by cases inspection, one can infer that $\Pi_2$ is just an instance of $1_i$, as it cannot be a rule acting on $!\Gamma_2, \Delta_2$ by the above hypothesis. So in particular $!\Gamma_2, \Delta_2$ is empty. We conclude as $M\{()/()\} = M$.

- Let us consider now the cases of the last rule $r$ of $\Pi_1$ acting on the subject $M$.
  - If $r$ is $v$, then $M = p = x$, $A = B$ and $!\Gamma_1, \Delta_1$ is empty. We conclude by taking the derivation $\Pi_2$ of $!\Gamma_2, \Delta_2 \vdash V : A$, as $M\{V/x\} = V$.
  - If $r$ is $!_e$, then $M = x$, $p = !x$, $!A = B$ and $!\Gamma_1, \Delta_1$ is empty.
    Notice that $V = !V'$ and by cases inspection, one can infer that the last rule of $\Pi_2$ is a $!_i$, as it cannot be a rule acting on $!\Gamma_2, \Delta_2$ by the above hypothesis. We conclude in particular that $\Delta_2$ is empty and the immediate subderivation of $\Pi_2$ is a derivation of $!\Gamma_2 \vdash V' : A$. We then conclude by taking this derivation as $M\{!V'/!x\} = V'$.
  - If $r$ is $\multimap_e$, so that $M = M_1 M_2$ and the immediate subderivations of $\Pi_1$ are $\Pi_{1,1}$ of $!\Gamma_{1,1}, \Delta_{1,1} \vdash M_1 : B' \multimap B$ and $\Pi_{1,2}$ of $!\Gamma_{1,2}, \Delta_{1,2} \vdash M_2 : B'$, such that: $!\Gamma_1, \Delta_1, p : A = !\Gamma_{1,1} \cup !\Gamma_{1,2}, \Delta_{1,1} \cup \Delta_{1,2}$ We should now split in sub-cases, depending in which environment $p$ occurs.
    * If $p \in !\Gamma_{1,1} \cap !\Gamma_{1,2}$. Then $p = !x$ and $A = !A'$ for a suitable $x : A'$. We then have $V = !V'$ and we can infer by cases inspection that the last rule of $\Pi_2$ is a $!_i$, as it cannot be a rule acting on $!\Gamma_2, \Delta_2$ by the above hypothesis. We conclude in particular that $\Delta_2$ is empty. We then apply the induction hypothesis separately on $\Pi_{1,1}, \Pi_2$ and on $\Pi_{1,2}, \Pi_2$ getting a derivation of $!\Gamma_{1,1} \cup !\Gamma_2, \Delta_{1,1} \vdash M_1\{V/p\} : B' \to B$ and one of $!\Gamma_{1,2} \cup !\Gamma_2, \Delta_{1,2} \vdash M_2\{V/p\} : B'$. We then conclude by a $\multimap_e$ rule $!\Gamma_1 \cup !\Gamma_2, \Delta_1 \vdash M_1 M_2\{V/p\} : B$.
    * The case $p \in (!\Gamma_{1,i}, \Delta_i) \setminus (!\Gamma_{1,3-i}, \Delta_{3-i})$ is a simpler variant of the above one, just applying the induction hypothesis on $\Pi_{1,i}$ and $\Pi_2$.
  - The case $r$ is $\otimes_i$ is analogous to $\multimap_e$.
  - The case $r$ is $\multimap_i$, $\&_i$ are immediate application of the induction hypothesis.
  - The cases $r$ is an initial rule $\top_i$, $Z$ are immediate.
  - The case $r$ is $1_i$ or $F_2$, $S$, $M$, $R$ are not possible as the pattern $p$ cannot appear in the environment: these cases are in fact specific instances of the previous cases of some rule acting on $p_1$.

□

**Remark 2.** Let us remark that the proof above of the substitution lemma (Lemma 3) uses the hypothesis that an exponential pattern (i.e. a pattern belonging to $!\Gamma$) must have the basic form $!x$ for a variable $x$. If we had relaxed our definition of patterns, allowing for e.g. $!(p_1, p_2)$, then substitution lemma would have failed (and hence subject reduction). In fact, in this case we would have for the terms $V = !((x, x'), (x, x'))$ and $M = ((x, z), (x', z))$ and the type for $B = (A \otimes A') \otimes (A \otimes A')$ the possible judgements:

1. $!(x, x') : !(A \otimes A') \vdash V : !B$
2. $!(x, x') : !(A \otimes A'), !z : !B \vdash M : (A \otimes B) \otimes (A' \otimes B)$

while the term $M\{V/!z\} = ((x, ((x, x'), (x, x'))), (x', ((x, x'), (x, x'))))$ could not be typed under the environment $!(x, x') : !(A \otimes A')$ because there is no possibility between the $x$'s and $(x, x')$.

Finally, we prove subject reduction by using the following lemma and the pattern substitution lemma above.

**Lemma 4.** If $!\Gamma, \Delta \vdash \lambda p.M : A \multimap B$ is derivable, then $!\Gamma, \Delta, p : A \vdash M : B$.

*Sketch Proof.* By induction on the size of the derivation $!\Gamma, \Delta, p : A \vdash M : B$. □

**Theorem 4** (Subject Reduction)**.** Let $!\Gamma, \Delta \vdash M : A$ and $M \to N$, then $!\Gamma, \Delta \vdash N : A$.

*Proof.* Let $\Pi$ be the derivation for $!\Gamma, \Delta \vdash M : A$, we proceed by induction on $s(\Pi)$. We split depending on the last derivation rule in $\Pi$.

- If the last rule of $\Pi$ is a rule $r$ among $\{!_w, \&_{ei}, \otimes_e, 1_e\}$ acting on a pattern in $!\Gamma, \Delta$, then the immediate subderivation of $\Pi$ is $\Pi'$ of $!\Gamma', \Delta' \vdash M : A$ for suitable $!\Gamma', \Delta'$, such that $FV(\Delta') \subseteq FV(\Delta)$ and $FV(!\Gamma', \Delta') \subseteq FV(!\Gamma, \Delta)$. We can then apply the induction hypothesis on $\Pi'$, getting a derivation of $!\Gamma', \Delta' \vdash N : A$. We can then conclude by possibly applying the rule $r$ obtaining $!\Gamma, \Delta \vdash N : A$.

- For the other cases, we can then suppose that the last rule of $\Pi$ is not a rule acting on $!\Gamma, \Delta$. We then proceed by analyzing the cases in which the last rule $r$ of $\Pi$ acts on the term $M$.

  – If $r$ is $\multimap_i$, then $M = \lambda p.M'$ and $A = B_1 \multimap B_2$.
    By hypothesis we have that $\Pi$ is the derivation for $!\Gamma, \Delta \vdash \lambda p.M' : B_1 \multimap B_2$. Moreover, the redex can be only inside $M'$, so $N$ is in the form $\lambda p.N'$. The immediate subderivation of $\Pi$ above $r$ is $\Pi'$ for $!\Gamma, \Delta, p : B_1 \vdash M' : B_2$.
    By induction hypothesis on $\Pi'$ we have a derivation for $!\Gamma, \Delta, p : B_1 \vdash N' : B_2$. We can conclude that $!\Gamma, \Delta \vdash \lambda p.N' : B_1 \multimap B_2$ is well-typed by using the rule $r$ and by induction hypothesis.

  – If $r$ is $\&_i$, then $M = \langle M_1, M_2 \rangle$ and $A = B_1 \& B_2$.
    By hypothesis we have that $\Pi$ is the derivation for $!\Gamma, \Delta \vdash \langle M_1, M_2 \rangle : B_1 \& B_2$. Therefore we have two subderivations $\Pi_1$ and $\Pi_2$ above $r$ for $!\Gamma, \Delta \vdash M_1 : B_1$ and $!\Gamma, \Delta \vdash M_2 : B_2$, respectively.
    Note that the redex can be either in $M_1$ or in $M_2$. Let us suppose that the redex is in $M_1$ and so $M_1 \to M_1'$ (the other case being similar), we have that $N = \langle M_1', M_2 \rangle$. By induction hypothesis on $\Pi_1$ we have a derivation for $!\Gamma, \Delta \vdash M_1' : B_1$. We can conclude that $!\Gamma, \Delta \vdash \langle M_1', M_2 \rangle : B_1 \& B_2$ is well-typed by using the rule $r$ and by using $\Pi_2$ for the derivation of $M_2$ and the induction hypothesis for the derivation of $M_1'$.

  – If $r$ is $\multimap_e$, then $M = M_1 M_2$ and $!\Gamma = !\Gamma_1 \cup !\Gamma_2$ and $\Delta = \Delta_1, \Delta_2$.
    By hypothesis we have that $\Pi$ is the derivation for $!\Gamma_1 \cup !\Gamma_2, \Delta_1, \Delta_2 \vdash M_1 M_2 : A$. Therefore we have two subderivations $\Pi_1$ and $\Pi_2$ above $r$ for $!\Gamma_1, \Delta_1 \vdash M_1 : B \multimap A$ and $!\Gamma_2, \Delta_2 \vdash M_2 : B$, respectively.
    Let's proceed by analyzing the following subcases related to the redex:

    * If the redex is inside $M_1$ and so $M_1 \to M_1'$, then $N = M_1' M_2$. By induction hypothesis on $\Pi_1$ we have a derivation for $!\Gamma_1, \Delta_1 \vdash M_1' : B \multimap A$. We can conclude that $!\Gamma_1 \cup !\Gamma_2, \Delta_1, \Delta_2 \vdash M_1' M_2 : A$ is well-typed by using the rule $r$ and by using $\Pi_2$ for the derivation of $M_2$ and the induction hypothesis for the derivation of $M_1'$.
    * The redex is inside $M_2$ and so $M_2 \to M_2'$, then $N = M_1 M_2'$. By induction hypothesis on $\Pi_2$ we have a derivation for $!\Gamma_1, \Delta_1 \vdash M_2' : B$. We can conclude that $!\Gamma_1 \cup !\Gamma_2, \Delta_1, \Delta_2 \vdash M_1 M_2' : A$ is well-typed by using the rule $r$ and by using $\Pi_1$ for the derivation of $M_1$ and the induction hypothesis for the derivation of $M_2'$.

* The redex is $M_1 M_2$, then we proceed by induction on the reduction step, so we analyze the cases of the reduction rules in Figure 3.4.
  - If the reduction rule is $\beta_\lambda$, then $M_1 = \lambda p.M_1'$ and $M_2$ is a value $V$ for the pattern $p$. Moreover, $N$ is in the form $M_1'\{V/p\}$.
    By hypothesis we have that $\Pi$ is the derivation for $!\Gamma_1 \cup !\Gamma_2, \Delta_1, \Delta_2 \vdash (\lambda p.M_1')V : A$. Therefore we have two subderivations $\Pi_1$ and $\Pi_2$ above $r$ for $!\Gamma_1, \Delta_1 \vdash \lambda p.M_1' : B \multimap A$ and $!\Gamma_2, \Delta_2 \vdash V : B$, respectively.
    By Lemma 4 on $!\Gamma_1, \Delta_1 \vdash \lambda p.M_1' : B \multimap A$ we have a derivation for $!\Gamma_1, \Delta_1, p : B \vdash M_1' : A$.
    We can conclude by applying Lemma 3, getting a derivation for the judgement $!\Gamma_1 \cup !\Gamma_2, \Delta_1, \Delta_2 \vdash M_1'\{V/p\} : A$.
  - If the reduction rule is $\beta_F$, $\beta_{\dot{*}}$ or $\beta_{\dot{+}}$, then the proof is simple and direct.

□

### 3.2.2 Progress

Progress property identifies a grammar to the $\beta$-normal forms and corresponds in proof-theory to the sub-formula property. We express here this grammar only for the closed terms as this is what we need and the generalisation to open terms is more involved.

**Proposition 1** (Progress)**.** The set of closed $\beta$-nf is given by:

$$W ::= \ \lambda p^A.M \mid (\,) \mid (W_1, W_2) \mid \langle\,\rangle \mid \langle W_1, W_2\rangle \mid !W \mid \underline{r} \mid \underline{f} \mid \dot{+} \mid \dot{*}(!\underline{r})$$

where $M$, $W$ or $W_i$ are $\beta$-nf ($W$ and $W_i$ are moreover closed).

*Sketch Proof.* By induction on $W$. □

### 3.2.3 Strong Normalization

In a rewriting system, strong normalization (SN) ensures that no term appears in an infinite reduction sequence. In $\lambda$LL, strong normalisation (Corollary 2) is obtained by using the notion of reducibility (see e.g. [53], Definition 1 and Corollary 1), which should be adapted in order to deal with the multiplicative connectives ($\otimes$, $\mathtt{1}$) without having the involutive negation $A^\perp$ of classical linear logic.

Let $\mathscr{T}_A$ be the set of terms of type $A$, for some typing environment. A typical way to define reducibility for linear logic formulas is by using orthogonality [6] — a map $\mathcal{X} \mapsto \mathcal{X}^\perp$ over sets of terms which formalises the notion of "passing a test". In this scenario, the reducibility for the type tensor $\mathtt{RED}_{A\otimes B}$ is defined as $\{(M, N) \subseteq \mathscr{T}_{A\otimes B} \mid M \subseteq \mathtt{RED}_A \wedge N \subseteq \mathtt{RED}_B\}^{\perp\perp}$. Unfortunately, we cannot apply this method immediately in $\lambda$LL, as we have not an involutive negation (a type operator $(\,)^\perp$ such that $A^{\perp\perp} = A$). However, we can overcome the difficulty by using let-expressions and the ground type $\mathbb{R}$.

**Definition 1** (Reducibility)**.** We define the sets Test and RED by mutual recursion as follows

$$\mathtt{Test}_{x:A} \stackrel{\text{def}}{=} \{\Delta, x : A \vdash N : \mathbb{R} \mid \forall M \in \mathtt{RED}_A.\ N\{{}^{M}\!/x\} \in \mathtt{RED}_\mathbb{R}\}$$

$$\mathtt{Test}_{!x:!A} \stackrel{\text{def}}{=} \{\Delta, !x :!A \vdash N : \mathbb{R} \mid \forall M \in \mathtt{RED}_A.\ \mathtt{let}\ !x = !M\ \mathtt{in}\ N \in \mathtt{RED}_\mathbb{R}\}$$

$$\mathtt{Test}_{():1} \stackrel{\text{def}}{=} \{\Delta \vdash N : \mathbb{R} \mid N \in \mathtt{RED}_\mathbb{R}\}$$

$$\mathtt{Test}_{(x_1,x_2):A_1\otimes A_2} \stackrel{\text{def}}{=} \{\Delta, (x_1, x_2) : A_1 \otimes A_2 \vdash N : \mathbb{R}.\ \mathtt{let}\ (x_1, x_2) = (V_1, V_2)\ \mathtt{in}\ N \in \mathtt{RED}_\mathbb{R}\}$$

$$\mathtt{Test}_{\langle x_1,x_2\rangle:A_1\&A_2} \stackrel{\text{def}}{=} \{\Delta, \langle x_1, x_2\rangle : A_1\&A_2 \vdash N : \mathbb{R} \text{ s.t. } \langle V_1, V_2\rangle \text{ typable.}$$

$$
\begin{aligned}
&\qquad\qquad\qquad \texttt{let}\ \langle x_1, x_2\rangle = \langle V_1, V_2\rangle\ \texttt{in}\ N \in \texttt{RED}_{\mathbb{R}}\}\\
\texttt{RED}_{\mathbb{R}} &\stackrel{\text{def}}{=} \{M \in \mathscr{T}_{\mathbb{R}} \mid M \text{ is SN}\}\\
\texttt{RED}_{1} &\stackrel{\text{def}}{=} \{M \in \mathscr{T}_{1} \mid \forall N \in \texttt{Test}_{():1}.\ \texttt{let}\ () = M\ \texttt{in}\ N \in \texttt{RED}_{\mathbb{R}}\}\\
\texttt{RED}_{A\otimes B} &\stackrel{\text{def}}{=} \{M \in \mathscr{T}_{A\otimes B} \mid \forall N \in \texttt{Test}_{(x_1,x_2):A\otimes B}.\ \texttt{let}\ (x_1, x_2) = M\ \texttt{in}\ N \in \texttt{RED}_{\mathbb{R}}\}\\
\texttt{RED}_{\top} &\stackrel{\text{def}}{=} \{M \in \mathscr{T}_{\top} \mid M \text{ is SN}\}\\
\texttt{RED}_{A\& B} &\stackrel{\text{def}}{=} \{M \in \mathscr{T}_{A\& B} \mid \forall N \in \texttt{Test}_{\langle x_1,x_2\rangle:A\& B}.\ \texttt{let}\ \langle x_1, x_2\rangle = M\ \texttt{in}\ N \in \texttt{RED}_{\mathbb{R}}\}\\
\texttt{RED}_{!A} &\stackrel{\text{def}}{=} \{M \in \mathscr{T}_{!A} \mid \forall N \in \texttt{Test}_{!x:!A}.\ \texttt{let}\ !x = M\ \texttt{in}\ N \in \texttt{RED}_{\mathbb{R}}\}\\
\texttt{RED}_{A\multimap B} &\stackrel{\text{def}}{=} \{M \in \mathscr{T}_{A\multimap B} \mid \forall N \in \texttt{RED}_{A}.\ MN \in \texttt{RED}_{B}\}
\end{aligned}
$$

We call *neutral* a term generated by the following grammar:

$$\mathcal{N} ::= \ x \mid MN \qquad \text{(Neutral)}$$

Let $\nu(M)$ be the number which bounds the length of every normalisation sequence beginning from $M$.

**Lemma 5** (Properties of Reducibility)**.** Given a type $A$, $\texttt{RED}_A$ enjoys the following properties:

(PR0) $\texttt{Test}_{x:A}$ is not empty.

(PR1) If $M \in \texttt{RED}_A$ then $M$ is SN.

(PR2) If $M \in \texttt{RED}_A$ and $M \to N$ then $N \in \texttt{RED}_A$.

(PR3) If $M$ is Neutral and $\forall N.M \to N$ implies $N \in \texttt{RED}_A$, then $M \in \texttt{RED}_A$.

*Proof.* By induction on $A$.

- Case $A = \mathbb{R}$:
  - **(PR0)**: In order to prove that $\texttt{Test}_{x:\mathbb{R}}$ is not empty we build a $\Delta, x : \mathbb{R} \vdash N : \mathbb{R}$ s.t. $\forall M \in \texttt{RED}_{\mathbb{R}}.N\{^M/_x\} \in \texttt{RED}_{\mathbb{R}}$.
    We take $N \stackrel{\text{def}}{=} x$ and we have to show that $\forall M \in \texttt{RED}_{\mathbb{R}}.x\{^M/_x\} \in \texttt{RED}_{\mathbb{R}}$. By substitution $x\{^M/_x\} = M$, so we can conclude as $M \in \texttt{RED}_{\mathbb{R}}$.
  - **(PR1)**: Direct by definition.
  - **(PR2)**: By hypothesis $M \in \texttt{RED}_{\mathbb{R}}$, so by definition $M$ is SN. Suppose that $M \to N$, we have to show that $N \in \texttt{RED}_{\mathbb{R}}$. More precisely, by definition of $\texttt{RED}_{\mathbb{R}}$ we have to prove that $N$ is SN. We know that $M$ is SN, so every term to which $M$ reduces to is SN, in particular $N$ is SN and we can conclude.
  - **(PR3)**: By hypothesis $\forall N.M \to N$ implies $N \in \texttt{RED}_{\mathbb{R}}$, so by definition of $\texttt{RED}_{\mathbb{R}}$ we have that $N$ is SN. More precisely, we have that $\nu(M) = \nu(N) + 1$ and since $N$ is SN we have that $\nu(N)$ is finite and so $\nu(M)$ is finite and $M$ is SN.
- Case $A = B \multimap C$:
  - **(PR0)**: In order to prove that $\texttt{Test}_{x:B\multimap C}$ is not empty we build $\Delta, x : B \multimap C \vdash N : \mathbb{R}$ such that
    $\forall M \in \texttt{RED}_{\mathbb{R}}.N\{^M/_x\} \in \texttt{RED}_{\mathbb{R}}$.
    By IH on $C$ (PR0) we have that $\texttt{Test}_{y:C}$ is not empty, so let us take $N_1 \in \texttt{Test}_{y:C}$ typed as $\Delta, y : C \vdash N_1 : \mathbb{R}$ and s.t. $\forall N_2 \in \texttt{RED}_{\mathbb{R}}.N\{^{N_2}/_y\} \in \texttt{RED}_{\mathbb{R}}$.

Let us consider a variable $z$ of type $B$, observe that $z$ is Neutral so by IH on $B$ (PR3) we have $z \in \mathtt{RED}_B$.
We construct $N \stackrel{\text{def}}{=} N_1\{^{xz}/_y\}$ and we proceed showing that $N \in \mathtt{Test}_{x:B\multimap C}$. Let us consider a generic term $Q \in \mathtt{RED}_{B\multimap C}$, by construction and by the fact that by definition of $\mathtt{RED}_{B\multimap C}$ then $Qz \in \mathtt{RED}_C$ we have $N\{^Q/_x\} = N_1\{^{Qz}/_y\}$. Since $Q$ is a generic term in $\mathtt{RED}_{B\multimap C}$ then $N$ is an element of $\mathtt{Test}_{x:B\multimap C}$, so we can conclude that $\mathtt{Test}_{x:B\multimap C}$ is not empty.

- **(PR1)**: Let us consider a variable $x$ of type $B$, observe that $x$ is Neutral so by IH on $B$ (PR3) we have $x \in \mathtt{RED}_B$.
  By hypothesis $M \in \mathtt{RED}_{B\multimap C}$ and by definition of $\mathtt{RED}_{B\multimap C}$ we have that $Mx \in \mathtt{RED}_C$. By IH on $C$ (PR1) we have that $Mx$ is SN and so all the subterms of $Mx$ are SN, in particular $M$ is SN and we can conclude.
- **(PR2)**: Let us consider a generic $P \in \mathtt{RED}_B$. By hypothesis we have that $M \in \mathtt{RED}_{B\multimap C}$ and by definition of $\mathtt{RED}_{B\multimap C}$ we have that $MP \in \mathtt{RED}_C$. Moreover, $MP \to NP$ as by hypothesis $M \to N$. By IH on $C$ (PR2) we have that $NP \in \mathtt{RED}_C$. Since $P$ is a generic term in $\mathtt{RED}_B$ then by definition of $\mathtt{RED}_{B\multimap C}$ we have that $N \in \mathtt{RED}_{B\multimap C}$ and we can conclude.
- **(PR3)**: Let us consider a generic $P \in \mathtt{RED}_B$ and we want to show that $MP \in \mathtt{RED}_C$. We proceed by induction on $\nu(P)$ to show that every term to which $MP$ reduces in one step is in $\mathtt{RED}_C$:
  * If $MP \to NP$ with $M \to N$. By hypothesis we know that $N \in \mathtt{RED}_{B\multimap C}$ so by definition $NP \in \mathtt{RED}_C$.
  * If $MP \to MP'$ with $P \to P'$. We have that $\nu(P') < \nu(P)$ so we can apply the IH on $\nu(P')$ and obtain that $MP' \in \mathtt{RED}_C$.

  By IH on $C$ (PR3) we have that $MP \in \mathtt{RED}_C$. Since $P$ is a generic term in $\mathtt{RED}_B$ we can conclude that $M \in \mathtt{RED}_{B\multimap C}$.

- Case $A = B \otimes C$:
  - **(PR0)**: In order to prove that $\mathtt{Test}_{x:B\otimes C}$ is not empty we build $\Delta, x : B\otimes C \vdash N : \mathbb{R}$ such that
    $\forall V_1 \in \mathtt{RED}_B.\forall V_2 \in \mathtt{RED}_C.\mathtt{let}\ (x_1, x_2) = (V_1, V_2)\ \mathtt{in}\ N \in \mathtt{RED}_{\mathbb{R}}$.
    By IH on $B$ (PR0) we have that $\mathtt{Test}_{x_1:B}$ is not empty, so let us take $N_1 \in \mathtt{Test}_{x_1:B}$ typed as $\Delta_1, x_1 : B \vdash N_1 : \mathbb{R}$ such that $\forall N_1' \in \mathtt{RED}_{\mathbb{R}}.N_1\{^{N_1'}/_{x_1}\} \in \mathtt{RED}_{\mathbb{R}}$.
    By IH on $C$ (PR0) we have that $\mathtt{Test}_{x_2:C}$ is not empty, so let us take $N_2 \in \mathtt{Test}_{x_2:C}$ typed as $\Delta_2, x_2 : C \vdash N_2 : \mathbb{R}$ such that $\forall N_2' \in \mathtt{RED}_{\mathbb{R}}.N_2\{^{N_2'}/_{x_2}\} \in \mathtt{RED}_{\mathbb{R}}$.
    We construct $N \stackrel{\text{def}}{=} N_1 * N_2$ and we have to prove that $N \in \mathtt{Test}_{x:B\otimes C}$. Let us consider a generic $W_1 \in \mathtt{RED}_B$ and a generic $W_2 \in \mathtt{RED}_C$. By construction we have that $\mathtt{let}\ (x_1, x_2) = (W_1, W_2)\ \mathtt{in}\ N = \mathtt{let}\ (x_1, x_2) = (W_1, W_2)\ \mathtt{in}\ N_1 * N_2$. We have to prove that $\mathtt{let}\ (x_1, x_2) = (W_1, W_2)\ \mathtt{in}\ N_1 * N_2 \in \mathtt{RED}_{\mathbb{R}}$, so by definition we have to prove that it is SN. We proceed by induction on $\mu(W_1) + \mu(W_2) + \mu(N_1) + \mu(N_2)$ to show that every term to which $\mathtt{let}\ (x_1, x_2) = (W_1, W_2)\ \mathtt{in}\ N_1 * N_2$ reduces in one step is SN:
    * If $\mathtt{let}\ (x_1, x_2) = (W_1, W_2)\ \mathtt{in}\ N_1 * N_2 \to (N_1 * N_2)\{^{(W_1,W_2)}/_{(x_1,x_2)}\}$. By substitution we have $(N_1 * N_2)\{^{(W_1,W_2)}/_{(x_1,x_2)}\} = (N_1 * N_2)\{^{W_1}/_{x_1}, ^{W_2}/_{x_2}\} = (N_1\{^{W_1}/_{x_1}\}) * (N_2\{^{W_2}/_{x_2}\})$.
      We have chosen $N_1 \in \mathtt{Test}_{x_1:B}$ so we have that $N_1\{^{W_1}/_{x_1}\} \in \mathtt{RED}_{\mathbb{R}}$, so by definition of $\mathtt{RED}_{\mathbb{R}}$ it is also SN. We have chosen $N_2 \in \mathtt{Test}_{x_2:C}$ so we have that

$N_2\{^{W_2}/_{x_2}\} \in \mathtt{RED}_{\mathbb{R}}$, so by definition of $\mathtt{RED}_{\mathbb{R}}$ it is also SN. All the subterms of $(N_1\{^{W_1}/_{x_1}\})*(N_2\{^{W_2}/_{x_2}\})$ are SN so it is SN and we can conclude.

  * If $\mathtt{let}\ (x_1, x_2) = (W_1, W_2)\ \mathtt{in}\ N_1 * N_2 \to \mathtt{let}\ (x_1, x_2) = (W_1', W_2)\ \mathtt{in}\ N_1 * N_2$ with $W_1 \to W_1'$. We have that $\nu(W_1') < \nu(W_1)$ so we can apply the IH and conclude that $\mathtt{let}\ (x_1, x_2) = (W_1', W_2)\ \mathtt{in}\ N_1 * N_2$ is SN.
  * All the other cases where $W_2 \to W_2'$ and $N_i \to N_i'$ for $2 \leq i \leq 1$ are similar to the previous case.

- **(PR1)**: Let us consider a generic $N \in \mathtt{Test}_{(x_1,x_2):B\otimes C}$ (by (PR0) this set is not empty). By hypothesis $M \in \mathtt{RED}_{B\otimes C}$ and by definition of $\mathtt{RED}_{B\otimes C}$ we have that $\mathtt{let}\ (x_1, x_2) = M\ \mathtt{in}\ N \in \mathtt{RED}_{\mathbb{R}}$. By definition of $\mathtt{RED}_{\mathbb{R}}$ we have that $\mathtt{let}\ (x_1, x_2) = M\ \mathtt{in}\ N$ is SN, so all the subterms of it are SN. In particular, $M$ is SN and we can conclude.
- **(PR2)**: Let us consider a generic $P \in \mathtt{Test}_{(x_1,x_2):B\otimes C}$ (by (PR0) this set is not empty). By hypothesis $M \in \mathtt{RED}_{B\otimes C}$ and by definition of $\mathtt{RED}_{B\otimes C}$ we have that $\mathtt{let}\ (x_1, x_2) = M\ \mathtt{in}\ P \in \mathtt{RED}_{\mathbb{R}}$. Moreover, by hypothesis $M \to N$ so we have that $\mathtt{let}\ (x_1, x_2) = M\ \mathtt{in}\ P \to \mathtt{let}\ (x_1, x_2) = N\ \mathtt{in}\ P$ and by IH on $\mathbb{R}$ (PR2) we have that $\mathtt{let}\ (x_1, x_2) = N\ \mathtt{in}\ P \in \mathtt{RED}_{\mathbb{R}}$. Since $P$ is a generic term in $\mathtt{Test}_{(x_1,x_2):B\otimes C}$ and $\mathtt{let}\ (x_1, x_2) = N\ \mathtt{in}\ P \in \mathtt{RED}_{\mathbb{R}}$ then we can conclude that $N \in \mathtt{RED}_{B\otimes C}$.
- **(PR3)**: Let us consider a generic $P \in \mathtt{Test}_{(x_1,x_2):B\otimes C}$ (by (PR0) this set is not empty), we want to show that $\mathtt{let}\ (x_1, x_2) = M\ \mathtt{in}\ P \in \mathtt{RED}_{\mathbb{R}}$. By definition of $\mathtt{RED}_{\mathbb{R}}$ we have to show that $\mathtt{let}\ (x_1, x_2) = M\ \mathtt{in}\ P$ is SN. We proceed by induction on $\nu(P)$ to show that every term to which $\mathtt{let}\ (x_1, x_2) = M\ \mathtt{in}\ P$ reduces in one step is SN:
  * If $\mathtt{let}\ (x_1, x_2) = M\ \mathtt{in}\ P \to \mathtt{let}\ (x_1, x_2) = N\ \mathtt{in}\ P$ with $M \to N$. By hypothesis we have that $N \in \mathtt{RED}_{B\otimes C}$ so by definition of $\mathtt{RED}_{B\otimes C}$ and since $P$ is a generic term in $\mathtt{Test}_{(x_1,x_2):B\otimes C}$ then we have that $\mathtt{let}\ (x_1, x_2) = N\ \mathtt{in}\ P \in \mathtt{RED}_{\mathbb{R}}$. By definition of $\mathtt{RED}_{\mathbb{R}}$ we have that $\mathtt{let}\ (x_1, x_2) = N\ \mathtt{in}\ P$ is SN and we can conclude.
  * If $\mathtt{let}\ (x_1, x_2) = M\ \mathtt{in}\ P \to \mathtt{let}\ (x_1, x_2) = N\ \mathtt{in}\ P'$ with $P \to P'$. We have that $\nu(P') < \nu(P)$ so we can apply the IH and conclude that $\mathtt{let}\ (x_1, x_2) = N\ \mathtt{in}\ P'$ is SN.

  All the terms to which $\mathtt{let}\ (x_1, x_2) = M\ \mathtt{in}\ P$ reduces in one step are SN and so it is SN. Moreover, by definition of $\mathtt{RED}_{\mathbb{R}}$ we have $\mathtt{let}\ (x_1, x_2) = M\ \mathtt{in}\ P \in \mathtt{RED}_{\mathbb{R}}$. Since $P$ is a generic term in $\mathtt{Test}_{(x_1,x_2):B\otimes C}$ and $\mathtt{let}\ (x_1, x_2) = M\ \mathtt{in}\ P \in \mathtt{RED}_{\mathbb{R}}$ then we can conclude that $M \in \mathtt{RED}_{B\otimes C}$.

- Case $A = B\&C$:

  - **(PR0)**: In order to prove that $\mathtt{Test}_{x:B\&C}$ is not empty we build $\Delta, x : B\&C \vdash N : \mathbb{R}$ such that
    $\forall V_1 \in \mathtt{RED}_B.\forall V_2 \in \mathtt{RED}_C.\mathtt{let}\ \langle x_1, x_2\rangle = \langle V_1, V_2\rangle\ \mathtt{in}\ N \in \mathtt{RED}_{\mathbb{R}}$.
    By IH on $B$ (PR0) we have that $\mathtt{Test}_{x_1:B}$ is not empty, so let us take $N_1 \in \mathtt{Test}_{x_1:B}$ typed as $\Delta, x_1 : B \vdash N_1 : \mathbb{R}$ such that $\forall N_1' \in \mathtt{RED}_{\mathbb{R}}.N_1\{^{N_1'}/_{x_1}\} \in \mathtt{RED}_{\mathbb{R}}$.
    By IH on $C$ (PR0) we have that $\mathtt{Test}_{x_2:C}$ is not empty, so let us take $N_2 \in \mathtt{Test}_{x_2:C}$ typed as $\Delta, x_2 : C \vdash N_2 : \mathbb{R}$ such that $\forall N_2' \in \mathtt{RED}_{\mathbb{R}}.N_2\{^{N_2'}/_{x_2}\} \in \mathtt{RED}_{\mathbb{R}}$.
    We construct $N \stackrel{\text{def}}{=} \dot{+}\langle N_1, N_2\rangle$ and we have to prove that $N \in \mathtt{Test}_{x:B\&C}$. Let us consider a generic $W_1 \in \mathtt{RED}_B$ and a generic $W_2 \in \mathtt{RED}_C$. By construction we have that $\mathtt{let}\ \langle x_1, x_2\rangle = \langle W_1, W_2\rangle\ \mathtt{in}\ N = \mathtt{let}\ \langle x_1, x_2\rangle = \langle W_1, W_2\rangle\ \mathtt{in}\ \dot{+}\langle N_1, N_2\rangle$. We have to prove that $\mathtt{let}\ \langle x_1, x_2\rangle = \langle W_1, W_2\rangle\ \mathtt{in}\ \dot{+}\langle N_1, N_2\rangle \in \mathtt{RED}_{\mathbb{R}}$, so by definition we have to prove that it is SN. We proceed by induction on $\mu(W_1)+\mu(W_2)+\mu(N_1)+\mu(N_2)$

to show that every term to which $\texttt{let}\ \langle x_1, x_2\rangle = \langle W_1, W_2\rangle\ \texttt{in}\ \dot{+}\langle N_1, N_2\rangle$ reduces in one step is SN:

* If $\texttt{let}\ \langle x_1, x_2\rangle = \langle W_1, W_2\rangle\ \texttt{in}\ \dot{+}\langle N_1, N_2\rangle \to (\dot{+}\langle N_1, N_2\rangle)\{^{\langle W_1, W_2\rangle}/_{\langle x_1, x_2\rangle}\}$. By substitution we have $(\dot{+}\langle N_1, N_2\rangle)\{^{\langle W_1, W_2\rangle}/_{\langle x_1, x_2\rangle}\} = (\dot{+}\langle N_1, N_2\rangle)\{^{W_1}/_{x_1}, ^{W_2}/_{x_2}\} = \dot{+}\langle N_1\{^{W_1}/_{x_1}, ^{W_2}/_{x_2}\}, N_2\{^{W_1}/_{x_1}, ^{W_2}/_{x_2}\}\rangle$.
  We have chosen $N_1 \in \texttt{Test}_{x_1:B}$ so we have that $N_1\{^{W_1}/_{x_1}\} \in \texttt{RED}_{\mathbb{R}}$, so by definition of $\texttt{RED}_{\mathbb{R}}$ it is also SN. We have chosen $N_2 \in \texttt{Test}_{x_2:C}$ so we have that $N_2\{^{W_2}/_{x_2}\} \in \texttt{RED}_{\mathbb{R}}$, so by definition of $\texttt{RED}_{\mathbb{R}}$ it is also SN. All the subterms of $\dot{+}\langle N_1\{^{W_1}/_{x_1}, ^{W_2}/_{x_2}\}, N_2\{^{W_1}/_{x_1}, ^{W_2}/_{x_2}\}\rangle$ are SN so it is SN and we can conclude.
* If $\texttt{let}\ \langle x_1, x_2\rangle = \langle W_1, W_2\rangle\ \texttt{in}\ \dot{+}\langle N_1, N_2\rangle \to \texttt{let}\ \langle x_1, x_2\rangle = (W_1', W_2)\ \texttt{in}\ \dot{+}\langle N_1, N_2\rangle$ with $W_1 \to W_1'$. We have that $\nu(W_1') < \nu(W_1)$ so we can apply the IH and conclude that $\texttt{let}\ \langle x_1, x_2\rangle = (W_1', W_2)\ \texttt{in}\ \dot{+}\langle N_1, N_2\rangle$ is SN.
* All the other cases where $W_2 \to W_2'$ and $N_i \to N_i'$ for $2 \leq i \leq 1$ are similar to the previous case.

- **(PR1)**: Let us consider a generic $N \in \texttt{Test}_{\langle x_1,x_2\rangle:B\&C}$ (by (PR0) this set is not empty). By hypothesis $M \in \texttt{RED}_{B\&C}$ and by definition of $\texttt{RED}_{B\&C}$ we have that $\texttt{let}\ \langle x_1, x_2\rangle = M\ \texttt{in}\ N \in \texttt{RED}_{\mathbb{R}}$. By definition of $\texttt{RED}_{\mathbb{R}}$ we have that $\texttt{let}\ \langle x_1, x_2\rangle = M\ \texttt{in}\ N$ is SN, so all the subterms of it are SN. In particular, $M$ is SN and we can conclude.
- **(PR2)**: Let us consider a generic $P \in \texttt{Test}_{\langle x_1,x_2\rangle:B\&C}$ (by (PR0) this set is not empty). By hypothesis $M \in \texttt{RED}_{B\&C}$ and by definition of $\texttt{RED}_{B\&C}$ we have that $\texttt{let}\ \langle x_1, x_2\rangle = M\ \texttt{in}\ P \in \texttt{RED}_{\mathbb{R}}$. Moreover, by hypothesis $M \to N$ so we have that $\texttt{let}\ \langle x_1, x_2\rangle = M\ \texttt{in}\ P \to \texttt{let}\ \langle x_1, x_2\rangle = N\ \texttt{in}\ P$ and by IH on $\mathbb{R}$ (PR2) we have that $\texttt{let}\ \langle x_1, x_2\rangle = N\ \texttt{in}\ P \in \texttt{RED}_{\mathbb{R}}$. Since $P$ is a generic term in $\texttt{Test}_{\langle x_1,x_2\rangle:B\&C}$ and $\texttt{let}\ \langle x_1, x_2\rangle = N\ \texttt{in}\ P \in \texttt{RED}_{\mathbb{R}}$ then we can conclude that $N \in \texttt{RED}_{B\&C}$.
- **(PR3)**: Let us consider a generic $P \in \texttt{Test}_{\langle x_1,x_2\rangle:B\&C}$ (by (PR0) this set is not empty), we want to show that $\texttt{let}\ \langle x_1, x_2\rangle = M\ \texttt{in}\ P \in \texttt{RED}_{\mathbb{R}}$. By definition of $\texttt{RED}_{\mathbb{R}}$ we have to show that $\texttt{let}\ \langle x_1, x_2\rangle = M\ \texttt{in}\ P$ is SN. We proceed by induction on $\nu(P)$ to show that every term to which $\texttt{let}\ \langle x_1, x_2\rangle = M\ \texttt{in}\ P$ reduces in one step is SN:

  * If $\texttt{let}\ \langle x_1, x_2\rangle = M\ \texttt{in}\ P \to \texttt{let}\ \langle x_1, x_2\rangle = N\ \texttt{in}\ P$ with $M \to N$. By hypothesis we have that $N \in \texttt{RED}_{B\&C}$ so by definition of $\texttt{RED}_{B\&C}$ and since $P$ is a generic term in $\texttt{Test}_{\langle x_1,x_2\rangle:B\&C}$ then we have that $\texttt{let}\ \langle x_1, x_2\rangle = N\ \texttt{in}\ P \in \texttt{RED}_{\mathbb{R}}$. By definition of $\texttt{RED}_{\mathbb{R}}$ we have that $\texttt{let}\ \langle x_1, x_2\rangle = N\ \texttt{in}\ P$ is SN and we can conclude.
  * If $\texttt{let}\ \langle x_1, x_2\rangle = M\ \texttt{in}\ P \to \texttt{let}\ \langle x_1, x_2\rangle = N\ \texttt{in}\ P'$ with $P \to P'$. We have that $\nu(P') < \nu(P)$ so we can apply the IH and conclude that $\texttt{let}\ \langle x_1, x_2\rangle = N\ \texttt{in}\ P'$ is SN.

  All the terms to which $\texttt{let}\ \langle x_1, x_2\rangle = M\ \texttt{in}\ P$ reduces in one step are SN and so it is SN. Moreover, by definition of $\texttt{RED}_{\mathbb{R}}$ we have $\texttt{let}\ \langle x_1, x_2\rangle = M\ \texttt{in}\ P \in \texttt{RED}_{\mathbb{R}}$. Since $P$ is a generic term in $\texttt{Test}_{\langle x_1,x_2\rangle:B\&C}$ and $\texttt{let}\ \langle x_1, x_2\rangle = M\ \texttt{in}\ P \in \texttt{RED}_{\mathbb{R}}$ then we can conclude that $M \in \texttt{RED}_{B\&C}$.

• Case $A = !B$:

- **(PR0)**: In order to prove that $\texttt{Test}_{!y:!B}$ is not empty we build $\Delta, !y :!B \vdash N : \mathbb{R}$ such that $\forall M \in \texttt{RED}_B.\texttt{let}\ !y =!M\ \texttt{in}\ N \in \texttt{RED}_{\mathbb{R}}$.
  By IH on $B$ (PR0) we have that $\texttt{Test}_{x:B}$ is not empty, so let us take $N_1 \in \texttt{Test}_{x:B}$ typed as $\Delta, x : B \vdash N_1 : \mathbb{R}$ such that $\forall N_1' \in \texttt{RED}_{\mathbb{R}}.N_1\{^{N_1'}/_x\} \in \texttt{RED}_{\mathbb{R}}$.
  We construct $N \stackrel{\text{def}}{=} \texttt{let}\ x = y\ \texttt{in}\ N_1$ and we have to prove that $N \in \texttt{Test}_{!y:!B}$. Let us consider a generic $Q \in \mathtt{RED}_B$. By substitution and by construction we have that $N\{^{!Q}/_{!y}\} = N\{^{Q}/_{y}\} = (\mathtt{let}\ x = y\ \mathtt{in}\ N_1)\{^{Q}/_{y}\} = \mathtt{let}\ x = Q\ \mathtt{in}\ N_1$. We have to prove that $\mathtt{let}\ x = Q\ \mathtt{in}\ N_1 \in \mathtt{RED}_{\mathbb{R}}$, so by definition we have to prove that it is SN. We proceed by induction on $\mu(Q) + \mu(N_1)$ to show that every term to which $\mathtt{let}\ x = Q\ \mathtt{in}\ N_1$ reduces in one step is SN:

  * $\mathtt{let}\ x = Q\ \mathtt{in}\ N_1 \to \mathtt{let}\ x = Q'\ \mathtt{in}\ N_1$ with $Q \to Q'$. We have that $\nu(Q') < \nu(Q)$ so we can apply the IH and conclude that $\mathtt{let}\ x = Q'\ \mathtt{in}\ N_1$ is SN.
  * $\mathtt{let}\ x = Q\ \mathtt{in}\ N_1 \to \mathtt{let}\ x = Q\ \mathtt{in}\ N_1'$ with $N_1 \to N_1'$. We have that $\nu(N_1') < \nu(N_1)$ so we can apply the IH and conclude that $\mathtt{let}\ x = Q\ \mathtt{in}\ N_1'$ is SN.
  * $\mathtt{let}\ x = Q\ \mathtt{in}\ N_1 \to N_1\{^{Q}/_{x}\}$. We have chosen $N_1 \in \mathtt{Test}_{x:B}$ so it is such that $\forall N_1' \in \mathtt{RED}_{\mathbb{R}}.N_1\{^{N_1'}/_{x}\} \in \mathtt{RED}_{\mathbb{R}}$. Since $Q$ is a generic term in $\mathtt{RED}_B$ then we have that $N_1\{^{Q}/_{x}\} \in \mathtt{RED}_{\mathbb{R}}$. By definition of $\mathtt{RED}_{\mathbb{R}}$ we have that $N_1\{^{Q}/_{x}\}$ is SN and we can conclude.

- **(PR1)**: Let us consider a generic $N \in \mathtt{Test}_{!x:!B}$ (by (PR0) this set is not empty). By hypothesis $M \in \mathtt{RED}_{!B}$ and by definition of $\mathtt{RED}_{!B}$ we have that $\mathtt{let}\ !x = M\ \mathtt{in}\ N \in \mathtt{RED}_{\mathbb{R}}$. By definition of $\mathtt{RED}_{\mathbb{R}}$ we have that $\mathtt{let}\ !x = M\ \mathtt{in}\ N$ is SN, so all the subterms of it are SN. In particular, $M$ is SN and we can conclude.
- **(PR2)**: Let us consider a generic $P \in \mathtt{Test}_{!x:!B}$ (by (PR0) this set is not empty). By hypothesis $M \in \mathtt{RED}_{!B}$ and by definition of $\mathtt{RED}_{!B}$ we have that $\mathtt{let}\ !x = M\ \mathtt{in}\ P \in \mathtt{RED}_{\mathbb{R}}$. Moreover, by hypothesis $M \to N$ so we have that $\mathtt{let}\ !x = M\ \mathtt{in}\ P \to \mathtt{let}\ !x = N\ \mathtt{in}\ P$ and by IH on $\mathbb{R}$ (PR2) we have that $\mathtt{let}\ !x = N\ \mathtt{in}\ P \in \mathtt{RED}_{\mathbb{R}}$. Since $P$ is a generic term in $\mathtt{Test}_{!x:!B}$ and $\mathtt{let}\ !x = N\ \mathtt{in}\ P \in \mathtt{RED}_{\mathbb{R}}$ then we can conclude that $N \in \mathtt{RED}_{!B}$.
- **(PR3)**: Let us consider a generic $P \in \mathtt{Test}_{!x:!B}$ (by (PR0) this set is not empty), we want to show that $\mathtt{let}\ !x = M\ \mathtt{in}\ P \in \mathtt{RED}_{\mathbb{R}}$. By definition of $\mathtt{RED}_{\mathbb{R}}$ we have to show that $\mathtt{let}\ !x = M\ \mathtt{in}\ P$ is SN. We proceed by induction on $\nu(P)$ to show that every term to which $\mathtt{let}\ !x = M\ \mathtt{in}\ P$ reduces in one step is SN:

  * If $\mathtt{let}\ !x = M\ \mathtt{in}\ P \to \mathtt{let}\ !x = N\ \mathtt{in}\ P$ with $M \to N$. By hypothesis we have that $N \in \mathtt{RED}_{!B}$ so by definition of $\mathtt{RED}_{!B}$ and since $P$ is a generic term in $\mathtt{Test}_{!x:!B}$ then we have that $\mathtt{let}\ !x = N\ \mathtt{in}\ P \in \mathtt{RED}_{\mathbb{R}}$. By definition of $\mathtt{RED}_{\mathbb{R}}$ we have that $\mathtt{let}\ !x = N\ \mathtt{in}\ P$ is SN and we can conclude.
  * If $\mathtt{let}\ !x = M\ \mathtt{in}\ P \to \mathtt{let}\ !x = M\ \mathtt{in}\ P'$ with $P \to P'$. We have that $\nu(P') < \nu(P)$ so we can apply the IH and conclude that $\mathtt{let}\ !x = M\ \mathtt{in}\ P'$ is SN.

  All the terms to which $\mathtt{let}\ !x = M\ \mathtt{in}\ P$ reduces in one step are SN and so it is SN. Moreover, by definition of $\mathtt{RED}_{\mathbb{R}}$ we have $\mathtt{let}\ !x = M\ \mathtt{in}\ P \in \mathtt{RED}_{\mathbb{R}}$. Since $P$ is a generic term in $\mathtt{Test}_{!x:!B}$ and $\mathtt{let}\ !x = M\ \mathtt{in}\ P \in \mathtt{RED}_{\mathbb{R}}$ then we can conclude that $M \in \mathtt{RED}_{!B}$.

• All the other cases are simple.

□

We proceed by using these properties to show that all terms in λLL are reducible (by proving Lemma 10) hence strong normalization follows. More precisely, in order to do this we need some auxiliary lemmas which we state below. The proofs follow a standard approach, using Definition 1 and the properties of reducibility listed in Lemma 5. For the sake of clarity of the manuscript we omit the detailed proofs which can however be found at the end of the chapter.

**Lemma 6.** Let $\Delta \vdash M : A$, if $\forall N \in \mathtt{Test}_{x:A}.N\{^{M}/_{x}\} \in \mathtt{RED}_{\mathbb{R}}$, then $M \in \mathtt{RED}_A$.

*Sketch Proof.* By induction on $A$. See page 59 for more details. □

**Lemma 7.** If $(M_1, M_2) \in \mathtt{RED}_{A_1 \otimes A_2}$, then $M_i \in \mathtt{RED}_{A_i}$ for $1 \leq i \leq 2$.

*Sketch Proof.* The detailed proof can be found at page 60. □

**Lemma 8.** If $!M \in \mathtt{RED}_{!A}$ then $M \in \mathtt{RED}_A$.

*Sketch Proof.* The detailed proof can be found at page 61. □

**Lemma 9.** If $\langle M_1, M_2\rangle \in \mathtt{RED}_{A_1 \& A_2}$, then $M_i \in \mathtt{RED}_{A_i}$ for $1 \leq i \leq 2$.

*Sketch Proof.* The detailed proof can be found at page 62. □

**Lemma 10.** Given $p_1 : A_1, \ldots, p_n : A_n \vdash M : B$, $\forall V_i \in \mathtt{RED}_{A_i}$. $\overline{M} = M\{V_1/p_1, \ldots, V_n/p_n\} \in \mathtt{RED}_B$, where $V_i$ value for the pattern $p_i$.

*Proof.* Let $\Pi$ be the derivation for the judgement $p_1 : A_1, \ldots, p_n : A_n \vdash M : B$, we proceed by induction on $s(\Pi)$. We split depending on the last derivation rule in $\Pi$.

- The last rule $r$ of $\Pi$ is a rule acting on a pattern in the set $\{p_1 : A_1, \ldots, p_n : A_n\}$. Let us suppose that $r$ acts on the pattern $p_1$, we then split in the following subcases:
  - If $r$ is of type $!_w$ and, since it acts on the pattern $p_1$, we have that $p_1 = !x : !A'$. By definition of value for a pattern we have that $V_1 = !N$ for some term $N$.
    The immediate subderivation $\Pi$ is $\Pi'$ for the judgement $p_2 : A_2, \ldots, p_n : A_n \vdash M : B$, so we have that $x \notin FV(M)$.
    We have to prove that $\overline{M} = M\{{}^{!N}\!/_{!x}, {}^{V_2}\!/_{p_2}, \ldots, {}^{V_n}\!/_{p_n}\} \in \mathtt{RED}_B$. Observe that $M\{{}^{!N}\!/_{!x}, {}^{V_2}\!/_{p_2}, \ldots, {}^{V_n}\!/_{p_n}\} = M\{{}^{V_2}\!/_{p_2}, \ldots, {}^{V_n}\!/_{p_n}\}$ since $x \notin FV(M)$. By IH on $\Pi'$ we have that $\forall V_i \in \mathtt{RED}_{A_i}$ for $2 \leq i \leq n$. $M\{{}^{V_2}\!/_{p_2}, \ldots, {}^{V_n}\!/_{p_n}\} \in \mathtt{RED}_B$ and so we can conclude.
  - If $r$ is of type $\&_{ei}$ and, since it acts on the pattern $p_1$, we have that $p_1 = \langle p_{1,1}, p_{1,2}\rangle : A_{1,1}\&A_{1,2}$. By definition of value for a pattern we have that $V_1 = \langle V_{1,1}, V_{1,2}\rangle$ where $V_{1,i}$ is a value for the pattern $p_{1,i}$ for $i \in \{1,2\}$.
    Let us suppose $i = 1$ (the other case being similar). The immediate subderivation $\Pi$ is $\Pi'$ for the judgement $p_{1,1} : A_{1,1}, p_2 : A_2, \ldots, p_n : A_n \vdash M : B$ by $\&_{e1}$-typing rule, so we have that $FV(p_{1,2})$ are not free in $M$.
    We have to prove that $\overline{M} = M\{{}^{\langle V_{1,1}, V_{1,2}\rangle}\!/_{\langle p_{1,1}, p_{1,2}\rangle}, {}^{V_2}\!/_{p_2}, \ldots, {}^{V_n}\!/_{p_n}\} \in \mathtt{RED}_B$. By substitution we have

    $$M\{{}^{\langle V_{1,1}, V_{1,2}\rangle}\!/_{\langle p_{1,1}, p_{1,2}\rangle}, {}^{V_2}\!/_{p_2}, \ldots, {}^{V_n}\!/_{p_n}\} = M\{{}^{V_{1,1}}\!/_{p_{1,1}}, {}^{V_{1,2}}\!/_{p_{1,2}}, {}^{V_2}\!/_{p_2}, \ldots, {}^{V_n}\!/_{p_n}\}$$

    and since $FV(p_{1,2})$ are not free in $M$ we have that $\overline{M} = M\{{}^{V_{1,1}}\!/_{p_{1,1}}, {}^{V_2}\!/_{p_2}, \ldots, {}^{V_n}\!/_{p_n}\}$. By hypothesis we have that $\langle V_{1,1}, V_{1,2}\rangle \in \mathtt{RED}_{A_{1,1}\&A_{1,2}}$, we can apply Lemma 9 we have that $V_{1,1} \in \mathtt{RED}_{A_{1,1}}$. By IH on $\Pi'$ we have that

    $$\forall V_i \in \mathtt{RED}_{A_i} \text{ for } 2 \leq i \leq n. \forall V_{1,1} \in \mathtt{RED}_{A_{1,1}}.\ M\{{}^{V_{1,1}}\!/_{p_{1,1}}, {}^{V_2}\!/_{p_2}, \ldots, {}^{V_n}\!/_{p_n}\} \in \mathtt{RED}_B$$

    and so we can conclude.

– If $r$ is of type $\otimes_e$ and, since it acts on the pattern $p_1$, we have that $p_1 = (p_{1,1}, p_{1,2}) : A_{1,1} \otimes A_{1,2}$. By definition of value for a pattern we have that $V_1 = (V_{1,1}, V_{1,2})$ where $V_{1,i}$ is a value for the pattern $p_{1,i}$ for $i \in \{1, 2\}$.
The immediate subderivation $\Pi$ is $\Pi'$ for the judgement $p_{1,1} : A_{1,1}, p_{1,2} : A_{1,2}, p_2 : A_2, \ldots, p_n : A_n \vdash M : B$.
We have to prove that $\overline{M} = M\{^{(V_{1,1},V_{1,2})}/_{(p_{1,1},p_{1,2})}, ^{V_2}/_{p_2}, \ldots, ^{V_n}/_{p_n}\} \in \mathtt{RED}_B$. By substitution we have

$$M\{^{(V_{1,1},V_{1,2})}/_{(p_{1,1},p_{1,2})}, ^{V_2}/_{p_2}, \ldots, ^{V_n}/_{p_n}\} = M\{^{V_{1,1}}/_{p_{1,1}}, ^{V_{1,2}}/_{p_{1,2}}, ^{V_{1,2}}/_{p_{1,2}}, ^{V_2}/_{p_2}, \ldots, ^{V_n}/_{p_n}\}$$

By hypothesis we have that $(V_{1,1}, V_{1,2}) \in \mathtt{RED}_{A_{1,1} \otimes A_{1,2}}$, we can apply Lemma 7 we have that $V_{1,1} \in \mathtt{RED}_{A_{1,1}}$ and $V_{1,2} \in \mathtt{RED}_{A_{1,2}}$. By IH on $\Pi'$ we have that

$$\forall V_i \in \mathtt{RED}_{A_i} \text{ for } 2 \leq i \leq n. \forall V_{1,j} \in \mathtt{RED}_{A_{1,j}} \text{ for } 2 \leq j \leq 1.$$
$$M\{^{V_{1,1}}/_{p_{1,1}}, ^{V_{1,2}}/_{p_{1,2}}, ^{V_2}/_{p_2}, \ldots, ^{V_n}/_{p_n}\} \in \mathtt{RED}_B$$

and so we can conclude.

– If $r$ is of type $1_e$ and, since it acts on the pattern $p_1$, we have that $p_1 = () : \mathtt{1}$. By definition of value for a pattern we have that $V_1 = ()$.
The immediate subderivation $\Pi$ is $\Pi'$ for the judgement $p_2 : A_2, \ldots, p_n : A_n \vdash M : B$.
We have to prove that $\overline{M} = M\{^{()}/_{()}, ^{V_2}/_{p_2}, \ldots, ^{V_n}/_{p_n}\} \in \mathtt{RED}_B$. By substitution we have
$M\{^{()}/_{()}, ^{V_2}/_{p_2}, \ldots, ^{V_n}/_{p_n}\} = M\{^{V_2}/_{p_2}, \ldots, ^{V_n}/_{p_n}\}$. By IH on $\Pi'$ we have that $\forall V_i \in \mathtt{RED}_{A_i}$ for $2 \leq i \leq n$. $M\{^{V_2}/_{p_2}, \ldots, ^{V_n}/_{p_n}\} \in \mathtt{RED}_B$ and so we can conclude.

- For the other cases we can suppose that the last rule $r$ of $\Pi$ is not acting an a pattern in the set $\{p_1 : A_1, \ldots, p_n : A_n\}$. We then consider the cases in which $r$ acts on the subject $M$ and we split in the following cases:

  – If $r$ is of type $v$ then $M = x : A_1$ and $p_1 = x : A_1$. Moreover, $B = A_1$ and $FV(p_i)$ are not free in $M$ for $2 \leq i \leq n$.
  In this case we have $\overline{M} = \{^{V_1}/_x, ^{V_2}/_{p_2}, \ldots, ^{V_n}/_{p_n}\}$. Since $FV(p_i)$ are not free in $M$ for $2 \leq i \leq n$ so by substitution we have that $\overline{M} = \{^{V_1}/_x\} = V_1$. Recall that in this case $B = A_1$ and by hypothesis $V_1 \in \mathtt{RED}_{A_1}$, so $\overline{M} \in \mathtt{RED}_{A_1}$ and we can conclude.

  – If $r$ is of type $!_e$ then $M = x : A'_1$ and $p_1 : !A'_1$. Moreover, $B = A'_1$ and $FV(p_i)$ are not free in $M$ for $2 \leq i \leq n$. By definition of value for a pattern we have that $V_1 = !N$ for some term $N$.
  In this case we have $\overline{M} = M\{^{!N}/_{!x}, ^{V_2}/_{p_2}, \ldots, ^{V_n}/_{p_n}\} = x\{^{!N}/_{!x}, ^{V_2}/_{p_2}, \ldots, ^{V_n}/_{p_n}\}$. By substitution and since $FV(p_i)$ are not free in $M$ for $2 \leq i \leq n$ we have that $\overline{M} = x\{^{!N}/_{!x}\} = x\{^N/_x\} = N$.
  Recall that in this case $B = A'_1$ and $\overline{M} = N$, by hypothesis we have that $!N \in \mathtt{RED}_{!A'_i}$ we can apply Lemma 8 obtaining that $N \in \mathtt{RED}_{A'_1}$, so $\overline{M} \in \mathtt{RED}_{A'_1}$ and we can conclude.

  – If $r$ is of type $\&_i$ then $M = \langle M_1, M_2 \rangle$ and $B = B_1 \& B_2$.
  The immediate subderivations of $\Pi$ are $\Pi_1$ and $\Pi_2$ for the judgements $p_1 : A_1, \ldots, p_n : A_n \vdash M_1 : B_1$ and $p_1 : A_1, \ldots, p_n : A_n \vdash M_2 : B_2$, respectively.
  By substitution we have that

  $$\overline{M} = \langle M_1, M_2 \rangle\{^{V_1}/_{p_1}, \ldots, ^{V_n}/_{p_n}\} = \langle M_1\{^{V_1}/_{p_1}, \ldots, ^{V_n}/_{p_n}\}, M_2\{^{V_1}/_{p_1}, \ldots, ^{V_n}/_{p_n}\}\rangle$$

  We have to prove that $\overline{M} = \langle M_1\{^{V_1}/_{p_1}, \ldots, ^{V_n}/_{p_n}\}, M_2\{^{V_1}/_{p_1}, \ldots, ^{V_n}/_{p_n}\}\rangle \in \mathtt{RED}_{B_1 \& B_2}$.
  We consider a generic $N \in \mathtt{Test}_{\langle x_1, x_2 \rangle : B_1 \& B_2}$ and by definition of $\mathtt{RED}_{B_1 \& B_2}$ we have to prove that $\mathtt{let}\ \langle x_1, x_2 \rangle = \langle M_1\{^{V_1}/_{p_1}, \ldots, ^{V_n}/_{p_n}\}, M_2\{^{V_1}/_{p_1}, \ldots, ^{V_n}/_{p_n}\}\rangle\ \mathtt{in}\ N \in \mathtt{RED}_{\mathbb{R}}$.
  More precisely, by definition of $\mathtt{RED}_{\mathbb{R}}$ we have to show that it is SN. Observe that:

* By IH on $\Pi_i$ we have that $\forall V_j \in \mathtt{RED}_{A_j}$ for $1 \leq i \leq n$. $\overline{M_i} = M_i\{^{V_1}/_{p_1}, \ldots, ^{V_n}/_{p_n}\} \in \mathtt{RED}_{B_i}$, so we can apply (PR1) of Lemma 5 obtaining that $\overline{M_i}$ is SN.
* Since $N \in \mathtt{Test}_{\langle x_1,x_2\rangle:B_1\&B_2}$ by definition of $\mathtt{Test}_{\langle x_1,x_2\rangle:B_1\&B_2}$ we have that $N$ is SN.

and so we can conclude.

– If $r$ is of type $\otimes_i$ then $M = (M_1, M_2)$ and $B = B_1 \otimes B_2$. Moreover, $!\Gamma_1, !\Gamma_2, \Delta_1, \Delta_2 = p_1 : A_1, \ldots, p_n : A_n$. The immediate subderivations of $\Pi$ are $\Pi_1$ and $\Pi_2$ for the judgements $!\Gamma_1, \Delta_1 \vdash M_1 : B_1$ and $!\Gamma_2, \Delta_2 \vdash M_2 : B_2$, respectively.
Let $\overrightarrow{W}$ be the value for $!\Gamma_1, !\Gamma_2, \Delta_1, \Delta_2$ and $\overrightarrow{W_i}$ be the value for $!\Gamma_i, \Delta_i$ with $i \in \{1, 2\}$.
By substitution we have that

$$\overline{M} = (M_1, M_2)\{\overrightarrow{W}/!\Gamma_1, !\Gamma_2, \Delta_1, \Delta_2\} = (M_1\{\overrightarrow{W_1}/!\Gamma_1, \Delta_1\}, M_2\{\overrightarrow{W_2}/!\Gamma_2, \Delta_2\})$$

We have to prove that $\overline{M} = (M_1\{\overrightarrow{W_1}/!\Gamma_1, \Delta_1\}, M_2\{\overrightarrow{W_2}/!\Gamma_2, \Delta_2\}) \in \mathtt{RED}_{B_1\otimes B_2}$. We consider a generic $N \in \mathtt{Test}_{(x_1,x_2):B_1\otimes B_2}$ and by definition of $\mathtt{RED}_{B_1\otimes B_2}$ we have to prove that $\mathtt{let}\ (x_1, x_2) = (M_1\{\overrightarrow{W_1}/!\Gamma_1, \Delta_1\}, M_2\{\overrightarrow{W_2}/!\Gamma_2, \Delta_2\})\ \mathtt{in}\ N \in \mathtt{RED}_{\mathbb{R}}$. More precisely, by definition of $\mathtt{RED}_{\mathbb{R}}$ we have to show that it is SN. Observe that:

* By IH on $\Pi_i$ we have that $\overline{M_i} = M_i\{\overrightarrow{W_i}/!\Gamma_i, \Delta_i\} \in \mathtt{RED}_{B_i}$, so we can apply (PR1) of Lemma 5 obtaining that $\overline{M_i}$ is SN.
* Since $N \in \mathtt{Test}_{(x_1,x_2):B_1\otimes B_2}$ by definition of $\mathtt{Test}_{(x_1,x_2):B_1\otimes B_2}$ we have that $N$ is SN.

and so we can conclude.

– If $r$ is of type $!_i$ then $M =!M_1$ and $B = !B_1$. The immediate subderivation $\Pi$ is $\Pi'$ for the judgement $p_1 : A_1, \ldots, p_n : A_n \vdash M_1 : B_1$.
By substitution we have that

$$\overline{M} = (!M_1)\{^{V_1}/_{p_1}, \ldots, ^{V_n}/_{p_n}\} =!M_1\{^{V_1}/_{p_1}, \ldots, ^{V_n}/_{p_n}\}$$

We have to prove that $\overline{M} =!M_1\{^{V_1}/_{p_1}, \ldots, ^{V_n}/_{p_n}\} \in \mathtt{RED}_{!B_1}$. We consider a generic $N \in \mathtt{Test}_{!x:!B_1}$ and by definition of $\mathtt{RED}_{!B_1}$ we have to prove that
$\mathtt{let}\ !x =!M_1\{^{V_1}/_{p_1}, \ldots, ^{V_n}/_{p_n}\}\ \mathtt{in}\ N \in \mathtt{RED}_{\mathbb{R}}$. More precisely, by definition of $\mathtt{RED}_{\mathbb{R}}$ we have to show that it is SN. Observe that:

* By IH on $\Pi'$ we have that $\forall V_i \in \mathtt{RED}_{A_i}$. $M_1\{^{V_1}/_{p_1}, \ldots, ^{V_n}/_{p_n}\} \in \mathtt{RED}_{B_1}$, so we can apply (PR1) of Lemma 5 on $B_1$ obtaining that $M_1\{^{V_1}/_{p_1}, \ldots, ^{V_n}/_{p_n}\}$ is SN.
* Since $N \in \mathtt{Test}_{!x:!B_1}$ by definition of $\mathtt{Test}_{!x:!B_1}$ we have that $N$ is SN.

and so we can conclude.

– If $r$ is of type $\multimap_i$ then $M = \lambda q.M_1$ and $B = B_1 \multimap B_2$.
The immediate subderivation of $\Pi$ is $\Pi'$ for the judgements $q : B_1, p_1 : A_1, \ldots, p_n : A_n \vdash M_1 : B_2$.
By substitution we have that

$$\overline{M} = (\lambda q.M_1)\{^{V_1}/_{p_1}, \ldots, ^{V_n}/_{p_n}\} = \lambda q.M_1\{^{V_1}/_{p_1}, \ldots, ^{V_n}/_{p_n}\}$$

We have to prove that $\overline{M} = \lambda q.M_1\{^{V_1}/_{p_1}, \ldots, ^{V_n}/_{p_n}\} \in \mathtt{RED}_{B_1\multimap B_2}$. We consider a generic $V \in \mathtt{RED}_{!B_1}$ such that $V$ value for $q$. By definition of $\mathtt{RED}_{B_1\multimap B_2}$ we have to show that
$(\lambda q.M_1\{^{V_1}/_{p_1}, \ldots, ^{V_n}/_{p_n}\})V \in \mathtt{RED}_{B_2}$. We proceed by induction on $\nu(M_1)$ to show that every term to which $(\lambda q.M_1\{^{V_1}/_{p_1}, \ldots, ^{V_n}/_{p_n}\})V$ reduces in one step is a term in $\mathtt{RED}_{B_2}$:

* $(\lambda q.M_1\{{}^{V_1}/_{p_1},\dots,{}^{V_n}/_{p_n}\})V \to (\lambda q.M_1'\{{}^{V_1}/_{p_1},\dots,{}^{V_n}/_{p_n}\})V$ with $M_1 \to M_1'$. We have that $\nu(M_1') < \nu(M_1)$ so by IH we have that $(\lambda q.M_1'\{{}^{V_1}/_{p_1},\dots,{}^{V_n}/_{p_n}\})V \in \mathtt{RED}_{B_2}$.
  * $(\lambda q.M_1\{{}^{V_1}/_{p_1},\dots,{}^{V_n}/_{p_n}\})V \to M_1\{{}^{V}/_{q},{}^{V_1}/_{p_1},\dots,{}^{V_n}/_{p_n}\}$. By IH on $\Pi'$ we have that $\forall V_i \in \mathtt{RED}_{A_i}.\forall V \in \mathtt{RED}_{B_1}.\ M_1\{{}^{V}/_{q},{}^{V_1}/_{p_1},\dots,{}^{V_n}/_{p_n}\} \in \mathtt{RED}_{B_2}$ and so we can conclude.

  Notice that $(\lambda q.M_1\{{}^{V_1}/_{p_1},\dots,{}^{V_n}/_{p_n}\})V$ is Neutral and by the reasoning above it always reduces in one step to terms in $\mathtt{RED}_{B_2}$, so we can apply (PR3) of Lemma 5 on $B_2$ obtaining that $(\lambda q.M_1\{{}^{V_1}/_{p_1},\dots,{}^{V_n}/_{p_n}\})V \in \mathtt{RED}_{B_2}$. Since $V$ is a generic term in $\mathtt{RED}_{B_1}$, by definition of $\mathtt{RED}_{B_1\multimap B_2}$ we can conclude that $\overline{M} = \lambda q.M_1\{{}^{V_1}/_{p_1},\dots,{}^{V_n}/_{p_n}\} \in \mathtt{RED}_{B_1\multimap B_2}$.

– All the other cases are immediate.

□

**Corollary 1.** If $M$ is a term in λLL then $M$ is reducible.

*Sketch Proof.* By induction the term $M$, we apply the properties of reducibility stated in Lemma 5. Specifically, the key is to demonstrate that every term $M$ can be reduced to some normal form, using the fact that the set of reducible terms is closed under reduction steps and is non-empty. □

**Corollary 2.** If $M$ is a term in λLL then $M$ is SN.

*Sketch Proof.* By induction the term $M$, using auxiliary Lemmas 6, 7, 8, 9 in the corresponding inductive cases. □

### 3.2.4 Confluence

We show that the confluence property holds for $\to$ in λLL.

**Theorem 5** (Confluence λLL)**.** If $M' \,{}^*\!\leftarrow M \to^* M''$ then there is $N$ such that $M' \to^* N \,{}^*\!\leftarrow M''$.

More precisely, the confluence property stated in Theorem 5 is derived by combining weak confluence (Lemma 14) with an application of Newman's Lemma [23] (Lemma 11). This strategy employs well-known results from rewriting theory to establish that the reduction relation in λLL is confluent, provided that certain conditions, namely strong normalization and weak confluence, are satisfied.

**Lemma 11** (Newman's Lemma λLL)**.** If the reduction relation $\to$ in λLL enjoys strong normalization and weak confluence, then it is confluent.

In our case, we have already demonstrated strong normalization of λLL in the previous subsection (Corollary 2). Weak confluence, on the other hand, is established in Lemma 14. More precisely, in order to prove weak confluence we proceed through several intermediate lemmas. The first key result establishes that reductions are preserved under substitution:

**Lemma 12.** If $M \to M'$ then $M\{V/p\} \to M'\{V/p\}$.

The next lemma extends this property above to sequences of reductions:

**Lemma 13.** If $V \to^* V'$ then $M\{V/p\} \to^* M\{V'/p\}$.

Finally, weak confluence is formalized as follows:

**Lemma 14** (Weak Confluence $\lambda$LL)**.** If $M' \leftarrow M \rightarrow M''$ then there exists $N$ such that $M' \rightarrow^* N$ and $M'' \rightarrow^* N$.

*Sketch Proof.* By induction on $M$. The case for $M = (\lambda p.M_1)V$ is proved by using Lemma 12 when $M' = M_1\{V/p\} \leftarrow M = (\lambda p.M_1)V \rightarrow (\lambda p.M_1')V = M''$ and Lemma 13 when $M' = M_1\{V/p\} \leftarrow M = (\lambda p.M_1)V \rightarrow (\lambda p.M_1)V' = M''$. □

## 3.3 Logical Equivalence $\sim$

The $\beta$-equivalence is too narrow to compare terms of complex types: our ultimate goal is to compute numeric functions and we are interested whether two terms can be interchanged in a program of ground type without changing the numeric function computed by this latter. A typing system offers a way of extending $\beta$-equivalence by firing the extensional behaviour over ground types, using the notion of logical relation.

**Definition 2** ($\sim_A$)**.** Given a type $A$, $\sim_A$ is a binary relation between closed terms of type $A$:

- $M \sim_{\mathbb{R}} N$ or $M \sim_{\mathbf{1}} N$ iff $M =_\beta N$,
- $M \sim_{A_1 \otimes A_2} N$, iff $M \rightarrow^* (M_1, M_2)$, $N \rightarrow^* (N_1, N_2)$ and $M_i \sim_{A_i} N_i$ for $i \in \{1, 2\}$,
- $M \sim_{!A} N$ iff $M \rightarrow^* !M$, $N \rightarrow^* !N$ and $M \sim_A N$,
- $M \sim_\top N$ always,
- $M \sim_{A_1 \& A_2} N$ iff $(\lambda\langle x_1, x_2\rangle.x_i)M \sim_{A_i} (\lambda\langle x_1, x_2\rangle.x_i)N$, for every $i \in \{1, 2\}$,
- $M \sim_{A \multimap B} N$ iff for all $M' \sim_A N'$, $MM' \sim_B NN'$.

Notice that thanks to the strong normalisation and progress properties (Corollary 2 and Proposition 1), the definition of $\sim$ for the $\top$ and $\&$ connectives is analogous to ones for the multiplicative connectives. We extend $\sim$ to open terms:

**Definition 3** ($\sim_{\Gamma \vdash A}$)**.** Let $\Gamma \vdash M : A$ and $\Gamma \vdash N : A$, with $\Gamma = p_1 : A_1, \ldots, p_n : A_n$, we set: $M \sim_{\Gamma \vdash A} N$ iff $\forall i \leq n, \forall V_i \sim_{A_i} V_i', M\{V_1/p_1, \ldots, V_n/p_n\} \sim_A N\{V_1'/p_1, \ldots, V_n'/p_n\}$.

Henceforth, we may omit type annotation on $\sim_A$ or $\sim_{\Gamma \vdash A}$ whenever clear from the context or irrelevant.

Moreover, we achieve the standard properties of the logical relation $\sim$ of Definition 2.

**Lemma 15.** Given $M =_\beta N$, then $M \sim N$.

*Sketch Proof.* First one prove the statement for closed term of a type $A$ by induction on $A$. The extension to open terms follows because $=_\beta$-equivalence is contextual. □

**Proposition 2.** The relation $\sim$ is an equivalence relation extending $=_\beta$ and context closed, i.e. $M \sim N$ implies $\gamma[M] \sim \gamma[N]$ for every $\gamma[\,]$.

*Sketch Proof.* Lemma 15 implies $=_\beta \subseteq \sim$ and hence reflexivity of $\sim$. The symmetry and transitive properties of $=_\beta$ are lifted to $\sim$ by induction on the definition of $\sim$. Context closure is proven by induction on $\gamma[\,]$. □

**Lemma 16.** Given a type $E$ in the grammar $\otimes$-sequence types (resp. a type $H$ in the grammar $\&$-sequence types) we have that $\sim_E$ (resp. $\sim_H$) coincides with $\beta$-equivalence of closed terms of that type $E$ (resp. $H$).

*Sketch Proof.* By Proposition 2, we need to prove only that $\sim_E$ (or $\sim_H$) is included in $=_\beta$. This follows easily by induction on the definition of $\sim$. □

The following lemma gives examples of $\sim$-equivalent terms which are not in general $\beta$-equivalent.

**Lemma 17.** Given terms $M$ and $N$ and an exponential safe context $\gamma[\,]$ (i.e. a context generated by the grammar in Figure 3.3 without $!\gamma[\,]$) s.t. no free variable of $M$ can be captured by binders in $\gamma[\,]$, as well as a pattern $p$ such that its free variables are not free in $\gamma[\,]$, we have: $(\lambda p.\gamma[N])M \sim \gamma[(\lambda p.N)M]$.

More precisely, the lemma above gives exactly the extension to $\beta$-reduction we need to achieve our results, summarised in the diagram of Figure 6.2. In fact, we can replace $\sim$ by extending $\beta$-equivalence with an adaptation of the $\sigma$-equivalences given in [95] and achieve the same results.

**Remark 3.** Notice that $\texttt{let}\ (p,q) = M\ \texttt{in}\ N$ (resp. $\texttt{let}\ (\,) = M\ \texttt{in}\ N$) is a special case of $MN$, so we have also that $\texttt{let}\ (p,q) = M\ \texttt{in}\ \gamma[N] \sim \gamma[\texttt{let}\ (p,q) = M\ \texttt{in}\ N]$ (resp. $\texttt{let}\ () = M\ \texttt{in}\ \gamma[N] \sim \gamma[\texttt{let}\ () = M\ \texttt{in}\ N]$).

## 3.4 &-sequence Types as Vector Spaces

Let us consider a &-sequence type $H$ (see grammar &-sequence types). Notice that the closed $\beta$-normal form of $H$ are nested tuples of real numbers, called numeral sequences in Chapter 4. In fact, the $\beta$-equivalence classes of $H$ define a real vector space of dimension equal to the number of occurrences of $\mathbb{R}$ in $H$: vector addition is given by $\dot{+}$ and scalar multiplication by $\dot{*}$. Normalisation and confluence assure that one can select $\beta$-normal form's as canonical representatives of the elements of this vector space, and the rewriting rules lift the algebraic properties of addition and multiplication over $\mathbb{R}$ to $H$, e.g. $M_1\dot{+}(M_2\dot{+}M_3) =_\beta (M_1\dot{+}M_2)\dot{+}M_3$, for closed terms of type $H$.

Moreover, this vector space $H$ is associated with a canonical base $\mathcal{B}_H$ that can be defined inductively on $H$ as follows:

$$\mathcal{B}_\mathbb{R} = \underline{1}\,, \qquad \mathcal{B}_\top = \{\langle\,\rangle\}\,, \qquad \mathcal{B}_{H_1\&H_2} = \{\langle V_1, \underline{0}\rangle, \text{ s.t. } V_1 \in \mathcal{B}_{H_1}\} \cup \{\langle \underline{0}, V_2\rangle \text{ s.t. } V_2 \in \mathcal{B}_{H_2}\}\,.$$

Similarly, one defines an inner product $\mathcal{I}_H$ as a closed term of type $H \otimes H \multimap \mathbb{R}$ by induction on $H$:

$$\mathcal{I}_\mathbb{R} \stackrel{\text{def}}{=} \lambda(h,h').\ h\dot{*}h'\,, \qquad \mathcal{I}_\top \stackrel{\text{def}}{=} \lambda(h,h').\ \underline{0}\,,$$

$$\mathcal{I}_{H_1\&H_2} \stackrel{\text{def}}{=} \lambda(\langle h_1,h_2\rangle, \langle h_1',h_2'\rangle).\ \mathcal{I}_{H_1}(h_1,h_1')\dot{+}\mathcal{I}_{H_2}(h_2,h_2')\,.$$

In this way, one can recover syntactically the isomorphism between an euclidean space and its dual, by $\text{dual}_H : H \multimap (H \multimap \mathbb{R})$ and $\overline{\text{dual}}_H : (H \multimap \mathbb{R}) \multimap H$:

$$\text{dual}_H = \lambda h.\lambda h'.\mathcal{I}_H(h,h')\,, \qquad \overline{\text{dual}}_H = \lambda f.\sum_{V\in\mathcal{B}_H}(f(V))\dot{*}_H V\,.$$

Section 5.3 will compare our transpose transformation with the one obtained by using this isomorphism.

Similar constructions are possible with a generic type $A$, but cannot be defined in general by syntactical terms, in fact the dimension of a vector space associated with an exponential type $!A$ may be infinite. Quantitative semantics (e.g. [43, 75]) or resource $\lambda$-calculus (e.g. [45]) provide more suitable frameworks for describing such spaces. However, we do not explore here these systems, as the transpose transformation is restricted to &-sequence types in the AD Systems under consideration.

## 3.5 Workload

We adapt the notion of workload $\mathcal{W}$ from [93, Section 4.3], as recalled in Subsection 2.1.2. The goal is to bound the number of numeric steps required to evaluate a term $M$, i.e., the $\beta_F$, $\beta_{\dot{+}}$, and $\beta_{\dot{*}}$ reduction steps. While this is complex for full $\beta$-reduction in $\lambda$LL, we identify a reduction strategy (safe reduction) and conditions on $M$ (Definition 4) that ensure $M$ reaches its $\beta$-normal form in at most $\mathcal{W}(M)$ numeric $\beta$-steps (Proposition 4). These conditions hold for the terms used in subsequent chapters to validate Autodiff transformations, providing quantitative soundness for our Autodiff encoding.

In Chapter 6 we will discuss a refined notion of workload based on a quantitative type system, which allows us to simplify Definition 4 (see Remark 5).

The set of *strong values* is defined as (for $\mathsf{c} \in \{\underline{r}, \underline{f}, \dot{+}, \dot{*}\}$):

$$W ::= \ x \mid \mathsf{c} \mid \lambda p.M \mid (\,) \mid (W_1, W_2) \mid \langle\,\rangle \mid \langle W_1, W_2\rangle \mid !W \mid *W \qquad \text{(Strong Values)}$$

The *safe reduction* ($s$-reduction in short) is a *call by closed strong value* reduction: we just replace $\beta_\lambda$ in Figure 3.4 with

$$\beta_s : \ (\lambda p.M)W \xrightarrow{s} M\{W/p\} \quad \text{for } W \text{ closed strong value}$$

The *workload* $\mathcal{W}(A)$ of a type $A$ is the number of occurrences of $\mathbb{R}$ not under the scope of a !, $\mathcal{W}(M)$ of a term $M$ is the number of numerical functions not under a ! as well as the number of possible numerals erased during a reduction, i.e.:

$$\mathcal{W}(\underline{f}) \overset{\text{def}}{=} \mathcal{W}(\dot{+}) \overset{\text{def}}{=} \mathcal{W}(\dot{*}) \overset{\text{def}}{=} 1$$

$$\mathcal{W}(x) \overset{\text{def}}{=} \mathcal{W}(!M) \overset{\text{def}}{=} \mathcal{W}((\,)) \overset{\text{def}}{=} \mathcal{W}(\langle\,\rangle) \overset{\text{def}}{=} \mathcal{W}(\underline{r}) \overset{\text{def}}{=} 0$$

$$\mathcal{W}(\lambda p.M) \overset{\text{def}}{=} \mathcal{W}(M) + \sum_{x:A \in FV(p)\setminus FV(M)} \mathcal{W}(A)$$

$$\mathcal{W}(MN) \overset{\text{def}}{=} \mathcal{W}(\langle M, N\rangle) \overset{\text{def}}{=} \mathcal{W}((M, N)) \overset{\text{def}}{=} \mathcal{W}(M) + \mathcal{W}(N)$$

In the definition above, the case for lambda abstraction includes a second summand that accounts for erased arguments, variables that are bound in the pattern but do not appear in the body of the abstraction. These arguments are discarded during $\beta$-reduction, but since they may represent numeric values or functions, their elimination can contribute to the overall cost. This potential cost is estimated based on the types of the erased variables. Moreover, it is worth noting that $\mathcal{W}(\sigma_{\mathcal{I}}) = 0$, and $\mathcal{W}(\dot{+}_H) = \mathcal{W}(\dot{*}_H) = \mathcal{W}(H)$.

A variable of type $A$ is *ground* if $A$ has no arrow, otherwise it is *higher-order*.

**Definition 4** (Safe Term)**.** A term $M$ is *safe* if:

(i) for any subterm $!M'$ in $M$, $\mathcal{W}(M') = 0$;

(ii) for any subterm $\langle M_1, M_2\rangle$ in $M$, $FV(M_1) \cap FV(M_2)$ has only ground variables.

Condition (ii) is very technical and will be discussed in Remark 5 at the end of this section.

Now our goal is to show that any safe closed term $M$ reaches its $\beta$-normal form in at most $\mathcal{W}(M)$ numeric $\beta$-steps by using safe reduction, to do this we need to prove some auxiliary lemmas. First, we show that strong values of ground type have a null workload

**Lemma 18.** Let $W$ be a strong value of ground type. We have $\mathcal{W}(W) = 0$ and all free variables of $W$ are ground.

*Sketch Proof.* By induction on $W$. Notice in particular that the strong value $W = \underline{*}W'$ has type $\mathbb{R} \multimap \mathbb{R}$, so it is not ground. □

Notice that the hypothesis $W$ be of ground type is important: for example $\dot{+}$ is a (closed) strong value with a non null workload.

The following lemmas help to formalize the connection between the different reduction strategies involved in our system, $\beta$-reduction and $s$-reduction, and the concept of strong values, offering insight into when a term reaches its final, irreducible state under these reduction strategies.

**Lemma 19.** If $M$ is a closed normal form for safe reduction, then $M$ is a closed strong value.

*Proof.* By induction on $M$.

- Case $M = M_1 M_2$:
  By hypothesis $M$ is a closed normal form for safe reduction, so $M_1$ and $M_2$ are closed normal forms for safe reduction. By induction hypotheses they can be supposed closed strong values. We split into sub-cases, depending on $M_1$. Let us consider the case in which $M_1$ is of arrow type. By typing $M_1$ cannot be a tuple (additive or multiplicative), neither an exponential !, nor a numeral. Moreover, since $M$ is closed by hypothesis we have that $M_1$ cannot be a free variable. Therefore, the remaining cases are abstraction, numeric function ($\underline{f}$, $\dot{+}$, $\dot{*}$) or $*W$. We details these cases as follows:

  – Subcase $M_1 = \lambda p.M_1'$:
    By inductive hypothesis $M_2$ is a closed strong value, so $M = (\lambda p.M_1')M_2$ is a $\beta_s$ redex, which is contrary to the hypothesis of $M$ be a $s$-normal form.

  – Subcase $M_1 = \underline{f}$ or $M_1 = \dot{+}$:
    By typing $M_2$ is a closed term of type $!\mathbb{R}$ (for unary $\underline{f}$) or $!\mathbb{R} \otimes !\mathbb{R}$ (for binary $\underline{f}$) or $\mathbb{R} \,\&\, \mathbb{R}$ (for $\dot{+}$). By induction hypothesis, $M_2$ is a closed strong value of type $!\mathbb{R}$ or $!\mathbb{R} \otimes !\mathbb{R}$ or $\mathbb{R} \,\&\, \mathbb{R}$. One can check that the only closed strong values of these types are tuples (multiplicative or additive) of numerals. Therefore, $M_1 M_2$ is a $\beta_s$ redex, which is contrary to the hypothesis $M$ is a normal form for the safe reduction.

  – Subcase $M_1 = \dot{*}$:
    By inductive hypothesis $M_2$ is a closed strong value, so $\dot{*}M_2$ is a closed strong value and we can conclude.

  – Subcase $M_1 = \dot{*}W$:
    By inductive hypothesis $M_2$ is a closed strong value and by typing it is of type $\mathbb{R}$, so it is a numeral. Moreover, $W$ also is a closed value of type $\mathbb{R}$, so $M_1 M_2$ is a numerical, hence safe redex, which is contrary to the hypothesis $M$ is a normal form for the safe reduction.

- All the other cases are similar or immediate.

□

**Lemma 20.** Let $M$ be a ground closed term. The following are equivalent:

1. $M$ is a closed strong value,
2. $M$ is a closed normal form for the whole reduction $\to$,
3. $M$ is a closed normal form for the safe reduction.

*Sketch Proof.* The implication (1) $\Rightarrow$ (2) is by induction on the grammar of Strong Values, remarking that the hypothesis of $M$ ground implies that $M$ is not an abstraction. The implication (2) $\Rightarrow$ (3) is immediate, as safe redexes are also $\beta_\lambda$-redexes. The implication (3) $\Rightarrow$ (1) is by Lemma 19. $\square$

The main auxiliary lemma is related to the properties retained by the substitution in the context of safe reduction. More precisely, the following statement is both qualitative, as it guarantees that safeness and typing are preserved during substitution, and quantitative, ensuring that the workload does not increase with respect to the sum of the cost related to the analysed term and the cost of the substituted value.

**Remark 4.** It is worth noting that in the statement of Safe Substitution we must require the strong value $W$ to be closed, since this is the only way to ensure that the substitution $M\{W/p\}$ is a safe term. Specifically, when $M = \langle M_1, M_2\rangle$, this requirement prevents the substitution from introducing free variables that could compromise the safeness of $M\{W/p\}$. Without assuming that $W$ is closed, the intersection $FV(M_1\{W/p\}) \cap FV(M_2\{W/p\})$ could contain non-ground variables, violating the safeness condition in item (ii) of Definition 4. The closure of $W$ is required solely to preserve the safeness of the substitution (see proof of Claim 1 at page 63).

**Lemma 21** (Safe Substitution)**.** Given a safe term $M$ such that $!\Gamma, \Delta, p : A \vdash M : B$ and a safe closed strong value $W$ for the pattern $p$ such that $\vdash W : A$, we have:

1. $M\{W/p\}$ is a safe term;
2. $!\Gamma, \Delta \vdash M\{W/p\} : B$
3. $\mathcal{W}(M\{W/p\}) \leq \mathcal{W}(W) + \mathcal{W}(M)$.

*Sketch Proof.* Claim 1 is proved by induction on $M$, using the properties of safe terms listed in Definition 4 and the definition of $\mathcal{W}(M)$. The two delicate cases are $M = !M'$ and $M = \langle M_1, M_2\rangle$, namely those related to the conditions in Definition 4.

Furthermore, we prove Claim 2 and Claim 3 by a similar approach we used for the Pattern Substitution Lemma (Lemma 3). More precisely, for any derivation $\Pi_1$ of $!\Gamma, \Delta, p : A \vdash M : B$ and $\Pi_2$ of $\vdash W : A$, we give a derivation of $!\Gamma, \Delta \vdash M\{W/p\} : B$ by induction on the lexicographically ordered pair $(s(\Pi_2), s(\Pi_1))$, where $s(\Pi_i)$ is the number of derivation rules of $\Pi_i$. We refer to page 63 for the detailed proof. $\square$

Moreover, we proceed by showing that the workload decreases along safe reduction (Proposition 3) and the safeness of a term is preserved along safe reduction (Lemma 23).

**Lemma 22.** If $M$ is safe, then there is no numerical operation (i.e. $\dot{+}$, $\dot{*}$ or $\underline{f}$) under a !.

*Sketch Proof.* We proceed by strengthening the statement, proving by induction on $M$ the following two claims:

1. If $M$ is safe, then there is not numerical operation under a !;
2. If moreover $\mathcal{W}(M) = 0$, then there is no numerical operation at all in $M$.

$\square$

**Proposition 3.** Let $M$ be a safe term. If $M \to N$ is a safe step, then $\mathcal{W}(N) \leq \mathcal{W}(M)$. If moreover the step is numerical, then $\mathcal{W}(N) < \mathcal{W}(M)$.

*Proof.* By induction on the evaluation context $\gamma[\,]$. The induction step splits according to the cases of Figure 3.3, while the base case splits according to Figure 3.4.

In the base case of induction, if the $\beta$-step is $\beta_s$ then we use the Safe Substitution Lemma (Lemma 21) and we conclude.

In the induction step, if $\gamma[\,] = !\gamma'[\,]$, so that $M = !\gamma'[M_0]$ and $N = !\gamma'[N_0]$, we then have $\mathcal{W}(N) = 0$ by definition and the two inequalities $\leq$ hold trivially. As for the strict inequality in case of numerical steps: since $M$ is safe, by Lemma 22 there is no numerical operator in $\gamma'[M_0]$, so the step $M \to N$ cannot be numerical. □

**Lemma 23** (Safeness Invariance)**.** If $M$ is safe and $M \xrightarrow{s} N$, then $N$ is safe too.

*Sketch Proof.* By induction on the evaluation context $\gamma[\,]$ of the reduction step $M \xrightarrow{s} N$.

In the base case of induction, if the $\beta$-step is $\beta_s$ then we conclude by using item 1 of Lemma 21.

One case of the induction step is subtle: if $\gamma[\,] = !\gamma'[\,]$, so $M = !\gamma'[M_0]$, $N = !\gamma'[N_0]$ and $\gamma'[M_0] \xrightarrow{s} \gamma'[N_0]$. By induction hypothesis $\gamma'[N_0]$ is safe. In order to prove that $!\gamma'[N_0]$ is safe too, we must prove that $\mathcal{W}(\gamma'[N_0]) = 0$. By Proposition 3, we have $\mathcal{W}(\gamma'[N_0]) \leq \mathcal{W}(\gamma'[M_0])$. By safeness of $M$ (Definition 4), we have $\mathcal{W}(\gamma'[M_0]) = 0$ and we can conclude. □

We are finally able to prove that a safe closed term $M$ normalizes by using safe reduction in at most $\mathcal{W}(M)$ steps. Formally, this is stated as follows

**Proposition 4.** A safe closed term $M$ reduces by any maximal safe-reduction sequence to a strong value $W$ in at most $\mathcal{W}(M)$ numeric steps. If moreover $M$ is of ground type, $W$ is a $\beta$-normal form.

*Proof.* Consider a maximal safe-reduction sequence $(M_i)_{i=0}^{n}$ starting from $M$, i.e. $M_0 = M$ and $M_n$ is a safe-normal form.

By Subject Reduction, all $M_i$'s are closed. In particular, Lemma 19 gives that $M_n$ is a closed strong value.

By Lemma 23, all $M_i$'s are safe too. So we can apply Proposition 3 to each reduction step and getting that the sequence $\mathcal{W}(M_0), \mathcal{W}(M_1), \dots$ is decreasing, moreover it strictly decreases if the step is numeric. We conclude that $\mathcal{W}(M)$ bounds the number of numeric steps of this sequence.

Since $M_n$ is a closed strong value of ground type, Lemma 20 assures that $M_n$ is also a $\beta$-normal form. □

**Remark 5.** Let us recall Condition (ii) in Definition 4 of a safe term $M$ as follows: for any subterm $\langle M_1, M_2\rangle$ in $M$, $FV(M_1) \cap FV(M_2)$ has only ground variables. Observe that this condition is fundamental as it enables the Proposition 4 by restricting additive duplication. The condition can be omitted with a more intricate definition of workload that accounts for additive duplication through an appropriate quantitative type system. This generalisation is detailed in Chapter 6, but we opt for simplicity in this section, as it suffices to validate Autodiff.

Moreover, let us state the following lemma that will be useful to show that our transformations are work preserving in Chapter 5.

**Lemma 24.** $\mathcal{W}(\S M) = \mathcal{W}(M)$.

*Proof.* By notational convention defined in Section 3.1 we have $\S M = \langle (\,), M\rangle$, so we can conclude as follows

$$\mathcal{W}(\S M) = \mathcal{W}(\langle (\,), M\rangle) = \mathcal{W}((\,)) + \mathcal{W}(M) = \mathcal{W}(M)$$

because by definition of workload $\mathcal{W}((\,)) = 0$. □

# Detailed Proofs

## Proofs Strong Normalization

**Lemma 6.** Let $\Delta \vdash M : A$, if $\forall N \in \texttt{Test}_{x:A}.N\{^{M}/_{x}\} \in \texttt{RED}_{\mathbb{R}}$, then $M \in \texttt{RED}_{A}$.

*Proof.* By induction on $A$.

- Case $A = \mathbb{R}$:
  Notice that $x$ of type $\mathbb{R}$ is such that $x \in \texttt{Test}_{x:\mathbb{R}}$, so $x\{^{M}/_{x}\} \in \texttt{RED}_{\mathbb{R}}$ and we can conclude $M \in \texttt{RED}_{\mathbb{R}}$.

- Case $A = B \multimap C$:
  By hypothesis we have that $\Delta \vdash M : B \multimap C$ and $\forall N \in \texttt{Test}_{x:B\multimap C}.N\{^{M}/_{x}\} \in \texttt{RED}_{\mathbb{R}}$ and we want to prove that $M \in \texttt{RED}_{B\multimap C}$. Observe that $N$ is a generic term in $\texttt{Test}_{x:B\multimap C}$ so $x$ of type $B \multimap C$ is free in $N$.
  By (PR0) of Lemma 5 on $C$ we have that $\texttt{Test}_{y:C}$ is not empty, so let us take $N_1 \in \texttt{Test}_{y:C}$ typed as $\Delta, y : C \vdash N_1 : \mathbb{R}$ such that $\forall N_2 \in \texttt{RED}_{C}.N_1\{^{N_2}/_{y}\} \in \texttt{RED}_{\mathbb{R}}$.
  Let us consider a generic term $P \in \texttt{RED}_{B}$, we construct $N \stackrel{\text{def}}{=} N_1\{^{xP}/_{y}\}$. and we have to show that $N \in \texttt{Test}_{x:B\multimap C}$. Let us consider a generic term $Q \in \texttt{RED}_{B\multimap C}$. By construction and by the fact that $QP \in \texttt{RED}_{C}$ by definition of $\texttt{RED}_{B\multimap C}$ we have that $N\{^{Q}/_{x}\} = N_1\{^{QP}/_{y}\}$. Moreover, we have chosen $N_1 \in \texttt{Test}_{y:C}$ so it is such that $\forall N_2 \in \texttt{RED}_{C}.N_1\{^{N_2}/_{y}\} \in \texttt{RED}_{\mathbb{R}}$. Since $QP \in \texttt{RED}_{C}$, $Q$ is a generic term in $\texttt{RED}_{B\multimap C}$ and $P$ is a generic term in $\texttt{RED}_{B}$ then we have that $N_1\{^{QP}/_{y}\} \in \texttt{RED}_{\mathbb{R}}$.
  By construction we have $N_1\{^{MP}/_{y}\} = N\{^{M}/_{x}\}$ and by hypothesis $N\{^{M}/_{x}\} \in \texttt{RED}_{\mathbb{R}}$, so also $N_1\{^{MP}/_{y}\} \in \texttt{RED}_{\mathbb{R}}$. We can apply the IH on $C$ obtaining that $MP \in \texttt{RED}_{C}$. Since $P$ is a generic term in $\texttt{RED}_{B}$ then we can conclude that $M \in \texttt{RED}_{B\multimap C}$.

- Case $A = B \otimes C$:
  By hypothesis we have that $\Delta \vdash M : B \otimes C$ and $\forall N \in \texttt{Test}_{x:B\otimes C}.N\{^{M}/_{x}\} \in \texttt{RED}_{\mathbb{R}}$ and we want to prove that $M \in \texttt{RED}_{B\otimes C}$. Observe that $N$ is a generic term in $\texttt{Test}_{x:B\otimes C}$ so $x$ of type $B \otimes C$ is free in $N$.
  By (PR0) of Lemma 5 on $B \otimes C$ we have that $\texttt{Test}_{(x_1,x_2):B\otimes C}$ is not empty, so let us take $N_1 \in \texttt{Test}_{(x_1,x_2):B\otimes C}$ typed as $\Delta, (x_1, x_2) : B \otimes C \vdash N_1 : \mathbb{R}$ such that $\forall V_1 \in \texttt{RED}_{B}.\forall V_2 \in \texttt{RED}_{C}.\texttt{let } (x_1, x_2) = (V_1, V_2) \texttt{ in } N_1 \in \texttt{RED}_{\mathbb{R}}$.
  We construct $N \stackrel{\text{def}}{=} \texttt{let } (x_1, x_2) = x \texttt{ in } N_1$ and we have to show that $N \in \texttt{Test}_{x:B\otimes C}$. Let us consider a generic term $P \in \texttt{RED}_{B\otimes C}$. By construction and by substitution we have that $N\{^{P}/_{x}\} = (\texttt{let } (x_1, x_2) = x \texttt{ in } N_1)\{^{P}/_{x}\} = \texttt{let } (x_1, x_2) = P \texttt{ in } N_1$. Moreover, we have chosen $N_1 \in \texttt{Test}_{(x_1,x_2):B\otimes C}$ and $P$ is a generic term in $\texttt{RED}_{B\otimes C}$, so by definition of $\texttt{RED}_{B\otimes C}$ we have that $\texttt{let } (x_1, x_2) = P \texttt{ in } N_1 \in \texttt{RED}_{\mathbb{R}}$. More precisely, we have also that $N\{^{P}/_{x}\} \in \texttt{RED}_{\mathbb{R}}$ and since $P$ is a generic term in $\texttt{RED}_{B\otimes C}$, by definition of $\texttt{Test}_{x:B\otimes C}$ we can conclude that $N \in \texttt{Test}_{x:B\otimes C}$.
  By construction we have $N\{^{M}/_{x}\} = \texttt{let } (x_1, x_2) = M \texttt{ in } N_1$ and by hypothesis $N\{^{M}/_{x}\} \in \texttt{RED}_{\mathbb{R}}$, so also $\texttt{let } (x_1, x_2) = M \texttt{ in } N_1 \in \texttt{RED}_{\mathbb{R}}$. Since $N_1$ is a generic term in $\texttt{Test}_{(x_1,x_2):B\otimes C}$ then we can conclude that $M \in \texttt{RED}_{B\otimes C}$.

- Case $A = B\&C$:
  By hypothesis we have that $\Delta \vdash M : B\&C$ and $\forall N \in \texttt{Test}_{x:B\&C}.N\{^{M}/_{x}\} \in \texttt{RED}_{\mathbb{R}}$ and we want to prove that $M \in \texttt{RED}_{B\&C}$. Observe that $N$ is a generic term in $\texttt{Test}_{x:B\&C}$ so $x$ of type $B\&C$ is free in $N$.
  By (PR0) of Lemma 5 on $B\&C$ we have that $\texttt{Test}_{\langle x_1,x_2\rangle:B\&C}$ is not empty, so let us take $N_1 \in \texttt{Test}_{\langle x_1,x_2\rangle:B\&C}$ typed as $\Delta, \langle x_1, x_2\rangle : B\&C \vdash N_1 : \mathbb{R}$ such that $\forall V_1 \in \texttt{RED}_{B}.\forall V_2 \in$

$\mathtt{RED}_C$ s.t. $\langle V_1, V_2\rangle$
typable. $\mathtt{let}\ \langle x_1, x_2\rangle = \langle V_1, V_2\rangle\ \mathtt{in}\ N_1 \in \mathtt{RED}_{\mathbb{R}}$.
We construct $N \stackrel{\text{def}}{=} \mathtt{let}\ \langle x_1, x_2\rangle = x\ \mathtt{in}\ N_1$ and we have to show that $N \in \mathtt{Test}_{x:B\&C}$. Let us consider a generic term $P \in \mathtt{RED}_{B\&C}$. By construction and by substitution we have that $N\{P/x\} = (\mathtt{let}\ \langle x_1, x_2\rangle = x\ \mathtt{in}\ N_1)\{P/x\} = \mathtt{let}\ \langle x_1, x_2\rangle = P\ \mathtt{in}\ N_1$. Moreover, we have chosen $N_1 \in \mathtt{Test}_{\langle x_1,x_2\rangle:B\&C}$ and $P$ is a generic term in $\mathtt{RED}_{B\&C}$, so by definition of $\mathtt{RED}_{B\&C}$ we have that $\mathtt{let}\ \langle x_1, x_2\rangle = P\ \mathtt{in}\ N_1 \in \mathtt{RED}_{\mathbb{R}}$. More precisely, we have also that $N\{P/x\} \in \mathtt{RED}_{\mathbb{R}}$ and since $P$ is a generic term in $\mathtt{RED}_{B\&C}$, by definition of $\mathtt{Test}_{x:B\&C}$ we can conclude that $N \in \mathtt{Test}_{x:B\&C}$.
By construction we have $N\{M/x\} = \mathtt{let}\ \langle x_1, x_2\rangle = M\ \mathtt{in}\ N_1$ and by hypothesis $N\{M/x\} \in \mathtt{RED}_{\mathbb{R}}$, so also $\mathtt{let}\ \langle x_1, x_2\rangle = M\ \mathtt{in}\ N_1 \in \mathtt{RED}_{\mathbb{R}}$. Since $N_1$ is a generic term in $\mathtt{Test}_{\langle x_1,x_2\rangle:B\&C}$ then we can conclude that $M \in \mathtt{RED}_{B\&C}$.

- Case $A = !B$:
  By hypothesis we have that $\Delta \vdash M : !B$ and $\forall N \in \mathtt{Test}_{x:!B}.N\{M/x\} \in \mathtt{RED}_{\mathbb{R}}$ and we want to prove that $M \in \mathtt{RED}_{!B}$. Observe that $N$ is a generic term in $\mathtt{Test}_{x:!B}$ so $x$ of type $!B$ is free in $N$.
  By (PR0) of Lemma 5 on $!B$ we have that $\mathtt{Test}_{!y:!B}$ is not empty, so let us take $N_1 \in \mathtt{Test}_{!y:!B}$ typed as $\Delta, !y : !B \vdash N_1 : \mathbb{R}$ such that $\forall N_2 \in \mathtt{RED}_B$ s.t. $!\Gamma \vdash N_2 : B$. $\mathtt{let}\ !y = !N_2\ \mathtt{in}\ N_1 \in \mathtt{RED}_{\mathbb{R}}$.
  We construct $N \stackrel{\text{def}}{=} \mathtt{let}\ !y = x\ \mathtt{in}\ N_1$ and we have to show that $N \in \mathtt{Test}_{x:!B}$. Let us consider a generic term $P \in \mathtt{RED}_{!B}$. By construction and by substitution we have that $N\{P/x\} = (\mathtt{let}\ !y = x\ \mathtt{in}\ N_1)\{P/x\} = \mathtt{let}\ !y = P\ \mathtt{in}\ N_1$. Moreover, we have chosen $N_1 \in \mathtt{Test}_{!y:!B}$ and $P$ is a generic term in $\mathtt{RED}_{!B}$, so by definition of $\mathtt{RED}_{!B}$ we have that $\mathtt{let}\ !y = P\ \mathtt{in}\ N_1 \in \mathtt{RED}_{\mathbb{R}}$. More precisely, we have also that $N\{P/x\} \in \mathtt{RED}_{\mathbb{R}}$ and since $P$ is a generic term in $\mathtt{RED}_{!B}$, by definition of $\mathtt{Test}_{x:!B}$ we can conclude that $N \in \mathtt{Test}_{x:!B}$.
  By construction we have $N\{M/x\} = \mathtt{let}\ !y = M\ \mathtt{in}\ N_1$ and by hypothesis $N\{M/x\} \in \mathtt{RED}_{\mathbb{R}}$, so also $\mathtt{let}\ !y = M\ \mathtt{in}\ N_1 \in \mathtt{RED}_{\mathbb{R}}$. Since $N_1$ is a generic term in $\mathtt{Test}_{!y:!B}$ then we can conclude that $M \in \mathtt{RED}_{!B}$.

□

**Lemma 7.** If $(M_1, M_2) \in \mathtt{RED}_{A_1\otimes A_2}$, then $M_i \in \mathtt{RED}_{A_i}$ for $1 \leq i \leq 2$.

*Proof.* By hypothesis $(M_1, M_2) \in \mathtt{RED}_{A_1\otimes A_2}$. By definition of $\mathtt{RED}_{A_1\otimes A_2}$ we have that $(M_1, M_2) \in \mathscr{T}_{A_1\otimes A_2}$ so it is well-typed as $!\Gamma_1 \cup !\Gamma_2, \Delta_1, \Delta_2 \vdash (M_1, M_2) : A_1 \otimes A_2$ and by $\otimes_i$-typing rule we have also that $!\Gamma_i, \Delta_i \vdash M_i : A_i$ for $1 \leq i \leq 2$.
We take a generic $N_1 \in \mathtt{Test}_{x_1:A_1}$ typed as $\Delta_1', x_1 : A_1 \vdash N_1 : \mathbb{R}$ and such that $\forall N_1' \in \mathtt{RED}_{A_1}.N_1\{N_1'/x_1\} \in \mathtt{RED}_{\mathbb{R}}$. We take a generic $N_2 \in \mathtt{Test}_{x_2:A_2}$ typed as $\Delta_2', x_2 : A_2 \vdash N_2 : \mathbb{R}$ and such that $\forall N_2' \in \mathtt{RED}_{A_2}.N_2\{N_2'/x_2\} \in \mathtt{RED}_{\mathbb{R}}$. We construct $N \stackrel{\text{def}}{=} N_1 {*} N_2$ and we want to prove that $N \in \mathtt{Test}_{(x_1,x_2):A_1\otimes A_2}$. Observe that $N$ is well-typed as $\Delta_1', \Delta_2', (x_1, x_2) : A_1 \otimes A_2 \vdash N : \mathbb{R}$. Let us consider a generic $W_1 \in \mathtt{RED}_{A_1}$ and a generic $W_2 \in \mathtt{RED}_{A_2}$. By construction we have that $\mathtt{let}\ (x_1, x_2) = (W_1, W_2)\ \mathtt{in}\ N = \mathtt{let}\ (x_1, x_2) = (W_1, W_2)\ \mathtt{in}\ N_1 {*} N_2$. We have to prove that $\mathtt{let}\ (x_1, x_2) = (W_1, W_2)\ \mathtt{in}\ N_1 {*} N_2 \in \mathtt{RED}_{\mathbb{R}}$, so by definition we have to prove that it is SN. We proceed by induction on $\nu(W_1) + \nu(W_2) + \nu(N_1) + \nu(N_2)$ to show that every term to which $\mathtt{let}\ (x_1, x_2) = (W_1, W_2)\ \mathtt{in}\ N_1 {*} N_2$ reduces in one step is SN:

- If $\mathtt{let}\ (x_1, x_2) = (W_1, W_2)\ \mathtt{in}\ N_1 {*} N_2 \to (N_1 {*} N_2)\{(W_1, W_2)/(x_1, x_2)\}$.
  By substitution we have $(N_1 {*} N_2)\{(W_1, W_2)/(x_1, x_2)\} = (N_1 {*} N_2)\{W_1/x_1, W_2/x_2\} = (N_1\{W_1/x_1\}){*}(N_2\{W_2/x_2\})$. We have chosen $N_1 \in \mathtt{Test}_{x_1:A_1}$ so we have that $N_1\{W_1/x_1\} \in \mathtt{RED}_{\mathbb{R}}$, so by definition of $\mathtt{RED}_{\mathbb{R}}$ it is also SN.

We have chosen $N_2 \in \texttt{Test}_{x_2:A_2}$ so we have that $N_2\{{}^{W_2}/_{x_2}\} \in \texttt{RED}_{\mathbb{R}}$, so by definition of $\texttt{RED}_{\mathbb{R}}$ it is also SN. All the subterms of $(N_1\{{}^{W_1}/_{x_1}\})\dot{*}(N_2\{{}^{W_2}/_{x_2}\})$ are SN so it is SN and we can conclude.

- If $\texttt{let } (x_1, x_2) = (W_1, W_2) \texttt{ in } N_1\dot{*}N_2 \to \texttt{let } (x_1, x_2) = (W_1', W_2) \texttt{ in } N_1\dot{*}N_2$ with $W_1 \to W_1'$. We have that $\nu(W_1') < \nu(W_1)$ so we can apply the IH and conclude that $\texttt{let } (x_1, x_2) = (W_1', W_2) \texttt{ in } N_1\dot{*}N_2$ is SN.

- All the other cases where $W_2 \to W_2'$ and $N_i \to N_i'$ for $2 \leq i \leq 1$ are similar to the previous case.

Summing up we have that $N \in \texttt{Test}_{(x_1,x_2):A_1\otimes A_2}$.
In order to apply Lemma 6 on $M_1$ and $M_2$ we have to show that $\forall N \in \texttt{Test}_{x_i:A_i}.N\{{}^{M_i}/_{x_i}\} \in \texttt{RED}_{\mathbb{R}}$, we proceed as follows:

1. We have shown that $N \in \texttt{Test}_{(x_1,x_2):A_1\otimes A_2}$, so by definition of $\texttt{Test}_{(x_1,x_2):A_1\otimes A_2}$ we have that $\texttt{let } (x_1, x_2) = (M_1, M_2) \texttt{ in } N \in \texttt{RED}_{\mathbb{R}}$. By construction we have that $\texttt{let } (x_1, x_2) = (M_1, M_2) \texttt{ in } N = \texttt{let } (x_1, x_2) = (M_1, M_2) \texttt{ in } N_1\dot{*}N_2$, so we have also that $\texttt{let } (x_1, x_2) = (M_1, M_2) \texttt{ in } N_1\dot{*}N_2 \in \texttt{RED}_{\mathbb{R}}$.

2. By reduction we have that $\texttt{let } (x_1, x_2) = (M_1, M_2) \texttt{ in } N_1\dot{*}N_2 \to (N_1\dot{*}N_2)\{{}^{(M_1,\,M_2)}/_{(x_1,\,x_2)}\}$ and by substitution we have that $(N_1\dot{*}N_2)\{{}^{(M_1,\,M_2)}/_{(x_1,\,x_2)}\} = (N_1\{{}^{M_1}/_{x_1}\})\dot{*}(N_2\{{}^{M_2}/_{x_2}\})$.

3. By item 1 we have that $\texttt{let } (x_1, x_2) = (M_1, M_2) \texttt{ in } N_1\dot{*}N_2 \in \texttt{RED}_{\mathbb{R}}$ and by item 2 it reduces to $(N_1\{{}^{M_1}/_{x_1}\})\dot{*}(N_2\{{}^{M_2}/_{x_2}\})$, so we can apply (PR2) of Lemma 5 on $\mathbb{R}$ obtaining that $(N_1\{{}^{M_1}/_{x_1}\})\dot{*}(N_2\{{}^{M_2}/_{x_2}\}) \in \texttt{RED}_{\mathbb{R}}$. By definition of $\texttt{RED}_{\mathbb{R}}$ we have that $(N_1\{{}^{M_1}/_{x_1}\})\dot{*}(N_2\{{}^{M_2}/_{x_2}\})$ is SN, so all the subterms of it are SN. More precisely $N_i\{{}^{M_i}/_{x_i}\}$ for $1 \leq i \leq 2$ are SN and by definition $N_i\{{}^{M_i}/_{x_i}\} \in \texttt{RED}_{\mathbb{R}}$.

Since $N_i$ is a generic term in $\texttt{Test}_{x_i:A_i}$ we can apply Lemma 6 and conclude. □

**Lemma 8.** If $!M \in \texttt{RED}_{!A}$ then $M \in \texttt{RED}_A$.

*Proof.* By hypothesis $!M \in \texttt{RED}_{!A}$. By definition of $\texttt{RED}_{!A}$ we have that $!M \in \mathscr{T}_{!A}$ so it is well-typed as $!\Gamma \vdash !M : !A$ and by $!_i$-typing rule we have also that $!\Gamma \vdash M : A$.
We take a generic $N_1 \in \texttt{Test}_{x:A}$ typed as $\Delta, x : A \vdash N_1 : \mathbb{R}$ and such that $\forall N_2 \in \texttt{RED}_A.N_1\{{}^{N_2}/_x\} \in \texttt{RED}_{\mathbb{R}}$. We construct $N \stackrel{\text{def}}{=} \texttt{let } x = y \texttt{ in } N_1$ and we want to prove that $N \in \texttt{Test}_{!y:!A}$. Let us consider a generic $Q \in \texttt{RED}_A$. By construction and substitution we have that $N_1\{{}^{!Q}/_{!y}\} = N_1\{{}^{Q}/_y\} = (\texttt{let } x = y \texttt{ in } N_1)\{{}^{Q}/_y\} = \texttt{let } x = Q \texttt{ in } N_1$. We have to prove that $\texttt{let } x = Q \texttt{ in } N_1 \in \texttt{RED}_{\mathbb{R}}$, so by definition we have to prove that it is SN. We proceed by induction on $\nu(Q) + \nu(N_1)$ to show that every term to which $\texttt{let } x = Q \texttt{ in } N_1$ reduces in one step is SN:

- $\texttt{let } x = Q \texttt{ in } N_1 \to \texttt{let } x = Q' \texttt{ in } N_1$ with $Q \to Q'$. We have that $\nu(Q') < \nu(Q)$ so we can apply the IH and conclude that $\texttt{let } x = Q' \texttt{ in } N_1$ is SN.

- $\texttt{let } x = Q \texttt{ in } N_1 \to \texttt{let } x = Q \texttt{ in } N_1'$ with $N_1 \to N_1'$. We have that $\nu(N_1') < \nu(N_1)$ so we can apply the IH and conclude that $\texttt{let } x = Q \texttt{ in } N_1'$ is SN.

- $\texttt{let } x = Q \texttt{ in } N_1 \to N_1\{{}^{Q}/_x\}$. We have chosen $N_1 \in \texttt{Test}_{x:A}$ so it is such that $\forall N_2 \in \texttt{RED}_A.N_1\{{}^{N_2}/_x\} \in \texttt{RED}_{\mathbb{R}}$. Since $Q$ is a generic term in $\texttt{RED}_A$ then we have that $N_1\{{}^{Q}/_x\} \in \texttt{RED}_{\mathbb{R}}$. By definition of $\texttt{RED}_{\mathbb{R}}$ we have that $N_1\{{}^{Q}/_x\}$ is SN and we can conclude.

Summing up we have that $N \in \texttt{Test}_{!y:!A}$.
By hypothesis $!M \in \texttt{RED}_{!A}$, so by definition of $\texttt{RED}_{!A}$ and by the fact that $N$ is a generic term in $\texttt{Test}_{!y:!A}$ we have that $\texttt{let}\ !y = !M\ \texttt{in}\ N \in \texttt{RED}_{\mathbb{R}}$. Moreover, by reduction we have that $\texttt{let}\ !y = !M\ \texttt{in}\ N \to N\{^{!M}/_{!y}\}$ so we can apply (PR2) of Lemma 5 on $\mathbb{R}$ obtaining $N\{^{!M}/_{!y}\} \in \texttt{RED}_{\mathbb{R}}$. Since $N_1$ is a term in $\texttt{Test}_{x:A}$ so it is such that $\forall N_2 \in \texttt{RED}_A.N_1\{^{N_2}/_x\} \in \texttt{RED}_{\mathbb{R}}$, so we can conclude that $M \in \texttt{RED}_A$. ☐

**Lemma 9.** If $\langle M_1, M_2\rangle \in \texttt{RED}_{A_1\&A_2}$, then $M_i \in \texttt{RED}_{A_i}$ for $1 \le i \le 2$.

*Proof.* By hypothesis $\langle M_1, M_2\rangle \in \texttt{RED}_{A_1\&A_2}$. By definition of $\texttt{RED}_{A_1\&A_2}$ we have that $\langle M_1, M_2\rangle \in \mathscr{T}_{A_1\&A_2}$ so it is well-typed as $\Delta \vdash \langle M_1, M_2\rangle : A_1\&A_2$ and by $\&_i$-typing rule we have also that $\Delta \vdash M_i : A_i$ for $1 \le i \le 2$.
We take a generic $N$ such that $\Gamma, \langle x_1, x_2\rangle : A_1\&A_2 \vdash N : \mathbb{R}$.
We want to prove that $N \in \texttt{Test}_{\langle x_1,x_2\rangle:A_1\&A_2}$. Observe that $N$ is well-typed as $\Delta_1', \Delta_2', \langle x_1, x_2\rangle : A_1\&A_2 \vdash N : \mathbb{R}$. Let us consider a generic $W_1 \in \texttt{RED}_{A_1}$ and a generic $W_2 \in \texttt{RED}_{A_2}$. We have to prove that $\texttt{let}\ \langle x_1, x_2\rangle = \langle W_1, W_2\rangle\ \texttt{in}\ N \in \texttt{RED}_{\mathbb{R}}$, so by definition we have to prove that it is SN. We proceed by induction on $\nu(W_1) + \nu(W_2) + \nu(N)$ to show that every term to which $\texttt{let}\ \langle x_1, x_2\rangle = \langle W_1, W_2\rangle\ \texttt{in}\ N$ reduces in one step are SN:

- If $\texttt{let}\ \langle x_1, x_2\rangle = \langle W_1, W_2\rangle\ \texttt{in}\ N \to N\{^{\langle W_1, W_2\rangle}/_{\langle x_1, x_2\rangle}\}$.
  By substitution we have $N\{^{\langle W_1, W_2\rangle}/_{\langle x_1, x_2\rangle}\} = N\{^{W_1}/_{x_1}\}\{^{W_2}/_{x_2}\}$.
  Notice that from $\Gamma, \langle x_1, x_2\rangle : A_1\&A_2 \vdash N : \mathbb{R}$ we can also get $\Gamma, x_i : A_i \vdash N : \mathbb{R}$ by $\&_{ei}$-typing rule for $1 \le i \le 2$. Since $W_i$ is a generic term in $\texttt{RED}_{A_i}$ we have that $N\{^{W_i}/_{x_i}\} \in \texttt{RED}_{\mathbb{R}}$, so by definition we have that $N \in \texttt{Test}_{x_i:A_i}$. We can conclude that $N\{^{W_1}/_{x_1}\}\{^{W_2}/_{x_2}\} \in \texttt{RED}_{\mathbb{R}}$ and so it is SN.

- If $\texttt{let}\ \langle x_1, x_2\rangle = \langle W_1, W_2\rangle\ \texttt{in}\ N \to \texttt{let}\ \langle x_1, x_2\rangle = \langle W_1', W_2\rangle\ \texttt{in}\ N$ with $W_1 \to W_1'$. We have that $\nu(W_1') < \nu(W_1)$ so we can apply the IH and conclude that $\texttt{let}\ \langle x_1, x_2\rangle = \langle W_1', W_2\rangle\ \texttt{in}\ N$ is SN.

- All the other cases where $W_2 \to W_2'$ and $N_i \to N_i'$ for $2 \le i \le 1$ are similar to the previous case.

Summing up we have that $N \in \texttt{Test}_{\langle x_1,x_2\rangle:A_1\&A_2}$.
In order to apply Lemma 6 on $M_1$ and $M_2$ we have to show that $\forall N \in \texttt{Test}_{x_i:A_i}.N\{^{M_i}/_{x_i}\} \in \texttt{RED}_{\mathbb{R}}$, we proceed as follows:

1. We have shown that $N \in \texttt{Test}_{\langle x_1,x_2\rangle:A_1\&A_2}$, so by definition of $\texttt{Test}_{\langle x_1,x_2\rangle:A_1\&A_2}$ we have that $\texttt{let}\ \langle x_1, x_2\rangle = \langle M_1, M_2\rangle\ \texttt{in}\ N \in \texttt{RED}_{\mathbb{R}}$.

2. By reduction we have that $\texttt{let}\ \langle x_1, x_2\rangle = \langle M_1, M_2\rangle\ \texttt{in}\ N \to N\{^{\langle M_1, M_2\rangle}/_{\langle x_1, x_2\rangle}\}$ and by substitution we have that $N\{^{\langle M_1, M_2\rangle}/_{\langle x_1, x_2\rangle}\} = N\{^{M_1}/_{x_1}\}\{^{M_2}/_{x_2}\}$.

3. By item 1 we have that $\texttt{let}\ \langle x_1, x_2\rangle = \langle M_1, M_2\rangle\ \texttt{in}\ N \in \texttt{RED}_{\mathbb{R}}$ and by item 2 it reduces to $N\{^{M_1}/_{x_1}\}\{^{M_2}/_{x_2}\}$, so we can apply (PR2) of Lemma 5 on $\mathbb{R}$ obtaining that $N\{^{M_1}/_{x_1}\}\{^{M_2}/_{x_2}\} \in \texttt{RED}_{\mathbb{R}}$. By definition of $\texttt{RED}_{\mathbb{R}}$ and by the fact that $\Gamma, \langle x_1, x_2\rangle : A_1\&A_2 \vdash N : \mathbb{R}$ and by $\&_{ei}$-typing rule $\Gamma, x_i : A_i \vdash N : \mathbb{R}\ 1 \le i \le 2$, we have that $N\{^{M_1}/_{x_1}\}\{^{M_2}/_{x_2}\}$ is SN and so it is in $\texttt{RED}_{\mathbb{R}}$. More precisely $N_i\{^{M_i}/_{x_i}\}$ for $1 \le i \le 2$ are SN and by definition $N_i\{^{M_i}/_{x_i}\} \in \texttt{RED}_{\mathbb{R}}$.

We can apply Lemma 6 and conclude. ☐

## Proofs Workload

**Lemma 21** (Safe Substitution)**.** Given a safe term $M$ such that $!\Gamma, \Delta, p : A \vdash M : B$ and a safe closed strong value $W$ for the pattern $p$ such that $\vdash W : A$, we have:

1. $M\{W/p\}$ is a safe term;
2. $!\Gamma, \Delta \vdash M\{W/p\} : B$
3. $\mathcal{W}(M\{W/p\}) \leq \mathcal{W}(W) + \mathcal{W}(M)$.

*Proof Claim 1.* By induction on $M$. The only non trivial cases are $M = !M'$ and $M = \langle M_1, M_2 \rangle$, we detail them as follows:

- Case $M = !M'$:
  By definition of substitution $M\{W/p\} = (!M')\{W/p\} = !(M'\{W/p\})$. Moreover, Definition 4 of safe term item $i$ we have to prove that $\mathcal{W}(M'\{W/p\}) = 0$ in order to conclude that $M\{W/p\}$ is safe.

  Recall that, by hypothesis $M = !M'$ is a safe term, so by item $i$ of Definition 4 we know that $\mathcal{W}(M') = 0$. Moreover, by typing the pattern $p$ is of exponential type this means that $p = x$ and $W = !W'$ for some strong value $W'$. Hence by definition of workload $\mathcal{W}(W) = \mathcal{W}(!W') = 0$.

  By item 3 of this lemma we have that $\mathcal{W}(M'\{W/p\}) \leq \mathcal{W}(W) + \mathcal{W}(M')$ which is equal to zero and so we can conclude.

- Case $M = \langle M_1, M_2 \rangle$:
  By definition of substitution $M\{W/p\} = (\langle M_1, M_2 \rangle)\{W/p\} = \langle M_1\{W/p\}, M_2\{W/p\} \rangle$. Moreover, Definition 4 of safe term item $ii$ we have to prove that $FV(M_1\{W/p\}) \cap FV(M_2\{W/p\})$ has only ground variables in order to conclude that $M\{W/p\}$ is safe.

  Recall that, by hypothesis $M = \langle M_1, M_2 \rangle$ is a safe term, so by item $ii$ of Definition 4 we know that $FV(M_1) \cap FV(M_2)$ has only ground variables. Moreover, by hypothesis $W$ is closed and substituting a closed value cannot introduce any new free variables, so we can conclude that $FV(M_1\{W/p\}) \cap FV(M_2\{W/p\})$ has only ground variables.

□

*Proof Claim 2 and 3.* For any derivation $\Pi_1$ of $!\Gamma, \Delta, p : A \vdash M : B$ and $\Pi_2$ of $\vdash W : A$, we give a derivation of $!\Gamma, \Delta \vdash M\{W/p\} : B$ by induction on the lexicographically ordered pair $(s(\Pi_2), s(\Pi_1))$, where $s(\Pi_i)$ is the number of derivation rules of $\Pi_i$.

We split depending on the last derivation rule in $\Pi_1$ or $\Pi_2$.

- If the last rule of $\Pi_1$ is a rule $r$ among $\{!_w, \&_{ei}, \otimes_e, 1_e\}$ acting on a pattern in $!\Gamma, \Delta$, then the immediate subderivation of $\Pi_1$ is $\Pi_1'$ of $!\Gamma', \Delta', p : A \vdash M : B$. We can conclude by induction hypothesis on $(s(\Pi_2), s(\Pi_1'))$ getting a) type derivation for $!\Gamma', \Delta' \vdash M\{W/p\} : B$; b) $\mathcal{W}(M\{W/p\}) \leq \mathcal{W}(M) + \mathcal{W}(W)$.

- For the other cases, we can then suppose that the last rules of $\Pi_1$ is not acting on $!\Gamma, \Delta$. We then split in further sub-cases depending if the last rules of $\Pi_1$ acts on the pattern $p : A$ or acts on the term $M$.

  Let us consider first the cases of a last rule $r$ of $\Pi_1$ acting on the pattern $p : A$.

– If $r$ is of type $!_w$, then $p = !x$ and $W = !W'$ for some safe closed strong value $W'$.
Notice that the subderivation $\Pi'_1$ above $r$ in $\Pi_1$ has conclusion $!\Gamma, \Delta \vdash M : B$.
By Lemma 1 $x \notin FV(M)$ and by applying Lemma 2 we have: $M\{!W'/!x\} = M$, so item 2 of the lemma holds by taking $\Pi'_1$.
By definition of workload we have $\mathcal{W}(W) = 0$ and so item 3 of the lemma holds as $\mathcal{W}(M\{W/p\}) = \mathcal{W}(M) \leq \mathcal{W}(M) + \mathcal{W}(W) = \mathcal{W}(M)$.

– If $r$ is of type $\&_{ei}$, then $p = \langle p_1, p_2\rangle : A_1 \,\&\, A_2$ and $W = \langle W_1, W_2\rangle$ for some safe strong values $W_i$ for $p_i$.
Notice that the subderivation $\Pi'_1$ above $r$ in $\Pi_1$ has conclusion $!\Gamma, \Delta, p_i : A_i \vdash M : B$.
By Lemma 1 $FV(p_{3-i}) \notin FV(M)$ so by Lemma 2, $M\{W_i/p_i\}\{W_{3-i}/p_{3-i}\} = M\{W_i/p_i\}$.
By cases inspection, one can infer that the last rule of $\Pi_2$ is a $\&_i$. Therefore we have a subderivation $\Pi'_2$ above such rule for the judgement $\vdash W_i : A_i$.
Let us suppose $i = 1$ (the other case being similar), so we have: $M\{W_1/p_1\}\{W_2/p_2\} = M\{W_1/p_1\}$.
By induction hypothesis on $(s(\Pi'_2), s(\Pi'_1))$ we have: a) a derivation for $!\Gamma, \Delta \vdash M\{W_1/p_1\} : B$; b) $\mathcal{W}(M\{W_1/p_1\}) \leq \mathcal{W}(M) + \mathcal{W}(W_1)$.
The item 2 of the lemma holds because $M\{W/p\} = M\{W_1/p_1\}$ and by point a of the induction hypothesis.
We show that item 3 of the lemma holds as follows

$$
\begin{aligned}
\mathcal{W}(M\{W/p\}) &= \mathcal{W}(M^{\langle W_1, W_2\rangle}/_{\langle p_1, p_2\rangle}) \\
&= \mathcal{W}(M\{W_1/p_1\}\{W_2/p_2\}) \\
&= \mathcal{W}(M\{W_1/p_1\}) \\
&\overset{\text{IH}}{\leq} \mathcal{W}(M) + \mathcal{W}(W_1) \\
&\leq \mathcal{W}(M) + \mathcal{W}(W) \\
&= \mathcal{W}(M) + \mathcal{W}(\langle W_1, W_2\rangle) \\
&= \mathcal{W}(M) + \mathcal{W}(W_1) + \mathcal{W}(W_2)
\end{aligned}
$$

– If $r$ is of type $\otimes_e$, then $p = (p_1, p_2) : A_1 \otimes A_2$ and $W = (W_1, W_2)$ for some values $W_i$ of $p_i$.
Notice that the subderivation $\Pi'_1$ above $r$ in $\Pi_1$ has conclusion $!\Gamma, \Delta, p_1 : A_1, p_2 : A_2 \vdash M : B$.
By cases inspection, one can infer that the last rule of $\Pi_2$ is a $\otimes_i$. Therefore we have two subderivations $\Pi_{2,1}$ and $\Pi_{2,2}$ above such rule for $\vdash W_1 : A_1$ and $\vdash W_2 : A_2$, respectively.
We have to prove that: 1) $M\{W/p\} = \mathcal{W}(M^{(W_1, W_2)}/_{(p_1, p_2)}) = M\{W_1/p_1\}\{W_2/p_2\}$ is well-typed as $!\Gamma, \Delta \vdash M\{W_1/p_1\}\{W_2/p_2\} : B$; 2) $\mathcal{W}(M\{W_1/p_1\}\{W_2/p_2\}) \leq \mathcal{W}(M) + \mathcal{W}(W_1) + \mathcal{W}(W_2)$
By induction hypothesis on $(s(\Pi_{2,1}), s(\Pi'_1))$ we have: a) a derivation for $!\Gamma, \Delta, p_2 : A_2 \vdash M\{W_1/p_1\} : B$; b) $\mathcal{W}(M\{W_1/p_1\}) \leq \mathcal{W}(M) + \mathcal{W}(W_1)$.
We proceed by applying the induction hypothesis on $(s(\Pi_{2,2}), s(\Pi'_1))$ obtaining: a) a derivation for $!\Gamma, \Delta \vdash M\{W_1/p_1\}\{W_2/p_2\} : B$; b) $\mathcal{W}((M\{W_1/p_1\})\{W_2/p_2\}) \leq \mathcal{W}(M\{W_1/p_1\}) + \mathcal{W}(W_2)$.
We can conclude as item 2 holds directly from item a of IH on $(s(\Pi_{2,2}), s(\Pi'_1))$ and item 3 holds as

$$
\mathcal{W}(M\{W_1/p_1\}) + \mathcal{W}(W_2) \overset{\text{IH on } (s(\Pi_{2,1}), s(\Pi'_1))}{\leq} \mathcal{W}(M) + \mathcal{W}(W_1) + \mathcal{W}(W_2)
$$

- If $r$ is of type $1_e$, then $p = ( ) : \mathtt{1}$ and $W = ( )$.
   Notice that the subderivation $\Pi_1'$ above $r$ in $\Pi_1$ has conclusion $!\Gamma, \Delta \vdash M : B$.
   By cases inspection, one can infer that the last rule of $\Pi_2$ is a $1_i$.
   By definition of substitution we have $M\{( )/( )\} = M$ and by definition of workload we have $\mathcal{W}(( )) = 0$, so we can conclude.

- Let us consider now the cases in which the last rule $r$ in $\Pi_1$ acts on the subject $M$.

 - If $r$ is of type $v$, then $M = p = x : A$. Moreover, $A = B$ and $!\Gamma, \Delta$ is empty.
   By definition of substitution $M\{W/p\} = W$.
   By definition of workload we have $\mathcal{W}(x) = 0$, so item 3 of the lemma holds.
   Moreover, item 2 of the lemma holds because by hypothesis we have a derivation $\Pi_2$ for $\vdash W : A$.
 - If $r$ is of type $!_e$, then $M = x$ and $p = !x$. Moreover, $B = !A$ and $!\Gamma, \Delta$ is empty.
   By definition of value for a pattern $W = !W'$ for some safe closed strong value $W'$.
   By hypothesis $W$ is safe and by item $i$ in Definition 4 of safeness we have $\mathcal{W}(W') = 0$.
   By cases inspection, one can infer that the last rule of $\Pi_2$ is a $!_i$. Therefore we have an immediate subderivation $\Pi_2'$ above such rule for the judgement $\vdash W' : A$.
   By definition of substitution $M\{W/p\} = W'$, so item 2 of the lemma holds by taking the derivation $\Pi_2'$.
   By definition od workload we have $\mathcal{W}(!W') = 0$. We can conclude that item 3 of the lemma holds as follows

$$
\begin{aligned}
\mathcal{W}(M\{W/p\}) &= \mathcal{W}(x\{!W'/!x\}) \\
&= \mathcal{W}(x\{W'/x\}) \\
&= \mathcal{W}(W') \\
&= 0 \\
&\leq \mathcal{W}(M) + \mathcal{W}(W) \\
&= \mathcal{W}(x) + \mathcal{W}(!W') \\
&= 0
\end{aligned}
$$

 - If $r$ is of type $\&_i$, then $M = \langle M_1, M_2 \rangle$ and $B = B_1 \& B_2$.
   The immediate subderivations of $\Pi_1$ are $\Pi_{1,1}$ and $\Pi_{1,2}$ above $r$ for $!\Gamma, \Delta, p : A \vdash M_1 : B_1$ and $!\Gamma, \Delta, p : A \vdash M_2 : B_2$, respectively.
   By induction hypothesis on $(s(\Pi_2), s(\Pi_{1,i}'))$ with $i \in \{1, 2\}$ we have: a) a type derivation for $!\Gamma, \Delta \vdash M_i\{W/p\} : B_i$; b) $\mathcal{W}(M_i\{W/p\}) = \mathcal{W}(M_i) + \mathcal{W}(W)$.
   By definition of substitution we have that $M\{W/p\} = \langle M_1, M_2 \rangle\{W/p\} = \langle M_1\{W/p\}, M_2\{W/p\} \rangle$.
   We can conclude that item 2 of the lemma holds by using $\&_i$ and the inductive hypotheses.
   By item 1 of the lemma $M\{W/p\}$ is safe, so in this case $\langle M_1, M_2 \rangle\{W/p\}$ is safe and by item $ii$ in Definition 4 of safe term we have that $FV(M_1) \cap FV(M_2)$ contains only ground variables. Moreover, $W$ is a strong value of ground type and by Lemma 18 we have $\mathcal{W}(W) = 0$.
   We show that item 3 of the lemma holds as follows

$$
\mathcal{W}(M\{W/p\}) = \mathcal{W}((\langle M_1, M_2 \rangle)\{W/p\})
$$

$$
\begin{aligned}
&= \mathcal{W}(\langle M_1\{W/p\}, M_2\{W/p\}\rangle) \\
&= \mathcal{W}(M_1\{W/p\}) + \mathcal{W}(M_2\{W/p\}) \\
&\overset{\text{IHs}}{\leq} \mathcal{W}(W) + \mathcal{W}(M_1) + \mathcal{W}(W) + \mathcal{W}(M_2) \\
&= \mathcal{W}(M_1) + \mathcal{W}(M_2) \\
&\leq \mathcal{W}(M) + \mathcal{W}(W) \\
&= \mathcal{W}(\langle M_1, M_2\rangle) + \mathcal{W}(W) \\
&= \mathcal{W}(M_1) + \mathcal{W}(M_2) + \mathcal{W}(W) \\
&= \mathcal{W}(M_1) + \mathcal{W}(M_2)
\end{aligned}
$$

– All other cases are similar or immediate.

□

Chapter 4

# Translation

In this chapter, we define a formal translation of Linear A expressions into terms of $\lambda$LL, establishing a type-directed correspondence that faithfully preserves the linearity constraints enforced by the type system of Linear A. This translation embeds Linear A types and programs into the type system of $\lambda$LL, which is governed by the resource-sensitive principles of Linear Logic, thereby enabling a logical interpretation of the type system of Linear A.

First, we give two translations $\mathtt{p}$ and $\mathtt{t}$ of Linear A types depending whether these latter refer to primal or tangent data:

$$\mathtt{p}(\mathbb{R}) = \mathbb{R}\,, \qquad \mathtt{p}(\mathtt{1}) = \mathtt{1}\,, \qquad \mathtt{p}(\tau \otimes \sigma) = !\mathtt{p}(\tau) \otimes !\mathtt{p}(\sigma)\,, \qquad \text{(Primal Translation)}$$
$$\mathtt{t}(\mathbb{R}) = \mathbb{R}\,, \qquad \mathtt{t}(\mathtt{1}) = \top\,, \qquad \mathtt{t}(\tau \otimes \sigma) = \mathtt{t}(\tau) \,\&\, \mathtt{t}(\sigma)\,. \qquad \text{(Tangent Translation)}$$

Observe that $\mathtt{p}(\tau)$ (resp. $\mathtt{t}(\tau)$) is generated by the grammar $\otimes$-sequence types (resp. &-sequence types).

**Remark 6.** One can prove by induction on a Linear A type $\tau$ that $\mathtt{p}(\tau)$ is a retraction of $!\mathtt{t}(\tau)$, namely $\mathtt{p}(\tau) \dashv\vdash !\mathtt{t}(\tau)$. In fact, in LL we have (see e.g. [84]) the two isomorphisms $\mathtt{1} \dashv\vdash !\top$ and $!A \otimes !B \dashv\vdash !(A \,\&\, B)$, as well as the retraction pair $!A \dashv\vdash !!A$. This shows that morally one can consider primal types as the exponential promotion of the tangent types, supposing that primal $\mathbb{R}$ is equivalent to the ! of tangent $\mathbb{R}$.

We extend $\mathtt{t}$ on $\otimes$-sequence types: $\mathtt{t}(\mathbb{R}) = \mathbb{R}, \mathtt{t}(\mathtt{1}) = \top, \mathtt{t}(!D \otimes !E) = \mathtt{t}(D) \& \mathtt{t}(E)$. Moreover, the notion of numerals extend to sequence types, in the spirit of Subsection 2.1.2. A *numeral sequence* $\underline{\vec{r}}$ for the type $\mathbb{R}$ (resp. $!\mathbb{R}$) is simply a numeral $\underline{p}$ (resp. the promotion $!\underline{p}$ of a numeral); a numeral sequence for $\top$ (resp. $\mathtt{1}$) is the empty sequence $\langle\ \rangle$ (resp. $(\,)$); a numeral sequence $\underline{\vec{r}}$ for $H\&H'$ (resp. $!(E \otimes E')$) is $\langle \underline{\vec{s}}, \underline{\vec{s}}' \rangle$ (resp. $!(\underline{\vec{s}}, \underline{\vec{s}}')$) for $\underline{\vec{s}}$, $\underline{\vec{s}}'$ numeral sequences for respectively $H$ and $H'$ (resp. $E$ and $E'$). Notice that the numeral sequences are closed strong values, so we can substitute a numeral sequence $\underline{\vec{r}}$ for $x$ in a term: $M\{\underline{\vec{r}}/x\}$.

We use the same metavariables $\underline{\vec{r}}$, $\underline{\vec{s}}$ to denote numeral sequences, both for Linear A and $\lambda$LL. We may silently suppose the immediate correspondence between the numeral sequences of a Linear A type $\tau$, and their $\lambda$LL siblings $\mathtt{t}(\tau)$ and $\mathtt{p}(\tau)$. Given an environment $\Gamma$ of sequence types, we write $\underline{\vec{r}} \in \Gamma$ for a function mapping every $x : A \in \Gamma$ to a numeral sequence $\underline{\vec{r}}_x$ for the type $A$.

We give two translations of Linear A expressions into $\lambda$LL, the translation $\delta$ defined on the top of the grammar of Linear A (Section 4.1), and the translation $\delta^{\mathsf{B}}$ dedicated to the grammar of Linear B (Section 4.2). This latter being a fragment of Linear A, $\delta$ and $\delta^{\mathsf{B}}$ basically coincide on Linear B (Proposition 8). The translation $\delta^{\mathsf{B}}$ however eases the notation as it takes advantage from the three-sorted grammar of Linear B.

## 4.1 Translation of Linear A into λLL

Consider a Linear A judgement $x_1 : \tau_1, \ldots, x_n : \tau_n; \dot{y}_1 : \sigma_1, \ldots, \dot{y}_m : \sigma_m \vdash^{\mathrm{Jax}} e : (\tau, \sigma)$. The idea of the translation $\delta(e)$ is to associate the "primal operators" of $e$ with the multiplicative operators of λLL and the "tangent operators" with the additive operators. There are however some subtleties. First, the "primal part" is scattered with exponential modalities, enabling the duplication/erasing of primal values, according to the call-by-value translation of λ-calculus into LL (see e.g. [82]). Namely, a free primal variable $x_i : \tau_i$ of $e$ is associated with a $!x_i$ pattern of type $!\mathtt{p}(\tau_i)$ in $\delta(e)$. Second, the "tangent part" of $e$ is represented as a linear map $(\mathtt{t}(\sigma_1) \,\&\, \cdots \,\&\, \mathtt{t}(\sigma_m)) \multimap \mathtt{t}(\sigma)$ incorporating the free tangent variables of $e$ as parameters of the map. This map is a kind of matrix representing the tangent part of $e$. Finally, this map is encapsulated by the affine modality (so getting a final type $\S((\mathtt{t}(\sigma_1) \,\&\, \cdots \,\&\, \mathtt{t}(\sigma_m)) \multimap \mathtt{t}(\sigma))$ for the "tangent part" of $e$) allowing for discharging it whenever not necessary. This is a useful hack to compactly represent very sparse matrices, as the ones hinted by Autodiff (see Remark 7).

Technically, the definition of $\delta(e)$ depends on a function $\rho$ associating the free primal variables in $e$ to λLL variables and an enumeration $\theta$ of the set $FV^t(e)$ of the free tangent variables in $e$. Let us ease the notation by adopting the convention of using the same name for the primal variables and their associated λLL variables, so that we can omit to explicit $\rho$ and simply write $\delta_\theta(e)$. Figure 4.1 gives the definition of $\delta_\theta(e)$ by structural induction on $e$, using the notational conventions of the previous sections. In particular, given the enumeration $\theta = (\dot{y_1} : \sigma_1, \ldots, \dot{y_m} : \sigma_m)$ of $FV^t(e)$, by a slight extension of the notation given in Section 3.1, we will denote by $\&\mathtt{t}(\theta)$ the type $\&(\mathtt{t}(\sigma_1), \ldots, \mathtt{t}(\sigma_m)) = \mathtt{t}(\sigma_1) \& \cdots \& \mathtt{t}(\sigma_m)$.

**Remark 7.** Tangent computations essentially consist into matrix multiplications, here implemented by the specialised sum $\dot{+}$ and product $\dot{*}$ on the elements of the &-sequence types. However, automatic differentiation has an essential feature that makes it different from just implementing matrix multiplication: the matrices considered are structured by blocks determined by the program structure and the multiplications do happen at the level of these blocks, not on the whole matrices. For instance, using the notation from the definition of $\delta_\theta(\mathtt{let}\ (x; \dot{y}) = e_1\ \mathtt{in}\ e_2)$ in Figure 4.1, we want that the variable $f$, which denotes the tangent computation corresponding to $e_1$, operates solely on the block of the additive tuple associated with the $\dot{y}$ input of $e_2$, rather than on all of its inputs. Moreover, it is worth noting that tangent computations, as the one referenced by $f$, are never duplicated. To reflect this invariant, Section 3.1 introduces a specialized affine modality $\S A \overset{\text{def}}{=} \mathtt{1} \,\&\, A$, which admits weakening but prohibits contraction. Accordingly, function types associated with tangent computations are wrapped with § rather than the standard exponential modality !. The definition of $\delta_\theta(\mathtt{let}\ (x; \dot{y}) = e_1\ \mathtt{in}\ e_2)$ in Figure 4.1 also highlights the necessity of the § modality on the type of $f$ for ensuring well-typedness. Specifically, the type derivation of the subterm $\langle f y_1, y_2 \rangle$ requires weakening the typing context of $y_2$ in order to introduce $\S f$, a step justified by the properties of the affine modality.

**Proposition 5** (Type $\delta$)**.** Given $x_1 : \tau_1, \ldots, x_n : \tau_n; \dot{y_1} : \sigma_1, \ldots, \dot{y_m} : \sigma_m \vdash^{\mathrm{Jax}} e : (\tau; \sigma)$ and an enumeration $\theta = (\dot{y_1} : \sigma_1, \ldots, \dot{y_m} : \sigma_m)$ of the set of the free tangent variables in $e$, then $\delta_\theta(e)$ is a well-typed term in λLL such that:

$$!x_1 : !\mathtt{p}(\tau_1), \ldots, !x_n : !\mathtt{p}(\tau_n) \vdash \delta_\theta(e) : !\mathtt{p}(\tau) \otimes \S((\&_{i=1}^{m} \mathtt{t}(\sigma_i)) \multimap \mathtt{t}(\sigma))$$

*Sketch Proof.* By structural induction on $e$. □

### 4.1.1 Soundness

The soundness of $\delta$ can be formally stated point-wise, by proving that $\delta(e)$ returns a term computing $!\underline{\vec{r}} \mapsto ![e]^{\mathtt{p}}_{\underline{\vec{r}}}$ and $!\underline{\vec{r}}, \underline{\vec{s}} \mapsto [e]^{\mathtt{t}}_{\underline{\vec{r}};\underline{\vec{s}}}$ for every numeral sequences $\underline{\vec{r}}$ and $\underline{\vec{s}}$ associated with, respectively, the primal and tangent free variables in $e$.

$$
\begin{aligned}
\delta_\theta((x;\dot{y})) &\overset{\text{def}}{=} (!x, \S(\lambda y^{\&\mathtt{t}(\theta)}.y)) \\
\delta_\theta(\mathtt{let}\ (x;\dot{y}) = e_1\ \mathtt{in}\ e_2) &\overset{\text{def}}{=} \mathtt{let}\ (!x, \S f) = \delta_{\theta \cap FV^t(e_1)}(e_1)\ \mathtt{in} \\
&\quad \mathtt{let}\ (!z, \S g) = \delta_{\dot{y}, \theta \cap FV^t(e_2)}(e_2)\ \mathtt{in} \\
&\quad \left( !z, \S \left( \lambda y^{\&\mathtt{t}(\theta)}. \begin{array}{l} \mathtt{let}\ \langle y_1, y_2 \rangle = \sigma^{\&\mathtt{t}(\theta)}_{FV^t(e_1)} y\ \mathtt{in} \\ g(\overline{\sigma}^{\&\mathtt{t}(\dot{y}, \theta \cap FV^t(e_2))}_{\dot{y}} \langle f y_1, y_2 \rangle) \end{array} \right) \right) \\
\delta_\theta(\otimes()) &\overset{\text{def}}{=} (!(\,), \S(\lambda y^\top.\langle\,\rangle)) \\
\delta_\theta(\otimes(x_1, x_2)) &\overset{\text{def}}{=} (!(!x_1, !x_2), \S(\lambda y^\top.\langle\,\rangle)) \\
\delta_\theta(\mathtt{let}\ \otimes() = z\ \mathtt{in}\ e) &\overset{\text{def}}{=} \mathtt{let}\ (\,) = z\ \mathtt{in}\ \delta_\theta(e) \\
\delta_\theta(\mathtt{let}\ \otimes(x_1, x_2) = z\ \mathtt{in}\ e) &\overset{\text{def}}{=} \mathtt{let}\ (!x_1, !x_2) = z\ \mathtt{in}\ \delta_\theta(e) \\
\delta_\theta(\dot{\otimes}()) &\overset{\text{def}}{=} (!(\,), \S(\lambda y^\top.\langle\,\rangle)) \\
\delta_\theta(\dot{\otimes}(\dot{x_1}, \dot{x_2})) &\overset{\text{def}}{=} (!(\,), \S(\lambda y^{\&\mathtt{t}(\theta)}.y)) \\
\delta_\theta(\mathtt{let}\ \dot{\otimes}() = \dot{z}\ \mathtt{in}\ e) &\overset{\text{def}}{=} \mathtt{let}\ (!x, \S f) = \delta_{\theta \setminus \dot{z}}(e)\ \mathtt{in} \\
&\quad (!x, \S(\lambda y^{\&\mathtt{t}(\theta)}.\mathtt{let}\ \langle z, y' \rangle = \sigma^{\&\mathtt{t}(\theta)}_{\{\theta(\dot{z})\}} y\ \mathtt{in}\ f y')) \\
\delta_\theta(\mathtt{let}\ \dot{\otimes}(\dot{x_1}, \dot{x_2}) = \dot{z}\ \mathtt{in}\ e) &\overset{\text{def}}{=} \mathtt{let}\ (!x, \S f) = \delta_{\dot{x_1}, \dot{x_2}, \theta \setminus \dot{z}}(e)\ \mathtt{in} \\
&\quad \left( !x, \S \left( \lambda y^{\&\mathtt{t}(\theta)}. \begin{array}{l} \mathtt{let}\ \langle \langle x_1, x_2 \rangle, y' \rangle = \sigma_{\{\theta(\dot{z})\}} y\ \mathtt{in} \\ f(\overline{\sigma}^{\&\mathtt{t}(\dot{x_1}, \dot{x_2}, \theta \setminus \dot{z})}_{\dot{x_1}, \dot{x_2}} \langle x_1, x_2, y' \rangle) \end{array} \right) \right) \\
\delta_\theta(\underline{r}) &\overset{\text{def}}{=} (!\underline{r}, \S(\lambda y^\top.\langle\,\rangle)) \\
\delta_\theta(\underline{f}(x_1, x_2)) &\overset{\text{def}}{=} (\underline{f}(!x_1, !x_2), \S(\lambda y^\top.\langle\,\rangle)) \\
\delta_\theta(\dot{0}_\sigma) &\overset{\text{def}}{=} (!(\,), \S(\lambda y^\top.\underline{0}_{\mathtt{t}(\sigma)})) \\
\delta_\theta(\dot{y_1} \dot{+} \dot{y_2}) &\overset{\text{def}}{=} (!(\,), \S(\lambda y^{\&\mathtt{t}(\theta)}.\dot{+}_{\&\mathtt{t}(\theta)} y)) \\
\delta_\theta(x \dot{*} \dot{y}) &\overset{\text{def}}{=} (!(\,), \S(\lambda y^{\&\mathtt{t}(\theta)}.\dot{*}_{\&\mathtt{t}(\theta)} x y)) \\
\delta_\theta(\mathrm{dup}(\dot{y})) &\overset{\text{def}}{=} (!(\,), \S(\lambda y^{\&\mathtt{t}(\theta)}.\langle y, y \rangle)) \\
\delta_\theta(\mathrm{drop}(e)) &\overset{\text{def}}{=} \mathtt{let}\ (!x, \S f) = \delta_\theta(e)\ \mathtt{in}\ (!(\,), \S(\lambda y^{\&\mathtt{t}(\theta)}.\mathtt{let}\ z = f y\ \mathtt{in}\ \langle\,\rangle))
\end{aligned}
$$

Figure 4.1: Translation $\delta_\theta(e)$ into λLL of a Linear A expression $\Gamma; \dot{\Gamma} \vdash^{\text{Jax}} e : (\tau, \sigma)$ given en enumeration $\theta$ of $\dot{\Gamma}$. Note that in this case $\overline{\sigma}$ can be the identity or neutrality, in case of identity we can omit it.

**Proposition 6** (Soundness $\delta$)**.** Given $\Gamma; \dot{\Sigma} \vdash^{\text{Jax}} e : (\tau; \sigma)$, an enumeration $\theta$ of the tangent variables in $\dot{\Sigma}$, then:

- $\forall \underline{\vec{r}}$ for $\Gamma$: $\delta_\theta(e)[!\underline{\vec{r}}/\mathtt{p}(\Gamma)] \to^* (!\llbracket e \rrbracket^{\mathtt{p}}_{\underline{\vec{r}}}, \S F)$,
- $\forall \underline{\vec{s}}$ for the type $\&\theta$: $F\underline{\vec{s}} \to^* \llbracket e \rrbracket^{\mathtt{t}}_{\underline{\vec{r}};\underline{\vec{s}}}$.

*Proof.* We proceed by induction on $e$. An interesting case is $e = \mathtt{let}\ (x; \dot{y}) = e_1\ \mathtt{in}\ e_2$ which is well-typed by the judgement $\Gamma_1 \cup \Gamma_2; \dot{\Gamma}_1, \dot{\Gamma}_2 \vdash^{\text{Jax}} \mathtt{let}\ (x;\dot{y}) = e_1\ \mathtt{in}\ e_2 : (\tau; \sigma)$. We fix a numeral sequence $!\underline{\vec{r}}$ for $\mathtt{p}(\Gamma_1), \mathtt{p}(\Gamma_2)$ and we can observe that $!\underline{\vec{r}}|FV(e_i)$ is a numeral sequence for $\mathtt{p}(\Gamma_i)$. We recall from Subsection 2.1.2 that

$$\llbracket \mathtt{let}\ (x;\dot{y}) = e_1\ \mathtt{in}\ e_2 \rrbracket^{\mathtt{p}}_{\underline{\vec{r}}} = \llbracket e_2 \rrbracket^{\mathtt{p}}_{\underline{\vec{r}}|FV(e_2), x \mapsto \llbracket e_1 \rrbracket^{\mathtt{p}}_{\underline{\vec{r}}|FV(e_1)}} \tag{4.1}$$

$$\llbracket \mathtt{let}\ (x;\dot{y}) = e_1\ \mathtt{in}\ e_2 \rrbracket^{\mathtt{t}}_{\underline{\vec{r}};\underline{\vec{s}}} = \llbracket e_2 \rrbracket^{\mathtt{t}}_{\underline{\vec{r}}|FV(e_2), x \mapsto \llbracket e_1 \rrbracket^{\mathtt{p}}_{\underline{\vec{r}}|FV(e_1)}; \underline{\vec{s}}|FV^t(e_2), \dot{y} \mapsto \llbracket e_1 \rrbracket^{\mathtt{t}}_{\underline{\vec{s}}|FV^t(e_1)}} \tag{4.2}$$

and we proceed as follows to prove the first claim

$$\delta_\theta(e)[!\underline{\vec{r}}/\mathtt{p}(\Gamma)] \stackrel{\text{def}}{=} \begin{pmatrix} \mathtt{let}\ (!x, \S f) = \delta_{\theta \cap FV^t(e_1)}(e_1)\ \mathtt{in} \\ \mathtt{let}\ (!z, \S g) = \delta_{\dot{y}, \theta \cap FV^t(e_2)}(e_2)\ \mathtt{in} \\ (!z, \S F) \end{pmatrix} [!\underline{\vec{r}}/\mathtt{p}(\Gamma)] \tag{4.3}$$

$$\begin{aligned} & \mathtt{let}\ (!x, \S f) = \delta_{\theta \cap FV^t(e_1)}(e_1)[!\underline{\vec{r}}|FV(e_1)/\mathtt{p}(\Gamma_1)]\ \mathtt{in} \\ = & \mathtt{let}\ (!z, \S g) = \delta_{\dot{y}, \theta \cap FV^t(e_2)}(e_2)[!\underline{\vec{r}}|FV(e_2)/\mathtt{p}(\Gamma_2)]\ \mathtt{in} \\ & (!z, \S F) \end{aligned} \tag{4.4}$$

$$\begin{aligned} & \mathtt{let}\ (!x, \S f) = (!\llbracket e_1 \rrbracket^{\mathtt{p}}_{\underline{\vec{r}}|FV(e_1)}, \S F_1)\ \mathtt{in} \\ \to^* & \mathtt{let}\ (!z, \S g) = (!\llbracket e_2 \rrbracket^{\mathtt{p}}_{\underline{\vec{r}}|FV(e_2)}, \S F_2)\ \mathtt{in} \\ & (!z, \S F) \end{aligned} \tag{4.5}$$

$$\to^* \begin{array}{l} \mathtt{let}\ (!x, \S f) = (!\llbracket e_1 \rrbracket^{\mathtt{p}}_{\underline{\vec{r}}|FV(e_1)}, \S F_1)\ \mathtt{in} \\ (!\llbracket e_2 \rrbracket^{\mathtt{p}}_{\underline{\vec{r}}|FV(e_2)}, \S F\{\S F_2/g\}) \end{array} \tag{4.6}$$

$$\to^* (!\llbracket e_2 \rrbracket^{\mathtt{p}}_{\underline{\vec{r}}|FV(e_2), x \mapsto \llbracket e_1 \rrbracket^{\mathtt{p}}_{\underline{\vec{r}}|FV(e_1)}}, \S F\{\S F_2/g, \S F_1/f\}) \tag{4.7}$$

where $\S F\{\S F_2/g, \S F_1/f\} = \S(\lambda y^{\&\mathtt{t}(\theta)}.\mathtt{let}\ \langle y_1, y_2 \rangle = \sigma^{\&\mathtt{t}(\theta)}_{FV^t(e_1)} y\ \mathtt{in}\ F_2\langle F_1 y_1, y_2 \rangle)$. The passage from line (4.4) to (4.5) is by induction hypothesis and the passage from line (4.5) (resp. (4.6)) to (4.6) (resp. (4.7)) is obtained by applying $\beta_\lambda$. We can conclude by observing that the first term of the tuple in (4.7) is equal to (4.1).

In order to prove the second claim, we fix $\underline{\vec{s}}$ for the type $\&\theta$ and and we can observe that $\underline{\vec{s}}|FV^t(e_i)$ is a numeral sequence for $\mathtt{t}(\dot{\Gamma}_i)$. Formally, we proceed as follows

$$(\lambda y^{\&\mathtt{t}(\theta)}.\mathtt{let}\ \langle y_1, y_2 \rangle = \sigma^{\&\mathtt{t}(\theta)}_{FV^t(e_1)} y\ \mathtt{in}\ F_2\langle F_1 y_1, y_2 \rangle)\underline{\vec{s}} \tag{4.8}$$

$$\to \mathtt{let}\ \langle y_1, y_2 \rangle = \sigma^{\&\mathtt{t}(\theta)}_{FV^t(e_1)} \underline{\vec{s}}\ \mathtt{in}\ F_2\langle F_1 y_1, y_2 \rangle \tag{4.9}$$

$$= \mathtt{let}\ \langle y_1, y_2 \rangle = (\lambda \langle x_i \rangle_{i=1}^n . \langle \langle x_i \rangle_{i \in FV^t(e_1)}, \langle x_i \rangle_{i \notin FV^t(e_1)} \rangle)\underline{\vec{s}}\ \mathtt{in}\ F_2\langle F_1 y_1, y_2 \rangle \tag{4.10}$$

$$= \mathtt{let}\ \langle y_1, y_2 \rangle = (\lambda \langle x_i \rangle_{i=1}^n . \langle \langle x_i \rangle_{i \in FV^t(e_1)}, \langle x_i \rangle_{i \in FV^t(e_2)} \rangle)\underline{\vec{s}}\ \mathtt{in}\ F_2\langle F_1 y_1, y_2 \rangle \tag{4.11}$$

$$\to \mathtt{let}\ \langle y_1, y_2 \rangle = \langle \underline{\vec{s}}|FV^t(e_1), \underline{\vec{s}}|FV^t(e_2) \rangle\ \mathtt{in}\ F_2\langle F_1 y_1, y_2 \rangle \tag{4.12}$$

$$\to^* F_2\langle F_1\underline{\vec{s}}|FV^t(e_1), \underline{\vec{s}}|FV^t(e_2)\rangle \tag{4.13}$$

$$\to^* F_2\langle [\![e_1]\!]^{\mathtt{t}}_{\underline{\vec{r}};\underline{\vec{s}}|FV^t(e_1)}, \underline{\vec{s}}|FV^t(e_2)\rangle \tag{4.14}$$

$$\to^* [\![e_2]\!]^{\mathtt{t}}_{\underline{\vec{r}}|FV(e_2),x\mapsto[\![e_1]\!]^{\mathtt{p}}_{\underline{\vec{r}}|FV(e_1)};\underline{\vec{s}}|FV^t(e_2),\dot{y}\mapsto[\![e_1]\!]^{\mathtt{t}}_{\underline{\vec{s}}|FV^t(e_1)}} \tag{4.15}$$

The passage from line (4.8) to (4.9) is obtained by applying $\beta_\lambda$.
The passage from line (4.9) to (4.10) is obtained by definition of splitting term $\sigma$.
The passage from line (4.10) to (4.11) is obtained by observing that a variable $x$ such that $x \in \mathtt{t}(\theta)$ and $x \notin FV^t(e_1)$ is a free tangent variable in $e_2$.
The passage from line (4.11) (resp. (4.12)) to (4.12) (resp. (4.13)) is obtained by applying $\beta_\lambda$.
The passage from line (4.13) (resp. (4.14)) to (4.14) (resp. (4.15)) is obtained by inductive hypothesis on $e_1$ (resp. $e_2$).
We can conclude by observing that (4.15) is equal to (4.2). □

### 4.1.2 Work Preservation

In Section 2.1.2, the workload of a JAX type $\tau$, denoted by $\mathcal{W}^{\mathtt{Jax}}(\tau)$, is defined as the number of occurrences of the base type $\mathbb{R}$ within $\tau$. The following lemma establishes a correspondence between the type translation $\mathtt{t}(\cdot)$ and the workload.

**Lemma 25.** Given $\tau$ in the grammar of JAX Types, then $\mathcal{W}^{\mathtt{Jax}}(\tau) = \mathcal{W}(\mathtt{t}(\tau))$.

*Proof.* By induction on $\tau$. □

We must also ensure that the translation $\delta(e)$ computes $[\![e]\!]^{\mathtt{p}}_{\underline{\vec{r}}}$ and $[\![e]\!]^{\mathtt{t}}_{\underline{\vec{r}};\underline{\vec{s}}}$ efficiently — specifically, without performing more flops than those required by the original Linear A expression $e$. This property is essential to guarantee that the translation preserves the computational efficiency. Notice that $\delta(e)$ satisfies the conditions of Definition 4, so by Proposition 4 we can use the workload of a term as a bound to the number of numeric steps.

**Proposition 7** (Workload $\delta$)**.** Given $\Gamma; \dot{\Gamma} \vdash^{\mathtt{Jax}} e : (\tau; \sigma)$, an enumeration $\theta$ of $\dot{\Gamma}$, then $\delta_\theta(e)$ is safe and $\mathcal{W}(\delta_\theta(e)) \leq \mathcal{W}^{\mathtt{Jax}}(e)$.

*Proof.* The safeness of $\delta_\theta(e)$ is easy to prove by induction on $e$ simply checking the items in Definition 4. Let us focus on the proof related to work preservation of $\delta_\theta(e)$, we proceed by induction on $e$.

- Case $e = \mathtt{let}\ (x; \dot{y}) = e_1\ \mathtt{in}\ e_2$:

$$\mathcal{W}^{\mathtt{Jax}}(e) = \mathcal{W}^{\mathtt{Jax}}(\mathtt{let}\ (x;\dot{y}) = e_1\ \mathtt{in}\ e_2) = \mathcal{W}^{\mathtt{Jax}}(e_1) + \mathcal{W}^{\mathtt{Jax}}(e_2)$$

$$\mathcal{W}(\delta_\theta(e)) = \mathcal{W}\begin{pmatrix} \mathtt{let}\ (!x, \S f) = \delta_{\theta\cap FV^t(e_1)}(e_1)\ \mathtt{in} \\ \mathtt{let}\ (!z, \S g) = \delta_{\dot{y},\theta\cap FV^t(e_2)}(e_2)\ \mathtt{in} \\ (!z, \S(\lambda y^{\&\mathtt{t}(\theta)}.\mathtt{let}\ \langle y_1, y_2\rangle = \sigma^{\&\mathtt{t}(\theta)}_{FV^t(e_1)} y\ \mathtt{in}\ g\langle f y_1, y_2\rangle)) \end{pmatrix}$$

  Observe that our workload essentially counts the number of numerical operations not under a ! and the number of possible numerals erased during a reduction. In this case nothing is erased, so the sums related to the workload for the two let-constructs and for the $\lambda$-abstraction are equal to zero. This means that $\mathcal{W}(\delta_\theta(e))$ is equal to $\mathcal{W}(\delta_{\theta\cap FV^t(e_1)}(e_1)) + \mathcal{W}(\delta_{\dot{y},\theta\cap FV^t(e_2)}(e_2))$ and we can conclude by inductive hypotheses.

- Case $e = \mathrm{drop}(e_1)$:
  By hypothesis $\Gamma; \dot{\Gamma} \vdash^{\text{Jax}} \mathrm{drop}(e_1) : (\mathtt{1}; \mathtt{1})$ and by typing $\Gamma; \dot{\Gamma} \vdash^{\text{Jax}} e_1 : (\tau_1; \sigma_1)$.
  By definition we have:

$$\mathcal{W}^{\mathtt{Jax}}(e) = \mathcal{W}^{\mathtt{Jax}}(\mathrm{drop}(e_1)) = \mathcal{W}^{\mathtt{Jax}}(e_1) + \mathcal{W}^{\mathtt{Jax}}(\tau_1) + \mathcal{W}^{\mathtt{Jax}}(\sigma_1)$$

$$\begin{aligned}\mathcal{W}(\delta_\theta(e)) &= \mathcal{W}(\mathtt{let}\ (!x, \S f) = \delta_\theta(e)\ \mathtt{in}\ ((\,), \S(\lambda y^{\&\mathtt{t}(\theta)}.\mathtt{let}\ z = fy\ \mathtt{in}\ \langle\ \rangle))) \\ &= \mathcal{W}(\delta_{\theta \cap FV^t(e_1)}(e_1)) + \mathcal{W}(\mathtt{t}(\sigma_1))\end{aligned}$$

  Observe that we do not count the workload associated to the primal output of $e_1$.
  By Lemma 25 we have that $\mathcal{W}^{\mathtt{Jax}}(\sigma_1) = \mathcal{W}(\mathtt{t}(\sigma_1))$ and we can conclude.

- Case $e = \underline{r}$:

$$\begin{aligned}\mathcal{W}^{\mathtt{Jax}}(e) &= \mathcal{W}^{\mathtt{Jax}}(\underline{r}) = 1 \\ \mathcal{W}(\delta_\theta(e)) &= \mathcal{W}((!\underline{r}, \S(\lambda y^{\top}.\langle\ \rangle))) = 0\end{aligned}$$

- Case $e = \dot{0}_\sigma$:

$$\begin{aligned}\mathcal{W}^{\mathtt{Jax}}(e) &= \mathcal{W}^{\mathtt{Jax}}(\dot{0}_\sigma) = 1 + \mathcal{W}^{\mathtt{Jax}}(\sigma) \\ \mathcal{W}(\delta_\theta(e)) &= \mathcal{W}((!(\,), \S(\lambda y^{\top}.\underline{0}_{\mathtt{t}(\sigma)}))) = 0\end{aligned}$$

- All the other cases are simple and direct.

□

It is worth noting that the cost of evaluating an expression is preserved by AD transformations in Autodiff (see Claim 2 of Theorems 1, 2, 3 in Section 2.2).

## 4.2 Translation of Linear B into $\lambda$LL

We can simplify the definition of $\delta$ on the fragment Linear B, by taking advantage of its three-sorted grammar. The three-sorted grammar of Linear B identifies a class of "purely primal expressions" $e^p$ in the subgrammar (Primal) and a class of "purely tangent expressions" $\dot{e}$ in the subgrammar (Tangent) and then the mixing $d$ of the two in the grammar (Linear B). One can take advantage of this structure in order to define a translation $\delta^{\mathsf{B}}$ of Linear B into $\lambda$LL that is more lightweight than the one provided by $\delta$. Specifically, $\delta^{\mathsf{B}}$ may be obtained by extracting a single component from the output pair produced by $\delta$.

Figure 4.2 defines the translation $\delta^{\mathsf{B}}_\theta(d)$ of a Linear B expression $d$, on the top of the definitions of $\delta^{\mathsf{B}}(e^p)$, given a primal expression $e^p$ (Figure 4.2a) and of $\delta^{\mathsf{B}}_\theta(\dot{e})$, given a tangent expression $\dot{e}$ (Figure 4.2b). Notice that $\delta^{\mathsf{B}}(e^p)$ omits the index $\theta$ on $e^p$ as primal expressions have no free tangent variables. The definition of $\delta^{\mathsf{B}}$ on purely primal expressions $e^p$ is simple and just commutes with all operators except on tuples where one has to manage exponentials (see Figure 4.2a). The cases for the purely tangent $\dot{e}$ expression are more involved and depends on an enumeration $\theta$ of the free tangent variables, exactly as $\delta$ (see Figure 4.2b)

The following proposition states the type of the translation $\delta^{\mathsf{B}}$ and relates translation $\delta$ with translation $\delta^{\mathsf{B}}$

$$
\begin{aligned}
\delta^{\mathsf{B}}(x) &\stackrel{\text{def}}{=} !x \\
\delta^{\mathsf{B}}(\mathtt{let}\ x = e_1^p\ \mathtt{in}\ e_2^p) &\stackrel{\text{def}}{=} \mathtt{let}\ !x = \delta^{\mathsf{B}}(e_1^p)\ \mathtt{in}\ \delta^{\mathsf{B}}(e_2^p) \\
\delta^{\mathsf{B}}(\otimes()) &\stackrel{\text{def}}{=} !(\,) \\
\delta^{\mathsf{B}}(\otimes(e_1^p, e_2^p)) &\stackrel{\text{def}}{=} !(\delta^{\mathsf{B}}(e_1^p), \delta^{\mathsf{B}}(e_2^p)) \\
\delta^{\mathsf{B}}(\mathtt{let}\ \otimes() = z\ \mathtt{in}\ e^p) &\stackrel{\text{def}}{=} \mathtt{let}\ (\,) = z\ \mathtt{in}\ \delta^{\mathsf{B}}(e^p) \\
\delta^{\mathsf{B}}(\mathtt{let}\ \otimes(x_1, x_2) = z\ \mathtt{in}\ e^p) &\stackrel{\text{def}}{=} \mathtt{let}\ (!x_1, !x_2) = z\ \mathtt{in}\ \delta^{\mathsf{B}}(e^p) \\
\delta^{\mathsf{B}}(\underline{r}) &\stackrel{\text{def}}{=} !\underline{r} \\
\delta^{\mathsf{B}}(\mathrm{f}(x_1, \dots, x_n)) &\stackrel{\text{def}}{=} \underline{f}(!x_1, \dots, !x_n) \\
\delta^{\mathsf{B}}(\mathrm{drop}(e^p)) &\stackrel{\text{def}}{=} \mathtt{let}\ !x = \delta^{\mathsf{B}}(e^p)\ \mathtt{in}\ !(\,)
\end{aligned}
$$

(a) Definition of $\delta^{\mathsf{B}}(e^p)$, given an expression $e^p$ in (Primal). The enumeration $\theta$ of $FV^t(e^p)$ is empty.

$$
\begin{aligned}
\delta^{\mathsf{B}}_\theta(\dot{x}) &\stackrel{\text{def}}{=} \lambda y^{\&\mathtt{t}(\theta)}.y \\
\delta^{\mathsf{B}}_\theta(\mathtt{let}\ \dot{y} = \dot{e_1}\ \mathtt{in}\ \dot{e_2}) &\stackrel{\text{def}}{=} \mathtt{let}\ \S f = \S\delta^{\mathsf{B}}_{\theta \cap FV^t(e_1)}(\dot{e_1})\ \mathtt{in} \\
&\qquad \mathtt{let}\ \S g = \S\delta^{\mathsf{B}}_{\dot{y}, \theta \cap FV^t(e_2)}(\dot{e_2})\ \mathtt{in} \\
&\qquad \lambda y^{\&\mathtt{t}(\theta)}.\mathtt{let}\ \langle y_1, y_2 \rangle = \sigma_{FV^t(e_1)} y\ \mathtt{in}\ g(\overline{\sigma}_{\dot{y}}^{\&\mathtt{t}(\dot{y}, \theta \cap FV^t(e_2))} \langle f y_1, y_2 \rangle) \\
\delta^{\mathsf{B}}_\theta(\dot{\otimes}()) &\stackrel{\text{def}}{=} \lambda y^{\top}.\langle\ \rangle \\
\delta^{\mathsf{B}}_\theta(\dot{\otimes}(\dot{e_1}, \dot{e_2})) &\stackrel{\text{def}}{=} \mathtt{let}\ \S f = \S\delta^{\mathsf{B}}_{\theta \cap FV^t(e_1)}(\dot{e_1})\ \mathtt{in} \\
&\qquad \mathtt{let}\ \S g = \S\delta^{\mathsf{B}}_{\theta \cap FV^t(e_2)}(\dot{e_2})\ \mathtt{in} \\
&\qquad \lambda y^{\&\mathtt{t}(\theta)}.\mathtt{let}\ \langle y_1, y_2 \rangle = \sigma_{FV^t(e_1)} y\ \mathtt{in}\ \langle f y_1, g y_2 \rangle \\
\delta^{\mathsf{B}}_\theta(\mathtt{let}\ \dot{\otimes}() = \dot{z}\ \mathtt{in}\ e) &\stackrel{\text{def}}{=} \mathtt{let}\ \S f = \S\delta^{\mathsf{B}}_{\theta \setminus \dot{z}}(e)\ \mathtt{in} \\
&\qquad \lambda y^{\&\mathtt{t}(\theta)}.\mathtt{let}\ \langle z, y' \rangle = \sigma_{\{\theta(\dot{z})\}} y\ \mathtt{in}\ f y' \\
\delta^{\mathsf{B}}_\theta(\mathtt{let}\ \dot{\otimes}(\dot{x_1}, \dot{x_2}) = \dot{z}\ \mathtt{in}\ e) &\stackrel{\text{def}}{=} \mathtt{let}\ \S f = \S\delta^{\mathsf{B}}_{\dot{x_1}, \dot{x_2}, \theta \setminus \dot{z}}(e)\ \mathtt{in} \\
&\qquad \lambda y^{\&\mathtt{t}(\theta)}.\mathtt{let}\ \langle \langle x_1, x_2 \rangle, y' \rangle = \sigma_{\{\theta(\dot{z})\}} y\ \mathtt{in}\ f(\overline{\sigma}_{\dot{x_1}, \dot{x_2}}^{\&\mathtt{t}(\dot{x_1}, \dot{x_2}, \theta \setminus \dot{z})} \langle x_1, x_2, y' \rangle) \\
\delta^{\mathsf{B}}_\theta(\mathrm{dup}(\dot{x})) &\stackrel{\text{def}}{=} \lambda y^{\&\mathtt{t}(\theta)}.\langle y, y \rangle \\
\delta^{\mathsf{B}}_\theta(\dot{0}_\sigma) &\stackrel{\text{def}}{=} \lambda y^{\top}.0_{\mathtt{t}(\sigma)} \\
\delta^{\mathsf{B}}_\theta(\dot{x_1} \dot{+} \dot{x_2}) &\stackrel{\text{def}}{=} \lambda y^{\&\mathtt{t}(\theta)}.\dot{+} y \\
\delta^{\mathsf{B}}_\theta(x \dot{*} \dot{y}) &\stackrel{\text{def}}{=} \lambda y^{\&\mathtt{t}(\theta)}.\dot{*}(\bar{!}x, y) \\
\delta^{\mathsf{B}}_\theta(\mathrm{drop}(\dot{e})) &\stackrel{\text{def}}{=} \mathtt{let}\ \S f = \S\delta^{\mathsf{B}}_\theta(\dot{e})\ \mathtt{in} \\
&\qquad \lambda y^{\&\mathtt{t}(\theta)}.\mathtt{let}\ z = f y\ \mathtt{in}\ \langle\ \rangle
\end{aligned}
$$

(b) Definition of $\delta^{\mathsf{B}}_\theta(\dot{e})$, given an expression $\dot{e}$ in Tangent and en enumeration $\theta$ of $FV^t(\dot{e})$.

$$
\begin{aligned}
\delta^{\mathsf{B}}_\theta((e^p; \dot{e})) &\stackrel{\text{def}}{=} (\delta^{\mathsf{B}}(e^p), \S\delta^{\mathsf{B}}_\theta(\dot{e})) \\
\delta^{\mathsf{B}}_\theta(\mathtt{let}\ x = e^p\ \mathtt{in}\ d) &\stackrel{\text{def}}{=} \mathtt{let}\ !x = \delta^{\mathsf{B}}(e^p)\ \mathtt{in}\ \delta^{\mathsf{B}}_\theta(d) \\
\delta^{\mathsf{B}}_\theta(\mathtt{let}\ \otimes() = z\ \mathtt{in}\ d) &\stackrel{\text{def}}{=} \mathtt{let}\ (\,) = z\ \mathtt{in}\ \delta^{\mathsf{B}}_\theta(d) \\
\delta^{\mathsf{B}}_\theta(\mathtt{let}\ \otimes(x_1, x_2) = z\ \mathtt{in}\ d) &\stackrel{\text{def}}{=} \mathtt{let}\ (!x_1, !x_2) = z\ \mathtt{in}\ \delta^{\mathsf{B}}_\theta(d)
\end{aligned}
$$

(c) Definition of $\delta^{\mathsf{B}}_\theta(d)$, given an expression $d$ in Linear B and en enumeration $\theta$ of $FV^t(d)$.

Figure 4.2: Translation $\delta^{\mathsf{B}}$ of Linear B expressions. Note that in this case $\overline{\sigma}$ can be the identity or neutrality, in case of identity we can omit it.

**Proposition 8** (Type $\delta^{\mathsf{B}}$)**.** Given a Linear B expression of type $x_1 : \tau_1, \ldots, x_n : \tau_n; \dot{y_1} : \sigma_1, \ldots, \dot{y_m} : \sigma_m \vdash^{\text{Jax}} d : (\tau; \sigma)$ and an enumeration $\theta = (\dot{y_1} : \sigma_1, \ldots, \dot{y_m} : \sigma_m)$ of the free tangent variables of $d$, then $\delta^{\mathsf{B}}_\theta(d) \in \lambda$LL such that:

$$!x_1 : !\mathtt{p}(\tau_1), ..., !x_n : !\mathtt{p}(\tau_n) \vdash \delta^{\mathsf{B}}_\theta(d) : !\mathtt{p}(\tau) \otimes \S(\,(\&_{i=1}^m \mathtt{t}(\sigma_i)) \multimap \mathtt{t}(\sigma))\,.$$

Moreover, $\delta_\theta(d) \sim \delta^{\mathsf{B}}_\theta(d)$.

In order to prove the proposition above we need the following two auxiliary lemmas on the translation $\delta^{\mathsf{B}}$ applied to (Primal) and to (Tangent).

**Lemma 26** (Type Primal $\delta^{\mathsf{B}}$)**.** Given a Primal Linear B expression of type $x_1 : \tau_1, \ldots, x_n : \tau_n; \vdash^{\text{Jax}} e^p : (\tau; \mathtt{1})$, then $\delta^{\mathsf{B}}(e^p)$ is a well-typed $\lambda$LL term such that: $!x_1 : !\mathtt{p}(\tau_1), \ldots, !x_n : !\mathtt{p}(\tau_n) \vdash \delta^{\mathsf{B}}(e^p) : !\mathtt{p}(\tau)$ . Moreover, $\delta_{()}(e^p) \sim (\delta^{\mathsf{B}}(e^p), \S(\lambda y^\top.\langle\,\rangle))$.

*Sketch Proof.* Notice that the definition of $\delta^{\mathsf{B}}(e^p)$ on a primal expression $e^p$ is basically the identity on almost all operators, but the proof of this lemma is not immediate as the left hand-side of the definition in Figure 4.2a uses JAX syntactical sugar, while on the right-hand side we have true $\lambda$LL terms. □

**Lemma 27** (Type Tangent $\delta^{\mathsf{B}}$)**.** Given a Tangent Linear B expression of type $x_1 : \tau_1, \ldots, x_n : \tau_n; \dot{y_1} : \sigma_1, \ldots, \dot{y_m} : \sigma_m \vdash^{\text{Jax}} \dot{e} : (\mathtt{1}; \sigma)$ and an enumeration $\theta = (\dot{y_1} : \sigma_1, \ldots, \dot{y_m} : \sigma_m)$ of the free tangent variables of $\dot{e}$, then $\delta^{\mathsf{B}}_\theta(\dot{e})$ is a well-typed expression in $\lambda$LL such that: $!x_1 : !\mathtt{p}(\tau_1), \ldots, !x_n : !\mathtt{p}(\tau_n) \vdash \delta^{\mathsf{B}}_\theta(\dot{e}) : \S(\,(\&_{i=1}^m \mathtt{t}(\sigma_i)) \multimap \mathtt{t}(\sigma))$ . Moreover, $\delta_\theta(\dot{e}) \sim (!(\,), \S\delta^{\mathsf{B}}_\theta(\dot{e}))$.

*Sketch Proof.* By induction on $\dot{e}$. □

Finally, the following lemma establishes a connection between $\delta^{\mathsf{B}}$ and the stack $E$ of primal let-definitions employed in the definition of the unzipping transformation for Autodiff (Subsection 2.2.2). This correspondence is instrumental in facilitating the proof of soundness for our unzipping transformation, which will be demonstrated in Section 5.2.

**Lemma 28.** We have that $\delta^{\mathsf{B}}(E[(e^p; \dot{e})]) = \delta^{\mathsf{B}}(E)[(\delta^{\mathsf{B}}(e^p), \delta^{\mathsf{B}}(\dot{e}))]$.

*Sketch Proof.* By immediate induction on $E[]$. □

The next two lemmas play a key role in the proof of the soundness theorem for the transpose transformation in our setting. The first lemma establishes a correspondence between the fusion expression in JAX and the fusion term in $\lambda$LL, mediated by the translation $\delta^{\mathsf{B}}$.

**Lemma 29.** Given $\theta = (\dot{x}_1, \ldots, \dot{x}_n)$ and two partitions $\theta_1$ and $\theta_2$ of $\theta$ such that $\dot{y}_1 : \otimes\theta_1$ and $\dot{y}_2 : \otimes\theta_2$, then $\delta^{\mathsf{B}}_{\dot{y}_1,\dot{y}_2}(\overline{\sigma}^{\text{Jax}}_{\dot{y}_1,\dot{y}_2;\theta}) \sim \overline{\sigma}^\theta_{\theta_1}$

Furthermore, since the transpose transformation in Autodiff is defined using the syntactic sugar introduced for Linear B, while $\delta^{\mathsf{B}}$ operates directly on the core grammar of Linear B (excluding syntactic sugar), it is important to establish a correspondence between these two formulations. This relationship is formalized in the following lemma.

**Lemma 30.** We have the following:

1. Given two tangent JAX expressions $\Gamma_1; \dot{\Gamma}_1 \vdash^{\text{Jax}} \dot{e}_1 : (\mathtt{1}; \tau_1 \dot{\otimes} \sigma_1)$ and $\Gamma_2; \dot{\Gamma}_2, \dot{x}_1 : \tau_1, \dot{x}_2 : \sigma_1 \vdash^{\text{Jax}} \dot{e}_2 : (\mathtt{1}; \tau_2 \dot{\otimes} \sigma_2)$, and $\theta$ is an enumeration of $\dot{\Gamma}_1, \dot{\Gamma}_2$, then

$$\delta^{\mathsf{B}}_\theta(\mathtt{let}\ \dot{\otimes}(\dot{x_1}, \dot{x_2}) = \dot{e}_1\ \mathtt{in}\ \dot{e}_2) \sim \begin{array}{l} \mathtt{let}\ \S f_1 = \S\delta^{\mathsf{B}}_{\theta \cap FV^t(\dot{e}_1)}(\dot{e}_1)\ \mathtt{in} \\ \mathtt{let}\ \S f_2 = \S\delta^{\mathsf{B}}_{\dot{x}_1, \dot{x}_2, \theta \cap FV^t(\dot{e}_2)}(\dot{e}_2)\ \mathtt{in} \\ \lambda u^{\&\mathtt{t}(\theta)}.\mathtt{let}\ \langle u_1, u_2\rangle = \sigma^\theta_{FV^t(\dot{e}_1)} u\ \mathtt{in}\ f_2(\overline{\sigma}^\theta_{(\dot{x}_1, \dot{x}_2)}\langle f_1 u_1, u_2\rangle) \end{array}$$

where $\S f_1$ and $\S f_2$ have the following types:

$$\S f_1 : \S(\&\mathtt{t}(\theta \cap FV^t(\dot{e}_1)) \multimap \mathtt{t}(\tau_1)\&\mathtt{t}(\sigma_1))$$
$$\S f_2 : \S(\&\mathtt{t}(\dot{x}_1, \dot{x}_2, \theta \cap FV^t(\dot{e}_2)) \multimap \mathtt{t}(\tau_2)\&\mathtt{t}(\sigma_2))$$

2. Given two tangent JAX expressions $\Gamma_1; \dot{\Gamma}_1 \vdash^{\text{Jax}} \dot{e}_1 : (\mathtt{1}; \tau_1 \dot{\otimes} \sigma_1)$ and $\Gamma_2; \dot{\Gamma}_2 \vdash^{\text{Jax}} \dot{e}_2 : (\mathtt{1}; \tau_2 \dot{\otimes} \sigma_2)$, and $\theta$ is an enumeration of $\dot{\Gamma}_1, \dot{\Gamma}_2$, then

$$\delta^{\mathsf{B}}_\theta(\dot{\otimes}(\dot{e}_1, \dot{e}_2)) \sim \begin{array}{l} \mathtt{let}\ \S f_1 = \S\delta^{\mathsf{B}}_{\theta \cap FV^t(\dot{e}_1)}(\dot{e}_1)\ \mathtt{in} \\ \mathtt{let}\ \S f_2 = \S\delta^{\mathsf{B}}_{\theta \cap FV^t(\dot{e}_2)}(\dot{e}_2)\ \mathtt{in} \\ \lambda u^{\&\mathtt{t}(\theta)}.\mathtt{let}\ \langle u_1, u_2\rangle = \sigma^{\theta}_{FV^t(\dot{e}_1)} u\ \mathtt{in}\ \langle f_1 u_1, f_2 u_2\rangle \end{array}$$

where $\S f_1$ and $\S f_2$ have the following types:

$$\S f_1 : \S(\&\mathtt{t}(\theta \cap FV^t(\dot{e}_1)) \multimap \mathtt{t}(\tau_1)\&\mathtt{t}(\sigma_1))$$
$$\S f_2 : \S(\&\mathtt{t}(\theta \cap FV^t(\dot{e}_2)) \multimap \mathtt{t}(\tau_2)\&\mathtt{t}(\sigma_2))$$

*Sketch Proof.* The proof is very syntactic and not particularly interesting, so we postpone the proof of Claim 1 to the end of the chapter (jump to page 78 for more details). □

### 4.2.1 Soundness

Similarly to $\delta$, soundness can be proved for $\delta^{\mathsf{B}}$. More precisely, soundness of $\delta^{\mathsf{B}}$ follows as a corollary from Proposition 8 and the soundness of $\delta$ (Proposition 6).

**Corollary 3** (Soundness of $\delta^{\mathsf{B}}$)**.** Given a Linear B expression of type $\Gamma; \dot{\Sigma} \vdash^{\text{Jax}} d : (\tau; \sigma)$, an enumeration $\theta$ of the tangent variables in $\dot{\Sigma}$, then:

- for every numeral sequence $\underline{\vec{r}}$ for $\Gamma$: $\delta^{\mathsf{B}}_\theta(d)[!\underline{\vec{r}}/\mathtt{p}(\Gamma)] \to^* (![d]^{\mathtt{p}}_{\underline{\vec{r}}}, \S F)$,
- and moreover, for every numeral sequence $\underline{\vec{s}}$ for the type $\&\theta$: $F\underline{\vec{s}} \to^* [e]^{\mathtt{t}}_{\underline{\vec{r}};\underline{\vec{s}}}$.

### 4.2.2 Work Preservation

Similarly to Proposition 7, one can check the workload preservation property for $\delta^{\mathsf{B}}$ as in Proposition 9 by first proving the workload preservation of $\delta^{\mathsf{B}}$ on Primal and Tangent.

**Lemma 31** (Workload Primal $\delta^{\mathsf{B}}$)**.** Given a Primal Linear B expression of type $x_1 : \tau_1, \ldots, x_n : \tau_n; \vdash^{\text{Jax}} e^p : (\tau; \mathtt{1})$, then $\delta^{\mathsf{B}}(e^p)$ is safe and $\delta^{\mathsf{B}}(e^p) \leq \mathcal{W}^{\mathtt{Jax}}(e^p)$.

*Sketch Proof.* By induction on $e^p$. □

**Lemma 32** (Workload Tangent $\delta^{\mathsf{B}}$)**.** Given a Tangent Linear B expression of type $x_1 : \tau_1, \ldots, x_n : \tau_n; \dot{y}_1 : \sigma_1, \ldots, \dot{y}_m : \sigma_m \vdash^{\text{Jax}} \dot{e} : (\mathtt{1}; \sigma)$ and an enumeration $\theta = (\dot{y}_1 : \sigma_1, \ldots, \dot{y}_m : \sigma_m)$ of the free tangent variables of $\dot{e}$, then $\delta^{\mathsf{B}}_\theta(\dot{e})$ is safe and $\delta^{\mathsf{B}}(\dot{e}) \leq \mathcal{W}^{\mathtt{Jax}}(\dot{e})$.

*Sketch Proof.* By induction on $\dot{e}$. □

**Proposition 9** (Workload $\delta^{\mathsf{B}}$)**.** Given a Linear B expression of type $\Gamma; \dot{\Gamma} \vdash^{\text{Jax}} d : (\tau; \sigma)$, an enumeration $\theta$ of the tangent variables in $\dot{\Gamma}$, then $\delta^{\mathsf{B}}_\theta(d)$ is safe and $\mathcal{W}(\delta^{\mathsf{B}}_\theta(d)) \leq \mathcal{W}^{\mathtt{Jax}}(d)$.

*Sketch Proof.* The safeness of $\delta^{\mathsf{B}}_\theta(d)$ is easy to prove by induction on $d$ simply checking the items in Definition 4. Let us focus on the proof related to workload preservation of $\delta^{\mathsf{B}}_\theta(d)$, we proceed by induction on $d$ and we use Lemma 31 and Lemma 32 in the case $d = (e^p; \dot{e})$. □

$$
\begin{aligned}
P, Q &::= \;!x \mid (\lambda !x.P)Q \mid \underline{f}(!x_1, !x_2) \mid !\underline{r} \mid !(\,) \mid !(P, Q) \mid \texttt{let}\ p^{\otimes} = z\ \texttt{in}\ P && (\lambda\mathrm{LL}^{\mathtt{P}})\\
U &::= \;\underline{0} \mid u \mid FU \mid \langle\,\rangle \mid \langle U_1, U_2\rangle && (\lambda\mathrm{LL}^{\mathtt{t}})\\
F, G &::= \;\lambda p^{\&}.U \mid f \mid \texttt{let}\ \S f = \S F\ \texttt{in}\ G \mid \dot{+} \mid \dot{*}x && (\lambda\mathrm{LL}^{\mathtt{f}})\\
R, S &::= \;(P, \S F) \mid \texttt{let}\ (!x, \S f) = S\ \texttt{in}\ R && (\lambda\mathrm{LL}^{\mathtt{A}})\\
&\quad \mid \texttt{let}\ \S f = \S F\ \texttt{in}\ R \mid \texttt{let}\ !x = P\ \texttt{in}\ R \mid \texttt{let}\ p^{\otimes} = z\ \texttt{in}\ R
\end{aligned}
$$

Figure 4.3: The 4-sorted grammar giving the fragment $\lambda\mathrm{LL}^{\mathtt{A}}$ of $\lambda$LL, together with the metavariables associated with each sort. The pattern $!x$ is supposed the exponential of a $\otimes$-sequence type, namely $!x_1$, $!x_2$ are of type $!\mathbb{R}$. The variable $u$ (resp. $f$) is supposed of a &-sequence (resp. tangent function) type. Metavariables $p^{\otimes}$, $p^{\&}$ vary over patterns of resp. $\otimes$-sequence, &-sequence types.

## 4.3 Dissecting Linear A into $\lambda$LL

The grammar of Linear A is projected into $\lambda$LL along $\delta$. The grammars in Figure 4.3 morally dissect the image sets of $\delta$ and split the purely tangent and purely primal terms. This fragment is constructed on the top of the fragments $\lambda\mathrm{LL}^{\mathtt{P}}$ and $\lambda\mathrm{LL}^{\mathtt{t}}$ basically giving, respectively, the purely primal and purely tangent part of the $\delta$ images. Notice that $\lambda\mathrm{LL}^{\mathtt{P}}$ corresponds to the restriction to ground types of the call-by-value translation of $\lambda$-calculus into linear logic [82]. The sort $\lambda\mathrm{LL}^{\mathtt{P}}$ identifies computations, while values are given by variables, numerals and the empty tuple.

The typing environments of the terms in $\lambda\mathrm{LL}^{\mathtt{A}}$ may contain exponential patterns $!x : !E$ associated with primal data, or patterns $\S f$ associated with some tangent computation, so of type $\S(L \multimap H)$ for some &-sequence types $L$ and $H$. Henceforth, we denote by $!\Sigma$ the environments of exponentiated $\otimes$-sequence patterns, i.e. $!\Sigma$ stands for, e.g. $!x_1 : !E_1, \ldots, !x_n : !E_n$, and by $\S\Phi$ the environments of the affine patterns of functions between &-sequences, i.e. $\S\Phi$ stands for, e.g. $\S f_1 : \S(L_1 \multimap H_1), \ldots, \S f_m : \S(L_m \multimap H_m)$. We call $!\Sigma$ as $!$-*environment* and $\S\Phi$ as $\S$-*environment*.

**Proposition 10.** We have the following:

1. $\forall P \in \lambda\mathrm{LL}^{\mathtt{P}}$, $!\Sigma \vdash P : !E$,
2. $\forall U \in \lambda\mathrm{LL}^{\mathtt{t}}$, $!\Sigma, \S\Phi, p^{\&} : L \vdash U : H$,
3. $\forall F \in \lambda\mathrm{LL}^{\mathtt{f}}$, $!\Sigma, \S\Phi \vdash F : L \multimap H$,
4. $\forall R \in \lambda\mathrm{LL}^{\mathtt{A}}$, $!\Sigma, \S\Phi \vdash R : !E \otimes \S(L \multimap H)$,

for suitable $!\Sigma$, $\S\Phi$, $p^{\&}$, $E, L, H$.

Moreover, the following proposition relates $\delta^{\mathtt{B}}$ on purely primal JAX expressions and $\lambda\mathrm{LL}^{\mathtt{P}}$.

**Proposition 11.** Given $e^p \in$ Primal, $\delta^{\mathtt{B}}(e^p) \in \lambda\mathrm{LL}^{\mathtt{P}}$.

Similarly, the following proposition relates $\delta$ and $\lambda\mathrm{LL}^{\mathtt{A}}$.

**Proposition 12.** Given $e \in$ Linear A and an enumeration $\theta$ of $FV^t(e)$, $\delta_\theta(e) \in \lambda\mathrm{LL}^{\mathtt{A}}$.

One can wonder whether the converse of Proposition 12 on $\lambda\text{LL}^{\mathtt{A}}$ holds, namely, whether all terms in $\lambda\text{LL}^{\mathtt{A}}$ are image of some Linear A expression via the translation $\delta$. This is not the case, in fact Linear A represents terms in a kind of administrative normal form, where all tuples and unpacking are restricted to variables. For instance, the term $\mathtt{let}\ (!x, \S f) = (!\underline{r}, \S(\lambda u.u))\ \mathtt{in}\ (!x, \S(\lambda u.f(fu)))$ is in $\lambda\text{LL}^{\mathtt{A}}$ but not in the image set of $\delta$. This example illustrates that $\lambda\text{LL}^{\mathtt{A}}$ is strictly more expressive than Linear A, and that the translation is not surjective.

Similarly, the converse of Proposition 11 does not hold. A counterexample is provided by the term $!((\lambda!x.!x)!x, !y)$, which is a term in $\lambda\text{LL}^{\mathtt{P}}$ but does not correspond to any purely primal expression in Linear B.

Finally, let us proceed with our toy example

**Illustrative Example 8** (Translation $\delta^{\mathtt{B}}$)**.** Consider the Linear B expression in Equation 2.2 computing our function $g$. We proceed by applying the translation $\delta^{\mathtt{B}}$ in Figure 4.2 and we obtain the following $\lambda\text{LL}^{\mathtt{P}}$ term $P$

$$
\begin{aligned}
P = \delta^{\mathtt{B}}(e^p) = \ & \mathtt{let}\ !v_1 = \underline{sin}\ !x\ \mathtt{in} \\
& \mathtt{let}\ !v_2 = !v_1\ \underline{*}\ !y\ \mathtt{in} \\
& \mathtt{let}\ !v_3 = \underline{cos}\ !x\ \mathtt{in} \\
& \mathtt{let}\ !v_4 = !v_2\ \underline{+}\ !v_3\ \mathtt{in} \\
& !v_4
\end{aligned} \tag{4.16}
$$

which is well-typed as $!x : !\mathbb{R}, !y : !\mathbb{R} \vdash P : !\mathbb{R}$.

# Detailed Proof

**Lemma 30.** We have the following:

1. Given two tangent JAX expressions $\Gamma_1; \dot{\Gamma}_1 \vdash^{\text{Jax}} \dot{e}_1 : (\mathtt{1}; \tau_1 \dot{\otimes} \sigma_1)$ and $\Gamma_2; \dot{\Gamma}_2, \dot{x}_1 : \tau_1, \dot{x}_2 : \sigma_1 \vdash^{\text{Jax}} \dot{e}_2 : (\mathtt{1}; \tau_2 \dot{\otimes} \sigma_2)$, and $\theta$ is an enumeration of $\dot{\Gamma}_1, \dot{\Gamma}_2$, then

$$\delta^{\mathsf{B}}_{\theta}(\mathtt{let}\ \dot{\otimes}(\dot{x_1}, \dot{x_2}) = \dot{e}_1\ \mathtt{in}\ \dot{e}_2) \sim \begin{array}{l} \mathtt{let}\ \S f_1 = \S\delta^{\mathsf{B}}_{\theta \cap FV^t(\dot{e}_1)}(\dot{e}_1)\ \mathtt{in} \\ \mathtt{let}\ \S f_2 = \S\delta^{\mathsf{B}}_{\dot{x}_1, \dot{x}_2, \theta \cap FV^t(\dot{e}_2)}(\dot{e}_2)\ \mathtt{in} \\ \lambda u^{\&\mathtt{t}(\theta)}.\mathtt{let}\ \langle u_1, u_2\rangle = \sigma^{\theta}_{FV^t(\dot{e}_1)} u\ \mathtt{in}\ f_2(\overline{\sigma}^{\theta}_{(\dot{x}_1, \dot{x}_2)}\langle f_1 u_1, u_2\rangle) \end{array}$$

   where $\S f_1$ and $\S f_2$ have the following types:

$$\begin{array}{l} \S f_1 : \S(\&\mathtt{t}(\theta \cap FV^t(\dot{e}_1)) \multimap \mathtt{t}(\tau_1)\&\mathtt{t}(\sigma_1)) \\ \S f_2 : \S(\&\mathtt{t}(\dot{x}_1, \dot{x}_2, \theta \cap FV^t(\dot{e}_2)) \multimap \mathtt{t}(\tau_2)\&\mathtt{t}(\sigma_2)) \end{array}$$

2. Given two tangent JAX expressions $\Gamma_1; \dot{\Gamma}_1 \vdash^{\text{Jax}} \dot{e}_1 : (\mathtt{1}; \tau_1 \dot{\otimes} \sigma_1)$ and $\Gamma_2; \dot{\Gamma}_2 \vdash^{\text{Jax}} \dot{e}_2 : (\mathtt{1}; \tau_2 \dot{\otimes} \sigma_2)$, and $\theta$ is an enumeration of $\dot{\Gamma}_1, \dot{\Gamma}_2$, then

$$\delta^{\mathsf{B}}_{\theta}(\dot{\otimes}(\dot{e_1}, \dot{e_2})) \sim \begin{array}{l} \mathtt{let}\ \S f_1 = \S\delta^{\mathsf{B}}_{\theta \cap FV^t(\dot{e}_1)}(\dot{e}_1)\ \mathtt{in} \\ \mathtt{let}\ \S f_2 = \S\delta^{\mathsf{B}}_{\theta \cap FV^t(\dot{e}_2)}(\dot{e}_2)\ \mathtt{in} \\ \lambda u^{\&\mathtt{t}(\theta)}.\mathtt{let}\ \langle u_1, u_2\rangle = \sigma^{\theta}_{FV^t(\dot{e}_1)} u\ \mathtt{in}\ \langle f_1 u_1, f_2 u_2\rangle \end{array}$$

   where $\S f_1$ and $\S f_2$ have the following types:

$$\begin{array}{l} \S f_1 : \S(\&\mathtt{t}(\theta \cap FV^t(\dot{e}_1)) \multimap \mathtt{t}(\tau_1)\&\mathtt{t}(\sigma_1)) \\ \S f_2 : \S(\&\mathtt{t}(\theta \cap FV^t(\dot{e}_2)) \multimap \mathtt{t}(\tau_2)\&\mathtt{t}(\sigma_2)) \end{array}$$

*Proof Claim 1.* Observe that $\mathtt{let}\ \dot{\otimes}(\dot{x_1}, \dot{x_2}) = \dot{e}_1\ \mathtt{in}\ \dot{e}_2$ is syntactic sugar for the expression $\mathtt{let}\ \dot{z} = \dot{e}_1\ \mathtt{in}\ \mathtt{let}\ \dot{\otimes}(\dot{x_1}, \dot{x_2}) = \dot{z}\ \mathtt{in}\ \dot{e}_2 \in$ Linear B.

$$\delta^{\mathsf{B}}_{\theta}(\mathtt{let}\ \dot{\otimes}(\dot{x_1}, \dot{x_2}) = \dot{e}_1\ \mathtt{in}\ \dot{e}_2) \approx \delta^{\mathsf{B}}_{\theta}(\mathtt{let}\ \dot{z} = \dot{e}_1\ \mathtt{in}\ \mathtt{let}\ \dot{\otimes}(\dot{x_1}, \dot{x_2}) = \dot{z}\ \mathtt{in}\ \dot{e}_2) \tag{4.17}$$

$$= \begin{array}{l} \mathtt{let}\ \S f_1 = \S\delta^{\mathsf{B}}_{\theta \cap FV^t(\dot{e}_1)}(\dot{e}_1)\ \mathtt{in} \\ \mathtt{let}\ \S g = \S\delta^{\mathsf{B}}_{\dot{z}, \theta \cap FV^t(\dot{e}_2)}(\mathtt{let}\ \dot{\otimes}(\dot{x_1}, \dot{x_2}) = \dot{z}\ \mathtt{in}\ \dot{e}_2)\ \mathtt{in} \\ \lambda u^{\&\mathtt{t}(\theta)}.\mathtt{let}\ \langle u_1, u_2\rangle = \sigma^{\theta}_{FV^t(\dot{e}_1)} u\ \mathtt{in}\ g(\overline{\sigma}^{\dot{z}, \theta \cap FV^t(\dot{e}_2)}_{\dot{z}}\langle f_1 u_1, u_2\rangle) \end{array} \tag{4.18}$$

$$\to \begin{array}{l} \mathtt{let}\ \S f_1 = \S\delta^{\mathsf{B}}_{\theta \cap FV^t(\dot{e}_1)}(\dot{e}_1)\ \mathtt{in} \\ \mathtt{let}\ \S g = \S\delta^{\mathsf{B}}_{\dot{z}, \theta \cap FV^t(\dot{e}_2)}(\mathtt{let}\ \dot{\otimes}(\dot{x_1}, \dot{x_2}) = \dot{z}\ \mathtt{in}\ \dot{e}_2)\ \mathtt{in} \\ \lambda u^{\&\mathtt{t}(\theta)}.\mathtt{let}\ \langle u_1, u_2\rangle = \sigma^{\theta}_{FV^t(\dot{e}_1)} u\ \mathtt{in}\ g\langle f_1 u_1, u_2\rangle \end{array} \tag{4.19}$$

$$= \begin{array}{l} \mathtt{let}\ \S f_1 = \S\delta^{\mathsf{B}}_{\theta \cap FV^t(\dot{e}_1)}(\dot{e}_1)\ \mathtt{in} \\ \mathtt{let}\ \S g = \S\left(\begin{array}{l} \mathtt{let}\ \S f_2 = \S\delta^{\mathsf{B}}_{\dot{x_1}, \dot{x_2}, \theta \cap FV^t(\dot{e}_2) \setminus \dot{z}}(e)\ \mathtt{in} \\ \lambda y^{\&\mathtt{t}(\theta \cap FV^t(\dot{e}_2))}. \\ \quad \mathtt{let}\ \langle\langle x_1, x_2\rangle, y'\rangle = \sigma_{\{\theta \cap FV^t(\dot{e}_2)(\dot{z})\}} y\ \mathtt{in}\ f_2(\overline{\sigma}_{\dot{x_1}, \dot{x_2}}\langle x_1, x_2, y'\rangle) \end{array}\right)\ \mathtt{in} \\ \lambda u^{\&\mathtt{t}(\theta)}.\mathtt{let}\ \langle u_1, u_2\rangle = \sigma^{\theta}_{FV^t(\dot{e}_1)} u\ \mathtt{in}\ g\langle f_1 u_1, u_2\rangle \end{array} \tag{4.20}$$

$$\begin{aligned}&\texttt{let}\ \S f_1 = \S\delta^{\mathsf{B}}_{\theta\cap FV^t(\dot e_1)}(\dot e_1)\ \texttt{in}\\ \to\ &\texttt{let}\ \S g = \S\begin{pmatrix}\texttt{let}\ \S f_2 = \S\delta^{\mathsf{B}}_{\dot x_1,\dot x_2,\theta\cap FV^t(\dot e_2)\setminus\dot z}(e)\ \texttt{in}\\ \lambda y^{\&\mathtt{t}(\theta\cap FV^t(\dot e_2))}.\\ \quad\texttt{let}\ \langle\langle x_1,x_2\rangle,y'\rangle = \sigma_{\{\theta\cap FV^t(\dot e_2)(\dot z)\}}y\ \texttt{in}\ f_2\langle x_1,x_2,y'\rangle\end{pmatrix}\texttt{in}\\ &\lambda u^{\&\mathtt{t}(\theta)}.\texttt{let}\ \langle u_1,u_2\rangle = \sigma^{\theta}_{FV^t(\dot e_1)}u\ \texttt{in}\ g\langle f_1u_1,u_2\rangle\end{aligned} \tag{4.21}$$

$$\begin{aligned}&\texttt{let}\ \S f_1 = \S\delta^{\mathsf{B}}_{\theta\cap FV^t(\dot e_1)}(\dot e_1)\ \texttt{in}\\ &\texttt{let}\ \S f_2 = \S\delta^{\mathsf{B}}_{\dot x_1,\dot x_2,\theta\cap FV^t(\dot e_2)}(\dot e_2)\ \texttt{in}\\ \sim\ &\texttt{let}\ \S g = \S(\lambda y^{\&\mathtt{t}(\theta\cap FV^t(\dot e_2))}.\texttt{let}\ \langle\langle x_1,x_2\rangle,y'\rangle = \sigma_{\{\theta\cap FV^t(\dot e_2)(\dot z)\}}y\ \texttt{in}\ f_2\langle x_1,x_2,y'\rangle)\texttt{in}\\ &\lambda u^{\&\mathtt{t}(\theta)}.\texttt{let}\ \langle u_1,u_2\rangle = \sigma^{\theta}_{FV^t(\dot e_1)}u\ \texttt{in}\ g\langle f_1u_1,u_2\rangle\end{aligned} \tag{4.22}$$

$$\begin{aligned}&\texttt{let}\ \S f_1 = \S\delta^{\mathsf{B}}_{\theta\cap FV^t(\dot e_1)}(\dot e_1)\ \texttt{in}\\ \to^*\ &\texttt{let}\ \S f_2 = \S\delta^{\mathsf{B}}_{\dot x_1,\dot x_2,\theta\cap FV^t(\dot e_2)}(\dot e_2)\ \texttt{in}\\ &\lambda u^{\&\mathtt{t}(\theta)}.\texttt{let}\ \langle u_1,u_2\rangle = \sigma^{\theta}_{FV^t(\dot e_1)}u\ \texttt{in}\ f_2(\overline{\sigma}^{\theta}_{(\dot x_1,\dot x_2)}\langle f_1u_1,u_2\rangle)\end{aligned} \tag{4.23}$$

where the passage from line (4.18) (resp. (4.20)) to line (4.19) (resp. (4.21)) is because by typing $\overline{\sigma}^{\dot z,\theta\cap FV^t(\dot e_2)}_{\dot z}$ (resp. $\overline{\sigma}^{\dot x_1,\dot x_2,\theta\cap FV^t(\dot e_2)}_{\dot x_1,\dot x_2}$) is the identity function and so we perform a step of $\beta_\lambda$. The passage from line (4.21) to line (4.22) is by Lemma 17. The last equation is obtained after several steps of $\beta_\lambda$. □

# Chapter 5

# AD System of $\lambda$LL

This chapter is devoted to defining the AD system of $\lambda$LL. In particular, we extend and adapt the core transformations previously introduced in the context of Autodiff to operate within the $\lambda$LL framework. The chapter begins by presenting the forward, unzipping, and transpose transformations applied to $\lambda$LL terms. For each of these transformations, we rigorously establish soundness by demonstrating that they commute with the $\delta$ translation, modulo the equivalence relation $\sim$ (refer to Theorems 7, 9, and 12). In addition to soundness, we also analyze the cost associated with these transformations, showing that they preserve the workload (see Theorems 8, 10, and Corollary 4). This property is crucial for ensuring that the AD system remains efficient.

Finally, Section 5.4 investigates how the unzipping transformation can be safely skipped in our setting. We leverage the fact that both unzipping and transpose transformations preserve the logical equivalence relation $\sim$ (as shown in Propositions 14 and 15) to argue that the intermediate unzipping step can be bypassed, without compromising the semantics of the transformed program and while preserving the overall workload. This optimization not only simplifies the AD system but also enhances the modularity and parallelism of the reverse-mode AD. These benefits are particularly significant in the context of programs composed of independent subroutines, where improved parallelism can lead to performance gains. In order to concretely illustrate these advantages, Subsection 5.4.1 presents a comparative example, highlighting the behavioural differences between Autodiff and the AD framework developed for $\lambda$LL. This example underscores the modular and compositional strengths of our approach.

## 5.1 Forward

As described in Chapter 1, forward mode AD is a program transformation that computes derivatives by propagating information from inputs to outputs, evaluating both the function and its derivatives in a single pass. In this section, we formalize this transformation for $\lambda$LL terms and establish its soundness in Subsection 5.1.1, as well as its preservation of workload in Subsection 5.1.2. Additionally, Example 9 illustrates the application of our forward transformation to the term $P$ introduced in the previous chapter (see Example 8).

Forward AD is defined in our setting as a program transformation $\mathcal{F}$ mapping a term $P \in \lambda\mathrm{LL}^{\mathtt{P}}$ of some type $E$ and an enumeration $\theta = (!x_1 : !E_1, \dots, !x_n : !E_n)$ of the free variables in $P$ into a term $\mathcal{F}_\theta(P)$ (Figure 5.1). The intuition is the same as for the $\mathcal{F}^{\mathtt{Jax}}$ transformation: the only difference is that now the tangent part is represented as a linear map taking in input the tangent siblings of the $P$ free variables. Moreover, Theorem 6 will prove that $\mathcal{F}_\theta(P)$ is a well-typed $\lambda\mathrm{LL}^{\mathtt{A}}$ term with the same free variables as $P$ and type $E \otimes \S\left((\&_{i=1}^{n}\mathtt{t}(E_i)) \multimap \mathtt{t}(E)\right)$, where $\mathtt{t}$ is the translation of types extended to ⊗-sequence types given in the previous chapter.

Formally, let $P \in \lambda LL^{\mathtt{P}}$ and $\theta = (!x_1 : !E_1, \ldots, !x_n : !E_n)$ be an enumeration of $FV(P)$, we define $\mathcal{F}_\theta(P)$ by structural induction on $P$ as in Figure 5.1.

$$\mathcal{F}_{(!x)}(!x) \stackrel{\text{def}}{=} (!x, \S(\lambda u.u))$$

$$\mathcal{F}_{(\,)}(!\underline{r}) \stackrel{\text{def}}{=} (!\underline{r}, \S(\lambda u.\underline{0}))$$

$$\mathcal{F}_{(\,)}(!(\,)) \stackrel{\text{def}}{=} (!(\,), \S(\lambda u.\langle\,\rangle))$$

$$\begin{aligned}\mathcal{F}_\theta(!(P,Q)) \stackrel{\text{def}}{=} & \mathtt{let}\ (!x, \S f) = \mathcal{F}_{\theta\cap FV(P)}(P)\ \mathtt{in} \\ & \mathtt{let}\ (!y, \S g) = \mathcal{F}_{\theta\cap FV(Q)}(Q)\ \mathtt{in} \\ & \left( !(!x,!y), \S \left( \lambda u^{\&\mathtt{t}(\theta)}. \begin{array}{l} \mathtt{let}\ \langle u_{PQ}, u' \rangle = \sigma_{FV(P)\cap FV(Q)} u\ \mathtt{in} \\ \mathtt{let}\ \langle u_P, u_Q \rangle = \sigma_{FV(P)\setminus FV(Q)} u'\ \mathtt{in} \\ \langle f\langle u_{PQ}, u_P\rangle, g\langle u_{PQ}, u_Q\rangle\rangle \end{array} \right)\right)\end{aligned}$$

$$\begin{aligned}\mathcal{F}_\theta((\lambda !x.P)Q) \stackrel{\text{def}}{=} & \mathtt{let}\ (!x, \S f) = \mathcal{F}_{\theta\cap FV(Q)}(Q)\ \mathtt{in} \\ & \mathtt{let}\ (!y, \S g) = \mathcal{F}_{x,\theta\cap FV(P)}(P)\ \mathtt{in} \\ & \left( !y, \S \left( \lambda u^{\&\mathtt{t}(\theta)}. \begin{array}{l} \mathtt{let}\ \langle u_{PQ}, u' \rangle = \sigma_{FV(P)\cap FV(Q)} u\ \mathtt{in} \\ \mathtt{let}\ \langle u_P, u_Q \rangle = \sigma_{(FV(P)\setminus\{!x\})\setminus FV(Q)} u'\ \mathtt{in} \\ g\langle f\langle u_{PQ}, u_Q\rangle, u_{PQ}, u_P\rangle \end{array} \right)\right)\end{aligned}$$

$$\begin{aligned}\mathcal{F}_{(!x_1,!x_2)}(\underline{f}(!x_1,!x_2)) \stackrel{\text{def}}{=} & \mathtt{let}\ !y_1 = \underline{\partial_1 f}(!x_1,!x_2)\ \mathtt{in} \\ & \mathtt{let}\ !y_2 = \underline{\partial_2 f}(!x_1,!x_2)\ \mathtt{in} \\ & (\underline{f}(!x_1,!x_2), \S(\lambda\langle u_1,u_2\rangle^{\mathtt{t}((!x_1,!x_2))}.(y_1 \dot{*} u_1) \dot{+} (y_2 \dot{*} (u_2))))\end{aligned}$$

$$\begin{aligned}\mathcal{F}_\theta(\mathtt{let}\ p^\otimes = z\ \mathtt{in}\ P) \stackrel{\text{def}}{=} & \mathtt{let}\ (!x, \S f) = (z, \S(\lambda u.u))\ \mathtt{in}\ \mathtt{let}\ (!x_1,!x_2) = x\ \mathtt{in} \\ & \mathtt{let}\ (!y, \S g) = \mathcal{F}_{FV(p^\otimes),\theta\cap FV(P)}(P)\ \mathtt{in} \\ & \left(!y, \S\left(\lambda u^{\&\mathtt{t}(\theta)}.\mathtt{let}\ \langle u_P, u'\rangle = \sigma_{FV(P)\setminus\{z\}} u\ \mathtt{in}\ g\langle f u', u_P\rangle\right)\right)\end{aligned}$$

Figure 5.1: Definition of the forward transformation $\mathcal{F}$.

**Theorem 6** (Type $\mathcal{F}$)**.** Given a judgment $!\Sigma \vdash P : !E$ of $\lambda LL^{\mathtt{P}}$ and an enumeration $\theta = (!x_1 : !E_1, \ldots, !x_n : !E_n)$ of $!\Sigma$, then $\mathcal{F}_\theta(P)$ is a $\lambda LL^{\mathtt{A}}$ term of type:

$$!\Sigma \vdash \mathcal{F}_\theta(P) : !E \otimes \S((\&_{i=1}^n \mathtt{t}(E_i)) \multimap \mathtt{t}(E)).$$

*Sketch Proof.* By induction on $P$. □

**Illustrative Example 9** (Forward Transformation $\lambda$LL)**.** Consider the $\lambda\mathrm{LL}^{\mathtt{P}}$ term $P$ in Equation 4.16, obtained by applying the translation $\delta^{\mathtt{B}}$ to the purely primal expression of Example 4, and an enumeration $\theta = (!x : !\mathbb{R}, !y : !\mathbb{R})$ of $FV(P)$. We proceed by applying our forward transformation to $P$ (Figure 5.1) and we obtain exactly the following $\lambda\mathrm{LL}^{\mathtt{A}}$ term which is well-typed as $!x : !\mathbb{R}, !y : !\mathbb{R} \vdash \mathcal{F}_\theta(P) : !\mathbb{R} \otimes \S((\mathbb{R}\&\mathbb{R}) \multimap \mathbb{R})$.

$$
\begin{aligned}
&\mathcal{F}_{(!x:!\mathbb{R},!y:!\mathbb{R})}(P) = \\
&\quad \mathtt{let}\ (!v_1, \S f_1^{\mathbb{R}\multimap\mathbb{R}}) = \begin{pmatrix} \mathtt{let}\ !w_1 = \underline{cos}\ !x\ \mathtt{in} \\ (\underline{sin}\ !x, \S(\lambda u.w_1 \dot{*} u)) \end{pmatrix} \mathtt{in} \\
&\quad \mathtt{let}\ (!v_2, \S f_2^{(\mathbb{R}\&\mathbb{R})\multimap\mathbb{R}}) = \begin{pmatrix} \mathtt{let}\ !w_2 = !y\ \mathtt{in}\ \mathtt{let}\ !w_3 = !v_1\ \mathtt{in} \\ (!v_1 \underline{*}\ !y, \S(\lambda\langle u_1, u_2\rangle.(w_2 \dot{*} u_1) \dot{+} (w_3 \dot{*} u_2))) \end{pmatrix} \mathtt{in} \\
&\quad \mathtt{let}\ (!v_3, \S f_3^{\mathbb{R}\multimap\mathbb{R}}) = \begin{pmatrix} \mathtt{let}\ !w_4 = \underline{-sin}\ !x\ \mathtt{in} \\ (\underline{cos}\ !x, \S(\lambda u.w_4 \dot{*} u)) \end{pmatrix} \mathtt{in} \\
&\quad \mathtt{let}\ (!v_4, \S f_4^{(\mathbb{R}\&\mathbb{R})\multimap\mathbb{R}}) = \begin{pmatrix} \mathtt{let}\ !w_5 = !\underline{1}\ \mathtt{in}\ \mathtt{let}\ !w_6 = !\underline{1}\ \mathtt{in} \\ (!v_2 \underline{+}\ !v_3, \S(\lambda\langle u_1, u_2\rangle.(w_5 \dot{*} u_1) \dot{+} (w_6 \dot{*} u_2))) \end{pmatrix} \mathtt{in} \\
&\quad \left(!v_4, \S\left(\lambda u^{\mathbb{R}\&\mathbb{R}}. \begin{array}{l} \mathtt{let}\ \langle dx, dy\rangle = u\ \mathtt{in} \\ f_4\langle f_2\langle f_1\ dx, dy\rangle, f_3\ dx\rangle \end{array}\right)\right)
\end{aligned}
\tag{5.1}
$$

where the part in black is computing out function $g(x, y) = (sin(x) * y) + cos(x)$ and the part in blue is computing the directional derivative of $g$ which is $\mathrm{D}(g)(x, \dot{dx}, y, \dot{dy}) = (cos(x) * y) * \dot{dx} + sin(x) * \dot{dy} - sin(x) * \dot{dx}$.

### 5.1.1 Soundness

The forward transformation in $\lambda$LL is proved to be sound (Theorem 7), by means of the following auxiliary lemmas

**Lemma 33.** Given $\Gamma, u : L \vdash M : H$ for some &-sequence types $L$ and $H$, then $\lambda u.M \sim_{\Gamma, L\multimap H} \lambda p.M\{p/u\}$ where $FV(p) \cap FV(M) = \emptyset$.

*Proof.* By Definition 2 of $\sim$ on open terms, we have to prove that

$$\forall i \leq n. \forall V_i \sim V_i'.\ (\lambda u.M)\{V_i/p_i\} \overset{?}{\sim}_{L\multimap H} (\lambda p.M\{p/u\})\{V_i'/p_i\}$$

where $V_i, V_i'$ are closed values for a pattern $p_i \in \Gamma$.
By definition of substitution we have $(\lambda u.M)\{V_i/p_i\} = \lambda u.M\{V_i/p_i\}$. By definition of substitution and by the hypothesis $FV(p) \cap FV(M) = \emptyset$, we have that $(\lambda p.M\{p/u\})\{V_i'/p_i\} = \lambda p.M\{V_i'/p_i\}\{p/u\}$.

Let $\overline{M} = M\{V_i/p_i\}$ and $\widehat{M} = M\{V_i'/p_i\}$, by Definition 2 of $\sim$ on closed terms we have to prove that

$$\forall N \sim_L N'.\ (\lambda u.\overline{M})N \overset{?}{\sim}_H (\lambda p.\widehat{M}\{p/u\})N'$$

By Progress Property (Proposition 1) we have that $(\lambda u.\overline{M})N =_\beta (\lambda u.\overline{M})W$ and $(\lambda p.\widehat{M}\{p/u\})N'$ $=_\beta (\lambda p.\widehat{M}\{p/u\})W'$ where $W$ and $W'$ are Strong Values. We proceed by induction on the pattern $p$ of &-sequence type as follows:

- If $p = u$ then $W$ is a value and we can conclude.

- If $p = \langle q_1, q_2\rangle$ then $L = L_1 \,\&\, L_2$. Moreover, by definition of value for a pattern $W = \langle W_1, W_2\rangle$ and $W' = \langle W_1', W_2'\rangle$.

  Recall that $W \sim_L W'$ so in this case $\langle W_1, W_2\rangle \sim_{L_1\&L_2} \langle W_1', W_2'\rangle$ and by definition $W_i \sim_{L_i} W_i'$ with $i \in \{1,2\}$.

  We have to prove that

$$(\lambda u.\overline{M})\langle W_1, W_2\rangle \overset{?}{\sim}_H (\lambda\langle q_1, q_2\rangle.\widehat{M}\{\langle q_1,q_2\rangle/u\})\langle W_1', W_2'\rangle$$

  By $\beta$-equivalence we have that

$$(\lambda u.\overline{M})\langle W_1, W_2\rangle =_\beta \overline{M}\{\langle W_1,W_2\rangle/u\}$$
$$(\lambda\langle q_1, q_2\rangle.\widehat{M}\{\langle q_1,q_2\rangle/u\})\langle W_1', W_2'\rangle =_\beta \widehat{M}\{\langle q_1,q_2\rangle/u\}\{\langle W_1',W_2'\rangle/u\}$$

  Moreover, by the hypothesis $FV(p) \cap FV(M) = \emptyset$, we have that $\widehat{M}\{\langle q_1,q_2\rangle/u\}\{\langle W_1',W_2'\rangle/u\} = \widehat{M}\{\langle W_1',W_2'\rangle/u\}$. We can conclude by reflexivity of $\sim$ obtaining that $\overline{M}\{\langle W_1,W_2\rangle/u\} \sim_H \widehat{M}\{\langle W_1',W_2'\rangle/u\}$.

□

More precisely, the following lemma $\sim$-relates the $\delta$ translation of $\mathrm{dup}(\dot{u})$ with the additive contraction of λLL terms, it will be useful to prove the soundness of our forward mode.

**Lemma 34.** Let $\Gamma; \dot{u_1} : \tau, \dot{u_2} : \tau, \dot{\Gamma} \vdash^{\mathtt{Jax}} e : (\sigma; \sigma')$, and let $\theta$ be an enumeration of the tangent variables in $\dot{u} : \tau, \dot{\Gamma}$, then:

$$\delta_\theta\begin{pmatrix}\texttt{let } \dot{a} = \mathrm{dup}(\dot{u}) \texttt{ in} \\ \texttt{let } \dot{\otimes}(\dot{u_1}, \dot{u_2}) = \dot{a} \texttt{ in } e\end{pmatrix} \sim \begin{array}{l}\texttt{let } (!x, \S f) = \delta_{\dot{u_1},\dot{u_2},\theta\backslash\dot{u}}(e) \texttt{ in} \\ \left(!x, \S\left(\lambda y^{\&\mathtt{t}(\theta)}.\ \begin{array}{l}\texttt{let } \langle\langle u_1, u_2\rangle, y'\rangle = \sigma_{\{\theta(\dot{u})\}} y \texttt{ in} \\ f(\overline{\sigma}^{\&\mathtt{t}(\dot{u_1},\dot{u_2},\theta\backslash\dot{u})}_{\dot{u_1},\dot{u_2}}\langle x_1, x_2, y'\rangle)\end{array}\right)\right)\end{array}$$

*Sketch Proof.* By applying the definition of $\delta$ and by using Lemma 17 and Proposition 2. □

Finally, we prove the soundness property for forward mode as follows

**Theorem 7** (Soundness $\mathcal{F}$)**.** Given a Linear B expression $e^p$ in (Primal), an enumeration $\theta = (x_1, \ldots, x_n)$ of the set $FV(e^p)$, a renaming $\phi = (x_1 \mapsto \dot{y_1}, \ldots, x_n \mapsto \dot{y_n})$ of $FV(e^p)$ into tangent JAX variables, and let $\theta' = (\dot{y_1}, \ldots, \dot{y_n})$ be the image of $\theta$ under $\phi$, we have:

$$\mathcal{F}_\theta(\delta^{\mathtt{B}}(e^p)) \sim \delta_{\theta'}(\mathcal{F}^{\mathtt{Jax}}_\phi(e^p))$$

*Proof.* We proceed by induction on $e^p$. The only delicate case is $e^p = \texttt{let } x = e_1^p \texttt{ in } e_2^p$. For the sake of simplicity we assume that $e_1^p$ and $e_2^p$ only share one primal variable, denoted by $z$.

$$\begin{aligned}\mathcal{F}_\theta(\delta^{\mathtt{B}}(e^p)) &= \mathcal{F}_\theta(\delta^{\mathtt{B}}(\texttt{let } x = e_1^p \texttt{ in } e_2^p)) \\ &= \mathcal{F}_\theta(\texttt{let } !x = \delta^{\mathtt{B}}(e_1^p) \texttt{ in } \delta^{\mathtt{B}}(e_2^p)) \approx \mathcal{F}_\theta((\lambda !x.\delta^{\mathtt{B}}(e_2^p))\delta^{\mathtt{B}}(e_1^p)) \\ &\phantom{=}\ \texttt{let } (!x, \S f) = \mathcal{F}_{\theta\cap FV(\delta^{\mathtt{B}}(e_1^p))}(\delta^{\mathtt{B}}(e_1^p)) \texttt{ in} \\ &\phantom{=}\ \texttt{let } (!y, \S g) = \mathcal{F}_{x,\theta\cap FV(\delta^{\mathtt{B}}(e_2^p))}(\delta^{\mathtt{B}}(e_2^p)) \texttt{ in} \\ &= \left(!y, \S\left(\lambda u^{\&\mathtt{t}(\theta)}.\ \begin{array}{l}\texttt{let } \langle u_{PQ}, u'\rangle = \sigma_{FV(\delta^{\mathtt{B}}(e_1^p))\cap FV(\delta^{\mathtt{B}}(e_2^p))} u \texttt{ in} \\ \texttt{let } \langle u_P, u_Q\rangle = \sigma_{(FV(\delta^{\mathtt{B}}(e_1^p))\backslash\{x\})\backslash FV(\delta^{\mathtt{B}}(e_2^p))} u' \texttt{ in} \\ g\langle f\langle u_{PQ}, u_Q\rangle, u_{PQ}, u_P\rangle\end{array}\right)\right)\end{aligned}$$

$$
\begin{aligned}
&\texttt{let }(!x, \S f) = \mathcal{F}_{\theta\cap FV(\delta^{\mathsf{B}}(e_1^p))}(\delta^{\mathsf{B}}(e_1^p))\texttt{ in}\\
&\texttt{let }(!y, \S g) = \mathcal{F}_{x,\theta\cap FV(\delta^{\mathsf{B}}(e_2^p))}(\delta^{\mathsf{B}}(e_2^p))\texttt{ in}\\
= &\left(!y, \S\left(\lambda u^{\&\mathtt{t}(\theta)}.\ \underbrace{\begin{array}{l}\texttt{let }\langle u_{PQ}, u'\rangle = \sigma_{FV(e_1^p)\cap FV(e_2^p)} u\texttt{ in}\\ \texttt{let }\langle u_P, u_Q\rangle = \sigma_{(FV(e_1^p)\setminus\{x\})\setminus FV(e_2^p)} u'\texttt{ in}\\ g\langle f\langle u_{PQ}, u_Q\rangle, u_{PQ}, u_P\rangle\end{array}}_{H}\right)\right)
\end{aligned}
$$

where the last line is obtained by observing that $FV(\delta^{\mathsf{B}}(e^p)) = FV(e^p)$ which is an immediate consequence of Lemma 26.

$$
\begin{aligned}
\delta_{\theta'}(\mathcal{F}_{\phi}^{\mathsf{Jax}}(e^p)) &= \delta_{\theta'}(\mathcal{F}_{\phi}^{\mathsf{Jax}}(\texttt{let }x = e_1^p\texttt{ in }e_2^p))\\
&\qquad \texttt{let }(!x, \S f) = \delta_{\theta'\cap FV^t(\mathcal{F}^{\mathsf{Jax}}_{\phi_1,\{z\mapsto \dot{w}_1\}}(e_1^p))}(\mathcal{F}^{\mathsf{Jax}}_{\phi_1,\{z\mapsto \dot{w}_1\}}(e_1^p))\texttt{ in}\\
&\qquad \texttt{let }(!y, \S g) = \delta_{\theta'\cap FV^t(\mathcal{F}^{\mathsf{Jax}}_{\phi_2,\{z\mapsto \dot{w}_2\}}(e_2^p))}(\mathcal{F}^{\mathsf{Jax}}_{\phi_2,\{z\mapsto \dot{w}_2\}}(e_2^p))\texttt{ in}\\
&\sim\ \ (!y, \S(\lambda y^{\&\mathtt{t}(\theta')}.\underbrace{\texttt{let }\langle y_1, y_2\rangle = \sigma^{\&\mathtt{t}(\theta)}_{\theta''\cap FV^t(\mathcal{F}^{\mathsf{Jax}}_{\phi_1,\{z\mapsto \dot{w}_1\}}(e_1^p))} y\texttt{ in }g\langle f y_1, y_2\rangle}_{H'}))
\end{aligned}
$$

where the last line is obtained by applying Lemma 34 and some $\beta$-steps.

We can observe that $\phi(FV(e)) = FV^t(e)$ and then by inductive hypotheses we have that

$$
\mathcal{F}_{\theta\cap FV(\delta^{\mathsf{B}}(e_i^p))}(\delta^{\mathsf{B}}(e_i^p)) \sim \delta_{\theta'\cap FV^t(\mathcal{F}^{\mathsf{Jax}}_{\phi_i,\{z\mapsto \dot{w}_i\}}(e_i^p))}(\mathcal{F}^{\mathsf{Jax}}_{\phi_i,\{z\mapsto \dot{w}_i\}}(e_i^p)) \qquad \text{with } i \in \{1, 2\}.
$$

In order to conclude the proof we have to show that $\lambda u^{\&\mathtt{t}(\theta)}.H \sim \lambda y^{\&\mathtt{t}(\theta')}.H'$. Let $p$ be a complete pattern of type $\&\mathtt{t}(\theta) = \&\mathtt{t}(\theta')$, then by Lemma 33 we have:

$$
\begin{aligned}
\lambda u^{\&\mathtt{t}(\theta)}.H &\sim \lambda p.H\{p/u\}\\
\lambda u^{\&\mathtt{t}(\theta)'}.H' &\sim \lambda p.H'\{p/u\}
\end{aligned}
$$

so it is easy to see that $\lambda p.H\{p/u\} \Rightarrow_\beta \lambda p.H'\{p/u\}$ and we can conclude.

□

### 5.1.2 Work Preservation

Theorem 8 shows that our forward transformation is work preserving up to a constant factor. In fact, $\mathcal{F}_\theta(P)$ introduces a constant number of numerical operations in case of $P$ numeric function. This observation was previously noted in the context of forward transformation in Autodiff (see Remark 1) and will be further elaborated in the subsequent proof.

**Theorem 8** (Workload $\mathcal{F}$)**.** There is a constant $c$ such that $\forall P \in \lambda\mathrm{LL}^{\mathsf{P}}$ and $\forall\theta$ enumeration of $FV(P)$, $\mathcal{W}(\mathcal{F}_\theta(P)) \leq c \cdot \mathcal{W}(P)$. If moreover $P$ is safe, then $\mathcal{F}_\theta(P)$ is safe too.

*Proof.* We proceed by induction on $P$. The part of the statement related to the safeness of $\mathcal{F}_\theta(P)$ is easy to prove by induction on $P$ simply checking the items in Definition 4. In contrast, the work preservation aspect of the statement requires a more careful analysis. Let us consider the two most interesting cases:

- Case $P = (\lambda !x.Q_1)Q_2$:

$$\begin{aligned}\mathcal{W}(P) &= \mathcal{W}((\lambda !x.Q_1)Q_2)\\ &= \mathcal{W}(\lambda !x.Q_1) + \mathcal{W}(Q_2)\\ &= \mathcal{W}(Q_1) + \mathcal{W}(Q_2)\end{aligned}$$

where the last line follows from the observation that $!x$ is an exponential pattern, meaning that all occurrences of $\mathbb{R}$ within its type appear under the scope of a bang modality (!) and are therefore excluded from the workload calculation.

$$\begin{aligned}\mathcal{W}(\mathcal{F}_\theta(P)) &= \mathcal{W}(\mathcal{F}_\theta((\lambda !x.Q_1)Q_2))\\ &= \mathcal{W}\begin{pmatrix}\texttt{let } (!x, \S f) = \mathcal{F}_{\theta\cap FV(Q_2)}(Q_2) \texttt{ in}\\ \texttt{let } (!y, \S g) = \mathcal{F}_{FV(!x),\theta\cap FV(Q_1)}(Q_1) \texttt{ in}\\ (!y, \S(\lambda u^{\&\texttt{t}(\theta)}.\texttt{let } \langle u_{1,2}, u_1, u_2\rangle = D_{Q_1,Q_2,!x}\ u \texttt{ in } g\langle f\langle u_{1,2}, u_1\rangle, u_{1,2}, u_2\rangle))\end{pmatrix}\end{aligned}$$

Observe that our workload accounts for the number of numerical operations not occurring under a ! modality, as well as the numerals potentially erased during reduction. In this case, no erasure occurs; consequently, the workload contributions from the two let-constructs and the $\lambda$-abstraction are equal to zero. Therefore, it follows that $\mathcal{W}(\mathcal{F}_\theta(P))$ is equal to $\mathcal{W}(\mathcal{F}_{\theta\cap FV(Q_2)}(Q_2)) + \mathcal{W}(\mathcal{F}_{FV(!x),\theta\cap FV(Q_1)}(Q_1))$.

We conclude by inductive hypotheses and by taking $c = 1$.

- Case $P = \underline{f}(!x_1, !x_2)$:

$$\mathcal{W}(\underline{f}(!x_1, !x_2)) = 1$$

$$\begin{aligned}\mathcal{W}(\mathcal{W}(\mathcal{F}_\theta(P))) &= \mathcal{W}(\mathcal{F}_{(!x_1,!x_2)}(\underline{f}(!x_1, !x_2)))\\ &= \mathcal{W}\begin{pmatrix}\texttt{let } !y_1 = \underline{\partial_1 f}(!x_1, !x_2) \texttt{ in}\\ \texttt{let } !y_2 = \underline{\partial_2 f}(!x_1, !x_2) \texttt{ in}\\ (\underline{f}(!x_1, !x_2), \S(\lambda\langle u_1, u_2\rangle.(y_1 \dot{*} u_1)\dot{+}(y_2 \dot{*}(u_2))))\end{pmatrix}\\ &= \sum_{i=1}^{2} \mathcal{W}(\underline{\partial_i f}(!x_1, !x_2)) + \mathcal{W}(\underline{f}(!x_1, !x_2)) + \mathcal{W}((y_1 \dot{*} u_1)\dot{+}(y_2 \dot{*}(u_2)))\\ &= 2 + 1 + 3 = 6\end{aligned}$$

We can conclude by taking $c = 6$.

Observe that this is the only case in which we use $c > 1$.

Moreover, it is interesting to details also the case of $n$-ary function $\underline{f}(!x_1, \ldots, !x_n)$ for which we have to suppose that the maximal arity of numeric function primitive of λLL is bounded by a constant $b$. We have to fix this constant because in that case we have:

$$\begin{aligned}&\mathcal{W}(\mathcal{F}_{(!x_1,\ldots,!x_n)}(\underline{f}(!x_1, \ldots, !x_n)))\\ &\quad = \sum_{i=1}^{n} \mathcal{W}(\underline{\partial_i f}(!x_1, !x_2)) + \mathcal{W}(\underline{f}(!x_1, \ldots, !x_n)) + n - 1 + \sum_{i=1}^{n} \mathcal{W}(y_1 \dot{*} u_1)\\ &\quad = n + 1 + n - 1 + n = 3n\end{aligned}$$

where $n$ is the arity of the numerical function $\underline{f}$ and the cost $n - 1$ is for the binary sums performed by the forward transformation. We take $c > 3b$ and we can conclude as $b$, unlike $n$, does not depend on the term but it is fixed once for the language.

□

## 5.2 Unzipping

The unzipping transformation, denoted by $\mathcal{U}$, is an endo-transformation on $\lambda\mathrm{LL}^{\mathtt{A}}$ terms that explicitly separates primal and tangent let-definitions. By exploiting the expressive power of the logical equivalence relation $\sim$, we demonstrate that, within our framework, this transformation can be understood as a sequence of commutations between `let` constructs (Proposition 13 and discussion in Remark 8). Adopting the same methodology as in the previous section, we establish that the unzipping transformation is both sound (Theorem 9) and work preserving (Theorem 10). In addition, we prove that it preserves the logical equivalence $\sim$ (Proposition 14), a property that will be instrumental in subsequent results. In order to illustrate the practical impact of this transformation, we proceed by applying the unzipping transformation to our running example (see Example 10).

$$\epsilon[\,] ::= [\,] \mid \mathtt{let}\ p^{\otimes} = z\ \mathtt{in}\ \epsilon[\,] \mid (\lambda !x.\epsilon[\,])Q$$

Figure 5.2: Grammar of context of exponential let-definitions. We will often use the let notation for the last case, i.e. $\mathtt{let}\ !x = Q\ \mathtt{in}\ \epsilon[\,]$.

Formally, the transformation is defined on the top of a structural decomposition $\mathcal{U}^{\bullet}(S)$ of $S$ into a triplet $(\epsilon[\,], P, F)$ of a context $\epsilon[\,]$ of exponential let-definitions generated by the grammar in Figure 5.2, a $\lambda\mathrm{LL}^{\mathtt{P}}$ term $P$ and a $\lambda\mathrm{LL}^{\mathtt{f}}$ term $F$, so that $\mathcal{U}(S) \stackrel{\text{def}}{=} \epsilon[(P, \S F)]$. The definition of $\mathcal{U}^{\bullet}(S)$ is given in Figure 5.3, and here below are the expected properties.

$$\begin{aligned}
\mathcal{U}^{\bullet}((P, \S F)) &\stackrel{\text{def}}{=} ([\,], P, F)\\
\mathcal{U}^{\bullet}(\mathtt{let}\ (!x, \S f) = S_1\ \mathtt{in}\ S_2) &\stackrel{\text{def}}{=} (\epsilon_1[\mathtt{let}\ !x = P_1\ \mathtt{in}\ \epsilon_2[]], P_2, (\mathtt{let}\ \S f = \S F_1\ \mathtt{in}\ F_2))\\
\mathcal{U}^{\bullet}(\mathtt{let}\ !x = P\ \mathtt{in}\ S_1) &\stackrel{\text{def}}{=} (\mathtt{let}\ !x = P\ \mathtt{in}\ \epsilon_1[], P_1, F_1)\\
\mathcal{U}^{\bullet}(\mathtt{let}\ f = \S F\ \mathtt{in}\ S_1) &\stackrel{\text{def}}{=} (\epsilon_1[], P_1, (\mathtt{let}\ f = \S F\ \mathtt{in}\ F_1))\\
\mathcal{U}^{\bullet}(\mathtt{let}\ p^{\otimes} = z\ \mathtt{in}\ S) &\stackrel{\text{def}}{=} (\mathtt{let}\ p^{\otimes} = z\ \mathtt{in}\ \epsilon_1[], P_1, F_1)
\end{aligned}$$

Figure 5.3: The unzipping $\mathcal{U}(S)$ is defined as $\mathcal{U}(S) \stackrel{\text{def}}{=} \epsilon[(P, \S F)]$, where $\epsilon[\,]$, $P$ and $F$ is given by the decomposition $\mathcal{U}^{\bullet}(S)$ above, with $\epsilon[\,]$ denoting contexts, i.e. terms with exactly one hole $[]$. In the inductive cases we suppose $\mathcal{U}^{\bullet}(S_i) = (\epsilon_i[], P_i, F_i)$.

The following proposition plays a key role in proving that the unzipping transformation preserves the logical equivalence $\sim$, and it will also be instrumental in Section 5.4 for justifying that this transformation can be safely omitted within our framework.

**Proposition 13.** Given $S \in \lambda\mathrm{LL}^{\mathtt{A}}$, we have: $S \sim \mathcal{U}(S)$, in particular they have the same type.

*Proof.* We proceed by induction on $S$. The only two delicate cases are for $\mathcal{U}(\mathtt{let}\ (!x, \S f) = S_1\ \mathtt{in}\ S_2)$ and for $\mathcal{U}(\mathtt{let}\ \S f = F\ \mathtt{in}\ S_1)$. Let us consider the former (the latter being an immediate variant). By definition $\mathcal{U}(S_i) \sim \epsilon_i[(P_i, \S F_i)]$, where $\mathcal{U}^{\bullet}(S_i) = (\epsilon_i[], P_i, F_i)$. Then:

$$\begin{aligned}
&\mathtt{let}\ (!x, \S f) = S_1\ \mathtt{in}\ S_2\\
&\sim \mathtt{let}\ (!x, \S f) = \epsilon_1[(P_1, F_1)]\ \mathtt{in}\ \epsilon_2[(P_2, F_2)]\\
&\sim \epsilon_1[\mathtt{let}\ (!x, \S f) = (P_1, F_1)\ \mathtt{in}\ \epsilon_2[(P_2, F_2)]]\\
&\sim \epsilon_1[\mathtt{let}\ !x = P_1\ \mathtt{in}\ \mathtt{let}\ \S f = F_1\ \mathtt{in}\ \epsilon_2[(P_2, F_2)]]
\end{aligned}$$

$$\sim \epsilon_1[\texttt{let}\ !x = P_1\ \texttt{in}\ \epsilon_2[(P_2, \texttt{let}\ \S f = F_1\ \texttt{in}\ F_2)]]$$

where the passages from the second-last line to the last line are correct as by Proposition 10 the context $\epsilon_2[]$ and $P_2$ have only free exponential variables, so in particular $f \notin FV(\epsilon_2[]) \cup FV(P_2)$ and Lemma 17 is applied. □

**Remark 8** (Unzipping as sequence of let-commutations)**.** It is worth noting that Proposition 13 shows that the unzipping transformation can be interpreted as a sequence of let-commutations. This conclusion stems from the fact that its proof depends solely on Lemma 17, which offers the minimal extension to $\beta$-reduction required for the result, and is applied exclusively to contexts $\epsilon[\,]$ involving exponential let-definitions (refer to the grammar in Figure 5.2).

**Illustrative Example 10** (Unzipping Transformation λLL)**.** Consider the $\lambda\text{LL}^{\texttt{A}}$ term $\mathcal{F}_\theta(P)$ in Equation 5.1, obtained by applying the forward mode to our program $P$. We proceed by applying our unzipping transformation $\mathcal{U}$ to $\mathcal{F}_\theta(P)$ as defined in Figure 5.3 and we obtain exactly the following $\lambda\text{LL}^{\texttt{A}}$ term $\mathcal{U}(\mathcal{F}_\theta(P))$ which has the same type as $\mathcal{F}_\theta(P)$.

$$\begin{aligned}
\mathcal{U}(\mathcal{F}_\theta(P)) = \;& \texttt{let}\ !v_1 = \underline{sin}\ !x\ \texttt{in}\ \texttt{let}\ !v_2 = !v_1\ \underline{*}\ !y\ \texttt{in} \\
& \texttt{let}\ !v_3 = \underline{cos}\ !x\ \texttt{in}\ \texttt{let}\ !v_4 = !v_2\ \underline{+}\ !v_3\ \texttt{in} \\
& \texttt{let}\ !w_1 = \underline{cos}\ !x\ \texttt{in}\ \texttt{let}\ !w_2 = !y\ \texttt{in} \\
& \texttt{let}\ !w_3 = !v_1\ \texttt{in}\ \texttt{let}\ !w_4 = \underline{-sin}\ !x\ \texttt{in} \\
& \texttt{let}\ !w_5 = \underline{1}\ \texttt{in}\ \texttt{let}\ !w_6 = \underline{1}\ \texttt{in} \\
& \left( !v_4, \S \left( \begin{array}{l}
\texttt{let}\ \S {f_1}^{\mathbb{R}\multimap\mathbb{R}} = \S(\lambda dx. w_1\ \dot{*}\ dx)\ \texttt{in} \\
\texttt{let}\ \S {f_2}^{(\mathbb{R}\&\mathbb{R})\multimap\mathbb{R}} = \S(\lambda\langle l_1, dy\rangle.(w_2\ \dot{*}\ l_1)\ \dot{+}\ (w_3\ \dot{*}\ dy))\ \texttt{in} \\
\texttt{let}\ \S {f_3}^{\mathbb{R}\multimap\mathbb{R}} = \S(\lambda dx. w_4\ \dot{*}\ dx)\ \texttt{in} \\
\texttt{let}\ \S {f_4}^{(\mathbb{R}\&\mathbb{R})\multimap\mathbb{R}} = \S(\lambda\langle l_2, l_3\rangle.(w_5\ \dot{*}\ l_2)\ \dot{+}\ (w_6\ \dot{*}\ l_3))\ \texttt{in} \\
\lambda u^{\mathbb{R}\&\mathbb{R}}.\texttt{let}\ \langle dx, dy\rangle = u\ \texttt{in}\ f_4\langle f_2\langle f_1\ dx, dy\rangle, f_3\ dx\rangle
\end{array} \right) \right)
\end{aligned} \tag{5.2}$$

We now examine the term above and observe that the first part of the term black part performs the primal computation, which consists of evaluating the primal output (stored in $!v_4$) along with a sequence of intermediate values ($!w_1$,$!w_2$,$!w_3$,$!w_4$,$!w_5$,$!w_6$) that influence certain computations performed in the tangent part. Comparing the purely tangent computation (part in blue) of the terms in Equation 5.1 and Equation 5.2 we can observe that applying the unzipping transformation results in a term where the tangent computation is placed entirely at the end of the term. Therefore, the transformation can be seen as let-commutations, according to Remark 8.

### 5.2.1 $\sim$ Preservation

We show that our unzipping transformation preserves the logical relation $\sim$. In other words, if two terms are logically equivalent before unzipping, then their unzipped forms remain logically equivalent. This property means that the unzipping transformation respects the logical relation $\sim$, maintaining the semantic equivalence of terms. This will be particularly important in Section 5.4, where we argue that unzipping can be safely omitted without altering the program's semantics.

**Proposition 14** (Unzipping preserves $\sim$)**.** Given $R, R' \in \lambda\text{LL}^{\texttt{A}}$, if $R \sim R'$ then $\mathcal{U}(R) \sim \mathcal{U}(R')$.

*Proof.* By Proposition 13 we have $R \sim \mathcal{U}(R)$ and $R' \sim \mathcal{U}(R')$, so we can conclude by transitivity of $\sim$ (Proposition 2). □

### 5.2.2 Soundness

The unzipping transformation in $\lambda$LL is shown to be sound by proving that it commutes with the $\delta$ translation, modulo the equivalence relation $\sim$.

**Theorem 9** (Soundness $\mathcal{U}$)**.** Given $\Gamma;\dot{\Gamma} \vdash^{\mathtt{Jax}} e:(\tau;\sigma)$ and an enumeration $\theta$ of $\dot{\Gamma}$, then

$$\mathcal{U}(\delta_\theta(e)) \;\sim\; \delta^{\mathsf{B}}_\theta(\mathcal{U}^{\mathtt{Jax}}(e)) \;\sim\; \delta_\theta(\mathcal{U}^{\mathtt{Jax}}(e)).$$

*Proof.* The equivalence $\delta^{\mathsf{B}}_\theta(\mathcal{U}^{\mathtt{Jax}}(e)) \sim \delta_\theta(\mathcal{U}^{\mathtt{Jax}}(e))$ is a consequence of Proposition 8. The first equivalence is proven by induction on $e$. Let us consider the two most delicate cases.

First, let $e = \mathtt{let}\ (x;\dot{y}) = e_1\ \mathtt{in}\ e_2$, so that

$$\delta_\theta(e) \stackrel{\text{def}}{=} \mathtt{let}\ (!x,\S f) = \delta_{\theta\cap FV^t(e_1)}(e_1)\ \mathtt{in}\ \mathtt{let}\ (!z,\S g) = \delta_{\dot{y},\theta\cap FV^t(e_2)}(e_2)\ \mathtt{in}\ (!z, F_{gf})$$

where $F_{gf} = \S(\lambda y^{\&\mathtt{t}(\theta)}.\mathtt{let}\ \langle y_1,y_2\rangle = \sigma^{\&\mathtt{t}(\theta)}_{FV^t(e_1)} y\ \mathtt{in}\ g\langle f y_1, y_2\rangle)$. By induction hypothesis, we have that: $\mathcal{U}(\delta_{\theta\cap FV^t(e_i)}(e_i)) \sim \delta^{\mathsf{B}}_{\theta\cap FV^t(e_i)}(\mathcal{U}^{\mathtt{Jax}}(e_i))$, for $i\in\{1,2\}$.

Let us write: $\mathcal{U}^{\mathtt{Jax}\bullet}(e_i) \stackrel{\text{def}}{=} (E_i[\,], e^p_i, \dot{e}_i)$ and $\mathcal{U}^\bullet(\delta_{\theta\cap FV^t(e_i)}(e_i)) \stackrel{\text{def}}{=} (\epsilon_i[], P_i, F_i)$. We have:

$$\mathcal{U}(\delta(e)) \tag{5.3}$$
$$\stackrel{\text{def}}{=} \epsilon_1[\mathtt{let}\ !x = P_1\ \mathtt{in}\ \epsilon_2[(P_2, \mathtt{let}\ \S f = F_1\ \mathtt{in}\ \mathtt{let}\ \S g = F_2\ \mathtt{in}\ F_{gf})]] \tag{5.4}$$
$$\sim \epsilon_1[\mathtt{let}\ !x = P_1\ \mathtt{in}\ \mathtt{let}\ \S f = F_1\ \mathtt{in}\ \epsilon_2[\mathtt{let}\ \S g = F_2\ \mathtt{in}\ (P_2, F_{gf})]] \tag{5.5}$$
$$=_\beta \epsilon_1[\mathtt{let}\ (!x,\S f) = (P_1,F_1)\ \mathtt{in}\ \epsilon_2[\mathtt{let}\ (!z,\S g) = (P_2,F_2)\ \mathtt{in}\ (!z,F_{gf})]] \tag{5.6}$$
$$\sim \mathtt{let}\ (!x,\S f) = \epsilon_1[(P_1,\S F_1)]\ \mathtt{in}\ \mathtt{let}\ (!z,\S g) = \epsilon_2[(P_2,\S F_2)]\ \mathtt{in}\ (!z,F_{gf}) \tag{5.7}$$
$$= \mathtt{let}\ (!x,\S f) = \mathcal{U}(\delta(e_1))\ \mathtt{in}\ \mathtt{let}\ (!z,\S g) = \mathcal{U}(\delta(e_2))\ \mathtt{in}\ (!z,F_{gf}) \tag{5.8}$$
$$\sim \mathtt{let}\ (!x,\S f) = \delta^{\mathsf{B}}(\mathcal{U}^{\mathtt{Jax}}(e_1))\ \mathtt{in}\ \mathtt{let}\ (!z,\S g) = \delta^{\mathsf{B}}(\mathcal{U}^{\mathtt{Jax}}(e_2))\ \mathtt{in}\ (!z,F_{gf}) \tag{5.9}$$
$$= \mathtt{let}\ (!x,\S f) = \delta^{\mathsf{B}}(E_1[(e^p_1,\dot{e}_1)])\ \mathtt{in}\ \mathtt{let}\ (!z,\S g) = \delta^{\mathsf{B}}(E_2[(e^p_2,\dot{e}_2)])\ \mathtt{in}\ (!z,F_{gf}) \tag{5.10}$$
$$\sim \mathtt{let}\ (!x,\S f) = \delta^{\mathsf{B}}(E_1)[(\delta^{\mathsf{B}}(e^p_1),\delta^{\mathsf{B}}(\dot{e}_1))]\ \mathtt{in}\ \mathtt{let}\ (!z,\S g) = \delta^{\mathsf{B}}(E_1)[(\delta^{\mathsf{B}}(e^p_2),\delta^{\mathsf{B}}(\dot{e}_2))]\ \mathtt{in}\ (!z,F_{gf}) \tag{5.11}$$
$$\sim \delta^{\mathsf{B}}(E_1)[\mathtt{let}\ !x = \delta^{\mathsf{B}}(e^p_1)\ \mathtt{in}\ \delta^{\mathsf{B}}(E_2)[(\delta^{\mathsf{B}}(e^p_2), \mathtt{let}\ \S f = \delta^{\mathsf{B}}(\dot{e}_1)\ \mathtt{in}\ \mathtt{let}\ \S g = \delta^{\mathsf{B}}(\dot{e}_2)\ \mathtt{in}\ F_{gf})]] \tag{5.12}$$
$$= \delta^{\mathsf{B}}(E_1)[\mathtt{let}\ !x = \delta^{\mathsf{B}}(e^p_1)\ \mathtt{in}\ \delta^{\mathsf{B}}(E_2)[(\delta^{\mathsf{B}}(e^p_2), \delta^{\mathsf{B}}(\mathtt{let}\ \dot{y} = \dot{e}_1\ \mathtt{in}\ \dot{e}_2))]] \tag{5.13}$$
$$= \delta^{\mathsf{B}}(E_1[\mathtt{let}\ !x = e^p_1\ \mathtt{in}\ E_2[(e^p_2, \mathtt{let}\ \dot{y} = \dot{e}_1\ \mathtt{in}\ \dot{e}_2)]]) \tag{5.14}$$
$$= \delta^{\mathsf{B}}_\theta(\mathcal{U}^{\mathtt{Jax}}(e)) \tag{5.15}$$

The passage from line (5.4) (resp. (5.6)) to (5.5) (resp.(5.7)) uses Proposition 10 and Lemma 17, and the line (5.8) to (5.9) is the induction hypothesis. At the end, the passage from line (5.10) (resp. (5.13)) to (5.11) (resp. (5.14)) uses Lemma 28.

We detail also the case is $e = \mathrm{drop}(e_1)$ too. Suppose $\mathcal{U}^{\mathtt{Jax}}(e_1) = E_1[(e^p_1,\dot{e}_1)]$ as well as $\mathcal{U}(\delta(e_1)) = \epsilon_1[(P_1,\S F_1)]$. We have:

$$\mathcal{U}(\delta(e)) \stackrel{\text{def}}{=} \mathcal{U}(\mathtt{let}\ (!x,\S f) = \delta(e_1)\ \mathtt{in}\ (!(\,),\S\lambda y.\lambda y.\mathtt{let}\ z = fy\ \mathtt{in}\ \langle\,\rangle)) \tag{5.16}$$
$$\to^* \mathcal{U}(\mathtt{let}\ (!x,\S f) = \delta(e_1)\ \mathtt{in}\ (!(\,),\S\langle\,\rangle)) \tag{5.17}$$
$$= \epsilon_1[\mathtt{let}\ !x = P_1\ \mathtt{in}\ (!(\,), \mathtt{let}\ \S f = F_1\ \mathtt{in}\ \S\lambda y.\langle\,\rangle)] \tag{5.18}$$
$$\sim \mathtt{let}\ (!x,\S f) = \epsilon_1[(P_1,F_1)]\ \mathtt{in}\ (!(\,),\S\lambda y.\langle\,\rangle) \tag{5.19}$$
$$= \mathtt{let}\ (!x,\S f) = \mathcal{U}(\delta(e_1))\ \mathtt{in}\ (!(\,),\S\lambda y.\langle\,\rangle) \tag{5.20}$$
$$\sim \mathtt{let}\ (!x,\S f) = \delta^{\mathsf{B}}(\mathcal{U}^{\mathtt{Jax}}(e_1))\ \mathtt{in}\ (!(\,),\S\lambda y.\langle\,\rangle) \tag{5.21}$$

$$= \texttt{let } (!x, \S f) = \delta^{\texttt{B}}(E_1[(e_1^p, \dot{e}_1)]) \texttt{ in } (!(\,), \S\lambda y.\langle\,\rangle) \tag{5.22}$$
$$= \texttt{let } (!x, \S f) = \delta^{\texttt{B}}(E_1)[(\delta^{\texttt{B}}(e_1^p), \delta^{\texttt{B}}(\dot{e}_1))] \texttt{ in } (!(\,), \S\lambda y.\langle\,\rangle) \tag{5.23}$$
$$\sim \delta^{\texttt{B}}(E_1)[(\texttt{let } !x = \delta^{\texttt{B}}(e_1^p) \texttt{ in } !(\,); \texttt{let } \S f = \delta^{\texttt{B}}(\dot{e}_1) \texttt{ in } \S\lambda y.\langle\,\rangle)] \tag{5.24}$$
$$= \delta^{\texttt{B}}(E_1)[(\delta^{\texttt{B}}(\mathrm{drop}(e_1^p)); \delta^{\texttt{B}}(\mathrm{drop}(\dot{e}_1)))] \tag{5.25}$$
$$= \delta^{\texttt{B}}(E_1[(\mathrm{drop}(e_1^p); \mathrm{drop}(\dot{e}_1))]) = \delta^{\texttt{B}}(\mathcal{U}^{\texttt{Jax}}(e)) \tag{5.26}$$

where the passage from line (5.18) to (5.19) uses Proposition 10 and Lemma 17, the passage from line (5.22) (resp. (5.25)) to line (5.23) (resp. (5.26)) is given by Lemma 28, and line (5.20) to (5.21) is the induction hypothesis. □

### 5.2.3 Work Preservation

We also establish that the unzipping transformation in λLL preserves the workload.

**Theorem 10** (Workload $\mathcal{U}$)**.** For $S \in \lambda\mathrm{LL}^{\texttt{A}}$, $\mathcal{W}(\mathcal{U}(S)) \leq \mathcal{W}(S)$. If moreover $S$ is safe, then $\mathcal{U}(S)$ is safe too.

*Proof.* The safeness of $\mathcal{U}$ is easy to prove by induction on $S$ simply checking the items in Definition 4. Let us focus on the proof related to work preservation of $\mathcal{U}$, we proceed by induction on $S$. The only two delicate cases:

- Case $S = (P, \S F)$:
  Let $\mathcal{U}^\bullet(S) = ([], P, F)$ and by definition $\mathcal{U}(S) = [(P, \S F)] = (P, \S F)$, so in this case we have that $S = \mathcal{U}(S)$ and we can conclude.
- Case $S = \texttt{let } (!x, \S f) = S_1 \texttt{ in } S_2$:
  Let $\mathcal{U}^\bullet(S_i) = (\epsilon_i[], P_i, F_i)$ and by definition $\mathcal{U}(S_i) = \epsilon_i[(P_i, \S F_i)]$.

  By induction hypothesis on $S_i$ we have that

  $$\mathcal{W}(\mathcal{U}(S_i)) = \mathcal{W}(\epsilon_i[(P_i, \S F_i)]) \leq \mathcal{W}(S_i).$$

  Moreover, $\mathcal{U}^\bullet(S) = (\epsilon_1[\texttt{let } !x = P_1 \texttt{ in } \epsilon_2[]], P_2, (\texttt{let } \S f = \S F_1 \texttt{ in } F_2))$ so we have:

  $$\begin{aligned}\mathcal{U}(S) &\stackrel{\text{def}}{=} \epsilon_1[\texttt{let } !x = P_1 \texttt{ in } \epsilon_2[(P_2, \texttt{let } \S f = \S F_1 \texttt{ in } \S F_2)]] \\ &\stackrel{\text{Prop. 10 + Lemma 17}}{\sim} \epsilon_1[\texttt{let } !x = P_1 \texttt{ in let } \S f = \S F_1 \texttt{ in } \epsilon_2[(P_2, \S F_2])] \\ &=_\beta \epsilon_1[\texttt{let } (!x, \S f) = (P_1, \S F_1) \texttt{ in } \epsilon_2[(P_2, \S F_2])] \\ &\stackrel{\text{Prop. 10 + Lemma 17}}{\sim} \texttt{let } (!x, \S f) = \epsilon_1[(P_1, \S F_1)] \texttt{ in } \epsilon_2[(P_2, \S F_2])\end{aligned} \tag{5.27}$$

  and we can conclude as follows:

  $$\begin{aligned}\mathcal{W}(\mathcal{U}(S)) &\stackrel{\text{Eq. 5.27}}{=} \mathcal{W}(\texttt{let } (!x, \S f) = \epsilon_1[(P_1, \S F_1)] \texttt{ in } \epsilon_2[(P_2, \S F_2])) \\ &\stackrel{\text{IHs}}{\leq} \mathcal{W}(\texttt{let } (!x, \S f) = S_1 \texttt{ in } S_2) \\ &= \mathcal{W}(S)\end{aligned}$$

□

Moreover, the lemma below follows directly form the work-preservation of the unzipping transformation (Theorem 10) and will be useful in the following section to prove that the transpose transformation is work preserving.

**Lemma 35.** Given $R \in \lambda\mathrm{LL}^{\texttt{A}}$ and $\mathcal{U}^\bullet(R) = (\epsilon[\,], P, F)$, we have: $\mathcal{W}(\epsilon[P]) \leq \mathcal{W}(R)$.

## 5.3 Transpose

Among the transformations considered, the transpose is the most intricate within our framework, as it necessitates careful handling to ensure preservation of the workload. This objective demands a more refined approach that systematically avoids the introduction of redundant operations, which could otherwise lead to an increase in the cost of the transformation. Consequently, the proof of work preservation involves a more technical and detailed analysis (see Lemma 41). The soundness of the transpose transformation is likewise formally established in Subsection 5.3.2.

Furthermore, it is important to recall that, in JAX, the transpose transformation is defined only over a restricted fragment of the codomain of the forward transformation — specifically, the Linear B fragment. This limitation necessitates an intermediate unzipping step between the forward and transpose phases. In contrast, our approach defines the transpose transformation directly on $\lambda\text{LL}^{\mathtt{A}}$, which also serves as the codomain of the forward transformation within $\lambda$LL. Additionally, we prove that our transpose transformation preserves the logical equivalence relation $\sim$ (see Proposition 15), further reinforcing its semantic robustness.

We define the transpose transformation $\mathcal{T}$ in Figure 5.4. The definition splits in three sub-definitions, giving, respectively, the action of $\mathcal{T}$ on the terms of $\lambda\text{LL}^{\mathtt{f}}$ (Figure 5.4a), on the terms of $\lambda\text{LL}^{\mathtt{t}}$ (Figure 5.4b) and finally on $\lambda\text{LL}^{\mathtt{A}}$ (Figure 5.4c). The first two definitions are mutually recursive, while $\mathcal{T}$ lifts to $\lambda\text{LL}^{\mathtt{A}}$ by a simple commutation with the exponential constructors.

The core of the definition is in the case of a term $U \in \lambda\text{LL}^{\mathtt{t}}$. Let us give some intuitions. By Proposition 10, $U$ has an &-sequence type $H$ and have at most one free &-sequence pattern $p^{\&} : L$. As described in Section 3.4, up to the logical equivalence $\sim$, the term $U$ can be interpreted as a linear map from the vector space corresponding to $L$ to the vector space associated with $H$. Linear maps are transposable: the term $\lambda f^{H \multimap \mathbb{R}} q^{\&^L}.fU$ is in fact of type $(H \multimap \mathbb{R}) \multimap (L \multimap \mathbb{R})$. The spaces associated with $L$ and $H$ have finite dimension, so they are isomorphic to their duals, via the terms dual and $\overline{\text{dual}}$ defined in Section 3.4. By composing all these terms we get (simplifying a bit with $\beta$-reduction):

$$\overleftarrow{U} = \overline{\text{dual}}_L \left(\lambda p^{\&}.\text{dual}_H(q^{\&})U\right) \tag{5.28}$$

which is a term of type $q^{\&} : H \vdash \overleftarrow{U} : L$, reversing $U$.

Why not simply defining $\mathcal{T}(U)$ as $\overleftarrow{U}$? Because it is highly inefficient with respect to the number of flops! In fact, $\overline{\text{dual}}_L \stackrel{\text{def}}{=} \lambda f^{L \multimap \mathbb{R}}. \sum_{V \in \mathcal{B}_L} (f(V)) *_L V$ is not a safe term as it replicates $f$ as many times as the dimension of the space associated with the input type $L$, which can be exceedingly large. The essence of AD lies in leveraging the structure of $U$ to derive a term extensionally equivalent to $\overleftarrow{U}$ but performing a number of flops comparable with $U$. Our goal is to show that this low-level manipulation is fully compatible with the Curry-Howard correspondence with linear logic. This is what is doing Figure 5.4, especially Figure 5.4b defining $\mathcal{T}(U)$ (the other sub-figures then extend $\mathcal{T}$ to the entire $\lambda\text{LL}^{\mathtt{A}}$ in a simple way) and in Lemma 38.

The definition of $\mathcal{T}(U)$ (Figure 5.4b) should take into account two crucial features of the &-pattern $p^{\&} : L$ in the typing environment of $U$: variables in $p^{\&}$ may occur several times in $U$ (because of additive contraction) or do not occur at all (because of & elimination). Different occurrences in $U$ of a variable in $p^{\&}$ should split into different variables in $\mathcal{T}(U)$. So for example, if $p^{\&} = \langle u, u' \rangle$ and $U = \langle \langle u, u \rangle, u' \rangle$, then $\mathcal{T}(U)$ is a term $\beta$-equivalent to $\langle u_1 + u_2, u' \rangle$ where $u_1$ and $u_2$ correspond to the two occurrences of $u$ in the term $U$. This splitting is implemented by a *variable renaming* $\alpha$ and of an action $\alpha[M]$ of the renaming on a term $M$, replacing any $u \in \text{Dom}(\alpha) \cap FV(M)$ with $\alpha(u)$.

Formally, we define the *variable renaming* $\alpha$ as a bijection between two sets of variables $\text{Dom}(\alpha)$ and $\text{Cod}(\alpha)$ preserving types (i.e. if $x : A$, then $\alpha(x) : A$). We denote by $\emptyset$ the renaming of empty domain. Given a term $M$ and a renaming $\alpha$, we define the *term renaming* $\alpha[M]$ as the term obtained by replacing all free occurrences in $M$ of variables $x \in \mathrm{Dom}(\alpha)$ by $\alpha(x)$, keeping the other variables untouched. On the contrast, giving a pattern $p^{\&}$ and a renaming $\alpha$, we define the *pattern renaming* $\alpha\langle p^{\&}\rangle$ as the action of renaming the occurrences in $p^{\&}$ of variables in $\mathrm{Dom}(\alpha)$ and of dropping the components of $p^{\&}$ having no variable in $\mathrm{Dom}(\alpha)$. More precisely, $\alpha\langle p^{\&}\rangle$ is defined by induction on $p^{\&}$:

$$\begin{aligned}
\alpha\langle u\rangle &\stackrel{\text{def}}{=} \alpha(u) && \text{if } u \in \mathrm{Dom}(\alpha),\\
\alpha\langle\langle p_1, p_2\rangle\rangle &\stackrel{\text{def}}{=} \langle \alpha\langle p_1\rangle, \alpha\langle p_2\rangle\rangle && \text{if } FV(p_1) \cap \mathrm{Dom}(\alpha) \neq \emptyset \text{ and } FV(p_2) \cap \mathrm{Dom}(\alpha) \neq \emptyset,\\
\alpha\langle\langle p_1, p_2\rangle\rangle &\stackrel{\text{def}}{=} \alpha\langle p_i\rangle && \text{if } FV(p_i) \cap \mathrm{Dom}(\alpha) \neq \emptyset \text{ and } FV(p_{3-1}) \cap \mathrm{Dom}(\alpha) = \emptyset\\
\alpha\langle p^{\&}\rangle &\stackrel{\text{def}}{=} t && \text{if } FV(p^{\&}) \cap \mathrm{Dom}(\alpha) = \emptyset,\ t \text{ fresh variable of type } t : \top
\end{aligned}$$

**Example 1** (Term Renaming vs Pattern Renaming)**.** Consider $\alpha = (x \mapsto x', y \mapsto y')$ and $M = p^{\&} = \langle\langle x, y\rangle, \langle x, \langle u, z\rangle\rangle\rangle$, then the term renaming applied to $M$ is $\alpha[M] = \langle\langle x', y'\rangle, \langle x', \langle u, z\rangle\rangle\rangle$, while the pattern renaming $\alpha\langle p^{\&}\rangle = \langle\langle x', y'\rangle, x'\rangle$.

**Lemma 36.** If $\Gamma, p^{\&} : L \vdash M : A$, then for every renaming $\alpha$ such that $FV(M) \cap FV(p^{\&}) \subseteq \mathrm{Dom}(\alpha) \subseteq FV(p^{\&})$, we have that:

1. $\Gamma, \alpha[p^{\&}] : L \vdash \alpha[M] : A$,
2. $\Gamma, \alpha\langle p^{\&}\rangle : L' \vdash \alpha[M] : A$, where $L'$ is the type of $\alpha\langle p^{\&}\rangle$,
3. $\mathcal{W}(\alpha[M]) = \mathcal{W}(M)$.

*Sketch Proof.* By induction on a derivation of $\Gamma, p^{\&} : L \vdash M : A$. The condition $\mathrm{Dom}(\alpha) \subseteq FV(p^{\&})$ is necessary to avoid the renaming in $\alpha[M]$ of variables in $\Gamma$. □

Moreover, the variables in $p^{\&}$ not occurring in $U$ will be associated with $\underline{0}$ terms. However, we must be parsimonious in adding such $\underline{0}$, as if they were summed with other terms, they would cost some useless numerical additions. Our notion of renaming is then partial, in the sense that $\mathrm{Dom}(\alpha)$ can be strictly smaller than the set $FV(p^{\&})$, in fact it will be $FV(p^{\&}) \cap FV(U)$.

More precisely, we implement all this by defining a closed term $\mu_{p^{\&},\alpha_1,\alpha_2}$ given a pattern $p^{\&} : L$ and two partial renamings $\alpha_i$ of the variables in $FV(p^{\&})$ which has the form:

$$\mu_{p^{\&},\alpha_1,\alpha_2} \stackrel{\text{def}}{=} \lambda\langle \alpha_1\langle p^{\&}\rangle, \alpha_2\langle p^{\&}\rangle\rangle.\nu(p^{\&}, \alpha_1, \alpha_2) \tag{5.29}$$

where $\alpha_i\langle p^{\&}\rangle$ is the pattern obtained from $p^{\&}$ by renaming its variables in $\mathrm{Dom}(\alpha_i)$ and by replacing the components not in $\mathrm{Dom}(\alpha_i)$ by a dummy variable $t : \top$, while $\nu(p^{\&}, \alpha_1, \alpha_2)$ is the $\lambda\mathrm{LL}^{\mathtt{t}}$ term obtained by exploring $p^{\&}$ and replacing its variables $u$ common to $\mathrm{Dom}(\alpha_1)$ and $\mathrm{Dom}(\alpha_2)$ by $\alpha_1(u)\dot{+}\alpha_2(u)$, those $u$ specific to $\mathrm{Dom}(\alpha_i) \setminus \mathrm{Dom}(\alpha_{3-i})$ by $\alpha_i(u)$ and the variables not in $\mathrm{Dom}(\alpha_1) \cup \mathrm{Dom}(\alpha_2)$ by zero terms $\underline{0}$.

**Example 2.** Consider a pattern $p^{\&} = \langle\langle x, y\rangle, \langle x, u\rangle\rangle$ and two partial renamings $\alpha_1 = (x \mapsto x_1, y \mapsto y_1)$ and $\alpha_2 = (x \mapsto x_2)$, we have $\mu_{p^{\&},\alpha_1,\alpha_2}$ equal to:

$$\lambda\langle\langle\langle x_1, y_1\rangle, x_1\rangle, \langle x_2, x_2\rangle\rangle.\langle\langle x_1\dot{+}x_2, y_1\rangle, \langle x_1\dot{+}x_2, \underline{0}\rangle\rangle\,.$$

Formally, given a pattern $p^{\&} : L$ and two renamings $\alpha_1$ and $\alpha_2$, the term $\nu(p^{\&}, \alpha_1, \alpha_2)$ is defined by induction on $p^{\&}$ as follows:

$$\nu(p^{\&}, \alpha_1, \alpha_2) \stackrel{\text{def}}{=} 0_L \qquad \text{if } FV(p^{\&}) \cap (\mathrm{Dom}(\alpha_1) \cup \mathrm{Dom}(\alpha_2)) = \emptyset,$$

$$\begin{aligned}
\nu(u,\alpha_1,\alpha_2) &\stackrel{\text{def}}{=} \alpha_1(u) +_L \alpha_2(u) && \text{if } u \in \mathrm{Dom}(\alpha_1) \cap \mathrm{Dom}(\alpha_2),\\
\nu(u,\alpha_1,\alpha_2) &\stackrel{\text{def}}{=} \alpha_i(u) && \text{if } u \in \mathrm{Dom}(\alpha_i) \setminus \mathrm{Dom}(\alpha_{3-i}),\\
\nu(\langle p_1,p_2\rangle,\alpha_1,\alpha_2) &\stackrel{\text{def}}{=} \langle \nu(p_1,\alpha_1,\alpha_2), \nu(p_2,\alpha_1,\alpha_2)\rangle && \text{otherwise.}
\end{aligned}$$

**Example 3.** Consider $\alpha$ and $p^{\&}$ as in the Example 1, and the renaming $\alpha'' = (x \mapsto x'')$, we have: $\nu(p^{\&},\alpha,\alpha'') = \langle\langle x' + x'', y'\rangle, \langle x' + x'', \underline{0}\rangle\rangle$.

**Lemma 37.** Let $p^{\&} : L$ be a pattern, $\alpha_1$ and $\alpha_2$ be two renamings with disjoint codomains and let $L_i$ be the type of $\alpha_i\langle p^{\&}\rangle$. We have that:

$$\lambda\langle \alpha_1\langle p^{\&}\rangle, \alpha_2\langle p^{\&}\rangle\rangle.\nu(p^{\&},\alpha_1,\alpha_2)$$

is a closed term of type $(L_1 \,\&\, L_2) \multimap L$ and of workload $\mathcal{W}(\mathrm{Dom}(\alpha_1) \cap \mathrm{Dom}(\alpha_2) \cap FV(p^{\&}))$.

*Sketch Proof.* Notice that we are supposing that the fresh variables of type $\top$ introduced by $\alpha_1\langle p^{\&}\rangle$ and $\alpha_2\langle p^{\&}\rangle$ are pairwise different, so that the hypothesis of $\mathrm{Cod}(\alpha_1) \cap \mathrm{Cod}(\alpha_2) = \emptyset$ guarantees that $\langle \alpha_1\langle p^{\&}\rangle, \alpha_2\langle p^{\&}\rangle\rangle$ is a well-defined pattern, i.e. there are no different occurrences of the same variable. We then prove by induction on $p^{\&}$ that:

- $\langle \alpha_1\langle p^{\&}\rangle, \alpha_2\langle p^{\&}\rangle\rangle : L_1 \,\&\, L_2 \vdash \nu(p^{\&},\alpha_1,\alpha_2) : L$ is derivable,
- any variable in $\alpha_1\langle p^{\&}\rangle$, $\alpha_2\langle p^{\&}\rangle$ of type different from $\top$ occurs free in $\nu(p^{\&},\alpha_1,\alpha_2)$,
- $\mathcal{W}(\nu(p^{\&},\alpha_1,\alpha_2)) = \mathcal{W}(\mathrm{Dom}(\alpha_1) \cap \mathrm{Dom}(\alpha_2) \cap FV(p^{\&}))$

□

Let's analyze the definition in Figure 5.4. Given $!\Sigma, \S\Phi \vdash F : L \multimap H$, Figure 5.4a defines the term $\mathcal{T}_{\S\overleftarrow{\Phi}}(F) : H \multimap L$, transposing $F$. The definition depends on a choice $\S\overleftarrow{\Phi}$ of new variables $\overleftarrow{f} : \S(H' \multimap L')$ for every variable $f : \S(L' \multimap H') \in \S\Phi$, which makes the transpose transformation compositional. So $!\Sigma, \S\overleftarrow{\Phi} \vdash \mathcal{T}_{\S\overleftarrow{\Phi}}(F) : H \multimap L$. The definition calls Figure 5.4b in the case of $F = \lambda p^{\&}.U$, with $!\Sigma, \S\Phi, p^{\&} : L \vdash U : H$. However, the transpose transformation applied to $U$ does not gives exactly a term of type $L$ as we cut off the components of $p^{\&}$ which have no free variable in $U$ so to avoid useless zero sums. So Figure 5.4b defines a term of type $!\Sigma, \S\overleftarrow{\Phi}, q^{\&} : H \vdash \mathcal{T}_{\S\overleftarrow{\Phi}, p^{\&}:L'}(U) : L'$, where $p^{\&} : L'$ is obtained by cutting off the useless components of $p^{\&} : L$ by using the identity renaming $\alpha$ restricted to $FV(U)$ and by defining $p^{\&} = \alpha\langle p^{\&}\rangle$. Then $\mathcal{T}_{\S\overleftarrow{\Phi}, p^{\&}:L'}(U) : L'$ is lifted to the right type $L$ in order to get $\mathcal{T}_{\S\overleftarrow{\Phi}}(F) : H \multimap L$ by adding zero on the components of $p^{\&} : L$ cut off in $p^{\&}$, by using the term $\nu(p^{\&},\alpha,\emptyset)$ (see first case of Figure 5.4a).

The additive pattern $q^{\&} : H$ in the typing environment of $\mathcal{T}_{\S\overleftarrow{\Phi}, p^{\&}:L'}(U) : L'$ can be given inductively. In the variable case ($U = u$), $q^{\&} = u$, in the cases of 0, $\langle\,\rangle$ and $FU'$, $q^{\&}$ can be any fresh pattern[1] of type $H$ and in case of the tuple $\langle U_1, U_2\rangle$, $q^{\&}$ is the tuple $\langle q_1^{\&}, q_2^{\&}\rangle$ of the additive patterns generated by the two recursive calls.

The definition of $\mathcal{T}_{\S\overleftarrow{\Phi}, p^{\&}}(\langle U_1, U_2\rangle)$ is crucial for having a reasonable workload. We consider in fact two renamings $\alpha_1$ and $\alpha_2$ of disjoint codomains so to split the different occurrences of variables in $p^{\&}$ which are additively contracted in $\langle U_1, U_2\rangle$. The fusion term $\nu(p^{\&},\alpha_1,\alpha_2)$ will then add the components of $p^{\&}$ that are split in $\langle \alpha_1[U_1], \alpha_2[U_2]\rangle$. The fact that $\mathrm{Dom}(\alpha_i) = FV(U_i) \cap FV(p^{\&})$ guarantees that there are not useless (but costly) zero sums in $\nu(p^{\&},\alpha_1,\alpha_2)$.

[1] For ease of writing, we avoid to explicit in the subscript of $\mathcal{T}(U)$ which fresh pattern has been chosen as this choice is irrelevant.

$$\mathcal{T}_{\S\overleftarrow{\Phi}}(\lambda p^{\&}.U) \stackrel{\text{def}}{=} \lambda q^{\&}.\mu_{p^{\&},\alpha,\emptyset}\langle \mathcal{T}_{\S\overleftarrow{\Phi},p^{\&}}(U), \langle\,\rangle\rangle$$

$$\mathcal{T}_{\S\overleftarrow{\Phi},\S\overleftarrow{f}}(f) \stackrel{\text{def}}{=} \overleftarrow{f}$$

$$\mathcal{T}_{\S\overleftarrow{\Phi}}(\texttt{let}\ \S f = \S F\ \texttt{in}\ G) \stackrel{\text{def}}{=} \begin{cases} \texttt{let}\ \S\overleftarrow{f} = \S\mathcal{T}_{\S\overleftarrow{\Phi}}(F)\ \texttt{in}\ \mathcal{T}_{\S\overleftarrow{\Phi},\S\overleftarrow{f}}(G) & \text{if } f \in FV(G), \\ \mathcal{T}_{\S\overleftarrow{\Phi}}(G) & \text{otherwise.} \end{cases}$$

$$\mathcal{T}_{\S\overleftarrow{\Phi}}(\dot{+}) \stackrel{\text{def}}{=} \lambda u.\langle u, u\rangle$$

$$\mathcal{T}_{\S\overleftarrow{\Phi}}(\dot{*}x) \stackrel{\text{def}}{=} \dot{*}x$$

(a) Definition of $\mathcal{T}$ on $\lambda\mathrm{LL}^{\mathtt{f}}$. If $!\Sigma, \S\Phi \vdash F : L \multimap H$, then $!\Sigma, \S\overleftarrow{\Phi} \vdash \mathcal{T}_{\S\overleftarrow{\Phi}}(F) : H \multimap L$. In the case of $\lambda p^{\&}.U$, $\alpha$ is the identity renaming restricted to $FV(p^{\&}) \cap FV(U)$, i.e. $\mathrm{Dom}(\alpha) = FV(p^{\&}) \cap FV(U)$ and $\alpha(u) = u$, while $\emptyset$ denotes the empty renaming, i.e. $\mathrm{Dom}(\emptyset) = \emptyset$; moreover $q^{\&}$ is the free pattern in $\mathcal{T}_{\S\overleftarrow{\Phi},p^{\&}}(U)$ and the term $\mu_{p^{\&},\alpha,\emptyset}$ is an instance of (5.29). In the case of $\texttt{let}\ \S f = \S F\ \texttt{in}\ G$, $\mathcal{T}$ simply eliminate $\S F$ if $f \notin FV(G)$, to avoid useless computation.

$$\mathcal{T}_{\S\overleftarrow{\Phi},p^{\&}}(u) \stackrel{\text{def}}{=} u \qquad \mathcal{T}_{\S\overleftarrow{\Phi},p^{\&}}(FU') \stackrel{\text{def}}{=} (\lambda q^{\&\prime}.\mathcal{T}_{\S\overleftarrow{\Phi},p^{\&}}(U'))(\mathcal{T}_{\S\overleftarrow{\Phi}}(F)q^{\&})$$

$$\mathcal{T}_{\S\overleftarrow{\Phi},p^{\&}}(\underline{0}) \stackrel{\text{def}}{=} \mathcal{T}_{\S\overleftarrow{\Phi},p^{\&}}(\langle\,\rangle) \stackrel{\text{def}}{=} \langle\,\rangle \quad \mathcal{T}_{\S\overleftarrow{\Phi},p^{\&}}(\langle U_1, U_2\rangle) \stackrel{\text{def}}{=} \mu_{p^{\&},\alpha_1,\alpha_2}\langle \mathcal{T}_{\S\overleftarrow{\Phi},\alpha_1[p^{\&}]}(\alpha_1[U_1]), \mathcal{T}_{\S\overleftarrow{\Phi},\alpha_2[p^{\&}]}(\alpha_2[U_2])\rangle$$

(b) Definition of $\mathcal{T}$ on $\lambda\mathrm{LL}^{\mathtt{t}}$. Given $!\Sigma, \S\Phi, p^{\&} : L \vdash U : H$, we have $!\Sigma, \S\overleftarrow{\Phi}, q^{\&} : H \vdash \mathcal{T}_{\S\overleftarrow{\Phi},p^{\&}}(U) : L'$, for a suitable pattern $q^{\&} : H$ and a type $L'$ erasing from $L$ the components of $p^{\&}$ not in $FV(U)$. In the $FU'$ case, $q^{\&\prime}$ is the pattern associated with $\mathcal{T}_{\S\overleftarrow{\Phi},p^{\&}}(U')$. In the $\langle U_1, U_2\rangle$ case, $\alpha_1$, $\alpha_2$ are two renamings of disjoint codomains s.t. $\mathrm{Dom}(\alpha_i) = FV(U_i) \cap FV(p^{\&})$, finally $\mu_{p^{\&},\alpha_1,\alpha_2}$ is given in (5.29).

$$\mathcal{T}_{\S\overleftarrow{\Phi}}((P, \S F)) \stackrel{\text{def}}{=} (P, \S\mathcal{T}_{\S\overleftarrow{\Phi}}(F))$$

$$\mathcal{T}_{\S\overleftarrow{\Phi}}(\texttt{let}\ (!x, \S f) = R\ \texttt{in}\ S) \stackrel{\text{def}}{=} \begin{cases} \texttt{let}\ !x = \epsilon[P]\ \texttt{in}\ \mathcal{T}_{\S\overleftarrow{\Phi}}(S),\ \text{for } \mathcal{U}^{\bullet}(R) = (\epsilon[], P, F) & \text{if } f \notin FV(S) \\ \texttt{let}\ (!x, \S\overleftarrow{f}) = \mathcal{T}_{\S\overleftarrow{\Phi}}(R)\ \texttt{in}\ \mathcal{T}_{\S\overleftarrow{\Phi},\S\overleftarrow{f}}(S) & \text{otherwise} \end{cases}$$

$$\mathcal{T}_{\S\overleftarrow{\Phi}}(\texttt{let}\ p^{\otimes} = z\ \texttt{in}\ S) \stackrel{\text{def}}{=} \texttt{let}\ p^{\otimes} = z\ \texttt{in}\ \mathcal{T}_{\S\overleftarrow{\Phi}}(S)$$

$$\mathcal{T}_{\S\overleftarrow{\Phi}}(\texttt{let}\ !x = P\ \texttt{in}\ S) \stackrel{\text{def}}{=} \texttt{let}\ !x = P\ \texttt{in}\ \mathcal{T}_{\S\overleftarrow{\Phi}}(S)$$

$$\mathcal{T}_{\S\overleftarrow{\Phi}}(\texttt{let}\ \S f = \S F\ \texttt{in}\ S) \stackrel{\text{def}}{=} \begin{cases} \texttt{let}\ \S\overleftarrow{f} = \S\mathcal{T}_{\S\overleftarrow{\Phi}}(F)\ \texttt{in}\ \mathcal{T}_{\S\overleftarrow{\Phi},\S\overleftarrow{f}}(S) & \text{if } f \notin FV(S) \\ \mathcal{T}_{\S\overleftarrow{\Phi},\S\overleftarrow{f}}(S) & \text{otherwise} \end{cases}$$

(c) Definition of $\mathcal{T}$ on $\lambda\mathrm{LL}^{\mathtt{A}}$. If $!\Sigma, \S\Phi \vdash S : !E \otimes \S(L \multimap H)$, then $!\Sigma, \S\overleftarrow{\Phi} \vdash \mathcal{T}_{\S\overleftarrow{\Phi}}(S) : !E \otimes \S(H \multimap L)$. As before, $\mathcal{T}_{\S\overleftarrow{\Phi}}(\texttt{let}\ \S f = \S F\ \texttt{in}\ S)$ supposes $f \in FV(S)$, otherwise $\mathcal{T}$ simply eliminate $\S F$.

Figure 5.4: Definition of the transpose transformation $\mathcal{T}$.

The following theorem states the typing preservation property of the transpose transformation:

**Theorem 11** (Type $\mathcal{T}$)**.** Let $!\Sigma, \S\Phi \vdash R : !E \otimes \S(L \multimap H)$ in $\lambda\text{LL}^{\mathtt{A}}$, then $\mathcal{T}_{\S\overleftarrow{\Phi}}(R)$ is well-typed as $!\Sigma, \S\overleftarrow{\Phi} \vdash \mathcal{T}_{\S\overleftarrow{\Phi}}(R) : !E \otimes \S(H \multimap L)$

*Sketch Proof.* By induction on $R$. □

Moreover, we show that our transpose transformation on $U \in \lambda\text{LL}^{\mathtt{t}}$ produces a term which is extensionally equivalent to $\overleftarrow{U}$ of Equation 5.28 but satisfying the condition of Proposition 4.

**Lemma 38.** Given a term $U \in \lambda\text{LL}^{\mathtt{t}}$ such that $p^{\&} : L \vdash U : H$ and let $q^{\&} : H$ be the free additive pattern in $\mathcal{T}_{p^{\&}}(U)$, then we have that

$$\overleftarrow{U} \sim_{q^{\&}:H\vdash L} \mu_{p^{\&},\alpha,\emptyset}\langle \mathcal{T}_{p^{\&}}(U), \langle\,\rangle\rangle$$

where $\alpha$ is the identity renaming restricted to $FV(p^{\&}) \cap FV(U)$.

*Sketch Proof.* By structural induction on $U$ we prove that

$$\overleftarrow{U}\{{}^{V}/_{q^{\&}}\} \sim_{q^{\&}:H\vdash L} (\lambda q^{\&}.\mu_{p^{\&},\alpha,\emptyset}\langle \mathcal{T}_{p^{\&}}(U), \langle\,\rangle\rangle)V'$$

for any value $V \sim_H V'$. This proof is very long and technical and we postpone it to the end of the chapter (jump to page 103 for a detailed discussion). □

Finally, we apply the transpose transformation in λLL on our running example.

**Illustrative Example 11** (Transpose Transformation λLL)**.** Consider the $\lambda\text{LL}^{\mathtt{A}}$ term $\mathcal{U}(\mathcal{F}_\theta(P))$ in Equation 5.2. We proceed by applying our transpose transformation $\mathcal{T}$ to it as defined in Figure 5.4 and we obtain a $\lambda\text{LL}^{\mathtt{A}}$ term $\mathcal{T}(\mathcal{U}(\mathcal{F}_\theta(P)))$ which is logical equivalent to the one in Equation 5.30 and it is well-typed as $!x : !\mathbb{R}, !y : !\mathbb{R} \vdash \mathcal{T}(\mathcal{U}(\mathcal{F}_\theta(P))) : !\mathbb{R} \otimes \S(\mathbb{R} \multimap (\mathbb{R}\&\mathbb{R}))$.

$$\begin{aligned}
&\mathcal{T}(\mathcal{U}(\mathcal{F}_\theta(P))) \sim \\
&\quad \texttt{let } !v_1 = \underline{sin}\ !x \texttt{ in let } !v_2 = !v_1 \mathbin{\underline{*}} !y \texttt{ in} \\
&\quad \texttt{let } !v_3 = \underline{cos}\ !x \texttt{ in let } !v_4 = !v_2 \mathbin{\underline{+}} !v_3 \texttt{ in} \\
&\quad \texttt{let } !w_1 = \underline{cos}\ !x \texttt{ in let } !w_2 = !y \texttt{ in} \\
&\quad \texttt{let } !w_3 = !v_1 \texttt{ in let } !w_4 = \underline{-sin}\ !x \texttt{ in} \\
&\quad \left( !v_4, \S \left( \begin{array}{l}
\texttt{let } \S\overleftarrow{f_1}^{\,\mathbb{R}\multimap\mathbb{R}} = \S(\lambda\bar{l}.w_1 \mathbin{\dot{*}} \bar{l}) \texttt{ in} \\
\texttt{let } \S\overleftarrow{f_2}^{\,\mathbb{R}\multimap(\mathbb{R}\&\mathbb{R})} = \S(\lambda\bar{l}.\langle w_2 \mathbin{\dot{*}} \bar{l}, w_3 \mathbin{\dot{*}} \bar{l}\rangle) \texttt{ in} \\
\texttt{let } \S\overleftarrow{f_3}^{\,\mathbb{R}\multimap\mathbb{R}} = \S(\lambda\bar{l}.w_4 \mathbin{\dot{*}} \bar{l}) \texttt{ in} \\
\texttt{let } \S\overleftarrow{f_4}^{\,\mathbb{R}\multimap(\mathbb{R}\&\mathbb{R})} = \S(\lambda\bar{l}.\langle \bar{l}, \bar{l}\rangle) \texttt{ in} \\
\lambda\bar{z}^{\,\mathbb{R}}. \left( \begin{array}{l}
(\lambda\langle\langle s_1, s_2\rangle, s_3\rangle.\langle s_1 \mathbin{\dot{+}} s_3, s_2\rangle) \\
\left( (\lambda\langle \overline{z'}, \overline{z''}\rangle.\langle(\lambda\langle \overline{z_1}, \overline{z_2}\rangle.\langle \overleftarrow{f_1}\overline{z_1}, \overline{z_2}\rangle)(\overleftarrow{f_2}\overline{z'}), \overleftarrow{f_3}\overline{z''}\rangle)(\overleftarrow{f_4}\bar{z}) \right)
\end{array} \right)
\end{array} \right) \right)
\end{aligned} \tag{5.30}$$

Let us analyze the term above. The primal computation begins with the forward computation, where primitive operations are computing step by step. More precisely, it first compute $\underline{sin}\ !x$ which is stored in $!v_1$, then multiply this by $!y$ to obtain $!v_2$. Next, $\underline{cos}\ !x$ is evaluated and assigned to $!v_3$, and finally added to $!v_2$, resulting in the expression $!v_4$. It is easy to see that $v_4$ is computing $g(x, y) = (sin(x) * y) + cos(x)$. Then, the primal computation proceed by computing the tape, a sequence of primal variables $(!w_1, !w_2, !w_3, !w_4)$ influencing certain tangent computations.

The computation proceeds with the reverse-mode differentiation pass. Let us focus on the part in red, especially the last line. The adjoint $\overline{z}$ is duplicated using a fanout operation (denoted in $\overleftarrow{f_4}$), allowing it to be passed to multiple components of the computational graph (recall Figure 1.3b from Chapter 1). One path traces the effect of $\underline{cos}\ !x$ back to $!x$ , while another flows through the $\underline{sin}\ !x \mathbin{\underline{*}} y$ term, contributing to both partial derivatives of $g$ with respect to $x$ and $y$. The linear functions $\overleftarrow{f_1}, \overleftarrow{f_2}, \overleftarrow{f_3}$ are each responsible for applying the chain rule locally. Finally, all these contributions are collected, summed where necessary, and scaled by the adjoint $\overline{z}$. This yields the resulting gradient vector: $\nabla g(x, y, \overline{z}) = (\overline{z} * (y * cos(x) - sin(x)), sin(x) * \overline{z})$.

### 5.3.1 ∼ Preservation

We show that our transpose transformation preserves the logical relation $\sim$, this will be useful to show that in our framework the unzipping transformation can be skipped.

More precisely, in order to prove that our transpose transformation preserves the logical relation $\sim$, we show the following auxiliary lemma using the fact that our transpose transformation on $U$ produces a term which is extensionally equivalent to $\overleftarrow{U}$.

**Lemma 39.** Given a well-typed term $R \in \lambda\mathrm{LL}^{\mathtt{A}}$ of type $!E \otimes \S(L \multimap H)$, we have that

$$\mathcal{T}_{\S\overleftarrow{\Phi}}(R) \sim \texttt{let}\ (!x, \S f) = R\ \texttt{in}\ \left(x, \S\left(\lambda q^{\&}.\overline{\mathrm{dual}}_L\left(\lambda p^{\&}.\mathrm{dual}_H(q^{\&})(fp^{\&})\right)\right)\right)$$

*Proof.* Let us denote by $\overleftarrow{R}$ the right-hand side of the equation in the statement. We first consider the case where $R$ is closed (and hence $\overleftarrow{R}$ is closed too).

By strong normalisation and Proposition 2, we can also consider that $R$ is a $\beta$-normal form. By the progress property (Proposition 1) we have that $R = (P, \S(\lambda p^{\&^H}.U))$ for $U$ a $\beta$-normal form of type $L$ with exactly one free pattern $p^{\&} : H$. We thus have that $\overleftarrow{R} \to^* (P, \overleftarrow{U})$, with $\overleftarrow{U}$ defined as in (5.28). By Lemma 38 and Proposition 2, we have that

$$\overleftarrow{R} \sim \left(P, \mu_{p^{\&},\alpha,\emptyset}\langle \mathcal{T}_{p^{\&}}(U), \langle\,\rangle\rangle\right) = (P, \mathcal{T}(\lambda p^{\&}.U)) = \mathcal{T}(R)$$

where $\alpha$ is the identity renaming restricted to $FV(p^{\&}) \cap FV(U)$. Therefore we can conclude.

In the case $R$ is open with a typing environment $\Gamma$, we have to prove $R[V_i/x_i] \sim \overleftarrow{R}[V_i'/x_i]$, for every $V_i \sim_{A_i} V_i'$ bunches of values for $x_i : A_i \in \Gamma$. Since $R[V_1/x_1, \ldots, V_n/x_n]$ is a closed term, we have already $R[V_1/x_1, \ldots, V_n/x_n] \sim \overleftarrow{R}[V_1/x_1, \ldots, V_n/x_n]$ by the previous reasoning. We then conclude by iterating $n-1$ times the context closure of $\sim$ (Proposition 2):

$$\overleftarrow{R}[V_1/x_1, \ldots, V_n/x_n] \sim \overleftarrow{R}[V_1'/x_1, V_2/x_2, \ldots, V_n/x_n] \cdots \sim \overleftarrow{R}[V_1'/x_1, \ldots, V_n'/x_n]$$

□

Finally, we are ready to show that our transpose transformation preserves the logical relation $\sim$ as follows

**Proposition 15** (Transpose preserves $\sim$)**.** Given $R, R' \in \lambda\mathrm{LL}^{\mathtt{A}}$, if $R \sim R'$ then $\mathcal{T}(R) \sim \mathcal{T}(R')$.

*Proof.* By using the context closure of $\sim$ (Proposition 2) and twice Lemma 39 as follows

$$\begin{aligned}\mathcal{T}_{\S\overleftarrow{\Phi}}(R) &\sim \texttt{let}\ (!x, \S f) = R\ \texttt{in}\ \left(x, \S\left(\lambda q^{\&}.\overline{\mathrm{dual}}_L\left(\lambda p^{\&}.\mathrm{dual}_H(q^{\&})(fp^{\&})\right)\right)\right)\\ &\sim \texttt{let}\ (!x, \S f) = R'\ \texttt{in}\ \left(x, \S\left(\lambda q^{\&}.\overline{\mathrm{dual}}_L\left(\lambda p^{\&}.\mathrm{dual}_H(q^{\&})(fp^{\&})\right)\right)\right)\\ &\sim \mathcal{T}_{\S\overleftarrow{\Phi}}(R')\end{aligned}$$

□

### 5.3.2 Soundness

In order to show that the soundness property for the transpose transformation holds we need to prove the following auxiliary lemma about the soundness of the transpose transformation on tangent expressions of Linear B.

**Lemma 40** (Soundness Transpose on Tangent)**.** Given a well-typed Tangent expression in Linear B $\Gamma; \dot{\Gamma} \vdash^{\text{Jax}} \dot{e} : (\mathbf{1}; \tau)$ and an enumeration $\theta$ for $\dot{\Gamma}$, then $\mathcal{T}(\delta^{\mathsf{B}}_{\theta}(\dot{e})) \sim \delta^{\mathsf{B}}_{\dot{u}:\tau}(\mathcal{T}^{\mathsf{Jax}}_{\theta;\dot{u}:\tau}(\dot{e}))$.

*Sketch Proof.* By induction on $\dot{e}$. The proof is very syntactic and for the sake of readability of the manuscript we postpone it to the end of the chapter (jump to page 107 for more details). □

**Theorem 12** (Soundness $\mathcal{T}$)**.** Given a well-typed expression in Linear B $\Gamma; \dot{\Gamma} \vdash^{\text{Jax}} d : (\sigma; \tau)$ and an enumeration $\theta$ of $\dot{\Gamma}$, then

$$\mathcal{T}(\delta_{\theta}(d)) \sim \mathcal{T}(\delta^{\mathsf{B}}_{\theta}(d)) \sim \delta^{\mathsf{B}}_{\dot{u}:\tau}(\mathcal{T}^{\mathsf{Jax}}_{\theta;\dot{u}:\tau}(d)) \sim \delta_{\dot{u}:\tau}(\mathcal{T}^{\mathsf{Jax}}_{\theta;\dot{u}:\tau}(d))$$

*Proof.* The first equivalence is a consequence of Proposition 15 and the last equivalence is by Proposition 8. We should then prove $\mathcal{T}(\delta^{\mathsf{B}}_{\theta}(d)) \sim \delta^{\mathsf{B}}_{\dot{u}:\tau}(\mathcal{T}^{\mathsf{Jax}}_{\theta;\dot{u}:\tau}(d))$ and we proceed by induction on $d$ using Lemma 40. □

### 5.3.3 Work Preservation

Given a finite set of variables $\mathcal{V}$, we will write $\mathcal{W}(\mathcal{V})$ for the sum $\sum_{x:A\in\mathcal{V}} \mathcal{W}(A)$. In the case of a set $\S\Phi$ of variables of type $f' : \S(L' \multimap H')$ and a term $M$, we also use the notation:

$$\mathcal{W}(\S\Phi^{\text{in}}_{M}) \stackrel{\text{def}}{=} \sum_{f:\S(L'\multimap H')\in\S\Phi\cap FV(M)} \mathcal{W}(L') \qquad \mathcal{W}(\S\Phi^{\text{out}}_{M}) \stackrel{\text{def}}{=} \sum_{f:\S(L'\multimap H')\in\S\Phi\cap FV(M)} \mathcal{W}(H')$$

Work preservation of the transpose transformation in $\lambda$LL follows directly as a corollary of the following lemma

**Lemma 41.** We have the following:

1. if $!\Sigma, \S\Phi, p^{\&} : L \vdash U : H$ and $\alpha$ is the identity renaming restricted to $FV(p^{\&}) \cap FV(U)$, then:

$$\mathcal{W}(\lambda q^{\&}.\mu_{p^{\&},\alpha,\emptyset}\langle \mathcal{T}_{\S\overleftarrow{\Phi},p^{\&}}(U), \langle\rangle\rangle) + \mathcal{W}(L) + \mathcal{W}(\S\overleftarrow{\Phi}^{\text{in}}_{\mathcal{T}_{\S\overleftarrow{\Phi},p^{\&}}(U)})$$
$$\leq \mathcal{W}(\lambda p^{\&}.U) + \mathcal{W}(H) + \mathcal{W}(\S\overleftarrow{\Phi}^{\text{out}}_{\mathcal{T}_{\S\overleftarrow{\Phi},p^{\&}}(U)})$$

2. if $!\Sigma, \S\Phi \vdash F : L \multimap H$, then:

$$\mathcal{W}(\mathcal{T}_{\S\overleftarrow{\Phi}}(F)) + \mathcal{W}(L) + \mathcal{W}(\S\overleftarrow{\Phi}^{\text{in}}_{\mathcal{T}_{\S\overleftarrow{\Phi}}(F)}) \leq \mathcal{W}(F) + \mathcal{W}(H) + \mathcal{W}(\S\overleftarrow{\Phi}^{\text{out}}_{\mathcal{T}_{\S\overleftarrow{\Phi}}(F)})$$

3. if $!\Sigma, \S\Phi \vdash R : !E \otimes \S(L \multimap H)$, then:

$$\mathcal{W}(\mathcal{T}_{\S\overleftarrow{\Phi}}(R)) + \mathcal{W}(L) + \mathcal{W}(\S\overleftarrow{\Phi}^{\text{in}}_{\mathcal{T}_{\S\overleftarrow{\Phi}}(R)}) \leq \mathcal{W}(R) + \mathcal{W}(H) + \mathcal{W}(\S\overleftarrow{\Phi}^{\text{out}}_{\mathcal{T}_{\S\overleftarrow{\Phi}}(R)})$$

The statement above is very technical, the core of the statement is in Claim 1 which is related to the work preservation of the transpose transformation on $\lambda\text{LL}^{\mathsf{t}}$ (defined in Figure 5.4b). Recall that the definition of transpose calls Figure 5.4b in the case of $F = \lambda p^{\&}.U$ where $!\Sigma, \S\Phi, p^{\&} : L \vdash U : H$. As we explained extensively before, the transpose transformation applied to $U$ return a term of type $L'$ obtained erasing from $L$ the components of $p^{\&}$ not in $FV(U)$ to avoid useless zero sums, obtaining an efficient transformation compared to $\overleftarrow{U}$. In order to restore the type $L$ of the hypothesis in Claim 1 we lifted $\mathcal{T}_{\S\overleftarrow{\Phi},p^{\&}}(U)$ to the right type $L$ by adding zero in the components of $p^{\&}$ which we cut off, exactly as we did in the first case of Figure 5.4a.

Moreover, we can observe also in our case that the workload needs to be amortized exactly as we have seen in Claim 2 of Theorem 3, which states the work preservation of transpose transformation in JAX.

Formally, the proof of Lemma 41 proceeds by induction on the term, analyzing the cases in Figure 5.4. In the following we try to give the intuition of the proof for each claim. However, the proof is very subtle and an interested reader can check the detailed proof at the end of the chapter.

*Proof Claim 1: Cases of $\mathcal{T}$ on $\lambda LL^{\mathtt{t}}$.* By typing of $\lambda\mathrm{LL}^{\mathtt{t}}$ we have that a term $U \in \lambda\mathrm{LL}^{\mathtt{t}}$ is well-typed as: $!\Sigma, \S\Phi, p^{\&} : L \vdash U : H$, so we are in the first case of the lemma.

By Equation 5.29 we have that

$$\lambda q^{\&}.\mu_{p^{\&},\alpha,\emptyset}\langle \mathcal{T}_{\S\overleftarrow{\Phi},p^{\&}}(U), \langle\,\rangle\rangle = \lambda q^{\&}.(\lambda\langle\alpha\langle p^{\&}\rangle, \emptyset\langle p^{\&}\rangle\rangle.\nu(p^{\&},\alpha,\emptyset))\langle \mathcal{T}_{\S\overleftarrow{\Phi},p^{\&}}(U), \langle\,\rangle\rangle$$

Observe that $FV(q^{\&}) \cap FV(\lambda\langle\alpha\langle p^{\&}\rangle, \emptyset\langle p^{\&}\rangle\rangle.\nu(p^{\&},\alpha,\emptyset)) = \emptyset$ so we have that

$$\begin{aligned}\mathcal{W}(\lambda q^{\&}.(\lambda\langle\alpha\langle p^{\&}\rangle, \emptyset\langle p^{\&}\rangle\rangle.\nu(p^{\&},\alpha,\emptyset))\langle \mathcal{T}_{\S\overleftarrow{\Phi},p^{\&}}(U), \langle\,\rangle\rangle)\\ = \mathcal{W}(\lambda\langle\alpha\langle p^{\&}\rangle, \emptyset\langle p^{\&}\rangle\rangle.\nu(p^{\&},\alpha,\emptyset)) + \mathcal{W}(\lambda q^{\&}.\langle \mathcal{T}_{\S\overleftarrow{\Phi},p^{\&}}(U), \langle\,\rangle\rangle)\end{aligned}$$

Moreover, by Lemma 37 we have $\mathcal{W}(\lambda\langle\alpha\langle p^{\&}\rangle, \emptyset\langle p^{\&}\rangle\rangle.\nu(p^{\&},\alpha,\emptyset)) = \mathcal{W}(\mathrm{Dom}(\alpha) \cap \emptyset \cap FV(p^{\&})) = \mathcal{W}(\emptyset) = 0$, so we have that the above equation is equal to $\mathcal{W}(\lambda q^{\&}.\langle \mathcal{T}_{\S\overleftarrow{\Phi},p^{\&}}(U), \langle\,\rangle\rangle)$ and can be simplified as $\mathcal{W}(\lambda q^{\&}.\mathcal{T}_{\S\overleftarrow{\Phi},p^{\&}}(U))$.

Summing up, in this case of the lemma it is enough to prove that

$$\mathcal{W}(\lambda q^{\&}.\mathcal{T}_{\S\overleftarrow{\Phi},p^{\&}}(U)) + \mathcal{W}(L) + \mathcal{W}(\S\overleftarrow{\Phi}^{\mathrm{in}}_{\mathcal{T}_{\S\overleftarrow{\Phi},p^{\&}}(U)}) \leq \mathcal{W}(\lambda p^{\&}.U) + \mathcal{W}(H) + \mathcal{W}(\S\overleftarrow{\Phi}^{\mathrm{out}}_{\mathcal{T}_{\S\overleftarrow{\Phi},p^{\&}}(U)})$$

and we proceed by analyzing the cases in Figure 5.4b. The proof of this Claim is very subtle and so for completeness we report the proof of all cases of induction at the end of the chapter (jump to page 111 for full details). □

*Proof Claim 2: Cases of $\mathcal{T}$ on $\lambda LL^{\mathtt{f}}$.* By typing of $\lambda\mathrm{LL}^{\mathtt{f}}$ we have that a term $F \in \lambda\mathrm{LL}^{\mathtt{f}}$ is well-typed as: $!\Sigma, \S\Phi \vdash F : L \multimap H$, so we are in the second case of the lemma and we want to prove that:

$$\mathcal{W}(\mathcal{T}_{\S\overleftarrow{\Phi}}(F)) + \mathcal{W}(L) + \mathcal{W}(\S\overleftarrow{\Phi}^{\mathrm{in}}_{\mathcal{T}_{\S\overleftarrow{\Phi}}(F)}) \leq \mathcal{W}(F) + \mathcal{W}(H) + \mathcal{W}(\S\overleftarrow{\Phi}^{\mathrm{out}}_{\mathcal{T}_{\S\overleftarrow{\Phi}}(F)}).$$

We proceed by analyzing the cases in Figure 5.4a. At the end of the chapter we will discuss the two most interesting cases: case $F = \lambda p^{\&}.U$ and composition (jump to page 117 for a detailed discussion). □

*Proof Claim 3: Cases of $\mathcal{T}$ on $\lambda LL^{\mathtt{A}}$.* By typing of $\lambda\mathrm{LL}^{\mathtt{A}}$ we have that a term $R \in \lambda\mathrm{LL}^{\mathtt{A}}$ is well-typed as: $!\Sigma, \S\Phi \vdash R : !E \otimes \S(L \multimap H)$, so we are in the third case of the lemma and we want to prove that:

$$\mathcal{W}(\mathcal{T}_{\S\overleftarrow{\Phi}}(R)) + \mathcal{W}(L) + \mathcal{W}(\S\overleftarrow{\Phi}^{\mathrm{in}}_{\mathcal{T}_{\S\overleftarrow{\Phi}}(R)}) \leq \mathcal{W}(R) + \mathcal{W}(H) + \mathcal{W}(\S\overleftarrow{\Phi}^{\mathrm{out}}_{\mathcal{T}_{\S\overleftarrow{\Phi}}(R)})$$

We proceed by analyzing the cases in Figure 5.4c. At the end of the chapter we will discuss the two most interesting cases: composition with $\lambda\mathrm{LL}^{\mathtt{f}}$ and composition (jump to page 120 for more details). □

**Corollary 4.** For every $\lambda LL^{\mathtt{A}}$ term with no free affine tangent variable $!\Sigma \vdash R : E \otimes \S(L \multimap H)$, we have $\mathcal{W}(\mathcal{T}(R)) + \mathcal{W}(L) \leq \mathcal{W}(R) + \mathcal{W}(H)$. If moreover $R$ is safe, then $\mathcal{T}(R)$ is safe too.

## 5.4 Skipping Unzipping

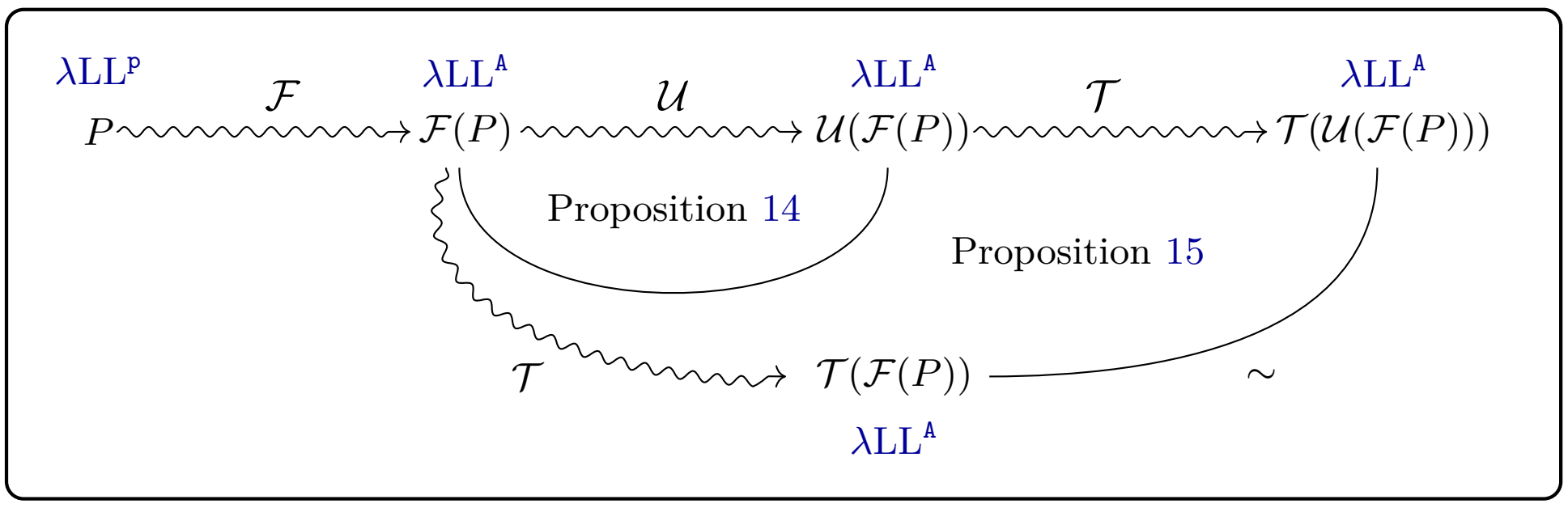

**AD System of $\lambda$LL**

Figure 5.5: How to Skip Unzipping in AD System of $\lambda$LL.

Recall from Section 2.2 that the JAX transpose is defined only on the fragment Linear B which is basically the image set of the JAX unzipping applied to Linear A, so that unzipping is a necessary step before the transpose. On the contrast, our $\lambda$LL transpose is defined on $\lambda LL^{\mathtt{A}}$, containing the whole image set of Linear A along the $\delta$ encoding (Proposition 12). One can then wonder whether applying or not the $\lambda$LL unzipping yields equivalent terms. This is in fact the case and can be achieved by a simple corollary (Corollary 5) from the fact that unzipping and transpose preserve the logical relation $\sim$ (Proposition 14 and Proposition 15, respectively), as described in Figure 5.5.

**Corollary 5** (Skipping Unzipping)**.** Given $R \in \lambda LL^{\mathtt{A}}$, then $\mathcal{T}(R) \sim \mathcal{T}(\mathcal{U}(R))$.

### 5.4.1 Benefits in Modularity and Parallelism

The implementation of the reverse mode as formalized in Linear A can obscure the parallel structure of a program due to the need for the unzipping transformation, which is not modular. By skipping unzipping as described above, we can preserve the program's inherent parallel structure. Let's illustrate this with an example. Consider the program $P = Q_1 * Q_2$ where $Q_1$ and $Q_2$ are two complex, independent subprograms of P, sharing only one input. Once a value for this latter is provided, $Q_1$ and $Q_2$ can be executed in parallel, needing to synchronise only at the end of their execution to perform the multiplication (seen as a numeric function, not the specialised $*$). We consider the two AD systems summarized in Figure 5.6. In order to keep the comparison between them more evident, we use $Q_1$ and $Q_2$ both as subexpressions in Linear A and as subterms in $\lambda$LL, even if technically we should translate them into the two languages.

**Autodiff Linear A.** We start by applying the AD system of JAX, called Autodiff, to the program with some syntactic simplifications for the sake of clarity. Recall that Autodiff decomposes reverse mode AD into three different transformations as described in Figure 5.6a.

We translate the program $P$ into a purely primal expression in Linear B and we obtain the expression $e$ in Figure 5.7a. In order to keep the comparison more evident we use the subprograms of $P$, namely $Q_1$ and $Q_2$, as subexpressions of $e$ without translating them.

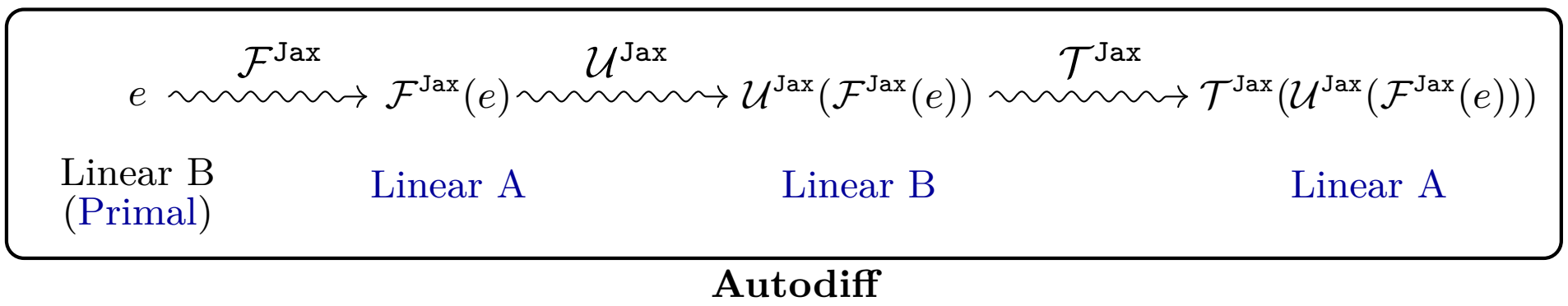

(a) Autodiff Linear A.

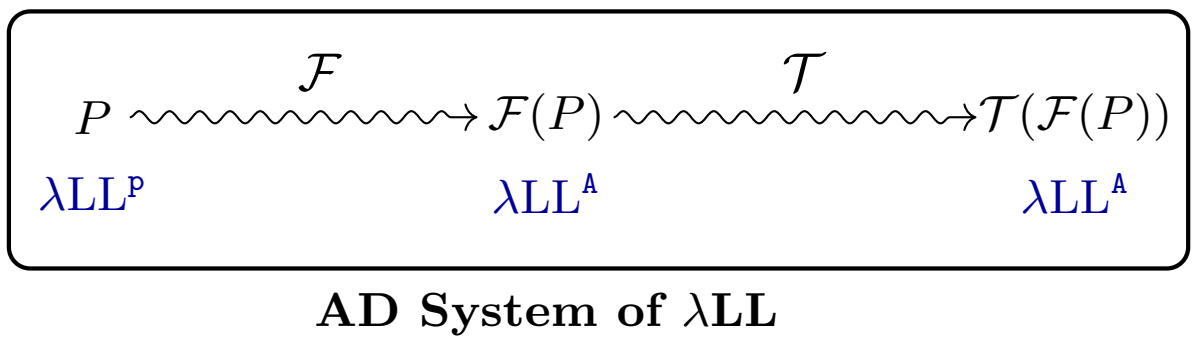

(b) AD System of λLL without Unzipping.

Figure 5.6: The two AD Systems.

We proceed by following the steps described in Figure 5.6a. First, we apply the transformation $\mathcal{F}^{\texttt{Jax}}$ (defined in Figure 2.3) to the expression $e$ and we obtain the expression in Figure 5.7b, where let $\dot{a} = \texttt{dup}(\dot{u})$ in $e'$ is syntactic sugar for let $(b; \dot{a}) = \texttt{dup}(\dot{u})$ in let $\otimes () = b$ in $e'$.

Note that we can not directly apply the transpose transformation to the expression $\mathcal{F}^{\texttt{Jax}}_{x \to \dot{u}}(e)$ in Figure 5.7b because it is an expression in Linear A not in Linear B and the transpose transformation is defined on Linear B. We proceed by applying the unzipping transformation $\mathcal{U}^{\texttt{Jax}}$ (given in Figure 2.5), assuming $\mathcal{U}^{\texttt{Jax}}(\mathcal{F}^{\texttt{Jax}}_{x \to \dot{v}_i}(Q_i)) = E_i$ in $(e^p_i, \dot{e}_i)$. After unzipping, we obtain an expression in Linear B, the latter is described in Figure 5.7c by using some simplifications.

Finally, we can apply the transpose transformation $\mathcal{T}^{\texttt{Jax}}$ (defined in Figure 2.7 and Figure 2.8) to the expression in Figure 5.7c. We obtain the Linear A expression defined in Figure 5.7d which computes the gradient of the initial program $P$ backward.

**AD System of λLL without Unzipping.** We apply the AD system of λLL as described in Figure 5.6b. We start by translating the program $P$ into a purely primal term in $\lambda LL^{\mathtt{P}}$ and we obtain the term $M$ in Figure 5.8a. As we did for JAX, we use the subprograms of $P$, namely $Q_1$ and $Q_2$, as subterms of $M$ without translating them.

Observe that by notational conventions defined in Section 3.1 we use the `let` notation for the application to an abstraction and the multiplication is a binary numeric function of type $!\mathbb{R} \otimes !\mathbb{R} \multimap !\mathbb{R}$, so we have that $M \approx (\lambda !y_1.(\lambda !y_2.\underline{*}(!y_1, !y_2))Q_2)Q_1$.

We proceed by following the steps described in Figure 5.6b. First, we apply the transformation $\mathcal{F}$ (given in Section 5.1 and in Figure 5.1) to the term $M$ and we obtain, after some $\beta$-steps and simplifications, the term in Figure 5.8b. Moreover, for the sake of readability, we reduce $\mathcal{F}_{x:!\mathbb{R}}(M)$ into the term $N$ in Figure 5.8c via $\beta$-reduction (defined in Figure 3.4). Then, we apply the transpose transformation $\mathcal{T}$ (given in Section 5.3 and in Figure 5.4) to the term $N$ and we obtain a term which is logical equivalent to the term in Figure 5.8d. This last step is very subtle, so at the end of the chapter we give a careful description of how to obtain the term in Figure 5.8d, starting from the term $N$ and applying the transpose transformation and the logical equivalence $\sim$ (jump to the page 124 for more details).

**Comparison.** An efficient approach to compute the gradient of $P$ would be to alternate the forward and transpose transformations by computing the independent computations related to $Q_1$ and $Q_2$ in parallel and only at the end compose the results.

Consider the programs in Figure 5.7d and in Figure 5.8d. The expression obtained by applying the reverse mode as formalized in Linear A requires to perform the two transformations

$$e = \begin{array}{l} \text{let } y_1 = Q_1 \text{ in} \\ \text{let } y_2 = Q_2 \text{ in} \\ y_1 \mathbin{\underline{*}} y_2 \end{array}$$

(a) Expression of (Primal) computing $P$.

$$\mathcal{F}^{\texttt{Jax}}_{x\to\dot{u}}(e) =$$
$$\begin{array}{l} \text{let } \dot{a} = \texttt{dup}(\dot{u}) \text{ in} \\ \text{let } \dot{\otimes}(\dot{v_1}, \dot{v_2}) = \dot{a} \text{ in} \\ \text{let } (y_1; \dot{z_1}) = \mathcal{F}^{\texttt{Jax}}_{x\to\dot{v_1}}(Q_1) \text{ in} \\ \text{let } (y_2; \dot{z_2}) = \mathcal{F}^{\texttt{Jax}}_{x\to\dot{v_2}}(Q_2) \text{ in} \\ (y_1 \mathbin{\underline{*}} y_2; (y_2 \dot{*} \dot{z_1}) \dot{+} (y_1 \dot{*} \dot{z_2})) \end{array}$$

(b) Forward Transformation of Autodiff applied to $e$

$$\mathcal{U}^{\texttt{Jax}}(\mathcal{F}^{\texttt{Jax}}_{x\to\dot{u}}(e)) =$$
$$\begin{array}{l} E_1 \text{ in let } y_1 = e^p_1 \text{ in} \\ E_2 \text{ in let } y_1 = e^p_2 \text{ in} \\ \left( y_1 \mathbin{\underline{*}} y_2;\ \begin{array}{l} \text{let } \dot{a} = \texttt{dup}(\dot{u}) \text{ in} \\ \text{let } \dot{\otimes}(\dot{v_1}, \dot{v_2}) = \dot{a} \text{ in} \\ \text{let } \dot{z_1} = \dot{e}_1 \text{ in} \\ \text{let } \dot{z_2} = \dot{e}_2 \text{ in} \\ (y_2 \dot{*} \dot{z_1}) \dot{+} (y_1 \dot{*} \dot{z_2}) \end{array} \right) \end{array}$$

(c) Unzipping Transformation of Autodiff applied to $\mathcal{F}^{\texttt{Jax}}_{x\to\dot{u}}(e)$ where $\mathcal{U}^{\texttt{Jax}}(\mathcal{F}^{\texttt{Jax}}_{x\to\dot{v}_i}(Q_i)) = E_i$ in $(e^p_i, \dot{e}_i)$

$$\mathcal{T}^{\texttt{Jax}}_{\dot{u}:\mathbb{R};\dot{w}:\mathbb{R}}(\mathcal{U}^{\texttt{Jax}}(\mathcal{F}^{\texttt{Jax}}_{x\to\dot{u}}(e))) =$$
$$\begin{array}{l} E_1 \text{ in let } y_1 = e^p_1 \text{ in} \\ E_2 \text{ in let } y_1 = e^p_2 \text{ in} \\ \left( y_1 \mathbin{\underline{*}} y_2;\ \begin{array}{l} \text{let } \dot{b} = \texttt{dup}(\dot{w}) \text{ in} \\ \text{let } \dot{\otimes}(\dot{b_1}, \dot{b_2}) = \dot{b} \text{ in} \\ \text{let } \dot{c_1} = y_2 \dot{*} \dot{b_1} \text{ in} \\ \text{let } \dot{d_1} = \mathcal{T}^{\texttt{Jax}}_{\dot{z_1}:\mathbb{R};\dot{c_1}:\mathbb{R}}(\dot{e}_1) \text{ in} \\ \text{let } \dot{c_2} = y_1 \dot{*} \dot{b_2} \text{ in} \\ \text{let } \dot{d_2} = \mathcal{T}^{\texttt{Jax}}_{\dot{z_2}:\mathbb{R};\dot{c_2}:\mathbb{R}}(\dot{e}_2) \text{ in} \\ \dot{d_1} \dot{+} \dot{d_2} \end{array} \right) \end{array}$$

(d) Transpose Transformation of Autodiff applied to $\mathcal{U}^{\texttt{Jax}}(\mathcal{F}^{\texttt{Jax}}_{x\to\dot{u}}(e))$

Figure 5.7: Autodiff Linear A applied to the program $P$ by skipping Unzipping.

sequentially. We can observe that by looking at the expression given in Figure 5.7d where the forward transformation, computing the primals $y_1$ and $y_2$, and the transpose, computing the co-tangents $d_1$ and $d_2$, are executed in a strict order. This requires performing the forward for both $Q_1$ and $Q_2$ before executing the transpose for both. However, since the subroutines $Q_1$ and $Q_2$ are independent, we should, in principle, be able to mix the transformations of $Q_1$ and $Q_2$ in various ways, resulting in an equivalent program. In our framework, we indeed have this flexibility. We can either apply unzipping to get a program similar to the one described above, or we can apply the transpose transformation without unzipping, resulting in the term described Figure 5.8d. In this case, the syntax of the term highlights that the two phases of $\mathcal{T}(\mathcal{F}(Q_1))$ are independent from those of $\mathcal{T}(\mathcal{F}(Q_2))$ and can be computed in parallel.

The two programs are extensionally equivalent as a consequence of Corollary 5, which exemplifies the kind of results our encoding enables. However, they can be implemented differently, with the parallel structure of the latter being more explicit than in the former.

$$M = \texttt{let } !y_1 = Q_1 \texttt{ in} \\ \texttt{let } !y_2 = Q_2 \texttt{ in} \\ !y_1 \,\underline{*}\, !y_2$$

(a) Term of $\lambda LL^{\mathtt{P}}$ computing $P$.

$$\begin{aligned}&\mathcal{F}_{x:!\mathbb{R}}(M) = \\ &\quad \texttt{let } (!y_1, \S f_1) = \mathcal{F}_{x:!\mathbb{R}}(Q_1) \texttt{ in} \\ &\quad \texttt{let } (!y_2, \S f_2) = \mathcal{F}_{x:!\mathbb{R}}(Q_2) \texttt{ in} \\ &\quad \texttt{let } !z_1 = y_2 \texttt{ in let } !z_2 = y_1 \texttt{ in} \\ &\quad \texttt{let } \S g = \S(\lambda\langle u_1, u_2\rangle.(z_1 \dot{*} u_1) \dot{+} (z_2 \dot{*} u_2)) \texttt{ in} \\ &\quad (!y_1 \,\underline{*}\, !y_2, \S(\lambda u^{\mathbb{R}}.\texttt{let } \langle u_1, u_2\rangle = \langle u, u\rangle \texttt{ in } g\langle f_1 u_1, f_2 u_2\rangle))\end{aligned}$$

(b) Forward Transformation in $\lambda LL^{\mathtt{A}}$ applied to $M$

$$\begin{aligned}N = \;& \texttt{let } (!y_1, \S f_1) = \mathcal{F}_{x:!\mathbb{R}}(Q_1) \texttt{ in} \\ & \texttt{let } (!y_2, \S f_2) = \mathcal{F}_{x:!\mathbb{R}}(Q_2) \texttt{ in} \\ & (!y_1 \,\underline{*}\, !y_2, \S(\lambda u^{\mathbb{R}}.\texttt{let } \langle u_1, u_2\rangle = \langle u, u\rangle \texttt{ in } (y_2 \dot{*} (f_1 u_1)) \dot{+} (y_1 \dot{*} (f_2 u_2))))\end{aligned}$$

(c) Term of $\lambda LL^{\mathtt{A}}$ s.t. $\mathcal{F}_{x:!\mathbb{R}}(M) \to^* N$.

$$\begin{aligned}\mathcal{T}_{\S\overleftarrow{\Phi}}(N) \sim \;& \texttt{let } (y_1, \overleftarrow{f_1}) = \mathcal{T}_{\S\overleftarrow{\Phi}}(\mathcal{F}_{x:!\mathbb{R}}(Q_1)) \texttt{ in} \\ & \texttt{let } (y_2, \overleftarrow{f_2}) = \mathcal{T}_{\S\overleftarrow{\Phi}, \overleftarrow{f_1}}(\mathcal{F}_{x:!\mathbb{R}}(Q_2)) \texttt{ in} \\ & (!y_1 \,\underline{*}\, !y_2, \S(\lambda h^{\mathbb{R}}.\texttt{let } \langle w_1, w_2\rangle = \langle h, h\rangle \texttt{ in } (y_2 \dot{*} (\overleftarrow{f_1} w_1)) \dot{+} (y_1 \dot{*} \overleftarrow{f_2} w_2)))\end{aligned}$$

(d) Transpose Transformation in $\lambda LL^{\mathtt{A}}$ applied to $N$.

Figure 5.8: AD System of λLL without unzipping applied to the program $P$.

# Detailed Proofs

## Bridging the Gap between $\overleftarrow{U}$ and $\mathcal{T}(U)$

In this Appendix we want to show that our transpose transformation on $U \in \lambda\mathrm{LL}^{\mathtt{t}}$ produces a term which is extensionally equivalent to $\overleftarrow{U}$ as defined in Equation 5.28 but satisfying the condition of Proposition 4. This is formalised in Lemma 38.

**Lemma 42.** Given a pattern $p^{\&} : L$ and $p^{\&} : L \vdash U_i : H_i$ we have that

$$\nu(p^{\&}, \alpha_1, \emptyset)\{^{U_1}/_{\alpha_1\langle U_1\rangle}\} \dot{+}_L \nu(p^{\&}, \alpha_2, \emptyset)\{^{U_2}/_{\alpha_2\langle U_2\rangle}\} \quad \sim_L \quad \nu(p^{\&}, \alpha'_1, \alpha'_2)\{^{\alpha'_i[U_i]}/_{\alpha'_i\langle\alpha'_i[U_i]\rangle}\}_{i=1}^{2}$$

where

- $\alpha_1$ and $\alpha_2$ are two identity renamings such that $\mathrm{Dom}(\alpha_i) = \mathrm{Cod}(\alpha_i) = FV(p^{\&}) \cap FV(U_i)$.
- $\alpha'_1$ and $\alpha'_2$ are two renamings such that $\mathrm{Dom}(\alpha'_i) = FV(p^{\&}) \cap FV(U_i)$ and $\mathrm{Cod}(\alpha'_1) \cap \mathrm{Cod}(\alpha'_2) = \emptyset$.

*Proof.* By induction on $p^{\&}$.

- Case $p^{\&} = \langle p_1^{\&}, p_2^{\&}\rangle$:
  In this case we have $L = L_1 \& L_2$.

$$\begin{aligned}
&\nu(p^{\&}, \alpha_1, \emptyset)\{^{U_1}/_{\alpha_1\langle U_1\rangle}\} \dot{+}_L \nu(p^{\&}, \alpha_2, \emptyset)\{^{U_2}/_{\alpha_2\langle U_2\rangle}\}\\
&\quad = \nu(\langle p_1^{\&}, p_2^{\&}\rangle, \alpha_1, \emptyset)\{^{U_1}/_{\alpha_1\langle U_1\rangle}\} \dot{+}_L \nu(\langle p_1^{\&}, p_2^{\&}\rangle, \alpha_2, \emptyset)\{^{U_2}/_{\alpha_2\langle U_2\rangle}\}\\
&\quad \overset{\text{Def }\nu}{=} \langle \nu(p_1^{\&}, \alpha_1, \emptyset), \nu(p_2^{\&}, \alpha_1, \emptyset)\rangle\{^{U_1}/_{\alpha_1\langle U_1\rangle}\} \dot{+}_L \langle \nu(p_1^{\&}, \alpha_2, \emptyset), \nu(p_2^{\&}, \alpha_2, \emptyset)\rangle\{^{U_2}/_{\alpha_2\langle U_2\rangle}\}\\
&\quad = \begin{array}{c}\langle \nu(p_1^{\&}, \alpha_1, \emptyset)\{^{U_1}/_{\alpha_1\langle U_1\rangle}\}, \nu(p_2^{\&}, \alpha_1, \emptyset)\{^{U_1}/_{\alpha_1\langle U_1\rangle}\}\rangle\\ \dot{+}_{L_1\&L_2}\\ \langle \nu(p_1^{\&}, \alpha_2, \emptyset)\{^{U_2}/_{\alpha_2\langle U_2\rangle}\}, \nu(p_2^{\&}, \alpha_2, \emptyset)\{^{U_2}/_{\alpha_2\langle U_2\rangle}\}\rangle\end{array}\\
&\quad = \langle \begin{array}{l}\nu(p_1^{\&}, \alpha_1, \emptyset)\{^{U_1}/_{\alpha_1\langle U_1\rangle}\} \dot{+}_{L_1} \nu(p_1^{\&}, \alpha_2, \emptyset)\{^{U_2}/_{\alpha_2\langle U_2\rangle}\},\\ \nu(p_2^{\&}, \alpha_1, \emptyset)\{^{U_1}/_{\alpha_1\langle U_1\rangle}\} \dot{+}_{L_2} \nu(p_2^{\&}, \alpha_2, \emptyset)\{^{U_2}/_{\alpha_2\langle U_2\rangle}\}\end{array} \rangle
\end{aligned} \tag{5.31}$$

$$\begin{aligned}
&\nu(p^{\&}, \alpha'_1, \alpha'_2)\{^{\alpha'_i[U_i]}/_{\alpha'_i\langle\alpha'_i[U_i]\rangle}\}_{i=1}^{2}\\
&\quad = \nu(\langle p_1^{\&}, p_2^{\&}\rangle, \alpha'_1, \alpha'_2)\{^{\alpha'_i[U_i]}/_{\alpha'_i\langle\alpha'_i[U_i]\rangle}\}_{i=1}^{2}\\
&\quad \overset{\text{Def }\nu}{=} \langle \nu(p_1^{\&}, \alpha'_1, \alpha'_2), \nu(p_2^{\&}, \alpha'_1, \alpha'_2)\rangle\{^{\alpha'_i[U_i]}/_{\alpha'_i\langle\alpha'_i[U_i]\rangle}\}_{i=1}^{2}\\
&\quad = \langle \nu(p_1^{\&}, \alpha'_1, \alpha'_2)\{^{\alpha'_i[U_i]}/_{\alpha'_i\langle\alpha'_i[U_i]\rangle}\}_{i=1}^{2}, \nu(p_2^{\&}, \alpha'_1, \alpha'_2)\{^{\alpha'_i[U_i]}/_{\alpha'_i\langle\alpha'_i[U_i]\rangle}\}_{i=1}^{2}\rangle
\end{aligned} \tag{5.32}$$

  By induction hypotheses we have that

$$\text{Eq. 5.31} \overset{\text{IHs}}{\sim} \langle \nu(p_1^{\&}, \alpha'_1, \alpha'_2)\{^{\alpha'_i[U_i]}/_{\alpha'_i\langle\alpha'_i[U_i]\rangle}\}_{i=1}^{2}, \nu(p_2^{\&}, \alpha'_1, \alpha'_2)\{^{\alpha'_i[U_i]}/_{\alpha'_i\langle\alpha'_i[U_i]\rangle}\}_{i=1}^{2}\rangle = \text{Eq. 5.32}$$

  and we can conclude.

- Case $p^{\&} = u$:
  In this case we have to analyze the cases in the definition of $\nu$. More precisely, we first split in subcases depending whether $u \in FV(U_1) \cup FV(U_2)$ or not:

– If $u \notin FV(U_1) \cup FV(U_2)$ then $u \notin \mathrm{Dom}(\alpha_i) = \{u\} \cap FV(U_i) = \emptyset$, so we have

$$\nu(p^{\&}, \alpha_1, \emptyset)\{{}^{U_1}/_{\alpha_1\langle U_1\rangle}\} \dotplus_L \nu(p^{\&}, \alpha_2, \emptyset)\{{}^{U_2}/_{\alpha_2\langle U_2\rangle}\} = 0_L \dotplus_L 0_L$$

Moreover, $u \notin \mathrm{Dom}(\alpha'_1) \cup \mathrm{Dom}(\alpha'_2) = \emptyset$ so we have

$$\nu(p^{\&}, \alpha'_1, \alpha'_2)\{{}^{\alpha'_i[U_i]}/_{\alpha'_i\langle\alpha'_i[U_i]\rangle}\}_{i=1}^2 = 0_L$$

We can conclude as $0_L \dotplus_L 0_L \sim_L 0_L$.

– If $u \in FV(U_1) \cup FV(U_2)$ then we have to split in subcases depending whether $u \in FV(U_1) \cap FV(U_2)$ or not:

  ∗ If $u \notin FV(U_1) \cap FV(U_2)$ then $u$ can be free in $U_1$ or in $U_2$. Let us consider the first case (the other being similar).
  In this case we have that $u \in \mathrm{Dom}(\alpha'_1)$ and $u \notin \mathrm{Dom}(\alpha'_2)$, so we have

$$\begin{aligned}\nu(p^{\&}, \alpha_1, \emptyset)\{{}^{U_1}/_{\alpha_1\langle U_1\rangle}\} \dotplus_L \nu(p^{\&}, \alpha_2, \emptyset)\{{}^{U_2}/_{\alpha_2\langle U_2\rangle}\} &= \alpha_1(u)\{{}^{U_1}/_{\alpha_1\langle U_1\rangle}\} \dotplus_L 0_L \\ &= u\{{}^{U_1}/_{\alpha_1\langle U_1\rangle}\} \dotplus_L 0_L \\ &= u\{{}^{U_1}/_{u}\} \dotplus_L 0_L \\ &= U_1 \dotplus_L 0_L \sim U_1\end{aligned}$$

  Moreover, $u \in \mathrm{Dom}(\alpha'_1) \setminus \mathrm{Dom}(\alpha'_2)$ and we have

$$\begin{aligned}\nu(p^{\&}, \alpha'_1, \alpha'_2)\{{}^{\alpha'_i[U_i]}/_{\alpha'_i\langle\alpha'_i[U_i]\rangle}\}_{i=1}^2 &= \alpha'_1(u)\{{}^{\alpha'_i[U_i]}/_{\alpha'_i\langle\alpha'_i[U_i]\rangle}\}_{i=1}^2 \\ &= \alpha'_1(u)\{{}^{\alpha'_1[U_1]}/_{\alpha'_i\langle\alpha'_1[U_1]\rangle}\} && (5.33) \\ &= \alpha'_1(u)\{{}^{\alpha'_1[U_1]}/_{\alpha'_1\langle u\rangle}\} && (5.34) \\ &= \alpha'_1(u)\{{}^{\alpha'_1[U_1]}/_{\alpha'_1(u)}\} && (5.35) \\ &\sim \alpha'_1[U_1]\end{aligned}$$

  where the passage from (5.33) to (5.34) is because by definition of $\alpha\langle p^{\&}\rangle$ we have that $\alpha'_i\langle\alpha'_i[U_i]\rangle = \alpha'_i\langle u\rangle$ as in this case by typing we have that $L = H_1$ and $U_1 = u$. Moreover, the passage from (5.34) to (5.35) is because by definition $\alpha\langle u\rangle = \alpha(u)$ when $u \in \mathrm{Dom}(\alpha)$.
  We can conclude by $\alpha$-renaming which is included in $\sim$.

  ∗ If $u \in FV(U_1) \cap FV(U_2)$ then $u \in \mathrm{Dom}(\alpha_i)$, so we have

$$\begin{aligned}&\nu(p^{\&}, \alpha_1, \emptyset)\{{}^{U_1}/_{\alpha_1\langle U_1\rangle}\} \dotplus_L \nu(p^{\&}, \alpha_2, \emptyset)\{{}^{U_2}/_{\alpha_2\langle U_2\rangle}\} \\ &\quad= \alpha_1(u)\{{}^{U_1}/_{\alpha_1\langle U_1\rangle}\} \dotplus_L \alpha_2(u)\{{}^{U_2}/_{\alpha_2\langle U_2\rangle}\} \\ &\quad= u\{{}^{U_1}/_{\alpha_1\langle U_1\rangle}\} \dotplus_L u\{{}^{U_2}/_{\alpha_2\langle U_2\rangle}\} \\ &\quad= u\{{}^{U_1}/_{u}\} \dotplus_L u\{{}^{U_2}/_{u}\} \\ &\quad= U_1 \dotplus_L U_2\end{aligned}$$

  Moreover, $u \in \mathrm{Dom}(\alpha'_1) \cap \mathrm{Dom}(\alpha'_2)$ so we have that

$$\begin{aligned}\nu(p^{\&}, \alpha'_1, \alpha'_2)\{{}^{\alpha'_i[U_i]}/_{\alpha'_i\langle\alpha'_i[U_i]\rangle}\}_{i=1}^2 &= (\alpha'_1(u) \dotplus_L \alpha'_2(u))\{{}^{\alpha'_i[U_i]}/_{\alpha'_i\langle\alpha'_i[U_i]\rangle}\}_{i=1}^2 \\ &= (\alpha'_1(u) \dotplus_L \alpha'_2(u))\{{}^{\alpha'_i[U_i]}/_{\alpha'_i\langle\alpha'_i[U_i]\rangle}\}_{i=1}^2 \\ &= \alpha'_1(u)\{{}^{\alpha'_i[U_i]}/_{\alpha'_i\langle\alpha'_i[U_i]\rangle}\}_{i=1}^2 \dotplus_L \alpha'_2(u)\{{}^{\alpha'_i[U_i]}/_{\alpha'_i\langle\alpha'_i[U_i]\rangle}\}_{i=1}^2 && (5.36) \\ &= \alpha'_1(u)\{{}^{\alpha'_i[U_i]}/_{\alpha'_i\langle u\rangle}\}_{i=1}^2 \dotplus_L \alpha'_2(u)\{{}^{\alpha'_i[U_i]}/_{\alpha'_i\langle u\rangle}\}_{i=1}^2 && (5.37) \\ &= \alpha'_1(u)\{{}^{\alpha'_i[U_i]}/_{\alpha'_i(u)}\}_{i=1}^2 \dotplus_L \alpha'_2(u)\{{}^{\alpha'_i[U_i]}/_{\alpha'_i(u)}\}_{i=1}^2 && (5.38)\end{aligned}$$

$$= \alpha_1'[U_1] \;\dot{+}_L\; \alpha_2'[U_2]$$

where the passage from (5.36) to (5.37) is because by definition of $\alpha\langle p^{\&}\rangle$ we have that $\alpha_i'\langle\alpha_i'[U_i]\rangle = \alpha_i'\langle u\rangle$ as in this case by typing we have that $L = H_i$ and $U_i = u$ for $i \in \{1, 2\}$. Moreover, the passage from (5.37) to (5.38) is because by definition $\alpha\langle u\rangle = \alpha(u)$ when $u \in \mathrm{Dom}(\alpha)$.
We can conclude by $\alpha$-renaming which is included in $\sim$.

□

**Corollary 6.** Given a pattern $p^{\&} : L$ we have that

$$(\mu_{p^{\&},\alpha_1,\emptyset}\langle U_1, \langle\rangle\rangle) \;\dot{+}_L\; (\mu_{p^{\&},\alpha_2,\emptyset}\langle U_2, \langle\rangle\rangle) \quad \sim_L \quad \mu_{p^{\&},\alpha_1',\alpha_2'}\langle \alpha_1'[U_1], \alpha_2'[U_2]\rangle$$

Finally, we prove Lemma 38 using Corollary 6 in the tuple case as follows

**Lemma 38.** Given a term $U \in \lambda\mathrm{LL}^{\mathsf{t}}$ such that $p^{\&} : L \vdash U : H$ and let $q^{\&} : H$ be the free additive pattern in $\mathcal{T}_{p^{\&}}(U)$, then we have that

$$\overleftarrow{U} \sim_{q^{\&}:H\vdash L} \mu_{p^{\&},\alpha,\emptyset}\langle \mathcal{T}_{p^{\&}}(U), \langle\,\rangle\rangle$$

where $\alpha$ is the identity renaming restricted to $FV(p^{\&}) \cap FV(U)$.

*Proof.* By structural induction on $U$ we prove that for any $V \sim_H V'$, we have

$$\overleftarrow{U}\{{}^{V}\!/_{q^{\&}}\} \sim_{q^{\&}:H\vdash L} \left(\lambda q^{\&}.\mu_{p^{\&},\alpha,\emptyset}\langle \mathcal{T}_{p^{\&}}(U), \langle\,\rangle\rangle\right) V'$$

where $\alpha$ is the identity renaming restricted to $FV(p^{\&}) \cap FV(U)$.

Let us consider the case $U = \langle U_1, U_2\rangle$, then we have $H = H_1 \& H_2$ and $q^{\&} = \langle q_1^{\&}, q_2^{\&}\rangle$.

By definition of value for a pattern we have $V = \langle V_1, V_2\rangle$. We proceed by analyzing $\overleftarrow{U}\{{}^{V}\!/_{q^{\&}}\}$ as follows

$$\begin{aligned}
\overleftarrow{U}\{{}^{V}\!/_{q^{\&}}\} &\overset{\text{Eq. 5.28}}{=} \left(\overline{\mathrm{dual}}_L\left(\lambda p^{\&}.\mathrm{dual}_H(q^{\&})U\right)\right)\{{}^{V}\!/_{q^{\&}}\} \\
&= \left(\overline{\mathrm{dual}}_L\left(\lambda p^{\&}.\mathrm{dual}_H(q^{\&})U\right)\right)\{{}^{V}\!/_{q^{\&}}\} \\
&\overset{\text{Def. }\overline{\mathrm{dual}}_L}{=} \left(\left(\lambda f.\sum_{W\in\mathcal{B}_L}(f(W))\dot{*}_L W\right)\left(\lambda p^{\&}.\mathrm{dual}_H(q^{\&})U\right)\right)\{{}^{V}\!/_{q^{\&}}\} \\
&\overset{\text{Def. }\overline{\mathrm{dual}}_H}{=} \left(\left(\lambda f.\sum_{W\in\mathcal{B}_L}(f(W))\dot{*}_L W\right)\left(\lambda p^{\&}.(\lambda h.\lambda h'.\mathcal{I}_{H_1\&H_2}(h,h'))(q^{\&})U\right)\right)\{{}^{V}\!/_{q^{\&}}\} \\
&= \left(\lambda f.\sum_{W\in\mathcal{B}_L}(f(W))\dot{*}_L W\right)\left(\lambda p^{\&}.(\lambda h.\lambda h'.\mathcal{I}_{H_1\&H_2}(h,h'))(V)U\right) \\
&= \left(\lambda f.\sum_{W\in\mathcal{B}_L}(f(W))\dot{*}_L W\right)\left(\lambda p^{\&}.(\lambda h.\lambda h'.\mathcal{I}_{H_1\&H_2}(h,h'))(\langle V_1, V_2\rangle)\langle U_1, U_2\rangle\right) \\
&\to^* \left(\lambda f.\sum_{W\in\mathcal{B}_L}(f(W))\dot{*}_L W\right)\left(\lambda p^{\&}.\mathcal{I}_{H_1\&H_2}(\langle V_1, V_2\rangle, \langle U_1, U_2\rangle)\right) \\
&\to \sum_{W\in\mathcal{B}_L}\left(\lambda p^{\&}.\mathcal{I}_{H_1\&H_2}(\langle V_1, V_2\rangle, \langle U_1, U_2\rangle)\right)(W)\dot{*}_L W
\end{aligned}$$

$$\to \sum_{W\in\mathcal{B}_L} \mathcal{I}_{H_1\&H_2}(\langle V_1,V_2\rangle,\langle U_1\{{}^{W}/_{p^\&}\},U_2\{{}^{W}/_{p^\&}\}\rangle)\dot{*}_L W$$

$$\overset{\text{Def. } \mathcal{I}_H \ +\beta\text{-red}}{\to^*} \sum_{W\in\mathcal{B}_L} (\mathcal{I}_{H_1}(V_1,U_1\{{}^{W}/_{p^\&}\}) \dotplus \mathcal{I}_{H_2}(V_2,U_2\{{}^{W}/_{p^\&}\}))\dot{*}_L W$$

$$\overset{\text{assoc}}{\sim} \left(\sum_{W\in\mathcal{B}_L} \mathcal{I}_{H_1}(V_1,U_1\{{}^{W}/_{p^\&}\})\dot{*}W\right) \dotplus_L \left(\sum_{W\in\mathcal{B}_L} \mathcal{I}_{H_2}(V_2,U_2\{{}^{W}/_{p^\&}\})\dot{*}W\right)$$

$$\sim \left(\left(\overline{\text{dual}}_L\left(\lambda p^\&.\text{dual}_{H_1}(q^\&)U_1\right)\right)\{{}^{V_1}/_{q_1^\&}\}\right) \dotplus_L \left(\left(\overline{\text{dual}}_L\left(\lambda p^\&.\text{dual}_{H_2}(q^\&)U_2\right)\right)\{{}^{V_2}/_{q_2^\&}\}\right)$$

$$= \overleftarrow{U_1}\{{}^{V_1}/_{q_1^\&}\} \dotplus_L \overleftarrow{U_2}\{{}^{V_2}/_{q_2^\&}\}$$

Summing up we have that

$$\overleftarrow{U}\{{}^{V}/_{q^\&}\} \sim \overleftarrow{U_1}\{{}^{V_1}/_{q_1^\&}\} \dotplus_L \overleftarrow{U_2}\{{}^{V_2}/_{q_2^\&}\} \tag{5.39}$$

By definition of value for a pattern we have $V' = \langle V_1', V_2'\rangle$. We proceed by analyzing $\left(\lambda q^\&.\mu_{p^\&,\alpha,\emptyset}\langle\mathcal{T}_{p^\&}(U),\langle\,\rangle\rangle\right)V'$ where $\alpha$ is the identity renaming restricted to $FV(p^\&)\cap(FV(U_1)\cup FV(U_2))$ as follows

$$\left(\lambda q^\&.\mu_{p^\&,\alpha,\emptyset}\langle\mathcal{T}_{p^\&}(U),\langle\,\rangle\rangle\right)V' = \left(\lambda q^\&.\mu_{p^\&,\alpha,\emptyset}\langle\mathcal{T}_{p^\&}(\langle U_1,U_2\rangle),\langle\,\rangle\rangle\right)V'$$

$$= \left(\lambda q^\&.\mu_{p^\&,\alpha,\emptyset}\langle\mu_{p^\&,\alpha_1',\alpha_2'}\langle\mathcal{T}_{\alpha_1'[p^\&]}(\alpha_1'[U_1]),\mathcal{T}_{\alpha_2'[p^\&]}(\alpha_2'[U_2])\rangle,\langle\,\rangle\rangle\right)V'$$

where $\alpha_1'$ and $\alpha_2'$ are two renamings with distinct codomains such that $\text{Dom}(\alpha_i') = FV(p^\&)\cap FV(U_i)$

$$\sim \left(\lambda q^\&.(\lambda\langle x,t\rangle.x)\langle\mu_{p^\&,\alpha_1',\alpha_2'}\langle\mathcal{T}_{\alpha_1'[p^\&]}(\alpha_1'[U_1]),\mathcal{T}_{\alpha_2'[p^\&]}(\alpha_2'[U_2])\rangle,\langle\,\rangle\rangle\right)V'$$

$$\sim \left(\lambda q^\&.\mu_{p^\&,\alpha_1',\alpha_2'}\langle\mathcal{T}_{\alpha_1'[p^\&]}(\alpha_1'[U_1]),\mathcal{T}_{\alpha_2'[p^\&]}(\alpha_2'[U_2])\rangle\right)V'$$

$$= \left(\lambda\langle q_1^\&,q_2^\&\rangle.\mu_{p^\&,\alpha_1',\alpha_2'}\langle\mathcal{T}_{\alpha_1'[p^\&]}(\alpha_1'[U_1]),\mathcal{T}_{\alpha_2'[p^\&]}(\alpha_2'[U_2])\rangle\right)\langle V_1',V_2'\rangle$$

Summing up we have that

$$\left(\lambda q^\&.\mu_{p^\&,\alpha,\emptyset}\langle\mathcal{T}_{p^\&}(U),\langle\,\rangle\rangle\right)V' \sim \left(\lambda\langle q_1^\&,q_2^\&\rangle.\mu_{p^\&,\alpha_1',\alpha_2'}\langle\mathcal{T}_{\alpha_1'[p^\&]}(\alpha_1'[U_1]),\mathcal{T}_{\alpha_2'[p^\&]}(\alpha_2'[U_2])\rangle\right)\langle V_1',V_2'\rangle \tag{5.40}$$

We can conclude that *Eq.* 5.39 $\sim$ *Eq.* 5.40 by applying induction hypotheses and Corollary 6 as follows

$$\overleftarrow{U}\{{}^{V}/_{q^\&}\} \overset{\text{Eq. 5.39}}{\sim} \overleftarrow{U_1}\{{}^{V_1}/_{q_1^\&}\} \dotplus_L \overleftarrow{U_2}\{{}^{V_2}/_{q_2^\&}\}$$

$$\overset{\text{IHs}}{\sim} \left(\lambda q_1^\&.\mu_{p^\&,\alpha_1,\emptyset}\langle\mathcal{T}_{p^\&}(U_1),\langle\,\rangle\rangle\right)V_1' \dotplus_L \left(\lambda q_2^\&.\mu_{p^\&,\alpha_2,\emptyset}\langle\mathcal{T}_{p^\&}(U_2),\langle\,\rangle\rangle\right)V_2'$$

where $\alpha_1$ and $\alpha_2$ are two identity renamings such that $\text{Dom}(\alpha_i) = FV(p^\&)\cap FV(U_i)$.

$$\sim \left(\lambda\langle q_1^\&,q_2^\&\rangle.\mu_{p^\&,\alpha_1,\emptyset}\langle\mathcal{T}_{p^\&}(U_1),\langle\,\rangle\rangle \dotplus_L \mu_{p^\&,\alpha_2,\emptyset}\langle\mathcal{T}_{p^\&}(U_2),\langle\,\rangle\rangle\right)\langle V_1',V_2'\rangle$$

$$\overset{\text{Cor. 6}}{\sim} \left(\lambda\langle q_1^\&,q_2^\&\rangle.\mu_{p^\&,\alpha_1',\alpha_2'}\langle\alpha_1'[\mathcal{T}_{p^\&}(U_1)],\alpha_2'[\mathcal{T}_{p^\&}(U_2)]\rangle\right)\langle V_1',V_2'\rangle$$

$$\sim \left(\lambda\langle q_1^\&,q_2^\&\rangle.\mu_{p^\&,\alpha_1',\alpha_2'}\langle\mathcal{T}_{\alpha_1'[p^\&]}(\alpha_1'[U_1]),\mathcal{T}_{\alpha_2'[p^\&]}(\alpha_2'[U_2])\rangle\right)\langle V_1',V_2'\rangle$$

$$\overset{\text{Eq. 5.40}}{\sim} \left(\lambda q^\&.\mu_{p^\&,\alpha,\emptyset}\langle\mathcal{T}_{p^\&}(U),\langle\,\rangle\rangle\right)V'$$

□

## Proof Soundness Transpose

**Lemma 40** (Soundness Transpose on Tangent)**.** Given a well-typed Tangent expression in Linear B $\Gamma;\dot{\Gamma} \vdash^{\text{Jax}} \dot{e} : (\mathtt{1};\tau)$ and an enumeration $\theta$ for $\dot{\Gamma}$, then $\mathcal{T}(\delta^{\mathsf{B}}_{\theta}(\dot{e})) \sim \delta^{\mathsf{B}}_{\dot{u}:\tau}(\mathcal{T}^{\mathsf{Jax}}_{\theta;\dot{u}:\tau}(\dot{e}))$.

*Proof.* We detail the proof for the composition below, so let's consider the case $\dot{e} = \mathtt{let}\ \dot{x} = \dot{e}_1\ \mathtt{in}\ \dot{e}_2$. Observe that by hypothesis we have $\Gamma;\dot{\Gamma} \vdash^{\text{Jax}} \mathtt{let}\ \dot{x} = \dot{e}_1\ \mathtt{in}\ \dot{e}_2 : (\mathtt{1};\tau)$ and by derived typing rules of JAX (Figure 2.2) we have $\Gamma = \Gamma_1 \cup \Gamma_2$ and $\dot{\Gamma} = \dot{\Gamma}_1, \dot{\Gamma}_2$ such that

$$\Gamma_1;\dot{\Gamma}_1 \vdash^{\text{Jax}} \dot{e}_1 : (\mathtt{1};\tau_1)$$
$$\Gamma_2;\dot{\Gamma}_2, \dot{x} : \tau_1 \vdash^{\text{Jax}} \dot{e}_2 : (\mathtt{1};\tau)$$

First, let's analyze $\mathcal{T}(\delta^{\mathsf{B}}_{\theta}(\dot{e}))$

$$\mathcal{T}(\delta^{\mathsf{B}}_{\theta}(\dot{e})) = \mathcal{T}(\delta^{\mathsf{B}}_{\theta}(\mathtt{let}\ \dot{x} = \dot{e}_1\ \mathtt{in}\ \dot{e}_2))$$
$$= \mathcal{T}\left(\begin{array}{l} \mathtt{let}\ \S f = \S\delta^{\mathsf{B}}_{\theta\cap FV^t(\dot{e}_1)}(\dot{e}_1)\ \mathtt{in} \\ \mathtt{let}\ \S g = \S\delta^{\mathsf{B}}_{\dot{x},\theta\cap FV^t(\dot{e}_2)}(\dot{e}_2)\ \mathtt{in} \\ \lambda u^{\&\mathtt{t}(\theta)}.\mathtt{let}\ \langle u_1, u_2\rangle = \sigma_{FV^t(\dot{e}_1)} u\ \mathtt{in}\ g(\overline{\sigma}^{\&\mathtt{t}(\dot{x},\theta\cap FV^t(\dot{e}_2))}_{\dot{u}}\langle f u_1, u_2\rangle) \end{array}\right)$$

where $\S f$ is of type $\S(\mathtt{t}(\theta\cap FV^t(\dot{e}_1)) \multimap \mathtt{t}(\tau_1))$, $\S g$ is of type $\S(\mathtt{t}(\dot{x},\theta\cap FV^t(\dot{e}_2)) \multimap \mathtt{t}(\tau))$ and $\delta^{\mathsf{B}}_{\theta}(\dot{e})$ is well-typed as $\mathtt{p}(\Gamma) \vdash \delta^{\mathsf{B}}_{\theta}(\dot{e}) : \&\mathtt{t}(\theta) \multimap \mathtt{t}(\tau)$.

By the definition of transpose transformation in $\lambda$LL (Figure 5.4) we have

$$= \begin{array}{l} \mathtt{let}\ \S\overleftarrow{f} = \S\mathcal{T}(\delta^{\mathsf{B}}_{\theta\cap FV^t(\dot{e}_1)}(\dot{e}_1))\ \mathtt{in} \\ \mathtt{let}\ \S\overleftarrow{g} = \S\mathcal{T}(\delta^{\mathsf{B}}_{\dot{x},\theta\cap FV^t(\dot{e}_2)}(\dot{e}_2))\ \mathtt{in} \\ \mathcal{T}(\lambda u^{\&\mathtt{t}(\theta)}.\mathtt{let}\ \langle u_1, u_2\rangle = \sigma_{FV^t(\dot{e}_1)} u\ \mathtt{in}\ g(\overline{\sigma}^{\&\mathtt{t}(\dot{x},\theta\cap FV^t(\dot{e}_2))}_{\dot{u}}\langle f u_1, u_2\rangle)) \end{array}$$

where $\S\overleftarrow{f}$ is of type $\S(\mathtt{t}(\tau_1) \multimap \mathtt{t}(\theta\cap FV^t(\dot{e}_1)))$, $\S g$ is of type $\S(\mathtt{t}(\tau) \multimap \mathtt{t}(\dot{x},\theta\cap FV^t(\dot{e}_2)))$ and $\mathcal{T}(\delta^{\mathsf{B}}_{\theta}(\dot{e}))$ is well-typed as $\mathtt{p}(\Gamma) \vdash \mathcal{T}(\delta^{\mathsf{B}}_{\theta}(\dot{e})) : \mathtt{t}(\tau) \multimap \&\mathtt{t}(\theta)$.

Now we want to compute $\mathcal{T}(\lambda u^{\&\mathtt{t}(\theta)}.U)$ where

$$U = \mathtt{let}\ \langle u_1, u_2\rangle = \sigma_{FV^t(\dot{e}_1)} u\ \mathtt{in}\ g(\overline{\sigma}^{\&\mathtt{t}(\dot{x},\theta\cap FV^t(\dot{e}_2))}_{\dot{u}}\langle f u_1, u_2\rangle)$$
$$\to \mathtt{let}\ \langle u_1, u_2\rangle = \sigma_{FV^t(\dot{e}_1)} u\ \mathtt{in}\ g(\langle f u_1, u_2\rangle) = U'$$

because by typing $\overline{\sigma}^{\&\mathtt{t}(\dot{x},\theta\cap FV^t(\dot{e}_2))}_{\dot{u}}$ is the identity and so we perform a $\beta_\lambda$ step. First we apply the case $\lambda p^{\&}.U$ of Figure 5.4a and we obtain that

$$\mathcal{T}(\lambda u^{\&\mathtt{t}(\theta)}.U') = \lambda v^{\mathtt{t}(\tau)}.\mu_{u,\alpha,\emptyset}\langle \mathcal{T}_{\S\overleftarrow{\Phi},u}(U'), \langle\,\rangle\rangle \tag{5.41}$$

where $\overleftarrow{\Phi} = \overleftarrow{f}, \overleftarrow{g}$. Moreover, $U'$ is syntactic sugar for the term $(\lambda\langle u_1, u_2\rangle.g\langle f u_1, u_2\rangle)(\sigma_{FV^t(\dot{e}_1)} u)$ and $\sigma_{FV^t(\dot{e}_1)}$ is of type $\&\mathtt{t}(\theta) \multimap (\&\mathtt{t}(\theta\cap FV^t(\dot{e}_1)))\&(\&\mathtt{t}(\theta\cap FV^t(\dot{e}_2)))$.

In order to make the proof clearer, we compute the transpose $\mathcal{T}(U')$ step by step, underlining at each step the subterm that we are going to analyze.

$$\begin{array}{l} \mathcal{T}((\lambda\langle u_1, u_2\rangle.g\langle f u_1, u_2\rangle)(\sigma_{FV^t(\dot{e}_1)} u)) \\ \quad = (\lambda v'.\underline{\mathcal{T}_{\overleftarrow{\Phi},u}(\sigma_{FV^t(\dot{e}_1)})})(\mathcal{T}_{\overleftarrow{\Phi}}(\lambda\langle u_1, u_2\rangle.g\langle f u_1, u_2\rangle)v) \\ \quad = (\lambda v'.\overline{\sigma}^{\&\mathtt{t}(\theta)} v')(\underline{\mathcal{T}_{\overleftarrow{\Phi}}(\lambda\langle u_1, u_2\rangle.g\langle f u_1, u_2\rangle)}v) \end{array}$$

$$\begin{aligned}
&\text{where } \overline{\sigma}^{\&\mathtt{t}(\theta)} \text{ has type } (\&\mathtt{t}(\theta \cap FV^t(\delta^{\mathsf{B}}(\dot{e_1}))))\&(\&\mathtt{t}(\theta \cap FV^t(\delta^{\mathsf{B}}(\dot{e_2})))) \multimap \&\mathtt{t}(\theta) \\
&= (\lambda v'.\overline{\sigma}^{\&\mathtt{t}(\theta)} v')((\lambda w^{\mathtt{t}(\tau)}.(\mu_{\langle u_1,u_2\rangle,\alpha',\emptyset} \underline{\mathcal{T}_{\overleftarrow{\Phi},\langle u_1,u_2\rangle}(g\langle f u_1, u_2\rangle)}))v) \qquad (5.42)
\end{aligned}$$

Let's compute $\mathcal{T}_{\overleftarrow{\Phi},\langle u_1,u_2\rangle}(g\langle f u_1, u_2\rangle)$ as follows

$$\begin{aligned}
&\mathcal{T}_{\overleftarrow{\Phi},\langle u_1,u_2\rangle}(g\langle f u_1, u_2\rangle) \\
&\qquad = (\lambda\langle z_1,z_2\rangle^{\mathtt{t}(\tau_1)\&(\&\mathtt{t}(\theta\cap FV^t(\delta^{\mathsf{B}}(\dot{e_2}))))}.\mathcal{T}_{\overleftarrow{\Phi},\langle u_1,u_2\rangle}(\langle f u_1, u_2\rangle))(\underline{\mathcal{T}_{\overleftarrow{\Phi}}(g)}w) \\
&\qquad = (\lambda\langle z_1,z_2\rangle.\underline{\mathcal{T}_{\overleftarrow{\Phi},\langle u_1,u_2\rangle}(\langle f u_1, u_2\rangle)})(\overleftarrow{g}\, w) \\
&\qquad = (\lambda\langle z_1,z_2\rangle.\mu_{\langle u_1,u_2\rangle,\alpha_1,\alpha_2}\langle \mathcal{T}_{\overleftarrow{\Phi},\alpha_1[u_1]}(\alpha_1[f u_1]), \mathcal{T}_{\overleftarrow{\Phi},\alpha_2[u_2]}(\alpha_2[u_2])\rangle)(\overleftarrow{g}\, w) \qquad (5.43)
\end{aligned}$$

Recall that in our transpose we use the renamings and the term $\mu$ to deal with variables in the pattern $p^{\&}$ occurring several times or does not occur, respectively. Observe that in this case the variables $u, u_1, u_2$ are used exactly once so we have that $\alpha_1$ and $\alpha_2$ are the identity renamings and $\mu$ is the identity term, so we have the following

$$\begin{aligned}
&= (\lambda\langle z_1,z_2\rangle.(\lambda w'.w')\langle \mathcal{T}_{\overleftarrow{\Phi},u_1}(f u_1), \mathcal{T}_{\overleftarrow{\Phi},u_2}(u_2)\rangle)(\overleftarrow{g}\, w) \\
&\to (\lambda\langle z_1,z_2\rangle.\langle \mathcal{T}_{\overleftarrow{\Phi},u_1}(f u_1), \underline{\mathcal{T}_{\overleftarrow{\Phi},u_2}(u_2)}\rangle)(\overleftarrow{g}\, w) \\
&= (\lambda\langle z_1,z_2\rangle.\langle \underline{\mathcal{T}_{\overleftarrow{\Phi},u_1}(f u_1)}, z_2\rangle)(\overleftarrow{g}\, w) \\
&= (\lambda\langle z_1,z_2\rangle.\langle (\lambda y.\mathcal{T}_{\overleftarrow{\Phi},u_1}(u_1))(\underline{\mathcal{T}_{\overleftarrow{\Phi}}(f)}z_1), z_2\rangle)(\overleftarrow{g}\, w) \\
&= (\lambda\langle z_1,z_2\rangle.\langle (\lambda y.\underline{\mathcal{T}_{\overleftarrow{\Phi},u_1}(u_1)})(\overleftarrow{f}\, z_1), z_2\rangle)(\overleftarrow{g}\, w) \\
&= (\lambda\langle z_1,z_2\rangle.\langle (\lambda y.y)(\overleftarrow{f}\, z_1), z_2\rangle)(\overleftarrow{g}\, w) \\
&\to (\lambda\langle z_1,z_2\rangle.\langle \overleftarrow{f}\, z_1, z_2\rangle)(\overleftarrow{g}\, w) \qquad (5.44)
\end{aligned}$$

Summing up Equations (5.42)-(5.44) we have the term

$$(\lambda v'.\overline{\sigma}^{\&\mathtt{t}(\theta)} v')((\lambda w^{\mathtt{t}(\tau)}.(\mu_{\langle u_1,u_2\rangle,\alpha',\emptyset}(\lambda\langle z_1,z_2\rangle.\langle \overleftarrow{f}\, z_1, z_2\rangle)(\overleftarrow{g}\, w)))v)$$

Observe that in this case we have that $\alpha'$ is the identity renaming and $\mu$ is the identity term, so we have

$$\begin{aligned}
&(\lambda v'.\overline{\sigma}^{\&\mathtt{t}(\theta)} v')((\lambda w^{\mathtt{t}(\tau)}.((\lambda\langle w',\langle\rangle\rangle.w')(\lambda\langle z_1,z_2\rangle.\langle \overleftarrow{f}\, z_1, z_2\rangle)(\overleftarrow{g}\, w)))v) \\
&\qquad \to (\lambda v'.\overline{\sigma}^{\&\mathtt{t}(\theta)} v')((\lambda w^{\mathtt{t}(\tau)}.((\lambda\langle z_1,z_2\rangle.\langle \overleftarrow{f}\, z_1, z_2\rangle)(\overleftarrow{g}\, w)))v) \qquad (5.45)
\end{aligned}$$

Summing up Equations (5.41)-(5.45) we have the term

$$\lambda v^{\mathtt{t}(\tau)}.\mu_{u,\alpha,\emptyset}\langle (\lambda v'.\overline{\sigma}^{\&\mathtt{t}(\theta)} v')((\lambda w^{\mathtt{t}(\tau)}.((\lambda\langle z_1,z_2\rangle.\langle \overleftarrow{f}\, z_1, z_2\rangle)(\overleftarrow{g}\, w)))v), \langle\rangle\rangle$$

Observe that in this case we have that $\alpha'$ is the identity renaming and $\mu$ is the term $\lambda\langle w', \langle\rangle\rangle.w'$, so we have

$$\begin{aligned}
&\lambda v^{\mathtt{t}(\tau)}.(\lambda\langle w',\langle\rangle\rangle.w')\langle (\lambda v'.\overline{\sigma}^{\&\mathtt{t}(\theta)} v')((\lambda w^{\mathtt{t}(\tau)}.((\lambda\langle z_1,z_2\rangle.\langle \overleftarrow{f}\, z_1, z_2\rangle)(\overleftarrow{g}\, w)))v), \langle\rangle\rangle \\
&\qquad \to \lambda v^{\mathtt{t}(\tau)}.(\lambda v'.\overline{\sigma}^{\&\mathtt{t}(\theta)} v')((\lambda w^{\mathtt{t}(\tau)}.((\lambda\langle z_1,z_2\rangle.\langle \overleftarrow{f}\, z_1, z_2\rangle)(\overleftarrow{g}\, w)))v) \\
&\qquad \to \lambda v^{\mathtt{t}(\tau)}.(\lambda v'.\overline{\sigma}^{\&\mathtt{t}(\theta)} v')((\lambda\langle z_1,z_2\rangle.\langle \overleftarrow{f}\, z_1, z_2\rangle)(\overleftarrow{g}\, v))
\end{aligned}$$

$$\approx \lambda v^{\mathtt{t}(\tau)}.(\lambda v'.\overline{\sigma}^{\&\mathtt{t}(\theta)}v')(\mathtt{let}\ \langle z_1, z_2\rangle = \overleftarrow{g}\, v\ \mathtt{in}\ \langle \overleftarrow{f}\, z_1, z_2\rangle)$$

Finally, we have that

$$\begin{aligned} & \mathtt{let}\ \S\overleftarrow{f} = \S\mathcal{T}(\delta^{\mathtt{B}}_{\theta\cap FV^t(\dot{e_1})}(\dot{e_1}))\ \mathtt{in} \\ \mathcal{T}(\delta^{\mathtt{B}}_{\theta}(\dot{e})) \sim\ & \mathtt{let}\ \S\overleftarrow{g} = \S\mathcal{T}(\delta^{\mathtt{B}}_{\dot{x},\theta\cap FV^t(\dot{e_2})}(\dot{e_2}))\ \mathtt{in} \\ & \lambda v^{\mathtt{t}(\tau)}.(\lambda v'.\overline{\sigma}^{\&\mathtt{t}(\theta)}v')(\mathtt{let}\ \langle z_1, z_2\rangle = \overleftarrow{g}\, v\ \mathtt{in}\ \langle \overleftarrow{f}\, z_1, z_2\rangle)\end{aligned}$$

Now we have to analyze $\delta^{\mathtt{B}}_{\dot{u}:\tau}(\mathcal{T}^{\mathtt{Jax}}_{\theta;\dot{u}:\tau}(\dot{e}))$. Recall that by derived typing rules of JAX (Figure 2.2) we have $\Gamma = \Gamma_1 \cup \Gamma_2$ and $\dot{\Gamma} = \dot{\Gamma}_1, \dot{\Gamma}_2$ such that $\Gamma_1; \dot{\Gamma}_1 \vdash^{\mathrm{Jax}} \dot{e}_1 : (\mathtt{1}; \tau_1)$ and $\Gamma_2; \dot{\Gamma}_2, \dot{x} : \tau_1 \vdash^{\mathrm{Jax}} \dot{e}_2 : (\mathtt{1}; \tau)$. Notice that $FV^t(\dot{e}_2) = \dot{\Gamma}_2 \cup \{\dot{x}\}$, so the transpose of $\dot{e}_2$ will return a tuple containing these two parts.

We have that

$$\begin{aligned}\delta^{\mathtt{B}}_{\dot{u}:\tau}(\mathcal{T}^{\mathtt{Jax}}_{\theta;\dot{u}:\tau}(\dot{e})) &= \delta^{\mathtt{B}}_{\dot{u}:\tau}(\mathcal{T}^{\mathtt{Jax}}_{\theta;\dot{u}:\tau}(\mathtt{let}\ \dot{x} = \dot{e}_1\ \mathtt{in}\ \dot{e}_2)) \\ &= \delta^{\mathtt{B}}_{\dot{u}:\tau}\begin{pmatrix}\mathtt{let}\ \dot{\otimes}(\dot{x}, \dot{u_2}) = \mathcal{T}^{\mathtt{Jax}}_{\dot{x}:\tau_1,\theta\cap FV^t(\dot{e_2});\dot{u}:\tau}(\dot{e_2})\ \mathtt{in} \\ \mathtt{let}\ \dot{u}_1 = \mathcal{T}^{\mathtt{Jax}}_{\theta\cap FV^t(\dot{e_1});\dot{x}:\tau_1}(\dot{e_1})\ \mathtt{in}\ \overline{\sigma}^{\mathrm{Jax}}_{\dot{u}_1,\dot{u}_2;\theta}\end{pmatrix}\end{aligned}$$

By typing of transpose transformation in JAX we have

$$\begin{aligned}\Gamma_2; \dot{u} : \tau &\vdash^{\mathrm{Jax}} \mathcal{T}^{\mathtt{Jax}}_{\dot{x}:\tau_1,\theta\cap FV^t(\dot{e_2});\dot{u}:\tau}(\dot{e_2}) : (\mathtt{1}; \tau_1 \otimes (\otimes(\theta \cap FV^t(\dot{e_2})))) \\ \Gamma_1; \dot{x} : \tau_1 &\vdash^{\mathrm{Jax}} \mathcal{T}^{\mathtt{Jax}}_{\theta\cap FV^t(\dot{e_1});\dot{x}:\tau_1}(\dot{e_1}) : (\mathtt{1}; \otimes(\theta \cap FV^t(\dot{e_1})))\end{aligned}$$

Notice that the transpose of $\dot{e_2}$ has only $\dot{u}$ ad its free tangent variable, so $FV^t(\mathcal{T}^{\mathtt{Jax}}(\dot{e_2})) = \{\dot{u}\}$ and it returns as output a tangent tuple of type $\tau_1$ (the type of the output of $\dot{e_1}$) and the $n$-ary tensor related to $\dot{\Gamma}_2$. We have also that $FV^t(\mathcal{T}^{\mathtt{Jax}}(\dot{e_1})) = \{\dot{x}\}$.

Moreover, $\mathtt{let}\ \dot{u}_1 = \mathcal{T}^{\mathtt{Jax}}_{\theta\cap FV^t(\dot{e_1});\dot{x}:\tau_1}(\dot{e_1})\ \mathtt{in}\ \overline{\sigma}^{\mathrm{Jax}}_{\dot{u}_1,\dot{u}_2;\theta}$ and $\overline{\sigma}^{\mathrm{Jax}}_{\dot{u}_1,\dot{u}_2;\theta}$ are well-typed as

$$\begin{aligned}\Gamma_1; \dot{x} : \tau_1, \dot{u}_2 : \otimes(\theta \cap FV^t(\dot{e_2})) &\vdash^{\mathrm{Jax}} \mathtt{let}\ \dot{u}_1 = \mathcal{T}^{\mathtt{Jax}}_{\theta\cap FV^t(\dot{e_1});\dot{x}:\tau_1}(\dot{e_1})\ \mathtt{in}\ \overline{\sigma}^{\mathrm{Jax}}_{\dot{u}_1,\dot{u}_2;\theta} : (\mathtt{1}; \otimes\theta) \\ ; \dot{u}_1 : \otimes(\theta \cap FV^t(\dot{e_1})), \dot{u}_2 : \otimes(\theta \cap FV^t(\dot{e_2})) &\vdash^{\mathrm{Jax}} \overline{\sigma}^{\mathrm{Jax}}_{\dot{u}_1,\dot{u}_2;\theta} : (\mathtt{1}; \otimes\theta)\end{aligned}$$

Observe that now we have to translate the let-construction in the form $\mathtt{let}\ \dot{\otimes}(\dot{x}_1, \dot{x}_2) = \dot{e_1}\ \mathtt{in}\ \dot{e_2}$ which is syntactic sugar and $\delta^{\mathtt{B}}$ is defined by induction on Linear B and not on the syntactic sugar, so we apply Lemma 30 obtaining the following term

$$\begin{aligned}&\mathtt{let}\ \S\overleftarrow{g} = \S\delta^{\mathtt{B}}_{\dot{u}}(\mathcal{T}^{\mathtt{Jax}}_{\dot{x}:\tau_1,\theta\cap FV^t(\dot{e_2});\dot{u}:\tau}(\dot{e_2}))\ \mathtt{in} \\ &\mathtt{let}\ \S f_2 = \S\delta^{\mathtt{B}}_{\dot{u}_1,\dot{x}}(\mathtt{let}\ \dot{u}_1 = \mathcal{T}^{\mathtt{Jax}}_{\theta\cap FV^t(\dot{e_1});\dot{x}:\tau_1}(\dot{e_1})\ \mathtt{in}\ \overline{\sigma}^{\mathrm{Jax}}_{\dot{u}_1,\dot{u}_2;\theta})\ \mathtt{in} \\ &\lambda u^{\mathtt{t}(\tau)}.\mathtt{let}\ \langle u_1, u_2\rangle = \sigma_{FV^t(\mathcal{T}^{\mathtt{Jax}}(\dot{e_2}))}u\ \mathtt{in}\ f_2\ \overline{\sigma}^{\&\mathtt{t}(\tau_1\otimes(\otimes(\theta\cap FV^t(\dot{e_2}))))}_{\dot{x}}\langle \overleftarrow{g}\, u_1, u_2\rangle\end{aligned}$$

where $FV^t(\mathcal{T}^{\mathtt{Jax}}(\dot{e_2})) = \{\dot{u}\}$ and $\dot{u}$ is of type $\tau$, so $\sigma_{FV^t(\mathcal{T}^{\mathtt{Jax}}(\dot{e_2}))}u = (\lambda\langle w, \langle\rangle\rangle.w)$. Moreover, by typing we have also that $\overline{\sigma}^{\&\mathtt{t}(\tau_1\otimes(\otimes(\theta\cap FV^t(\dot{e_2}))))}_{\dot{x}}$ is the identity term.

$$\begin{aligned}&\mathtt{let}\ \S\overleftarrow{g} = \S\delta^{\mathtt{B}}_{\dot{u}}(\mathcal{T}^{\mathtt{Jax}}_{\dot{x}:\tau_1,\theta\cap FV^t(\dot{e_2});\dot{u}:\tau}(\dot{e_2}))\ \mathtt{in} \\ \to^*\ &\mathtt{let}\ \S f_2 = \S\delta^{\mathtt{B}}_{\dot{u}_1,\dot{x}}(\mathtt{let}\ \dot{u}_1 = \mathcal{T}^{\mathtt{Jax}}_{\theta\cap FV^t(\dot{e_1});\dot{x}:\tau_1}(\dot{e_1})\ \mathtt{in}\ \overline{\sigma}^{\mathrm{Jax}}_{\dot{u}_1,\dot{u}_2;\theta})\ \mathtt{in} \\ &\lambda u^{\mathtt{t}(\tau)}.f_2 \overleftarrow{g}\, u\end{aligned}$$

$$
\begin{aligned}
& \texttt{let } \S\overleftarrow{g} = \S\delta^{\mathsf{B}}_{\dot u}(\mathcal{T}^{\mathsf{Jax}}_{\dot x:\tau_1,\theta\cap FV^t(\dot{e_2});\dot u:\tau}(\dot{e_2}))\texttt{ in} \\
= \; & \texttt{let } \S f_2 = \S\left(\begin{array}{l}\texttt{let } \S\overleftarrow{f} = \S\delta^{\mathsf{B}}_{\dot x}(\mathcal{T}^{\mathsf{Jax}}_{\theta\cap FV^t(\dot{e_1});\dot x:\tau_1}(\dot{e_1}))\texttt{ in} \\ \texttt{let } \S g' = \S\delta^{\mathsf{B}}_{\dot u_1,\dot u_2}(\overline{\sigma}^{\mathsf{Jax}}_{\dot u_1,\dot u_2;\theta})\texttt{ in} \\ \lambda v.\left(\begin{array}{l}\texttt{let } \langle v_1,v_2\rangle = \sigma_{FV^t(\mathcal{T}^{\mathsf{Jax}}(\dot{e_1}))} v\texttt{ in} \\ g'\overline{\sigma}^{(\&\mathtt{t}(\theta\cap FV^t(\dot{e_1})))\&(\&\mathtt{t}(\theta\cap FV^t(\dot{e_2})))}_{\dot u_1}\langle\overleftarrow{f}\, v_1, v_2\rangle\end{array}\right)\end{array}\right)\texttt{ in} \\
& \lambda u^{\mathtt{t}(\tau)}.f_2\,\overleftarrow{g}\,u
\end{aligned}
$$

$$
\begin{aligned}
& \texttt{let } \S\overleftarrow{g} = \S\delta^{\mathsf{B}}_{\dot u}(\mathcal{T}^{\mathsf{Jax}}_{\dot x:\tau_1,\theta\cap FV^t(\dot{e_2});\dot u:\tau}(\dot{e_2}))\texttt{ in} \\
\overset{\text{Lemma 29}}{=} \; & \texttt{let } \S f_2 = \S\left(\begin{array}{l}\texttt{let } \S\overleftarrow{f} = \S\delta^{\mathsf{B}}_{\dot x}(\mathcal{T}^{\mathsf{Jax}}_{\theta\cap FV^t(\dot{e_1});\dot x:\tau_1}(\dot{e_1}))\texttt{ in} \\ \texttt{let } \S g' = \S(\lambda w.\overline{\sigma}^{\&\mathtt{t}(\theta)} w)\texttt{ in} \\ \lambda v.\left(\begin{array}{l}\texttt{let } \langle v_1,v_2\rangle = \sigma_{FV^t(\mathcal{T}^{\mathsf{Jax}}(\dot{e_1}))} v\texttt{ in} \\ g'\overline{\sigma}^{(\&\mathtt{t}(\theta\cap FV^t(\dot{e_1})))\&(\&\mathtt{t}(\theta\cap FV^t(\dot{e_2})))}_{\dot u_1}\langle\overleftarrow{f}\, v_1, v_2\rangle\end{array}\right)\end{array}\right)\texttt{ in} \\
& \lambda u^{\mathtt{t}(\tau)}.f_2\,\overleftarrow{g}\,u
\end{aligned}
$$

where $v$ is of type $\&\mathtt{t}(\tau_1 \otimes (\otimes(\theta \cap FV^t(\dot{e_2})))) = \mathtt{t}(\tau_1)\&(\&\mathtt{t}(\theta \cap FV^t(\dot{e_2})))$ and $FV^t(\mathcal{T}^{\mathsf{Jax}}(\dot{e_1})) = \{\dot x\}$ and $\dot x$ is of type $\tau_1$. By typing we have that $\sigma_{FV^t(\mathcal{T}^{\mathsf{Jax}}(\dot{e_1}))} v$ and $\overline{\sigma}^{(\&\mathtt{t}(\theta\cap FV^t(\dot{e_1})))\&(\&\mathtt{t}(\theta\cap FV^t(\dot{e_2})))}_{\dot u_1}$ are the identity.

$$
\begin{aligned}
& \texttt{let } \S\overleftarrow{g} = \S\delta^{\mathsf{B}}_{\dot u}(\mathcal{T}^{\mathsf{Jax}}_{\dot x:\tau_1,\theta\cap FV^t(\dot{e_2});\dot u:\tau}(\dot{e_2}))\texttt{ in} \\
\to^* \; & \texttt{let } \S f_2 = \S\left(\begin{array}{l}\texttt{let } \S\overleftarrow{f} = \S\delta^{\mathsf{B}}_{\dot x}(\mathcal{T}^{\mathsf{Jax}}_{\theta\cap FV^t(\dot{e_1});\dot x:\tau_1}(\dot{e_1}))\texttt{ in} \\ \lambda v.\texttt{let } \langle v_1,v_2\rangle = v\texttt{ in } (\lambda w.\overline{\sigma}^{\&\mathtt{t}(\theta)} w)\langle\overleftarrow{f}\, v_1, v_2\rangle\end{array}\right)\texttt{ in} \\
& \lambda u^{\mathtt{t}(\tau)}.f_2\,\overleftarrow{g}\,u
\end{aligned}
$$

$$
\begin{aligned}
& \texttt{let } \S\overleftarrow{f} = \S\delta^{\mathsf{B}}_{\dot x}(\mathcal{T}^{\mathsf{Jax}}_{\theta\cap FV^t(\dot{e_1});\dot x:\tau_1}(\dot{e_1}))\texttt{ in} \\
\overset{\text{Lemma 17}}{\sim} \; & \texttt{let } \S\overleftarrow{g} = \S\delta^{\mathsf{B}}_{\dot u}(\mathcal{T}^{\mathsf{Jax}}_{\dot x:\tau_1,\theta\cap FV^t(\dot{e_2});\dot u:\tau}(\dot{e_2}))\texttt{ in} \\
& \texttt{let } \S f_2 = \S\left(\lambda v.\texttt{let } \langle v_1,v_2\rangle = v\texttt{ in } (\lambda w.\overline{\sigma}^{\&\mathtt{t}(\theta)} w)\langle\overleftarrow{f}\, v_1, v_2\rangle\right)\texttt{ in} \\
& \lambda u^{\mathtt{t}(\tau)}.f_2\,\overleftarrow{g}\,u
\end{aligned}
$$

$$
\begin{aligned}
& \texttt{let } \S\overleftarrow{f} = \S\delta^{\mathsf{B}}_{\dot x}(\mathcal{T}^{\mathsf{Jax}}_{\theta\cap FV^t(\dot{e_1});\dot x:\tau_1}(\dot{e_1}))\texttt{ in} \\
\to \; & \texttt{let } \S\overleftarrow{g} = \S\delta^{\mathsf{B}}_{\dot u}(\mathcal{T}^{\mathsf{Jax}}_{\dot x:\tau_1,\theta\cap FV^t(\dot{e_2});\dot u:\tau}(\dot{e_2}))\texttt{ in} \\
& \lambda u^{\mathtt{t}(\tau)}.(\lambda v.\texttt{let } \langle v_1,v_2\rangle = v\texttt{ in } (\lambda w.\overline{\sigma}^{\&\mathtt{t}(\theta)} w)\langle\overleftarrow{f}\, v_1, v_2\rangle)(\overleftarrow{g}\,u)
\end{aligned}
$$

$$
\begin{aligned}
& \texttt{let } \S\overleftarrow{f} = \S\delta^{\mathsf{B}}_{\dot x}(\mathcal{T}^{\mathsf{Jax}}_{\theta\cap FV^t(\dot{e_1});\dot x:\tau_1}(\dot{e_1}))\texttt{ in} \\
\to \; & \texttt{let } \S\overleftarrow{g} = \S\delta^{\mathsf{B}}_{\dot u}(\mathcal{T}^{\mathsf{Jax}}_{\dot x:\tau_1,\theta\cap FV^t(\dot{e_2});\dot u:\tau}(\dot{e_2}))\texttt{ in} \\
& \lambda u^{\mathtt{t}(\tau)}.\texttt{let } \langle v_1,v_2\rangle = (\overleftarrow{g}\,u)\texttt{ in } (\lambda w.\overline{\sigma}^{\&\mathtt{t}(\theta)} w)\langle\overleftarrow{f}\, v_1, v_2\rangle
\end{aligned}
$$

$$
\begin{aligned}
& \texttt{let } \S\overleftarrow{f} = \S\delta^{\mathsf{B}}_{\dot x}(\mathcal{T}^{\mathsf{Jax}}_{\theta\cap FV^t(\dot{e_1});\dot x:\tau_1}(\dot{e_1}))\texttt{ in} \\
\overset{\text{Lemma 17}}{\sim} \; & \texttt{let } \S\overleftarrow{g} = \S\delta^{\mathsf{B}}_{\dot u}(\mathcal{T}^{\mathsf{Jax}}_{\dot x:\tau_1,\theta\cap FV^t(\dot{e_2});\dot u:\tau}(\dot{e_2}))\texttt{ in} \\
& \lambda u^{\mathtt{t}(\tau)}.(\lambda w.\overline{\sigma}^{\&\mathtt{t}(\theta)} w)(\texttt{let } \langle v_1,v_2\rangle = (\overleftarrow{g}\,u)\texttt{ in } \langle\overleftarrow{f}\, v_1, v_2\rangle)
\end{aligned}
$$

Summing up we have

$$\mathcal{T}(\delta^{\mathtt{B}}_{\theta}(\dot{e})) \sim \begin{array}{l}\mathtt{let}\ \S\overleftarrow{f} = \S\mathcal{T}(\delta^{\mathtt{B}}_{\theta\cap FV^t(\dot{e}_1)}(\dot{e}_1))\ \mathtt{in}\\ \mathtt{let}\ \S\overleftarrow{g} = \S\mathcal{T}(\delta^{\mathtt{B}}_{\dot{x},\theta\cap FV^t(\dot{e}_2)}(\dot{e}_2))\ \mathtt{in}\\ \lambda v^{\mathtt{t}(\tau)}.(\lambda v'.\overline{\sigma}^{\&\mathtt{t}(\theta)}v')(\mathtt{let}\ \langle z_1, z_2\rangle = \overleftarrow{g}\,v\ \mathtt{in}\ \langle \overleftarrow{f}\,z_1, z_2\rangle)\end{array}$$

$$\delta^{\mathtt{B}}_{\dot{u}:\tau}(\mathcal{T}^{\mathtt{Jax}}_{\theta;\dot{u}:\tau}(\dot{e})) \sim \begin{array}{l}\mathtt{let}\ \S\overleftarrow{f} = \S\delta^{\mathtt{B}}_{\dot{x}}(\mathcal{T}^{\mathtt{Jax}}_{\theta\cap FV^t(\dot{e}_1);\dot{x}:\tau_1}(\dot{e}_1))\ \mathtt{in}\\ \mathtt{let}\ \S\overleftarrow{g} = \S\delta^{\mathtt{B}}_{\dot{u}}(\mathcal{T}^{\mathtt{Jax}}_{\dot{x}:\tau_1,\theta\cap FV^t(\dot{e}_2);\dot{u}:\tau}(\dot{e}_2))\ \mathtt{in}\\ \lambda u^{\mathtt{t}(\tau)}.(\lambda w.\overline{\sigma}^{\theta}w)(\mathtt{let}\ \langle v_1, v_2\rangle = (\overleftarrow{g}\,u)\ \mathtt{in}\ \langle \overleftarrow{f}\,v_1, v_2\rangle)\end{array}$$

and we can conclude by induction hypothesis and $\alpha$-renaming. □

## Proof Work Preservation Transpose

**Lemma 41.** We have the following:

1. if $!\Sigma, \S\Phi, p^{\&} : L \vdash U : H$ and $\alpha$ is the identity renaming restricted to $FV(p^{\&}) \cap FV(U)$, then:

$$\mathcal{W}(\lambda q^{\&}.\mu_{p^{\&},\alpha,\emptyset}\langle \mathcal{T}_{\S\overleftarrow{\Phi},p^{\&}}(U), \langle\,\rangle\rangle) + \mathcal{W}(L) + \mathcal{W}(\S\overleftarrow{\Phi}^{\mathrm{in}}_{\mathcal{T}_{\S\overleftarrow{\Phi},p^{\&}}(U)}) \leq \mathcal{W}(\lambda p^{\&}.U) + \mathcal{W}(H) + \mathcal{W}(\S\overleftarrow{\Phi}^{\mathrm{out}}_{\mathcal{T}_{\S\overleftarrow{\Phi},p^{\&}}(U)})$$

2. if $!\Sigma, \S\Phi \vdash F : L \multimap H$, then:

$$\mathcal{W}(\mathcal{T}_{\S\overleftarrow{\Phi}}(F)) + \mathcal{W}(L) + \mathcal{W}(\S\overleftarrow{\Phi}^{\mathrm{in}}_{\mathcal{T}_{\S\overleftarrow{\Phi}}(F)}) \leq \mathcal{W}(F) + \mathcal{W}(H) + \mathcal{W}(\S\overleftarrow{\Phi}^{\mathrm{out}}_{\mathcal{T}_{\S\overleftarrow{\Phi}}(F)})$$

3. if $!\Sigma, \S\Phi \vdash R : !E \otimes \S(L \multimap H)$, then:

$$\mathcal{W}(\mathcal{T}_{\S\overleftarrow{\Phi}}(R)) + \mathcal{W}(L) + \mathcal{W}(\S\overleftarrow{\Phi}^{\mathrm{in}}_{\mathcal{T}_{\S\overleftarrow{\Phi}}(R)}) \leq \mathcal{W}(R) + \mathcal{W}(H) + \mathcal{W}(\S\overleftarrow{\Phi}^{\mathrm{out}}_{\mathcal{T}_{\S\overleftarrow{\Phi}}(R)})$$

*Proof Claim 1: Cases of $\mathcal{T}$ on $\lambda LL^{\mathtt{t}}$.* By the reasoning described above in this case it is enough to prove that

$$\mathcal{W}(\lambda q^{\&}.\mathcal{T}_{\S\overleftarrow{\Phi},p^{\&}}(U)) + \mathcal{W}(L) + \mathcal{W}(\S\overleftarrow{\Phi}^{\mathrm{in}}_{\mathcal{T}_{\S\overleftarrow{\Phi},p^{\&}}(U)}) \leq \mathcal{W}(\lambda p^{\&}.U) + \mathcal{W}(H) + \mathcal{W}(\S\overleftarrow{\Phi}^{\mathrm{out}}_{\mathcal{T}_{\S\overleftarrow{\Phi},p^{\&}}(U)})$$

and we proceed by analyzing the cases in Figure 5.4b.

- Case $U = \langle U_1, U_2\rangle$:
  In this case we have that $H = H_1 \& H_2$ and $q^{\&} = \langle q_1^{\&}, q_2^{\&}\rangle$. By hypothesis we have $!\Sigma, \S\Phi, p^{\&} : L \vdash \langle U_1, U_2\rangle : H_1 \& H_2$, so by &- typing rule we have $!\Sigma, \S\Phi, p^{\&} : L \vdash U_i : H_i$.

  Moreover, by item 1 of Lemma 36 we have $!\Sigma, \S\Phi, \alpha[p^{\&}] : L \vdash \alpha[U_i] : H_i$.

  By inductive hypothesis on $U_i$ we have

$$\mathcal{W}(\lambda q_i^{\&}.\mathcal{T}_{\S\overleftarrow{\Phi},p^{\&}}(U_i)) + \mathcal{W}(L) + \mathcal{W}(\S\overleftarrow{\Phi}^{\mathrm{in}}_{\mathcal{T}_{\S\Phi,p^{\&}}(U_i)}) \leq \mathcal{W}(\lambda\alpha[p^{\&}].\alpha[U_i]) + \mathcal{W}(H_i) + \mathcal{W}(\S\overleftarrow{\Phi}^{\mathrm{out}}_{\mathcal{T}_{\S\Phi,p^{\&}}(U_i)})$$

  We have:

$$\mathcal{W}(\lambda q^{\&}.\mathcal{T}_{\S\overleftarrow{\Phi},p^{\&}}(U)) = \mathcal{W}(\lambda\langle q_1^{\&}, q_2^{\&}\rangle.\mathcal{T}_{\S\overleftarrow{\Phi},p^{\&}}(\langle U_1, U_2\rangle))$$

$$= \mathcal{W}(\lambda\langle q_1^{\&}, q_2^{\&}\rangle.\mu_{p,\alpha_1,\alpha_2}\langle \mathcal{T}_{\overleftarrow{\S\Phi},\alpha_1[p^{\&}]}(\alpha_1[U_1]), \mathcal{T}_{\overleftarrow{\S\Phi},\alpha_2[p^{\&}]}(\alpha_2[U_2])\rangle) \tag{5.46}$$

$$= \mathcal{W}(\mu_{p,\alpha_1,\alpha_2}) + \sum_{i=1}^{2} \mathcal{W}(\lambda q_i^{\&}.\mathcal{T}_{\overleftarrow{\S\Phi},\alpha_i[p^{\&}]}(\alpha_i[U_i])) \tag{5.47}$$

$$= \mathcal{W}(\mu_{p,\alpha_1,\alpha_2}) + \sum_{i=1}^{2} \mathcal{W}(\lambda q_i^{\&}.\mathcal{T}_{\overleftarrow{\S\Phi},p^{\&}}(U_i)) \tag{5.48}$$

$$= \mathcal{W}(FV(U_1) \cap FV(U_2) \cap FV(p^{\&})) + \sum_{i=1}^{2} \mathcal{W}(\lambda q_i^{\&}.\mathcal{T}_{\overleftarrow{\S\Phi},p^{\&}}(U_i)) \tag{5.49}$$

where:

- The passage from line (5.46) to line (5.47) is because by typing we know that $FV(q_{3-i}^{\&}) \cap FV(\lambda q_i^{\&}.\mathcal{T}_{\overleftarrow{\S\Phi},\alpha_i[p^{\&}]}(\alpha_i[U_i])) = \emptyset$.
- The passage from line (5.47) to line (5.48) is obtained by applying item 3 of Lemma 36 which states that $\mathcal{W}(\alpha[M]) = \mathcal{W}(M)$.
- Recall that $\mathrm{Dom}(\alpha_i) = FV(U_i) \cap FV(p^{\&})$ and that by Equation 5.29 we have $\mu_{p,\alpha_1,\alpha_2} = \lambda\langle\alpha_1\langle p\rangle, \alpha_2\langle p\rangle\rangle.\nu(p, \alpha_1, \alpha_2)$. The passage from line (5.48) to line (5.49) is by Lemma 37 obtaining that $\mathcal{W}(\mu_{p,\alpha_1,\alpha_2}) = \mathcal{W}(\mathrm{Dom}(\alpha_1) \cap \mathrm{Dom}(\alpha_2) \cap FV(p)) = \mathcal{W}(FV(U_1) \cap FV(U_2) \cap FV(p^{\&}))$. More precisely, $\mathcal{W}(FV(U_1) \cap FV(U_2) \cap FV(p^{\&}))$ is the number of sums performed by $\nu(p, \alpha_1, \alpha_2)$.

We have also that

$$\begin{aligned} \mathcal{W}(\lambda p^{\&}.U) &= \mathcal{W}(\lambda p^{\&}.\langle U_1, U_2\rangle) \\ &= \mathcal{W}(FV(p^{\&}) \setminus (FV(U_1) \cup FV(U_2))) + \sum_{i=1}^{2} \mathcal{W}(U_i) \end{aligned}$$

where the last line is obtained by applying the definition of workload for the lambda abstraction.

Finally, we show that

$$\mathcal{W}(\lambda q^{\&}.\mathcal{T}_{\overleftarrow{\S\Phi},p^{\&}}(U)) + \mathcal{W}(L) + \mathcal{W}(\S\overleftarrow{\Phi}^{\text{in}}_{\mathcal{T}_{\S\Phi,p^{\&}}(U)}) \leq \mathcal{W}(\lambda p^{\&}.U) + \mathcal{W}(H) + \mathcal{W}(\S\overleftarrow{\Phi}^{\text{out}}_{\mathcal{T}_{\S\Phi,p^{\&}}(U)})$$

by using the following remark.

**Remark 9.** We need to analyze the quantity $\mathcal{W}(FV(U_1) \cap FV(U_2) \cap FV(p^{\&})) + \mathcal{W}(FV(p^{\&}) \setminus FV(U_1)) + \mathcal{W}(FV(p^{\&}) \setminus FV(U_2))$.

Let us consider the colours in Figure 5.9.

The quantity we are analyzing is the following

$$\mathcal{W}(\,FV(U_1) \cap FV(U_2) \cap FV(p^{\&})\,) + \mathcal{W}(\,FV(p^{\&}) \setminus FV(U_1)\,) + \mathcal{W}(\,FV(p^{\&}) \setminus FV(U_2)\,).$$

By observing the figure we have that:

$$\begin{aligned} FV(p^{\&}) \setminus FV(U_1) &= FV(p^{\&}) \setminus (FV(U_1) \cup FV(U_2)) \\ &\cup (FV(p^{\&}) \cap FV(U_2)) \setminus (FV(U_1) \cap FV(U_2)) \end{aligned} \tag{5.50}$$

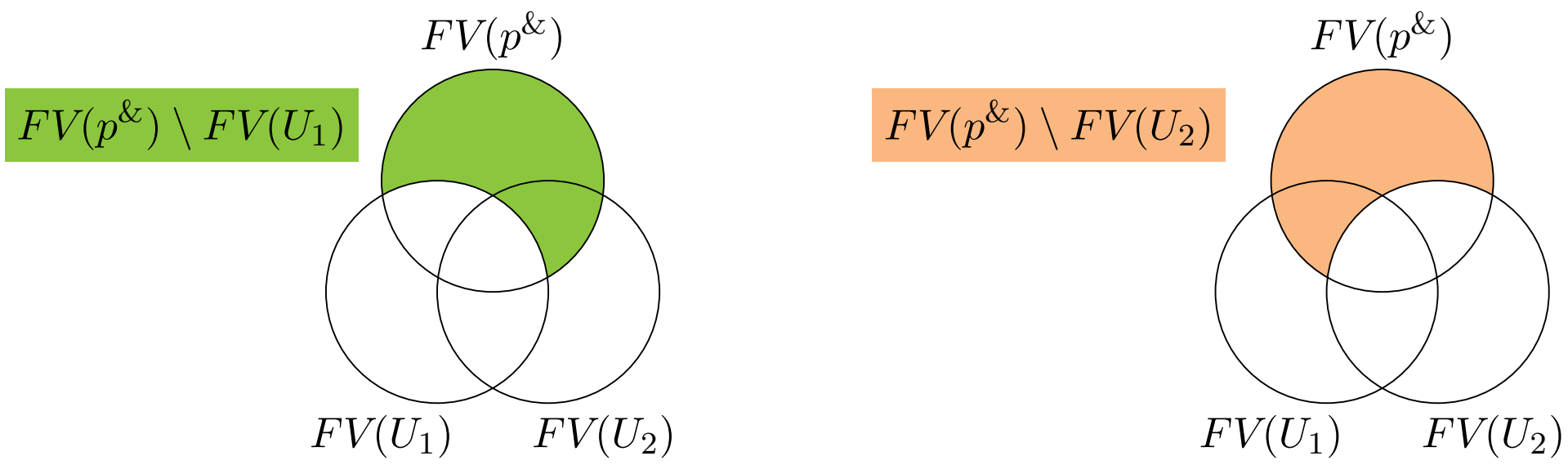

(a) Venn Diagram for $FV(p^{\&}) \setminus FV(U_1)$ and $FV(p^{\&}) \setminus FV(U_2)$

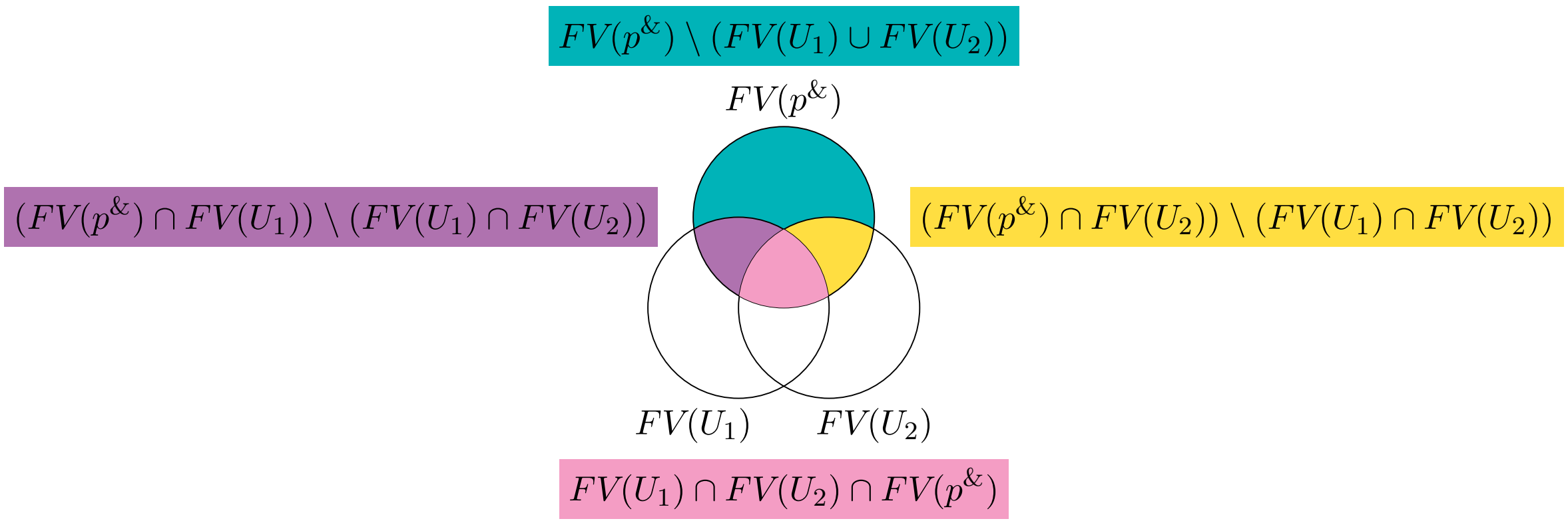

(b) Venn Diagram for the components of $FV(p^{\&})$

Figure 5.9: Venn Diagram for Work Preservation Transpose

$$\begin{aligned} FV(p^{\&}) \setminus FV(U_2) &= FV(p^{\&}) \setminus (FV(U_1) \cup FV(U_2)) \\ &\quad \cup (FV(p^{\&}) \cap FV(U_1)) \setminus (FV(U_1) \cap FV(U_2)) \end{aligned} \tag{5.51}$$

$$\begin{aligned} FV(p^{\&}) &= FV(p^{\&}) \setminus (FV(U_1) \cup FV(U_2)) \cup FV(U_1) \cap FV(U_2) \cap FV(p^{\&}) \cup \\ &\quad \cup (FV(p^{\&}) \cap FV(U_1)) \setminus (FV(U_1) \cap FV(U_2)) \\ &\quad \cup (FV(p^{\&}) \cap FV(U_2)) \setminus (FV(U_1) \cap FV(U_2)) \end{aligned} \tag{5.52}$$

We have:

$$\begin{aligned} &\mathcal{W}( FV(U_1) \cap FV(U_2) \cap FV(p^{\&}) ) + \mathcal{W}( FV(p^{\&}) \setminus FV(U_1) ) + \mathcal{W}( FV(p^{\&}) \setminus FV(U_2) ) \\ &\overset{\text{Eq. 5.50, 5.51}}{=} \mathcal{W}( FV(p^{\&}) \setminus (FV(U_1) \cup FV(U_2)) )+ \\ &\quad + \mathcal{W}( FV(p^{\&}) \setminus (FV(U_1) \cup FV(U_2)) ) + \mathcal{W}( FV(U_1) \cap FV(U_2) \cap FV(p^{\&}) )+ \\ &\quad + \mathcal{W}( (FV(p^{\&}) \cap FV(U_1)) \setminus (FV(U_1) \cap FV(U_2)) )+ \\ &\quad + \mathcal{W}( (FV(p^{\&}) \cap FV(U_2)) \setminus (FV(U_1) \cap FV(U_2)) ) \\ &\overset{\text{Eq. 5.52}}{=} \mathcal{W}( FV(p^{\&}) \setminus (FV(U_1) \cup FV(U_2)) ) + \mathcal{W}(FV(p^{\&})) \end{aligned}$$

Moreover, if $FV(p^{\&})$ is of type $L$ then we can conclude that the analyzed quantity is equal to

$$\mathcal{W}(FV(p^{\&}) \setminus (FV(U_1) \cup FV(U_2))) + \mathcal{W}(L)$$

More precisely, we proceed as follows

$$\begin{aligned}
&\mathcal{W}(\lambda q^{\&}.\mathcal{T}_{\S\overleftarrow{\Phi},p^{\&}}(U)) + \mathcal{W}(L) + \mathcal{W}(\S\overleftarrow{\Phi}^{\text{in}}_{\mathcal{T}_{\S\Phi,p^{\&}}(U)}) \\
&\quad = \mathcal{W}(FV(U_1) \cap FV(U_2) \cap FV(p^{\&})) + \sum_{i=1}^{2} \Big(\mathcal{W}(\lambda q_i^{\&}.\mathcal{T}_{\S\overleftarrow{\Phi},p^{\&}}(U_i))\Big) \\
&\qquad\qquad + \mathcal{W}(L) + \mathcal{W}(\S\overleftarrow{\Phi}^{\text{in}}_{\mathcal{T}_{\S\Phi,p^{\&}}(U)}) \\
&\quad = \mathcal{W}(FV(U_1) \cap FV(U_2) \cap FV(p^{\&})) + \mathcal{W}(L) \\
&\qquad\qquad + \sum_{i=1}^{2} \mathcal{W}(\lambda q_i^{\&}.\mathcal{T}_{\S\overleftarrow{\Phi},p^{\&}}(U_i)) + \mathcal{W}(\S\overleftarrow{\Phi}^{\text{in}}_{\mathcal{T}_{\S\Phi,p^{\&}}(U_i)}) \\
&\quad \overset{\text{IHs}}{\leq} \mathcal{W}(FV(U_1) \cap FV(U_2) \cap FV(p^{\&})) - \mathcal{W}(L) \\
&\qquad\qquad + \sum_{i=1}^{2} \mathcal{W}(\lambda\alpha[p^{\&}].\alpha[U_i]) + \mathcal{W}(H_i) + \mathcal{W}(\S\overleftarrow{\Phi}^{\text{out}}_{\mathcal{T}_{\S\Phi,p^{\&}}(U_i)}) \\
&\quad \overset{\text{Lemma 36}}{=} \mathcal{W}(FV(U_1) \cap FV(U_2) \cap FV(p^{\&})) - \mathcal{W}(L) \\
&\qquad\qquad + \sum_{i=1}^{2} \mathcal{W}(\lambda p^{\&}.U_i) + \mathcal{W}(H_i) + \mathcal{W}(\S\overleftarrow{\Phi}^{\text{out}}_{\mathcal{T}_{\S\Phi,p^{\&}}(U_i)}) \\
&\quad = \mathcal{W}(FV(U_1) \cap FV(U_2) \cap FV(p^{\&})) \\
&\qquad\qquad - \mathcal{W}(L) + \sum_{i=1}^{2} \mathcal{W}(U_i) + \mathcal{W}(FV(p^{\&}) \setminus FV(U_i)) + \mathcal{W}(H_i) + \mathcal{W}(\S\overleftarrow{\Phi}^{\text{out}}_{\mathcal{T}_{\S\Phi,p^{\&}}(U_i)}) \\
&\quad = \mathcal{W}(FV(U_1) \cap FV(U_2) \cap FV(p^{\&})) + \mathcal{W}(FV(p^{\&}) \setminus FV(U_1)) + \mathcal{W}(FV(p^{\&}) \setminus FV(U_2)) \\
&\qquad\qquad - \mathcal{W}(L) + \sum_{i=1}^{2} \mathcal{W}(U_i) + \mathcal{W}(H_i) + \mathcal{W}(\S\overleftarrow{\Phi}^{\text{out}}_{\mathcal{T}_{\S\Phi,p^{\&}}(U_i)}) \\
&\quad \overset{\text{Remark 9}}{=} \mathcal{W}(FV(p^{\&}) \setminus (FV(U_1) \cup FV(U_2))) + \mathcal{W}(L) \\
&\qquad\qquad - \mathcal{W}(L) + \sum_{i=1}^{2} \mathcal{W}(U_i) + \mathcal{W}(H_i) + \mathcal{W}(\S\overleftarrow{\Phi}^{\text{out}}_{\mathcal{T}_{\S\Phi,p^{\&}}(U_i)}) \\
&\quad = \mathcal{W}(FV(p^{\&}) \setminus (FV(U_1) \cup FV(U_2))) + \sum_{i=1}^{2} \mathcal{W}(U_i) + \mathcal{W}(H_i) + \mathcal{W}(\S\overleftarrow{\Phi}^{\text{out}}_{\mathcal{T}_{\S\Phi,p^{\&}}(U_i)}) \\
&\quad = \mathcal{W}(\lambda p^{\&}.U) + \mathcal{W}(H) + \mathcal{W}(\S\overleftarrow{\Phi}^{\text{out}}_{\mathcal{T}_{\S\Phi,p^{\&}}(U)})
\end{aligned}$$

and so we can conclude.

- Case $U = FU'$:
  By hypothesis we have $!\Sigma, \S\Phi, p^{\&} : L \vdash FU' : H$, so by typing we have

$$\S\Phi \vdash F : L_0 \multimap H$$

$$!\Sigma, \S\Phi, p^{\&} : L \vdash U' : L_0$$

By inductive hypothesis on $F$ (case 2 of the lemma) we have

$$\mathcal{W}(\mathcal{T}_{\S\Phi}(F)) + \mathcal{W}(L) + \mathcal{W}(\S\overleftarrow{\Phi}^{\text{in}}_{\mathcal{T}_{\S\Phi}(F)}) \leq \mathcal{W}(F) + \mathcal{W}(H) + \mathcal{W}(\S\overleftarrow{\Phi}^{\text{out}}_{\mathcal{T}_{\S\Phi}(F)})$$

By inductive hypothesis on $U'$ (case 1 of the lemma) we have

$$\mathcal{W}(\lambda q^{\&}.\mathcal{T}_{\S\overleftarrow{\Phi},p^{\&}}(U'))+\mathcal{W}(L)+\mathcal{W}(\S\overleftarrow{\Phi}^{\text{in}}_{\mathcal{T}_{\S\Phi,p^{\&}}(U')}) \leq \mathcal{W}(\lambda p^{\&}.U')+\mathcal{W}(L_0)+\mathcal{W}(\S\overleftarrow{\Phi}^{\text{out}}_{\mathcal{T}_{\S\Phi,p^{\&}}(U')})$$

We have:

$$\begin{aligned}\mathcal{W}(\lambda q^{\&}.\mathcal{T}_{\S\overleftarrow{\Phi},p^{\&}}(U)) &= \mathcal{W}(\lambda q^{\&}.\mathcal{T}_{\S\overleftarrow{\Phi},p^{\&}}(FU'))\\ &= \mathcal{W}((\lambda q^{\&\prime}.\mathcal{T}_{\S\overleftarrow{\Phi},p^{\&}}(U'))(\mathcal{T}_{\S\overleftarrow{\Phi}}(F)q^{\&}))\\ &= \mathcal{W}(\lambda q^{\&\prime}.\mathcal{T}_{\S\overleftarrow{\Phi},p^{\&}}(U'))+\mathcal{W}(\mathcal{T}_{\S\overleftarrow{\Phi}}(F))\end{aligned}$$

$$\begin{aligned}\mathcal{W}(\lambda p^{\&}.U) &= \mathcal{W}(\lambda p^{\&}.FU')\\ &= \mathcal{W}(F(\lambda p^{\&}.U'))=\mathcal{W}(F)+\mathcal{W}(\lambda p^{\&}.U')\end{aligned}$$

where the last line is obtained by observing that by typing we have that $p^{\&}$ is free only in $U'$.

We show that

$$\mathcal{W}(\lambda q^{\&}.\mathcal{T}_{\S\overleftarrow{\Phi},p^{\&}}(U))+\mathcal{W}(L)+\mathcal{W}(\S\overleftarrow{\Phi}^{\text{in}}_{\mathcal{T}_{\S\Phi,p^{\&}}(U)}) \leq \mathcal{W}(\lambda p^{\&}.U)+\mathcal{W}(H)+\mathcal{W}(\S\overleftarrow{\Phi}^{\text{out}}_{\mathcal{T}_{\S\Phi,p^{\&}}(U)})$$

as follows

$$\begin{aligned}&\mathcal{W}(\lambda q^{\&}.\mathcal{T}_{\S\overleftarrow{\Phi},p^{\&}}(FU'))+\mathcal{W}(L)+\mathcal{W}(\S\overleftarrow{\Phi}^{\text{in}}_{\mathcal{T}_{\S\Phi,p^{\&}}(U)})\\ &\quad= \mathcal{W}(\lambda q^{\&\prime}.\mathcal{T}_{\S\overleftarrow{\Phi},p^{\&}}(U'))+\mathcal{W}(\mathcal{T}_{\S\overleftarrow{\Phi}}(F))+\mathcal{W}(L)+\mathcal{W}(\S\overleftarrow{\Phi}^{\text{in}}_{\mathcal{T}_{\S\Phi,p^{\&}}(U')})+\mathcal{W}(\S\overleftarrow{\Phi}^{\text{in}}_{\mathcal{T}_{\S\Phi}(F)})\\ &\quad\overset{\text{IH on U'}}{\leq} \mathcal{W}(\lambda p^{\&}.U')+\mathcal{W}(L_0)+\mathcal{W}(\S\overleftarrow{\Phi}^{\text{out}}_{\mathcal{T}_{\S\Phi,p^{\&}}(U')})+\mathcal{W}(\mathcal{T}_{\S\overleftarrow{\Phi}}(F))+\mathcal{W}(\S\overleftarrow{\Phi}^{\text{in}}_{\mathcal{T}_{\S\Phi}(F)})\\ &\quad\overset{\text{IH on F}}{\leq} \mathcal{W}(\lambda p^{\&}.U')+\mathcal{W}(\S\overleftarrow{\Phi}^{\text{out}}_{\mathcal{T}_{\S\Phi,p^{\&}}(U')})+\mathcal{W}(F)+\mathcal{W}(H)+\mathcal{W}(\S\overleftarrow{\Phi}^{\text{out}}_{\mathcal{T}_{\S\Phi}(F)})\\ &\quad= \mathcal{W}(\lambda p^{\&}.U')+\mathcal{W}(F)+\mathcal{W}(H)+\mathcal{W}(\S\overleftarrow{\Phi}^{\text{out}}_{\mathcal{T}_{\S\Phi,p^{\&}}(U)})\\ &\quad= \mathcal{W}(\lambda p^{\&}.U)+\mathcal{W}(H)+\mathcal{W}(\S\overleftarrow{\Phi}^{\text{out}}_{\mathcal{T}_{\S\Phi,p^{\&}}(U)})\end{aligned}$$

so we can conclude.

- Case $U = u$:
  Observe that by typing

$$L = H, \quad \mathcal{W}(\S\overleftarrow{\Phi}^{\text{in}}_{\mathcal{T}_{\S\Phi,p^{\&}}(u)}) = \mathcal{W}(\S\overleftarrow{\Phi}^{\text{out}}_{\mathcal{T}_{\S\Phi,p^{\&}}(u)}) = 0 \quad \text{and} \quad q^{\&} = p^{\&} = u. \tag{5.53}$$

We have:

$$\begin{aligned}&\mathcal{W}(\lambda q^{\&}.\mathcal{T}_{\S\overleftarrow{\Phi},p^{\&}}(u))+\mathcal{W}(L)+\mathcal{W}(\S\overleftarrow{\Phi}^{\text{in}}_{\mathcal{T}_{\S\Phi,p^{\&}}(U)})\\ &\quad= \mathcal{W}(\lambda q^{\&}.u)+\mathcal{W}(L)+\mathcal{W}(\S\overleftarrow{\Phi}^{\text{in}}_{\mathcal{T}_{\S\Phi,p^{\&}}(U)})\\ &\quad\overset{\text{Eq. 5.53}}{=} \mathcal{W}(\lambda u.u)+\mathcal{W}(L)\\ &\quad= \mathcal{W}(L)\\ &\quad\leq \mathcal{W}(\lambda p^{\&}.u)+\mathcal{W}(H)+\mathcal{W}(\S\overleftarrow{\Phi}^{\text{out}}_{\mathcal{T}_{\S\Phi,p^{\&}}(U)})\end{aligned}$$

$$\overset{\text{Eq. 5.53}}{=} \mathcal{W}(\lambda u.u) + \mathcal{W}(H)$$
$$= \mathcal{W}(H)$$

and we can conclude because by typing we know that $L = H$.

- Case $U = \underline{0}$:
  By typing $H = \mathbb{R}$, so $q^{\&}$ is of type $\mathbb{R}$. We have:

$$\mathcal{W}(\lambda q^{\&}.\mathcal{T}_{\S\overleftarrow{\Phi},p^{\&}}(U)) = \mathcal{W}(\lambda q^{\&}.\mathcal{T}_{\S\overleftarrow{\Phi},p^{\&}}(\underline{0})) = \mathcal{W}(\lambda q^{\&}.\langle\ \rangle) = \mathcal{W}(\mathbb{R}) = 1$$

$$\mathcal{W}(\lambda p^{\&}.U) = \mathcal{W}(\lambda p^{\&}.\underline{0}) = \mathcal{W}(\underline{0}) + \mathcal{W}(L) = \mathcal{W}(L)$$

By definition we have that

$$\mathcal{W}(\S\overleftarrow{\Phi}^{\text{in}}_{\mathcal{T}_{\S\Phi,p^{\&}}(\underline{0})}) = \mathcal{W}(\S\overleftarrow{\Phi}^{\text{out}}_{\mathcal{T}_{\S\Phi,p^{\&}}(\underline{0})}) = 0 \tag{5.54}$$

because $\mathcal{T}_{\S\Phi,p^{\&}}(\underline{0})$ has only free variables of ground type.

We have:

$$\mathcal{W}(\lambda q^{\&}.\mathcal{T}_{\S\overleftarrow{\Phi},p^{\&}}(\underline{0})) + \mathcal{W}(L) + \mathcal{W}(\S\overleftarrow{\Phi}^{\text{in}}_{\mathcal{T}_{\S\Phi,p^{\&}}(\underline{0})})$$
$$= 1 + \mathcal{W}(L) + \mathcal{W}(\S\overleftarrow{\Phi}^{\text{in}}_{\mathcal{T}_{\S\Phi,p^{\&}}(\underline{0})})$$
$$\overset{\text{Eq. 5.54}}{=} 1 + \mathcal{W}(L)$$
$$\leq \mathcal{W}(\lambda p^{\&}.\underline{0}) + \mathcal{W}(H) + \mathcal{W}(\S\overleftarrow{\Phi}^{\text{out}}_{\mathcal{T}_{\S\Phi,p^{\&}}(\underline{0})})$$
$$= \mathcal{W}(L) + \mathcal{W}(H) + \mathcal{W}(\S\overleftarrow{\Phi}^{\text{out}}_{\mathcal{T}_{\S\Phi,p^{\&}}(\underline{0})})$$
$$\overset{\text{Eq. 5.54}}{=} \mathcal{W}(L) + \mathcal{W}(H)$$
$$= \mathcal{W}(L) + \mathcal{W}(\mathbb{R})$$
$$= \mathcal{W}(L) + 1$$

- Case $U = \langle\ \rangle$:
  By typing $H = \top$, so $q^{\&}$ is of type $\top$. We have:

$$\mathcal{W}(\lambda q^{\&}.\mathcal{T}_{\S\overleftarrow{\Phi},p^{\&}}(U)) = \mathcal{W}(\lambda q^{\&}.\mathcal{T}_{\S\overleftarrow{\Phi},p^{\&}}(\langle\ \rangle)) = \mathcal{W}(\lambda q^{\&}.\langle\ \rangle) = \mathcal{W}(\top) = 0$$

$$\mathcal{W}(\lambda p^{\&}.U) = \mathcal{W}(\lambda p^{\&}.\langle\ \rangle) = \mathcal{W}(\langle\ \rangle) + \mathcal{W}(L) = \mathcal{W}(L)$$

By definition we have that

$$\mathcal{W}(\S\overleftarrow{\Phi}^{\text{in}}_{\mathcal{T}_{\S\Phi,p^{\&}}(\langle\ \rangle)}) = \mathcal{W}(\S\overleftarrow{\Phi}^{\text{out}}_{\mathcal{T}_{\S\Phi,p^{\&}}(\langle\ \rangle)}) = 0 \tag{5.55}$$

because $\mathcal{T}_{\S\Phi,p^{\&}}(\langle\ \rangle)$ has only free variables of ground type.

We have:

$$\mathcal{W}(\lambda q^{\&}.\mathcal{T}_{\S\overleftarrow{\Phi},p^{\&}}(\langle\ \rangle)) + \mathcal{W}(L) + \mathcal{W}(\S\overleftarrow{\Phi}^{\text{in}}_{\mathcal{T}_{\S\Phi,p^{\&}}(\langle\ \rangle)})$$
$$= 0 + \mathcal{W}(L) + \mathcal{W}(\S\overleftarrow{\Phi}^{\text{in}}_{\mathcal{T}_{\S\Phi,p^{\&}}(\langle\ \rangle)})$$

$$
\begin{aligned}
&\overset{\text{Eq. 5.55}}{=} \mathcal{W}(L)\\
&\leq \mathcal{W}(\lambda p^{\&}.\langle\,\rangle) + \mathcal{W}(H) + \mathcal{W}(\S\overleftarrow{\Phi}^{\text{out}}_{\mathcal{T}_{\S\Phi,p^{\&}}(\langle\,\rangle)})\\
&= \mathcal{W}(L) + \mathcal{W}(H) + \mathcal{W}(\S\overleftarrow{\Phi}^{\text{out}}_{\mathcal{T}_{\S\Phi,p^{\&}}(\langle\,\rangle)})\\
&\overset{\text{Eq. 5.55}}{=} \mathcal{W}(L) + \mathcal{W}(H)\\
&= \mathcal{W}(L) + \mathcal{W}(\top)\\
&= \mathcal{W}(L)
\end{aligned}
$$

□

*Proof Claim 2: Cases of $\mathcal{T}$ on $\lambda LL^{\mathsf{f}}$.* Recall that we have to show

$$\mathcal{W}(\mathcal{T}_{\S\overleftarrow{\Phi}}(F)) + \mathcal{W}(L) + \mathcal{W}(\S\overleftarrow{\Phi}^{\text{in}}_{\mathcal{T}_{\S\overleftarrow{\Phi}}(F)}) \leq \mathcal{W}(F) + \mathcal{W}(H) + \mathcal{W}(\S\overleftarrow{\Phi}^{\text{out}}_{\mathcal{T}_{\S\overleftarrow{\Phi}}(F)}).$$

We proceed by analyzing the cases in Figure 5.4a.

- Case $F = \lambda p^{\&}.U$:
  By hypothesis we have $!\Sigma, \S\Phi \vdash \lambda p^{\&}.U : L \multimap H$, so by typing we have $!\Sigma, \S\Phi, p^{\&} : L \vdash U : H$

  By inductive hypothesis on $U$ (case 1 of the lemma) we have

  $$\mathcal{W}(\lambda q^{\&}.\mu_{p^{\&},\alpha,\emptyset}\langle \mathcal{T}_{\S\overleftarrow{\Phi},p^{\&}}(U), \langle\,\rangle\rangle) + \mathcal{W}(L) + \mathcal{W}(\S\overleftarrow{\Phi}^{\text{in}}_{\mathcal{T}_{\S\overleftarrow{\Phi},p^{\&}}(U)}) \leq \mathcal{W}(\lambda p^{\&}.U) + \mathcal{W}(H) + \mathcal{W}(\S\overleftarrow{\Phi}^{\text{out}}_{\mathcal{T}_{\S\overleftarrow{\Phi},p^{\&}}(U)})$$

  We have:

  $$\mathcal{W}(\mathcal{T}_{\S\overleftarrow{\Phi}}(F)) = \mathcal{W}(\mathcal{T}_{\S\overleftarrow{\Phi}}(\lambda p^{\&}.U)) = \mathcal{W}(\lambda q^{\&}.\mu_{p^{\&},\alpha,\emptyset}\langle \mathcal{T}_{\S\overleftarrow{\Phi},p^{\&}}(U), \langle\,\rangle\rangle)$$

  $$\mathcal{W}(F) = \mathcal{W}(\lambda p^{\&}.U)$$

  We show that

  $$\mathcal{W}(\mathcal{T}_{\S\overleftarrow{\Phi}}(\lambda p^{\&}.U)) + \mathcal{W}(L) + \mathcal{W}(\S\overleftarrow{\Phi}^{\text{in}}_{\mathcal{T}_{\S\overleftarrow{\Phi}}(\lambda p^{\&}.U)}) \leq \mathcal{W}(F) + \mathcal{W}(H) + \mathcal{W}(\S\overleftarrow{\Phi}^{\text{out}}_{\mathcal{T}_{\S\overleftarrow{\Phi}}(\lambda p^{\&}.U)})$$

  as follows

  $$
  \begin{aligned}
  &\mathcal{W}(\mathcal{T}_{\S\overleftarrow{\Phi}}(\lambda p^{\&}.U)) + \mathcal{W}(L) + \mathcal{W}(\S\overleftarrow{\Phi}^{\text{in}}_{\mathcal{T}_{\S\overleftarrow{\Phi}}(\lambda p^{\&}.U)})\\
  &\qquad = \mathcal{W}(\lambda q^{\&}.\mu_{p^{\&},\alpha,\emptyset}\langle \mathcal{T}_{\S\overleftarrow{\Phi},p^{\&}}(U), \langle\,\rangle\rangle) + \mathcal{W}(L) + \mathcal{W}(\S\overleftarrow{\Phi}^{\text{in}}_{\mathcal{T}_{\S\overleftarrow{\Phi},p^{\&}}(U)})\\
  &\qquad \overset{\text{IH on U}}{\leq} \mathcal{W}(\lambda p^{\&}.U) + \mathcal{W}(H) + \mathcal{W}(\S\overleftarrow{\Phi}^{\text{out}}_{\mathcal{T}_{\S\overleftarrow{\Phi},p^{\&}}(U)})
  \end{aligned}
  $$

  and so we can conclude because in this case $\S\overleftarrow{\Phi}^{\text{out}}_{\mathcal{T}_{\S\overleftarrow{\Phi}}(\lambda p^{\&}.U)} = \S\overleftarrow{\Phi}^{\text{out}}_{\mathcal{T}_{\S\overleftarrow{\Phi},p^{\&}}(U)}$.

- Case $F = \texttt{let}\ \S f^{\S(L_0 \multimap H_0)} = \S G_1\ \texttt{in}\ G_2$:
  By hypothesis we have $!\Sigma, \S\Phi \vdash \texttt{let}\ \S f^{\S(L_0 \multimap H_0)} = \S G_1\ \texttt{in}\ G_2 : L \multimap H$, so by typing we have:

  $$!\Sigma, \S\Phi \vdash G_1 : L_0 \multimap H_0$$
  $$!\Sigma, \S\Phi, f : L_0 \multimap H_0 \vdash G_2 : L \multimap H$$

  Observe that $f$ may not be free in $G_2$, so we have to analyze the following subcases:

– Subcase $f \in FV(G_2)$:
By inductive hypothesis on $G_1$ (case 2 of the lemma) we have

$$\mathcal{W}(\mathcal{T}_{\S\overleftarrow{\Phi}}(G_1)) + \mathcal{W}(L_0) + \mathcal{W}(\S\overleftarrow{\Phi}^{\text{in}}_{\mathcal{T}_{\S\overleftarrow{\Phi}}(G_1)}) \leq \mathcal{W}(F) + \mathcal{W}(H_0) + \mathcal{W}(\S\overleftarrow{\Phi}^{\text{out}}_{\mathcal{T}_{\S\overleftarrow{\Phi}}(G_1)}).$$

By inductive hypothesis on $G_2$ (case 2 of the lemma) we have

$$\begin{aligned}\mathcal{W}(\mathcal{T}_{\S\overleftarrow{\Phi},\S\overleftarrow{f}}(G_2)) + \mathcal{W}(L) + \mathcal{W}(\S\overleftarrow{\Phi}^{\text{in}}_{\mathcal{T}_{\S\overleftarrow{\Phi},\S\overleftarrow{f}}(G_2)}) + \boxed{\mathcal{W}(H_0)} \\ \leq \mathcal{W}(F) + \mathcal{W}(H) + \mathcal{W}(\S\overleftarrow{\Phi}^{\text{out}}_{\mathcal{T}_{\S\overleftarrow{\Phi},\S\overleftarrow{f}}(G_2)}) + \boxed{\mathcal{W}(L_0)} \\ \text{because } \overleftarrow{f} : \S(H_0 \multimap L_0) \in FV(\mathcal{T}_{\S\overleftarrow{\Phi},\S\overleftarrow{f}}(G_2))\end{aligned}$$

We have:

$$\begin{aligned}\mathcal{W}(\mathcal{T}_{\S\overleftarrow{\Phi}}(F)) &= \mathcal{W}(\mathcal{T}_{\S\overleftarrow{\Phi}}(\texttt{let } \S f^{\S(L_0 \multimap H_0)} = \S G_1 \texttt{ in } G_2)) \\ &= \mathcal{W}(\texttt{let } \overleftarrow{f}^{\S(H_0 \multimap L_0)} = \S\mathcal{T}_{\S\overleftarrow{\Phi},\S\overleftarrow{f}}(G_2) \texttt{ in } \mathcal{T}_{\S\overleftarrow{\Phi}}(G_1)) \\ &\approx \mathcal{W}((\lambda f.\mathcal{T}_{\S\overleftarrow{\Phi}}(G_1))\S\mathcal{T}_{\S\overleftarrow{\Phi},\S\overleftarrow{f}}(G_2)) \\ &= \mathcal{W}(\mathcal{T}_{\S\overleftarrow{\Phi}}(G_1)) + \mathcal{W}(\S\mathcal{T}_{\S\overleftarrow{\Phi},\S\overleftarrow{f}}(G_2)) \\ &\overset{\text{Lemma 24}}{=} \mathcal{W}(\mathcal{T}_{\S\overleftarrow{\Phi}}(G_1)) + \mathcal{W}(\mathcal{T}_{\S\overleftarrow{\Phi},\S\overleftarrow{f}}(G_2))\end{aligned}$$

$$\begin{aligned}\mathcal{W}(F) &= \mathcal{W}(\texttt{let } \S f^{\S(L_0 \multimap H_0)} = \S G_1 \texttt{ in } G_2) \\ &= \mathcal{W}(\S G_1) + \mathcal{W}(G_2) \\ &\overset{\text{Lemma 24}}{=} \mathcal{W}(G_1) + \mathcal{W}(G_2)\end{aligned}$$

We can conclude as follows

$$\begin{aligned}&\mathcal{W}(\mathcal{T}_{\S\overleftarrow{\Phi}}(F)) + \mathcal{W}(L) + \mathcal{W}(\S\overleftarrow{\Phi}^{\text{in}}_{\mathcal{T}_{\S\overleftarrow{\Phi}}(F)}) \\ &= \mathcal{W}(\mathcal{T}_{\S\overleftarrow{\Phi}}(G_1)) + \mathcal{W}(\mathcal{T}_{\S\overleftarrow{\Phi},\S\overleftarrow{f}}(G_2)) + \mathcal{W}(L) + \mathcal{W}(\S\overleftarrow{\Phi}^{\text{in}}_{\mathcal{T}_{\S\overleftarrow{\Phi}}(F)}) \\ &= \mathcal{W}(\mathcal{T}_{\S\overleftarrow{\Phi}}(G_1)) + \underline{\mathcal{W}(\mathcal{T}_{\S\overleftarrow{\Phi},\S\overleftarrow{f}}(G_2))} + \underline{\mathcal{W}(L)} \\ &\qquad + \mathcal{W}(\S\overleftarrow{\Phi}^{\text{in}}_{\mathcal{T}_{\S\overleftarrow{\Phi}}(G_1)}) + \underline{\mathcal{W}(\S\overleftarrow{\Phi}^{\text{in}}_{\mathcal{T}_{\S\overleftarrow{\Phi},\S\overleftarrow{f}}(G_2)})} \\ &\overset{\text{IH on } G_2}{\leq} \underline{\mathcal{W}(\mathcal{T}_{\S\overleftarrow{\Phi}}(G_1))} + \underline{\mathcal{W}(\S\overleftarrow{\Phi}^{\text{in}}_{\mathcal{T}_{\S\overleftarrow{\Phi}}(G_1)})} - \boxed{\mathcal{W}(H_0)} + \\ &\qquad + \mathcal{W}(G_2) + \mathcal{W}(H) + \mathcal{W}(\S\overleftarrow{\Phi}^{\text{out}}_{\mathcal{T}_{\S\overleftarrow{\Phi},\S\overleftarrow{f}}(G_2)}) + \underline{\boxed{\mathcal{W}(L_0)}} \\ &\overset{\text{IH on } G_1}{\leq} \mathcal{W}(G_1) + \mathcal{W}(H_0) + \mathcal{W}(\S\overleftarrow{\Phi}^{\text{out}}_{\mathcal{T}_{\S\overleftarrow{\Phi}}(G_1)}) - \boxed{\mathcal{W}(H_0)} \\ &\qquad + \mathcal{W}(G_2) + \mathcal{W}(H) + \mathcal{W}(\S\overleftarrow{\Phi}^{\text{out}}_{\mathcal{T}_{\S\overleftarrow{\Phi},\S\overleftarrow{f}}(G_2)}) \\ &= \mathcal{W}(G_1) + \mathcal{W}(\S\overleftarrow{\Phi}^{\text{out}}_{\mathcal{T}_{\S\overleftarrow{\Phi}}(G_1)}) + \mathcal{W}(G_2) + \mathcal{W}(H) + \mathcal{W}(\S\overleftarrow{\Phi}^{\text{out}}_{\mathcal{T}_{\S\overleftarrow{\Phi},\S\overleftarrow{f}}(G_2)}) \\ &= \mathcal{W}(G_1) + \mathcal{W}(G_2) + \mathcal{W}(H) + \mathcal{W}(\S\overleftarrow{\Phi}^{\text{out}}_{\mathcal{T}_{\S\overleftarrow{\Phi}}(F)}) \\ &\leq \mathcal{W}(F) + \mathcal{W}(H) + \mathcal{W}(\S\overleftarrow{\Phi}^{\text{out}}_{\mathcal{T}_{\S\overleftarrow{\Phi}}(F)})\end{aligned}$$

$$= \mathcal{W}(G_1) + \mathcal{W}(G_2) + \mathcal{W}(H) + \mathcal{W}(\S\overleftarrow{\Phi}^{\text{out}}_{\mathcal{T}_{\S\overleftarrow{\Phi}}(F)})$$

– Subcase $f \notin FV(G_2)$:
By typing we have $!\Sigma, \S\Phi \vdash G_2 : L \multimap H$.
By inductive hypothesis on $G_2$ (case 2 of the lemma) we have

$$\mathcal{W}(\mathcal{T}_{\S\overleftarrow{\Phi}}(G_2)) + \mathcal{W}(L) + \mathcal{W}(\S\overleftarrow{\Phi}^{\text{in}}_{\mathcal{T}_{\S\overleftarrow{\Phi}}(G_2)}) \leq \mathcal{W}(F) + \mathcal{W}(H) + \mathcal{W}(\S\overleftarrow{\Phi}^{\text{out}}_{\mathcal{T}_{\S\overleftarrow{\Phi}}(G_2)})$$

We have:

$$\mathcal{W}(\mathcal{T}_{\S\overleftarrow{\Phi}}(F)) = \mathcal{W}(\mathcal{T}_{\S\overleftarrow{\Phi}}(\texttt{let } \S f^{\S(L_0 \multimap H_0)} = \S G_1 \texttt{ in } G_2)) = \mathcal{W}(\mathcal{T}_{\S\overleftarrow{\Phi}}(G_2))$$

$$\begin{aligned}
\mathcal{W}(F) &= \mathcal{W}(\texttt{let } \S f^{\S(L_0 \multimap H_0)} = \S G_1 \texttt{ in } G_2) \\
&\approx \mathcal{W}((\lambda \S f.G_2)\S G_1) \\
&= \mathcal{W}(\S G_1) + \mathcal{W}(G_2) + \mathcal{W}(FV(f) \setminus FV(G_2)) \\
&= \mathcal{W}(\S G_1) + \mathcal{W}(G_2) + \mathcal{W}(f) \\
&= \mathcal{W}(\S G_1) + \mathcal{W}(G_2) + \mathcal{W}(\S(L_0 \multimap H_0)) \\
&\approx \mathcal{W}(\S G_1) + \mathcal{W}(G_2) + \mathcal{W}(\mathtt{1}\&(L_0 \multimap H_0)) \\
&= \mathcal{W}(\S G_1) + \mathcal{W}(G_2) + \mathcal{W}(\mathtt{1}) + \mathcal{W}(L_0) + \mathcal{W}(H_0) \\
&= \mathcal{W}(\S G_1) + \mathcal{W}(G_2) + \mathcal{W}(L_0) + \mathcal{W}(H_0) \\
&\overset{\text{Lemma 24}}{=} \mathcal{W}(G_1) + \mathcal{W}(G_2) + \mathcal{W}(L_0) + \mathcal{W}(H_0)
\end{aligned}$$

Observe that in this case we have

$$\S\overleftarrow{\Phi}^{\text{in}}_{\mathcal{T}_{\S\overleftarrow{\Phi}}(F)} = \S\overleftarrow{\Phi}^{\text{in}}_{\mathcal{T}_{\S\overleftarrow{\Phi}}(G_2)} \quad \text{and} \quad \S\overleftarrow{\Phi}^{\text{out}}_{\mathcal{T}_{\S\overleftarrow{\Phi}}(F)} = \S\overleftarrow{\Phi}^{\text{out}}_{\mathcal{T}_{\S\overleftarrow{\Phi}}(G_2)} \tag{5.56}$$

because $\mathcal{T}_{\S\overleftarrow{\Phi}}(F) = \mathcal{T}_{\S\overleftarrow{\Phi}}(G_2)$.
We can conclude as follows

$$\begin{aligned}
&\mathcal{W}(\mathcal{T}_{\S\overleftarrow{\Phi}}(F)) + \mathcal{W}(L) + \mathcal{W}(\S\overleftarrow{\Phi}^{\text{in}}_{\mathcal{T}_{\S\overleftarrow{\Phi}}(F)}) \\
&\quad = \mathcal{W}(\mathcal{T}_{\S\overleftarrow{\Phi}}(G_2)) + \mathcal{W}(L) + \mathcal{W}(\S\overleftarrow{\Phi}^{\text{in}}_{\mathcal{T}_{\S\overleftarrow{\Phi}}(F)}) \\
&\quad \overset{\text{Eq. 5.56}}{=} \mathcal{W}(\mathcal{T}_{\S\overleftarrow{\Phi}}(G_2)) + \mathcal{W}(L) + \mathcal{W}(\S\overleftarrow{\Phi}^{\text{in}}_{\mathcal{T}_{\S\overleftarrow{\Phi}}(G_2)}) \\
&\quad \overset{\text{IH on } G_2}{\leq} \mathcal{W}(G_2) + \mathcal{W}(H) + \mathcal{W}(\S\overleftarrow{\Phi}^{\text{out}}_{\mathcal{T}_{\S\overleftarrow{\Phi}}(G_2)}) \\
&\quad \leq \mathcal{W}(F) + \mathcal{W}(H) + \mathcal{W}(\S\overleftarrow{\Phi}^{\text{out}}_{\mathcal{T}_{\S\overleftarrow{\Phi}}(F)}) \\
&\quad = \mathcal{W}(G_1) + \mathcal{W}(G_2) + \mathcal{W}(L_0) + \mathcal{W}(H_0) + \mathcal{W}(H) + \mathcal{W}(\S\overleftarrow{\Phi}^{\text{out}}_{\mathcal{T}_{\S\overleftarrow{\Phi}}(F)}) \\
&\quad \overset{\text{Eq. 5.56}}{=} \mathcal{W}(G_1) + \mathcal{W}(G_2) + \mathcal{W}(L_0) + \mathcal{W}(H_0) + \mathcal{W}(H) + \mathcal{W}(\S\overleftarrow{\Phi}^{\text{out}}_{\mathcal{T}_{\S\overleftarrow{\Phi}}(G_2)})
\end{aligned}$$

□

*Proof Claim 3: Cases of $\mathcal{T}$ on $\lambda LL^{\mathtt{A}}$.* Recall that we have to show

$$\mathcal{W}(\mathcal{T}_{\S\overleftarrow{\Phi}}(F)) + \mathcal{W}(L) + \mathcal{W}(\S\overleftarrow{\Phi}^{\text{in}}_{\mathcal{T}_{\S\overleftarrow{\Phi}}(F)}) \leq \mathcal{W}(F) + \mathcal{W}(H) + \mathcal{W}(\S\overleftarrow{\Phi}^{\text{out}}_{\mathcal{T}_{\S\overleftarrow{\Phi}}(F)}).$$

We proceed by analyzing the cases in Figure 5.4a.

- Case $R = \mathtt{let}\ \S f^{\S(L_0 \multimap H_0)} = \S F\ \mathtt{in}\ S$:
  By hypothesis we have $!\Sigma, \S\Phi \vdash \mathtt{let}\ \S f^{\S(L_0 \multimap H_0)} = \S F\ \mathtt{in}\ S : !E \otimes \S(L \multimap H)$ and by typing we have:

  $$!\Sigma, \S\Phi \vdash F : L_0 \multimap H_0$$
  $$!\Sigma, \S\Phi, f : \S(L_0 \multimap H_0) \vdash S : !E \otimes \S(L \multimap H)$$

  Observe that $f$ may not be free in $S$, so we have to analyze the following subcases (similar to what we did in the case of composition of $\lambda LL^{\mathtt{f}}$):

  - Subcase $f \in FV(S)$:
    By inductive hypothesis on $F$ (case 2 of the lemma) we have:

    $$\mathcal{W}(\mathcal{T}_{\S\overleftarrow{\Phi}}(F)) + \mathcal{W}(L) + \mathcal{W}(\S\overleftarrow{\Phi}^{\text{in}}_{\mathcal{T}_{\S\overleftarrow{\Phi}}(F)}) \leq \mathcal{W}(F) + \mathcal{W}(H) + \mathcal{W}(\S\overleftarrow{\Phi}^{\text{out}}_{\mathcal{T}_{\S\overleftarrow{\Phi}}(F)}).$$

    By inductive hypothesis on $S$ (case 3 of the lemma) we have:

    $$\begin{aligned} &\mathcal{W}(\mathcal{T}_{\S\overleftarrow{\Phi},\S\overleftarrow{f}}(S)) + \mathcal{W}(L) + \mathcal{W}(\S\overleftarrow{\Phi}^{\text{in}}_{\mathcal{T}_{\S\overleftarrow{\Phi},\S\overleftarrow{f}}(S)}) + \mathcal{W}(H_0) \\ &\qquad \leq \mathcal{W}(S) + \mathcal{W}(H) + \mathcal{W}(\S\overleftarrow{\Phi}^{\text{out}}_{\mathcal{T}_{\S\overleftarrow{\Phi},\S\overleftarrow{f}}(S)}) + \mathcal{W}(L_0) \\ &\qquad \text{because } \overleftarrow{f} : \S(H_0 \multimap L_0) \in FV(\mathcal{T}_{\S\overleftarrow{\Phi},\S\overleftarrow{f}}(S)) \end{aligned}$$

    We have:

    $$\begin{aligned} \mathcal{W}(\mathcal{T}_{\S\overleftarrow{\Phi}}(R)) &= \mathcal{W}(\mathcal{T}_{\S\overleftarrow{\Phi}}(\mathtt{let}\ \S f^{\S(L_0 \multimap H_0)} = \S F\ \mathtt{in}\ S)) \\ &= \mathcal{W}(\mathtt{let}\ \overleftarrow{f}^{\S(H_0 \multimap L_0)} = \S\mathcal{T}_{\S\overleftarrow{\Phi},\S\overleftarrow{f}}(S)\ \mathtt{in}\ \mathcal{T}_{\S\overleftarrow{\Phi}}(F)) \\ &\approx \mathcal{W}((\lambda f.\mathcal{T}_{\S\overleftarrow{\Phi}}(F))\S\mathcal{T}_{\S\overleftarrow{\Phi},\S\overleftarrow{f}}(S)) \\ &= \mathcal{W}(\mathcal{T}_{\S\overleftarrow{\Phi}}(F)) + \mathcal{W}(\S\mathcal{T}_{\S\overleftarrow{\Phi},\S\overleftarrow{f}}(S)) \\ &\overset{\text{Lemma 24}}{=} \mathcal{W}(\mathcal{T}_{\S\overleftarrow{\Phi}}(F)) + \mathcal{W}(\mathcal{T}_{\S\overleftarrow{\Phi},\S\overleftarrow{f}}(S)) \end{aligned}$$

    $$\begin{aligned} \mathcal{W}(R) &= \mathcal{W}(\mathtt{let}\ \S f^{\S(L_0 \multimap H_0)} = \S F\ \mathtt{in}\ S) \\ &= \mathcal{W}(\S F) + \mathcal{W}(S) \\ &\overset{\text{Lemma 24}}{=} \mathcal{W}(F) + \mathcal{W}(S) \end{aligned}$$

    We can conclude as follows

    $$\begin{aligned} &\mathcal{W}(\mathcal{T}_{\S\overleftarrow{\Phi}}(R)) + \mathcal{W}(L) + \mathcal{W}(\S\overleftarrow{\Phi}^{\text{in}}_{\mathcal{T}_{\S\overleftarrow{\Phi}}(R)}) \\ &\quad = \mathcal{W}(\mathcal{T}_{\S\overleftarrow{\Phi}}(F)) + \mathcal{W}(\mathcal{T}_{\S\overleftarrow{\Phi},\S\overleftarrow{f}}(S)) + \mathcal{W}(L) + \mathcal{W}(\S\overleftarrow{\Phi}^{\text{in}}_{\mathcal{T}_{\S\overleftarrow{\Phi}}(R)}) \\ &\quad = \mathcal{W}(\mathcal{T}_{\S\overleftarrow{\Phi}}(F)) + \underline{\mathcal{W}(\mathcal{T}_{\S\overleftarrow{\Phi},\S\overleftarrow{f}}(S))} + \underline{\mathcal{W}(L)} \end{aligned}$$

$$
\begin{aligned}
&\quad + \mathcal{W}(\S\overleftarrow{\Phi}^{\text{in}}_{\mathcal{T}_{\S\overleftarrow{\Phi}}(F)}) + \underline{\mathcal{W}(\S\overleftarrow{\Phi}^{\text{in}}_{\mathcal{T}_{\S\overleftarrow{\Phi},\S\overleftarrow{f}}(S)})} \\
&\overset{\text{IH on } S}{\leq} \underline{\mathcal{W}(\mathcal{T}_{\S\overleftarrow{\Phi}}(F))} + \underline{\mathcal{W}(\S\overleftarrow{\Phi}^{\text{in}}_{\mathcal{T}_{\S\overleftarrow{\Phi}}(F)})} - \mathcal{W}(H_0) + \\
&\quad + \mathcal{W}(S) + \mathcal{W}(H) + \mathcal{W}(\S\overleftarrow{\Phi}^{\text{out}}_{\mathcal{T}_{\S\overleftarrow{\Phi},\S\overleftarrow{f}}(S)}) + \underline{\mathcal{W}(L_0)} \\
&\overset{\text{IH on } F}{\leq} \mathcal{W}(F) + \mathcal{W}(H_0) + \mathcal{W}(\S\overleftarrow{\Phi}^{\text{out}}_{\mathcal{T}_{\S\overleftarrow{\Phi}}(F)}) - \mathcal{W}(H_0) + \mathcal{W}(S) + \mathcal{W}(H) \\
&\quad + \mathcal{W}(\S\overleftarrow{\Phi}^{\text{out}}_{\mathcal{T}_{\S\overleftarrow{\Phi},\S\overleftarrow{f}}(S)}) \\
&= \mathcal{W}(F) + \mathcal{W}(\S\overleftarrow{\Phi}^{\text{out}}_{\mathcal{T}_{\S\overleftarrow{\Phi}}(F)}) + \mathcal{W}(S) + \mathcal{W}(H) + \mathcal{W}(\S\overleftarrow{\Phi}^{\text{out}}_{\mathcal{T}_{\S\overleftarrow{\Phi},\S\overleftarrow{f}}(S)}) \\
&= \mathcal{W}(F) + \mathcal{W}(S) + \mathcal{W}(H) + \mathcal{W}(\S\overleftarrow{\Phi}^{\text{out}}_{\mathcal{T}_{\S\overleftarrow{\Phi}}(R)}) \\
&\leq \mathcal{W}(R) + \mathcal{W}(H) + \mathcal{W}(\S\overleftarrow{\Phi}^{\text{out}}_{\mathcal{T}_{\S\overleftarrow{\Phi}}(R)}) \\
&= \mathcal{W}(F) + \mathcal{W}(S) + \mathcal{W}(H) + \mathcal{W}(\S\overleftarrow{\Phi}^{\text{out}}_{\mathcal{T}_{\S\overleftarrow{\Phi}}(R)})
\end{aligned}
$$

– Subcase $f \notin FV(S)$:
By typing we have $!\Sigma, \S\Phi \vdash S : !E \otimes \S(L \multimap H)$.
By inductive hypothesis on $S$ (case 3 of the lemma) we have:

$$\mathcal{W}(\mathcal{T}_{\S\overleftarrow{\Phi}}(S)) + \mathcal{W}(L) + \mathcal{W}(\S\overleftarrow{\Phi}^{\text{in}}_{\mathcal{T}_{\S\overleftarrow{\Phi}}(S)}) \leq \mathcal{W}(S) + \mathcal{W}(H) + \mathcal{W}(\S\overleftarrow{\Phi}^{\text{out}}_{\mathcal{T}_{\S\overleftarrow{\Phi}}(S)})$$

We have:

$$\mathcal{W}(\mathcal{T}_{\S\overleftarrow{\Phi}}(R)) = \mathcal{W}(\mathcal{T}_{\S\overleftarrow{\Phi}}(\texttt{let } \S f^{\S(L_0 \multimap H_0)} = \S F \texttt{ in } S)) = \mathcal{W}(\mathcal{T}_{\S\overleftarrow{\Phi}}(S))$$

$$
\begin{aligned}
\mathcal{W}(R) &= \mathcal{W}(\texttt{let } \S f^{\S(L_0 \multimap H_0)} = \S F \texttt{ in } S) \\
&\approx \mathcal{W}((\lambda \S f.S)\S F) \\
&= \mathcal{W}(\S F) + \mathcal{W}(S) + \mathcal{W}(FV(f) \setminus FV(S)) \\
&= \mathcal{W}(\S F) + \mathcal{W}(S) + \mathcal{W}(L_0 \multimap H_0) \\
&= \mathcal{W}(\S F) + \mathcal{W}(S) + \mathcal{W}(L_0) + \mathcal{W}(H_0) \\
&\overset{\text{Lemma 24}}{=} \mathcal{W}(F) + \mathcal{W}(S) + \mathcal{W}(L_0) + \mathcal{W}(H_0)
\end{aligned}
$$

Observe that in this case we have

$$\S\overleftarrow{\Phi}^{\text{in}}_{\mathcal{T}_{\S\overleftarrow{\Phi}}(R)} = \S\overleftarrow{\Phi}^{\text{in}}_{\mathcal{T}_{\S\overleftarrow{\Phi}}(S)} \quad \text{and} \quad \S\overleftarrow{\Phi}^{\text{out}}_{\mathcal{T}_{\S\overleftarrow{\Phi}}(R)} = \S\overleftarrow{\Phi}^{\text{out}}_{\mathcal{T}_{\S\overleftarrow{\Phi}}(S)} \tag{5.57}$$

because $\mathcal{T}_{\S\overleftarrow{\Phi}}(R) = \mathcal{T}_{\S\overleftarrow{\Phi}}(S)$.
We can conclude as follows

$$
\begin{aligned}
&\mathcal{W}(\mathcal{T}_{\S\overleftarrow{\Phi}}(R)) + \mathcal{W}(L) + \mathcal{W}(\S\overleftarrow{\Phi}^{\text{in}}_{\mathcal{T}_{\S\overleftarrow{\Phi}}(R)}) \\
&\quad = \mathcal{W}(\mathcal{T}_{\S\overleftarrow{\Phi}}(S)) + \mathcal{W}(L) + \mathcal{W}(\S\overleftarrow{\Phi}^{\text{in}}_{\mathcal{T}_{\S\overleftarrow{\Phi}}(R)}) \\
&\quad \overset{\text{Eq. 5.57}}{=} \mathcal{W}(\mathcal{T}_{\S\overleftarrow{\Phi}}(S)) + \mathcal{W}(L) + \mathcal{W}(\S\overleftarrow{\Phi}^{\text{in}}_{\mathcal{T}_{\S\overleftarrow{\Phi}}(S)})
\end{aligned}
$$

$$\begin{aligned}
&\overset{\text{IH on } S}{\leq} \mathcal{W}(S) + \mathcal{W}(H) + \mathcal{W}(\S\overleftarrow{\Phi}^{\text{out}}_{\mathcal{T}_{\S\overleftarrow{\Phi}}(S)}) \\
&\leq \mathcal{W}(R) + \mathcal{W}(H) + \mathcal{W}(\S\overleftarrow{\Phi}^{\text{out}}_{\mathcal{T}_{\S\overleftarrow{\Phi}}(R)}) \\
&= \mathcal{W}(F) + \mathcal{W}(S) + \mathcal{W}(L_0) + \mathcal{W}(H_0) + \mathcal{W}(H) + \mathcal{W}(\S\overleftarrow{\Phi}^{\text{out}}_{\mathcal{T}_{\S\overleftarrow{\Phi}}(R)}) \\
&\overset{\text{Eq. 5.57}}{=} \mathcal{W}(F) + \mathcal{W}(S) + \mathcal{W}(L_0) + \mathcal{W}(H_0) + \mathcal{W}(H) + \mathcal{W}(\S\overleftarrow{\Phi}^{\text{out}}_{\mathcal{T}_{\S\overleftarrow{\Phi}}(S)})
\end{aligned}$$

- Case $R = \texttt{let } (!x, \S f^{\S(L_0 \multimap H_0)}) = S_1 \texttt{ in } S_2$:
  By hypothesis we have $!\Sigma, \S\Phi \vdash \texttt{let } (!x, \S f^{\S(L_0 \multimap H_0)}) = S_1 \texttt{ in } S_2 : !E \otimes \S(L \multimap H)$ and by typing we have:

$$!\Sigma, \S\Phi \vdash S_1 : !E_0 \otimes \S(L_0 \multimap H_0)$$
$$!\Sigma, !x : !E_0, \S\Phi, f : \S(L_0 \multimap H_0) \vdash S_2 : !E \otimes \S(L \multimap H)$$

  Observe that $f$ may not be free in $S_2$, so we have to analyze the following subcases:

  - Subcase $f \in FV(S_2)$:
    By inductive hypothesis on $S_1$ (case 3 of the lemma) we have:

$$\mathcal{W}(\mathcal{T}_{\S\overleftarrow{\Phi}}(S_1)) + \mathcal{W}(L) + \mathcal{W}(\S\overleftarrow{\Phi}^{\text{in}}_{\mathcal{T}_{\S\overleftarrow{\Phi}}(S_1)}) \leq \mathcal{W}(S_1) + \mathcal{W}(H) + \mathcal{W}(\S\overleftarrow{\Phi}^{\text{out}}_{\mathcal{T}_{\S\overleftarrow{\Phi}}(S_1)})$$

    By inductive hypothesis on $S_2$ (case 3 of the lemma) we have:

$$\begin{aligned}
&\mathcal{W}(\mathcal{T}_{\S\overleftarrow{\Phi},\S\overleftarrow{f}}(S_2)) + \mathcal{W}(L) + \mathcal{W}(\S\overleftarrow{\Phi}^{\text{in}}_{\mathcal{T}_{\S\overleftarrow{\Phi},\S\overleftarrow{f}}(S_2)}) + \mathcal{W}(H_0) \\
&\qquad \leq \mathcal{W}(S_2) + \mathcal{W}(H) + \mathcal{W}(\S\overleftarrow{\Phi}^{\text{out}}_{\mathcal{T}_{\S\overleftarrow{\Phi},\S\overleftarrow{f}}(S_2)}) + \mathcal{W}(L_0) \\
&\qquad \text{because } \overleftarrow{f} : \S(H_0 \multimap L_0) \in FV(\mathcal{T}_{\S\overleftarrow{\Phi},\S\overleftarrow{f}}(S_2))
\end{aligned}$$

    We have:

$$\begin{aligned}
\mathcal{W}(\mathcal{T}_{\S\overleftarrow{\Phi}}(R)) &= \mathcal{W}(\mathcal{T}_{\S\overleftarrow{\Phi}}(\texttt{let } (!x, \S f) = S_1 \texttt{ in } S_2)) \\
&= \mathcal{W}(\texttt{let } (!x, \S\overleftarrow{f}) = \mathcal{T}_{\S\overleftarrow{\Phi}}(S_1) \texttt{ in } \mathcal{T}_{\S\overleftarrow{\Phi},\S\overleftarrow{f}}(S_2)) \\
&\approx \mathcal{W}((\lambda f.\mathcal{T}_{\S\overleftarrow{\Phi}}(S_1))\mathcal{T}_{\S\overleftarrow{\Phi},\S\overleftarrow{f}}(S_2)) \\
&= \mathcal{W}(\mathcal{T}_{\S\overleftarrow{\Phi}}(S_1)) + \mathcal{W}(\mathcal{T}_{\S\overleftarrow{\Phi},\S\overleftarrow{f}}(S_2))
\end{aligned}$$

$$\begin{aligned}
\mathcal{W}(R) &= \mathcal{W}(\texttt{let } (!x, \S f) = S_1 \texttt{ in } S_2) \\
&= \mathcal{W}(S_1) + \mathcal{W}(S_2)
\end{aligned}$$

    We can conclude as follows

$$\begin{aligned}
&\mathcal{W}(\mathcal{T}_{\S\overleftarrow{\Phi}}(R)) + \mathcal{W}(L) + \mathcal{W}(\S\overleftarrow{\Phi}^{\text{in}}_{\mathcal{T}_{\S\overleftarrow{\Phi}}(R)}) \\
&\quad = \mathcal{W}(\mathcal{T}_{\S\overleftarrow{\Phi}}(S_1)) + \mathcal{W}(\mathcal{T}_{\S\overleftarrow{\Phi},\S\overleftarrow{f}}(S_2)) + \mathcal{W}(L) + \mathcal{W}(\S\overleftarrow{\Phi}^{\text{in}}_{\mathcal{T}_{\S\overleftarrow{\Phi}}(R)}) \\
&\quad = \mathcal{W}(\mathcal{T}_{\S\overleftarrow{\Phi}}(S_1)) + \underline{\mathcal{W}(\mathcal{T}_{\S\overleftarrow{\Phi},\S\overleftarrow{f}}(S_2))} + \underline{\mathcal{W}(L)}
\end{aligned}$$

$$
\begin{aligned}
& + \mathcal{W}(\S\overleftarrow{\Phi}^{\text{in}}_{\mathcal{T}_{\S\overleftarrow{\Phi}}(S_1)}) + \underline{\mathcal{W}(\S\overleftarrow{\Phi}^{\text{in}}_{\mathcal{T}_{\S\overleftarrow{\Phi},\S\overleftarrow{f}}(S_2)})} \\
\overset{\text{IH on } S_2}{\leq}\ & \underline{\mathcal{W}(\mathcal{T}_{\S\overleftarrow{\Phi}}(S_1))} + \underline{\mathcal{W}(\S\overleftarrow{\Phi}^{\text{in}}_{\mathcal{T}_{\S\overleftarrow{\Phi}}(S_1)})} - \mathcal{W}(H_0) + \\
& + \mathcal{W}(S_2) + \mathcal{W}(H) + \mathcal{W}(\S\overleftarrow{\Phi}^{\text{out}}_{\mathcal{T}_{\S\overleftarrow{\Phi},\S\overleftarrow{f}}(S_2)}) + \underline{\mathcal{W}(L_0)} \\
\overset{\text{IH on } S_1}{\leq}\ & \mathcal{W}(S_1) + \mathcal{W}(H_0) + \mathcal{W}(\S\overleftarrow{\Phi}^{\text{out}}_{\mathcal{T}_{\S\overleftarrow{\Phi}}(S_1)}) - \mathcal{W}(H_0) \\
& + \mathcal{W}(S_2) + \mathcal{W}(H) + \mathcal{W}(\S\overleftarrow{\Phi}^{\text{out}}_{\mathcal{T}_{\S\overleftarrow{\Phi},\S\overleftarrow{f}}(S_2)}) \\
& = \mathcal{W}(S_1) + \mathcal{W}(\S\overleftarrow{\Phi}^{\text{out}}_{\mathcal{T}_{\S\overleftarrow{\Phi}}(S_1)}) + \mathcal{W}(S_2) + \mathcal{W}(H) + \mathcal{W}(\S\overleftarrow{\Phi}^{\text{out}}_{\mathcal{T}_{\S\overleftarrow{\Phi},\S\overleftarrow{f}}(S_2)}) \\
& = \mathcal{W}(S_1) + \mathcal{W}(S_2) + \mathcal{W}(H) + \mathcal{W}(\S\overleftarrow{\Phi}^{\text{out}}_{\mathcal{T}_{\S\overleftarrow{\Phi}}(R)}) \\
& \leq \mathcal{W}(R) + \mathcal{W}(H) + \mathcal{W}(\S\overleftarrow{\Phi}^{\text{out}}_{\mathcal{T}_{\S\overleftarrow{\Phi}}(R)}) \\
& = \mathcal{W}(S_1) + \mathcal{W}(S_2) + \mathcal{W}(H) + \mathcal{W}(\S\overleftarrow{\Phi}^{\text{out}}_{\mathcal{T}_{\S\overleftarrow{\Phi}}(R)})
\end{aligned}
$$

– Subcase $f \notin FV(S_2)$:
By typing we have $!\Sigma, !x : !E_0, \S\Phi \vdash S_2 : !E \otimes \S(L \multimap H)$.
By inductive hypothesis on $S_2$ (case 3 of the lemma) we have:

$$
\mathcal{W}(\mathcal{T}_{\S\overleftarrow{\Phi}}(S_2)) + \mathcal{W}(L) + \mathcal{W}(\S\overleftarrow{\Phi}^{\text{in}}_{\mathcal{T}_{\S\overleftarrow{\Phi}}(S_2)}) \leq \mathcal{W}(S_2) + \mathcal{W}(H) + \mathcal{W}(\S\overleftarrow{\Phi}^{\text{out}}_{\mathcal{T}_{\S\overleftarrow{\Phi}}(S_2)})
$$

We have:

$$
\begin{aligned}
\mathcal{W}(\mathcal{T}_{\S\overleftarrow{\Phi}}(R)) &= \mathcal{W}(\mathcal{T}_{\S\overleftarrow{\Phi}}(\texttt{let } (!x, \S f) = S_1 \texttt{ in } S_2)) \\
&= \mathcal{W}(\texttt{let } !x = \epsilon_1[P_1] \texttt{ in } \mathcal{T}_{\S\overleftarrow{\Phi}}(S_2)) \quad \text{for } \mathcal{U}^{\bullet}(S_1) = (\epsilon_1[], P_1, F_1) \\
&= \mathcal{W}(\epsilon_1[P_1]) + \mathcal{W}(\mathcal{T}_{\S\overleftarrow{\Phi}}(S_2))
\end{aligned}
$$

$$
\begin{aligned}
\mathcal{W}(R) &= \mathcal{W}(\texttt{let } (!x, \S f) = S_1 \texttt{ in } S_2) \\
&\approx \mathcal{W}((\lambda(!x, \S f).S_2)S_1) \\
&= \mathcal{W}(S_1) + \mathcal{W}(S_2) + \mathcal{W}(FV((!x, \S f)) \setminus FV(S_2)) \\
&= \mathcal{W}(S_1) + \mathcal{W}(S_2) + \mathcal{W}(f) \\
&= \mathcal{W}(S_1) + \mathcal{W}(S_2) + \mathcal{W}(L_0 \multimap H_0) \\
&= \mathcal{W}(S_1) + \mathcal{W}(S_2) + \mathcal{W}(L_0) + \mathcal{W}(H_0)
\end{aligned}
$$

Observe that in this case we have that $\mathcal{W}(\S\overleftarrow{\Phi}^{\text{in}}_{\mathcal{T}_{\S\overleftarrow{\Phi}}(R)}) = \mathcal{W}(\S\overleftarrow{\Phi}^{\text{in}}_{\mathcal{T}_{\S\overleftarrow{\Phi}}(S_2)}) + \mathcal{W}(\S\overleftarrow{\Phi}^{\text{in}}_{!\epsilon_1[P_1]})$ because $\mathcal{T}_{\S\overleftarrow{\Phi}}(R) = \texttt{let } !x = \epsilon_1[P_1] \texttt{ in } \mathcal{T}_{\S\overleftarrow{\Phi}}(S_2)$. Moreover, $\epsilon_1[P_1] \in \lambda\mathrm{LL}^{\mathrm{P}}$ and so $\mathcal{W}(\S\overleftarrow{\Phi}^{\text{in}}_{\epsilon_1[P_1]}) = 0$. The same reasoning can be applied to $\mathcal{W}(\S\overleftarrow{\Phi}^{\text{out}}_{\mathcal{T}_{\S\overleftarrow{\Phi}}(R)})$. Summing up we have that:

$$
\mathcal{W}(\S\overleftarrow{\Phi}^{\text{in}}_{\mathcal{T}_{\S\overleftarrow{\Phi}}(R)}) = \mathcal{W}(\S\overleftarrow{\Phi}^{\text{in}}_{\mathcal{T}_{\S\overleftarrow{\Phi}}(S_2)}) \quad \text{and} \quad \mathcal{W}(\S\overleftarrow{\Phi}^{\text{out}}_{\mathcal{T}_{\S\overleftarrow{\Phi}}(R)}) = \mathcal{W}(\S\overleftarrow{\Phi}^{\text{out}}_{\mathcal{T}_{\S\overleftarrow{\Phi}}(S_2)}) \tag{5.58}
$$

We can conclude as follows

$$
\begin{aligned}
&\mathcal{W}(\mathcal{T}_{\S\overleftarrow{\Phi}}(R)) + \mathcal{W}(L) + \mathcal{W}(\S\overleftarrow{\Phi}^{\text{in}}_{\mathcal{T}_{\S\overleftarrow{\Phi}}(R)}) \\
&\qquad = \mathcal{W}(\epsilon_1[P_1]) + \mathcal{W}(\mathcal{T}_{\S\overleftarrow{\Phi}}(S_2)) + \mathcal{W}(L) + \mathcal{W}(\S\overleftarrow{\Phi}^{\text{in}}_{\mathcal{T}_{\S\overleftarrow{\Phi}}(R)}) \\
&\qquad \overset{\text{Eq. 5.58}}{=} \mathcal{W}(\epsilon_1[P_1]) + \underline{\mathcal{W}(\mathcal{T}_{\S\overleftarrow{\Phi}}(S_2))} + \underline{\mathcal{W}(L)} + \underline{\mathcal{W}(\S\overleftarrow{\Phi}^{\text{in}}_{\mathcal{T}_{\S\overleftarrow{\Phi}}(S_2)})} \\
&\qquad \overset{\text{IH on } S_2}{\leq} \mathcal{W}(\epsilon_1[P_1]) + \mathcal{W}(S_2) + \mathcal{W}(H) + \mathcal{W}(\S\overleftarrow{\Phi}^{\text{out}}_{\mathcal{T}_{\S\overleftarrow{\Phi}}(S_2)}) \\
&\qquad \leq \mathcal{W}(R) + \mathcal{W}(H) + \mathcal{W}(\S\overleftarrow{\Phi}^{\text{out}}_{\mathcal{T}_{\S\overleftarrow{\Phi}}(R)}) \\
&\qquad = \mathcal{W}(S_1) + \mathcal{W}(S_2) + \mathcal{W}(L_0) + \mathcal{W}(H_0) + \mathcal{W}(H) + \mathcal{W}(\S\overleftarrow{\Phi}^{\text{out}}_{\mathcal{T}_{\S\overleftarrow{\Phi}}(R)}) \\
&\qquad \overset{\text{Eq. 5.58}}{=} \mathcal{W}(S_1) + \mathcal{W}(S_2) + \mathcal{W}(L_0) + \mathcal{W}(H_0) + \mathcal{W}(H) + \mathcal{W}(\S\overleftarrow{\Phi}^{\text{out}}_{\mathcal{T}_{\S\overleftarrow{\Phi}}(S_2)})
\end{aligned}
$$

because by Lemma 35 $\mathcal{W}(\epsilon_1[P_1]) \leq \mathcal{W}(S_1)$.

□

## Details Example Modularity

In this appendix we consider the term $N$ in Figure which is defined as:

$$
\begin{aligned}
N = \;& \texttt{let } (!y_1, \S f_1) = \mathcal{F}_{x:!\mathbb{R}}(Q_1) \texttt{ in} \\
& \texttt{let } (!y_2, \S f_2) = \mathcal{F}_{x:!\mathbb{R}}(Q_2) \texttt{ in} \\
& (!y_1 \underline{*} !y_2, \S(\lambda u^{\mathbb{R}}.\texttt{let } \langle u_1, u_2\rangle = \langle u, u\rangle \texttt{ in } (y_2 \dot{*} (f_1 u_1)) \dot{+} (y_1 \dot{*} (f_2 u_2))))
\end{aligned}
$$

In order to make the example clearer, we perform the transpose step by step on the tangent part $\S(\lambda u^{\mathbb{R}}.\texttt{let } \langle u_1, u_2\rangle = \langle u, u\rangle \texttt{ in } (y_2 \dot{*} (f_1 u_1)) \dot{+} (y_1 \dot{*} (f_2 u_2)))$, underlining at each step the subterm that we are going to analyze.

Let $U = \texttt{let } \langle u_1, u_2\rangle = \langle u, u\rangle \texttt{ in } (y_2 \dot{*} (f_1 u_1)) \dot{+} (y_1 \dot{*} (f_2 u_2))$, we analyze $\mathcal{T}_{\S\overleftarrow{\Phi}, \S\overleftarrow{f_1}, \S\overleftarrow{f_2}}(\lambda u^{\mathbb{R}}.U)$ by applying case $\lambda p^{\&}.U$ of Figure 5.4a and we simplify it as follows

$$
\begin{aligned}
\mathcal{T}_{\S\overleftarrow{\Phi}, \S\overleftarrow{f_1}, \S\overleftarrow{f_2}}(\lambda u^{\mathbb{R}}.U) &= \lambda h^{\mathbb{R}}.\mu_{u,\alpha,\emptyset}\langle \mathcal{T}_{\S\overleftarrow{\Phi}, \S\overleftarrow{f_1}, \S\overleftarrow{f_2}, u}(U), \langle\,\rangle\rangle \\
&\qquad \text{where } \mathrm{Dom}(\alpha) = \{u\} \text{ and } \alpha(u) = z. \\
&\overset{\text{Eq. 5.29}}{=} \lambda h^{\mathbb{R}}.(\lambda\langle \alpha\langle u\rangle, \emptyset\langle u\rangle\rangle.\nu(u, \alpha, \emptyset))\langle \mathcal{T}_{\S\overleftarrow{\Phi}, \S\overleftarrow{f_1}, \S\overleftarrow{f_2}, u}(U), \langle\,\rangle\rangle \\
&\overset{\text{Def. } \alpha\langle p^{\&}\rangle}{=} \lambda h^{\mathbb{R}}.(\lambda\langle \alpha(u), t\rangle.\nu(u, \alpha, \emptyset))\langle \mathcal{T}_{\S\overleftarrow{\Phi}, \S\overleftarrow{f_1}, \S\overleftarrow{f_2}, u}(U), \langle\,\rangle\rangle \\
&\qquad \text{where } t \text{ is a fresh variable of type } \top. \\
&\overset{\text{Def. } \nu(p^{\&}, \alpha_1, \alpha_2)}{=} \lambda h^{\mathbb{R}}.(\lambda\langle \alpha(u), t\rangle.\alpha(u))\langle \mathcal{T}_{\S\overleftarrow{\Phi}, \S\overleftarrow{f_1}, \S\overleftarrow{f_2}, u}(U), \langle\,\rangle\rangle \\
&\overset{\alpha(u) = z}{=} \lambda h^{\mathbb{R}}.(\lambda\langle z, t\rangle.z)\langle \underline{\mathcal{T}_{\S\overleftarrow{\Phi}, \S\overleftarrow{f_1}, \S\overleftarrow{f_2}, u}(U)}, \langle\,\rangle\rangle
\end{aligned}
\tag{5.59}
$$

Now we want to analyze $\mathcal{T}_{\S\overleftarrow{\Phi}, \S\overleftarrow{f_1}, \S\overleftarrow{f_2}, u}(U)$. Observe that by notational conventions defined in Section 3.1 we have

$$
\begin{aligned}
U &= \texttt{let } \langle u_1, u_2\rangle = \langle u, u\rangle \texttt{ in } (y_2 \dot{*} (f_1 u_1)) \dot{+} (y_1 \dot{*} (f_2 u_2)) \\
&\approx (\underbrace{\lambda\langle u_1, u_2\rangle.(y_2 \dot{*} (f_1 u_1)) \dot{+} (y_1 \dot{*} (f_2 u_2))}_{F})\underbrace{\langle u, u\rangle}_{U'}
\end{aligned}
$$

Therefore we have $\mathcal{T}_{\overleftarrow{\S\Phi},\overleftarrow{\S f_1},\overleftarrow{\S f_2},u}(U) = \mathcal{T}_{\overleftarrow{\S\Phi},\overleftarrow{\S f_1},\overleftarrow{\S f_2},u}(FU')$ and we apply case $FU'$ of Figure 5.4b obtaining

$$\mathcal{T}_{\overleftarrow{\S\Phi},\overleftarrow{\S f_1},\overleftarrow{\S f_2},u}(FU') = (\lambda\langle v_1, v_2\rangle.\underline{\mathcal{T}_{\overleftarrow{\S\Phi},\overleftarrow{\S f_1},\overleftarrow{\S f_2},u}(U')})(\underline{\mathcal{T}_{\overleftarrow{\S\Phi},\overleftarrow{\S f_1},\overleftarrow{\S f_2}}(F)h}) \tag{5.60}$$

First, we proceed by analyzing $\mathcal{T}_{\overleftarrow{\S\Phi},\overleftarrow{\S f_1},\overleftarrow{\S f_2},u}(U')$. More precisely, we apply case $\langle U_1, U_2\rangle$ of Figure 5.4b and we simplify it as follows

$$\begin{aligned}
\mathcal{T}_{\overleftarrow{\S\Phi},\overleftarrow{\S f_1},\overleftarrow{\S f_2},u}(U') &= \mathcal{T}_{\overleftarrow{\S\Phi},\overleftarrow{\S f_1},\overleftarrow{\S f_2},u}(\langle u, u\rangle) \\
&\overset{\text{Fig. 5.4b}}{=} \mu_{u,\alpha_1,\alpha_2}\langle \mathcal{T}_{\overleftarrow{\S\Phi},\overleftarrow{\S f_1},\overleftarrow{\S f_2},\alpha_1[u]}(\alpha_1[u]), \mathcal{T}_{\overleftarrow{\S\Phi},\overleftarrow{\S f_1},\overleftarrow{\S f_2},\alpha_2[u]}(\alpha_2[u])\rangle \\
&\quad \text{where } \mathrm{Dom}(\alpha_1) = \mathrm{Dom}(\alpha_2) = \{u\} \text{ and } \alpha_i(u) = v_i \text{ with } i \in \{1,2\}. \\
&\overset{\text{Eq. 5.29}}{=} (\lambda\langle \alpha_1\langle u\rangle, \alpha_2\langle u\rangle\rangle.\nu(u,\alpha_1,\alpha_2))\langle \mathcal{T}_{\overleftarrow{\S\Phi},\overleftarrow{\S f_1},\overleftarrow{\S f_2},\alpha_1[u]}(\alpha_1[u]), \mathcal{T}_{\overleftarrow{\S\Phi},\overleftarrow{\S f_1},\overleftarrow{\S f_2},\alpha_2[u]}(\alpha_2[u])\rangle \\
&\overset{\text{Def. }\alpha\langle p^{\&}\rangle}{=} (\lambda\langle \alpha_1(u), \alpha_2(u)\rangle.\nu(u,\alpha_1,\alpha_2))\langle \mathcal{T}_{\overleftarrow{\S\Phi},\overleftarrow{\S f_1},\overleftarrow{\S f_2},\alpha_1[u]}(\alpha_1[u]), \mathcal{T}_{\overleftarrow{\S\Phi},\overleftarrow{\S f_1},\overleftarrow{\S f_2},\alpha_2[u]}(\alpha_2[u])\rangle \\
&\overset{\alpha_i(u) = v_i}{=} (\lambda\langle v_1, v_2\rangle.\nu(u,\alpha_1,\alpha_2))\langle \mathcal{T}_{\overleftarrow{\S\Phi},\overleftarrow{\S f_1},\overleftarrow{\S f_2},\alpha_1[u]}(\alpha_1[u]), \mathcal{T}_{\overleftarrow{\S\Phi},\overleftarrow{\S f_1},\overleftarrow{\S f_2},\alpha_2[u]}(\alpha_2[u])\rangle \\
&\overset{\text{Def. }\nu(p^{\&},\alpha_1,\alpha_2)}{=} (\lambda\langle v_1, v_2\rangle.v_1 \dot{+} v_2)\langle \mathcal{T}_{\overleftarrow{\S\Phi},\overleftarrow{\S f_1},\overleftarrow{\S f_2},\alpha_1[u]}(\alpha_1[u]), \mathcal{T}_{\overleftarrow{\S\Phi},\overleftarrow{\S f_1},\overleftarrow{\S f_2},\alpha_2[u]}(\alpha_2[u])\rangle \\
&\overset{\alpha\text{-renaming}}{=} (\lambda\langle v_1', v_2'\rangle.v_1' \dot{+} v_2')\langle \mathcal{T}_{\overleftarrow{\S\Phi},\overleftarrow{\S f_1},\overleftarrow{\S f_2},\alpha_1[u]}(\alpha_1[u]), \mathcal{T}_{\overleftarrow{\S\Phi},\overleftarrow{\S f_1},\overleftarrow{\S f_2},\alpha_2[u]}(\alpha_2[u])\rangle \\
&\overset{\text{Def. }\alpha[p^{\&}]}{=} (\lambda\langle v_1', v_2'\rangle.v_1' \dot{+} v_2')\langle \mathcal{T}_{\overleftarrow{\S\Phi},\overleftarrow{\S f_1},\overleftarrow{\S f_2},u}(\alpha_1[u]), \mathcal{T}_{\overleftarrow{\S\Phi},\overleftarrow{\S f_1},\overleftarrow{\S f_2},u}(\alpha_2[u])\rangle \\
&\overset{\text{Def. }\alpha[M]}{=} (\lambda\langle v_1', v_2'\rangle.v_1' \dot{+} v_2')\langle \mathcal{T}_{\overleftarrow{\S\Phi},\overleftarrow{\S f_1},\overleftarrow{\S f_2},u}(v_1), \mathcal{T}_{\overleftarrow{\S\Phi},\overleftarrow{\S f_1},\overleftarrow{\S f_2},u}(v_2)\rangle \\
&\overset{\text{Fig. 5.4b}}{=} (\lambda\langle v_1', v_2'\rangle.v_1' \dot{+} v_2')\langle v_1, v_2\rangle \\
&\to_\beta \; v_1 \dot{+} v_2
\end{aligned} \tag{5.61}$$

Finally, we analyze $\mathcal{T}_{\overleftarrow{\S\Phi},\overleftarrow{\S f_1},\overleftarrow{\S f_2}}(F)$ where $F$ is equal to $\lambda\langle u_1, u_2\rangle.(y_2 \dot{*} (f_1 u_1)) \dot{+} (y_1 \dot{*} (f_2 u_2))$. Notice that by notational conventions defined in Section 3.1 we have

$$\begin{aligned}
F \approx \lambda\langle u_1, u_2\rangle.&\texttt{let } j_1 = f_1 u_1 \texttt{ in let } j_2 = f_2 u_2 \texttt{ in} \\
&\texttt{let } s_1 = y_2 \dot{*} j_1 \texttt{ in let } s_1 = y_1 \dot{*} j_2 \texttt{ in} \\
&s_1 \dot{+} s_2
\end{aligned}$$

After many steps of transpose, $\beta$-steps and simplifications we get

$$\lambda d.(\lambda\langle w_1, w_2\rangle.\langle y_2 \dot{*} (\overleftarrow{f_1} w_1), y_1 \dot{*} (\overleftarrow{f_2} w_2)\rangle)\langle d, d\rangle \tag{5.62}$$

Summing up the tangent part of the last line in $\mathcal{T}(N)$ is equal to

$$\begin{aligned}
\mathcal{T}_{\overleftarrow{\S\Phi},\overleftarrow{\S f_1},\overleftarrow{\S f_2}}(\lambda u^{\mathbb{R}}.U) &\overset{\text{Eq. 5.59}}{=} \lambda h^{\mathbb{R}}.(\lambda\langle z, t\rangle.z)\langle \underline{\mathcal{T}_{\overleftarrow{\S\Phi},\overleftarrow{\S f_1},\overleftarrow{\S f_2},u}(U)}, \langle\,\rangle\rangle \\
&= \lambda h^{\mathbb{R}}.(\lambda\langle z, t\rangle.z)\langle \underline{\mathcal{T}_{\overleftarrow{\S\Phi},\overleftarrow{\S f_1},\overleftarrow{\S f_2},u}(FU')}, \langle\,\rangle\rangle \\
&\overset{\text{Eq. 5.60}}{=} \lambda h^{\mathbb{R}}.(\lambda\langle z, t\rangle.z)\langle(\lambda\langle v_1, v_2\rangle.\underline{\mathcal{T}_{\overleftarrow{\S\Phi},\overleftarrow{\S f_1},\overleftarrow{\S f_2},u}(U')})(\mathcal{T}_{\overleftarrow{\S\Phi},\overleftarrow{\S f_1},\overleftarrow{\S f_2}}(F)h), \langle\,\rangle\rangle \\
&\overset{\text{Eq. 5.61}}{=} \lambda h^{\mathbb{R}}.(\lambda\langle z, t\rangle.z)\langle(\lambda\langle v_1, v_2\rangle.v_1 \dot{+} v_2)(\underline{\mathcal{T}_{\overleftarrow{\S\Phi},\overleftarrow{\S f_1},\overleftarrow{\S f_2}}(F)h}), \langle\,\rangle\rangle \\
&\overset{\text{Eq. 5.62}}{=} \lambda h^{\mathbb{R}}.(\lambda\langle z, t\rangle.z)\langle(\lambda\langle v_1, v_2\rangle.v_1 \dot{+} v_2)((\lambda d.(\lambda\langle w_1, w_2\rangle.\langle y_2 \dot{*} (\overleftarrow{f_1} w_1), y_1 \dot{*} (\overleftarrow{f_2} w_2)\rangle)\langle d, d\rangle)h), \langle\,\rangle\rangle
\end{aligned}$$

For readability we can simplify the last line above as

$$\lambda h^{\mathbb{R}}.\texttt{let}\ \langle w_1, w_2\rangle = \langle h, h\rangle\ \texttt{in}\ (y_2 \dot{*} (\overleftarrow{f_1} w_1))\ \dot{+}\ (y_1 \dot{*} (\overleftarrow{f_2} w_2))$$

and we obtain the term in Figure 5.8d.

Chapter 6

# Quantitative Framework of $\lambda$LL

Let us recall the definition of safe term exactly as given in Section 3.5:

**Definition 4** (Safe Term)**.** A term $M$ is *safe* if:

(i) for any subterm $!M'$ in $M$, $\mathcal{W}(M') = 0$;

(ii) for any subterm $\langle M_1, M_2 \rangle$ in $M$, $FV(M_1) \cap FV(M_2)$ has only ground variables.

As outlined in Remark 5 of Section 3.5, condition (ii) in the definition above is very technical and plays a central role in proving that a safe closed term $M$ normalizes through safe reduction in at most $\mathcal{W}(M)$ steps (see Proposition 3). This condition is essential because it limits additive duplication of subterms with higher order type. However, it is possible to omit this condition by shifting focus from the syntactic structure of terms to their type derivations. Specifically, we can define the notion of workload directly on the type derivation of a term $M$ using a quantitative type system which carries how many times a subterm will be additively duplicated. This approach enables a more refined control over duplication, particularly of variables associated with function types between &-sequence types, without relying on the syntactic restriction in condition (ii).

In this chapter, we pursue this alternative path. We begin by presenting the definition of the quantitative type system which take carefully care of duplication of variables, particularly those related to function between &-sequence types, while analyzing the workload of the typed term. This approach allows us to revisit the definition of a safe term, as presented in Definition 6 below, which turns out to be more general than the one given in Definition 4. In Section 6.2, we present the quantitative version of Subject Reduction and we explore the corresponding adjustments to the cost model of $\lambda$LL within the quantitative setting. Finally, in Section 6.3 we provide quantitative soundness for the AD System of $\lambda$LL.

## 6.1 Quantitative Type System

Let us recall the terminology introduced in Section 3.5: a variable is *ground* if its type has no arrow, otherwise it is *higher-order*. We define a decoration of the type system such that it takes into account how many times an higher order variable is duplicated by means of an additive duplication. More precisely, the decorated types are defined by the following grammar

$$\mathcal{A}, \mathcal{B}, \mathcal{C} ::= \mathbb{R} \mid \mathcal{A} \overset{\mathrm{k}}{\multimap} \mathcal{B} \mid \mathtt{1} \mid \mathcal{A} \otimes \mathcal{B} \mid \top \mid \mathcal{A} \& \mathcal{B} \mid !\mathcal{A} \qquad \text{(Decorated Types)}$$

where $k \in \mathbb{N}^{>0}$ in the linear implication indicates that the input, or part of it, may be subjected to at most $k$ additive duplications. In order to relate decorated types to their undecorated counterparts, we define a forgetful map which erases all quantitative annotations. This map is defined inductively over the structure of decorated types and yields the corresponding standard type by systematically discarding the duplication bounds as follows:

$$|\mathbb{R}| = \mathbb{R}\,, \qquad |\mathtt{1}| = \mathtt{1}\,, \qquad |\top| = \top\,, \qquad |!\mathcal{A}| = !|\mathcal{A}|\,,$$
$$|\mathcal{A} \otimes \mathcal{B}| = |\mathcal{A}| \otimes |\mathcal{B}|\,, \quad |\mathcal{A} \& \mathcal{B}| = |\mathcal{A}| \& |\mathcal{B}|\,,$$
$$|\mathcal{A} \overset{\mathrm{k}}{\multimap} \mathcal{B}| = |\mathcal{A}| \multimap |\mathcal{B}|.$$

We define a decoration of a typing environment $\Delta$ as a function that assigns to each pattern $p$ of type $A$ in $\Delta$ a pair consisting of an *additive degree* $k \in \mathbb{N}^{>0}$ and a *decorated type* $\mathcal{A}$, satisfying $|\mathcal{A}| = A$, where $|\cdot|$ denotes the erasure of decorations defined above. The additive degree $k$ provides a upper bound on the number of times that at least one variable occurring freely in $p$, i.e., in $FV(p)$, is duplicated via additive duplication. This assignment captures, at the level of the typing environment, the quantitative behaviour of variable additive duplication.

**Definition 5** (Decoration of Typing Environment)**.** Given a type environment $\Delta$, we define $\mathcal{D}(\Delta)$ as the following function

$$\mathcal{D}(\Delta) = p : A \in \Delta \mapsto (k, \mathcal{A}) \text{ s.t. } k \in \mathbb{N}^{>0}, |\mathcal{A}| = A \text{ if } A = !A' \text{ or ground type, then } k = 1.$$

We will write $p^k : \mathcal{A}$ as an alternative notation for $p : (k, \mathcal{A})$.

Two operations on decorated typing environments are introduced to manipulate the additive degrees associated with variable patterns. Let $\mathcal{D}(\Delta)$ denote a decorated typing environment. Given a natural number $k \in \mathbb{N}^{>0}$, we define the scalar multiplication of $k$ with a decorated typing environment $\mathcal{D}(\Delta)$, denoted by $k \circledast \mathcal{D}(\Delta)$, as follows:

$$k \circledast \mathcal{D}(\Delta) = p : A \in \Delta \mapsto \begin{cases} (1, \mathcal{A}) & \text{if } A = !A' \text{ or ground type} \\ (k * k', \mathcal{A}) & \text{otherwise let } \mathcal{D}(\Delta)(p) = (k', \mathcal{A}) \end{cases}$$

This operation scales the additive degree of each variable by a factor of $k$, except in the case of types that are structurally immune to additive duplication, namely ! modality, $\mathbb{R}$, and $\mathtt{1}$, for which the additive degree remains fixed at 1.

Given two decorated typing environments $\mathcal{D}_1(\Delta)$ and $\mathcal{D}_2(\Delta)$ over the same typing environment $\Delta$, we define their pointwise addition, denoted by $\mathcal{D}_1(\Delta) \boxplus \mathcal{D}_2(\Delta)$, as follows:

$$\mathcal{D}_1(\Delta) \boxplus \mathcal{D}_2(\Delta) = p : A \in \Delta \mapsto \begin{cases} (1, \mathcal{A}) & \text{if } A = !A' \text{ or ground type} \\ (k_1 + k_2, \mathcal{A}) & \text{otherwise let } \mathcal{D}_i(\Delta)(p) = (k_i, \mathcal{A}) \text{ for } i \in \{1, 2\} \end{cases}$$

This operation aggregates the additive degrees of variables across two decorated environments, consistently assigning an additive degree of 1 to ground types and the ! modality, reflecting their exemption from additive duplication.

In addition, we enrich the typing judgment with a quantitative annotation in the form of a natural number $m \in \mathbb{N}^{\geq 0}$, which we refer to as the *workload* of the typed term. This annotation keeps track of the number of numerical operation not under the scope of a bang modality as well as the number of possible numerals erased during the reduction of the typed term.

By decorating typing judgments with such a workload annotation, we obtain a fine-grained, type-based mechanism for statically estimating the cost of a term in λLL. This approach enables the development of a quantitative linear type system that not only enforces usage constraints on variables but also provides upper bounds on the cost of evaluation, thereby supporting cost-sensitive reasoning within the logical framework.

Summing up, the decorated typing judgment takes the form

$$\mathcal{D}(\Delta) \vdash^{m} M : \mathcal{A}$$

where

- $M$ is the typed term.
- $\mathcal{D}(\Delta)$ is a decoration of the typing environment which takes into account how many times a pattern $p$ in $\Delta$ or a part of it is duplicated by means of an additive duplication, by associating to each pattern a number $k \in \mathbb{N}^{>0}$.
- $m \in \mathbb{N}^{\geq 0}$ is the workload which keeps track of the number of numerical operation not under the scope of a bang as well as the number of possible numerals erased during the reduction of the typed term.
- $\mathcal{A}$ is a decorated type in the grammar Decorated Types.

The quantitative type system is given in Figure 6.1. In the weakening rule $!_w$ the workload is equal to zero because we do not consider the number of numerical operation under the scope of a bang.

The typing rules governing the linear implication connective require particular care due to their interaction with variable duplication. In the introduction rule for linear implication, denoted $\multimap_i$, the abstraction $\lambda p.M$ is assigned the decorated type $\mathcal{A} \overset{k}{\multimap} \mathcal{B}$, where the pattern $p$ is annotated with an additive degree of $k$. This annotation enforces a bound on the number of additive duplications permitted for the variables introduced by $p$ during the evaluation of the term. Moreover, the cost associated with the abstraction is not determined solely by the cost $m$ of deriving the type of the body $M$. It must also account for the erasure of variables of type $\mathbb{R}$ that are introduced by the pattern $p$ but do not occur freely in $M$. These variables are instantiated during term application and subsequently discarded, incurring a cost due to erasure. To capture this additional cost, we define the total workload of the abstraction as the sum of the body's cost $m$ and the erasure cost $\mathcal{W}(FV(p) \setminus FV(M))$. Here, $FV(p)$ denotes the set of variables bound by the pattern $p$, and the cost of the difference $FV(p) \setminus FV(M)$ identifies those variables of type $\mathbb{R}$ that are unused in the body of the abstraction and hence contribute to the reduction cost through erasure. The elimination rule $\multimap_e$ describes how the types and the workloads of the function term $M$ and its argument $N$ combine to determine the type and the workload of the result. Formally, suppose we have a function term $M$ with decorated type $\mathcal{A} \overset{k}{\multimap} \mathcal{B}$ and workload $m_1$, and an argument term $N$ with decorated type $\mathcal{A}$ and workload $m_2$. Then, the application $MN$ is assigned the decorated type $\mathcal{B}$, and the resulting workload is given by $m_1 + k \cdot m_2$. This expression reflects the cost incurred by the potential additive duplication of the argument $N$ up to $k$ times, as permitted by the type of the function term. In particular, if $N$ contains numerical operations not guarded by a bang, then these may be duplicated in each application instance, justifying the multiplicative cost. Additionally, in order to preserve accurate accounting of variable usage, the additive degrees of the patterns in the decorated environment typing $N$ must also be scaled accordingly. That is, each pattern in the environment $\Delta_2$ that types the argument $N$ must have its additive degree multiplied by $k$. This transformation is captured formally by the operation: $k \circledast \mathcal{D}(\Delta_2)$, which appears in the conclusion of the $\multimap_e$ rule. This ensures that the total duplication impact of the argument is correctly reflected in the environment of the resulting term.

$$\frac{}{x^k : \mathcal{A} \vdash^{0} x : \mathcal{A}}\ v \qquad \frac{}{!x^1 : !\mathcal{A} \vdash^{m} x : \mathcal{A}}\ !_e$$

$$\frac{\mathcal{D}(!\Gamma) \vdash^{m} M : \mathcal{A}}{\mathcal{D}(!\Gamma) \vdash^{m} !M : !\mathcal{A}}\ !_i \qquad \frac{\mathcal{D}(\Delta) \vdash^{m} M : \mathcal{B}}{!x^1 : !\mathcal{A}, \mathcal{D}(\Delta) \vdash^{0} M : \mathcal{B}}\ !_w$$

$$\frac{p^k : \mathcal{A}, \mathcal{D}(\Delta) \vdash^{m} M : \mathcal{B}}{\mathcal{D}(\Delta) \vdash^{m+\mathcal{W}(FV(p)\setminus FV(M))} \lambda p.M : \mathcal{A} \overset{\mathrm{k}}{\multimap} \mathcal{B}}\ \multimap_i$$

$$\frac{\mathcal{D}(!\Gamma_1), \mathcal{D}(\Delta_1) \vdash^{m_1} M : \mathcal{A} \overset{\mathrm{k}}{\multimap} \mathcal{B} \qquad \mathcal{D}(!\Gamma_2), \mathcal{D}(\Delta_2) \vdash^{m_2} N : A}{\mathcal{D}(!\Gamma_1) \cup \mathcal{D}(!\Gamma_2), \mathcal{D}(\Delta_1), k \circledast \mathcal{D}(\Delta_2) \vdash^{m_1+k*m_2} MN : \mathcal{B}}\ \multimap_e$$

$$\frac{}{\vdash^{0} (\,) : \mathtt{1}}\ 1_i \qquad \frac{\mathcal{D}(\Delta) \vdash^{m} M : \mathcal{B}}{(\,)^1 : \mathtt{1}, \mathcal{D}(\Delta) \vdash^{m} M : \mathcal{B}}\ 1_e$$

$$\frac{p^{k_1} : \mathcal{A}, q^{k_2} : \mathcal{B}, \mathcal{D}(\Delta) \vdash^{m} M : \mathcal{C}}{(p,q)^{max(k_1,k_2)} : \mathcal{A} \otimes \mathcal{B}, \mathcal{D}(\Delta) \vdash^{m} M : \mathcal{C}}\ \otimes_e$$

$$\frac{\mathcal{D}(!\Gamma_1), \mathcal{D}(\Delta_1) \vdash^{m_2} M : \mathcal{A} \qquad \mathcal{D}(!\Gamma_2), \mathcal{D}(\Delta_2) \vdash^{m_1} N : \mathcal{B}}{\mathcal{D}(!\Gamma_1) \cup \mathcal{D}(!\Gamma_2), \mathcal{D}(\Delta_1), \mathcal{D}(\Delta_2) \vdash^{m_1+m_2} (M,N) : \mathcal{A} \otimes \mathcal{B}}\ \otimes_i$$

$$\frac{\mathcal{D}_1(\Delta) \vdash^{m_1} M_1 : \mathcal{A}_1 \qquad \mathcal{D}_2(\Delta) \vdash^{m_2} M_2 : \mathcal{A}_2}{\mathcal{D}_1(\Delta) \boxplus \mathcal{D}_2(\Delta) \vdash^{m_1+m_2} \langle M_1, M_2 \rangle : \mathcal{A}_1 \mathbin{\&} \mathcal{A}_2}\ \&_i$$

$$\frac{p_i^k : \mathcal{A}_i, \mathcal{D}(\Delta) \vdash^{m} M : \mathcal{B}}{\langle p_1, p_2 \rangle^k : \mathcal{A}_1 \mathbin{\&} \mathcal{A}_2, \mathcal{D}(\Delta) \vdash^{m} M : \mathcal{B}}\ \&_{ei},\ i \in \{1,2\} \qquad \frac{}{\mathcal{D}(\Delta) \vdash^{0} \langle\,\rangle : \top}\ \top$$

$$\frac{r \text{ real number}}{\vdash^{0} \underline{r} : \mathbb{R}}\ R \qquad \frac{}{\mathcal{D}(\Delta) \vdash^{0} \underline{0} : \mathbb{R}}\ Z$$

$$\frac{f \text{ binary map}}{\vdash^{1} \underline{f} : !\mathbb{R} \otimes !\mathbb{R} \overset{1}{\multimap} !\mathbb{R}}\ F_2 \qquad \frac{}{\vdash^{1} \dot{+} : \mathbb{R} \mathbin{\&} \mathbb{R} \overset{1}{\multimap} \mathbb{R}}\ S \qquad \frac{}{\vdash^{1} \dot{*} : \mathbb{R} \overset{1}{\multimap} \mathbb{R} \overset{1}{\multimap} \mathbb{R}}\ M$$

Figure 6.1: Quantitative Type System of λLL.

In the typing rule for additive conjunction, denoted $\&_i$, we analyze the two components of the conjunction independently, each under its own decorated typing environment. Specifically, we consider two decorations, $\mathcal{D}_1$ and $\mathcal{D}_2$, of the same underlying environment $\Delta$, corresponding respectively to the derivations of the left and right components of the additive pair. In the conclusion of the rule, the decoration of the environment reflects the combined usage of each pattern across both branches. For each pattern $p \in \Delta$, the additive degree assigned in the conclusion is obtained by summing the degrees assigned to $p$ in $\mathcal{D}_1$ and $\mathcal{D}_2$, respectively. This aggregation is formally expressed by the pointwise addition operation: $\mathcal{D}_1(\Delta) \boxplus \mathcal{D}_2(\Delta)$, which yields a new decoration of $\Delta$ where each pattern's degree reflects its total duplication across both components of the conjunction. This construction ensures that the cost analysis accurately accounts for the cumulative impact of variable usage in both branches of the additive conjunction.

Finally, we are ready to revise the definition of safe term as follows

**Definition 6** (Safe Term Quantitative)**.** A term $M$ is safe if any subterm $!M'$ in $M$ is such that $\mathcal{D}(\Delta) \vdash^0 M' : \mathcal{A}$.

## 6.2 Subject Reduction and Cost Model

Subject reduction is a fundamental property of type systems, ensuring that the type of a term is preserved under reduction. In a non-quantitative setting, subject reduction guarantees that if a term is well-typed and it reduces to another term, the resulting term remains well-typed under the same type. In a quantitative setting, however, subject reduction not only considers the preservation of types but also tracks workload associated with the terms.

Our calculus typed as in Figure 6.1 enjoys a quantitative version of subject reduction (Theorem 13) which ensures that the term's type remains consistent across *safe reductions* and that the workload strictly decreases or remains unchanged at each evaluation step. Quantitative Subject Reduction is proved by means of a quantitative pattern substitution lemma (Lemma 44).

**Remark 10.** It is important to observe that the quantitative version of subject reduction (Theorem 13) is restricted to safe reduction, namely the *call-by-strong value strategy* introduced in Section 3.2. This restriction is required since the workload of a safe term may increase when reduced using general $\beta$-reduction. For example, consider a safe term $M = (\lambda x.\langle x, x\rangle)(\dot{+}\langle \underline{3}, \underline{2}\rangle)$ with the judgment $\vdash^1 M : \mathbb{R}\&\mathbb{R}$. The argument $\dot{+}\langle \underline{3}, \underline{2}\rangle$ is not a strong value. If we reduce $M$ using standard $\beta$-reduction, we obtain $M \xrightarrow{\beta_\lambda} N = \langle \dot{+}\langle \underline{3}, \underline{2}\rangle, \dot{+}\langle \underline{3}, \underline{2}\rangle\rangle$. The term $N$ is still well-typed, with a derivation $\vdash^2 N : \mathbb{R}\&\mathbb{R}$. However, the workload has increased: while the workload of $M$ is 1, the workload of $N$ is 2.

In the quantitative type system we embed the reasoning about the workload of a term within its type derivation while maintaining the syntax of λLL unaltered. As a consequence, in order to express the statement of pattern substitution lemma in the quantitative setting we have to merge the statement of the pattern substitution lemma discussed in Subsection 3.2.1 (see Lemma 3) and the statement of the safe substitution lemma (see Lemma 21 given in Section 3.5).

Since the quantitative version of subject reduction is restricted to $s$-reduction, as discussed in Remark 10, the quantitative version of the pattern substitution lemma (Lemma 44) must also be restricted to strong values. This is in contrast to Lemma 3, which considers generic values.

**Remark 11.** It is worth noting that Lemma 44 does not require the strong value $W$ to be closed. As discussed in Remark 4 for Lemma 21, closure is necessary in the non-quantitative setting only to ensure the safeness of $M\{W/p\}$, due to condition (ii) of Definition 4. However, in the quantitative setting, this condition has been removed from the definition of safe terms (Definition 6), and as a result, the closure of $W$ is no longer required.

We prove the quantitative pattern substitution lemma (Lemma 44) using Lemma 2 and the following auxiliary lemma

**Lemma 43.** If $\mathcal{D}(\Delta) \vdash^m M : \mathcal{A}$, then $FV(M) \subseteq FV(\mathcal{D}(\Delta))$.

*Sketch Proof.* By induction on a derivation of $\mathcal{D}(\Delta) \vdash^m M : \mathcal{A}$. □

**Lemma 44** (Quantitative Pattern Substitution)**.** Given two derivable judgements $\mathcal{D}(!\Gamma_1), \mathcal{D}(\Delta_1), p^k : \mathcal{A} \vdash^m M : \mathcal{B}$ and $\mathcal{D}(!\Gamma_2), \mathcal{D}(\Delta_2) \vdash^n W : \mathcal{A}$ such that

(a) $W$ is a strong value for the pattern $p$,

(b) $FV(\mathcal{D}(!\Gamma_i, \Delta_i)) \cap FV(\mathcal{D}(\Delta_{3-i})) = \emptyset$ for $i \in \{1, 2\}$,

then we have that the judgement $\mathcal{D}(!\Gamma_1) \cup \mathcal{D}(!\Gamma_2), \mathcal{D}(\Delta_1), k \circledast \mathcal{D}(\Delta_2) \vdash^{m'} M\{W/p\} : \mathcal{B}$ is derivable. Moreover, if $M$ and $W$ are safe we have

1. $M\{W/p\}$ is a safe term (according to Definition 6)
2. $m' \leq k * n + m$

*Proof.* For any derivation $\Pi_1$ for $\mathcal{D}(!\Gamma_1), \mathcal{D}(\Delta_1), p^k : \mathcal{A} \vdash^m M : \mathcal{B}$ and $\Pi_2$ for $\mathcal{D}(!\Gamma_2), \mathcal{D}(\Delta_2) \vdash^n W : \mathcal{A}$, we give a derivation $\mathcal{D}(!\Gamma_1) \cup \mathcal{D}(!\Gamma_2), \mathcal{D}(\Delta_1), k \circledast \mathcal{D}(\Delta_2) \vdash^{m'} M\{W/p\} : \mathcal{B}$ by induction on the lexicographically ordered pair $(s(\Pi_2), s(\Pi_1))$. We split depending on the last derivation rule in $\Pi_1$ or $\Pi_2$.

- If the last rule of $\Pi_1$ is a rule $r$ among $\{!_w, \&_{ei}, \otimes_e, 1_e\}$ acting on a pattern in $\mathcal{D}(!\Gamma_1), \mathcal{D}(\Delta_1)$, then the immediate subderivation of $\Pi_1$ is $\Pi_1'$ of $\mathcal{D}(!\Gamma_1'), \mathcal{D}(\Delta_1'), p^k : \mathcal{A} \vdash^m M : \mathcal{B}$ for suitable $\mathcal{D}(!\Gamma_1'), \mathcal{D}(\Delta_1')$ such that $FV(\mathcal{D}(\Delta_1')) \subseteq FV(\mathcal{D}(\Delta_1))$ and $FV(\mathcal{D}(!\Gamma_1'), \mathcal{D}(\Delta_1')) \subseteq FV(\mathcal{D}(!\Gamma_1), \mathcal{D}(\Delta_1))$.

  We then have: $FV(\mathcal{D}(!\Gamma_1'), \mathcal{D}(\Delta_1')) \cap FV(\mathcal{D}(\Delta_2)) \subseteq FV(\mathcal{D}(!\Gamma_1), \mathcal{D}(\Delta_1)) \cap FV(\mathcal{D}(\Delta_2))$ where the latter is equal to $\emptyset$ by item b of the hypothesis.

  We can apply the induction hypothesis on $(s(\Pi_1'), s(\Pi_2))$, obtaining

  – A derivation for $\mathcal{D}(!\Gamma_1') \cup \mathcal{D}(!\Gamma_2), \mathcal{D}(\Delta_1'), k \circledast \mathcal{D}(\Delta_2) \vdash^{m'} M\{W/p\} : \mathcal{B}$.
  – If $M$ and $W$ are safe we have also that
    1. $M\{W/p\}$ is a safe term
    2. $m' \leq k * n + m$

- The cases where the last rule of $\Pi_2$ is a rule $r$ among $\{!_w, \&_{ei}, \otimes_e, 1_e\}$ acting on a pattern in $\mathcal{D}(!\Gamma_2), \mathcal{D}(\Delta_2)$ are analogous to the previous case.

- For the other cases we can suppose that the last rules of $\Pi_1$ and $\Pi_2$ are not rules acting on $\mathcal{D}(!\Gamma_i), \mathcal{D}(\Delta_i)$ for $1 \leq i \leq 2$. We then split in further subcases depending if the last rule of $\Pi_1$ acts on the pattern $p : A$ or acts on the term $M$.

  Let us consider first the cases related to the last rule $r$ of $\Pi_1$ acting on the pattern $p^k : \mathcal{A}$.

  – If $r$ is $!_w$, then $p = !x$ and $W = !W'$. Moreover, $k = 1$. By cases inspection, one can infer that the last rule of $\Pi_2$ is a $!_i$, as it cannot be a rule acting on $\mathcal{D}(!\Gamma_2), \mathcal{D}(\Delta_2)$ by the above hypothesis. In particular, we can conclude that $\mathcal{D}(\Delta_2)$ is empty. Notice that the subderivation $\Pi_1'$ above $\Pi_1$ has conclusion $\mathcal{D}(!\Gamma_1), \mathcal{D}(\Delta_1) \vdash^m M : \mathcal{B}$, by Lemma 43 we have that $!x \notin FV(M)$ and so by Lemma 2 we have that $M = M\{W/p\}$. We can apply the induction hypothesis and conclude.

- If $r$ is $\&_{ei}$, then $p = \langle p_1, p_2\rangle : \mathcal{A}_1 \& \mathcal{A}_2$ and $W = \langle W_1, W_2\rangle$ for some strong values $W_i$ for $p_i$.

  Notice that the subderivation $\Pi'_1$ above $r$ has conclusion $\mathcal{D}(!\Gamma_1), \mathcal{D}(\Delta_1), p_i^k : \mathcal{A}_i \vdash^m M : \mathcal{B}$.

  By cases inspection, one can infer that the last rule of $\Pi_2$ is a $\&_i$, as it cannot be a rule acting on $\mathcal{D}(!\Gamma_2), \mathcal{D}(\Delta_2)$ by the above hypothesis. Therefore, we have two subderivations $\Pi_{2,i}$ above such a rule of the judgement $\mathcal{D}_i(!\Gamma_2), \mathcal{D}_i(\Delta_2) \vdash^{n_i} W_i : \mathcal{A}$ such that $\mathcal{D}(!\Gamma_2, \Delta_2) = \mathcal{D}_1(!\Gamma_2, \Delta_2) \boxplus \mathcal{D}_2(!\Gamma_2, \Delta_2)$ and we have also that

$$n = n_1 + n_2. \tag{6.1}$$

  By Lemma 43 we have also that $FV(p_{3-i}) \cap FV(M) = \emptyset$ and so by Lemma 2 and by substitution we have $M\{W/p\} = M\{W_i/p_i\}\{W_{3-i}/p_{3-i}\} = M\{W_i/p_i\}$.

  Let us suppose $i = 1$ (the other being similar), so we have $M\{W/p\} = M\{W_1/p_1\}$.

  We can then apply the induction hypothesis on $(s(\Pi'_1), s(\Pi_{2,1}))$, obtaining

  * A derivation $\Pi_c$ for $\mathcal{D}(!\Gamma_1) \cup \mathcal{D}(!\Gamma_2), \mathcal{D}(\Delta_1), k \circledast \mathcal{D}(\Delta_2) \vdash^{m'} M\{W_1/p_1\} : \mathcal{B}$.
  * If $M$ and $W_1$ are safe then we have also that
    1. $M\{W_1/p_1\}$ is a safe term
    2. $m' \leq k * n_1 + m$

  We can conclude that $\mathcal{D}(!\Gamma_1) \cup \mathcal{D}(!\Gamma_2), \mathcal{D}(\Delta_1), k \circledast \mathcal{D}(\Delta_2) \vdash^{m'} M\{W/p\} : \mathcal{B}$ is well-typed because we know that $M\{W/p\} = M\{W_1/p_1\}$ and by IH we have a derivation $\Pi_c$ for it.

  Moreover, if $M$ and $W$ are safe then $M\{W/p\}$ is safe too because $M\{W/p\} = M\{W_1/p_1\}$ and if $W$ is safe then also $W_1$ is safe and by item 1 of IH we know that $M\{W_1/p_1\}$ is a safe term.

  Finally, supposing that $M$ and $W$ are safe, we show that $m' \leq k * n + m$ as follows

$$\begin{aligned} m' &\overset{\text{IH}}{\leq} k * n_1 + m \\ &\leq k * n + m \\ &\overset{\text{Eq. 6.1}}{\leq} k * (n_1 + n_2) + m \end{aligned}$$

- If $r$ is $\otimes_e$, then $p = (p_1, p_2) : \mathcal{A}_1 \otimes \mathcal{A}_2$ and $W = (W_1, W_2)$ for some strong values $W_i$ of $p_i$ and

$$k = max(k_1, k_2). \tag{6.2}$$

  Notice that the subderivation $\Pi'_1$ above $r$ has conclusion $\mathcal{D}(!\Gamma_1), \mathcal{D}(\Delta_1), p_1^{k_1} : \mathcal{A}_1, p_2^{k_2} : \mathcal{A}_2 \vdash^m M : \mathcal{B}$.

  Also, by cases inspection, one can infer that the last rule of $\Pi_2$ is a $\otimes_i$, as it cannot be a rule acting on $\mathcal{D}(!\Gamma_2), \mathcal{D}(\Delta_2)$ by the above hypothesis. Therefore we have two subderivations $\Pi_{2,1}$ and $\Pi_{2,2}$ above such a rule of the judgements, resp., $\mathcal{D}(!\Gamma_{2,1}), \mathcal{D}(\Delta_{2,1}) \vdash W_1 : A_1$ and $\mathcal{D}(!\Gamma_{2,2}), \mathcal{D}(\Delta_{2,2}) \vdash W_2 : A_2$ such that $\mathcal{D}(!\Gamma_2) = \mathcal{D}(!\Gamma_{2,1}) \cup \mathcal{D}(!\Gamma_{2,2})$ and $\mathcal{D}(\Delta_2) = \mathcal{D}(\Delta_{2,1}), \mathcal{D}(\Delta_{2,2})$. Observe that

$$n = n_1 + n_2. \tag{6.3}$$

  By definition of substitution we have $M\{W/p\} = M\{(W_1, W_2)/(p_1, p_2)\} = M\{W_1/p_1\}\{W_2/p_2\}$.

  We can then apply the induction hypothesis on $(s(\Pi'_1), s(\Pi_{2,1}))$, obtaining

* A derivation $\Pi'$ for $\mathcal{D}(!\Gamma_1) \cup \mathcal{D}(!\Gamma_{2,1}), \mathcal{D}(\Delta_1), k \circledast \mathcal{D}(\Delta_{2,1}) \vdash^{m_1} M\{W_1/p_1\} : \mathcal{B}$.
* If $M$ and $W_1$ are safe then we have also that
  1. $M\{W_1/p_1\}$ is a safe term
  2. $m' \leq k * n_1 + m$

Let us call the above induction hypothesis IH1.

We can apply the induction hypothesis on the latter derivation $\Pi_{2,2}$ (notice that $s(\Pi_{2,2}) < s(\Pi_2)$ and so we have $(s(\Pi_{2,2}), k) < (s(\Pi_2), s(\Pi_1))$ for any $k$). So we apply the induction hypothesis on $(s(\Pi'), s(\Pi_{2,2}))$, obtaining

* A derivation $\Pi_c$ for $\mathcal{D}(!\Gamma_1) \cup \mathcal{D}(!\Gamma_{2,1}) \cup \mathcal{D}(!\Gamma_{2,2}), \mathcal{D}(\Delta_1), k \circledast (\mathcal{D}(\Delta_{2,1}), \mathcal{D}(\Delta_{2,2})) \vdash^{m_2} M\{W_1/p_1\}\{W_2/p_2\} : \mathcal{B}$.
* If $M$ and $W_2$ are safe then we have also that
  1. $M\{W_1/p_1\}\{W_2/p_2\}$ is a safe term
  2. $m_2 \leq k_2 * n_2 + m_1$

Let us call the above induction hypothesis IH2.

Observe that in this case we have

$$m' = m_2. \tag{6.4}$$

We can conclude that $\mathcal{D}(!\Gamma_1) \cup \mathcal{D}(!\Gamma_2), \mathcal{D}(\Delta_1), k \circledast \mathcal{D}(\Delta_2) \vdash^{m'} M\{W/p\} : \mathcal{B}$ is well-typed because we know that $M\{W/p\} = M\{W_1/p_1\}\{W_2/p_2\}$. Additionally, $\mathcal{D}(!\Gamma_2)$ and $\mathcal{D}(\Delta_2)$ are such that: $\mathcal{D}(!\Gamma_2) = \mathcal{D}(!\Gamma_{2,1}) \cup \mathcal{D}(!\Gamma_{2,2})$ and $\mathcal{D}(\Delta_2) = \mathcal{D}(\Delta_{2,1}), \mathcal{D}(\Delta_{2,2})$, so by IH2 we have a derivation $\Pi_c$ for it.

Moreover, if $M$ and $W$ are safe then $M\{W/p\}$ is safe too because by substitution $M\{W/p\} = M\{W_1/p_1\}\{W_2/p_2\}$ and if $W$ is safe then also $W_1$ and $W_2$ are safe and by item 1 of IH2 we know that $M\{W_1/p_1\}\{W_2/p_2\}$ is a safe term.

Finally, supposing that $M$ and $W$ are safe, we show that $m' \leq k * n + m$ as follows

$$\begin{aligned} m' &\overset{\text{Eq. 6.4}}{=} m_2 \\ &\overset{\text{IH2}}{\leq} k_2 * n_2 + m_1 \\ &\overset{\text{IH1}}{\leq} k_2 * n_2 + k_1 * n_1 + m \\ &\leq k * n + m \\ &\overset{\text{Eq. 6.2}}{=} max(k_1, k_2) * n + m \\ &\overset{\text{Eq. 6.3}}{=} max(k_1, k_2) * (n_1 + n_2) + m \\ &= max(k_1, k_2) * n_1 + max(k_1, k_2) * n_2 + m \end{aligned}$$

We can conclude because $max(k_1, k_2) \geq k_i$ for $i \in \{1, 2\}$.

– If $r$ is $1_e$, then $p = (\,) : \mathtt{1}$ and $W = (\,)$. Moreover,

$$k = 1. \tag{6.5}$$

Notice that the subderivation $\Pi'_1$ above $r$ has conclusion $\mathcal{D}(!\Gamma_1), \mathcal{D}(\Delta_1) \vdash^m M : \mathcal{B}$.

Also, by cases inspection, one can infer that $\Pi_2$ is just an instance of $1_i$, as it cannot be a rule acting on $\mathcal{D}(!\Gamma_2), \mathcal{D}(\Delta_2)$ by the above hypothesis. So in particular $\mathcal{D}(!\Gamma_2), \mathcal{D}(\Delta_2)$ is empty and

$$n = 0. \tag{6.6}$$

By definition of substitution we have that $M\{W/p\} = M\{()/()\} = M$.
We have to show that $\mathcal{D}(!\Gamma_1) \cup \mathcal{D}(!\Gamma_2), \mathcal{D}(\Delta_1), k \circledast \mathcal{D}(\Delta_2) \vdash^{m'} M\{W/p\} : \mathcal{B}$ is well-typed. By substitution we have $M\{W/p\} = M$ and $\mathcal{D}(!\Gamma_2), \mathcal{D}(\Delta_2)$ are empty. Therefore, by $\Pi'_1$ is a derivation for it and we can conclude by observing that

$$m' = m. \tag{6.7}$$

Moreover, if $M$ and $W$ are safe then $M\{W/p\}$ is safe too because we have that $M\{W/p\} = M$.
Finally, supposing that $M$ and $W$ are safe, we show that $m' \leq k * n + m$ as follows

$$\begin{aligned} m' &\overset{\text{Eq. 6.4}}{=} m \\ &\leq k * n + m \\ &\overset{\text{Eq. 6.2}}{=} 1 * n + m \\ &\overset{\text{Eq. 6.3}}{=} 1 * 0 + m \\ &= m \end{aligned}$$

- Let us consider now the cases of the last rule $r$ of $\Pi_1$ acting on the subject $M$.
  - If $r$ is $v$, then $M = p = x$, $\mathcal{A} = \mathcal{B}$ and $\mathcal{D}(!\Gamma_1), \mathcal{D}(\Delta_1)$ is empty. Moreover, in this case we have also that

$$m = 0 \tag{6.8}$$

$$k = 1 \tag{6.9}$$

    By definition of substitution $M\{W/p\} = x\{W/x\} = W$.
    We can conclude that $\mathcal{D}(!\Gamma_1) \cup \mathcal{D}(!\Gamma_2), \mathcal{D}(\Delta_1), k \circledast \mathcal{D}(\Delta_2) \vdash^{m'} M\{W/p\} : \mathcal{B}$ is well-typed because by substitution we have that $M\{W/p\} = W$. Observe that

$$m' = n. \tag{6.10}$$

    Moreover, if $M$ and $W$ are safe then $M\{W/p\}$ is safe too because we have that $M\{W/p\} = W$.
    Finally, supposing that $M$ and $W$ are safe, we show that $m' \leq k * n + m$ as follows

$$\begin{aligned} m' &\overset{\text{Eq. 6.10}}{=} n \\ &\leq k * n + m \\ &\overset{\text{Eq. 6.8}}{=} k * n + 0 \\ &\overset{\text{Eq. 6.9}}{=} 1 * n \\ &= n \end{aligned}$$

  - If $r$ is $!_e$, then $M = x : \mathcal{B}$ and $p = !x : !\mathcal{B}$ so $\mathcal{A} = !\mathcal{B}$ and $\mathcal{D}(!\Gamma_1), \mathcal{D}(\Delta_1)$ is empty. Moreover, we have also that

$$k = 1. \tag{6.11}$$

    Observe that by definition of value for a pattern $W = !W'$ for some strong value $W'$, so $\Pi_2$ is a derivation for $\mathcal{D}(!\Gamma_2), \mathcal{D}(\Delta_2) \vdash^{n} !W' : !\mathcal{B}$
    By cases inspection, one can infer that the last rule of $\Pi_2$ is $!_i$, as it cannot be a rule acting on $\mathcal{D}(!\Gamma_2), \mathcal{D}(\Delta_2)$ by the above hypothesis. In particular $\mathcal{D}(\Delta_2)$ is empty and

the immediate subderivation of $\Pi_2$ is a derivation $\Pi_2'$ of $\mathcal{D}(!\Gamma_2) \vdash^{n'} W' : \mathcal{B}$. Observe that

$$n = 0. \tag{6.12}$$

By definition of substitution $M\{W/p\} = x\{!W'/!x\} = W'$.
We can conclude that $\mathcal{D}(!\Gamma_1) \cup \mathcal{D}(!\Gamma_2), \mathcal{D}(\Delta_1), k \circledast \mathcal{D}(\Delta_2) \vdash^{m'} M\{W/p\} : \mathcal{B}$ is well-typed because by substitution we have that $M\{W/p\} = W'$. Moreover, $\mathcal{D}(!\Gamma_1), \mathcal{D}(\Delta_1)$ are empty and $\mathcal{D}(\Delta_2)$ is empty so $\Pi_2'$ is a derivation for it. Observe that

$$m' = n'. \tag{6.13}$$

Moreover, if $M$ and $W$ are safe then $M\{W/p\} = W'$ is safe too because $W = !W'$ and so also $W'$ is safe. By definition of safeness of $W$ we have that

$$n' = 0. \tag{6.14}$$

Finally, supposing that $M$ and $W$ are safe, we show that $m' \leq k * n + m$ as follows

$$\begin{aligned} m' &\overset{\text{Eq. 6.13}}{=} n' \\ &\overset{\text{Eq. 6.14}}{=} 0 \\ &\leq k * n + m \\ &\overset{\text{Eq. 6.11}}{=} 1 * n + m \\ &\overset{\text{Eq. 6.12}}{=} 1 * 0 + m \\ &= m \end{aligned}$$

We can conclude as $m \in \mathbb{N}^{\geq 0}$.

– If $r$ is $\multimap_e$, then $M = M_1 M_2$. By $\multimap_e$ typing rule we have that the immediate subderivation of $\Pi_1$ are $\Pi_{1,1}$ for $\mathcal{D}(!\Gamma_{1,1}), \mathcal{D}(\Delta_{1,1}) \vdash^{m_1} M_1 : \mathcal{B}' \overset{k'}{\multimap} \mathcal{B}$ and $\Pi_{1,2}$ for $\mathcal{D}(!\Gamma_{1,2}), \mathcal{D}(\Delta_{1,2}) \vdash^{m_2} M_2 : \mathcal{B}'$ such that

$$\begin{aligned} \mathcal{D}(\Gamma_1) &= \mathcal{D}(\Gamma_{1,1}) \cup \mathcal{D}(\Gamma_{1,2}) \\ \mathcal{D}(\Delta_1) &= \mathcal{D}(\Delta_{1,1}), k' \circledast \mathcal{D}(\Delta_{1,2}) \end{aligned}$$

and we have also that

$$m = k' * m_2 + m_1. \tag{6.15}$$

We should now split in sub-cases, depending in which environment $p : \mathcal{A}$ occurs

* If $p \in \mathcal{D}(\Gamma_{1,1}) \cap \mathcal{D}(\Gamma_{1,2})$, then $p = !x$ and $\mathcal{A} = !\mathcal{A}'$ for a suitable $x : \mathcal{A}'$.
  By definition of value for a pattern $W = !W'$ for some strong value $W'$.
  Observe that $\mathcal{A} = !\mathcal{A}'$, so $\Pi_2'$ is a derivation for $\mathcal{D}(!\Gamma_2), \mathcal{D}(\Delta_2) \vdash^{n} !W' : !\mathcal{A}'$.
  By cases inspection, one can infer that the last rule of $\Pi_2$ is $!_i$, as it cannot be a rule acting on $\mathcal{D}(!\Gamma_2), \mathcal{D}(\Delta_2)$ by the above hypothesis. In particular $\mathcal{D}(\Delta_2)$ is empty and the immediate subderivation of $\Pi_2$ is a derivation $\Pi_2'$ of $\mathcal{D}(!\Gamma_2) \vdash^{n'} W' : \mathcal{A}'$. Observe that

  $$n = 0. \tag{6.16}$$

  By definition of substitution $M\{W/p\} = (M_1 M_2)\{W/p\} = (M_1\{W/p\})(M_2\{W/p\})$.
  We can apply the induction hypothesis on $(s(\Pi_{1,1}), s(\Pi_2))$, obtaining

  · A derivation $\Pi'$ for $(\mathcal{D}(!\Gamma_{1,1}) \cup \mathcal{D}(!\Gamma_2)) \setminus p, \mathcal{D}(\Delta_{1,1}) \vdash^{m_1'} M_1\{W/p\} : \mathcal{B}' \overset{k'}{\multimap} \mathcal{B}$.

- If $M_1$ and $W$ are safe we have also that
  1. $M_1\{W/p\}$ is a safe term
  2. $m'_1 \leq k * n + m_1$

Let us call the above induction hypothesis IH1.
We can apply the induction hypothesis on $(s(\Pi_{1,2}), s(\Pi_2))$, obtaining

- A derivation $\Pi''$ for $(\mathcal{D}(!\Gamma_{1,2}) \cup \mathcal{D}(!\Gamma_2)) \setminus p, \mathcal{D}(\Delta_{1,2}) \vdash^{m'_2} M_2\{W/p\} : \mathcal{B}'$.
- If $M_2$ and $W$ are safe we have also that
  1. $M_2\{W/p\}$ is a safe term
  2. $m'_2 \leq k * n + m_2$

Let us call the above induction hypothesis IH2.
We can conclude that $(\mathcal{D}(!\Gamma_1) \cup \mathcal{D}(!\Gamma_2)) \setminus p, \mathcal{D}(\Delta_1), k \circledast \mathcal{D}(\Delta_2) \vdash^{m'} M\{W/p\} : \mathcal{B}$ is well-typed because by substitution we have that $M\{W/p\} = (M_1\{W/p\})(M_2\{W/p\})$. Moreover, $\mathcal{D}(\Delta_2)$ is empty and so we apply $\multimap_e$ and we use the derivations $\Pi'$ of IH1 and $\Pi''$ of IH2. Observe that

$$m' = k' * m'_2 + m'_1. \tag{6.17}$$

Moreover, if $M$ and $W$ are safe then $M\{W/p\}$ is safe too because by substitution we have that $M\{W/p\} = (M_1\{W/p\})(M_2\{W/p\})$ and by induction hypotheses $M_1\{W/p\}$ and $M_2\{W/p\}$ are safe terms.
Finally, supposing that $M$ and $W$ are safe, we show that $m' \leq k*n+m$ as follows

$$\begin{aligned}
m' &\overset{\text{Eq. 6.17}}{=} k' * m'_2 + m'_1 \\
&\overset{\text{IH1}}{\leq} k' * m'_2 + k * n + m_1 \\
&\overset{\text{IH2}}{\leq} k' * (k * n + m_2) + k * n + m_1 \\
&\overset{\text{Eq. 6.16}}{=} k' * (k * 0 + m_2) + k * 0 + m_1 \\
&= k' * m_2 + m_1 \\
&\leq k * n + m \\
&\overset{\text{Eq. 6.16}}{=} k * 0 + m \\
&= m \\
&\overset{\text{Eq. 6.15}}{=} k' * m_2 + m_1
\end{aligned}$$

* If $p \in \mathcal{D}(\Gamma_{1,i}) \setminus \mathcal{D}(\Gamma_{1,(3-i)})$ for $i \in \{1, 2\}$, then this is a simpler variant of the sub-case above, just applying the induction hypothesis on $(s(\Pi_{1,i}), s(\Pi_2))$.
* If $p \in \mathcal{D}(\Delta_{1,2})$, then we have that $FV(p) \cap FV(M_1) = \emptyset$.
  By definition of substitution $M\{W/p\} = (M_1M_2)\{W/p\} = M_1(M_2\{W/p\})$.
  We can apply the induction hypothesis on $(s(\Pi_{1,2}), s(\Pi_2))$, obtaining
  - A derivation $\Pi''$ for $\mathcal{D}(!\Gamma_{1,2}) \cup \mathcal{D}(!\Gamma_2), \mathcal{D}(\Delta_{1,2}) \setminus p, k \circledast \mathcal{D}(\Delta_2) \vdash^{m'_2} M_2\{W/p\} : \mathcal{B}'$.
  - If $M_2$ and $W$ are safe we have also that
    1. $M_2\{W/p\}$ is a safe term
    2. $m'_2 \leq k * n + m_2$

  We can conclude that $\mathcal{D}(!\Gamma_1) \cup \mathcal{D}(!\Gamma_2), \mathcal{D}(\Delta_1) \setminus p, k \circledast \mathcal{D}(\Delta_2) \vdash^{m'} M\{W/p\} : \mathcal{B}$ is well-typed because by substitution we have that $M\{W/p\} = M_1(M_2\{W/p\})$ and so we apply $\multimap_e$ and we use the derivations $\Pi_{1,1}$ and $\Pi'$, where the latter is obtained by induction hypothesis. Observe that

$$m' = k' * m_2' + m_1. \tag{6.18}$$

Moreover, if $M$ and $W$ are safe then $M\{W/p\}$ is safe too because by substitution we have that $M\{W/p\} = M_1(M_2\{W/p\})$ and $M_1$ is safe as $M = M_1M_2$ is safe and moreover by induction hypothesis $M_2\{W/p\}$ is safe.
Finally, supposing that $M$ and $W$ are safe, we show that $m' \leq (k' * k) * n + m$ as follows

$$\begin{aligned} m' &\overset{\text{Eq. 6.18}}{=} k' * m_2' + m_1 \\ &\overset{\text{IH}}{\leq} k' * (k * n + m_2) + m_1 \\ &\leq (k' * k) * n + m \\ &\overset{\text{Eq. 6.15}}{=} (k' * k) * n + k' * m_2 + m_1 \\ &= k' * (k * n + m_2) + m_1 \end{aligned}$$

* If $p \in \mathcal{D}(\Delta_{1,1})$, then this case is simpler a variant of the sub-case above.

– If $r$ is $\multimap_i$, then $M = \lambda q.M_1$ and $\mathcal{B} = \mathcal{B}_1 \overset{\text{k'}}{\multimap} \mathcal{B}_2$.
Note that the subderivation $\Pi_1$ above $r$ is $\Pi_1'$ and it has conclusion $\mathcal{D}(!\Gamma_1), \mathcal{D}(\Delta_1), q^{k'} : \mathcal{B}_1 \vdash^{m_1} M_1 : \mathcal{B}_2$ and

$$m = m_1 + \mathcal{W}(FV(q) \setminus FV(M_1)). \tag{6.19}$$

By definition of substitution $M\{W/p\} = (\lambda q.M_1)\{W/p\} = \lambda q.M_1\{W/p\}$.
We can apply the induction hypothesis on $(s(\Pi_1'), s(\Pi_2))$, obtaining

* A derivation $\Pi'$ for $\mathcal{D}(!\Gamma_1) \cup \mathcal{D}(!\Gamma_2), \mathcal{D}(\Delta_1), k \circledast \mathcal{D}(\Delta_2), q^{k'} : \mathcal{B}_1 \vdash^{m_1'} M_1\{W/p\} : \mathcal{B}_2$.
* If $M_1$ and $W$ are safe we have also that
  1. $M_1\{W/p\}$ is a safe term
  2. $m_1' \leq k * n + m_1$

We can conclude that $\mathcal{D}(!\Gamma_1) \cup \mathcal{D}(!\Gamma_2), \mathcal{D}(\Delta_1), k \circledast \mathcal{D}(\Delta_2) \vdash^{m'} M\{W/p\} : \mathcal{B}_1 \overset{\text{k'}}{\multimap} \mathcal{B}_2$ is well-typed because by substitution we have that $M\{W/p\} = \lambda q.M_1\{W/p\}$ and so we apply $\multimap_i$ and we use the derivation $\Pi'$ obtained by induction hypothesis. Observe that

$$m' = m_1' + \mathcal{W}(FV(q) \setminus FV(M_1)). \tag{6.20}$$

Moreover, if $M$ and $W$ are safe then $M\{W/p\}$ is safe too, because by substitution we have that $M\{W/p\} = \lambda q.M_1\{W/p\}$ and by induction hypothesis $M_1\{W/p\}$ is a safe term.
Finally, supposing that $M$ and $W$ are safe, we show that $m' \leq k * n + m$ as follows

$$\begin{aligned} m' &\overset{\text{Eq. 6.20}}{=} m_1' + \mathcal{W}(FV(q) \setminus FV(M_1)) \\ &\overset{\text{IH}}{\leq} k * n + m_1 + \mathcal{W}(FV(q) \setminus FV(M_1)) \\ &\leq k * n + m \\ &\overset{\text{Eq. 6.19}}{=} k * n + m_1 + \mathcal{W}(FV(q) \setminus FV(M_1)) \end{aligned}$$

– If $r$ is $\&_i$, then $M = \langle M_1, M_2 \rangle$ and $\mathcal{B} = \mathcal{B}_1 \& \mathcal{B}_2$.
Notice that the immediate subderivations of $\Pi_1$ above $r$ are $\Pi_{1,1}$ for $\mathcal{D}_1(!\Gamma_1), \mathcal{D}_1(\Delta_1) \vdash^{m_1} M_1 : \mathcal{B}_1$ and $\Pi_{1,2}$ for $\mathcal{D}_2(!\Gamma_1), \mathcal{D}_2(\Delta_1) \vdash^{m_2} M_2 : \mathcal{B}_2$ such that

$$\mathcal{D}(!\Gamma_1) = \mathcal{D}_1(!\Gamma_1) \boxplus \mathcal{D}_2(!\Gamma_1)$$
$$\mathcal{D}(\Delta_1) = \mathcal{D}_1(\Delta_1) \boxplus \mathcal{D}_2(\Delta_1)$$

Moreover, by definition of decoration $\mathcal{D}$ and the operation $\boxplus$ we have also that $\mathcal{D}_i(!\Gamma_1) = \mathcal{D}(!\Gamma_1)$ for $i \in \{1, 2\}$. Observe that

$$m = m_1 + m_2. \tag{6.21}$$

We should now split in sub-cases, depending in which environment $p : \mathcal{A}$ occurs

∗ If $p \in \mathcal{D}(!\Gamma_1)$, then $p = !x$ and $\mathcal{A} = !\mathcal{A}'$. Moreover, in this case we have also that $k = 1$. This case is simple to prove.

∗ If $p \in \mathcal{D}(\Delta_1)$, then $\mathcal{D}_1(\Delta_1) = \mathcal{D}_1(\Delta_1'), \mathcal{D}_1(p)$ and $\mathcal{D}_2(\Delta_1) = \mathcal{D}_2(\Delta_1'), \mathcal{D}_2(p)$. Now we have to split depending on the cases in the definition of $\boxplus$. If $\mathcal{A} = !\mathcal{A}'$ or ground then by definition of $\boxplus$ we have $\mathcal{D}(p) \mapsto (1, \mathcal{A})$ and the proof is simple. Otherwise, let us suppose that $\mathcal{D}_i(p) \mapsto (k_i, \mathcal{A})$ for $i \in \{1, 2\}$, so by definition of $\boxplus$ we have that $\mathcal{D}(p) \mapsto (k_1 + k_2, \mathcal{A})$. More precisely, we have that

$$k = k_1 + k_2. \tag{6.22}$$

By definition of substitution $M\{W/p\} = (\langle M_1, M_2 \rangle)\{W/p\} = \langle M_1\{W/p\}, M_2\{W/p\} \rangle$. We can apply the induction hypothesis on $(s(\Pi_{1,i}), s(\Pi_2))$ for $i \in \{1, 2\}$, obtaining

· A derivation $\Pi_i'$ for $\mathcal{D}(!\Gamma_1) \cup \mathcal{D}(!\Gamma_2), \mathcal{D}_i(\Delta_1'), k \circledast \mathcal{D}(\Delta_2) \vdash^{m_i'} M_i\{W/p\} : \mathcal{B}_i$.

· If $M_i$ and $W$ are safe we have also that
  1. $M_i\{W/p\}$ is a safe term
  2. $m_i' \leq k_i * n + m_i$

We can conclude that $\mathcal{D}(!\Gamma_1) \cup \mathcal{D}(!\Gamma_2), \mathcal{D}(\Delta_1'), k \circledast \mathcal{D}(\Delta_2) \vdash^{m'} M\{W/p\} : \mathcal{B}_1 \& \mathcal{B}_2$ is well-typed because by substitution $M\{W/p\} = \langle M_1\{W/p\}, M_2\{W/p\} \rangle$ and $\mathcal{D}(\Delta_1') = \mathcal{D}_1(\Delta_1') \boxplus \mathcal{D}_2(\Delta_1')$, so we use $\&_i$ and the derivations $\Pi_1'$ and $\Pi_2'$ of the induction hypotheses. Observe that

$$m' = m_1' + m_2'. \tag{6.23}$$

Moreover, if $M$ and $W$ are safe then $M\{W/p\}$ is safe too, because by substitution $M\{W/p\} = \langle M_1\{W/p\}, M_2\{W/p\} \rangle$ and by induction hypotheses $M_1\{W/p\}$ and $M_2\{W/p\}$ are safe.
Finally, supposing that $M$ and $W$ are safe, we show that $m' \leq k*n+m$ as follows

$$\begin{aligned} m' &\overset{\text{Eq. 6.23}}{=} m_1' + m_2' \\ &\overset{\text{IHs}}{\leq} k_1 * n + m_1 + k_2 * n + m_2 \\ &\leq k * n + m \\ &\overset{\text{Eq. 6.22}}{=} (k_1 + k_2) * n + m \\ &\overset{\text{Eq. 6.21}}{=} (k_1 + k_2) * n + m_1 + m_2 \end{aligned}$$

– If $r$ is $\otimes_i$, then we proceed similarly to the case $\multimap_e$

- If $r$ is $\top$ or $r$ is $Z$, then the proof are immediate.
- The cases in which $r$ is in $\{1_e, F_2, S, M, R\}$ are not possible as the pattern $p$ cannot appear in the environment.

□

Similarly, also the quantitative subject reduction stated in Theorem 13 merges the statement of Subject Reduction Theorem (see Theorem 4) and the statement of Measure Decreasing Proposition (see Proposition 3)

**Lemma 45.** If $\mathcal{D}(!\Gamma), \mathcal{D}(\Delta) \vdash^{m} \lambda p.M : \mathcal{B} \overset{\text{k}}{\multimap} \mathcal{A}$ is derivable, then $\mathcal{D}(!\Gamma), \mathcal{D}(\Delta), p^{k} : \mathcal{B} \vdash^{m'} M : \mathcal{A}$ and $m' \leq m$.

*Sketch Proof.* By induction on the size of the derivation $\mathcal{D}(!\Gamma), \mathcal{D}(\Delta), p^{k} : \mathcal{B} \vdash^{m'} M : \mathcal{A}$. □

**Theorem 13** (Quantitative Subject Reduction)**.** If $\mathcal{D}(!\Gamma), \mathcal{D}(\Delta) \vdash^{m} M : \mathcal{A}$ and $M \overset{s}{\to} N$, then $\mathcal{D}(!\Gamma), \mathcal{D}(\Delta) \vdash^{n} N : \mathcal{A}$. Moreover, if $M$ is safe then $n \leq m$, and if the step is numerical we have $n < m$.

*Proof.* Let $\Pi$ be the derivation for $\mathcal{D}(!\Gamma), \mathcal{D}(\Delta) \vdash^{m} M : \mathcal{A}$, we proceed by induction on $s(\Pi)$. We split depending on the last derivation rule of $\Pi$.

- If the last rule of $\Pi$ is a rule $r$ among $\{!_w, \&_{ei}, \otimes_e, 1_e\}$ acting on a pattern in $\mathcal{D}(!\Gamma_1), \mathcal{D}(\Delta_1)$, then the immediate subderivation of $\Pi$ is $\Pi'$ of $\mathcal{D}(!\Gamma'), \mathcal{D}(\Delta') \vdash^{m'} M : \mathcal{A}$ for suitable $\mathcal{D}(!\Gamma_1'), \mathcal{D}(\Delta_1')$ such that $FV(\mathcal{D}(\Delta_1')) \subseteq FV(\mathcal{D}(\Delta_1))$ and $FV(\mathcal{D}(!\Gamma_1'), \mathcal{D}(\Delta_1')) \subseteq FV(\mathcal{D}(!\Gamma_1), \mathcal{D}(\Delta_1))$. Observe that
$$m = m'. \tag{6.24}$$
We can apply the induction hypothesis on $\Pi'$ obtaining the derivation for $\mathcal{D}(!\Gamma'), \mathcal{D}(\Delta') \vdash^{n'} N : \mathcal{A}$. Moreover, if $M$ is safe term we have also $n' \leq m'$.
We can conclude $\mathcal{D}(!\Gamma), \mathcal{D}(\Delta) \vdash^{n} N : \mathcal{A}$ is well-type by applying the rule $r$ and by using the derivation $\Pi'$ of induction hypothesis. We have also that
$$n' = n. \tag{6.25}$$
Moreover, if $M$ is safe then we show that $n \leq m$ as follows
$$n \overset{\text{Eq. 6.25}}{=} n' \overset{\text{IH}}{\leq} m' \overset{\text{Eq. 6.24}}{=} m$$
- For the other cases, we can suppose that the last rule of $\Pi$ is not a rule acting on $\mathcal{D}(!\Gamma_1), \mathcal{D}(\Delta_1)$. We then proceed by analyzing the cases in which the last rule $r$ of $\Pi$ acts on the term $M$.
  - If $r$ is $\multimap_e$, then $M = \lambda p.M'$ and $\mathcal{A} = \mathcal{B}_1 \overset{\text{k'}}{\multimap} \mathcal{B}_2$.
    In this case $\Pi$ is a derivation for $\mathcal{D}(!\Gamma), \mathcal{D}(\Delta) \vdash^{m} \lambda p.M' : \mathcal{B}_1 \overset{\text{k'}}{\multimap} \mathcal{B}_2$.
    The redex can be only inside $M'$, so $N$ is in the form $\lambda p.N'$. The immediate subderivations of $\Pi$ is $\Pi'$ for $\mathcal{D}(!\Gamma), \mathcal{D}(\Delta), p^{k'} : \mathcal{B}_1 \vdash^{m} M' : \mathcal{B}_2$. Observe that
    $$m = m' + \mathcal{W}(FV(p) \setminus FV(M')). \tag{6.26}$$
    We can apply the induction hypothesis on $s(\Pi')$ obtaining a derivation $\Pi''$ for $\mathcal{D}(!\Gamma), \mathcal{D}(\Delta), p^{k'} : \mathcal{B}_1 \vdash^{n'} N' : \mathcal{B}_2$, Moreover, if $M$ is safe then $n' \leq m'$.

We can conclude that $\mathcal{D}(!\Gamma), \mathcal{D}(\Delta) \vdash^{n} \lambda p.N' : \mathcal{B}_1 \overset{k'}{\multimap} \mathcal{B}_2$ is well-typed by applying the rule $\multimap_e$ and using the derivation $\Pi''$ obtained by induction hypothesis. We have also that

$$n = n' + \mathcal{W}(FV(p) \setminus FV(N')). \tag{6.27}$$

Observe that by typing we know that

$$\mathcal{W}(FV(p) \setminus FV(N')) = \mathcal{W}(FV(p) \setminus FV(M')). \tag{6.28}$$

Moreover, if $M$ is safe then we show that $n \leq m$ as follows

$$\begin{aligned} n &\overset{\text{Eq. 6.27}}{=} n' + \mathcal{W}(FV(p) \setminus FV(N')) \\ &\overset{\text{IH}}{\leq} m' + \mathcal{W}(FV(p) \setminus FV(N')) \\ &\overset{\text{Eq. 6.28}}{=} m' + \mathcal{W}(FV(p) \setminus FV(M')) \overset{\text{Eq. 6.26}}{=} m \end{aligned}$$

– If $r$ is $\&_i$, then $M = \langle M_1, M_2 \rangle$ and $\mathcal{A} = \mathcal{B}_1 \& \mathcal{B}_2$.
In this case $\Pi$ is a derivation for $\mathcal{D}(!\Gamma), \mathcal{D}(\Delta) \vdash^{m} \langle M_1, M_2 \rangle : \mathcal{B}_1 \& \mathcal{B}_2$. Notice that the immediate subderivations of $\Pi$ above $r$ are $\Pi_1$ for $\mathcal{D}_1(!\Gamma_1), \mathcal{D}_1(\Delta_1) \vdash^{m_1} M_1 : \mathcal{B}_1$ and $\Pi_2$ for $\mathcal{D}_2(!\Gamma_1), \mathcal{D}_2(\Delta_1) \vdash^{m_2} M_2 : \mathcal{B}_2$ such that

$$\begin{aligned} \mathcal{D}(!\Gamma_1) &= \mathcal{D}_1(!\Gamma_1) \boxplus \mathcal{D}_2(!\Gamma_1) \\ \mathcal{D}(\Delta_1) &= \mathcal{D}_1(\Delta_1) \boxplus \mathcal{D}_2(\Delta_1) \end{aligned}$$

Moreover, by definition of decoration $\mathcal{D}$ and the operation $\boxplus$ we have also that $\mathcal{D}_i(!\Gamma_1) = \mathcal{D}(!\Gamma_1)$ for $i \in \{1, 2\}$. Observe that

$$m = m_1 + m_2. \tag{6.29}$$

The redex can be either in $M_1$ or in $M_2$. Let us suppose that the redex is in $M_1$ and so $M_1 \overset{s}{\to} M_1'$ (the other case being similar), we have that $N = \langle M_1', M_2 \rangle$.
We can apply the induction hypothesis on $s(\Pi_1)$ obtaining a derivation $\Pi_1'$ for $\mathcal{D}_1(!\Gamma_1), \mathcal{D}_1(\Delta_1) \vdash^{n_1} M_1' : \mathcal{B}_1$. Moreover, if $M$ is safe then $n_1 \leq m_1$.
We can conclude that $\mathcal{D}(!\Gamma), \mathcal{D}(\Delta) \vdash^{n} \langle M_1', M_2 \rangle : \mathcal{B}_1 \& \mathcal{B}_2$ is well-typed by applying the rule $\&_i$ and using the derivations $\Pi_2$ and $\Pi_1'$, the latter obtained by induction hypothesis. We have also that

$$n = n_1 + m_2. \tag{6.30}$$

Moreover, if $M$ is safe then we show that $n \leq m$ as follows

$$n \overset{\text{Eq. 6.30}}{=} n_1 + m_2 \overset{\text{IH}}{\leq} m_1 + m_2 \overset{\text{Eq. 6.29}}{=} m$$

– If $r$ is $\multimap_e$, then $M = M_1 M_2$. In this case $\Pi$ is a derivation for $\mathcal{D}(!\Gamma), \mathcal{D}(\Delta) \vdash^{m} M_1 M_2 : \mathcal{A}$. By $\multimap_e$ typing rule we have that the immediate subderivation of $\Pi$ are $\Pi_1$ for $\mathcal{D}(!\Gamma_1), \mathcal{D}(\Delta_1) \vdash^{m_1} M_1 : \mathcal{B} \overset{k}{\multimap} \mathcal{A}$ and $\Pi_2$ for $\mathcal{D}(!\Gamma_2), \mathcal{D}(\Delta_2) \vdash^{m_2} M_2 : \mathcal{B}$ such that

$$\begin{aligned} \mathcal{D}(\Gamma) &= \mathcal{D}(\Gamma_1) \cup \mathcal{D}(\Gamma_2) \\ \mathcal{D}(\Delta) &= \mathcal{D}(\Delta_1), k' \circledast \mathcal{D}(\Delta_2) \end{aligned}$$

and we have also that

$$m = k' * m_2 + m_1. \tag{6.31}$$

We proceed by analyzing the following sub-cases related to the redex:

* If the redex is $M_1 M_2$, then we proceed by induction on the reduction step. More precisely, we analyze the rules in Figure 3.4.
   · If the reduction rule is $\beta_s$, then $M_1 = \lambda p.M_1'$ and $M_2$ is a strong value $W$ for the pattern $p$. Moreover, $N$ is in the form $M_1'\{W/p\}$.
   We recall that in this case $\Pi$ is a derivation for $\mathcal{D}(!\Gamma), \mathcal{D}(\Delta) \vdash^m (\lambda p.M_1')W : \mathcal{A}$. Therefore, in this case the two subderivations above $r$ are $\Pi_1$ for $\mathcal{D}(!\Gamma_1), \mathcal{D}(\Delta_1) \vdash^{m_1} \lambda p.M_1' : \mathcal{B} \overset{\mathrm{k}}{\multimap} \mathcal{A}$ and $\Pi_2$ for $\mathcal{D}(!\Gamma_2), \mathcal{D}(\Delta_2) \vdash^{m_2} W : \mathcal{B}$
   By Lemma 45 on $\mathcal{D}(!\Gamma_1), \mathcal{D}(\Delta_1) \vdash^{m_1} \lambda p.M_1' : \mathcal{B} \overset{\mathrm{k}}{\multimap} \mathcal{A}$ we have a derivation for $\mathcal{D}(!\Gamma_1), \mathcal{D}(\Delta_1), p^k : \mathcal{B} \vdash^{m_1'} M_1' : \mathcal{A}$ and
$$m_1' \leq m_1. \tag{6.32}$$
   By Lemma 44 we obtain a derivation $\Pi_c$ for $\mathcal{D}(!\Gamma), \mathcal{D}(\Delta), k \circledast \mathcal{D}(\Delta_2) \vdash^{m'} M_1'\{W/p\} : \mathcal{A}$. Moreover, if $M_1'$ and $W$ are safe we have
   1. $M_1'\{W/p\}$ is a safe term
   2. $m' \leq k * m_2 + m_1'$

   We can conclude that $\mathcal{D}(!\Gamma_1) \cup \mathcal{D}(!\Gamma_2), \mathcal{D}(\Delta_1), k \circledast \mathcal{D}(\Delta_2) \vdash^n M_1'\{W/p\} : \mathcal{A}$ because we know that $\mathcal{D}(\Gamma) = \mathcal{D}(\Gamma_1) \cup \mathcal{D}(\Gamma_2)$ so $\Pi_c$ is a derivation for it. We have also that
$$n = m' \tag{6.33}$$
   Moreover, if $M$ is safe then we show that $n \leq m$ as follows
$$n \overset{\text{Eq. 6.33}}{=} m' \overset{\text{IH}}{\leq} k * m_2 + m_1' \overset{\text{Eq. 6.32}}{\leq} k * m_2 + m_1 \overset{\text{Eq. 6.31}}{=} m$$
   · If the reduction rule is $\beta_{\dot{+}}$, then $M_1 = \dot{+}$ and $M_2 = \langle \underline{r_1}, \underline{r_2} \rangle$. Moreover, by typing $\mathcal{D}(!\Gamma), \mathcal{D}(\Delta)$ is empty and
$$m_1 = 1 \quad \text{and} \quad m_2 = 2. \tag{6.34}$$
   By $\beta_{\dot{+}}$ rule we have that $N$ is in the form $\langle \underline{r_1}, \underline{r_2} \rangle$ and
$$n = 1. \tag{6.35}$$
   It is easy to see that $N$ is well-typed. Moreover, if $M$ is safe then we show that $n < m$ because the rule is arithmetical. More precisely, we do it as follows
$$n \overset{\text{Eq. 6.36}}{=} 1 < 3 = 1 + 2 \overset{\text{Eq. 6.34}}{=} m_1 + m_2 \overset{\text{Eq. 6.31}}{=} m$$
   · If the reduction rule is $\beta_{\dot{*}}$ or $\beta_F$, then the proof is simple and direct as in the case for $\beta_{\dot{+}}$.
* If the redex is in $M_1$ and so $M_1 \overset{s}{\to} M_1'$, then $N = M_1' M_2$.
We can apply the induction hypothesis on $\Pi_1$ and obtain a derivation $\Pi_1'$ for $\mathcal{D}(!\Gamma_1), \mathcal{D}(\Delta_1) \vdash^{n_1} M_1' : \mathcal{B} \overset{\mathrm{k}}{\multimap} \mathcal{A}$. Moreover, if $M$ is safe $n_1 \leq m_1$.
We can conclude that $\mathcal{D}(!\Gamma), \mathcal{D}(\Delta) \vdash^n M_1 M_2' : \mathcal{A}$ is well-typed by applying the rule $\multimap_e$ and using the derivation $\Pi_2$ and $\Pi_1'$, the latter obtained by induction hypothesis. We have also that
$$n = k * m_2 + n_1 \tag{6.36}$$
Moreover, if $M$ is safe then we show that $n \leq m$ as follows
$$n \overset{\text{Eq. 6.36}}{=} k * m_2 + n_1 \overset{\text{IH}}{\leq} k * m_2 + m_1 \overset{\text{Eq. 6.29}}{=} m$$

* If the redex is in $M_2$ and so $M_2 \xrightarrow{s} M_2'$, then $N = M_1 M_2'$.
  We can apply the induction hypothesis on $\Pi_2$ and obtain a derivation $\Pi_2'$ for $\mathcal{D}(!\Gamma_2), \mathcal{D}(\Delta_2) \vdash^{n_2} M_2' : \mathcal{B}$. Moreover, if $M$ is safe $n_2 \leq m_2$.
  We can conclude that $\mathcal{D}(!\Gamma), \mathcal{D}(\Delta) \vdash^{n} M_1 M_2' : \mathcal{A}$ is well-typed by applying the rule $\multimap_e$ and using the derivation $\Pi_1$ and $\Pi_2'$, the latter obtained by induction hypothesis. We have also that

$$n = k * n_2 + m_1 \tag{6.37}$$

  Moreover, if $M$ is safe then we show that $n \leq m$ as follows

$$n \overset{\text{Eq. 6.37}}{=} k * n_2 + m_1 \overset{\text{IH}}{\leq} k * m_2 + m_1 \overset{\text{Eq. 6.29}}{=} m$$

□

Now we want to check that the safeness of a term as in Definition 6 is preserved along safe reduction. The proof of safeness invariance (Lemma 23 in Section 3.5) in the quantitative setting is modified according to the new definition of safe term. Let us state again this property and show the proof for the quantitative type system as follows

**Lemma 46** (Quantitative Safeness Invariance)**.** If $M$ is safe (according to Definition 6) and $M \xrightarrow{s} N$, then $N$ is safe too.

*Proof.* By induction on the evaluation context $\gamma[\,]$ of the reduction step $M \xrightarrow{s} N$. The induction step splits according to the cases in Figure 3.3, while the base case splits according to Figure 3.4.

In the base case of induction, if the $\beta$-step is $\beta_s$ then we conclude by using Lemma 44.

The induction step is subtle when $\gamma[\,] = !\gamma'[\,]$ because in this case we have $M = !\gamma'[M_0]$, $N = !\gamma'[N_0]$ and $\gamma'[M_0] \xrightarrow{s} \gamma'[N_0]$. By induction hypothesis $\gamma'[N_0]$ is safe. Let $\gamma'[M_0]$ be typed as $\mathcal{D}(!\Gamma), \mathcal{D}(\Delta) \vdash^{m} \gamma'[M_0] : \mathcal{A}$. By Quantitative Subject Reduction (Theorem 13) we have that $N = \gamma'[N_0]$ is well-typed as $\mathcal{D}(!\Gamma), \mathcal{D}(\Delta) \vdash^{n} \gamma'[N_0] : \mathcal{A}$ and

$$n \leq m, \tag{6.38}$$

since by hypothesis $M$ is safe. Moreover, By definition of safeness of $M$ (according to Definition 6) we have that

$$m = 0. \tag{6.39}$$

We can conclude because to prove that $N = \gamma'[N_0]$ is safe we have to show that $n = 0$, we do it as follows

$$n \overset{\text{Eq. 6.38}}{\leq} m \overset{\text{Eq. 6.39}}{=} 0$$

and by observing that $n \in \mathbb{N}^{\geq 0}$.

□

Finally, we want to show that the number of numerical step of a closed safe term $M$ is bounded by the workload of $M$. More precisely, the statement and the proof of Proposition 4 are adapted to the new definition of workload as follows

**Proposition 16.** A safe closed term $M$ such that $\vdash^{m} M : \mathcal{A}$ reduces by any maximal safe-reduction sequence to a strong value $W$ in at most $m$ numeric steps. Moreover, if $M$ is of ground type then $W$ is a $\beta$-normal form.

*Proof.* Consider a maximal safe-reduction sequence $(M_i)_{i=0}^n$ starting from $M$, so we have

$$M = M_0 \to M_1 \to \ldots \to M_n$$

and $M_n$ is a normal form.

By Quantitative Safeness Invariance (Lemma 46) we have that all $M_i$'s are safe. We can apply Quantitative Subject Reduction (Theorem 13) to each reduction step obtaining a derivation for $\vdash^{m_{i+1}} M_{i+1} : \mathcal{A}$ and since all $M_i$'s are safe we have also that $m_{i+1} \leq m_i$ for $i \in \{0, \ldots, n\}$. More precisely, we obtain that the sequence $m = m_0, m_1, \ldots, m_n$ is decreasing. Moreover, it strictly decrease when the step is numerical.

Moreover, we can observe that by Quantitative Subject Reduction we have that all $M_i$'s are closed. In particular $M_n$ is a closed normal form for safe reduction and by Lemma 19 we have that $M_n$ is a closed strong value.

We can conclude that $m$ bounds the number of numerical steps of this sequence. If moreover $\mathcal{A}$ is ground, by reasoning above we have that $M_n$ is a closed strong value and by Lemma 20 (precisely (1)⇒(2)) we can conclude that $M_n$ is a $\beta$-normal form. □

We also adapt the statement of Lemma 24 in Section 3.5 which will be useful in the next section to prove that our transformations are work preserving in the quantitative setting as well.

**Lemma 47.** Given $\mathcal{D}(!\Gamma), \mathcal{D}(\S\Sigma) \vdash^{m} \S M : \S A$ and $\mathcal{D}(!\Gamma), \mathcal{D}(\S\Sigma) \vdash^{m'} M : A$, then $m = m'$.

*Proof.* Observe that $\S M = \langle (\,), M \rangle$ and $\S A = 1\&A$. Therefore, we have the following type derivation

$$\dfrac{\vdash^{0} (\,) : 1 \qquad \overset{\text{By hypothesis}}{\mathcal{D}(!\Gamma), \mathcal{D}(\S\Sigma) \vdash^{m'} M : A}}{\mathcal{D}(!\Gamma), \mathcal{D}(\S\Sigma) \vdash^{0+m'} \langle (\,), M \rangle : 1\&A}\ \&_i$$

so $m = 0 + m' = m'$ and we can conclude. □

## 6.3 Work-Preservation of AD Transformations

The work-preservation statements of the AD transformations hold in the quantitative setting as well by replacing $\mathcal{W}(M)$ with the number $m$ such that $\mathcal{D}(\Delta) \vdash^{m} M : \mathcal{A}$. Nonetheless, extending these work-preservation properties to the quantitative setting introduces nontrivial subtleties, arising due to the introduction of workload and additive duplication annotations to the type derivation of a term in λLL. In order to rigorously address these challenges, the following subsections systematically revisit the work-preservation theorems for the AD transformations, adapting them to the quantitative type system. For each transformation, we restate the relevant theorems, aligning them with the refined notion of workload. Proofs are provided for the same representative cases discussed in Chapter 5, with careful attention paid to how the transformations interact with the enriched typing and cost-tracking mechanisms. This requires a detailed analysis of how each transformation's workload is affected in the quantitative setting, ensuring that workload preservation is upheld under the revised definitions.

In the following we will use $\mathcal{L}$ and $\mathcal{H}$ as &-decorated sequence types and $\mathcal{E}$ as ⊗-decorated sequence types.

### 6.3.1 Quantitative Work-Preservation Forward

Let $\mathtt{t}$ be the function defined in Chapter 4, which maps $\otimes$-sequence types to &-sequence types. We extend this function to $\otimes$-decorated sequence types in the natural way. As discussed in Subsection 5.1.2, our forward transformation is work-preserving up to a constant factor. This is because the definition of $\mathcal{F}_\theta(P)$, as given in Figure 5.1, introduces a constant number of numerical operations in the case where $P$ is a numeric function. More formally, the statement of Theorem 8 in the quantitative setting changes as follows

**Theorem 14** (Quantitative Workload $\mathcal{F}$)**.** There is a constant $c$ such that $\forall P \in \lambda\mathrm{LL}^{\mathsf{P}}$ typed as $\mathcal{D}(!\Sigma) \vdash^m P : !\mathcal{E}$ and $\forall\theta$ enumeration of $FV(P)$ such that $\mathcal{D}(!\Sigma) \vdash^n \mathcal{F}_\theta(P) : !\mathcal{E} \otimes (\&\mathtt{t}(\theta) \overset{1}{\multimap} \mathtt{t}(\mathcal{E}))$, we have $n \leq c \cdot m$. If moreover $P$ is safe, then $\mathcal{F}_\theta(P)$ is safe too.

*Proof.* We proceed by induction on $P$. The part of the statement related to the safeness of $\mathcal{F}_\theta(P)$ is easy to prove by induction on $P$ simply checking Definition 6. Instead, the work preservation part of the statement is more delicate.

Let us consider the two most delicate cases:

- Case $P = (\lambda !x.Q_1)Q_2$:
  By hypothesis we have $\mathcal{D}(!\Sigma) \vdash^m (\lambda !x.Q_1)Q_2 : !\mathcal{E}$. We analyze the type derivation for the latter judgement as follows
  - By $\multimap_e$-rule we have $\mathcal{D}(!\Sigma) = \mathcal{D}(!\Sigma_1) \cup \mathcal{D}(!\Sigma_2)$ such that

    $$\mathcal{D}(!\Sigma_1) \vdash^{m_1} \lambda !x.Q_1 : !\mathcal{E}_0 \overset{1}{\multimap} !\mathcal{E}$$
    $$\mathcal{D}(!\Sigma_2) \vdash^{m_2} Q_2 : !\mathcal{E}_0$$

    and $m = m_2 + m_1$.
  - By $\multimap_i$-rule applied to $\mathcal{D}(!\Sigma_1) \vdash^{m_1} \lambda !x.Q_1 : !\mathcal{E}_0 \overset{1}{\multimap} !\mathcal{E}$ we have

    $$\mathcal{D}(!\Sigma_1), !x^1 : !\mathcal{E}_0 \vdash^{m'_1} Q_1 : !\mathcal{E}$$

    and $m_1 = m'_1 + \mathcal{W}(FV(!x) \setminus FV(Q_1))$. Moreover, $\mathcal{W}(FV(!x) \setminus FV(Q_1)) = 0$ because if $x \in FV(Q_1)$ then $\mathcal{W}(FV(!x) \setminus FV(Q_1)) = \mathcal{W}(\emptyset)$, otherwise $\mathcal{W}(FV(!x) \setminus FV(Q_1)) = \mathcal{W}(!\mathcal{E}_0)$ which is equal to 0 because in our definition of workload of a type we do not count the occurrences of $\mathbb{R}$ under the scope of a !. Therefore, we have that $m_1 = m'_1$.

  Summing up, we have that
  $$m = m'_1 + m_2 \tag{6.40}$$

  By definition of forward we have that

  $$\mathcal{F}_\theta((\lambda !x.Q_1)Q_2) = \begin{array}{l} \mathtt{let}\ (!x, \S f) = \mathcal{F}_{\theta \cap FV(Q_2)}(Q_2)\ \mathtt{in} \\ \mathtt{let}\ (!y, \S g) = \mathcal{F}_{FV(!x), \theta \cap FV(Q_1)}(Q_1)\ \mathtt{in} \\ (!y, \S(\lambda u^{\&\mathtt{t}(\theta)}.\mathtt{let}\ \langle u_{1,2}, u_1, u_2\rangle = D_{Q_1,Q_2,!x}\ u\ \mathtt{in}\ g\langle f\langle u_{1,2}, u_1\rangle, u_{1,2}, u_2\rangle)) \end{array}$$

  Observe that our workload essentially counts the number of numerical operations not under a ! and the number of possible numerals erased during a reduction. In this case nothing is erased so the sums related to the workload for the two let-constructs and for the $\lambda$-abstraction are equal to zero. This means that by typing we have

  $$\mathcal{D}(!\Sigma_1), !x^1 : !\mathcal{E}_0 \vdash^{n_1} \mathcal{F}_{x, \theta \cap FV(Q_1)}(Q_1) : !\mathcal{E} \otimes \S(\&\mathtt{t}(x, \theta \cap FV(Q_1)) \overset{1}{\multimap} \mathtt{t}(\mathcal{E}))$$

$$\mathcal{D}(!\Sigma_1) \vdash^{n_2} \mathcal{F}_{\theta \cap FV(Q_2)}(Q_2) : !\mathcal{E}_0 \otimes \S(\&\mathtt{t}(\theta \cap FV(Q_2)) \overset{1}{\multimap} \mathtt{t}(\mathcal{E}_0))$$

and

$$n = n_1 + n_2 \tag{6.41}$$

By inductive hypothesis on $Q_1$ we have $n_1 \leq c \cdot m_1'$.

By inductive hypothesis on $Q_2$ we have $n_2 \leq c \cdot m_2$.

Finally, we show that $n \leq c \cdot m$ as follows

$$\begin{aligned} n &\overset{\text{Eq. 6.41}}{=} n_1 + n_2 \\ &\overset{\text{IH on } Q_1}{\leq} c \cdot m_1' + n_2 \\ &\overset{\text{IH on } Q_2}{\leq} c \cdot m_1' + c \cdot m_2 \\ &= c \cdot (m_1' + m_2) \\ &\overset{\text{Eq. 6.40}}{=} c \cdot m \end{aligned}$$

- Case $P = \underline{f}(!x_1, !x_2)$:
  By hypothesis we have $\mathcal{D}(!\Sigma) \vdash^{m} \underline{f}(!x_1, !x_2) : !\mathbb{R}$ and by typing we have $\mathcal{D}(!\Sigma) = \{!x_1{}^1 : !\mathbb{R}, !x_2{}^1 : !\mathbb{R}\}$. We analyze the type derivation for the latter judgement as follows

  – By $\multimap_e$-rule we have

  $$\vdash^{m_1} \underline{f} : !\mathbb{R} \otimes !\mathbb{R} \overset{1}{\multimap} !\mathbb{R}$$
  $$!x_1{}^1 : !\mathbb{R}, !x_2{}^1 : !\mathbb{R} \vdash^{m_2} (!x_1, !x_2) : !\mathbb{R} \otimes !\mathbb{R}$$

  and $m = m_2 + m_1$. Moreover, by $F2$-typing rule we have that $m_1 = 1$.

  – By $\otimes_e$-rule applied to $!x_1{}^1 : !\mathbb{R}, !x_2{}^1 : !\mathbb{R} \vdash^{m_2} (!x_1, !x_2) : !\mathbb{R} \otimes !\mathbb{R}$ we have

  $$!x_1{}^1 : !\mathbb{R} \vdash^{m_2'} !x_1 : !\mathbb{R}$$
  $$!x_2{}^1 : !\mathbb{R} \vdash^{m_2'} !x_2 : !\mathbb{R}$$

  and $m_2 = m_2' + m_2''$.

  – By $!_i$ we have that $m_2' = m_2'' = 0$.

  Summing up, we have

  $$m = 1 \tag{6.42}$$

  By definition of forward we have that

  $$\mathcal{F}_\theta(\underline{f}(!x_1, !x_2)) = \begin{array}{l} \mathtt{let}\ !y_1 = \underline{\partial_1 f}(!x_1, !x_2)\ \mathtt{in} \\ \mathtt{let}\ !y_2 = \underline{\partial_2 f}(!x_1, !x_2)\ \mathtt{in} \\ (\underline{f}(!x_1, !x_2), \S(\lambda\langle u_1, u_2\rangle.(y_1 \dot{*} u_1) \dot{+} (y_2 \dot{*}(u_2)))) \end{array}$$

  By hypothesis we have $!x_1{}^1 : !\mathbb{R}, !x_2{}^1 : !\mathbb{R} \vdash^{n} \mathcal{F}_\theta(P) : !\mathbb{R} \otimes (\mathbb{R}\&\mathbb{R} \overset{1}{\multimap} \mathbb{R})$. We analyze the type derivation for the latter judgement as follows

– After some steps of $\multimap_e, \multimap_i, \otimes_e, !_i$ we have

$$\begin{aligned}
&\vdash^1 \underline{\partial_1 f} : !\mathbb{R} \otimes !\mathbb{R} \overset{1}{\multimap} !\mathbb{R} \\
&\vdash^1 \underline{\partial_2 f} : !\mathbb{R} \otimes !\mathbb{R} \overset{1}{\multimap} !\mathbb{R} \\
&\vdash^1 \underline{f} : !\mathbb{R} \otimes !\mathbb{R} \overset{1}{\multimap} !\mathbb{R} \\
!{x_1}^1 : !\mathbb{R}, !{x_2}^1 : !\mathbb{R} &\vdash^0 (!x_1, !x_2) : !\mathbb{R} \otimes !\mathbb{R} \\
!{y_1}^1 : !\mathbb{R}, !{y_2}^1 : !\mathbb{R}, {u_1}^1 : \mathbb{R}, {u_2}^1 : \mathbb{R} &\vdash^0 (y_1 \dot{*} u_1) \dot{+} (y_2 \dot{*} (u_2)) : \mathbb{R}
\end{aligned}$$

and $n = 3 + n'$.

– By typing rules $S$ and $M$ we have $n' = 3$.

Summing up, we have

$$n = 6 \tag{6.43}$$

Finally, we show that $n \leq c \cdot m$ as follows

$$n \overset{\text{Eq. 6.43}}{=} 6 \leq c \cdot m \overset{\text{Eq. 6.42}}{=} c \cdot 1$$

and we can conclude by taking $c = 6$.

□

### 6.3.2 Quantitative Work-Preservation Unzipping

In the quantitative setting, the statement of Theorem 10 changes as follows

**Theorem 15** (Quantitative Workload $\mathcal{U}$)**.** Given $\mathcal{D}(!\Sigma), \mathcal{D}(\S\Phi) \vdash^m S : !\mathcal{E} \otimes (\mathcal{L} \overset{1}{\multimap} \mathcal{H})$ and $\mathcal{D}(!\Sigma), \mathcal{D}(\S\Phi) \vdash^n \mathcal{U}(S) : !\mathcal{E} \otimes (\mathcal{L} \overset{1}{\multimap} \mathcal{H})$, we have that $n \leq m$. If moreover $S$ is safe, then $\mathcal{U}(S)$ is safe too.

*Proof.* The safeness of $\mathcal{U}$ is easy to prove by induction on $S$ simply checking Definition 6. Let us focus on the proof related to work preservation of $\mathcal{U}$, we proceed by induction on $S$. The only two delicate cases are:

- Case $S = (P, \S F)$:
  Let $\mathcal{U}^\bullet(S) = ([], P, F)$ and by definition $\mathcal{U}(S) = [(P, \S F)] = (P, \S F)$, so in this case we have that $S = \mathcal{U}(S)$. Therefore, $m = n$ and we can conclude.

- Case $S = \texttt{let } (!x, \S f) = S_2 \texttt{ in } S_1$:
  Recall that $\texttt{let } (!x^{!\mathcal{E}_0}, \S f^{\S(\mathcal{L}_0 \overset{1}{\multimap} \mathcal{H}_0)}) = S_2 \texttt{ in } S_1$ is a syntactic sugar for $(\lambda(!x^{!\mathcal{E}_0}, \S f^{\S(\mathcal{L}_0 \overset{1}{\multimap} \mathcal{H}_0)}).S_1)S_2$.

  Let us analyze the type derivation for

  $$\mathcal{D}(!\Sigma), \mathcal{D}(\S\Phi) \vdash^m (\lambda(!x^{!\mathcal{E}_0}, \S f^{\S(\mathcal{L}_0 \overset{1}{\multimap} \mathcal{H}_0)}).S_1)S_2 : !\mathcal{E} \otimes (\mathcal{L} \overset{1}{\multimap} \mathcal{H})$$

  as follows

- By $\multimap_e$-rule we have

$$\mathcal{D}(!\Sigma) = \mathcal{D}(!\Sigma_1) \cup \mathcal{D}(!\Sigma_2)$$
$$\mathcal{D}(\S\Phi) = \mathcal{D}(\S\Phi_1), k \circledast \mathcal{D}(\S\Phi_2)$$

such that

$$\mathcal{D}(!\Sigma_1), \mathcal{D}(\S\Phi_1) \vdash^{m_1} \lambda(!x^{!\mathcal{E}_0}, \S f^{\S(\mathcal{L}_0 \overset{1}{\multimap} \mathcal{H}_0)}).S_1 : (!\mathcal{E}_0 \otimes (\mathcal{L}_0 \overset{1}{\multimap} \mathcal{H}_0)) \overset{k}{\multimap} (!\mathcal{E} \otimes (\mathcal{L} \overset{1}{\multimap} \mathcal{H}))$$
$$\mathcal{D}(!\Sigma_2), \mathcal{D}(\S\Phi_2) \vdash^{m_2} S_2 : !\mathcal{E}_0 \otimes (\mathcal{L}_0 \overset{1}{\multimap} \mathcal{H}_0)$$

and $m = k * m_2 + m_1$.

- By $\multimap_i$-rule applied to

$$\mathcal{D}(!\Sigma_1), \mathcal{D}(\S\Phi_1) \vdash^{m_1} \lambda(!x^{!\mathcal{E}_0}, \S f^{\S(\mathcal{L}_0 \overset{1}{\multimap} \mathcal{H}_0)}).S_1 : (!\mathcal{E}_0 \otimes (\mathcal{L}_0 \overset{1}{\multimap} \mathcal{H}_0)) \overset{k}{\multimap} (!\mathcal{E} \otimes (\mathcal{L} \overset{1}{\multimap} \mathcal{H}))$$

we have

$$\mathcal{D}(!\Sigma_1), \mathcal{D}(\S\Phi_1), (!x, \S f)^k : !\mathcal{E}_0 \otimes (\mathcal{L}_0 \overset{1}{\multimap} \mathcal{H}_0) \vdash^{m_1'} S_1 : !\mathcal{E} \otimes (\mathcal{L} \overset{1}{\multimap} \mathcal{H})$$

and $m_1 = m_1' + \mathcal{W}(FV((!x, \S f)) \setminus FV(S_1)) = m_1' + \mathcal{W}(FV(f) \setminus FV(S_1))$.

Summing up, we have that

$$m = k * m_2 + m_1' + \mathcal{W}(FV(f) \setminus FV(S_1)) \tag{6.44}$$

Let $\mathcal{U}^\bullet(S_i) = (\epsilon_i[], P_i, F_i)$ and by definition $\mathcal{U}(S_i) = \epsilon_i[(P_i, \S F_i)]$.
Moreover, $\mathcal{U}^\bullet(S) = (\epsilon_1[\texttt{let } !x = P_1 \texttt{ in } \epsilon_2[]], P_2, (\texttt{let } \S f = \S F_1 \texttt{ in } F_2))$ so we have:

$$\begin{aligned}
\mathcal{U}(S) &\overset{\text{def}}{=} \epsilon_2[\texttt{let } !x = P_2 \texttt{ in } \epsilon_1[(P_1, \texttt{let } \S f = \S F_2 \texttt{ in } \S F_1)]] \\
&\overset{\text{Prop. 10 + Lemma 17}}{\sim} \epsilon_2[\texttt{let } !x = P_2 \texttt{ in let } \S f = \S F_2 \texttt{ in } \epsilon_1[(P_1, \S F_1])] \\
&=_\beta \epsilon_2[\texttt{let } (!x, \S f) = (P_2, \S F_2) \texttt{ in } \epsilon_1[(P_1, \S F_1])] \\
&\overset{\text{Prop. 10 + Lemma 17}}{\sim} \texttt{let } (!x, \S f) = \epsilon_2[(P_2, \S F_2)] \texttt{ in } \epsilon_1[(P_1, \S F_1])] \\
&= \texttt{let } (!x, \S f) = \mathcal{U}(S_2) \texttt{ in } \mathcal{U}(S_1)
\end{aligned} \tag{6.45}$$

Recall that $\texttt{let } (!x, \S f) = \mathcal{U}(S_2) \texttt{ in } \mathcal{U}(S_1)$ is a syntactic sugar for $(\lambda(!x^{!\mathcal{E}_0}, \S f^{\S(\mathcal{L}_0 \overset{1}{\multimap} \mathcal{H}_0)}).\mathcal{U}(S_1))\mathcal{U}(S_2)$.
Let us analyze the type derivation for

$$\mathcal{D}(!\Sigma), \mathcal{D}(\S\Phi) \vdash^{n} (\lambda(!x^{!\mathcal{E}_0}, \S f^{\S(\mathcal{L}_0 \overset{1}{\multimap} \mathcal{H}_0)}).\mathcal{U}(S_1))\mathcal{U}(S_2) : !\mathcal{E} \otimes (\mathcal{L} \overset{1}{\multimap} \mathcal{H})$$

as follows

- By $\multimap_e$-rule we have

$$\mathcal{D}(!\Sigma) = \mathcal{D}(!\Sigma_1) \cup \mathcal{D}(!\Sigma_2)$$
$$\mathcal{D}(\S\Phi) = \mathcal{D}(\S\Phi_1), k \circledast \mathcal{D}(\S\Phi_2)$$

such that

$$\mathcal{D}(!\Sigma_1), \mathcal{D}(\S\Phi_1) \vdash^{n_1} \lambda(!x^{!\mathcal{E}_0}, \S f^{\S(\mathcal{L}_0 \overset{1}{\multimap} \mathcal{H}_0)}).\mathcal{U}(S_1) : (!\mathcal{E}_0 \otimes (\mathcal{L}_0 \overset{1}{\multimap} \mathcal{H}_0)) \overset{k}{\multimap} (!\mathcal{E} \otimes (\mathcal{L} \overset{1}{\multimap} \mathcal{H}))$$
$$\mathcal{D}(!\Sigma_2), \mathcal{D}(\S\Phi_2) \vdash^{n_2} \mathcal{U}(S_2) : !\mathcal{E}_0 \otimes (\mathcal{L}_0 \overset{1}{\multimap} \mathcal{H}_0)$$

and $n = k * n_2 + n_1$.

– By $\multimap_i$-rule applied to

$$\mathcal{D}(!\Sigma_1), \mathcal{D}(\S\Phi_1) \vdash^{n_1} \lambda(!x^{!\mathcal{E}_0}, \S f^{\S(\mathcal{L}_0 \overset{1}{\multimap} \mathcal{H}_0)}).\mathcal{U}(S_1) : (!\mathcal{E}_0 \otimes (\mathcal{L}_0 \overset{1}{\multimap} \mathcal{H}_0)) \overset{k}{\multimap} (!\mathcal{E} \otimes (\mathcal{L} \overset{1}{\multimap} \mathcal{H}))$$

we have

$$\mathcal{D}(!\Sigma_1), \mathcal{D}(\S\Phi_1), (!x, \S f)^k : !\mathcal{E}_0 \otimes (\mathcal{L}_0 \overset{1}{\multimap} \mathcal{H}_0) \vdash^{n_1'} \mathcal{U}(S_1) : !\mathcal{E} \otimes (\mathcal{L} \overset{1}{\multimap} \mathcal{H})$$

and $n_1 = n_1' + \mathcal{W}(FV((!x, \S f)) \setminus FV(\mathcal{U}(S_1))) = n_1' + \mathcal{W}(FV(f) \setminus FV(\mathcal{U}(S_1))) = n_1' + \mathcal{W}(FV(f) \setminus FV(S_1))$ where the last passage is because by typing $FV(\mathcal{U}(S_1)) = FV(S_1)$.

$$n = k * n_2 + n_1' + \mathcal{W}(FV(f) \setminus FV(S_1)) \tag{6.46}$$

By inductive hypothesis on $S_1$ we have $n_1' \leq m_1'$.

By inductive hypothesis on $S_2$ we have $n_2 \leq m_2$.

Finally, we have tho show that $n \leq m$ as follows

$$\begin{aligned} n &\overset{\text{Eq. 6.46}}{=} k * n_2 + n_1' + \mathcal{W}(FV(f) \setminus FV(S_1)) \\ &\overset{\text{IH on } S_1}{\leq} k * n_2 + m_1' + \mathcal{W}(FV(f) \setminus FV(S_1)) \\ &\overset{\text{IH on } S_2}{\leq} k * m_2 + m_1' + \mathcal{W}(FV(f) \setminus FV(S_1)) \\ &\overset{\text{Eq. 6.44}}{=} m \end{aligned}$$

□

It is convenient to adapt also the statement of Lemma 35 in Chapter 5.2 as it will be useful to prove the work-preservation of our transpose transformation in the quantitative setting.

**Lemma 48.** *Given $R \in \lambda\mathrm{LL}^{\mathsf{A}}$ such that $\mathcal{D}(!\Sigma), \mathcal{D}(\S\Phi) \vdash^m R : !\mathcal{E} \otimes (\mathcal{L} \overset{1}{\multimap} \mathcal{H})$ and let $\mathcal{U}^\bullet(R) = (\epsilon[], P, F)$ such that $\mathcal{D}(!\Sigma') \vdash^n \epsilon[P] : !\mathcal{E}$ we have: $n \leq m$.*

*Sketch Proof.* Immediate consequence of the definition $\mathcal{U}^\bullet(R)$. □

### 6.3.3 Quantitative Work-Preservation Transpose

Recall from Section 5.3 that the definition of transpose, presented in Figure 5.4, relies on a sophisticated system of renamings designed to ensure that the transformation preserves the workload. The assertions concerning the typing of this system must be revised to align with the quantitative type system introduced in this chapter. First, we adapt Lemma 36, which outlines the properties of renamings, as follows:

**Lemma 49.** *If $\mathcal{D}(\Gamma), p^{\&} : \mathcal{L} \vdash^m M : \mathcal{A}$, then for every renaming $\alpha$ such that $FV(M) \cap FV(p^{\&}) \subseteq \mathrm{Dom}(\alpha) \subseteq FV(p^{\&})$, we have that:*

1. *$\mathcal{D}(\Gamma), \alpha[p^{\&}] : \mathcal{L} \vdash^{m'} \alpha[M] : \mathcal{A}$,*
2. *$\mathcal{D}(\Gamma), \alpha\langle p^{\&}\rangle : \mathcal{L}' \vdash^{m''} \alpha[M] : \mathcal{A}$, where $\mathcal{L}'$ is the type of $\alpha\langle p^{\&}\rangle$,*
3. *$m = m' = m''$.*

Moreover, Lemma 37, which states the properties of $\mu$, is adapted to the quantitative setting as follows

**Lemma 50.** Let $p^\& : \mathcal{L}$ be a pattern, $\alpha_1$ and $\alpha_2$ be two renamings with disjoint codomains and let $\mathcal{L}_i$ be the type of $\alpha_i\langle p^\&\rangle$. We have that $\lambda\langle\alpha_1\langle p^\&\rangle, \alpha_2\langle p^\&\rangle\rangle.\nu(p^\&, \alpha_1, \alpha_2)$ is well-typed as follows

$$\vdash^{\mathcal{W}(\mathrm{Dom}(\alpha_1)\cap\mathrm{Dom}(\alpha_2)\cap FV(p^\&))} \lambda\langle\alpha_1\langle p^\&\rangle, \alpha_2\langle p^\&\rangle\rangle.\nu(p^\&, \alpha_1, \alpha_2) : (\mathcal{L}_1 \& \mathcal{L}_2) \overset{1}{\multimap} \mathcal{L}$$

As described in Subsection 5.3.3, the workload preservation property of the transpose transformation in λLL follows directly as a corollary of the technical Lemma 41. We adopt the same strategy here, highlighting the differences introduced by the quantitative type system.

The definition of $\mathcal{W}(\mathcal{V})$ have to be adapted to the decoration of the environment $\Phi$. More precisely, given a set $\S\mathcal{D}(\Phi)$ of variables of type $f^k : \S(\mathcal{L} \overset{1}{\multimap} \mathcal{H})$ and a term $M$ we use the notation:

$$\mathcal{W}(\S\mathcal{D}(\Phi)^{\mathrm{in}}_M) \overset{\mathrm{def}}{=} \sum_{f^k : \S(\mathcal{L} \overset{1}{\multimap} \mathcal{H}) \in \S\Phi \cap FV(M)} k * \mathcal{W}(\mathcal{L})$$

$$\mathcal{W}(\S\mathcal{D}(\Phi)^{\mathrm{out}}_M) \overset{\mathrm{def}}{=} \sum_{f^k : \S(\mathcal{L} \overset{1}{\multimap} \mathcal{H}) \in \S\Phi \cap FV(M)} k * \mathcal{W}(\mathcal{H})$$

where the workload $\mathcal{W}(\mathcal{A})$ of a decorated type $\mathcal{A}$ is the number of occurrences of $\mathbb{R}$ not under the scope of a !.

Work preservation of the transpose transformation in the quantitative setting of λLL follows directly as a corollary of the following lemma, which is the quantitative adaptation of Lemma 41.

**Lemma 51.** We have the following:

1. If $\mathcal{D}(!\Sigma), \mathcal{D}(\S\Phi), p^{\&1} : \mathcal{L} \vdash^m U : \mathcal{H}$, $\alpha$ is the identity renaming restricted to $FV(p^\&) \cap FV(U)$ and we have:

$$\mathcal{D}(!\Sigma), \mathcal{D}(\S\Phi) \vdash^{m+\mathcal{W}(FV(p^\&)\setminus FV(U))} \lambda p^\&.U : \mathcal{L} \overset{1}{\multimap} \mathcal{H}$$
$$\mathcal{D}(!\Sigma), \mathcal{D}(\S\overleftarrow{\Phi}) \vdash^n \lambda q^\&.\mu_{p^\&,\alpha,\emptyset}\langle\mathcal{T}_{\S\mathcal{D}(\overleftarrow{\Phi}),p^\&}(U), \langle\,\rangle\rangle : \mathcal{H} \overset{1}{\multimap} \mathcal{L}$$

   then:

$$n + \mathcal{W}(\mathcal{L}) + \mathcal{W}\left(\S\mathcal{D}(\overleftarrow{\Phi})^{\mathrm{in}}_{\mathcal{T}_{\S\mathcal{D}(\overleftarrow{\Phi}),p^\&}(U)}\right) \leq m + \mathcal{W}(FV(p^\&) \setminus FV(U)) + \mathcal{W}(\mathcal{H}) + \mathcal{W}\left(\S\mathcal{D}(\overleftarrow{\Phi})^{\mathrm{out}}_{\mathcal{T}_{\S\mathcal{D}(\overleftarrow{\Phi}),p^\&}(U)}\right)$$

2. If $\mathcal{D}(!\Sigma), \mathcal{D}(\S\Phi) \vdash^m F : \mathcal{L} \overset{1}{\multimap} \mathcal{H}$ and $\mathcal{D}(!\Sigma), \mathcal{D}(\S\overleftarrow{\Phi}) \vdash^n \mathcal{T}_{\S\mathcal{D}(\overleftarrow{\Phi})}(F) : \mathcal{H} \overset{1}{\multimap} \mathcal{L}$, then we have

$$n + \mathcal{W}(\mathcal{L}) + \mathcal{W}\left(\S\mathcal{D}(\overleftarrow{\Phi})^{\mathrm{in}}_{\mathcal{T}_{\S\mathcal{D}(\overleftarrow{\Phi})}(F)}\right) \leq m + \mathcal{W}(\mathcal{H}) + \mathcal{W}\left(\S\mathcal{D}(\overleftarrow{\Phi})^{\mathrm{out}}_{\mathcal{T}_{\S\mathcal{D}(\overleftarrow{\Phi})}(F)}\right)$$

3. If $\mathcal{D}(!\Sigma), \mathcal{D}(\S\Phi) \vdash^m R : !\mathcal{E} \otimes (\mathcal{L} \overset{1}{\multimap} \mathcal{H})$ and $\mathcal{D}(!\Sigma), \mathcal{D}(\S\overleftarrow{\Phi}) \vdash^n \mathcal{T}_{\mathcal{D}(\S\overleftarrow{\Phi})}(R) : !\mathcal{E} \otimes (\mathcal{H} \overset{1}{\multimap} \mathcal{L})$, then we have

$$n + \mathcal{W}(\mathcal{L}) + \mathcal{W}\left(\S\mathcal{D}(\overleftarrow{\Phi})^{\mathrm{in}}_{\mathcal{T}_{\mathcal{D}(\S\overleftarrow{\Phi})}(R)}\right) \leq m + \mathcal{W}(\mathcal{H}) + \mathcal{W}\left(\S\mathcal{D}(\overleftarrow{\Phi})^{\mathrm{out}}_{\mathcal{T}_{\mathcal{D}(\S\overleftarrow{\Phi})}(R)}\right)$$

Formally, the proof of the above lemma proceeds by induction on the structure of the term, analyzing the cases presented in Figure 5.4. As in the non-quantitative setting, we aim to convey the underlying intuition behind each claim of the statement. However, the proof is particularly intricate so we refer interested readers to the detailed version provided at the end of the chapter.

*Proof Quantitative Claim 1: Cases of $\mathcal{T}$ on $\lambda LL^{\mathtt{t}}$.* By typing of $\lambda\mathrm{LL}^{\mathtt{t}}$ we have that a term $U \in \lambda\mathrm{LL}^{\mathtt{t}}$ is well-typed as: $\mathcal{D}(!\Sigma), \mathcal{D}(\S\Phi), p^{\&^1} : \mathcal{L} \vdash^{m} U : \mathcal{H}$, so we are in the first case of the lemma. By Equation 5.29 we have that

$$\lambda q^{\&}.\mu_{p^{\&},\alpha,\emptyset}\langle \mathcal{T}_{\S\mathcal{D}(\overleftarrow{\Phi}),p^{\&}}(U), \langle\,\rangle\rangle = \lambda q^{\&}.(\lambda\langle \alpha\langle p^{\&}\rangle, \emptyset\langle p^{\&}\rangle\rangle.\nu(p^{\&},\alpha,\emptyset))\langle \mathcal{T}_{\S\mathcal{D}(\overleftarrow{\Phi}),p^{\&}}(U), \langle\,\rangle\rangle$$

Observe that $FV(q^{\&}) \cap FV(\lambda\langle \alpha\langle p^{\&}\rangle, \emptyset\langle p^{\&}\rangle\rangle.\nu(p^{\&},\alpha,\emptyset)) = \emptyset$ so we have that

$$\vdash^{n'} \lambda\langle \alpha\langle p^{\&}\rangle, \emptyset\langle p^{\&}\rangle\rangle.\nu(p^{\&},\alpha,\emptyset) : \mathcal{L}\&\top \overset{1}{\multimap} \mathcal{H}$$
$$\mathcal{D}(!\Sigma), \mathcal{D}(\S\overleftarrow{\Phi}) \vdash^{n''} \lambda q^{\&}.\langle \mathcal{T}_{\S\mathcal{D}(\overleftarrow{\Phi}),p^{\&}}(U), \langle\,\rangle\rangle : \mathcal{H} \overset{1}{\multimap} \mathcal{L}\&\top$$

and $n = n' + n''$. Moreover, by Lemma 50 we have $n' = \mathcal{W}(\mathrm{Dom}(\alpha) \cap \emptyset \cap FV(p^{\&})) = \mathcal{W}(\emptyset) = 0$, so we have that

$$n = n''. \tag{6.47}$$

By quantitative typing (Figure 6.1) we have that

$$\mathcal{D}(!\Sigma), \mathcal{D}(\S\overleftarrow{\Phi}) \vdash^{\overline{n}} \lambda q^{\&}.\mathcal{T}_{\S\mathcal{D}(\overleftarrow{\Phi}),p^{\&}}(U) : \mathcal{H}$$

and

$$n'' = \overline{n}. \tag{6.48}$$

Summing up, we have that $n \overset{\text{Eq. 6.47}}{=} n'' \overset{\text{Eq. 6.48}}{=} \overline{n}$ and in this case of the lemma is enough to prove that

$$\overline{n}+\mathcal{W}(\mathcal{L})+\mathcal{W}\left(\S\mathcal{D}(\overleftarrow{\Phi})^{\mathrm{in}}_{\mathcal{T}_{\S\mathcal{D}(\overleftarrow{\Phi}),p^{\&}}(U)}\right) \leq m+\mathcal{W}(FV(p^{\&})\backslash FV(U))+\mathcal{W}(\mathcal{H})+\mathcal{W}\left(\S\mathcal{D}(\overleftarrow{\Phi})^{\mathrm{out}}_{\mathcal{T}_{\S\mathcal{D}(\overleftarrow{\Phi}),p^{\&}}(U)}\right)$$

and we proceed by analyzing the cases in Figure 5.4b. The proof of this claim is very subtle and so for completeness we show all cases of induction at the end of the chapter (jump to page 153 for full details). □

*Proof Quantitative Claim 2: Cases of $\mathcal{T}$ on $\lambda LL^{\mathtt{f}}$.* By typing of $\lambda\mathrm{LL}^{\mathtt{f}}$ we have that a term $F \in \lambda\mathrm{LL}^{\mathtt{f}}$ is well-typed as $\mathcal{D}(!\Sigma), \mathcal{D}(\S\Phi) \vdash^{m} F : \mathcal{L} \overset{1}{\multimap} \mathcal{H}$, so we are in the second case of the lemma and we have also that $\mathcal{D}(!\Sigma), \mathcal{D}(\S\overleftarrow{\Phi}) \vdash^{n} \mathcal{T}_{\S\mathcal{D}(\overleftarrow{\Phi})}(F) : \mathcal{H} \overset{1}{\multimap} \mathcal{L}$. We want to prove that:

$$n + \mathcal{W}(\mathcal{L}) + \mathcal{W}\left(\S\mathcal{D}(\overleftarrow{\Phi})^{\mathrm{in}}_{\mathcal{T}_{\S\mathcal{D}(\overleftarrow{\Phi})}(F)}\right) \leq m + \mathcal{W}(\mathcal{H}) + \mathcal{W}\left(\S\mathcal{D}(\overleftarrow{\Phi})^{\mathrm{out}}_{\mathcal{T}_{\S\mathcal{D}(\overleftarrow{\Phi})}(F)}\right)$$

We proceed by analyzing the cases in Figure 5.4a. The two most interesting cases: case $F = \lambda p^{\&}.U$ and composition, jump to page 161 for a detailed discussion. □

*Proof Quantitative Claim 3: Cases of $\mathcal{T}$ on $\lambda LL^{\mathtt{A}}$.* By typing of $\lambda\mathrm{LL}^{\mathtt{A}}$ we have that a term $R \in \lambda\mathrm{LL}^{\mathtt{A}}$ is well-typed as $\mathcal{D}(!\Sigma), \mathcal{D}(\S\Phi) \vdash^{m} R : !\mathcal{E} \otimes (\mathcal{L} \overset{1}{\multimap} \mathcal{H})$, so we are in the third case of the lemma and we have also that $\mathcal{D}(!\Sigma), \mathcal{D}(\S\overleftarrow{\Phi}) \vdash^{n} \mathcal{T}_{\mathcal{D}(\S\overleftarrow{\Phi})}(R) : !\mathcal{E} \otimes (\mathcal{H} \overset{1}{\multimap} \mathcal{L})$. We want to prove that:

$$n + \mathcal{W}(\mathcal{L}) + \mathcal{W}\left(\S\mathcal{D}(\overleftarrow{\Phi})^{\text{in}}_{\mathcal{T}_{\mathcal{D}(\S\overleftarrow{\Phi})}(R)}\right) \leq m + \mathcal{W}(\mathcal{H}) + \mathcal{W}\left(\S\mathcal{D}(\overleftarrow{\Phi})^{\text{out}}_{\mathcal{T}_{\mathcal{D}(\S\overleftarrow{\Phi})}(R)}\right)$$

We proceed by analyzing the cases in Figure 5.4c. At the end of the chapter we will discuss the most interesting case related to composition (more details are given from page 166). □

**Corollary 7** (Quantitative Workload $\mathcal{T}$)**.** For every $\lambda\text{LL}^{\mathtt{A}}$ term with no free affine tangent variable $\mathcal{D}(!\Sigma) \vdash^{m} R : !\mathcal{E} \otimes (\mathcal{L} \overset{1}{\multimap} \mathcal{H})$ and $\mathcal{D}(!\Sigma) \vdash^{n} \mathcal{T}(R) : !\mathcal{E} \otimes (\mathcal{H} \overset{1}{\multimap} \mathcal{L})$, we have $n + \mathcal{W}(L) \leq m + \mathcal{W}(H)$. If moreover $R$ is safe, then $\mathcal{T}(R)$ is safe too.

# Proof Quantitative Work Preservation Transpose

**Lemma 51.** We have the following:

1. If $\mathcal{D}(!\Sigma), \mathcal{D}(\S\Phi), p^{\&^1} : \mathcal{L} \vdash^{m} U : \mathcal{H}$, $\alpha$ is the identity renaming restricted to $FV(p^{\&}) \cap FV(U)$ and we have:

$$\mathcal{D}(!\Sigma), \mathcal{D}(\S\Phi) \vdash^{m+\mathcal{W}(FV(p^{\&})\setminus FV(U))} \lambda p^{\&}.U : \mathcal{L} \overset{1}{\multimap} \mathcal{H}$$
$$\mathcal{D}(!\Sigma), \mathcal{D}(\S\overleftarrow{\Phi}) \vdash^{n} \lambda q^{\&}.\mu_{p^{\&},\alpha,\emptyset}\langle \mathcal{T}_{\S\mathcal{D}(\overleftarrow{\Phi}),p^{\&}}(U), \langle\,\rangle\rangle : \mathcal{H} \overset{1}{\multimap} \mathcal{L}$$

then:

$$n + \mathcal{W}(\mathcal{L}) + \mathcal{W}\Big(\S\mathcal{D}(\overleftarrow{\Phi})^{\text{in}}_{\mathcal{T}_{\S\mathcal{D}(\overleftarrow{\Phi}),p^{\&}}(U)}\Big) \leq m + \mathcal{W}(FV(p^{\&}) \setminus FV(U)) + \mathcal{W}(\mathcal{H}) + \mathcal{W}\Big(\S\mathcal{D}(\overleftarrow{\Phi})^{\text{out}}_{\mathcal{T}_{\S\mathcal{D}(\overleftarrow{\Phi}),p^{\&}}(U)}\Big)$$

2. If $\mathcal{D}(!\Sigma), \mathcal{D}(\S\Phi) \vdash^{m} F : \mathcal{L} \overset{1}{\multimap} \mathcal{H}$ and $\mathcal{D}(!\Sigma), \mathcal{D}(\S\overleftarrow{\Phi}) \vdash^{n} \mathcal{T}_{\S\mathcal{D}(\overleftarrow{\Phi})}(F) : \mathcal{H} \overset{1}{\multimap} \mathcal{L}$, then we have

$$n + \mathcal{W}(\mathcal{L}) + \mathcal{W}\Big(\S\mathcal{D}(\overleftarrow{\Phi})^{\text{in}}_{\mathcal{T}_{\S\mathcal{D}(\overleftarrow{\Phi})}(F)}\Big) \leq m + \mathcal{W}(\mathcal{H}) + \mathcal{W}\Big(\S\mathcal{D}(\overleftarrow{\Phi})^{\text{out}}_{\mathcal{T}_{\S\mathcal{D}(\overleftarrow{\Phi})}(F)}\Big)$$

3. If $\mathcal{D}(!\Sigma), \mathcal{D}(\S\Phi) \vdash^{m} R : !\mathcal{E} \otimes (\mathcal{L} \overset{1}{\multimap} \mathcal{H})$ and $\mathcal{D}(!\Sigma), \mathcal{D}(\S\overleftarrow{\Phi}) \vdash^{n} \mathcal{T}_{\mathcal{D}(\S\overleftarrow{\Phi})}(R) : !\mathcal{E} \otimes (\mathcal{H} \overset{1}{\multimap} \mathcal{L})$, then we have

$$n + \mathcal{W}(\mathcal{L}) + \mathcal{W}\Big(\S\mathcal{D}(\overleftarrow{\Phi})^{\text{in}}_{\mathcal{T}_{\mathcal{D}(\S\overleftarrow{\Phi})}(R)}\Big) \leq m + \mathcal{W}(\mathcal{H}) + \mathcal{W}\Big(\S\mathcal{D}(\overleftarrow{\Phi})^{\text{out}}_{\mathcal{T}_{\mathcal{D}(\S\overleftarrow{\Phi})}(R)}\Big)$$

*Proof Quantitative Claim 1: Cases of $\mathcal{T}$ on $\lambda LL^{t}$.* By the reasoning described above in this case it is enough to prove that

$$\overline{n} + \mathcal{W}(\mathcal{L}) + \mathcal{W}\Big(\S\mathcal{D}(\overleftarrow{\Phi})^{\text{in}}_{\mathcal{T}_{\S\mathcal{D}(\overleftarrow{\Phi}),p^{\&}}(U)}\Big) \leq m + \mathcal{W}(FV(p^{\&}) \setminus FV(U)) + \mathcal{W}(\mathcal{H}) + \mathcal{W}\Big(\S\mathcal{D}(\overleftarrow{\Phi})^{\text{out}}_{\mathcal{T}_{\S\mathcal{D}(\overleftarrow{\Phi}),p^{\&}}(U)}\Big)$$

and we proceed by analyzing the cases in Figure 5.4b.

- Case $U = \langle U_1, U_2 \rangle$:
  In this case we have that $\mathcal{H} = \mathcal{H}_1 \& \mathcal{H}_2$ and $q^{\&} = \langle q_1^{\&}, q_2^{\&} \rangle$.
  We have also $\lambda p^{\&}.U = \lambda p^{\&}.\langle U_1, U_2 \rangle$. Let us analyze the type derivation for

$$\mathcal{D}(!\Sigma), \mathcal{D}(\S\Phi) \vdash^{m+\mathcal{W}(FV(p^{\&})\setminus FV(U))} \lambda p^{\&}.\langle U_1, U_2 \rangle : \mathcal{L} \overset{1}{\multimap} (\mathcal{H}_1 \& \mathcal{H}_2)$$

  as follows

  - By $\multimap_i$-rule we have

$$\mathcal{D}(!\Sigma), \mathcal{D}(\S\Phi), p^{\&^1} : \mathcal{L} \vdash^{m} \langle U_1, U_2 \rangle : \mathcal{H}_1 \& \mathcal{H}_2$$

– By $\&_i$-rule we have that

$$\mathcal{D}(!\Sigma) = \mathcal{D}_1(!\Sigma) \boxplus \mathcal{D}_2(!\Sigma)$$
$$\mathcal{D}(\S\Phi) = \mathcal{D}_1(\S\Phi) \boxplus \mathcal{D}_2(\S\Phi)$$

Moreover, by definition of decoration $\mathcal{D}$ and the operation $\boxplus$ we have also that $\mathcal{D}_i(!\Sigma) = \mathcal{D}(!\Sigma)$ for $i \in \{1, 2\}$. Therefore, the subterms $U_1$ and $U_2$ are typed as

$$\mathcal{D}(!\Sigma), \mathcal{D}_1(\S\Phi), p^{\&^1} : \mathcal{L} \vdash^{m_1} U_1 : \mathcal{H}_1$$
$$\mathcal{D}(!\Sigma), \mathcal{D}_2(\S\Phi), p^{\&^1} : \mathcal{L} \vdash^{m_2} U_2 : \mathcal{H}_2$$

and

$$m = m_1 + m_2. \tag{6.49}$$

By definition of transpose we have

$$\lambda\langle q_1^{\&}, q_2^{\&}\rangle.\mathcal{T}_{\S\mathcal{D}(\overleftarrow{\Phi}),p^{\&}}(\langle U_1, U_2\rangle)$$
$$= \lambda\langle q_1^{\&}, q_2^{\&}\rangle.(\mu_{p,\alpha_1,\alpha_2}\langle \mathcal{T}_{\S\mathcal{D}(\overleftarrow{\Phi}),\alpha_1[p^{\&}]}(\alpha_1[U_1]), \mathcal{T}_{\S\mathcal{D}(\overleftarrow{\Phi}),\alpha_2[p^{\&}]}(\alpha_2[U_2])\rangle)$$

where $\alpha_1$ and $\alpha_2$ are two renamings such that

$$\mathrm{Cod}(\alpha_1) \cap \mathrm{Cod}(\alpha_2) = \emptyset \quad \text{and} \quad \mathrm{Dom}(\alpha_i) = FV(U_i) \cap FV(p^{\&}).$$

Moreover, we know that $FV(\mu_{p,\alpha_1,\alpha_2}) \cap FV(\langle q_1^{\&}, q_2^{\&}\rangle) = \emptyset$ and so by typing we have

$$FV(q_{3-i}^{\&}) \cap FV\left(\mathcal{T}_{\S\mathcal{D}(\overleftarrow{\Phi}),p^{\&}}(U_i)\right) = FV(q_{3-i}^{\&}) \cap FV\left(\mathcal{T}_{\S\mathcal{D}(\overleftarrow{\Phi}),\alpha_i[p^{\&}]}(\alpha_i[U_i])\right) = \emptyset$$

Therefore, we analyze the type derivation for

$$\mathcal{D}(!\Sigma), \mathcal{D}(\S\overleftarrow{\Phi}) \vdash^{\overline{n}} \mu_{p,\alpha_1,\alpha_2}\langle \lambda q_1^{\&}.\mathcal{T}_{\S\mathcal{D}(\overleftarrow{\Phi}),\alpha_1[p^{\&}]}(\alpha_1[U_1]), \lambda q_2^{\&}.\mathcal{T}_{\S\mathcal{D}(\overleftarrow{\Phi}),\alpha_2[p^{\&}]}(\alpha_2[U_2])\rangle : \mathcal{H}_1 \& \mathcal{H}_2$$

as follows:

– By $\multimap_e$-rule we type the terms $\mu_{p,\alpha_1,\alpha_2}$ and the term $\langle \lambda q_1^{\&}.\mathcal{T}_{\S\mathcal{D}(\overleftarrow{\Phi}),\alpha_1[p^{\&}]}(\alpha_1[U_1]), \lambda q_2^{\&}.\mathcal{T}_{\S\mathcal{D}(\overleftarrow{\Phi}),\alpha_2[p^{\&}]}(\alpha_2[U_2])\rangle$. By Definition of $\mu$ in Equation 5.29 and Lemma 50 we have that $\mu_{p,\alpha_1,\alpha_2}$ is typed as

$$\vdash^{\mathcal{W}(\mathrm{Dom}(\alpha_1)\cap\mathrm{Dom}(\alpha_2)\cap FV(p^{\&}))} \mu_{p,\alpha_1,\alpha_2} : (\mathcal{L}_1 \& \mathcal{L}_2) \overset{1}{\multimap} \mathcal{L}.$$

By typing we have

$$\mathcal{D}(!\Sigma), \mathcal{D}(\S\overleftarrow{\Phi}) \vdash^{n'} : (\mathcal{H}_1 \overset{1}{\multimap} \mathcal{L}_1) \overset{1}{\multimap} (\mathcal{H}_2 \overset{1}{\multimap} \mathcal{L}_2)$$

where

$$\overline{n} = n' + \mathcal{W}(\mathrm{Dom}(\alpha_1) \cap \mathrm{Dom}(\alpha_2) \cap FV(p^{\&})) \tag{6.50}$$

– By $\&_i$-rule we have that

$$\mathcal{D}(!\Sigma) = \mathcal{D}_1(!\Sigma) \boxplus \mathcal{D}_2(!\Sigma)$$
$$\mathcal{D}(\S\Phi) = \mathcal{D}_1(\S\Phi) \boxplus \mathcal{D}_2(\S\Phi)$$

Moreover, by definition of decoration $\mathcal{D}$ and the operation $\boxplus$ we have also that $\mathcal{D}_i(!\Sigma) = \mathcal{D}(!\Sigma)$ for $i \in \{1,2\}$. Therefore, we have that

$$\mathcal{D}(!\Sigma), \mathcal{D}_1(\S\overleftarrow{\Phi}) \vdash^{n'_1} \lambda q_1^{\&}.\mathcal{T}_{\S\mathcal{D}(\overleftarrow{\Phi}),\alpha_1[p^{\&}]}(\alpha_1[U_1]) : \mathcal{H}_1 \overset{1}{\multimap} \mathcal{L}_1$$
$$\mathcal{D}(!\Sigma), \mathcal{D}_2(\S\overleftarrow{\Phi}) \vdash^{n'_2} \lambda q_2^{\&}.\mathcal{T}_{\S\mathcal{D}(\overleftarrow{\Phi}),\alpha_2[p^{\&}]}(\alpha_2[U_2]) : \mathcal{H}_2 \overset{1}{\multimap} \mathcal{L}_2$$

and

$$n' = n'_1 + n'_2. \tag{6.51}$$

Moreover, by Lemma 49 we have

$$\mathcal{D}(!\Sigma), \mathcal{D}_1(\S\overleftarrow{\Phi}) \vdash^{n_1} \lambda q_1^{\&}.\mathcal{T}_{\S\mathcal{D}(\overleftarrow{\Phi}),p^{\&}}(U_1) : \mathcal{H}_1 \overset{1}{\multimap} \mathcal{L}_1$$
$$\mathcal{D}(!\Sigma), \mathcal{D}_2(\S\overleftarrow{\Phi}) \vdash^{n_2} \lambda q_2^{\&}.\mathcal{T}_{\S\mathcal{D}(\overleftarrow{\Phi}),p^{\&}}(U_2) : \mathcal{H}_2 \overset{1}{\multimap} \mathcal{L}_2$$

where

$$n'_1 = n_1 \quad \text{and} \quad n'_2 = n_2. \tag{6.52}$$

Summing up we have that

$$\begin{aligned}
\overline{n} &\overset{\text{Eq. } 6.50}{=} n' + \mathcal{W}(\mathrm{Dom}(\alpha_1) \cap \mathrm{Dom}(\alpha_2) \cap FV(p^{\&}))\\
&\overset{\text{Eq. } 6.51}{=} n'_1 + n'_2 + \mathcal{W}(\mathrm{Dom}(\alpha_1) \cap \mathrm{Dom}(\alpha_2) \cap FV(p^{\&}))\\
&\overset{\text{Eq. } 6.52}{=} n_1 + n_2 + \mathcal{W}(\mathrm{Dom}(\alpha_1) \cap \mathrm{Dom}(\alpha_2) \cap FV(p^{\&}))
\end{aligned}$$

so we have

$$\overline{n} = n_1 + n_2 + \mathcal{W}(\mathrm{Dom}(\alpha_1) \cap \mathrm{Dom}(\alpha_2) \cap FV(p^{\&})). \tag{6.53}$$

By inductive hypothesis on $U_i$ we have

$$\begin{aligned}
&n_i + \mathcal{W}(\mathcal{L}) + \mathcal{W}\Big(\S\mathcal{D}_i(\overleftarrow{\Phi})^{\text{in}}_{\mathcal{T}_{\S\mathcal{D}(\overleftarrow{\Phi}),p^{\&}}(U_i)}\Big)\\
&\qquad \leq m_i + \mathcal{W}(FV(p^{\&}) \setminus FV(U_i)) + \mathcal{W}(\mathcal{H}_i) + \mathcal{W}\Big(\S\mathcal{D}_i(\overleftarrow{\Phi})^{\text{out}}_{\mathcal{T}_{\S\mathcal{D}(\overleftarrow{\Phi}),p^{\&}}(U_i)}\Big)
\end{aligned}$$

Finally, we show that

$$\begin{aligned}
&\overline{n} + \mathcal{W}(\mathcal{L}) + \mathcal{W}\Big(\S\mathcal{D}(\overleftarrow{\Phi})^{\text{in}}_{\mathcal{T}_{\S\mathcal{D}(\overleftarrow{\Phi}),p^{\&}}(U)}\Big)\\
&\qquad \leq m + \mathcal{W}(FV(p^{\&}) \setminus FV(U)) + \mathcal{W}(\mathcal{H}) + \mathcal{W}\Big(\S\mathcal{D}(\overleftarrow{\Phi})^{\text{out}}_{\mathcal{T}_{\S\mathcal{D}(\overleftarrow{\Phi}),p^{\&}}(U)}\Big)
\end{aligned}$$

as follows

$$\begin{aligned}
&\overline{n} + \mathcal{W}(\mathcal{L}) + \mathcal{W}\Big(\S\mathcal{D}(\overleftarrow{\Phi})^{\text{in}}_{\mathcal{T}_{\S\mathcal{D}(\overleftarrow{\Phi}),p^{\&}}(U)}\Big)\\
&\overset{\text{Eq. } 6.53}{=} n_1 + n_2 + \mathcal{W}(\mathrm{Dom}(\alpha_1) \cap \mathrm{Dom}(\alpha_2) \cap FV(p^{\&})) + \mathcal{W}(\mathcal{L}) + \mathcal{W}\Big(\S\mathcal{D}_i(\overleftarrow{\Phi})^{\text{in}}_{\mathcal{T}_{\S\mathcal{D}(\overleftarrow{\Phi}),p^{\&}}(U)}\Big)\\
&= \mathcal{W}(\mathrm{Dom}(\alpha_1) \cap \mathrm{Dom}(\alpha_2) \cap FV(p^{\&})) + \mathcal{W}(\mathcal{L}) + \sum_{i=1}^{n} n_i + \mathcal{W}\Big(\S\mathcal{D}_i(\overleftarrow{\Phi})^{\text{in}}_{\mathcal{T}_{\S\mathcal{D}(\overleftarrow{\Phi}),p^{\&}}(U_i)}\Big)
\end{aligned}$$

$$\overset{\text{IHs}}{\leq} \mathcal{W}(\mathrm{Dom}(\alpha_1) \cap \mathrm{Dom}(\alpha_2) \cap FV(p^{\&})) - \mathcal{W}(\mathcal{L}) +$$
$$+ \sum_{i=1}^{n} m_i + \mathcal{W}(\mathcal{H}_i) + \mathcal{W}(FV(p^{\&}) \setminus FV(U_i)) + \mathcal{W}\left( \S\mathcal{D}_i(\overleftarrow{\Phi})^{\text{out}}_{\mathcal{T}_{\S\mathcal{D}(\overleftarrow{\Phi}),p^{\&}}(U_i)} \right)$$
$$= \mathcal{W}(\mathrm{Dom}(\alpha_1) \cap \mathrm{Dom}(\alpha_2) \cap FV(p^{\&})) - \mathcal{W}(\mathcal{L}) + \mathcal{W}\left( \S\mathcal{D}(\overleftarrow{\Phi})^{\text{out}}_{\mathcal{T}_{\S\mathcal{D}(\overleftarrow{\Phi}),p^{\&}}(U)} \right) + \mathcal{W}(\mathcal{H}) +$$
$$+ \sum_{i=1}^{n} m_i + \mathcal{W}(FV(p^{\&}) \setminus FV(U_i))$$
$$= m_1 + m_2 + \mathcal{W}(\mathcal{H}) - \mathcal{W}(\mathcal{L}) + \mathcal{W}\left( \S\mathcal{D}(\overleftarrow{\Phi})^{\text{out}}_{\mathcal{T}_{\S\mathcal{D}(\overleftarrow{\Phi}),p^{\&}}(U)} \right) +$$
$$+ \mathcal{W}(\mathrm{Dom}(\alpha_1) \cap \mathrm{Dom}(\alpha_2) \cap FV(p^{\&}))$$
$$+ \mathcal{W}(FV(p^{\&}) \setminus FV(U_1)) + \mathcal{W}(FV(p^{\&}) \setminus FV(U_2))$$
$$\overset{\text{Remark 9}}{=} m_1 + m_2 + \mathcal{W}(\mathcal{H}) - \mathcal{W}(\mathcal{L}) + \mathcal{W}\left( \S\mathcal{D}(\overleftarrow{\Phi})^{\text{out}}_{\mathcal{T}_{\S\mathcal{D}(\overleftarrow{\Phi}),p^{\&}}(U)} \right) +$$
$$+ \mathcal{W}(FV(p^{\&}) \setminus (FV(U_1) \cup FV(U_2))) + \mathcal{W}(\mathcal{L})$$
$$= m_1 + m_2 + \mathcal{W}(\mathcal{H}) + \mathcal{W}\left( \S\mathcal{D}(\overleftarrow{\Phi})^{\text{out}}_{\mathcal{T}_{\S\mathcal{D}(\overleftarrow{\Phi}),p^{\&}}(U)} \right) + \mathcal{W}(FV(p^{\&}) \setminus (FV(U_1) \cup FV(U_2)))$$
$$\leq m + \mathcal{W}(FV(p^{\&}) \setminus FV(U)) + \mathcal{W}(\mathcal{H}) + \mathcal{W}\left( \S\mathcal{D}(\overleftarrow{\Phi})^{\text{out}}_{\mathcal{T}_{\S\mathcal{D}(\overleftarrow{\Phi}),p^{\&}}(U)} \right)$$
$$\overset{\text{Eq. 6.49}}{=} m_1 + m_2 + \mathcal{W}(FV(p^{\&}) \setminus FV(U)) + \mathcal{W}(\mathcal{H}) + \mathcal{W}\left( \S\mathcal{D}(\overleftarrow{\Phi})^{\text{out}}_{\mathcal{T}_{\S\mathcal{D}(\overleftarrow{\Phi}),p^{\&}}(U)} \right)$$
$$= m_1 + m_2 + \mathcal{W}(FV(p^{\&}) \setminus (FV(U_1) \cup FV(U_2))) + \mathcal{W}(\mathcal{H}) + \mathcal{W}\left( \S\mathcal{D}(\overleftarrow{\Phi})^{\text{out}}_{\mathcal{T}_{\S\mathcal{D}(\overleftarrow{\Phi}),p^{\&}}(U)} \right)$$

- Case $U = FU_1$:
  Observe that by typing of $\lambda\mathrm{LL}^{\mathrm{f}}$ we have $FV(p^{\&}) \cap FV(F) = \emptyset$, so we analyze the type derivation for
  $$\mathcal{D}(!\Sigma), \mathcal{D}(\S\Phi) \vdash^{\widetilde{m}} F(\lambda p^{\&}.U_1) : \mathcal{H}$$
  and
  $$m + \mathcal{W}(FV(p^{\&}) \setminus FV(U)) = \widetilde{m} \tag{6.54}$$
  By $\multimap_e$ we have $\mathcal{D}(\S\Phi) = \mathcal{D}(\S\Phi_1), k' \circledast \mathcal{D}(\S\Phi_2)$ such that $\mathrm{Dom}(\mathcal{D}(\S\Phi_1)) \cap \mathrm{Dom}(\mathcal{D}(\S\Phi_2)) = \emptyset$ and $\mathcal{D}(!\Sigma) = \mathcal{D}(!\Sigma_1) \cup \mathcal{D}(!\Sigma_2)$. Therefore, we have that
  $$\mathcal{D}(!\Sigma_1), \mathcal{D}(\S\Phi_1) \vdash^{m_F} F : \mathcal{L}_0 \overset{\text{k'}}{\multimap} \mathcal{H}$$
  $$\mathcal{D}(!\Sigma_2), \mathcal{D}(\S\Phi_2) \vdash^{m'} \lambda p^{\&}.U_1 : \mathcal{L} \overset{1}{\multimap} \mathcal{L}_0$$
  and $\widetilde{m} = m_F + \overline{k} * m'$. Moreover, observe that $k' = 1$ because $\mathcal{L}_0$ is &-decorated sequence type and so $\mathcal{O}(\mathcal{L}_0) = 0$, so we have
  $$\widetilde{m} = m_F + m' \tag{6.55}$$
  By definition of transpose we have $\mathcal{T}_{\S\mathcal{D}(\overleftarrow{\Phi}),p^{\&}}(FU_1) = (\lambda q^{\&'}.\mathcal{T}_{\S\mathcal{D}(\overleftarrow{\Phi}),p^{\&}}(U_1))(\mathcal{T}_{\S\mathcal{D}(\overleftarrow{\Phi})}(F)q^{\&})$. In this case we have to analyze
  $$\lambda q^{\&}.\left((\lambda q^{\&'}.\mathcal{T}_{\S\mathcal{D}(\overleftarrow{\Phi}),p^{\&}}(U_1))(\mathcal{T}_{\S\mathcal{D}(\overleftarrow{\Phi})}(F)q^{\&})\right) = (\lambda q^{\&'}.\mathcal{T}_{\S\mathcal{D}(\overleftarrow{\Phi}),p^{\&}}(U_1))(\lambda q^{\&}.\mathcal{T}_{\S\mathcal{D}(\overleftarrow{\Phi})}(F)q^{\&})$$

More precisely, we to analyze the derivation for the judgement

$$\mathcal{D}(!\Sigma_1) \cup \mathcal{D}(!\Sigma_2), \mathcal{D}(\S\overleftarrow{\Phi}_1), \mathcal{D}(\S\overleftarrow{\Phi}_2) \vdash^{\overline{n}} (\lambda q^{\&'}.\mathcal{T}_{\S\mathcal{D}(\overleftarrow{\Phi}_2),p^\&}(U_1))(\lambda q^\&.\mathcal{T}_{\S\mathcal{D}(\overleftarrow{\Phi}_1)}(F)q^\&) : \mathcal{L}$$

as follows

- By $\multimap_e$ we have

$$\mathcal{D}(!\Sigma_2), \mathcal{D}(\S\overleftarrow{\Phi_2}) \vdash^{n_1} \lambda q^{\&'}.\mathcal{T}_{\S\mathcal{D}(\overleftarrow{\Phi}_2),p^\&}(U_1) : \mathcal{L}_0 \overset{\text{k''}}{\multimap} \mathcal{L}$$
$$\mathcal{D}(!\Sigma_1), \mathcal{D}(\S\overleftarrow{\Phi_1}) \vdash^{n_2} \lambda q^\&.\mathcal{T}_{\S\mathcal{D}(\overleftarrow{\Phi}_1)}(F)q^\& : \mathcal{H} \overset{\text{k'''}}{\multimap} \mathcal{L}_0$$

  and $\overline{n} = k'' * n_2 + n_1$. Moreover, observe that $k'' = 1$ because $\mathcal{L}_0$ is &-decorated sequence type and so $\mathcal{O}(\mathcal{L}_0) = 0$ and also $k''' = 1$, so we have

$$\overline{n} = n_2 + n_1 \tag{6.56}$$

- By $\multimap_i$ we have

$$\mathcal{D}(!\Sigma_1), \mathcal{D}(\S\overleftarrow{\Phi_1}), q^{\&^1} : \mathcal{H} \vdash^{n_2'} \mathcal{T}_{\S\mathcal{D}(\overleftarrow{\Phi}_1)}(F)q^\& : \mathcal{L}_0$$

  and $n_2 = n_2' + \mathcal{W}(FV(q^\&) \setminus FV(\mathcal{T}_{\S\mathcal{D}(\overleftarrow{\Phi}_1)}(F)q^\&)) = n_2' + \mathcal{W}(\emptyset) = n_2'$, so we have

$$n_2 = n_2'. \tag{6.57}$$

- By $\multimap_e$ we have

$$q^{\&^1} : \mathcal{H} \vdash^{0} q^\& : \mathcal{H}$$
$$\mathcal{D}(!\Sigma_1), \mathcal{D}(\S\overleftarrow{\Phi_1}) \vdash^{n_2''} \mathcal{T}_{\S\mathcal{D}(\overleftarrow{\Phi}_1)}(F) : \mathcal{H} \overset{1}{\multimap} \mathcal{L}_0$$

  and $n_2' = 1 * n_2'' + 0$, so we have

$$n_2' = n_2''. \tag{6.58}$$

Summing up we have

$$\overline{n} \overset{\text{Eq. 6.56}}{=} n_1 + n_2 \overset{\text{Eq. 6.57}}{=} n_1 + n_2' \overset{\text{Eq. 6.58}}{=} n_1 + n_2''$$

so we have

$$\overline{n} = n_1 + n_2'' \tag{6.59}$$

By inductive hypothesis on $F$ (case 2 of the lemma) we have

$$n_2'' + \mathcal{W}(\mathcal{L}_0) + \mathcal{W}\Big(\S\mathcal{D}(\overleftarrow{\Phi}_1)^{\text{in}}_{\mathcal{T}_{\S\overleftarrow{\Phi}_1}(F)}\Big) \leq m_F + \mathcal{W}(\mathcal{H}) + \mathcal{W}\Big(\S\mathcal{D}(\overleftarrow{\Phi}_1)^{\text{out}}_{\mathcal{T}_{\S\overleftarrow{\Phi}_1}(F)}\Big)$$

By inductive hypothesis on $U_1$ (case 1 of the lemma) we have

$$n_1 + \mathcal{W}(\mathcal{L}) + \mathcal{W}\Big(\S\mathcal{D}(\overleftarrow{\Phi}_2)^{\text{in}}_{\mathcal{T}_{\S\mathcal{D}(\overleftarrow{\Phi}_2),p^\&}(U_1)}\Big) \leq m' + \mathcal{W}(\mathcal{L}_0) + \mathcal{W}\Big(\S\mathcal{D}(\overleftarrow{\Phi}_2)^{\text{out}}_{\mathcal{T}_{\S\mathcal{D}(\overleftarrow{\Phi}_2),p^\&}(U_1)}\Big)$$

Recall that by typing $\mathrm{Dom}(\mathcal{D}(\S\Phi_1)) \cap \mathrm{Dom}(\mathcal{D}(\S\Phi_2)) = \emptyset$ and so we have

$$\begin{aligned}\mathcal{W}\Big(\S\mathcal{D}(\overleftarrow{\Phi})^{\mathrm{in}}_{\mathcal{T}_{\S\mathcal{D}(\overleftarrow{\Phi}),p^{\&}}}(U)\Big) &= \mathcal{W}\Big(\S\mathcal{D}(\overleftarrow{\Phi}_2)^{\mathrm{in}}_{\mathcal{T}_{\S\mathcal{D}(\overleftarrow{\Phi}_2),p^{\&}}}(U_1)\Big) + \mathcal{W}\Big(\S\mathcal{D}(\overleftarrow{\Phi}_1)^{\mathrm{in}}_{\mathcal{T}_{\S\overleftarrow{\Phi}_1}}(F)\Big)\\ \mathcal{W}\Big(\S\mathcal{D}(\overleftarrow{\Phi})^{\mathrm{out}}_{\mathcal{T}_{\S\mathcal{D}(\overleftarrow{\Phi}),p^{\&}}}(U)\Big) &= \mathcal{W}\Big(\S\mathcal{D}(\overleftarrow{\Phi}_2)^{\mathrm{out}}_{\mathcal{T}_{\S\mathcal{D}(\overleftarrow{\Phi}_2),p^{\&}}}(U_1)\Big) + \mathcal{W}\Big(\S\mathcal{D}(\overleftarrow{\Phi}_1)^{\mathrm{out}}_{\mathcal{T}_{\S\overleftarrow{\Phi}_1}}(F)\Big)\end{aligned} \tag{6.60}$$

Finally, we show that

$$\begin{aligned}&\overline{n} + \mathcal{W}(\mathcal{L}) + \mathcal{W}\Big(\S\mathcal{D}(\overleftarrow{\Phi})^{\mathrm{in}}_{\mathcal{T}_{\S\mathcal{D}(\overleftarrow{\Phi}),p^{\&}}}(U)\Big)\\ &\qquad\leq m + \mathcal{W}(FV(p^{\&}) \setminus FV(U)) + \mathcal{W}(\mathcal{H}) + \mathcal{W}\Big(\S\mathcal{D}(\overleftarrow{\Phi})^{\mathrm{out}}_{\mathcal{T}_{\S\mathcal{D}(\overleftarrow{\Phi}),p^{\&}}}(U)\Big)\end{aligned}$$

by using Remark 12 and Remark 13 as follows

$$\begin{aligned}&\overline{n} + \mathcal{W}(\mathcal{L}) + \mathcal{W}\Big(\S\mathcal{D}(\overleftarrow{\Phi})^{\mathrm{in}}_{\mathcal{T}_{\S\mathcal{D}(\overleftarrow{\Phi}),p^{\&}}}(U)\Big)\\
&\overset{\text{Eq. 6.59}}{=} n_1 + n_2'' + \mathcal{W}(\mathcal{L}) + \mathcal{W}\Big(\S\mathcal{D}(\overleftarrow{\Phi})^{\mathrm{in}}_{\mathcal{T}_{\S\mathcal{D}(\overleftarrow{\Phi}),p^{\&}}}(U)\Big)\\
&\overset{\text{Eq. 6.60}}{=} n_1 + n_2'' + \mathcal{W}(\mathcal{L}) + \mathcal{W}\Big(\S\mathcal{D}(\overleftarrow{\Phi}_2)^{\mathrm{out}}_{\mathcal{T}_{\S\mathcal{D}(\overleftarrow{\Phi}_2),p^{\&}}}(U_1)\Big) + \mathcal{W}\Big(\S\mathcal{D}(\overleftarrow{\Phi}_1)^{\mathrm{out}}_{\mathcal{T}_{\S\overleftarrow{\Phi}_1}}(F)\Big)\\
&\overset{\text{IH on } U_1}{\leq} n_2'' + \mathcal{W}\Big(\S\mathcal{D}(\overleftarrow{\Phi}_1)^{\mathrm{out}}_{\mathcal{T}_{\S\overleftarrow{\Phi}_1}}(F)\Big) + m' + \mathcal{W}(\mathcal{L}_0) + \mathcal{W}\Big(\S\mathcal{D}(\overleftarrow{\Phi}_2)^{\mathrm{out}}_{\mathcal{T}_{\S\mathcal{D}(\overleftarrow{\Phi}_2),p^{\&}}}(U_1)\Big)\\
&\overset{\text{IH on } F}{\leq} m_F + \mathcal{W}(\mathcal{H}) + \mathcal{W}\Big(\S\mathcal{D}(\overleftarrow{\Phi}_1)^{\mathrm{out}}_{\mathcal{T}_{\S\overleftarrow{\Phi}_1}}(F)\Big) + m' + \mathcal{W}\Big(\S\mathcal{D}(\overleftarrow{\Phi}_2)^{\mathrm{out}}_{\mathcal{T}_{\S\mathcal{D}(\overleftarrow{\Phi}_2),p^{\&}}}(U_1)\Big)\\
&\overset{\text{Eq. 6.60}}{=} m_F + m' + \mathcal{W}(\mathcal{H}) + \mathcal{W}\Big(\S\mathcal{D}(\overleftarrow{\Phi})^{\mathrm{in}}_{\mathcal{T}_{\S\mathcal{D}(\overleftarrow{\Phi}),p^{\&}}}(U)\Big)\\
&\overset{\text{Remark 12}}{=} m_F + m'' + \mathcal{W}(FV(p^{\&}) \setminus FV(U_1)) + \mathcal{W}(\mathcal{H}) + \mathcal{W}\Big(\S\mathcal{D}(\overleftarrow{\Phi})^{\mathrm{in}}_{\mathcal{T}_{\S\mathcal{D}(\overleftarrow{\Phi}),p^{\&}}}(U)\Big)\\
&\leq m + \mathcal{W}(FV(p^{\&}) \setminus FV(U)) + \mathcal{W}(\mathcal{H}) + \mathcal{W}\Big(\S\mathcal{D}(\overleftarrow{\Phi})^{\mathrm{out}}_{\mathcal{T}_{\S\mathcal{D}(\overleftarrow{\Phi}),p^{\&}}}(U)\Big)\\
&\overset{\text{Eq. 6.54}}{=} \widetilde{m} + \mathcal{W}(\mathcal{H}) + \mathcal{W}\Big(\S\mathcal{D}(\overleftarrow{\Phi})^{\mathrm{out}}_{\mathcal{T}_{\S\mathcal{D}(\overleftarrow{\Phi}),p^{\&}}}(U)\Big)\\
&\overset{\text{Remark 13}}{=} m_F + m'' + \mathcal{W}(FV(p^{\&}) \setminus FV(U_1)) + \mathcal{W}(\mathcal{H}) + \mathcal{W}\Big(\S\mathcal{D}(\overleftarrow{\Phi})^{\mathrm{out}}_{\mathcal{T}_{\S\mathcal{D}(\overleftarrow{\Phi}),p^{\&}}}(U)\Big)\end{aligned}$$

**Remark 12.** Recall that $m'$ is such that $\mathcal{D}(!\Sigma_2), \mathcal{D}(\S\Phi_2) \vdash^{m'} \lambda p^{\&}.U_1 : \mathcal{L} \overset{1}{\multimap} \mathcal{L}_0$, so by typing ($\multimap_i$-rule) we have $\mathcal{D}(!\Sigma_2), \mathcal{D}(\S\Phi_2), p^{\&^1} : \mathcal{L}_0 \vdash^{m''} U_1 : \mathcal{L}_0$ and $m' = m'' + \mathcal{W}(FV(p^{\&}) \setminus FV(U_1))$

**Remark 13.** Recall that by Equation 6.55 we have $\widetilde{m} = m_F + m'$ and by Remark 12 $\widetilde{m} = m_F + m'' + \mathcal{W}(FV(p^{\&}) \setminus FV(U_1))$.

- Case $U = u$:
Observe that by typing we have $\mathcal{L} = \mathcal{H}$ and $p^{\&} = q^{\&} = u$. Therefore, we have

$$\mathcal{W}\Big(\S\mathcal{D}(\overleftarrow{\Phi})^{\mathrm{in}}_{\mathcal{T}_{\S\mathcal{D}(\overleftarrow{\Phi}),p^{\&}}}(u)\Big) = \mathcal{W}\Big(\S\mathcal{D}(\overleftarrow{\Phi})^{\mathrm{out}}_{\mathcal{T}_{\S\mathcal{D}(\overleftarrow{\Phi}),p^{\&}}}(u)\Big) = 0 \tag{6.61}$$

More precisely, we have the following type derivations

$$\vdash^{m+\mathcal{W}(FV(p^{\&})\setminus FV(U))} \lambda u.u : \mathcal{L} \overset{1}{\multimap} \mathcal{H}$$
$$\vdash^{\overline{n}} \lambda u.u : \mathcal{H} \overset{1}{\multimap} \mathcal{L}$$

Observe that $\mathcal{W}(FV(p^{\&}) \setminus FV(U)) = \mathcal{W}(\emptyset) = 0$ because $p^{\&} = u$ and $U = u$. Moreover, by typing and previous observation it is easy to see that

$$\overline{n} = m + \mathcal{W}(FV(p^{\&}) \setminus FV(U)) = m = 0 \tag{6.62}$$

Finally, we show that

$$\overline{n} + \mathcal{W}(\mathcal{L}) + \mathcal{W}\left(\S\mathcal{D}(\overleftarrow{\Phi})^{\text{in}}_{\mathcal{T}_{\S\mathcal{D}(\overleftarrow{\Phi}),p^{\&}}(U)}\right) \leq m + \mathcal{W}(FV(p^{\&}) \setminus FV(U)) + \mathcal{W}(\mathcal{H}) + \mathcal{W}\left(\S\mathcal{D}(\overleftarrow{\Phi})^{\text{out}}_{\mathcal{T}_{\S\mathcal{D}(\overleftarrow{\Phi}),p^{\&}}(U)}\right)$$

as follows

$$\begin{aligned}
&\overline{n} + \mathcal{W}(\mathcal{L}) + \mathcal{W}\left(\S\mathcal{D}(\overleftarrow{\Phi})^{\text{in}}_{\mathcal{T}_{\S\mathcal{D}(\overleftarrow{\Phi}),p^{\&}}(U)}\right)\\
&\quad = \overline{n} + \mathcal{W}(\mathcal{L}) + \mathcal{W}\left(\S\mathcal{D}(\overleftarrow{\Phi})^{\text{in}}_{\mathcal{T}_{\S\mathcal{D}(\overleftarrow{\Phi}),p^{\&}}(u)}\right)\\
&\quad \overset{\text{Eq. 6.61}}{=} \overline{n} + \mathcal{W}(\mathcal{L})\\
&\quad \overset{\text{Eq. 6.62}}{=} \mathcal{W}(\mathcal{L})\\
&\quad \leq m + \mathcal{W}(FV(p^{\&}) \setminus FV(U)) + \mathcal{W}(\mathcal{H}) + \mathcal{W}\left(\S\mathcal{D}(\overleftarrow{\Phi})^{\text{out}}_{\mathcal{T}_{\S\mathcal{D}(\overleftarrow{\Phi}),p^{\&}}(U)}\right)\\
&\quad = m + \mathcal{W}(FV(p^{\&}) \setminus FV(U)) + \mathcal{W}(\mathcal{H}) + \mathcal{W}\left(\S\mathcal{D}(\overleftarrow{\Phi})^{\text{out}}_{\mathcal{T}_{\S\mathcal{D}(\overleftarrow{\Phi}),p^{\&}}(u)}\right)\\
&\quad \overset{\text{Eq. 6.61}}{=} m + \mathcal{W}(FV(p^{\&}) \setminus FV(U)) + \mathcal{W}(\mathcal{H})\\
&\quad \overset{\text{Eq. 6.62}}{=} \mathcal{W}(\mathcal{H})
\end{aligned}$$

and we can conclude as in this case $\mathcal{L} = \mathcal{H}$.

- Case $U = \underline{0}$:
  Observe that by typing $\mathcal{H} = \mathbb{R}$ so $q^{\&}$ is of type $\mathbb{R}$ and

$$\mathcal{W}\left(\S\mathcal{D}(\overleftarrow{\Phi})^{\text{in}}_{\mathcal{T}_{\S\mathcal{D}(\overleftarrow{\Phi}),p^{\&}}(\underline{0})}\right) = \mathcal{W}\left(\S\mathcal{D}(\overleftarrow{\Phi})^{\text{out}}_{\mathcal{T}_{\S\mathcal{D}(\overleftarrow{\Phi}),p^{\&}}(\underline{0})}\right) = 0 \tag{6.63}$$

By definition of transpose we have that $\mathcal{T}_{\S\mathcal{D}(\overleftarrow{\Phi}),p^{\&}}(\underline{0}) = \langle\rangle$. Therefore, in this case we have the following type derivations

$$\vdash^{m+\mathcal{W}(FV(p^{\&})\setminus FV(\underline{0}))} \lambda p^{\&}.\underline{0} : \mathcal{L} \overset{1}{\multimap} \mathbb{R}$$
$$\vdash^{\overline{n}} \lambda q^{\&}.\langle\rangle : \mathbb{R} \overset{1}{\multimap} \mathcal{L}$$

where $\overline{n} = \overline{n'} + \mathcal{W}(FV(q^{\&}) \setminus FV(\langle\rangle))$.

Moreover, we have:

- $\mathcal{W}(FV(q^{\&}) \setminus FV(\langle\rangle)) = \mathcal{W}(FV(q^{\&})) = \mathcal{W}(\mathcal{H}) = \mathcal{W}(\mathbb{R}) = 1$
- $\mathcal{W}(FV(p^{\&}) \setminus FV(\underline{0})) = \mathcal{W}(FV(p^{\&})) = \mathcal{W}(\mathcal{L})$
- $\overline{n'} = 0$ by $\top$-typing rule
- $m = 0$ by $Z$-typing rule

Summing up, we have

$$m + \mathcal{W}(FV(p^{\&}) \setminus FV(\underline{0})) = \mathcal{W}(\mathcal{L}) \tag{6.64}$$

$$\overline{n} = 1 \tag{6.65}$$

Finally, we show that

$$\overline{n} + \mathcal{W}(\mathcal{L}) + \mathcal{W}\left(\S\mathcal{D}(\overleftarrow{\Phi})^{\text{in}}_{\mathcal{T}_{\S\mathcal{D}(\overleftarrow{\Phi}),p^{\&}}(U)}\right) \leq m + \mathcal{W}(FV(p^{\&}) \setminus FV(U)) + \mathcal{W}(\mathcal{H}) + \mathcal{W}\left(\S\mathcal{D}(\overleftarrow{\Phi})^{\text{out}}_{\mathcal{T}_{\S\mathcal{D}(\overleftarrow{\Phi}),p^{\&}}(U)}\right)$$

as follows

$$\begin{aligned}
&\overline{n} + \mathcal{W}(\mathcal{L}) + \mathcal{W}\left(\S\mathcal{D}(\overleftarrow{\Phi})^{\text{in}}_{\mathcal{T}_{\S\mathcal{D}(\overleftarrow{\Phi}),p^{\&}}(U)}\right)\\
&= \overline{n} + \mathcal{W}(\mathcal{L}) + \mathcal{W}\left(\S\mathcal{D}(\overleftarrow{\Phi})^{\text{in}}_{\mathcal{T}_{\S\mathcal{D}(\overleftarrow{\Phi}),p^{\&}}(\underline{0})}\right)\\
&\overset{\text{Eq. 6.63}}{=} \overline{n} + \mathcal{W}(\mathcal{L})\\
&\overset{\text{Eq. 6.65}}{=} 1 + \mathcal{W}(\mathcal{L})\\
&\leq m + \mathcal{W}(FV(p^{\&}) \setminus FV(U)) + \mathcal{W}(\mathcal{H}) + \mathcal{W}\left(\S\mathcal{D}(\overleftarrow{\Phi})^{\text{out}}_{\mathcal{T}_{\S\mathcal{D}(\overleftarrow{\Phi}),p^{\&}}(U)}\right)\\
&= m + \mathcal{W}(FV(p^{\&}) \setminus FV(U)) + \mathcal{W}(\mathcal{H}) + \mathcal{W}\left(\S\mathcal{D}(\overleftarrow{\Phi})^{\text{out}}_{\mathcal{T}_{\S\mathcal{D}(\overleftarrow{\Phi}),p^{\&}}(\underline{0})}\right)\\
&\overset{\text{Eq. 6.63}}{=} m + \mathcal{W}(FV(p^{\&}) \setminus FV(U)) + \mathcal{W}(\mathcal{H})\\
&\overset{\text{Eq. 6.64}}{=} \mathcal{W}(\mathcal{L}) + \mathcal{W}(\mathcal{H})\\
&= \mathcal{W}(\mathcal{L}) + \mathcal{W}(\mathbb{R})\\
&= \mathcal{W}(\mathcal{L}) + 1
\end{aligned}$$

- Case $U = \langle\rangle$:
  Observe that by typing $\mathcal{H} = \top$ so $q^{\&}$ is of type $\top$ and

$$\mathcal{W}\left(\S\mathcal{D}(\overleftarrow{\Phi})^{\text{in}}_{\mathcal{T}_{\S\mathcal{D}(\overleftarrow{\Phi}),p^{\&}}(\langle\rangle)}\right) = \mathcal{W}\left(\S\mathcal{D}(\overleftarrow{\Phi})^{\text{out}}_{\mathcal{T}_{\S\mathcal{D}(\overleftarrow{\Phi}),p^{\&}}(\langle\rangle)}\right) = 0 \tag{6.66}$$

By definition of transpose we have that $\mathcal{T}_{\S\mathcal{D}(\overleftarrow{\Phi}),p^{\&}}(\langle\rangle) = \langle\rangle$. Therefore, in this case we have the following type derivations

$$\vdash^{m+\mathcal{W}(FV(p^{\&})\setminus FV(\langle\rangle))} \lambda p^{\&}.\langle\rangle : \mathcal{L} \overset{1}{\multimap} \top$$

$$\vdash^{\overline{n}} \lambda q^{\&}.\langle\rangle : \top \overset{1}{\multimap} \mathcal{L}$$

where $\overline{n} = \overline{n'} + \mathcal{W}(FV(q^{\&}) \setminus FV(\langle\rangle))$.

Moreover, we have:

- $\mathcal{W}(FV(q^{\&}) \setminus FV(\langle\rangle)) = \mathcal{W}(FV(q^{\&})) = \mathcal{W}(\mathcal{H}) = \mathcal{W}(\top) = 0$
- $\mathcal{W}(FV(p^{\&}) \setminus FV(\langle\rangle)) = \mathcal{W}(FV(p^{\&})) = \mathcal{W}(\mathcal{L})$
- $\overline{n'} = 0$ by $\top$-typing rule
- $m = 0$ by $Z$-typing rule

Summing up, we have

$$m + \mathcal{W}(FV(p^{\&}) \setminus FV(\underline{0})) = \mathcal{W}(\mathcal{L}) \tag{6.67}$$

$$\overline{n} = 1 \tag{6.68}$$

Finally, we show that

$$\overline{n} + \mathcal{W}(\mathcal{L}) + \mathcal{W}\Big(\S\mathcal{D}(\overleftarrow{\Phi})^{\text{in}}_{\mathcal{T}_{\S\mathcal{D}(\overleftarrow{\Phi}),p^{\&}}(U)}\Big) \leq m + \mathcal{W}(FV(p^{\&}) \setminus FV(U)) + \mathcal{W}(\mathcal{H}) + \mathcal{W}\Big(\S\mathcal{D}(\overleftarrow{\Phi})^{\text{out}}_{\mathcal{T}_{\S\mathcal{D}(\overleftarrow{\Phi}),p^{\&}}(U)}\Big)$$

as follows

$$\begin{aligned}
&\overline{n} + \mathcal{W}(\mathcal{L}) + \mathcal{W}\Big(\S\mathcal{D}(\overleftarrow{\Phi})^{\text{in}}_{\mathcal{T}_{\S\mathcal{D}(\overleftarrow{\Phi}),p^{\&}}(U)}\Big)\\
&= \overline{n} + \mathcal{W}(\mathcal{L}) + \mathcal{W}\Big(\S\mathcal{D}(\overleftarrow{\Phi})^{\text{in}}_{\mathcal{T}_{\S\mathcal{D}(\overleftarrow{\Phi}),p^{\&}}(\langle\rangle)}\Big)\\
&\overset{\text{Eq. 6.66}}{=} \overline{n} + \mathcal{W}(\mathcal{L})\\
&\overset{\text{Eq. 6.68}}{=} \mathcal{W}(\mathcal{L})\\
&\leq m + \mathcal{W}(FV(p^{\&}) \setminus FV(U)) + \mathcal{W}(\mathcal{H}) + \mathcal{W}\Big(\S\mathcal{D}(\overleftarrow{\Phi})^{\text{out}}_{\mathcal{T}_{\S\mathcal{D}(\overleftarrow{\Phi}),p^{\&}}(U)}\Big)\\
&= m + \mathcal{W}(FV(p^{\&}) \setminus FV(U)) + \mathcal{W}(\mathcal{H}) + \mathcal{W}\Big(\S\mathcal{D}(\overleftarrow{\Phi})^{\text{out}}_{\mathcal{T}_{\S\mathcal{D}(\overleftarrow{\Phi}),p^{\&}}(\langle\rangle)}\Big)\\
&\overset{\text{Eq. 6.66}}{=} m + \mathcal{W}(FV(p^{\&}) \setminus FV(U)) + \mathcal{W}(\mathcal{H})\\
&\overset{\text{Eq. 6.67}}{=} \mathcal{W}(\mathcal{L}) + \mathcal{W}(\mathcal{H})\\
&= \mathcal{W}(\mathcal{L}) + \mathcal{W}(\top)\\
&= \mathcal{W}(\mathcal{L})
\end{aligned}$$

□

*Proof Quantitative Claim 2: Cases of $\mathcal{T}$ on $\lambda LL^{\mathbf{f}}$.* Recall that we have to show

$$n + \mathcal{W}(\mathcal{L}) + \mathcal{W}\Big(\S\mathcal{D}(\overleftarrow{\Phi})^{\text{in}}_{\mathcal{T}_{\S\mathcal{D}(\overleftarrow{\Phi})}(F)}\Big) \leq m + \mathcal{W}(\mathcal{H}) + \mathcal{W}\Big(\S\mathcal{D}(\overleftarrow{\Phi})^{\text{out}}_{\mathcal{T}_{\S\mathcal{D}(\overleftarrow{\Phi})}(F)}\Big)$$

We proceed by analyzing the cases in Figure 5.4a.

- Case $F = \lambda p^{\&}.U$:
  By hypothesis we have $\mathcal{D}(!\Sigma), \mathcal{D}(\S\Phi) \vdash^{m} \lambda p^{\&}.U : \mathcal{L} \overset{1}{\multimap} \mathcal{H}$, so by $\multimap_i$-rule we have $\mathcal{D}(!\Sigma), \mathcal{D}(\S\Phi), p^{\&^1} : \mathcal{L} \vdash^{m'} U : \mathcal{H}$ and

$$m = m' + \mathcal{W}(FV(p^{\&}) \setminus FV(U)) \tag{6.69}$$

By definition of transpose we have that

$$\mathcal{T}_{\S\mathcal{D}(\overleftarrow{\Phi})}(\lambda p^{\&}.U) = \lambda q^{\&}.\mu_{p^{\&},\alpha,\emptyset}\langle \mathcal{T}_{\S\overleftarrow{\Phi},p^{\&}}(U), \langle\rangle\rangle$$

which is typed as

$$\mathcal{D}(!\Sigma), \mathcal{D}(\S\overleftarrow{\Phi}) \vdash^{n} \lambda q^{\&}.\mu_{p^{\&},\alpha,\emptyset}\langle \mathcal{T}_{\S\overleftarrow{\Phi},p^{\&}}(U), \langle\rangle\rangle : \mathcal{H} \overset{1}{\multimap} \mathcal{L}$$

By inductive hypothesis on $U$ (case 1 of the lemma) we have

$$\begin{aligned} n + \mathcal{W}(\mathcal{L}) + \mathcal{W}\Big(\S\mathcal{D}(\overleftarrow{\Phi})^{\text{in}}_{\mathcal{T}_{\S\mathcal{D}(\overleftarrow{\Phi}),p^{\&}}(U)}\Big) \\ \leq m' + \mathcal{W}(FV(p^{\&}) \setminus FV(U)) + \mathcal{W}(\mathcal{H}) + \mathcal{W}\Big(\S\mathcal{D}(\overleftarrow{\Phi})^{\text{out}}_{\mathcal{T}_{\S\mathcal{D}(\overleftarrow{\Phi}),p^{\&}}(U)}\Big) \end{aligned}$$

Finally, we show that

$$n + \mathcal{W}(\mathcal{L}) + \mathcal{W}\Big(\S\mathcal{D}(\overleftarrow{\Phi})^{\text{in}}_{\mathcal{T}_{\S\mathcal{D}(\overleftarrow{\Phi})}(F)}\Big) \leq m + \mathcal{W}(\mathcal{H}) + \mathcal{W}\Big(\S\mathcal{D}(\overleftarrow{\Phi})^{\text{out}}_{\mathcal{T}_{\S\mathcal{D}(\overleftarrow{\Phi})}(F)}\Big)$$

as follows

$$\begin{aligned} & n + \mathcal{W}(\mathcal{L}) + \mathcal{W}\Big(\S\mathcal{D}(\overleftarrow{\Phi})^{\text{in}}_{\mathcal{T}_{\S\mathcal{D}(\overleftarrow{\Phi})}(F)}\Big) \\ & \quad = n + \mathcal{W}(\mathcal{L}) + \mathcal{W}\Big(\S\mathcal{D}(\overleftarrow{\Phi})^{\text{in}}_{\mathcal{T}_{\S\mathcal{D}(\overleftarrow{\Phi})}(U)}\Big) \\ & \quad \overset{\text{IH on U}}{\leq} m' + \mathcal{W}(FV(p^{\&}) \setminus FV(U)) + \mathcal{W}(\mathcal{H}) + \mathcal{W}\Big(\S\mathcal{D}(\overleftarrow{\Phi})^{\text{out}}_{\mathcal{T}_{\S\mathcal{D}(\overleftarrow{\Phi}),p^{\&}}(U)}\Big) \\ & \quad \leq m + \mathcal{W}(\mathcal{H}) + \mathcal{W}\Big(\S\mathcal{D}(\overleftarrow{\Phi})^{\text{out}}_{\mathcal{T}_{\S\mathcal{D}(\overleftarrow{\Phi})}(F)}\Big) \\ & \quad = m + \mathcal{W}(\mathcal{H}) + \mathcal{W}\Big(\S\mathcal{D}(\overleftarrow{\Phi})^{\text{out}}_{\mathcal{T}_{\S\mathcal{D}(\overleftarrow{\Phi})}(U)}\Big) \\ & \quad \overset{\text{Eq. 6.69}}{=} m' + \mathcal{W}(FV(p^{\&}) \setminus FV(U)) + \mathcal{W}(\mathcal{H}) + \mathcal{W}\Big(\S\mathcal{D}(\overleftarrow{\Phi})^{\text{out}}_{\mathcal{T}_{\S\mathcal{D}(\overleftarrow{\Phi})}(U)}\Big) \end{aligned}$$

- Case $F = \texttt{let}\ \S f^{\ \S(\mathcal{L}_0 \overset{1}{\multimap} \mathcal{H}_0)} = \S G_2\ \texttt{in}\ G_1$:

  Recall that $\texttt{let}\ \S f^{\ \S(\mathcal{L}_0 \overset{1}{\multimap} \mathcal{H}_0)} = \S G_2\ \texttt{in}\ G_1$ is syntactic sugar for $(\lambda \S f.G_1)\S G_2$. Let us analyze the type derivation for $\mathcal{D}(!\Sigma), \mathcal{D}(\S\Phi) \vdash^{m} (\lambda \S f^{\ \S(\mathcal{L}_0 \overset{1}{\multimap} \mathcal{H}_0)}.G_1)\S G_2 : \mathcal{L} \overset{1}{\multimap} \mathcal{H}$ as follows

  - By $\multimap_e$ we have

    $$\begin{aligned} \mathcal{D}(!\Sigma) &= \mathcal{D}(!\Sigma_1) \cup \mathcal{D}(!\Sigma_2) \\ \mathcal{D}(\S\Phi) &= \mathcal{D}(\S\Phi_1), k \circledast \mathcal{D}(\S\Phi_2) \end{aligned}$$

    such that

    $$\begin{aligned} &\mathcal{D}(!\Sigma_1), \mathcal{D}(\S\Phi_1) \vdash^{m_1} \lambda \S f^{\ \S(\mathcal{L}_0 \overset{1}{\multimap} \mathcal{H}_0)}.G_1 : (\mathcal{L}_0 \overset{1}{\multimap} \mathcal{H}_0) \overset{k}{\multimap} (\mathcal{L} \overset{1}{\multimap} \mathcal{H}) \\ &\mathcal{D}(!\Sigma_2), \mathcal{D}(\S\Phi_2) \vdash^{m_2} \S G_2 : \S(\mathcal{L}_0 \overset{1}{\multimap} \mathcal{H}_0) \end{aligned}$$

    and $m = k * m_2 + m_1$.

– By $\S_r$-derived typing rule we have $\mathcal{D}(!\Sigma_2), \mathcal{D}(\S\Phi_2) \vdash^{m_2'} G_2 : \mathcal{L}_0 \overset{1}{\multimap} \mathcal{H}_0$. By Lemma 47 we have that $m_2 = m_2'$ and so

$$m = k * m_2' + m_1 \tag{6.70}$$

Observe that $f$ may not be free in $G_1$, so we have to analyze the following subcases:

– Subcase $f \in FV(G_1)$:

Let us analyze the judgement $\mathcal{D}(!\Sigma_1), \mathcal{D}(\S\Phi_1) \vdash^{m_1} \lambda\S f^{\ \S(\mathcal{L}_0 \overset{1}{\multimap} \mathcal{H}_0)}.G_1 : (\mathcal{L}_0 \overset{1}{\multimap} \mathcal{H}_0) \overset{k}{\multimap} (\mathcal{L} \overset{1}{\multimap} \mathcal{H})$, by $\multimap_i$ we have

$$\mathcal{D}(!\Sigma_1), \mathcal{D}(\S\Phi_1), f^k : \mathcal{L}_0 \overset{1}{\multimap} \mathcal{H}_0 \vdash^{m_1'} G_1 : \mathcal{L} \overset{1}{\multimap} \mathcal{H}$$

and $m_1 = m_1' + \mathcal{W}(FV(f) \setminus FV(G_1))$. Moreover, $\mathcal{W}(FV(f) \setminus FV(G_1)) = \mathcal{W}(\emptyset) = 0$ because in this subcase $f \in FV(G_1)$, so we have

$$m_1 = m_1' \tag{6.71}$$

By definition of transpose we have that

$$\mathcal{T}_{\S\mathcal{D}(\overleftarrow{\Phi})}(F) = \texttt{let } \S\overleftarrow{f}^{\ \S(\mathcal{H}_0 \overset{1}{\multimap} \mathcal{L}_0)} = \S\mathcal{T}_{\S\mathcal{D}(\overleftarrow{\Phi_2})}(G_2) \texttt{ in } \mathcal{T}_{\S\mathcal{D}(\overleftarrow{\Phi_1})}(G_1)$$

which is a syntactic sugar for $(\lambda\overleftarrow{f}^{\ \S(\mathcal{H}_0 \overset{1}{\multimap} \mathcal{L}_0)}.\mathcal{T}_{\S\mathcal{D}(\overleftarrow{\Phi_1})}(G_1))\S\mathcal{T}_{\S\mathcal{D}(\overleftarrow{\Phi_2})}(G_2)$.

Let us analyze the type derivation for

$$\mathcal{D}(!\Sigma), \mathcal{D}(\S\overleftarrow{\Phi}) \vdash^{n} (\lambda\overleftarrow{f}^{\ \S(\mathcal{H}_0 \overset{1}{\multimap} \mathcal{L}_0)}.\mathcal{T}_{\S\mathcal{D}(\overleftarrow{\Phi_1})}(G_1))\S\mathcal{T}_{\S\mathcal{D}(\overleftarrow{\Phi_2})}(G_2) : \mathcal{H} \overset{1}{\multimap} \mathcal{L}$$

as follows

* By $\multimap_e$ we have

$$\mathcal{D}(!\Sigma) = \mathcal{D}(!\Sigma_1) \cup \mathcal{D}(!\Sigma_2)$$
$$\mathcal{D}(\S\overleftarrow{\Phi}) = \mathcal{D}(\S\overleftarrow{\Phi_1}), k \circledast \mathcal{D}(\S\overleftarrow{\Phi_2})$$

  such that

$$\mathcal{D}(!\Sigma_1), \mathcal{D}(\S\overleftarrow{\Phi_1}) \vdash^{n_1} \lambda\overleftarrow{f}^{\ \S(\mathcal{H}_0 \overset{1}{\multimap} \mathcal{L}_0)}.\mathcal{T}_{\S\mathcal{D}(\overleftarrow{\Phi_1})}(G_1) : (\mathcal{H}_0 \overset{1}{\multimap} \mathcal{L}_0) \overset{k}{\multimap} (\mathcal{H} \overset{1}{\multimap} \mathcal{L})$$
$$\mathcal{D}(!\Sigma_2), \mathcal{D}(\S\overleftarrow{\Phi_2}) \vdash^{n_2} \S\mathcal{T}_{\S\mathcal{D}(\overleftarrow{\Phi_2})}(G_2) : \S(\mathcal{H}_0 \overset{1}{\multimap} \mathcal{L}_0)$$

  and $n = k * n_2 + n_1$.

* By $\S_r$-derived typing rule we have $\mathcal{D}(!\Sigma_2), \mathcal{D}(\S\Phi_2) \vdash^{n_2'} \mathcal{T}_{\S\mathcal{D}(\overleftarrow{\Phi_2})}(G_2) : \mathcal{H}_0 \overset{1}{\multimap} \mathcal{L}_0$. By Lemma 47 we have that $n_2 = n_2'$ and so

$$n = k * n_2' + n_1$$

* By $\multimap_i$ we have

$$\mathcal{D}(!\Sigma_1), \mathcal{D}(\S\overleftarrow{\Phi_1}), \overleftarrow{f}^k : \mathcal{H}_0 \overset{1}{\multimap} \mathcal{L}_0 \vdash^{n_1'} \S\mathcal{T}_{\S\mathcal{D}(\overleftarrow{\Phi_1})}(G_1) : \mathcal{H} \overset{1}{\multimap} \mathcal{L}$$

  and $n_1 = n_1' + \mathcal{W}(FV(f) \setminus FV(\mathcal{T}_{\S\mathcal{D}(\overleftarrow{\Phi_1})}(G_1)))$.

  Moreover, $\mathcal{W}(FV(\overleftarrow{f}) \setminus FV(\mathcal{T}_{\S\mathcal{D}(\overleftarrow{\Phi_1})}(G_1))) = \mathcal{W}(\emptyset) = 0$ because in this subcase $\overleftarrow{f} \in FV(\mathcal{T}_{\S\mathcal{D}(\overleftarrow{\Phi_1})}(G_1))$, so we have $n_1 = n_1'$.

Summing up, we have

$$n = k * n_2' + n_1' \tag{6.72}$$

Observe that $\mathcal{T}_{\S\mathcal{D}(\overleftarrow{\Phi})}(F)$ is well-typed as

$$\mathcal{D}(!\Sigma_1) \cup \mathcal{D}(!\Sigma_2), \mathcal{D}(\S\overleftarrow{\Phi_1}), \mathcal{D}(\S\overleftarrow{\Phi_2}) \vdash^{n} \mathcal{T}_{\S\mathcal{D}(\overleftarrow{\Phi})}(F) : \mathcal{H} \overset{1}{\multimap} \mathcal{L}$$

so we have that

$$\begin{aligned}
\mathcal{W}\Big(\S\mathcal{D}(\overleftarrow{\Phi})^{\mathrm{in}}_{\mathcal{T}_{\S\mathcal{D}(\overleftarrow{\Phi})}(F)}\Big) &= \mathcal{W}\Big(\S\mathcal{D}(\overleftarrow{\Phi_1})^{\mathrm{in}}_{\mathcal{T}_{\S\mathcal{D}(\overleftarrow{\Phi_1})}(G_1)}\Big) + k * \mathcal{W}\Big(\S\mathcal{D}(\overleftarrow{\Phi_2})^{\mathrm{in}}_{\mathcal{T}_{\S\mathcal{D}(\overleftarrow{\Phi_2})}(G_2)}\Big) \\
\mathcal{W}\Big(\S\mathcal{D}(\overleftarrow{\Phi})^{\mathrm{out}}_{\mathcal{T}_{\S\mathcal{D}(\overleftarrow{\Phi})}(F)}\Big) &= \mathcal{W}\Big(\S\mathcal{D}(\overleftarrow{\Phi_1})^{\mathrm{out}}_{\mathcal{T}_{\S\mathcal{D}(\overleftarrow{\Phi_1})}(G_1)}\Big) + k * \mathcal{W}\Big(\S\mathcal{D}(\overleftarrow{\Phi_2})^{\mathrm{out}}_{\mathcal{T}_{\S\mathcal{D}(\overleftarrow{\Phi_2})}(G_2)}\Big)
\end{aligned} \tag{6.73}$$

By inductive hypothesis on $G_1$ (case 2 of the lemma) we have

$$\begin{aligned}
&n_1' + \mathcal{W}(\mathcal{L}) + \mathcal{W}\Big(\S\mathcal{D}(\overleftarrow{\Phi_1})^{\mathrm{in}}_{\mathcal{T}_{\S\mathcal{D}(\overleftarrow{\Phi_1})}(G_1)}\Big) + k * \mathcal{W}(\mathcal{H}_0) \\
&\qquad\qquad \leq m_1' + \mathcal{W}(\mathcal{H}) + \mathcal{W}\Big(\S\mathcal{D}(\overleftarrow{\Phi_1})^{\mathrm{out}}_{\mathcal{T}_{\S\mathcal{D}(\overleftarrow{\Phi_1})}(G_1)}\Big) + k * \mathcal{W}(\mathcal{L}_0)
\end{aligned}$$

By inductive hypothesis on $G_2$ (case 2 of the lemma) we have

$$n_2' + \mathcal{W}(\mathcal{L}_0) + \mathcal{W}\Big(\S\mathcal{D}(\overleftarrow{\Phi_2})^{\mathrm{in}}_{\mathcal{T}_{\S\mathcal{D}(\overleftarrow{\Phi_2})}(G_2)}\Big) \leq m_2' + \mathcal{W}(\mathcal{H}_0) + \mathcal{W}\Big(\S\mathcal{D}(\overleftarrow{\Phi_2})^{\mathrm{out}}_{\mathcal{T}_{\S\mathcal{D}(\overleftarrow{\Phi_2})}(G_2)}\Big)$$

Finally, we show that

$$n + \mathcal{W}(\mathcal{L}) + \mathcal{W}\Big(\S\mathcal{D}(\overleftarrow{\Phi})^{\mathrm{in}}_{\mathcal{T}_{\S\mathcal{D}(\overleftarrow{\Phi})}(F)}\Big) \leq m + \mathcal{W}(\mathcal{H}) + \mathcal{W}\Big(\S\mathcal{D}(\overleftarrow{\Phi})^{\mathrm{out}}_{\mathcal{T}_{\S\mathcal{D}(\overleftarrow{\Phi})}(F)}\Big)$$

as follows

$$\begin{aligned}
&n + \mathcal{W}(\mathcal{L}) + \mathcal{W}\Big(\S\mathcal{D}(\overleftarrow{\Phi})^{\mathrm{in}}_{\mathcal{T}_{\S\mathcal{D}(\overleftarrow{\Phi})}(F)}\Big) \\
&\overset{\text{Eq. 6.72}}{=} k * n_2' + n_1' + \mathcal{W}(\mathcal{L}) + \mathcal{W}\Big(\S\mathcal{D}(\overleftarrow{\Phi})^{\mathrm{in}}_{\mathcal{T}_{\S\mathcal{D}(\overleftarrow{\Phi})}(F)}\Big) \\
&\overset{\text{Eq. 6.73}}{=} k * n_2' + n_1' + \mathcal{W}(\mathcal{L}) + \mathcal{W}\Big(\S\mathcal{D}(\overleftarrow{\Phi_1})^{\mathrm{in}}_{\mathcal{T}_{\S\mathcal{D}(\overleftarrow{\Phi_1})}(G_1)}\Big) + k * \mathcal{W}\Big(\S\mathcal{D}(\overleftarrow{\Phi_2})^{\mathrm{in}}_{\mathcal{T}_{\S\mathcal{D}(\overleftarrow{\Phi_2})}(G_2)}\Big) \\
&= k * \Big(\underline{n_2' + \mathcal{W}\Big(\S\mathcal{D}(\overleftarrow{\Phi_2})^{\mathrm{in}}_{\mathcal{T}_{\S\mathcal{D}(\overleftarrow{\Phi_2})}(G_2)}\Big)}\Big) + n_1' + \mathcal{W}(\mathcal{L}) + \mathcal{W}\Big(\S\mathcal{D}(\overleftarrow{\Phi_1})^{\mathrm{in}}_{\mathcal{T}_{\S\mathcal{D}(\overleftarrow{\Phi_1})}(G_1)}\Big) \\
&\overset{\text{IH on } G_2}{\leq} k * \Big(m_2' + \mathcal{W}(\mathcal{H}_0) + \mathcal{W}\Big(\S\mathcal{D}(\overleftarrow{\Phi_2})^{\mathrm{out}}_{\mathcal{T}_{\S\mathcal{D}(\overleftarrow{\Phi_2})}(G_2)}\Big) - \mathcal{W}(\mathcal{L}_0)\Big) + \\
&\qquad + n_1' + \mathcal{W}(\mathcal{L}) + \mathcal{W}\Big(\S\mathcal{D}(\overleftarrow{\Phi_1})^{\mathrm{in}}_{\mathcal{T}_{\S\mathcal{D}(\overleftarrow{\Phi_1})}(G_1)}\Big) \\
&= k * m_2' + \underline{k * \mathcal{W}(\mathcal{H}_0)} + k * \mathcal{W}\Big(\S\mathcal{D}(\overleftarrow{\Phi_2})^{\mathrm{out}}_{\mathcal{T}_{\S\mathcal{D}(\overleftarrow{\Phi_2})}(G_2)}\Big) - k * \mathcal{W}(\mathcal{L}_0) \\
&\qquad + \underline{n_1' + \mathcal{W}(\mathcal{L}) + \mathcal{W}\Big(\S\mathcal{D}(\overleftarrow{\Phi_1})^{\mathrm{in}}_{\mathcal{T}_{\S\mathcal{D}(\overleftarrow{\Phi_1})}(G_1)}\Big)}
\end{aligned}$$

$$\overset{\text{IH on } G_1}{\leq} k * m_2' + k * \mathcal{W}\left(\S\mathcal{D}(\overleftarrow{\Phi_2})^{\text{out}}_{\mathcal{T}_{\S\mathcal{D}(\overleftarrow{\Phi_2})}(G_2)}\right) - k * \mathcal{W}(\mathcal{L}_0) +$$

$$+ m_1' + \mathcal{W}(\mathcal{H}) + \mathcal{W}\left(\S\mathcal{D}(\overleftarrow{\Phi_1})^{\text{out}}_{\mathcal{T}_{\S\mathcal{D}(\overleftarrow{\Phi_1})}(G_1)}\right) + k * \mathcal{W}(\mathcal{L}_0)$$

$$= k * m_2' + m_1' + \mathcal{W}(\mathcal{H}) + \mathcal{W}\left(\S\mathcal{D}(\overleftarrow{\Phi_1})^{\text{out}}_{\mathcal{T}_{\S\mathcal{D}(\overleftarrow{\Phi_1})}(G_1)}\right) + k * \mathcal{W}\left(\S\mathcal{D}(\overleftarrow{\Phi_2})^{\text{out}}_{\mathcal{T}_{\S\mathcal{D}(\overleftarrow{\Phi_2})}(G_2)}\right)$$

$$\overset{\text{Eq. 6.73}}{=} k * m_2' + m_1' + \mathcal{W}(\mathcal{H}) + \mathcal{W}\left(\S\mathcal{D}(\overleftarrow{\Phi})^{\text{in}}_{\mathcal{T}_{\S\mathcal{D}(\overleftarrow{\Phi})}(F)}\right)$$

$$\leq m + \mathcal{W}(\mathcal{H}) + \mathcal{W}\left(\S\mathcal{D}(\overleftarrow{\Phi})^{\text{out}}_{\mathcal{T}_{\S\mathcal{D}(\overleftarrow{\Phi})}(F)}\right)$$

$$\overset{\text{Eq. 6.70}}{=} k * m_2' + m_1 + \mathcal{W}(\mathcal{H}) + \mathcal{W}\left(\S\mathcal{D}(\overleftarrow{\Phi})^{\text{out}}_{\mathcal{T}_{\S\mathcal{D}(\overleftarrow{\Phi})}(F)}\right)$$

$$\overset{\text{Eq. 6.71}}{=} k * m_2' + m_1' + \mathcal{W}(\mathcal{H}) + \mathcal{W}\left(\S\mathcal{D}(\overleftarrow{\Phi})^{\text{out}}_{\mathcal{T}_{\S\mathcal{D}(\overleftarrow{\Phi})}(F)}\right)$$

– Subcase $f \notin FV(G_1)$:

Let us analyze the judgement $\mathcal{D}(!\Sigma_1), \mathcal{D}(\S\Phi_1) \vdash^{m_1} \lambda\S f^{\ \S(\mathcal{L}_0 \overset{1}{\multimap} \mathcal{H}_0)}.G_1 : (\mathcal{L}_0 \overset{1}{\multimap} \mathcal{H}_0) \overset{k}{\multimap} (\mathcal{L} \overset{1}{\multimap} \mathcal{H})$, by $\multimap_i$ we have

$$\mathcal{D}(!\Sigma_1), \mathcal{D}(\S\Phi_1), f^k : \mathcal{L}_0 \overset{1}{\multimap} \mathcal{H}_0 \vdash^{m_1'} G_1 : \mathcal{L} \overset{1}{\multimap} \mathcal{H}$$

and $m_1 = m_1' + \mathcal{W}(FV(f) \setminus FV(G_1))$. Moreover, $\mathcal{W}(FV(f) \setminus FV(G_1)) = \mathcal{W}(f^1 : \mathcal{L}_0 \overset{1}{\multimap} \mathcal{H}_0) = \mathcal{W}(\mathcal{L}_0) + \mathcal{W}(\mathcal{H}_0)$ because in this subcase $f \notin FV(G_1)$, so we have

$$m_1 = m_1' + \mathcal{W}(\mathcal{L}_0) + \mathcal{W}(\mathcal{H}_0) \tag{6.74}$$

By definition of transpose we have that

$$\mathcal{T}_{\S\mathcal{D}(\overleftarrow{\Phi})}(F) = \mathcal{T}_{\S\mathcal{D}(\overleftarrow{\Phi_1})}(G_1)$$

and $\mathcal{D}(!\Sigma_1), \mathcal{D}(\S\overleftrightarrow{\Phi}) \vdash^{n} \mathcal{T}_{\S\mathcal{D}(\overleftarrow{\Phi_1})}(G_1) : \mathcal{H} \overset{1}{\multimap} \mathcal{L}$ is a derivation for it.
Observe that by typing

$$\begin{aligned}
\mathcal{W}\left(\S\mathcal{D}(\overleftarrow{\Phi})^{\text{in}}_{\mathcal{T}_{\S\mathcal{D}(\overleftarrow{\Phi})}(F)}\right) &= \mathcal{W}\left(\S\mathcal{D}(\overleftarrow{\Phi_1})^{\text{in}}_{\mathcal{T}_{\S\mathcal{D}(\overleftarrow{\Phi_1})}(G_1)}\right) \\
\mathcal{W}\left(\S\mathcal{D}(\overleftarrow{\Phi})^{\text{out}}_{\mathcal{T}_{\S\mathcal{D}(\overleftarrow{\Phi})}(F)}\right) &= \mathcal{W}\left(\S\mathcal{D}(\overleftarrow{\Phi_1})^{\text{out}}_{\mathcal{T}_{\S\mathcal{D}(\overleftarrow{\Phi_1})}(G_1)}\right)
\end{aligned} \tag{6.75}$$

By inductive hypothesis on $G_1$ (case 2 of the lemma) we have

$$n + \mathcal{W}(\mathcal{L}) + \mathcal{W}\left(\S\mathcal{D}(\overleftarrow{\Phi_1})^{\text{in}}_{\mathcal{T}_{\S\mathcal{D}(\overleftarrow{\Phi_1})}(G_1)}\right) \leq m_1' + \mathcal{W}(\mathcal{H}) + \mathcal{W}\left(\S\mathcal{D}(\overleftarrow{\Phi_1})^{\text{out}}_{\mathcal{T}_{\S\mathcal{D}(\overleftarrow{\Phi_1})}(G_1)}\right)$$

Finally, we show that

$$n + \mathcal{W}(\mathcal{L}) + \mathcal{W}\left(\S\mathcal{D}(\overleftarrow{\Phi})^{\text{in}}_{\mathcal{T}_{\S\mathcal{D}(\overleftarrow{\Phi})}(F)}\right) \leq m + \mathcal{W}(\mathcal{H}) + \mathcal{W}\left(\S\mathcal{D}(\overleftarrow{\Phi})^{\text{out}}_{\mathcal{T}_{\S\mathcal{D}(\overleftarrow{\Phi})}(F)}\right)$$

as follows

$$n + \mathcal{W}(\mathcal{L}) + \mathcal{W}\left(\S\mathcal{D}(\overleftarrow{\Phi})^{\text{in}}_{\mathcal{T}_{\S\mathcal{D}(\overleftarrow{\Phi})}(F)}\right)$$

$$\stackrel{\text{Eq. 6.75}}{=} n + \mathcal{W}(\mathcal{L}) + \mathcal{W}\left(\S\mathcal{D}(\overleftarrow{\Phi_1})^{\text{in}}_{\mathcal{T}_{\S\mathcal{D}(\overleftarrow{\Phi_1})}}(G_1)\right)$$
$$\stackrel{\text{IH on } G_1}{\leq} m_1' + \mathcal{W}(\mathcal{H}) + \mathcal{W}\left(\S\mathcal{D}(\overleftarrow{\Phi_1})^{\text{out}}_{\mathcal{T}_{\S\mathcal{D}(\overleftarrow{\Phi_1})}}(G_1)\right)$$
$$\stackrel{\text{Eq. 6.75}}{=} m_1' + \mathcal{W}(\mathcal{H}) + \mathcal{W}\left(\S\mathcal{D}(\overleftarrow{\Phi})^{\text{in}}_{\mathcal{T}_{\S\mathcal{D}(\overleftarrow{\Phi})}}(F)\right)$$
$$\stackrel{\text{Eq. 6.74}}{\leq} m_1 + \mathcal{W}(\mathcal{H}) + \mathcal{W}\left(\S\mathcal{D}(\overleftarrow{\Phi})^{\text{in}}_{\mathcal{T}_{\S\mathcal{D}(\overleftarrow{\Phi})}}(F)\right)$$
$$\leq m + \mathcal{W}(\mathcal{H}) + \mathcal{W}\left(\S\mathcal{D}(\overleftarrow{\Phi})^{\text{out}}_{\mathcal{T}_{\S\mathcal{D}(\overleftarrow{\Phi})}}(F)\right)$$
$$\stackrel{\text{Eq. 6.70}}{=} k * m_2' + m_1 + \mathcal{W}(\mathcal{H}) + \mathcal{W}\left(\S\mathcal{D}(\overleftarrow{\Phi})^{\text{out}}_{\mathcal{T}_{\S\mathcal{D}(\overleftarrow{\Phi})}}(F)\right)$$

and we can conclude as $k \in \mathbb{N}^{>0}$.

□

*Proof Quantitative Claim 3: Cases of $\mathcal{T}$ on $\lambda LL^{\mathtt{A}}$.* Recall that we have to show

$$n + \mathcal{W}(\mathcal{L}) + \mathcal{W}\left(\S\mathcal{D}(\overleftarrow{\Phi})^{\text{in}}_{\mathcal{T}_{\mathcal{D}(\S\overleftarrow{\Phi})}}(R)\right) \leq m + \mathcal{W}(\mathcal{H}) + \mathcal{W}\left(\S\mathcal{D}(\overleftarrow{\Phi})^{\text{out}}_{\mathcal{T}_{\mathcal{D}(\S\overleftarrow{\Phi})}}(R)\right)$$

We proceed by analyzing the cases in Figure 5.4a.

More precisely, we consider the case $R = \mathtt{let}\ (!x, \S f^{\S(\mathcal{H}_0 \overset{1}{\multimap} \mathcal{L}_0)}) = S_2\ \mathtt{in}\ S_1$.

Recall that $\mathtt{let}\ (!x^{!\mathcal{E}_0}, \S f^{\S(\mathcal{L}_0 \overset{1}{\multimap} \mathcal{H}_0)}) = S_2\ \mathtt{in}\ S_1$ is a syntactic sugar for $(\lambda(!x^{!\mathcal{E}_0}, \S f^{\S(\mathcal{L}_0 \overset{1}{\multimap} \mathcal{H}_0)}).S_1)S_2$. Let us analyze the type derivation for

$$\mathcal{D}(!\Sigma), \mathcal{D}(\S\Phi) \vdash^{m} (\lambda(!x^{!\mathcal{E}_0}, \S f^{\S(\mathcal{L}_0 \overset{1}{\multimap} \mathcal{H}_0)}).S_1)S_2 : !\mathcal{E} \otimes (\mathcal{L} \overset{1}{\multimap} \mathcal{H})$$

as follows

- By $\multimap_e$-rule we have

  $$\mathcal{D}(!\Sigma) = \mathcal{D}(!\Sigma_1) \cup \mathcal{D}(!\Sigma_2)$$
  $$\mathcal{D}(\S\Phi) = \mathcal{D}(\S\Phi_1), k \circledast \mathcal{D}(\S\Phi_2)$$

  such that

  $$\mathcal{D}(!\Sigma_1), \mathcal{D}(\S\Phi_1) \vdash^{m_1} \lambda(!x^{!\mathcal{E}_0}, \S f^{\S(\mathcal{L}_0 \overset{1}{\multimap} \mathcal{H}_0)}).S_1 : (!\mathcal{E}_0 \otimes (\mathcal{L}_0 \overset{1}{\multimap} \mathcal{H}_0)) \overset{k}{\multimap} (!\mathcal{E} \otimes (\mathcal{L} \overset{1}{\multimap} \mathcal{H}))$$
  $$\mathcal{D}(!\Sigma_2), \mathcal{D}(\S\Phi_2) \vdash^{m_2} S_2 : !\mathcal{E}_0 \otimes (\mathcal{L}_0 \overset{1}{\multimap} \mathcal{H}_0)$$

  and $m = k * m_2 + m_1$.

- By $\multimap_i$-rule applied to

  $$\mathcal{D}(!\Sigma_1), \mathcal{D}(\S\Phi_1) \vdash^{m_1} \lambda(!x^{!\mathcal{E}_0}, \S f^{\S(\mathcal{L}_0 \overset{1}{\multimap} \mathcal{H}_0)}).S_1 : (!\mathcal{E}_0 \otimes (\mathcal{L}_0 \overset{1}{\multimap} \mathcal{H}_0)) \overset{k}{\multimap} (!\mathcal{E} \otimes (\mathcal{L} \overset{1}{\multimap} \mathcal{H}))$$

  we have

  $$\mathcal{D}(!\Sigma_1), \mathcal{D}(\S\Phi_1), (!x, \S f)^k : !\mathcal{E}_0 \otimes (\mathcal{L}_0 \overset{1}{\multimap} \mathcal{H}_0) \vdash^{m_1'} S_1 : !\mathcal{E} \otimes (\mathcal{L} \overset{1}{\multimap} \mathcal{H})$$

  and $m_1 = m_1' + \mathcal{W}(FV((!x, \S f)) \setminus FV(S_1)) = m_1' + \mathcal{W}(FV(f) \setminus FV(S_1))$

- By $\otimes_e$-rule we have

$$\mathcal{D}(!\Sigma_1), \mathcal{D}(\S\Phi_1), x^{k_1} : !\mathcal{E}_0, f^{k_2} : \mathcal{L}_0 \overset{1}{\multimap} \mathcal{H}_0 \vdash^{m'_1} S_1 : !\mathcal{E} \otimes (\mathcal{L} \overset{1}{\multimap} \mathcal{H})$$

and $k = max(k_1, k_2)$. Observe that $k_1 = 1$ because by definition of decoration of pattern with exponential type, so $k = max(1, k_2) = k_2$ as $k_2 \in \mathbb{N}^{>0}$.

Summing up, we have that

$$m = k * m_2 + m'_1 + \mathcal{W}(FV(f) \setminus FV(S_1)) \tag{6.76}$$

$$k = k_2 \tag{6.77}$$

Observe that $f$ may not be free in $S_1$, so we have to analyze the following subcases:

- Subcase $f \in FV(S_1)$:
By definition of transpose we have:

$$\mathcal{T}_{\S\overleftarrow{\Phi}}(\texttt{let } (!x, \S f) = S_2 \texttt{ in } S_1) = \texttt{let } (!x, \S \overleftarrow{f}) = \mathcal{T}_{\S\overleftarrow{\Phi_2}}(S_2) \texttt{ in } \mathcal{T}_{\S\overleftarrow{\Phi_1}}(S_1)$$

which is a syntactic sugar for $(\lambda(!x^{!\mathcal{E}_0}, \S \overleftarrow{f}^{\S(\mathcal{H}_0 \overset{1}{\multimap} \mathcal{L}_0)}).\mathcal{T}_{\S\overleftarrow{\Phi_1}}(S_1))\mathcal{T}_{\S\overleftarrow{\Phi_2}}(S_2)$.

Let us analyze the type derivation for

$$\mathcal{D}(!\Sigma), \mathcal{D}(\S\overleftarrow{\Phi}) \vdash^{n} (\lambda(!x^{!\mathcal{E}_0}, \S \overleftarrow{f}^{\S(\mathcal{H}_0 \overset{1}{\multimap} \mathcal{L}_0)}).\mathcal{T}_{\S\overleftarrow{\Phi_1}}(S_1))\mathcal{T}_{\S\overleftarrow{\Phi_2}}(S_2) : !\mathcal{E} \otimes (\mathcal{H} \overset{1}{\multimap} \mathcal{L})$$

as follows

  - By $\multimap_e$-rule we have

$$\mathcal{D}(!\Sigma) = \mathcal{D}(!\Sigma_1) \cup \mathcal{D}(!\Sigma_2)$$
$$\mathcal{D}(\S\overleftarrow{\Phi}) = \mathcal{D}(\S\overleftarrow{\Phi}_1), k \circledast \mathcal{D}(\S\overleftarrow{\Phi}_2)$$

  such that

$$\mathcal{D}(!\Sigma_1), \mathcal{D}(\S\overleftarrow{\Phi_1}) \vdash^{n_1} \lambda(!x^{!\mathcal{E}_0}, \S \overleftarrow{f}^{\S(\mathcal{H}_0 \overset{1}{\multimap} \mathcal{L}_0)}).\mathcal{T}_{\S\overleftarrow{\Phi_1}}(S_1) : (!\mathcal{E}_0 \otimes (\mathcal{H}_0 \overset{1}{\multimap} \mathcal{L}_0)) \overset{k}{\multimap} (!\mathcal{E} \otimes (\mathcal{H} \overset{1}{\multimap} \mathcal{L}))$$

$$\mathcal{D}(!\Sigma_2), \mathcal{D}(\S\overleftarrow{\Phi_2}) \vdash^{n_2} \mathcal{T}_{\S\overleftarrow{\Phi_2}}(S_2) : !\mathcal{E}_0 \otimes (\mathcal{H}_0 \overset{1}{\multimap} \mathcal{L}_0)$$

  and $n = k * n_2 + n_1$.

  - By $\multimap_i$-rule applied to

$$\mathcal{D}(!\Sigma_1), \mathcal{D}(\S\overleftarrow{\Phi_1}) \vdash^{n_1} \lambda(!x^{!\mathcal{E}_0}, \S \overleftarrow{f}^{\S(\mathcal{H}_0 \overset{1}{\multimap} \mathcal{L}_0)}).\mathcal{T}_{\S\overleftarrow{\Phi_1}}(S_1) : (!\mathcal{E}_0 \otimes (\mathcal{H}_0 \overset{1}{\multimap} \mathcal{L}_0)) \overset{k}{\multimap} (!\mathcal{E} \otimes (\mathcal{H} \overset{1}{\multimap} \mathcal{L}))$$

  we have

$$\mathcal{D}(!\Sigma_1), \mathcal{D}(\S\overleftarrow{\Phi_1}), (!x, \S \overleftarrow{f})^{k} : !\mathcal{E}_0 \otimes (\mathcal{H}_0 \overset{1}{\multimap} \mathcal{L}_0) \vdash^{n'_1} \mathcal{T}_{\S\overleftarrow{\Phi_1}}(S_1) : !\mathcal{E} \otimes (\mathcal{H} \overset{1}{\multimap} \mathcal{L})$$

  and $n_1 = n'_1 + \mathcal{W}(FV((!x, \S \overleftarrow{f})) \setminus FV(S_1))$. Moreover, $\mathcal{W}(FV((!x, \S \overleftarrow{f})) \setminus FV(S_1)) = \mathcal{W}(FV(\S \overleftarrow{f}) \setminus FV(S_1)) = \mathcal{W}(\emptyset) = 0$ since in this subcase we are supposing $f \in FV(S_1)$ so we have also that $\overleftarrow{f} \in FV(S_1)$. Therefore, in this subcase we have $n_1 = n'_1$.

– By $\otimes_e$-rule we have

$$\mathcal{D}(!\Sigma_1), \mathcal{D}(\S\overleftarrow{\Phi_1}), !x^{k_1} : !\mathcal{E}_0, \S f^{k_2} : \S(\mathcal{H}_0 \overset{1}{\multimap} \mathcal{L}_0) \vdash^{m'_1} \mathcal{T}_{\S\overleftarrow{\Phi_1}}(S_1) : !\mathcal{E} \otimes (\mathcal{H} \overset{1}{\multimap} \mathcal{L})$$

and $k = max(k_1, k_2)$. Observe that $k_1 = 1$ because by definition of decoration of pattern with exponential type, so $k = max(1, k_2) = k_2$ as $k_2 \in \mathbb{N}^{>0}$.

Summing up, we have that

$$n = k * n_2 + n'_1 \tag{6.78}$$

$$k = k_2 \tag{6.79}$$

Furthermore, recall that

$$m \overset{\text{Eq. 6.76}}{=} k * m_2 + m'_1 + \mathcal{W}(FV(f) \setminus FV(S_1)) = k * m_2 + m'_1$$

since in this subcase we are supposing $f \in FV(S_1)$. Therefore, we have

$$m = k * m_2 + m'_1 \tag{6.80}$$

Observe that $\mathcal{T}_{\S\overleftarrow{\Phi}}(R)$ is well-typed as

$$\mathcal{D}(!\Sigma_1) \cup \mathcal{D}(!\Sigma_2), \mathcal{D}(\S\overleftarrow{\Phi_1}), k \circledast \mathcal{D}(\S\overleftarrow{\Phi_2}) \vdash^{n} \mathcal{T}_{\S\overleftarrow{\Phi}}(R) : !\mathcal{E} \otimes (\mathcal{H} \overset{1}{\multimap} \mathcal{L})$$

and by Equation 6.79 we have $k = k_2$. Therefore, we have that

$$\begin{aligned} \mathcal{W}\Big(\S\mathcal{D}(\overleftarrow{\Phi})^{\text{in}}_{\mathcal{T}_{\S\mathcal{D}(\overleftarrow{\Phi})}(F)}\Big) &= \mathcal{W}\Big(\S\mathcal{D}(\overleftarrow{\Phi_1})^{\text{in}}_{\mathcal{T}_{\S\mathcal{D}(\overleftarrow{\Phi_1})}(S_1)}\Big) + k_2 * \mathcal{W}\Big(\S\mathcal{D}(\overleftarrow{\Phi_2})^{\text{in}}_{\mathcal{T}_{\S\mathcal{D}(\overleftarrow{\Phi_2})}(S_2)}\Big) \\ \mathcal{W}\Big(\S\mathcal{D}(\overleftarrow{\Phi})^{\text{out}}_{\mathcal{T}_{\S\mathcal{D}(\overleftarrow{\Phi})}(F)}\Big) &= \mathcal{W}\Big(\S\mathcal{D}(\overleftarrow{\Phi_1})^{\text{out}}_{\mathcal{T}_{\S\mathcal{D}(\overleftarrow{\Phi_1})}(S_1)}\Big) + k_2 * \mathcal{W}\Big(\S\mathcal{D}(\overleftarrow{\Phi_2})^{\text{out}}_{\mathcal{T}_{\S\mathcal{D}(\overleftarrow{\Phi_2})}(S_2)}\Big) \end{aligned} \tag{6.81}$$

By inductive hypothesis on $S_1$ (case 2 of the lemma) we have

$$\begin{aligned} & n'_1 + \mathcal{W}(\mathcal{L}) + \mathcal{W}\Big(\S\mathcal{D}(\overleftarrow{\Phi_1})^{\text{in}}_{\mathcal{T}_{\S\mathcal{D}(\overleftarrow{\Phi_1})}(S_1)}\Big) + k_2 * \mathcal{W}(\mathcal{H}_0) \\ & \qquad \leq m'_1 + \mathcal{W}(\mathcal{H}) + \mathcal{W}\Big(\S\mathcal{D}(\overleftarrow{\Phi_1})^{\text{out}}_{\mathcal{T}_{\S\mathcal{D}(\overleftarrow{\Phi_1})}(S_1)}\Big) + k_2 * \mathcal{W}(\mathcal{L}_0) \end{aligned}$$

By inductive hypothesis on $S_2$ (case 2 of the lemma) we have

$$n_2 + \mathcal{W}(\mathcal{L}_0) + \mathcal{W}\Big(\S\mathcal{D}(\overleftarrow{\Phi_2})^{\text{in}}_{\mathcal{T}_{\S\mathcal{D}(\overleftarrow{\Phi_2})}(S_2)}\Big) \leq m_2 + \mathcal{W}(\mathcal{H}_0) + \mathcal{W}\Big(\S\mathcal{D}(\overleftarrow{\Phi_2})^{\text{out}}_{\mathcal{T}_{\S\mathcal{D}(\overleftarrow{\Phi_2})}(S_2)}\Big)$$

Finally, we show that

$$n + \mathcal{W}(\mathcal{L}) + \mathcal{W}\Big(\S\mathcal{D}(\overleftarrow{\Phi})^{\text{in}}_{\mathcal{T}_{\mathcal{D}(\S\overleftarrow{\Phi})}(R)}\Big) \leq m + \mathcal{W}(\mathcal{H}) + \mathcal{W}\Big(\S\mathcal{D}(\overleftarrow{\Phi})^{\text{out}}_{\mathcal{T}_{\mathcal{D}(\S\overleftarrow{\Phi})}(R)}\Big)$$

as follows

$$n + \mathcal{W}(\mathcal{L}) + \mathcal{W}\Big(\S\mathcal{D}(\overleftarrow{\Phi})^{\text{in}}_{\mathcal{T}_{\S\mathcal{D}(\overleftarrow{\Phi})}(R)}\Big)$$

$$
\begin{aligned}
&\overset{\text{Eq. 6.78}}{=} k * n_2 + n_1' + \mathcal{W}(\mathcal{L}) + \mathcal{W}\Big(\S\mathcal{D}(\overleftarrow{\Phi})^{\text{in}}_{\mathcal{T}_{\S\mathcal{D}(\overleftarrow{\Phi})}}(R)\Big) \\
&\overset{\text{Eq. 6.79}}{=} k_2 * n_2 + n_1' + \mathcal{W}(\mathcal{L}) + \mathcal{W}\Big(\S\mathcal{D}(\overleftarrow{\Phi})^{\text{in}}_{\mathcal{T}_{\S\mathcal{D}(\overleftarrow{\Phi})}}(R)\Big) \\
&\overset{\text{Eq. 6.81}}{=} k_2 * n_2 + n_1' + \mathcal{W}(\mathcal{L}) + \mathcal{W}\Big(\S\mathcal{D}(\overleftarrow{\Phi_1})^{\text{in}}_{\mathcal{T}_{\S\mathcal{D}(\overleftarrow{\Phi_1})}}(S_1)\Big) \\
&\qquad\qquad + k_2 * \mathcal{W}\Big(\S\mathcal{D}(\overleftarrow{\Phi_2})^{\text{in}}_{\mathcal{T}_{\S\mathcal{D}(\overleftarrow{\Phi_2})}}(S_2)\Big) \\
&= k_2 * \Big(\underline{n_2 + \mathcal{W}\Big(\S\mathcal{D}(\overleftarrow{\Phi_2})^{\text{in}}_{\mathcal{T}_{\S\mathcal{D}(\overleftarrow{\Phi_2})}}(S_2)\Big)}\Big) + n_1' + \mathcal{W}(\mathcal{L}) + \mathcal{W}\Big(\S\mathcal{D}(\overleftarrow{\Phi_1})^{\text{in}}_{\mathcal{T}_{\S\mathcal{D}(\overleftarrow{\Phi_1})}}(S_1)\Big) \\
&\overset{\text{IH on } S_2}{\leq} k_2 * \Big(m_2 + \mathcal{W}(\mathcal{H}_0) + \mathcal{W}\Big(\S\mathcal{D}(\overleftarrow{\Phi_2})^{\text{out}}_{\mathcal{T}_{\S\mathcal{D}(\overleftarrow{\Phi_2})}}(S_2)\Big) - \mathcal{W}(\mathcal{L}_0)\Big) + \\
&\qquad\quad + n_1' + \mathcal{W}(\mathcal{L}) + \mathcal{W}\Big(\S\mathcal{D}(\overleftarrow{\Phi_1})^{\text{in}}_{\mathcal{T}_{\S\mathcal{D}(\overleftarrow{\Phi_1})}}(S_1)\Big) \\
&= k_2 * m_2 + \underline{k_2 * \mathcal{W}(\mathcal{H}_0)} + k_2 * \mathcal{W}\Big(\S\mathcal{D}(\overleftarrow{\Phi_2})^{\text{out}}_{\mathcal{T}_{\S\mathcal{D}(\overleftarrow{\Phi_2})}}(S_2)\Big) - k_2 * \mathcal{W}(\mathcal{L}_0) + \\
&\qquad \underline{+ n_1' + \mathcal{W}(\mathcal{L}) + \mathcal{W}\Big(\S\mathcal{D}(\overleftarrow{\Phi_1})^{\text{in}}_{\mathcal{T}_{\S\mathcal{D}(\overleftarrow{\Phi_1})}}(S_1)\Big)} \\
&\overset{\text{IH on } S_1}{\leq} k_2 * m_2 + k_2 * \mathcal{W}\Big(\S\mathcal{D}(\overleftarrow{\Phi_2})^{\text{out}}_{\mathcal{T}_{\S\mathcal{D}(\overleftarrow{\Phi_2})}}(S_2)\Big) - k_2 * \mathcal{W}(\mathcal{L}_0) + \\
&\qquad\quad + m_1' + \mathcal{W}(\mathcal{H}) + \mathcal{W}\Big(\S\mathcal{D}(\overleftarrow{\Phi_1})^{\text{out}}_{\mathcal{T}_{\S\mathcal{D}(\overleftarrow{\Phi_1})}}(S_1)\Big) + k_2 * \mathcal{W}(\mathcal{L}_0) \\
&= k_2 * m_2 + k_2 * \mathcal{W}\Big(\S\mathcal{D}(\overleftarrow{\Phi_2})^{\text{out}}_{\mathcal{T}_{\S\mathcal{D}(\overleftarrow{\Phi_2})}}(S_2)\Big) + m_1' + \mathcal{W}(\mathcal{H}) \\
&\qquad\quad + \mathcal{W}\Big(\S\mathcal{D}(\overleftarrow{\Phi_1})^{\text{out}}_{\mathcal{T}_{\S\mathcal{D}(\overleftarrow{\Phi_1})}}(S_1)\Big) \\
&\overset{\text{Eq. 6.81}}{=} k_2 * m_2 + m_1' + \mathcal{W}(\mathcal{H}) + \mathcal{W}\Big(\S\mathcal{D}(\overleftarrow{\Phi})^{\text{out}}_{\mathcal{T}_{\S\mathcal{D}(\overleftarrow{\Phi})}}(R)\Big) \\
&\leq m + \mathcal{W}(\mathcal{H}) + \mathcal{W}\Big(\S\mathcal{D}(\overleftarrow{\Phi})^{\text{out}}_{\mathcal{T}_{\mathcal{D}(\S\overleftarrow{\Phi})}}(R)\Big) \\
&\overset{\text{Eq. 6.80}}{=} k * m_2 + m_1' + \mathcal{W}(\mathcal{H}) + \mathcal{W}\Big(\S\mathcal{D}(\overleftarrow{\Phi})^{\text{out}}_{\mathcal{T}_{\mathcal{D}(\S\overleftarrow{\Phi})}}(R)\Big) \\
&\overset{\text{Eq. 6.79}}{=} k_2 * m_2 + m_1' + \mathcal{W}(\mathcal{H}) + \mathcal{W}\Big(\S\mathcal{D}(\overleftarrow{\Phi})^{\text{out}}_{\mathcal{T}_{\mathcal{D}(\S\overleftarrow{\Phi})}}(R)\Big)
\end{aligned}
$$

- Subcase $f \notin FV(S_1)$:
  By definition of transpose we have:

  $$\mathcal{T}_{\S\overleftarrow{\Phi}}(\texttt{let}\ (!x, \S f) = S_2\ \texttt{in}\ S_1) = \texttt{let}\ !x = \epsilon[P]\ \texttt{in}\ \mathcal{T}_{\S\overleftarrow{\Phi}}(S_2) \qquad \text{for } \mathcal{U}^{\bullet}(S_2) = (\epsilon_1[], P, F)$$

  which is a syntactic sugar for $(\lambda !x^{!\mathcal{E}_0}.\mathcal{T}_{\S\overleftarrow{\Phi}}(S_1))\ \epsilon[P]$.

  Let us analyze the type derivation for

  $$\mathcal{D}(!\Sigma), \mathcal{D}(\S\overleftarrow{\Phi}) \vdash^{n} (\lambda !x^{!\mathcal{E}_0}.\mathcal{T}_{\S\overleftarrow{\Phi}}(S_1))\ \epsilon[P] : !\mathcal{E} \otimes (\mathcal{H} \overset{1}{\multimap} \mathcal{L})$$

  as follows

– By $\multimap_e$-rule we have

$$\mathcal{D}(!\Sigma) = \mathcal{D}(!\Sigma_1) \cup \mathcal{D}(!\Sigma_2)$$

such that

$$\mathcal{D}(!\Sigma_1), \mathcal{D}(\S\overleftarrow{\Phi}) \vdash^{n_1} \lambda !x^{!\mathcal{E}_0}.\mathcal{T}_{\S\overleftarrow{\Phi}}(S_1) : !\mathcal{E}_0 \overset{k'}{\multimap} (!\mathcal{E} \otimes (\mathcal{H} \overset{1}{\multimap} \mathcal{L}))$$
$$\mathcal{D}(!\Sigma_2), \mathcal{D}(\S\overleftarrow{\Phi}) \vdash^{n_2} \epsilon[P] : !\mathcal{E}_0$$

and

$$n = k' * n_2 + n_1 \tag{6.82}$$

Moreover, by Lemma 48 we have that

$$n_2 \leq m_2 \tag{6.83}$$

– By $\multimap_i$-rule applied to $\mathcal{D}(!\Sigma_1), \mathcal{D}(\S\overleftarrow{\Phi}) \vdash^{n_1} \lambda !x^{!\mathcal{E}_0}.\mathcal{T}_{\S\overleftarrow{\Phi}}(S_1) : !\mathcal{E}_0 \overset{k'}{\multimap} (!\mathcal{E} \otimes (\mathcal{H} \overset{1}{\multimap} \mathcal{L}))$ we have

$$\mathcal{D}(!\Sigma_1), !x^{k'} : !\mathcal{E}_0, \mathcal{D}(\S\overleftarrow{\Phi}) \vdash^{n'_1} \lambda !x^{!\mathcal{E}_0}.\mathcal{T}_{\S\overleftarrow{\Phi}}(S_1) : !\mathcal{E} \otimes (\mathcal{H} \overset{1}{\multimap} \mathcal{L})$$

and $n_1 = n'_1 + \mathcal{W}(FV(!x) \setminus FV(\mathcal{T}_{\S\overleftarrow{\Phi}}(S_1)))$. Moreover, we have that $\mathcal{W}(FV(!x) \setminus FV(\mathcal{T}_{\S\overleftarrow{\Phi}}(S_1))) = \mathcal{W}(!x : !\mathcal{E}) = 0$ because in our definition of workload of a type we do not count the occurrences of $\mathbb{R}$ under the scope of a !. Therefore, we have that

$$n_1 = n'_1 \tag{6.84}$$

Observe also that by definition of decoration of a pattern with exponential type we have

$$k' = 1 \tag{6.85}$$

Summing up, we have that

$$n \overset{\text{Eq. 6.82}}{=} k' * n_2 + n_1 \overset{\text{Eq. 6.85}}{=} n_2 + n_1 \overset{\text{Eq. 6.84}}{=} n_2 + n'_1$$

so we have

$$n = n'_1 + n_2 \tag{6.86}$$

Observe that $\mathcal{T}_{\S\overleftarrow{\Phi}}(R)$ is well-typed as

$$\mathcal{D}(!\Sigma_1) \cup \mathcal{D}(!\Sigma_2), \mathcal{D}(\S\overleftarrow{\Phi}) \vdash^{n} \mathcal{T}_{\S\overleftarrow{\Phi}}(R) : !\mathcal{E} \otimes (\mathcal{H} \overset{1}{\multimap} \mathcal{L})$$

so by typing we have that

$$\begin{aligned} \mathcal{W}\Big(\S\mathcal{D}(\overleftarrow{\Phi})^{\text{in}}_{\mathcal{T}_{\S\mathcal{D}(\overleftarrow{\Phi})}(F)}\Big) &= \mathcal{W}\Big(\S\mathcal{D}(\overleftarrow{\Phi})^{\text{in}}_{\mathcal{T}_{\S\mathcal{D}(\overleftarrow{\Phi})}(S_1)}\Big) \\ \mathcal{W}\Big(\S\mathcal{D}(\overleftarrow{\Phi})^{\text{out}}_{\mathcal{T}_{\S\mathcal{D}(\overleftarrow{\Phi})}(F)}\Big) &= \mathcal{W}\Big(\S\mathcal{D}(\overleftarrow{\Phi})^{\text{out}}_{\mathcal{T}_{\S\mathcal{D}(\overleftarrow{\Phi})}(S_1)}\Big) \end{aligned} \tag{6.87}$$

Recall that

$$m \overset{\text{Eq. 6.76}}{=} k * m_2 + m'_1 + \mathcal{W}(FV(f) \setminus FV(S_1)) \overset{\text{Eq. 6.77}}{=} k_2 * m_2 + m'_1 + \mathcal{W}(FV(f) \setminus FV(S_1))$$

Moreover, $\mathcal{W}(FV(f)\backslash FV(S_1)) = \mathcal{W}(f : \mathcal{L}_0 \overset{1}{\multimap} \mathcal{H}_0) = \mathcal{W}(\mathcal{L}_0)+\mathcal{W}(\mathcal{H}_0)$ since in this subcase we are supposing $f \notin FV(S_1)$. Therefore, in this case we have

$$m = k_2 * m_2 + m_1' + \mathcal{W}(\mathcal{L}_0) + \mathcal{W}(\mathcal{H}_0) \tag{6.88}$$

By inductive hypothesis on $S_1$ (case 2 of the lemma) we have

$$n_1' + \mathcal{W}(\mathcal{L}) + \mathcal{W}\Big(\S\mathcal{D}(\overleftarrow{\Phi_1})^{\text{in}}_{\mathcal{T}_{\S\mathcal{D}(\overleftarrow{\Phi_1})}(S_1)}\Big) \leq m_1' + \mathcal{W}(\mathcal{H}) + \mathcal{W}\Big(\S\mathcal{D}(\overleftarrow{\Phi_1})^{\text{out}}_{\mathcal{T}_{\S\mathcal{D}(\overleftarrow{\Phi_1})}(S_1)}\Big)$$

Finally, we show that

$$n + \mathcal{W}(\mathcal{L}) + \mathcal{W}\Big(\S\mathcal{D}(\overleftarrow{\Phi})^{\text{in}}_{\mathcal{T}_{\mathcal{D}(\S\overleftarrow{\Phi})}(R)}\Big) \leq m + \mathcal{W}(\mathcal{H}) + \mathcal{W}\Big(\S\mathcal{D}(\overleftarrow{\Phi})^{\text{out}}_{\mathcal{T}_{\mathcal{D}(\S\overleftarrow{\Phi})}(R)}\Big)$$

as follows

$$\begin{aligned}
& n + \mathcal{W}(\mathcal{L}) + \mathcal{W}\Big(\S\mathcal{D}(\overleftarrow{\Phi})^{\text{in}}_{\mathcal{T}_{\mathcal{D}(\S\overleftarrow{\Phi})}(R)}\Big) \\
& \overset{\text{Eq. 6.86}}{=} n_1' + n_2 + \mathcal{W}(\mathcal{L}) + \mathcal{W}\Big(\S\mathcal{D}(\overleftarrow{\Phi})^{\text{in}}_{\mathcal{T}_{\mathcal{D}(\S\overleftarrow{\Phi})}(R)}\Big) \\
& \overset{\text{Eq. 6.87}}{=} n_1' + n_2 + \mathcal{W}(\mathcal{L}) + \mathcal{W}\Big(\S\mathcal{D}(\overleftarrow{\Phi})^{\text{in}}_{\mathcal{T}_{\mathcal{D}(\S\overleftarrow{\Phi})}(S_1)}\Big) \\
& \overset{\text{IH on } S_1}{\leq} n_2 + m_1' + \mathcal{W}(\mathcal{H}) + \mathcal{W}\Big(\S\mathcal{D}(\overleftarrow{\Phi})^{\text{in}}_{\mathcal{T}_{\mathcal{D}(\S\overleftarrow{\Phi})}(S_1)}\Big) \\
& \overset{\text{Eq. 6.83}}{\leq} m_2 + m_1' + \mathcal{W}(\mathcal{H}) + \mathcal{W}\Big(\S\mathcal{D}(\overleftarrow{\Phi})^{\text{in}}_{\mathcal{T}_{\mathcal{D}(\S\overleftarrow{\Phi})}(S_1)}\Big) \\
& \leq m + \mathcal{W}(\mathcal{H}) + \mathcal{W}\Big(\S\mathcal{D}(\overleftarrow{\Phi})^{\text{out}}_{\mathcal{T}_{\mathcal{D}(\S\overleftarrow{\Phi})}(R)}\Big) \\
& \overset{\text{Eq. 6.88}}{=} k_2 * m_2 + m_1' + \mathcal{W}(\mathcal{L}_0) + \mathcal{W}(\mathcal{H}_0) + \mathcal{W}(\mathcal{H}) + \mathcal{W}\Big(\S\mathcal{D}(\overleftarrow{\Phi})^{\text{out}}_{\mathcal{T}_{\mathcal{D}(\S\overleftarrow{\Phi})}(R)}\Big)
\end{aligned}$$

and we can conclude as $k_2 \in \mathbb{N}^{>0}$.

□

# Conclusion and Future Perspectives of ADLL

Figure 6.2 summarises our main contributions. We have a linear $\lambda$-calculus $\lambda$LL with well-behaved $\beta$-reduction and a logical relation $\sim$ that compares programs with respect to their extensional behaviour on the ground types. We have defined a $\lambda$LL encoding $\delta$ of the Linear A system and the three transformations (forward $\mathcal{F}^{\mathtt{Jax}}$, unzipping $\mathcal{U}^{\mathtt{Jax}}$, transpose $\mathcal{T}^{\mathtt{Jax}}$) formalising AD implementation in libraries like JAX [93].

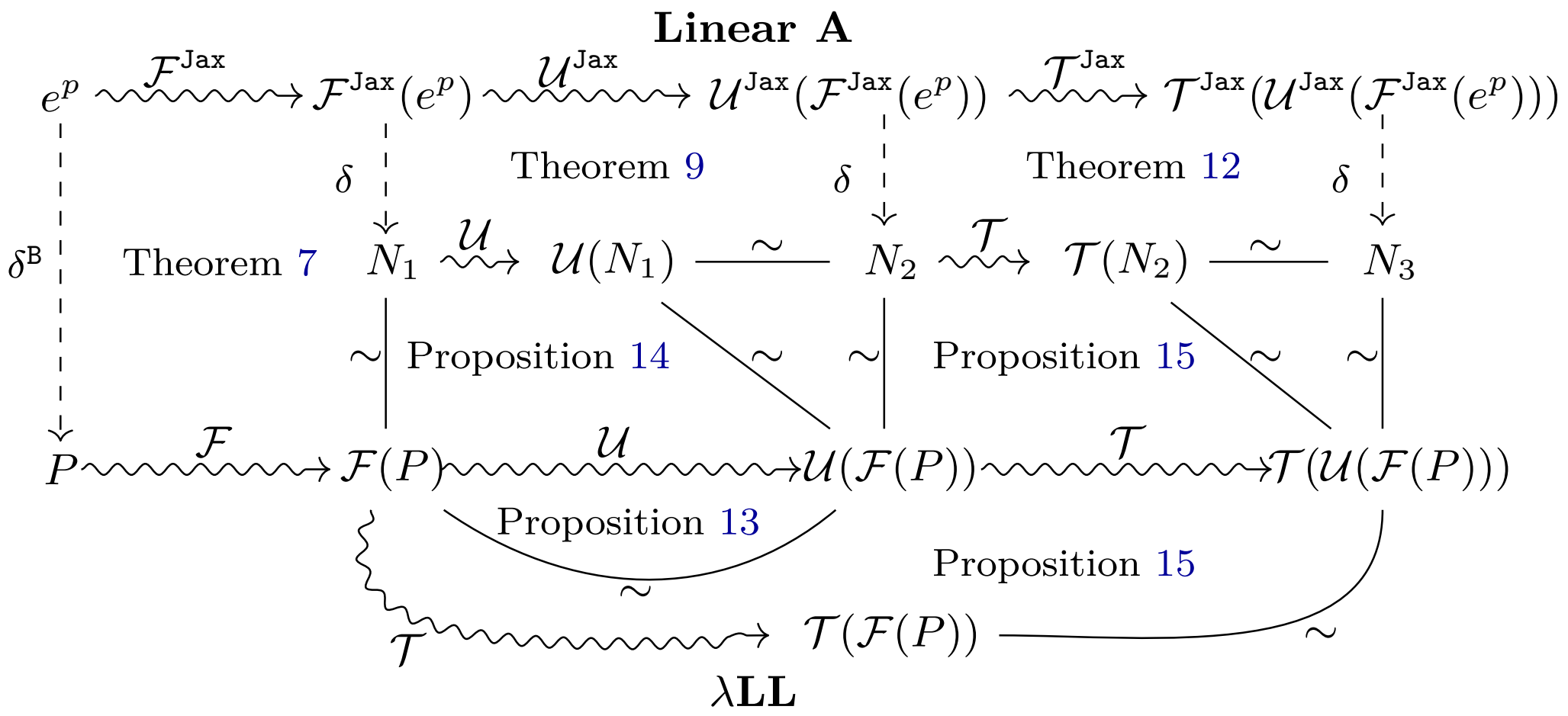

Figure 6.2: Comparing Linear A and $\lambda$LL

We have proven the soundness both qualitatively and quantitatively: qualitatively, because all transformations commute with the translation $\delta$ modulo our extensional equivalence $\sim$ (Theorems 7, 9, 12); quantitatively, because they all preserve the workload of the original ones (Theorems 8, 10 and Corollary 4 and their quantitative counterparts Theorems 14, 15 and Corollary 7). We have also shown that the unzipping and transpose transformations preserve our extensional equivalence $\sim$ (Propositions 14, 15). Moreover, the unzipping transformation in $\lambda$LL can be seen as let-commutation (Proposition 13 and Remark 8) and it can be skipped, providing more modularity and improving parallelism.

The three transformations $\mathcal{F}$, $\mathcal{U}$, and $\mathcal{T}$ have been defined for the fragment $\lambda\mathrm{LL}^{\mathtt{A}}$ of $\lambda$LL, which encompasses the $\delta$ encoding of Linear A. Our main concern was to compare our encoding with the original transformations in Linear A. However, we are considering to which extent one can generalise these transformations to generic terms of $\lambda$LL. This is a complex task if one aims to maintain a reasonable workload for the transformations (see discussion in Section 5.3).

This work is theoretical, and we have not implemented it yet. More precisely, our work can be implemented in two ways: either by making a toy implementation from scratch or by injecting our transformations into the JAX Autodiff code. We are more interested in the latter but it is more complex. Below we will try to give an idea of how we plan to proceed in this direction, hoping to clarify the benefits of this work from a practical point of view.

In JAX Autodiff, AD transformations manipulate `Jaxpr` objects, which essentially represent expressions of Linear A. The application of AD transformations is managed through a process called *tracing*. The key idea of tracing is that when you apply AD to a complex function composed of multiple primitive operations, JAX traces the primitive operations and constructs a computational graph of the function by creating the corresponding Jaxpr object. In the JAX tracing process, the unzipping transformation plays a central role to efficiently compute the gradient backward. We show that this unzipping transformation can be skipped using $\lambda$LL by gaining in modularity while preserving efficiency. Therefore, the practical contribution would be to inject our transformations into the JAX tracing process in such a way that the computational graph is build using objects representing terms of $\lambda$LL. Furthermore, this modification of the tracing process could be exploited to improve parallelism.

As mentioned in Chapter 1, many formalisations of AD have recently been introduced, raising questions about their interrelation. In particular, some of these systems are based on (some form of) the linear $\lambda$-calculus, e.g. [26] and [71]. We plan to investigate whether $\lambda$LL or an extension of it can be used as a common framework to compare these formalisations.

# Part II

# CryptoBLL

## Cryptography, Bounded Linear Logic and Lambda Calculus

Chapter 7

# Introduction to Computational Cryptography

The security of cryptographic protocols is concerned with ensuring that core communication goals, such as confidentiality, integrity, and authenticity, are achieved even in the presence of malicious entities. These entities may range from passive eavesdroppers, who silently observe network traffic in an attempt to extract sensitive information, to active adversaries capable of intercepting, modifying, injecting, or replaying messages to disrupt communication or deceive participants. In order to rigorously analyze whether a protocol can withstand such threats, researchers define security through precise formal models that specify both the capabilities of these adversaries and the properties the protocol must guarantee. These models form the basis for proving security properties and depend on underlying assumptions. While some models aim for information-theoretic security, which holds against adversaries with unlimited resources, most rely on computational assumptions, accepting negligible failure probabilities in exchange for practical efficiency. In this part of the thesis, we will focus on the second approach, grounded in *computational cryptography*. In this framework, protocols are typically constructed using well-defined *cryptographic schemes*, such as encryption, digital signatures, or key exchange, which are formalized independently with specific algorithms and security guarantees. These schemes serve as the building blocks of the protocol, whose overall security is proven by analyzing how these components interact under the defined threat model.

Computational cryptography relies on two important relaxations of *perfect secrecy* [103]. The latter requires that, when considering an encryption scheme, no information about an encrypted message be disclosed even to adversaries with unlimited computational power. In other words, learning the ciphertext gives the adversary no advantage at all in guessing the plaintext. However, while perfect secrecy is ideal in theory, it turns out to be too strict for most practical uses. It usually requires keys that are as long as the messages and can only be used once, which makes it impractical and unusable for interesting security proofs (see, e.g., [70, Section 2.3] for more details). In contrast, security definitions in the *computational model* take into account computational limits on the adversary and allow for a small probability of success. In other words, the cryptographic scheme may technically be vulnerable in some extremely rare cases, but the chance of such a failure is so small that it becomes irrelevant in practice.

There are several ways to formalize these two relaxations. Here, we consider the *asymptotic approach*, which is rooted in complexity theory and introduces a security parameter, denoted by $n$, shared by all the components involved in the scheme. Specifically, this parameter controls the size and structure of all elements in the cryptographic scheme, including the key space $\mathcal{K}$, message space $\mathcal{M}$, as well as the computational complexity of the scheme's algorithms. In this scenario, an efficient adversary is an algorithm working in probabilistic polynomial time in $n$ (PPT in the following) and the goal is to ensure that no such adversary can break the scheme with any meaningful advantage. More precisely, the probability with which the adversary can succeed is bounded by a negligible function in $n$. Formally, a function $\epsilon(n)$ is negligible if it decreases faster than the inverse of any polynomial. Intuitively, this means that the probability of a successful attack becomes insignificantly small as the security parameter increases, providing a strong, scalable notion of security that aligns with practical computational constraints.

When could we say that a scheme is secure in the computational sense? Among the many equivalent definitions in the literature, one of the handiest is the one based on *indistinguishability*, which states that a cryptographic scheme is secure if any PPT adversary succeeds in breaking it with at most negligible probability. Once the desired security property has been formally defined through a *cryptographic experiment*, the security of a scheme is proved using a technique known as *proof by reduction*. The $n$-security of a given cryptographic scheme $\Pi_1$ is reduced to the $n$-security of another scheme $\Pi_2$. Since $\Pi_2$ is assumed to be secure, the scheme $\Pi_1$ is hard to break. Formally, a proof by reduction is a proof by contraposition: the negation of the thesis implies the negation of the hypothesis. In other words, if an adversary $A$ is able to break the scheme $\Pi_1$ then there exists another adversary $B$ able to solve some presumably hard computational problem exploited by $\Pi_2$, with a similar effort.

In the literature there are several attempts aimed at modelling cryptographic constructions and proofs according to the computational model.

As outlined in the Introduction, the Universal Composability (UC) framework [28, 27] offers a rigorous foundation for reasoning about security under composition, ensuring that protocols remain secure even when combined with others. Despite its strengths, UC introduces significant technical complexity and struggles with higher-order settings involving dynamic behaviors such as runtime protocol generation [98, 72]. These limitations have driven efforts aimed at determining if it is possible to either extend the UC framework or to capture it by way of a calculus or process algebra (e.g. [29, 79, 73, 16]). In all the aforementioned works, a tension is evident between the need to be expressive, so as to capture UC proofs, and the need to keep the model simple enough, masking the details of probability and complexity as much as possible.

This work, however, focuses on the stand-alone setting, where even isolated protocol proofs present substantial challenges. Formalizing security in this context often involves intricate simulations, subtle reasoning about adversarial capabilities, and meticulous management of computational assumptions and probabilistic behavior. These factors contribute to the complexity of security proofs, posing significant challenges for their construction, validation, and especially mechanization. The broader aim is to bring greater structure and modularity to this process, drawing on ideas from formal methods to improve both clarity and rigor in cryptographic reasoning within the computational model of modern cryptography.

In this chapter, we begin by introducing the essential cryptographic preliminaries and then outline how this part of the thesis contributes to the broader effort of formally reasoning about cryptographic security in higher-order settings, specifically within the computational model of modern cryptography. While our approach is conceptually distant from the UC framework, it is motivated by similar foundational concerns. Specifically, we aim to develop a formal calculus tailored to the higher-order setting that addresses the longstanding tension between expressiveness and simplicity. Rather than formalizing UC directly, our goal is to provide an alternative framework that facilitates the reasoning about the security of higher-order cryptographic constructions in a modular and abstract way.

## 7.1 Private Encryption and CPA security

In order to stay self-contained, in this section we introduce the necessary cryptographic preliminaries. We first focus on one of the fundamental primitives in symmetric cryptography:

private-key encryption. We formally define the components of a private-key encryption scheme, which consists of algorithms for key generation, encryption, and decryption. Next, we describe a threat model in which the adversary is allowed to choose plaintexts and observe their corresponding ciphertexts. This scenario gives rise to the concept of chosen-plaintext attack (CPA) security, an essential notion that captures the confidentiality guarantees of encryption schemes in the presence of active adversaries. More precisely, the CPA security definition is formalized through an experiment based on indistinguishability, where the adversary's ability to distinguish encryptions of chosen messages is bounded by a negligible function in the security parameter.

Finally, in Subsection 7.1.1, we discuss the standard proof by reduction that establishes CPA security for the encryption scheme $\Pi_F$, induced by a pseudorandom function $F$, following the presentation in [70]. This involves defining pseudorandom functions, explaining how they give rise to encryption schemes, and proving their security under chosen-plaintext attacks.

The presentation provided here is partial and specifically focused on the aspects that will be relevant to this part of the thesis. While we concentrate on CPA security for private-key encryption, since it will serve as example in our framework, it is worth noting that many other cryptographic settings, such as Message Authentication Codes (MACs), also exhibit higher-order behaviour and could similarly be analyzed within our framework.

In the setting of private-key encryption, two parties share a key and use it to communicate secretly. Formally, a private-key encryption scheme $\Pi = (Gen, Enc, Dec)$ is defined by specifying a message space $\mathcal{M}$, a key space $\mathcal{K}$, and three algorithms: a procedure for generating keys ($Gen$), a procedure for encrypting ($Enc$), and a procedure for decrypting ($Dec$). An encryption scheme must satisfy the following correctness requirement: for every key $k \in \mathcal{K}$ output by $Gen$ and every message $m \in \mathcal{M}$, it holds that $Dec(k, Enc(k, m)) = m$.

$$
\begin{aligned}
&\mathtt{PrivK}^{CPA}_{A,\Pi}(n) :\\
&\quad k \leftarrow Gen(1^n)\\
&\quad m_0, m_1 \leftarrow A^{Enc_k(\cdot)}(1^n)\\
&\quad b \leftarrow \{0,1\}\\
&\quad c \leftarrow Enc(k, m_b)\\
&\quad g \leftarrow A(c, Enc_k(\cdot))\\
&\quad \mathtt{return}\ (b = g)
\end{aligned}
$$

Figure 7.1: Pseudocode for $\mathtt{PrivK}^{CPA}$ Experiment

In chosen-plaintext attacks, the adversary can obtain the ciphertext corresponding to plaintexts of its choice. Formally, this is modelled by giving the adversary $A$ access to an *encryption oracle*, denoted by $Enc_k(\cdot)$, viewed as a black box that encrypts messages of $A$'s choice using a key $k$ that is unknown to $A$. The adversary is allowed to interact with the encryption oracle a polynomial number of times in the security parameter $n$. Let us consider two messages $m_0, m_1$ chosen by the adversary and a ciphertext $c \leftarrow Enc(k, m_b)$, where $b$ is chosen at random. Security against chosen-plaintext attacks means that the adversary cannot guess the value of $b$ with probability significantly higher than $1/2$. This scenario is formalized by the cryptographic experiment $\mathtt{PrivK}^{CPA}_{A,\Pi}$, reported in Figure 7.1 exactly in the form it has in [70]. The security against chosen-plaintext attacks based on indistinguishability, also known as CPA security, is formalized as follows:

**Definition 7** (CPA security)**.** A private-key encryption scheme $\Pi$ is CPA-secure, if for all probabilistic polynomial-time adversaries $A$ there is a negligible function $\varepsilon$ such that

$$Pr[\mathtt{PrivK}^{CPA}_{A,\Pi}(n) = 1] \leq \frac{1}{2} + \varepsilon(n)$$

where the probability is taken over the randomness used by $A$, as well as the randomness used in the experiment.

### 7.1.1 CPA-Secure Encryption Scheme from Pseudorandom Functions

This subsection is devoted to the presentation of the game-based proof of security against active attacks for the encryption scheme $\Pi_F$ induced by a pseudorandom function $F$, exactly as presented in [70]. First, we define what a pseudorandom function is along with its key properties. Next, we proceed by defining an encryption scheme based on a pseudorandom function and we conclude by proving that the latter is secure with respect to chosen-plaintext attacks.

#### Pseudorandom Functions

A *pseudorandom function* [56, 70] is a function computed by any deterministic polytime algorithm taking two strings in input, and producing a string as output, in such a way that when the first of the two parameters is picked at random, the unary function obtained through currying is indistinguishable from a random one, all this to the eyes of adversaries working in probabilistic polynomial time. The notion of a pseudorandom function is closely related to that of a secure block-cipher. The formal definition of pseudorandom function requires the following preliminary notions:

- A *binary partial function* is a partial function from $\{0,1\}^* \times \{0,1\}^*$ to $\{0,1\}^*$.
- A binary partial function $F$ is *length-preserving* iff $F(k,x)$ is defined iff $|k| = |x|$ and in that case $|F(k,x)| = |x|$.
- A binary partial function is *efficient* when there exists a deterministic polytime algorithm that computes it.

Given a length-preserving binary partial function $F$, we denote by $F_k$ the total function from $\{0,1\}^{|k|}$ to $\{0,1\}^{|k|}$ defined in the natural way: $F_k(m) = F(k,m)$. Now we are able to state the formal definition of pseudorandom function as follows

**Definition 8.** Given a binary partial function $F$, which is length-preserving and efficient, we say that $F$ is a pseudorandom function (PRF) iff for every PPT distinguisher $D$, there exists a negligible function $\varepsilon$ such that

$$\left| Pr[D^{F_k(\cdot)}(1^n) = 1] - Pr[D^{f(\cdot)}(1^n) = 1] \right| \leq \varepsilon(n)$$

where $k$ is chosen among all strings of length $n$ randomly and $f(\cdot)$ is chosen among all functions from $\{0,1\}^n$ to $\{0,1\}^n$ randomly. Here, $D^{F_k(\cdot)}$ (resp. $D^{f(\cdot)}$) stands for the computation the distinguisher $D$ gives rise to when querying the function $F_k(\cdot)$ (resp. $f(\cdot)$) as an oracle and tries to decide whether the function is truly random or pseudorandom based on the outputs of the function.

### Constructing an Encryption Scheme from Pseudorandom Functions

We consider the encryption scheme $\Pi_F = (Gen^F, Enc^F, Dec^F)$ induced by a pseudorandom function $F$, described in [70, Construction 3.30]. This scheme uses the operator $\oplus$, namely the bitwise exclusive-or operation of two binary strings with the same length and $n$ as the security parameter. More precisely, the algorithms of $\Pi_F$ are defined as follows:

- On input $1^n$, the key generation algorithm $Gen^F$ samples $k$ from $\{0,1\}^n$ uniformly at random and returns it.
- The encryption algorithm $Enc^F$ takes a secret key $k$ and the message $m$ we want to encrypt. The encryption process make use of a $n$-bits random value $r$ and combines it with the key $k$ via the pseudorandom function $F$ to produce a ciphertext. The result is a pair $\langle r, F(k,r) \oplus m\rangle$ which consists of the random value and the encrypted message.
- The decryption algorithm $Dec^F$ takes a secret key $k$ and the pair $\langle r, s\rangle$, where $r$ is the random value and $s$ is the encrypted part, and uses the key $k$ and the random value $r$ to recover the original message $m$ from the encrypted part by computing $F(k,r) \oplus s$.

Observe that the use of a random value $r$ ensures that even if the same message is encrypted multiple times using the same key, the ciphertext will not necessarily be equal, due to the randomness of $r$.

### Proof of CPA Security

Before moving to the details of the proof, we highlight a common template that is used by most proofs of security for constructions based on a pseudorandom function $F$. Essentially, the goal is to reduce the problem of breaking the encryption scheme's security to the problem of distinguishing a pseudorandom function from a random function, which is assumed to be computationally hard. The first step of such proofs is to consider a idealized version of $\Pi_F$ in which the pseudorandom function is replaced with a random function $f$, we call this version $\widehat{\Pi}$. The proof proceeds by demonstrating that the scheme $\widehat{\Pi}$ is (unconditionally) secure, so the probability of any efficient adversary successfully breaking $\widehat{\Pi}$ is negligible. Since $F$ is a pseudorandom function, by definition $F$ is computationally indistinguishable from a random function to any efficient adversary, therefore $\widehat{\Pi}$ and $\Pi_F$ cannot behave in a significantly different way. Hence, the adversary's ability to break $\Pi_F$ is not substantially affected by the change from $F$ to $f$. In other words, if the adversary cannot break the security of $\widehat{\Pi}$, then it also cannot break the security of $\Pi_F$, thus establishing that $\Pi_F$ is secure under the assumption that the pseudorandom function $F$ is indistinguishable from a random function. Formally, the proof of security of $\Pi_F$ with respect to chosen-plaintext attacks is described in the following theorem.

**Theorem 16** (CPA Security of $\Pi_F$)**.** If $F$ is a pseudorandom function, then $\Pi_F$ is a CPA-secure private-key encryption scheme for messages of length $n$.

*Proof.* In Chapter 10 we will refer to specific steps of this proof so, to remain as self-contained as possible, we give the details of the proof below exactly as in the textbook [70]. We proceed with the steps described above:

1. Let $\widehat{\Pi} = (Gen^f, Enc^f, Dec^f)$ be an encryption scheme that is exactly the same as $\Pi_F$ except that a truly random function $f$ is used in place of $F_k$. What we want to prove is that for every PPT adversary $A$, there is a negligible function $\varepsilon$ such that:

$$\left|Pr[\mathtt{PrivK}^{CPA}_{A,\Pi_F}(n) = 1] - Pr[\mathtt{PrivK}^{CPA}_{A,\widehat{\Pi}}(n) = 1]\right| \leq \varepsilon(n) \tag{7.1}$$

   We prove this by reduction as follows

- From any adversary $A$ we construct a distinguisher $D_A$ for the pseudorandom function $F$. The distinguisher $D_A$ is given oracle access to some function $O$, and its goal is to determine whether this function is pseudorandom or random. To do this, $D_A$ emulates the experiment $\mathtt{PrivK}^{CPA}$ in such a way that $D_A^{F_k(\cdot)}$ behaves like $\mathtt{PrivK}^{CPA}_{A,\Pi^F}$ while $D_A^{f(\cdot)}$ behaves like $\mathtt{PrivK}^{CPA}_{A,\widehat{\Pi}}$. In detail the distinguisher $D_A^O$ is defined as follows:

  $D_A^O$ :

  1. Run $A(1^n)$. Whenever $A$ queries its encryption oracle on a message $m \in \{0,1\}^n$, answer this query in the following way:
     a) Choose a uniform $r \in \{0,1\}^n$.
     b) Query $O(r)$ and obtain response $y$.
     c) Return the ciphertext $\langle r, y \oplus m \rangle$ to $A$.
  2. When $A$ outputs messages $m_0, m_1 \in \{0,1\}^n$, choose a uniform bit $b \in \{0,1\}$ and then:
     a) Choose a uniform $r \in \{0,1\}^n$.
     b) Query $O(r)$ and obtain response $y$.
     c) Return the challenge ciphertext $\langle r, y \oplus m \rangle$ to $A$.
  3. Continue answering encryption-oracle queries of A as before, until $A$ outputs a bit $g$.
  4. Output 1 if $b = g$, and 0 otherwise.

  Note that $D_A^O$ runs in polynomial time since $A$ does.

- We analyze the following two cases:
  - Case $O = F_k$: If $D_A$'s oracle is a pseudorandom function, then the view of $A$ when run as a subroutine by $D_A$ is distributed identically to the view of $A$ in experiment $\mathtt{PrivK}^{CPA}_{A,\Pi_F}(n)$:

$$Pr_{k \leftarrow \{0,1\}^n}[D_A^{F_k(\cdot)}(1^n) = 1] = Pr[\mathtt{PrivK}^{CPA}_{A,\Pi_F}(n) = 1] \tag{7.2}$$

  - Case $O = f$: If $D_A$'s oracle is a random function, then the view of $A$ when run as a subroutine by $D_A$ is distributed identically to the view of $A$ in experiment $\mathtt{PrivK}^{CPA}_{A,\widehat{\Pi}}(n)$:

$$Pr_{f \leftarrow \mathbf{Func_n}}[D_A^{f(\cdot)}(1^n) = 1] = Pr[\mathtt{PrivK}^{CPA}_{A,\widehat{\Pi}}(n) = 1] \tag{7.3}$$

  where $\mathbf{Func_n}$ is the set of all functions mapping n-bit strings to n-bit strings.

- By the assumption that $F$ is a pseudorandom function (and since $D_A$ is efficient), there exists a negligible function $\varepsilon$ for which

$$|Pr_{k \leftarrow \{0,1\}^n}[D_A^{F_k(\cdot)}(1^n) = 1] - Pr_{f \leftarrow \mathbf{Func_n}}[D_A^{f(\cdot)}(1^n) = 1]| \leq \varepsilon(n)$$

  We can then conclude that Equation 7.1 holds.

2. We also show that

$$Pr[\mathtt{PrivK}^{CPA}_{A,\widehat{\Pi}}(n) = 1] \leq \frac{1}{2} + \frac{q(n)}{2^n} \tag{7.4}$$

where $q(n)$ is a bound on the number of encryption queries made by $A$. Formally, Equation 7.4 is proved by analyzing a probability event. More precisely, notice that every time a message $m$ is encrypted in $\texttt{PrivK}^{CPA}_{A,\widehat{\Pi}}$, whether by the encryption oracle or when creating the challenge ciphertext, a random string $r$ is chosen uniformly at random from $\{0,1\}^n$. Then, the ciphertext is formed as the pair $\langle r, f(r) \oplus m\rangle$, where $f$ is the random function since we are analyzing the experiment based on the encryption scheme $\widehat{\Pi}$. Let's call the random string used in the challenge ciphertext $r^\star$, so the challenge ciphertext looks like $\langle r^\star, f(r^\star) \oplus m_b\rangle$. Now, there are two possibilities:

- The value $r^\star$ is never used when answering any of $A$'s encryption-oracle queries. Consequently, $A$ gains no information about $f(r^\star)$ through its interaction with the encryption oracle, since $f$ is modeled as a truly random function. From $A$'s perspective, the string obtained by performing the xor operation between $f(r^\star)$ and $m_b$ in the challenge ciphertext is therefore uniformly distributed and independent of the rest of the experiment. Hence, the probability that $A$ correctly guesses $b$ in this case is exactly $\frac{1}{2}$.
- The value $r^\star$ is used in answering at least one of $A$'s encryption-oracle queries. In this case, $A$ can potentially distinguish which message, $m_0$ or $m_1$, was encrypted. This occurs because if the encryption oracle ever returns a ciphertext $\langle r^\star, s\rangle$ in response to $A$'s query to encrypt some message $m$, then the adversary $A$ learns $f(r^\star) = s \oplus m$. However, since $A$ makes at most $q(n)$ queries to the encryption oracle, hence at most $q(n)$ different values of $r$ are used in response to $A$'s queries. Given that $r^\star$ is chosen uniformly at random from $\{0,1\}^n$, the probability that $r^\star$ coincides with any of these previously used values is at most $\frac{q(n)}{2^n}$.

Let us consider $\texttt{Repeat}$ as the event in which $r^\star$ is used by the encryption oracle in answering at least one of $A$'s queries. As previously described, the probability that the event $\texttt{Repeat}$ occurs is upper bounded by $\frac{q(n)}{2^n}$. Moreover, conditioned on the event that $\texttt{Repeat}$ does not occur, denoted by $\overline{\texttt{Repeat}}$, the adversary $A$ succeeds in the $\texttt{PrivK}^{CPA}_{A,\widehat{\Pi}}$ experiment with probability exactly $\frac{1}{2}$. Formally, we have that

$$Pr[\texttt{Repeat}] = \frac{q(n)}{2^n} \quad \text{and} \quad Pr[\texttt{PrivK}^{CPA}_{A,\widehat{\Pi}}(n) = 1 \land \overline{\texttt{Repeat}}] = \frac{1}{2} \tag{7.5}$$

Therefore, we have that

$$\begin{aligned} Pr[\texttt{PrivK}^{CPA}_{A,\widehat{\Pi}}(n) = 1] &= Pr[\texttt{PrivK}^{CPA}_{A,\widehat{\Pi}}(n) = 1 \land \texttt{Repeat}] + Pr[\texttt{PrivK}^{CPA}_{A,\widehat{\Pi}}(n) = 1 \land \overline{\texttt{Repeat}}] \\ &\leq Pr[\texttt{Repeat}] + Pr[\texttt{PrivK}^{CPA}_{A,\widehat{\Pi}}(n) = 1 \land \overline{\texttt{Repeat}}] \\ &\overset{\text{Eq. 7.5}}{=} \frac{1}{2} + \frac{q(n)}{2^n} \end{aligned}$$

and we can conclude that Equation 7.4 holds.

3. By putting together point 1. and 2. above we can conclude that $\Pi_F$ is CPA-secure. More precisely, by combining Equation 7.4 with Equation 7.1, we see that there is a negligible function $\varepsilon$ such that $Pr[\texttt{PrivK}^{CPA}_{A,\Pi_F}(n) = 1] = \frac{1}{2} + \frac{q(n)}{2^n} + \varepsilon(n)$. Observe that, since $q(n)$ is a polynomial, by definition we have that $\frac{1}{2} + \frac{q(n)}{2^n}$ is negligible. Moreover, the sum of two negligible functions is negligible, and thus there exists a negligible function $\varepsilon'$ such that $Pr[\texttt{PrivK}^{CPA}_{A,\Pi_F}(n) = 1] = \frac{1}{2} + \varepsilon'(n)$, completing the proof.

□

## 7.2 Goals of CryptoBLL

The two predominant models in modern cryptography, namely the computational [57] and the symbolic [42] models, have had very different fates with respect to the application of language-based verification techniques to them. In the symbolic model, which does not account for complexity nor for probability, the application of classic verification methodologies (e.g. model checking [48], rewriting [87] and abstract interpretation [1]) is natural and has been extensively done. In the computational model, instead, all this is notoriously more problematic.

An interesting line of work, which has given rise to an increasing number of contributions in the last 25 years (see, e.g., [68, 85, 60, 37]), consists in the application of classical program equivalence theories to programming languages specifically designed to capture the reference notion of complexity in the computational model, namely that of a probabilistic polynomial-time algorithm. Once this is done, the gold standard notion of equivalence in cryptography, namely *computational indistinguishability* [70, 55], becomes a form of observational equivalence, thus paving the way towards the study of computational indistinguishability via standard tools from programming language theory, like logical relations [107, 106] and applicative bisimilarity [5, 101], which are sound by construction (although not necessarily complete) for observational equivalence.

This is precisely the direction we explore in this part of the thesis; our objective is to define a typed $\lambda$-calculus with references and probabilistic choice able to naturally capture the complexity constraints mentioned above through a form of graded modality. In order to reason about program equivalence in this framework, we develop a logical relation that is proved sound for an approximate observational equivalence capturing the essence of computational indistinguishability.

### 7.2.1 Motivations

We introduce a typed $\lambda$-calculus, called $\lambda$BLL, able to solve some expressiveness problems of our previous approach based on $\pi$-calculus and session types [37, 74]. At the level of terms, we enrich the pure $\lambda$-calculus with function symbols computing probabilistic polytime functions, namely the basic building blocks of any cryptographic protocol, and with global references that are essential for modelling experiments with a shared state between the components of the protocol. At the level of types, $\lambda$BLL has a linear type system with a correspondence to Bounded Linear Logic [52, 63, 38] which is able to model the computational complexity constraints of attackers. This calculus offers a flexible and expressive formalism that accurately captures and structures cryptographic constructions and experiments within the computational model, thanks to its integration of probabilistic polytime primitives, shared state through global references, and a resource-aware type system grounded in BLL.

This work establishes a logical relation that is proved sound for an approximate form of observational equivalence, precisely capturing computational indistinguishability in $\lambda$BLL and contributing to the formal foundations of cryptographic reasoning. A central feature of this relation is that it is defined via a logical metric, resulting in a relation that quantitatively measures the behavioural distance between terms. This approximate nature is crucial for cryptographic reasoning, where indistinguishability permits small behavioural differences between programs, as long as no efficient adversary can exploit them with more than negligible advantage. By embedding this approximation into the semantics, the framework provides a rigorous foundation for formalizing security properties and enables *equational reasoning* over terms in $\lambda$BLL. Specifically, if two terms are related by the logical relation, their behaviours are computationally indistinguishable to any PPT adversary. As a result, cryptographic proofs can be systematically structured as a series of equational steps, each justified by the logical relation to guarantee that the overall transformation preserves computational indistinguishability.

Concretely, we exemplify the successful application of our approach in the higher-order setting by providing an equational formalization of the standard proof by reduction that establishes CPA security for an encryption scheme built from a pseudorandom function. This formalization is particularly interesting because it leverages higher-order features, such as the use of an oracle, which allows for dynamic behaviours that are challenging to capture in first-order models present in the literature.

### 7.2.2 Outline of CryptoBLL

In Chapter 8, we introduce a typed $\lambda$-calculus with references and probabilistic choice, designed to naturally capture complexity of attackers via a form of graded modality, at the same time allowing to easily express primitives, experiments and reductions, fundamental ingredients in game-based proofs. The language, called $\lambda$BLL, is derived from Bounded Linear Logic [52, 63, 38], and its syntax, typing rules, and operational semantics are presented in detail. Special emphasis is placed on how linear types and modalities interact to control resource usage and how probabilistic effects and mutable references are incorporated in a coherent and type-safe manner.

Chapter 9 defines a logical relation for $\lambda$BLL that is sound with respect to an approximate observational equivalence capturing computational indistinguishability. The relation is based on a logical metric, enabling reasoning about programs that exhibit similar but not identical behaviour from a computational perspective. This chapter is a joint contribution with Zeinab Galal, who developed the logical relation framework. In order to stay self-contained and accessible, the key definitions and conceptual explanations are provided, while the detailed proofs are left out.

At the end, Chapter 10 explores both the practical expressiveness of $\lambda$BLL and its application to the computational model of modern cryptography. It first introduces a collection of function symbols and equations in $\lambda$BLL that encapsulate key computational patterns such as randomness, bitstring manipulation, and memory operations. The chapter then turns to a formalization of the standard proof by reduction that establishes CPA security for an encryption scheme built from a pseudorandom function, previously introduced in this chapter. In the $\lambda$BLL framework, the components of this proof are modelled as terms, and the proof itself is carried out through equational reasoning, guided by the equations introduced earlier. This example demonstrates the ability of $\lambda$BLL to encode cryptographic constructions and support a structured equational reasoning about their security properties.

### 7.2.3 Related Works

Although the literature regarding formal methods for the security analysis of protocols and primitives is much more abundant in the symbolic model than in the computational one, it certainly cannot be said that the latter has not been the subject of attention by the research community. The work on probabilistic relational Hoare logic which gave rise to the `EasyCrypt` tool [29], must certainly be mentioned. The result of Bana and Comon Lundt on inconsistency proofs as security proofs [14], which in turn gave rise to the `Squirrel` tool [11], is another pertinent example. In both cases, the model provides for the possibility of higher-order constructions, which however are not fully-fledged. In particular, managing complexity aspects and higher-order functions at the same time turns out to be hard.

This last direction is the one followed by the work on CSLR and its formalization [89]. In this case we find ourselves faced with a $\lambda$-calculus for polynomial time and its application to the study of cryptographic primitives. There are two differences with this work. First of all, the greater expressiveness of $\lambda$BLL allows to capture PPT even for second-order constructions. Furthermore, the logical relations introduced here effectively give rise to a notion of metric, while in CSLR the underlying equational theories are exact, even though a notion of observational equivalence similar to ours has been introduced.

In the literature devoted to capturing the UC framework via a formal calculus, it is worth mentioning the work by Liao et al. [79]. This work introduces the *Interactive Lambda Calculus (ILC)*, a formal calculus designed to model cryptographic protocols in a composable setting. The ILC combines aspects of $\lambda$-calculus and $\pi$-calculus to express complex protocol interactions while maintaining the composability guarantees that are central to the Universal Composability (UC) framework. This approach represents a significant step forward in capturing UC security in a formal, calculus-based system, addressing the challenges of modelling and reasoning about complex cryptographic behaviours such as protocol instantiation and interaction at runtime. While their approach is rooted in capturing UC security, our goal is more distant: we focus on equational reasoning about cryptographic constructions in higher-order settings, without aiming to formalize UC or its ideal/real execution paradigm. However, our objective differs significantly as it emphasizes equational reasoning for cryptographic constructions in higher-order settings, without seeking to formalize UC security or adopt the ideal/real paradigm.

Another attempt that goes in the same direction as ours is the work by Mitchell et al. [86], who introduced a process algebra in the style of Milner's CCS capable of modelling cryptographic protocols. Unlike ours, the resulting calculus is concurrent and this gives rise to a series of complications. Once again, despite the underlying notion of observational equivalence being approximate and therefore adhering to computational indistinguishability, the proposed notion of bisimulation is exact and as such much finer.

Logical relations [107, 88] are a powerful tool for relational reasoning about higher-order terms. They are known to work well in calculi with effects and in particular in presence of probabilistic choice effects [59, 24, 8]. It is also known that metric versions of logical relations can be given, and that they are useful for sensitivity analysis [94, 36]. The possibility of applying logical relations to calculi such as the cryptographic $\lambda$-calculus is well-known [60], but the underlying calculus turns out to be fundamentally different from ours, being in the tradition of the symbolic model and abstracting away from probabilistic effects and complexity constraints.

Chapter 8

# $\lambda$BLL

In this chapter we define a $\lambda$-calculus based language, called $\lambda$BLL, which is expressive enough to capture cryptographic constructions and experiments within the computational model in higher-order settings.

First, we define the syntax and type system of $\lambda$BLL in Section 8.1, illustrating how it enables the representation of complex interactions between adversaries and oracles. In Section 8.2, we explore how the calculus integrates probabilistic behaviour and mutable state through the combination of the distribution and state monads, providing a structured foundation for reasoning about cryptographic effects. Section 8.3 sets out the operational semantics of the calculus, giving precise rules for the execution of effectful computations. Finally, Section 8.4 demonstrates that $\lambda$BLL accurately characterizes probabilistic polynomial-time computation, ensuring that the calculus is not only expressive but also faithfully reflects the computational constraints relevant to cryptographic applications.

## 8.1 Syntax and Type System

In this section, we introduce $\lambda$BLL, a linear $\lambda$-calculus enriched with function symbols representing probabilistic polynomial-time computations and with mutable references for modelling shared state. The calculus is expressive enough to capture complex cryptographic experiments, including higher-order interactions between adversaries and oracles. Its type system, rooted in Bounded Linear Logic, employs a graded modality that tracks polynomial bounds, thereby enabling precise control over resource usage and reflecting the computational constraints of attackers in the realm of computational cryptography.

**Types** At the level of types, $\lambda$BLL has a linear type system with a correspondence to Bounded Linear Logic (BLL) [52] and graded-calculi [94], designed to enforce quantitative control over resource usage. In contrast to standard linear logic, where resources must be used exactly once and the bang modality (!) enables unrestricted duplication, BLL refines this modality by introducing explicit bounds. These bounds regulate how many times a resource can be duplicated, enabling more fine-grained control. In $\lambda$BLL, this is realized through a graded comonadic modality $!_p A$, where the index $p$ is a polynomial that governs how many times a term of type $A$ can be duplicated. These bounds are used to track and constrain the computational complexity of adversaries, which is essential in cryptographic reasoning. The polynomials used for indexing are built from positive natural numbers ($\mathbb{N}_{\geq 1}$), addition and multiplication, but also contain a polynomial variable $i$, representing the security parameter. Formally, these polynomials are

$$G ::= \mathbb{U} \mid \mathbb{B} \mid \mathbb{S}[p] \qquad \text{(Ground Types)}$$
$$P ::= G \mid P \otimes P \mid !_p A \qquad \text{(Positive Types)}$$
$$A ::= P \mid P \multimap A \qquad \text{(Types)}$$

Figure 8.1: Types of λBLL

generated by the following grammar:

$$p ::= 1 \mid i \mid p + p \mid p \times p \qquad \text{(Polynomials)}$$

It is important to observe that this structure allows λBLL to express and reason about families of types and terms indexed by the security parameter, aligning with the quantitative spirit of BLL and enabling formal reasoning about asymptotic resource bounds.

Figure 8.1 introduces the grammar of types in λBLL, which is designed to support fine-grained reasoning about resource usage in cryptographic computations. The system distinguishes between positive types and general types in the CBPV style [78, 46]. The base of the system is formed by *ground types*, which include the unit type ($\mathbb{U}$), boolean type ($\mathbb{B}$), and the type of fixed-length binary strings ($\mathbb{S}[p]$), where the length is determined by a polynomial $p$. These represent the fundamental data types commonly used in cryptographic constructions. Building on these, *positive types* include ground types, pairs formed with the tensor product ($P \otimes P$), and the modal type $!_p A$. As mentioned above, this modality is inspired by Bounded Linear Logic and captures controlled resource duplication, which is critical for modelling operations like bounded adversarial queries. Finally, *general types* include linear function types ($P \multimap A$), capturing the restriction, drawn from CBPV, that only positive types, representing values, may appear as function arguments.

**Syntax** At the term level, λBLL extends the pure λ-calculus with two key features essential for modelling cryptographic protocols: function symbols representing probabilistic polynomial-time computations, namely the basic building blocks of any cryptographic protocol, and global references, which enable the representation of shared mutable state across different components of a protocol or experiment. More precisely, we define a set of function symbols $\mathcal{F}$, where each function symbol $f$ in $\mathcal{F}$ comes equipped with:

- a type denoted $\mathtt{typeof}(f)$ of the form $G_1 \times \cdots \times G_m \to G$ where $G_1, \ldots, G_m$ and $G$ are ground types;
- for every polynomial $p$ in $\mathbb{N}_{\geq 1}[i]$, a term constructor $f_p$ of arity $m$.

For example, we will consider the function symbol `random` with $\mathtt{typeof}(\mathtt{random}) = \mathbb{S}[i]$ and arity 0 interpreted as a map randomly generating a string in $\{0,1\}^i$, and the function symbol `xor` with $\mathtt{typeof}(\mathtt{xor}) = \mathbb{S}[i] \times \mathbb{S}[i] \to \mathbb{S}[i]$ and arity 2 interpreted as a map computing the *bitwise exclusive-or* of binary strings.

The grammar of λBLL, as shown in Figure 8.2, explicitly separates values and computations, following the principles of CBPV.

Values represent data that can be stored, passed, or paired, but not directly executed, and they serve as the static components within a computation. At the base level, *ground values* provide the primitive data types of the language: the unit value $\star$, boolean values $\mathbf{t}$ and $\mathbf{f}$, and fixed-length binary strings $s$, which are central in representing cryptographic inputs such as keys

$$W ::= \star \mid \mathbf{t} \mid \mathbf{f} \mid s \qquad \text{(Ground Values)}$$

$$Z ::= x \mid W \mid \langle Z, Z \rangle \mid !M \qquad \text{(Positive Values)}$$

$$\mathcal{V} \ni V ::= Z \mid \lambda x.M \qquad \text{(Values)}$$

$$\begin{aligned} \Lambda \ni M ::= {} & \mathtt{return}\, V \mid \mathtt{der}(Z) \mid M Z \\ & \mid \mathtt{let}\, x = N \,\mathtt{in}\, M \mid \mathtt{let}\, \langle x, y \rangle = Z \,\mathtt{in}\, M \\ & \mid \mathtt{if}\, Z \,\mathtt{then}\, M \,\mathtt{else}\, N \mid \mathtt{loop}\, V\, p\, \mathtt{times}\, \mathtt{from}\, M \qquad \text{(Computations)} \\ & \mid \mathtt{set}\, r\, Z \mid \mathtt{get}\, r \\ & \mid f_p(Z_1, \ldots, Z_m) \end{aligned}$$

Figure 8.2: Syntax of $\lambda$BLL

and messages. *Positive values* include variables (e.g., $x$), pairs of values denoted as $\langle Z, Z \rangle$, and suspended computations of the form $!M$, governed by the graded comonadic modality.

In $\lambda$BLL, computations express the dynamic behaviour of the language, especially those involving probabilistic operations and shared mutable state. A central feature of the language is its treatment of memory references, which enable components of a protocol or adversary to interact through shared state. These references are restricted to store ground values, such as unit, booleans, and fixed-length bitstrings, thereby maintaining precise control over the flow of information. They are manipulated through two primitives: $\mathtt{set}\, r\, V$, which writes the value $V$ to the memory location $r$, and $\mathtt{get}\, r$, which retrieves the current content of $r$. In addition to memory operations, computations in $\lambda$BLL adopt standard constructs from the CBPV paradigm. Computations include $\mathtt{return}\, V$ for embedding values, $\mathtt{der}(Z)$ to force suspended computations, and $M Z$ for the application of a computation to a value. The language also supports control flow mechanisms such as conditional branching ($\mathtt{if}\, Z \,\mathtt{then}\, M \,\mathtt{else}\, N$), let-bindings for sequencing, and pattern matching on product values. Function symbols, introduced earlier, appear in computations as applications of the form $f_p(Z_1, \ldots, Z_m)$, representing probabilistic polynomial-time primitives. A distinctive feature of $\lambda$BLL is its support for bounded iteration via the construct $\mathtt{loop}\, V\, p\, \mathtt{times}\, \mathtt{from}\, M$, which executes $M$ up to $p$ times, where $p$ is a polynomial in the security parameter. This mechanism is especially well-suited for capturing cryptographic scenarios involving adversaries that interact repeatedly with an oracle.

The set of free variables $FV(M)$ of a term $M$ is defined by induction as usual. Moreover, we classify a variable as *ground* if its type contains no function type constructor ($\multimap$); otherwise, it is considered *higher-order*. Based on this distinction, we define the set $HOFV(M)$ as the subset of $FV(M)$ consisting only of higher-order variables.

**Type System** We consider two kinds of contexts to distinguish between term variables $x$ and memory references $r$. These contexts are formally defined as follows:

$$\Gamma ::= \varnothing \mid x : P, \Gamma \qquad \text{(Variable Context)}$$

$$\Theta ::= \varnothing \mid r : G, \Theta \qquad \text{(Reference Context)}$$

For a fixed reference context $\Theta = r_1 : G_1, \ldots, r_n : G_n$ assigning ground types to memory references, we have two kinds of typing judgements:

$$\Gamma \vdash_v^{\Theta} V : A \quad \text{and} \quad \Gamma \vdash_c^{\Theta} M : A$$

for values and computations respectively, where $\Gamma = x_1 : P_1, \ldots, x_n : P_n$ is a context assigning positive types to term variables. The operation of polynomial addition induces a binary partial operation $\boxplus$ on positive types defined by induction below:

$$G \boxplus G := G$$
$$(P \otimes Q) \boxplus (R \otimes S) := (P \boxplus R) \otimes (Q \boxplus S)$$
$$(!_p A) \boxplus (!_q A) := !_{p+q} A.$$

In order to account for polynomial multiplication, we also define for every polynomial $p \in \mathbb{N}_{\geq 1}[i]$, a total unary operation on positive types by induction:

$$p * G := G$$
$$p * (P \otimes Q) := (p * P) \otimes (p * Q)$$
$$p * (!_q A) := !_{p \times q} A.$$

It is important to note that, for ground types, the equalities $G \boxplus G := G$ and $p * G := G$ indicate that ground values, namely unit $\star$, booleans $\mathbf{t}$ and $\mathbf{f}$, and binary strings $s \in \{0,1\}^*$, can be freely duplicated. In contrast, we explicitly track polytime complexity for higher-order applications and effectful computations, following an approach similar to the one outlined in [40].

The partial operation $\boxplus$ on positive types can be extended to a total operation on variable contexts:

$$\varnothing \boxplus \varnothing := \varnothing$$
$$(x : P, \Gamma) \boxplus \Delta := \begin{cases} x : P, \Gamma \boxplus \Delta & \text{if } x \text{ does not occur in } \Delta \\ x : P \boxplus Q, \Gamma \boxplus \Sigma & \text{if } \Delta = x : Q, \Sigma \end{cases}$$

We also extend the operation $p * (-)$ on positive types to a total operation on term variables contexts:

$$p * \varnothing := \varnothing$$
$$p * (x : P, \Gamma) := (x : p * P), p * \Gamma$$

For a polynomial $p$ in $\mathbb{N}_{\geq 1}[i]$ and a type $A$, we write $Ap$ for the type $A[p/i]$ where we substitute all the occurrences of the security parameter $i$ by $p$. Similarly, for a term $M$, we write $Mp$ for the term $M[p/i]$.

The typing rules for λBLL, presented in Figure 8.3, formalize its type system through a set of rules that assign types to terms while respecting linearity and resource constraints. Notably, some of these rules are specifically designed to handle constructs essential for modelling cryptographic protocols. We proceed by describing these rules in detail, highlighting how they manage resource usage and support these critical constructs.

The function symbol rule types applications of probabilistic polynomial-time functions $f_p$. Given that $f$ has a known type signature from ground types $\text{typeof}(f) = G_1 \times \cdots \times G_m \to G$, the rule requires that each argument $Z_k$ is typed under environments where the security parameter $i$ in the types is substituted by the polynomial $p$. This substitution reflects how the type, and thus resource bounds, depends quantitatively on the security parameter. The typed term $\vdash_c^\Theta f_p(Z_1, \ldots, Z_m)$ is assigned the output type $Gp$, consistent with this parameterized interpretation. The variable contexts from each argument are combined using the operation $\boxplus$, which merges resource usage at the level of types to ensure that the total resource consumption is properly accounted for in the typing environment. These function symbols capture cryptographic primitives such as random bitstring generation or bitwise operations, enabling the precise modelling of probabilistic polynomial-time computations essential for cryptographic protocols.

The loop rule types the construct $\texttt{loop}\, V\, p\, \texttt{times}\, \texttt{from}\, M$, where a value $V : P \multimap P$ is iterated $p$ times over a computation $M : P$. The typing context for the iterated body $V$ is scaled by $p$ to capture that its environment is effectively duplicated for each iteration, ensuring the linear type discipline is preserved even under bounded repetition. This rule allows the system to express controlled loops with a polynomial bound on the number of iterations, which is critical for representing interactive adversaries that make a bounded number of queries.

The conditional rule types the standard branching construct $\texttt{if}\, Z\, \texttt{then}\, M\, \texttt{else}\, N$, requiring the condition $Z$ to be boolean, and both branches $M$ and $N$ to have the same type $A$. Importantly, the rule requires that $M$ and $N$ are typed under the same variable context, ensuring a consistent tracking of resources regardless of which branch is taken. The environment of the condition $Z$ is combined via $\boxplus$ with the environment used by the branches, reflecting the total resources needed for the entire conditional expression. This construct is essential for expressing decision-making in protocols, enabling computations to branch based on boolean tests while maintaining precise and sound resource accounting.

The set and get rules manage global references. Writing to a reference with $\texttt{set}\, r\, Z$ requires the value $Z$ to be typed with the ground type $G$ assigned to the reference $r$ by the reference context $\Theta$, and it is typed by the unit type $\mathbb{U}$. Reading from a reference $\texttt{get}\, r$ is typed by the ground type $G$, and requires the reference $r$ of type $G$ to be declared in the global reference context $\Theta$. These rules enable mutable shared state, a necessity for modelling oracles or adversaries that maintain prior knowledge.

## 8.2 Effects

Our calculus incorporates probabilistic effects with references by combining the distribution monad and the state monad. The distribution monad models probabilistic behaviour, which is essential for cryptographic protocols that rely on randomness, such as key generation or message selection. It allows us to represent computations that return values with specific probabilities, enabling simulations of adversarial behaviour. The state monad manages mutable state, crucial for protocols involving shared memory, such as oracles or adversaries maintaining prior knowledge. It provides structured access to memory, allowing for updates and retrievals in a controlled manner. By combining these monads, we can model cryptographic systems where randomness and mutable state interact, such as adversaries querying oracles while affecting system state. This integrated approach gives us the tools to reason about the behaviour and security of cryptographic protocols in a precise and formal way.

**Probability Distributions** Recall that for a set $X$, a *(finite) probability distribution* is a function $\mu : X \to [0,1]$ with finite support, i.e. the set $\mathbf{supp}(\mu) := \{x \in X \mid \mu(x) > 0\}$ is finite, and such that $\sum_{x \in X} \mu(x) = 1$. We denote by $\delta_x : X \to [0,1]$ the *Dirac distribution* mapping an element $y$ in $X$ to 1 if $y = x$ and to 0 otherwise. Any probability distribution $\mu$ is then equal to

$$\sum_{1 \le k \le m} a_k \delta_{x_k} \text{ where } \{x_1, \dots, x_m\} = \mathbf{supp}(\mu) \text{ and } a_k = \mu(x_k)$$

for $1 \le k \le m$. We denote by $\mathbf{D}(X)$ the set of all probability distributions over $X$. It induces a monad $(\mathbf{D}, \eta_{\mathbf{D}}, \gg\!=_{\mathbf{D}})$ on the category $\mathbf{Set}$ of sets and functions (we will omit the subscripts if there is no ambiguity). The unit has components $\eta_X : x \mapsto \delta_x$ given by Dirac distributions and the bind operator

$$\gg\!= : \mathbf{D}(X) \times \mathbf{Set}(X, \mathbf{D}(Y)) \to \mathbf{D}(Y)$$

maps a distribution $\mu = \sum_k a_k \delta_{x_k} \in \mathbf{D}(X)$ and a function $f : X \to \mathbf{D}(Y)$ to the pushforward distribution $\mu \gg\!= f := \sum_k a_k f(x_k)$.

$$\frac{}{x : P \vdash_v^\Theta x : P}\ \textsc{var} \qquad \frac{}{\vdash_v^\Theta \mathbf{t} : \mathbb{B}}\ \textsc{true} \qquad \frac{}{\vdash_v^\Theta \mathbf{f} : \mathbb{B}}\ \textsc{false}$$

$$\frac{\mathrm{typeof}(f) = G_1 \times \cdots \times G_m \to G \quad (\Gamma_k p \vdash_v^\Theta Z_k : G_k p)_{1\le k\le m} \quad p \in \mathbb{N}_{\ge 1}[i]}{\boxplus_k \Gamma_k p \vdash_c^\Theta f_p(Z_1,\dots,Z_m) : Gp}\ \textsc{fun}$$

$$\frac{s \in \{0,1\}^c \quad p : i \mapsto c \text{ is a constant polynomial}}{\vdash_v^\Theta s : \mathbb{S}[p]}\ \textsc{string} \qquad \frac{\Gamma \vdash_v^\Theta Z_1 : P \quad \Delta \vdash_v^\Theta Z_2 : Q}{\Gamma \boxplus \Delta \vdash_v^\Theta \langle Z_1, Z_2\rangle : P \otimes Q}\ \textsc{tensor}$$

$$\frac{\Gamma \vdash_v^\Theta Z : P\otimes Q \quad x:P, y:Q, \Delta \vdash_c^\Theta M : A}{\Gamma\boxplus\Delta \vdash_c^\Theta \mathtt{let}\,\langle x,y\rangle = Z\,\mathtt{in}\,M : A}\ \textsc{let_pair} \qquad \frac{}{\vdash_v^\Theta \star : \mathbb{U}}\ \textsc{unit}$$

$$\frac{\Gamma \vdash_c^\Theta M : A}{p * \Gamma \vdash_v^\Theta\ !M : !_p A}\ \textsc{bang} \qquad \frac{\Gamma \vdash_v^\Theta Z : !_1 A}{\Gamma \vdash_c^\Theta \mathtt{der}(Z) : A}\ \textsc{der} \qquad \frac{\Gamma \vdash_c^\Theta M : P \multimap A \quad \Delta \vdash_v^\Theta Z : P}{\Gamma\boxplus\Delta \vdash_c^\Theta MZ : A}\ \textsc{app}$$

$$\frac{\Gamma, x : P \vdash_c^\Theta M : A}{\Gamma \vdash_v^\Theta \lambda x.M : P \multimap A}\ \textsc{lam} \qquad \frac{\Gamma \vdash_v^\Theta V : A}{\Gamma \vdash_c^\Theta \mathtt{return}\,V : A}\ \textsc{eta} \qquad \frac{\Gamma \vdash_c^\Theta N : P \quad x : P, \Delta \vdash_c^\Theta M : A}{\Gamma\boxplus\Delta \vdash_c^\Theta \mathtt{let}\,x = N\,\mathtt{in}\,M : A}\ \textsc{let}$$

$$\frac{\Gamma \vdash_v^\Theta V : P \multimap P \quad \Delta \vdash_c^\Theta M : P}{(p * \Gamma) \boxplus \Delta \vdash_c^\Theta \mathtt{loop}\,V\,p\,\mathtt{times}\,\mathtt{from}\,M : P}\ \textsc{loop} \qquad \frac{\Gamma \vdash_v^\Theta Z : G \quad r : G \in \Theta}{\Gamma \vdash_c^\Theta \mathtt{set}\,r\,Z : \mathbb{U}}\ \textsc{set}$$

$$\frac{r : G \in \Theta}{\vdash_c^\Theta \mathtt{get}\,r : G}\ \textsc{get} \qquad \frac{\Gamma \vdash_v^\Theta Z : \mathbb{B} \quad \Delta \vdash_c^\Theta M : A \quad \Delta \vdash_c^\Theta N : A}{\Gamma\boxplus\Delta \vdash_c^\Theta \mathtt{if}\,Z\,\mathtt{then}\,M\,\mathtt{else}\,N : A}\ \textsc{case}$$

$$\frac{\Gamma \vdash_v^\Theta V : A \quad x \text{ not free in } \Gamma}{\Gamma, x : P \vdash_v^\Theta V : A}\ \textsc{weak} \qquad \frac{\Gamma \vdash_c^\Theta M : A \quad x \text{ not free in } \Gamma}{\Gamma, x : P \vdash_c^\Theta M : A}\ \textsc{weak}$$

Figure 8.3: Type System of λBLL

**Combining Probability with References** The general idea is that a *store* is a map from memory reference variables to values that preserves typing. More precisely, for a fixed closed reference context $\Theta = r_1 : G_1, \dots, r_m : G_m$ (meaning that the security parameter variable $i$ does not occur in the types $G_1, \dots G_m$), we denote by $\mathrm{St}_\Theta$ the set of functions $e : \{r_1, \dots, r_m\} \to \mathcal{V}$ such that $e(r_j) \in \{V \in \mathcal{V} \mid \cdot \vdash_v^\Theta V : G_j\}$ for all $1 \le j \le m$.

We associate to every closed $\Theta$ a corresponding monad $(\mathbf{T}_\Theta, \eta_\Theta, \gg\!=_\Theta)$ on **Set** given by the tensor product [67] of the distribution monad with the state monad $\mathbf{T}_\Theta := (\mathbf{D}(- \times \mathrm{St}_\Theta))^{\mathrm{St}_\Theta}$, similarly to [9]. The unit of $\mathbf{T}_\Theta$ has components $X \to \mathbf{D}(X \times \mathrm{St}_\Theta)^{\mathrm{St}_\Theta}$ mapping $x \in X$ and $e \in \mathrm{St}_\Theta$ to the Dirac distribution $\delta_{(x,e)}$. The bind operator

$$\gg\!=_\Theta : \mathbf{T}_\Theta X \times \mathbf{Set}(X, \mathbf{T}_\Theta Y) \to \mathbf{T}_\Theta Y$$

takes $\varphi \in \mathbf{T}_\Theta X$ and $f : X \to \mathbf{T}_\Theta Y$ to the map $\boldsymbol{\lambda} e.(\varphi(e) \gg\!=_{\mathbf{D}} \mathbf{eval} \circ (f \times \mathrm{id}_{\mathrm{St}_\Theta}))$ where $\boldsymbol{\lambda}$ and **eval** are respectively the Currying operator and the evaluation map induced by the Cartesian closed structure of **Set**. Explicitly, $\varphi \gg\!=_\Theta f$ maps a store $e \in \mathrm{St}_\Theta$ to the following distribution in $\mathbf{D}(Y \times \mathrm{St}_\Theta)$:

$$(y, e_1) \mapsto \sum_{(x,e_2)} \varphi(e)(x, e_2) \cdot f(x)(e_2)(y, e_1).$$

$\mathbf{T}_\Theta$ is a *strong monad* with respect to Cartesian products in **Set** and for sets $X, Y$, the strength has components $\mathrm{str}_{X,Y} : X \times \mathbf{T}_\Theta Y \to \mathbf{T}_\Theta(X \times Y)$ where for $x \in X$, $\varphi \in \mathbf{T}_\Theta Y = (\mathbf{D}(Y \times \mathrm{St}_\Theta))^{\mathrm{St}_\Theta}$ and $e \in \mathrm{St}_\Theta$, $\mathrm{str}_{X,Y}(x, \varphi)(e)$ is the distribution in $\mathbf{D}(X \times Y \times \mathrm{St}_\Theta)$ defined by:

$$\mathrm{str}_{X,Y}(x, \varphi)(e) : (x', y, e') \mapsto \delta_x(x') \times \varphi(e)(y, e').$$

**From Sets to Indexed Families** In order to work with general term sequents where the security parameter $i$ *may occur freely*, we generalize the discussion above from sets to families of sets. Let **ISet** be the category whose objects are families $X = \{X_n\}_{n \geq 1}$ of sets indexed by $\mathbb{N}_{\geq 1}$ and a morphism from $X = \{X_n\}_{n\geq 1}$ to $Y = \{Y_n\}_{n \geq 1}$ is a family of functions $\{f_n : X_n \to Y_n\}_{n \geq 1}$.

In our calculus, probabilistic effects are generated via the function symbols in $\mathcal{F}$. For each $f \in \mathcal{F}$ with $\mathsf{typeof}(f) = G_1 \times \cdots \times G_m \to G$, we assume that:

- there is a family $[\![f]\!] = \{[\![f]\!]_n\}_{n \geq 1}$ of set-functions $[\![f]\!]_n : [\![G_1]\!]_n \times \cdots \times [\![G_m]\!]_n \to \mathbf{D}([\![G]\!]_n)$ indexed over the security parameter $n \geq 1$ where $[\![\mathbb{S}[p]]\!]_n := \{0, 1\}^{p(n)}$, $[\![\mathbb{B}]\!]_n := \{\mathbf{t}, \mathbf{f}\}$ and $[\![\mathbb{U}]\!]_n := \{\star\}$.
- these functions can be evaluated in probabilistic polynomial time: there exists a PPT algorithm $\mathsf{alg}(f)$ such that for every $n \geq 1$, if $\mathsf{alg}(f)$ is fed with input $1^n$ and a tuple $t \in [\![G_1]\!]_n \times \cdots \times [\![G_m]\!]_n$, it returns $x \in [\![G]\!]_n$ with probability $[\![f]\!]_n(t)(x)$. This can be achieved by taking function symbols from a language guaranteeing the aforementioned complexity bounds [85, 41]. A very small amount of these would however be sufficient for completeness.

Now, for a general reference context $\Theta$ (whose types may contain $i$), we define a monad on **ISet** mapping an indexed family $X = \{X_n\}_{n \geq 1}$ to the family

$$\{\mathbf{T}_{\Theta n}(X_n)\}_{n \geq 1} = \{(\mathbf{D}(X_n \times \mathrm{St}_{\Theta n}))^{\mathrm{St}_{\Theta n}}\}_{n \geq 1}$$

which we will use for the operational semantics of $\lambda$BLL.

## 8.3 Operational Semantics

Given a variable context $\Gamma$, reference context $\Theta$ and type $A$, we define indexed families $\Lambda(\Gamma; \Theta; A) = \{\Lambda_n(\Gamma; \Theta; A)\}_{n \geq 1}$ and $\mathcal{V}(\Gamma; \Theta; A) = \{\mathcal{V}_n(\Gamma; \Theta; A)\}_{n \geq 1}$ of typable terms and values respectively as

$$\Lambda_n(\Gamma; \Theta; A) := \{M \in \Lambda \mid \Gamma n \vdash_c^{\Theta n} M : An\} \quad \text{and} \quad \mathcal{V}_n(\Gamma; \Theta; A) := \{V \in \mathcal{V} \mid \Gamma n \vdash_v^{\Theta n} V : An\}.$$

If the variable context $\Gamma$ is empty, we write $\Lambda_n(\Theta; A)$ and $\mathcal{V}_n(\Theta; A)$ for $\Lambda_n(\varnothing; \Theta; A)$ and $\mathcal{V}_n(\varnothing; \Theta; A)$ respectively.

For a fixed reference context $\Theta$ and type $A$, the small step operational semantics (defined in Figure 8.4) is an indexed family of maps

$$\{\mathbf{sm}_n^{\Theta;A} : \Lambda_n(\Theta; A) \times \mathrm{St}_{\Theta n} \longrightarrow \mathbf{D}(\Lambda_n(\Theta; A) \times \mathrm{St}_{\Theta n})\}_{n \geq 1}.$$

where for the case $\mathtt{set}\, r\, V$, $e[V/r]$ denotes the store mapping a reference $r'$ to $e(r')$ if $r' \neq r$ and to $V$ if $r' = r$.

$$
\begin{aligned}
\mathbf{sm}_n^{\Theta;A}(\mathtt{let}\, x = \mathtt{return}\, V \,\mathtt{in}\, M, e) &:= (\delta(M[V/x], e)) \\
\mathbf{sm}_n^{\Theta;A}(MZ, e) &:= (\mathbf{sm}_n^{\Theta;P\multimap A}(M, e)) \gg\!=_{\mathbf{D}} ((N, e') \mapsto \delta(NZ, e')) \\
\mathbf{sm}_n^{\Theta;A}(\mathtt{let}\, x = N \,\mathtt{in}\, M, e) &:= (\mathbf{sm}_N^{\Theta;P}(N, e)) \gg\!=_{\mathbf{D}} ((N', e') \mapsto \delta(\mathtt{let}\, x = N' \,\mathtt{in}\, M, e')) \\
\mathbf{sm}_n^{\Theta;A}(\mathtt{let}\, \langle x, y\rangle = \langle V, W\rangle \,\mathtt{in}\, M, e) &:= \delta(M[V/x, W/y], e) \\
\mathbf{sm}_n^{\Theta;A}(\mathtt{der}(!M), e)) &:= \delta(M, e) \\
\mathbf{sm}_n^{\Theta;A}((\mathtt{return}\, \lambda x.M)Z, e) &:= \delta(M[Z/x], e) \\
\mathbf{sm}_n^{\Theta;A}(\mathtt{if}\, \mathbf{t} \,\mathtt{then}\, M \,\mathtt{else}\, N, e) &:= \delta(M, e) \\
\mathbf{sm}_n^{\Theta;A}(\mathtt{if}\, \mathbf{f} \,\mathtt{then}\, M \,\mathtt{else}\, N, e) &:= \delta(N, e) \\
\mathbf{sm}_n^{\Theta;A}(\mathtt{set}\, r\, V, e) &:= \delta(\mathtt{return}\, \star, e[V/r]) \\
\mathbf{sm}_n^{\Theta;A}(\mathtt{get}\, r, e) &:= \delta(\mathtt{return}\, e(r), e) \\
\mathbf{sm}_n^{\Theta;A}(f_p(Z_1, \ldots, Z_m), e) &:= (M, e') \mapsto (\delta(e)(e')[\![f]\!]_{p(n)}(Z_1, \ldots, Z_m)(M), e') \\
\mathbf{sm}_n^{\Theta;A}(\mathtt{loop}\, \lambda x.M\, 1 \,\mathtt{times}\,\mathtt{from}\, N, e) &:= \delta(\mathtt{let}\, x = N \,\mathtt{in}\, M, e) \\
\mathbf{sm}_n^{\Theta;A}(\mathtt{loop}\, \lambda x.M\, k+1 \,\mathtt{times}\,\mathtt{from}\, N, e) &:= \delta(\mathtt{let}\, x = (\mathtt{loop}\, (\lambda x.M)\, k \,\mathtt{times}\,\mathtt{from}\, N) \,\mathtt{in}\, M, e)
\end{aligned}
$$

Figure 8.4: Small step semantics of λBLL

Moreover, our calculus is strongly normalizing and in addition to the small step operational semantics for one step reductions, we also provide a final or big step semantics for the convergence behaviour of terms. For a fixed reference context $\Theta$ and a type $A$, the final semantics $(\!|-|\!)^{\Theta,A}$ for closed λBLL-terms is a map in **ISet** corresponding to the indexed family of functions

$$\{(\!|-|\!)_n^{\Theta,A} : \Lambda_n(\Theta; A) \to \mathbf{D}(\mathcal{V}_n(\Theta; A) \times \mathrm{St}_{\Theta n})^{\mathrm{St}_{\Theta n}}\}_{n\geq 1}$$

obtained in two steps:

1. We first use the fact that each monad $\mathbf{T}_\Theta$ on **Set** extends to the category of $\omega$-complete partial orders with a bottom element ($\omega$-cppo) and Scott-continuous morphisms (it follows easily from the fact that both the distribution and the state monads extend to $\omega$-cppo's). It allows us to define inductively a family $(\!|-|\!)_n^{\Theta,A} : \Lambda_n(\Theta; A) \to \mathbf{T}_{\Theta n}^{\perp}(\mathcal{V}_n(\Theta; A))$ where for a set $X$, $\mathbf{T}_{\Theta n}^{\perp} := \mathbf{T}_{\Theta n}(X \uplus \{\perp\}, \leq)$ is the image of the flat ordering ($\perp \leq x$ for all $x \in X$) under $\mathbf{T}_{\Theta n}$ (the bottom element $\perp$ is added to account for computations which are possibly non-terminating). Similarly to [77], each map $(\!|-|\!)_n^{\Theta,A}$ is obtained as the supremum $\bigvee_{k\in\omega} (\!|-|\!)_{n,k}^{\Theta,A}$ where $(\!|-|\!)_{n,k}^{\Theta,A}$ is defined inductively below:

$$
\begin{aligned}
(\!|M|\!)^{\Theta,A}_{n,0} &:= \bot \\
(\!|\texttt{return}\, V|\!)^{\Theta,A}_{n,k+1} &:= \eta_{\mathcal{V}_n(\Theta;A)}(V) \\
(\!|\texttt{if}\, \mathbf{t}\, \texttt{then}\, M\, \texttt{else}\, N|\!)^{\Theta,A}_{n,k+1} &:= (\!|M|\!)^{\Theta,A}_{n,k} \\
(\!|\texttt{if}\, \mathbf{f}\, \texttt{then}\, M\, \texttt{else}\, N|\!)^{\Theta,A}_{n,k+1} &:= (\!|N|\!)^{\Theta,A}_{n,k} \\
(\!|\texttt{der}(!M)|\!)^{\Theta,A}_{n,k+1} &:= (\!|M|\!)^{\Theta,A}_{n,k} \\
(\!|(\texttt{return}\, \lambda x.M)Z|\!)^{\Theta,A}_{n,k+1} &:= (\!|M[Z/x]|\!)^{\Theta,A}_{n,k} \\
(\!|\texttt{set}\, r\, Z|\!)^{\Theta,A}_{n,k+1}(e) &:= \delta_{(\star,e[Z/r])} \\
(\!|\texttt{get}\, r|\!)^{\Theta,A}_{n,k+1}(e) &:= \delta_{(e(r),e)} \\
(\!|\texttt{let}\, \langle x,y\rangle = \langle Z,Z'\rangle\, \texttt{in}\, M|\!)^{\Theta,A}_{n,k+1} &:= (\!|M[Z/x,Z'/y]|\!)^{\Theta,A}_{n,k} \\
(\!|\texttt{let}\, x = N\, \texttt{in}\, M|\!)^{\Theta,A}_{n,k+1} &:= (\!|N|\!)^{\Theta,P}_{n,k} \gg\!= (U \mapsto (\!|M[U/x]|\!)^{\Theta,A}_{n,k}) \\
(\!|MZ|\!)^{\Theta,A}_{n,k+1} &:= (\!|M|\!)^{\Theta,P \overset{\Theta}{\multimap} A}_{n,k} \gg\!= (\lambda x.N \mapsto (\!|N[Z/x]|\!)^{\Theta,A}_{n,k})
\end{aligned}
$$

For the case $(\!|\texttt{loop}\, \lambda x.M\, m\, \texttt{times}\, \texttt{from}\, N|\!)^{\Theta,A}_{n,k+1}$, if $m = 1$, we define it to be

$$(\!|N|\!)^{\Theta,A}_{n,k} \gg\!= (U \mapsto (\!|M[U/x]|\!)^{\Theta,A}_{n,k})$$

and if $m > 1$, we take

$$(\!|\texttt{loop}\, \lambda x.M\, (m-1)\, \texttt{times}\, \texttt{from}\, N|\!)^{\Theta,A}_{n,k} \gg\!= (U \mapsto (\!|M[U/x]|\!)^{\Theta,A}_{n,k}).$$

For a function symbol $f$ with $\mathsf{typeof}(f) = G_1 \times \cdots \times G_m \to G$ and a polynomial $p$ in $\mathbb{N}_{\geq 1}[i]$, $(\!|f_p(W_1,\ldots,W_m)|\!)^{\Theta,A}_{n,k+1}(e)$ is the mapping

$$(V,e') \mapsto \delta_e(e')[\![f]\!]_{p(n)}(W_1,\ldots,W_m)(V).$$

2. We prove that for any $M \in \Lambda_n(\Theta;A)$ and $e \in \mathrm{St}_{\Theta n}$, $(M,e)$ reduces to some distribution $\mathscr{D}$ in polynomial time. As a corollary, we obtain that the final semantics map $(\!|-|\!)^{\Theta,A}_n$ can in fact be restricted to $\Lambda_n(\Theta;A) \to \mathbf{T}_{\Theta n}(\mathcal{V}_n(\Theta;A))$ since $\lambda$BLL is strongly normalizing.

In order to express the standard correspondence between the final (big step) semantics and the transitive closure of the small step semantics, we define the iterated small step reduction which for all $k \in \mathbb{N}$, is an indexed family of partial functions

$$\{\mathbf{sm}^{\Theta;A}_{n,k} : \Lambda_n(\Theta;A) \times \mathrm{St}_{\Theta n} \longrightarrow \mathbf{D}(\Lambda_n(\Theta;A) \times \mathrm{St}_{\Theta n})\}_{n \geq 1}.$$

defined as follows:

$$\mathbf{sm}^{\Theta;A}_{n,0}(M,e) := \begin{cases} \delta(V,e) & \text{if } M = \texttt{return}\, V \\ \bot & \text{otherwise} \end{cases}$$

$$\frac{\mathbf{sm}^{\Theta;A}_{n}(M,e) = \sum_k a_k \delta(M_k,e_k) \qquad \mathbf{sm}^{\Theta;A}_{n,l_k}(M_k,e_k) = \mathscr{D}_k}{\mathbf{sm}^{\Theta;A}_{n,1+\max_k l_k}(M,e) = \sum_k a_k \mathscr{D}_k}$$

and formulate the result as follows:

**Lemma 52.** For a fixed reference context $\Theta$, type $A$ and security parameter $n$, the following are equivalent for any term $M \in \Lambda_n(\Theta;A)$, store $e \in \mathrm{St}_{\Theta n}$ and distribution $\mathscr{D} \in \mathbf{D}(\mathcal{V}_n(\Theta;A) \times \mathrm{St}_{\Theta n})$:

$$(\!|M|\!)^{\Theta,A}_n(e) = \mathscr{D} \quad \Leftrightarrow \quad \exists k \in \mathbb{N},\ (\!|M|\!)^{\Theta,A}_{n,k}(e) = \mathscr{D} \quad \Leftrightarrow \quad \exists m \in \mathbb{N}, \mathbf{sm}^{\Theta;A}_{n,m}(M,e) = \mathscr{D}$$

### Progress

Up to this point, we have considered indexed families $\Lambda(\Gamma;\Theta;A)$ and $\mathcal{V}(\Gamma;\Theta;A)$, consisting of terms that are closed with respect to the security parameter variable $i$. In contrast, the progress property involves reasoning about terms in which the security parameter remains free. This distinction is crucial, as the progress property must hold uniformly for all instantiations of the security parameter. As a result, in order to express and verify this property, we must consider terms where the security parameter has not yet been fixed.

For a variable context $\Gamma$, a location context $\Theta$ and a type $A$, let $\Lambda_o(\Gamma;\Theta;A)$ and $\mathcal{V}_o(\Gamma;\Theta;A)$ be the sets of derivable computation terms and values respectively which are open for the security parameter variable $i$:

$$\Lambda_o(\Gamma;\Theta;A) := \{M \in \Lambda \mid \Gamma \vdash_c^\Theta M : A\} \qquad \mathcal{V}_o(\Gamma;\Theta;A) := \{V \in \mathcal{V} \mid \Gamma \vdash_v^\Theta V : A\}.$$

In particular, it requires that for every positive natural number $n$, the term obtained by substituting $n$ for the security parameter behaves correctly according to the semantics. This is formalized as follows

**Theorem 17** (Progress)**.** For every $M \in \Lambda_o(\varnothing;\Theta;A)$ (i.e., $\vdash_c^\Theta M : A$ is derivable), then either

1. $M = \texttt{return}\, V$ for some value $V \in \mathcal{V}_o(\varnothing;\Theta;A)$ or
2. $\forall n \geq 1$, $\forall e \in \mathrm{St}_{\Theta n}$, $\exists D \in \mathbf{D}(\Lambda_n(\Theta;A) \times \mathrm{St}_{\Theta n})$ such that $\mathbf{sm}_n^{\Theta;A}(M) = D$.

*Sketch Proof.* The first case is immediate. For the second case, we proceed by induction on the derivation of the typing judgment $\vdash_c^{\Theta n} M : An$. Since $M$ is open with respect to the security parameter, we analyze its behavior uniformly for each $n \geq 1$, ensuring that it either is a return value or makes a step in the operational semantics under any suitable store $e \in \mathrm{St}_{\Theta n}$. See page 205 for more details. □

## 8.4 Soundness and Completeness for Polynomial Time

A calculus like λBLL is especially meaningful when considered in the context of cryptographic applications, which we will explore in Chapter 10. Its relevance hinges on establishing a clear correspondence with the notion of probabilistic polynomial time (PPT), a foundational concept in computational cryptography. Ensuring this correspondence is crucial, as it provides theoretical justification for the use of λBLL as a sound and practical framework for modelling and reasoning about the computational aspects of protocol security. This section is therefore devoted to giving evidence supporting that such a correspondence indeed holds, laying the groundwork for the subsequent applications and results discussed later.

Before moving on to the description of soundness and completeness, however, it is worth outlining what is meant in this context by probabilistic polynomial time. In fact, what we mean by a PPT function can be deduced from how we defined function symbols in $\mathcal{F}$: these are families of functions, indexed on natural numbers, which possibly return a distribution and are computable by a probabilistic Turing machine working in polynomial time on the value of the underlying parameter. That basic functions are PPT holds by hypothesis, but that the same remains true for any term definable in the calculus needs to be proved. Moreover, the fact that any such function can be represented in λBLL has to be proved as well.

### 8.4.1 Soundness for PPT

In order to prove polynomial-time soundness, we must show that there exists a polynomial bound on the length of reduction sequences for any term in $\lambda$BLL, and that term size does not increase too much during reduction.

Let $\Lambda_n(\Theta; A)$ denote the set of well-typed terms of type $A$ under reference context $\Theta$, evaluated with respect to a security parameter $n \in \mathbb{N}_{\geq 1}$. Let $\mathrm{St}_{\Theta n}$ denote the set of stores admissible for $\Theta$ under parameter $n$, used to interpret reference values.

**Definition 9** (Cost)**.** For a term $M \in \Lambda_n(\Theta; A)$ and a store $e \in \mathrm{St}_{\Theta n}$, we define the evaluation cost $\mathrm{cost}_n^{A,\Theta}(M, e) \in \mathbb{N} \cup \{+\infty\}$ as

$$\mathrm{cost}_n^{A,\Theta}(M, e) := \inf_k \{\mathbf{sm}_{n,k}^{\Theta;A}(M, e) \neq \bot\}$$

The definition above captures the maximal number of steps required for the term to reduce to a normal form. Our goal is to show that, for every well-typed term $M$, this quantity is polynomially bounded in $n$, uniformly across all valid stores. In order to prove the latter, we follow the structure used in Section 4.2 in [37]. Specifically, we proceed by associating to each typing derivation $\pi$ for a term $M$ a polynomial $\mathit{poly}(\pi) \in \mathbb{N}_{\geq 1}[i]$ that serves as a bound on the computational complexity of evaluating $M$. This polynomial is defined so that it satisfies two key properties: first, it upper-bounds the evaluation cost as in Definition 9; and second, it dominates the size of the term. These properties ensure that evaluation proceeds within a number of steps that is polynomial in the size of the input term, and that simulating each such step is not too expensive. As a result, we conclude that all typable terms in $\lambda$BLL can be evaluated in polynomial time, and thus the language is sound for polynomial-time computation.

First, we formally define the size of terms and values in $\lambda$BLL by mutual induction, as presented in Figure 8.5. This mutual inductive definition reflects the layered structure of $\lambda$BLL syntax, distinguishing ground values, positive values, general values, and computations. It largely follows the classical approach to defining the syntactic size of terms: each constructor contributes a constant, and the total size is obtained by summing this constant with the sizes of its immediate subterms. Although $\lambda$BLL extends the standard linear lambda calculus with additional constructs, such as control flow operators, memory operations, and function symbols, the size of these additional constructs is defined consistently using the same syntactic measure.

Moreover, we assign a polynomial to each type in $\lambda$BLL by mutual induction in Figure 8.6, reflecting the structure of the type system. Ground types such as $\mathbb{U}$ and $\mathbb{B}$ have a fixed size equal to 1, while string types $\mathbb{S}[p]$ contribute a size polynomial equal to the parameter $p$. The composite types, such as tensor products, exponentials, and function types, are assigned polynomials formed by combining the polynomials of their components. In particular, the size of exponential types corresponds to a polynomial in the resource parameter, capturing the quantitative nature of resource usage in $\lambda$BLL. This polynomial assignment provides a precise and compositional measure of type complexity, supporting inductive reasoning and complexity analysis alongside the size definitions for terms.

For convenience, in this section we consider conditionals of the form $\texttt{if}\, Z\, \texttt{then}\, N_1\, \texttt{else}\, N_2$ where both branches $N_1$ and $N_2$ are closed terms. This restriction entails no loss of generality since any conditional with open branches can be equivalently represented using $\lambda$-abstractions, preserving typing and semantics. This simplification enables a modular complexity analysis, allowing the evaluation and cost of each branch to be considered independently, without affecting the expressiveness of the language.

We define the polynomial cost $\mathit{poly}(\cdot)$ associated with each typing derivation $\pi$ by induction on the typing rules, as detailed in Figure 8.7. For basic constructs such as variables and

$$\begin{aligned}
size_W(\star) &:= 1 \\
size_W(\mathbf{t}) &:= 1 \\
size_W(\mathbf{f}) &:= 1 \\
size_W(s) &:= |s|
\end{aligned}$$

(a) Size definition for ground values.

$$\begin{aligned}
size_Z(x) &:= 1 \\
size_Z(W) &:= size_W(W) \\
size_Z(\langle Z_1, Z_2 \rangle) &:= 1 + size_Z(Z_1) + size_Z(Z_2) \\
size_Z(!M) &:= 1 + size_M(M)
\end{aligned}$$

(b) Size definition for positive values.

$$\begin{aligned}
size_V(Z) &:= size_Z(Z) \\
size_V(\lambda x.M) &:= 1 + size_M(M)
\end{aligned}$$

(c) Size definition for values.

$$\begin{aligned}
size_M(\texttt{return}\, V) &:= 1 + size_V(V) \\
size_M(\texttt{der}(Z)) &:= 1 + size_Z(Z) \\
size_M(MZ) &:= 1 + size_M(M) + size_Z(Z) \\
size_M(\texttt{let}\, x = N \,\texttt{in}\, M) &:= 1 + size_M(N) + size_M(M) \\
size_M(\texttt{let}\, \langle x, y \rangle = Z \,\texttt{in}\, N) &:= 1 + size_Z(Z) + size_M(M) \\
size_M(\texttt{if}\, Z \,\texttt{then}\, M \,\texttt{else}\, N) &:= 1 + size_Z(Z) + size_M(M) + size_M(N) \\
size_M(\texttt{loop}\, V\, p \,\texttt{times}\,\texttt{from}\, M) &:= 1 + size_V(V) + size_M(M) \\
size_M(\texttt{set}\, r\, Z) &:= 1 + size_Z(Z) \\
size_M(\texttt{get}\, r) &::= 1 \\
size_M(f_p(Z_1, \ldots, Z_m)) &:= 1 + \sum_{i=1}^{m} size_Z(Z_i)
\end{aligned}$$

(d) Size definition for computations.

Figure 8.5: Definition of the size of a term in λBLL

$$\begin{aligned}
size_G(\mathbb{U}) &:= 1 \\
size_G(\mathbb{B}) &:= 1 \\
size_G(\mathbb{S}[p]) &:= p
\end{aligned}$$

(a) Size definition for ground types.

$$\begin{aligned}
size_P(G) &:= size_G(G) \\
size_P(P_1 \otimes P_2) &:= size_P(P_1) + size_P(P_2) \\
size_Z(!_p A) &:= p \times size_A(A)
\end{aligned}$$

(b) Size definition for positive values.

$$\begin{aligned}
size_A(P) &:= size_A(P) \\
size_A(P \multimap A) &:= 1 + size_P(P) + size_A(A)
\end{aligned}$$

(c) Size definition for generic types.

Figure 8.6: Definition of the size of a type in λBLL

ground values (booleans, unit, strings), the polynomial is given by the size of the type, reflecting their minimal complexity. For composite values, like pairs, abstractions, and applications, the polynomial cost is calculated by summing the polynomials of their sub-derivations and adding a constant overhead. The promotion rule introduces multiplicative cost: promoting a computation $M$ to a value of type $!_p A$ multiplies its cost by the resource polynomial $p$, reflecting potential duplication.

In the fragment related to computations, evaluation cost grows with the structure of the program. Function application, let-bindings, and pattern matching accumulate cost additively from their subderivations, with a fixed overhead. The return and dereliction rules similarly add one to the cost of their inner value. For conditional branches, the total cost includes the cost of evaluating the condition, plus the cost of both branches. While at runtime only one branch executes, both are accounted for to ensure a sound worst-case bound in the presence of branching. The loop construct $\texttt{loop}\, V\, p\, \texttt{times}\, \texttt{from}\, M$ reflects bounded iteration. Its cost includes a setup overhead, the cost of evaluating the initial argument $M$, and the cost of executing the loop body $V$ multiplied by the polynomial $p$ which bounds the number of iterations. This captures the total cost of repeated application and maintains polynomial boundedness. For mutable operations, reading a value (`get`) incurs a fixed cost depending on the ground type size, while writing (`set`) includes the cost of the value being stored and a constant overhead. Finally, the rule for function symbols $f_p(Z_1, \dots, Z_m)$ aggregates the cost of evaluating all its arguments and adds the size of the result type.

As we mentioned above, the polynomial cost $\mathit{poly}(\cdot)$ associated with a typing derivation is designed to upper-bound the evaluation cost of the corresponding term and to dominate its syntactic size. This latter property is formalized by the following lemma.

**Lemma 53.** *For every type derivation $\pi \triangleright \vdash_c^\Theta M : A$ we have that $\forall n \geq 1.\ \mathit{poly}(\pi)n \geq \mathit{size}_M(M)$. Similarly for values, for every type derivation $\rho \triangleright \vdash_c^\Theta V : A$ we have that $\forall n \geq 1.\ \mathit{poly}(\pi)n \geq \mathit{size}_V(V)$.*

The cost function $\mathit{poly}(\cdot)$ is designed to be monotone under substitution. Specifically, when a value is substituted for a variable within a term or another value, the overall cost increases by at most the cost associated with the substituted value. This cost of the substituted value is captured by the function $\underline{\mathit{poly}}(\cdot)$, which is defined to be always less than or equal to $\mathit{poly}(\cdot)$, providing a refined bound on the complexity contribution of values. Consequently, the cost of a substituted term can be bounded by the sum of the original term's cost and the $\underline{\mathit{poly}}$-measure of the substituted value. The use of such a refined cost measure $\underline{\mathit{poly}}(\cdot)$, consistently bounded above by the main cost function $\mathit{poly}(\cdot)$, is a common and standard pattern in complexity analyses of typed languages. This approach enables precise control over the complexity contribution of substituted values during the substitution process.

Formally, the refined measure $\underline{\mathit{poly}}(\cdot)$ is defined in Figure 8.8 and deviates from $\mathit{poly}(\cdot)$ primarily by assigning zero cost to base values such as booleans, unit, and constant strings. Moreover, it is worth mentioning that in the rule for the tensor product, $\underline{\mathit{poly}}(\cdot)$ sums the costs of the components directly without adding the constant overhead present in $\mathit{poly}(\cdot)$. The relationship between the two measures is captured formally by the following monotonicity property:

**Lemma 54** (Monotonicity of $\underline{\mathit{poly}}(\cdot)$ wrt $\mathit{poly}(\cdot)$)**.** *For every type derivation $\pi \triangleright \vdash_c^\Theta M : A$ we have that $\forall n \geq 1.\ \underline{\mathit{poly}}(\pi)n \leq \mathit{poly}(\pi)n$.*
*Similarly for values, for every type derivation $\rho \triangleright \vdash_c^\Theta V : A$ we have that $\forall n \geq 1.\ \underline{\mathit{poly}}(\rho)n \leq \mathit{poly}(\rho)n$.*

*Sketch Proof.* By mutual induction on $\pi$ and $\rho$.
The result follows directly from inspecting the definitions in Figures 8.7 and 8.8. For every

$$poly\left(\frac{}{x : P \vdash^{\Theta}_{v} x : P}\ \textsc{var}\right) := size_P(P) \qquad poly\left(\frac{}{\vdash^{\Theta}_{v} \mathbf{t} : \mathbb{B}}\ \textsc{true}\right) := 1 \qquad poly\left(\frac{}{\vdash^{\Theta}_{v} \mathbf{f} : \mathbb{B}}\ \textsc{false}\right) := 1$$

$$poly\left(\frac{}{\vdash^{\Theta}_{v} * : \mathbb{U}}\ \textsc{unit}\right) := 1 \qquad poly\left(\frac{s \in \{0,1\}^c\ p : i \mapsto c \text{ is a const. polynomial}}{\vdash^{\Theta}_{v} s : \mathbb{S}[p]}\ \textsc{string}\right) := p$$

$$poly\left(\frac{\pi_1 \triangleright \Gamma \vdash^{\Theta}_{v} Z_1 : P\ \pi_2 \triangleright \Delta \vdash^{\Theta}_{v} Z_2 : Q}{\Gamma \boxplus \Delta \vdash^{\Theta}_{v} \langle Z_1, Z_2 \rangle : P \otimes Q}\ \textsc{tensor}\right) := 1 + poly(\pi_1) + poly(\pi_2)$$

$$poly\left(\frac{\pi' \triangleright \Gamma \vdash^{\Theta}_{c} M : A}{p * \Gamma \vdash^{\Theta}_{v} !M : !_p A}\ \textsc{bang}\right) := p \times poly(\pi') \qquad poly\left(\frac{\pi' \triangleright \Gamma, x : P \vdash^{\Theta}_{c} M : A}{\Gamma \vdash^{\Theta}_{v} \lambda x.M : P \multimap A}\ \textsc{lam}\right) := 1 + poly(\pi')$$

$$poly\left(\frac{\pi' \triangleright \Gamma \vdash^{\Theta}_{v} V : A\ x \text{ not free in } \Gamma}{\Gamma, x : P \vdash^{\Theta}_{v} V : A}\ \textsc{weak}\right) := poly(\pi')$$

(a) Polynomial Cost Assignment for Values.

$$poly\left(\frac{\pi' \triangleright \Gamma \vdash^{\Theta}_{v} V : A}{\Gamma \vdash^{\Theta}_{c} \mathtt{return}\, V : A}\ \textsc{eta}\right) := 1 + poly(\pi') \qquad poly\left(\frac{\pi' \triangleright \Gamma \vdash^{\Theta}_{v} Z : !_1 A}{\Gamma \vdash^{\Theta}_{c} \mathtt{der}(Z) : A}\ \textsc{der}\right) := 1 + poly(\pi')$$

$$poly\left(\frac{\pi_1 \triangleright \Gamma \vdash^{\Theta}_{c} M : P \multimap A \qquad \pi_2 \triangleright \Delta \vdash^{\Theta}_{v} Z : P}{\Gamma \boxplus \Delta \vdash^{\Theta}_{c} MZ : A}\ \textsc{app}\right) := 1 + poly(\pi_1) + poly(\pi_2)$$

$$poly\left(\frac{\pi_1 \triangleright \Gamma \vdash^{\Theta}_{c} N : P \qquad \pi_2 \triangleright x : P, \Delta \vdash^{\Theta}_{c} M : A}{\Gamma \boxplus \Delta \vdash^{\Theta}_{c} \mathtt{let}\, x = N \,\mathtt{in}\, M : A}\ \textsc{let}\right) := 1 + poly(\pi_1) + poly(\pi_2)$$

$$poly\left(\frac{\pi_1 \triangleright \Gamma \vdash^{\Theta}_{v} Z : P \otimes Q \qquad x : P, y : Q, \pi_2 \triangleright \Delta \vdash^{\Theta}_{c} M : A}{\Gamma \boxplus \Delta \vdash^{\Theta}_{c} \mathtt{let}\, \langle x, y \rangle = Z \,\mathtt{in}\, M : A}\ \textsc{let_pair}\right) := 1 + poly(\pi_1) + poly(\pi_2)$$

$$poly\left(\frac{\pi_g \triangleright \Gamma \vdash^{\Theta}_{v} Z : \mathbb{B} \qquad \pi_1 \triangleright \Delta \vdash^{\Theta}_{c} M : A \qquad \pi_2 \triangleright \Delta \vdash^{\Theta}_{c} N : A}{\Gamma \boxplus \Delta \vdash^{\Theta}_{c} \mathtt{if}\, Z \,\mathtt{then}\, M \,\mathtt{else}\, N : A}\ \textsc{case}\right) := \begin{array}{l} 1 + poly(\pi_g) \\ + poly(\pi_1) + poly(\pi_2) \end{array}$$

$$poly\left(\frac{\pi_1 \triangleright \Gamma \vdash^{\Theta}_{v} V : P \multimap P \qquad \pi_2 \triangleright \Delta \vdash^{\Theta}_{c} M : P}{(p * \Gamma) \boxplus \Delta \vdash^{\Theta}_{c} \mathtt{loop}\, V\, p \,\mathtt{times}\,\mathtt{from}\, M : P}\ \textsc{loop}\right) := 1 + p + p \times poly(\pi_1) + poly(\pi_2)$$

$$poly\left(\frac{\pi' \triangleright \Gamma \vdash^{\Theta}_{v} Z : G \qquad r : G \in \Theta}{\Gamma \vdash^{\Theta}_{c} \mathtt{set}\, r\, Z : \mathbb{U}}\ \textsc{set}\right) := 2 + poly(\pi')$$

$$poly\left(\frac{r : G \in \Theta}{\vdash^{\Theta}_{c} \mathtt{get}\, r : G}\ \textsc{get}\right) := 2 + size_G(G)$$

$$poly\left(\frac{(\pi_k \triangleright \Gamma_k p \vdash^{\Theta}_{v} Z_k : G_k p)_{1 \leq k \leq m} \qquad p \in \mathbb{N}_{\geq 1}[i]}{\boxplus_k \Gamma_k p \vdash^{\Theta}_{c} f_p(Z_1, \ldots, Z_m) : Gp}\ \textsc{fun}\right) := 1 + size_G(G) + \sum_{k=1}^{m} poly(\pi_k)$$

(b) Polynomial Cost Assignment for Computations.

Figure 8.7: Polynomial Cost Assignment for each Typing Rule.

rule, the cost assigned by $\underline{poly}(\cdot)$ is either strictly smaller or equal to the corresponding cost assigned by $poly(\cdot)$, with differences arising only in the treatment of base values and tensor introduction. $\square$

The proof of substitution in the context of polynomial cost assignment and typing relies on several key auxiliary lemmas that establish fundamental properties.

The following lemmas establish key foundational properties about values of ground types. Lemma 55 ensures that every value in the semantic interpretation of a ground type is typable by a derivation in an empty reference context. Lemma 56 shows that the cost assigned by $poly(\cdot)$ to the derivation with reference context $\Theta$ for a closed value of ground type matches exactly the size of the ground type, reflecting their minimal complexity. Finally, Lemma 57 shows that the refined cost measure $\underline{poly}(\cdot)$ assigns zero cost to such derivations.

**Lemma 55.** For any ground type $G$, $\forall n \geq 1.\forall V \in [\![G]\!]_n$ then $\vdash_v^{\varnothing} V : Gn$.

*Sketch Proof.* By cases inspection on $G$. See page 207 for more details. $\square$

**Lemma 56.** For any ground type $G$ and value $V$ with derivation $\pi \triangleright \vdash_v^{\Theta} V : G$, we have that $poly(\pi) = size_G(G)$.

*Sketch Proof.* We proceed by induction on the last rule of $\pi$. See page 207 for more details. $\square$

**Lemma 57.** For any ground type $G$ and value $V$ with derivation $\pi \triangleright \Gamma \vdash_v^{\Theta} V : G$, we have that $\underline{poly}(\pi) = 0$.

*Sketch Proof.* We proceed by induction on the last rule of $\pi$. See page 208 for more details. $\square$

Moreover, Lemma 58 ensures that the polynomial cost assignment $\underline{poly}(\cdot)$ is coherent with the structure of positive types built using the binary operation $P \boxplus Q$ and the unary operation $p * P$ on positive types. In both cases, the lemma formally establishes that the polynomial assignment $\underline{poly}(\cdot)$ aligns with the inductive structure of positive types. This alignment is essential for reasoning modularly about cost during substitution and composition of terms.

**Lemma 58.**

A. For a value $V$, positive types $P$ and $Q$, if $\pi \triangleright \vdash_v^{\Theta} V : P \boxplus Q$, then there exist derivations $\pi_1 \triangleright \vdash_v^{\Theta} V : P$ and $\pi_2 \triangleright \vdash_v^{\Theta} V : Q$ such that $\underline{poly}(\pi) = \underline{poly}(\pi_1) + \underline{poly}(\pi_2)$.

B. For a positive value $Z$, positive type $P$ and polynomial $p$ in $\mathbb{N}_{\geq 1}[i]$, if $\pi \triangleright \vdash_v^{\Theta} Z : p * P$, then there exists a derivation $\pi' \triangleright \vdash_v^{\Theta} Z : P$ such that $\underline{poly}(\pi) = p \times \underline{poly}(\pi')$.

*Sketch Proof.* The first case is proved by induction on the definition of $P \boxplus Q$ and the second case is by induction on the definition of $p * P$. See page 208 for more details. $\square$

Building on these foundational cost decompositions, we now state the substitution lemma. It ensures that substituting a well-typed value for a variable within a term or another value preserves typing and adds at most the cost of the substituted value to the overall cost. As described earlier, this additional cost is measured using the refined cost function $\underline{poly}(\cdot)$, which provides a tighter bound on the complexity contribution of values.

$$\underline{poly}\left(\frac{}{x : P \vdash^{\Theta}_{v} x : P}\ \text{VAR}\right) := size_P(P) \qquad \underline{poly}\left(\frac{}{\vdash^{\Theta}_{v} \mathbf{t} : \mathbb{B}}\ \text{TRUE}\right) := 0 \qquad \underline{poly}\left(\frac{}{\vdash^{\Theta}_{v} \mathbf{f} : \mathbb{B}}\ \text{FALSE}\right) := 0$$

$$\underline{poly}\left(\frac{}{\vdash^{\Theta}_{v} * : \mathbb{U}}\ \text{UNIT}\right) := 0 \qquad \underline{poly}\left(\frac{s \in \{0,1\}^c\ p : i \mapsto c \text{ is a const. polynomial}}{\vdash^{\Theta}_{v} s : \mathbb{S}[p]}\ \text{STRING}\right) := 0$$

$$\underline{poly}\left(\frac{\pi_1 \triangleright \Gamma \vdash^{\Theta}_{v} Z_1 : P\ \pi_2 \triangleright \Delta \vdash^{\Theta}_{v} Z_2 : Q}{\Gamma \boxplus \Delta \vdash^{\Theta}_{v} \langle Z_1, Z_2 \rangle : P \otimes Q}\ \text{TENSOR}\right) := \underline{poly}(\pi_1) + \underline{poly}(\pi_2)$$

$$\underline{poly}\left(\frac{\pi' \triangleright \Gamma \vdash^{\Theta}_{c} M : A}{p * \Gamma \vdash^{\Theta}_{v} !M : !_p A}\ \text{BANG}\right) := p \times \underline{poly}(\pi') \qquad \underline{poly}\left(\frac{\pi' \triangleright \Gamma, x : P \vdash^{\Theta}_{c} M : A}{\Gamma \vdash^{\Theta}_{v} \lambda x.M : P \multimap A}\ \text{LAM}\right) := 1 + \underline{poly}(\pi')$$

$$\underline{poly}\left(\frac{\pi' \triangleright \Gamma \vdash^{\Theta}_{v} V : A\ x \text{ not free in } \Gamma}{\Gamma, x : P \vdash^{\Theta}_{v} V : A}\ \text{WEAK}\right) := \underline{poly}(\pi')$$

(a) $\underline{poly}(\cdot)$ Cost Assignment for Values.

$$\underline{poly}\left(\frac{\pi' \triangleright \Gamma \vdash^{\Theta}_{v} V : A}{\Gamma \vdash^{\Theta}_{c} \texttt{return}\, V : A}\ \text{ETA}\right) := 1 + \underline{poly}(\pi') \qquad \underline{poly}\left(\frac{\pi' \triangleright \Gamma \vdash^{\Theta}_{v} Z : !_1 A}{\Gamma \vdash^{\Theta}_{c} \texttt{der}(Z) : A}\ \text{DER}\right) := 1 + \underline{poly}(\pi')$$

$$\underline{poly}\left(\frac{\pi_1 \triangleright \Gamma \vdash^{\Theta}_{c} M : P \multimap A \qquad \pi_2 \triangleright \Delta \vdash^{\Theta}_{v} Z : P}{\Gamma \boxplus \Delta \vdash^{\Theta}_{c} MZ : A}\ \text{APP}\right) := 1 + \underline{poly}(\pi_1) + \underline{poly}(\pi_2)$$

$$\underline{poly}\left(\frac{\pi_1 \triangleright \Gamma \vdash^{\Theta}_{c} N : P \qquad \pi_2 \triangleright x : P, \Delta \vdash^{\Theta}_{c} M : A}{\Gamma \boxplus \Delta \vdash^{\Theta}_{c} \texttt{let}\, x = N \,\texttt{in}\, M : A}\ \text{LET}\right) := 1 + \underline{poly}(\pi_1) + \underline{poly}(\pi_2)$$

$$\underline{poly}\left(\frac{\pi_1 \triangleright \Gamma \vdash^{\Theta}_{v} Z : P \otimes Q \qquad x : P, y : Q, \pi_2 \triangleright \Delta \vdash^{\Theta}_{c} M : A}{\Gamma \boxplus \Delta \vdash^{\Theta}_{c} \texttt{let}\, \langle x, y \rangle = Z \,\texttt{in}\, M : A}\ \text{LET_PAIR}\right) := 1 + \underline{poly}(\pi_1) + \underline{poly}(\pi_2)$$

$$\underline{poly}\left(\frac{\pi_g \triangleright \Gamma \vdash^{\Theta}_{v} Z : \mathbb{B} \qquad \pi_1 \triangleright \Delta \vdash^{\Theta}_{c} M : A \qquad \pi_2 \triangleright \Delta \vdash^{\Theta}_{c} N : A}{\Gamma \boxplus \Delta \vdash^{\Theta}_{c} \texttt{if}\, Z \,\texttt{then}\, M \,\texttt{else}\, N : A}\ \text{CASE}\right) := \begin{array}{l} 1 + \underline{poly}(\pi_g) \\ + \underline{poly}(\pi_1) + \underline{poly}(\pi_2) \end{array}$$

$$\underline{poly}\left(\frac{\pi_1 \triangleright \Gamma \vdash^{\Theta}_{v} V : P \multimap P \qquad \pi_2 \triangleright \Delta \vdash^{\Theta}_{c} M : P}{(p * \Gamma) \boxplus \Delta \vdash^{\Theta}_{c} \texttt{loop}\, V\, p \,\texttt{times from}\, M : P}\ \text{LOOP}\right) := 1 + p + p \times \underline{poly}(\pi_1) + \underline{poly}(\pi_2)$$

$$\underline{poly}\left(\frac{\pi' \triangleright \Gamma \vdash^{\Theta}_{v} Z : G \qquad r : G \in \Theta}{\Gamma \vdash^{\Theta}_{c} \texttt{set}\, r\, Z : \mathbb{U}}\ \text{SET}\right) := 2 + \underline{poly}(\pi')$$

$$\underline{poly}\left(\frac{r : G \in \Theta}{\vdash^{\Theta}_{c} \texttt{get}\, r : G}\ \text{GET}\right) := 2 + size_G(G)$$

$$\underline{poly}\left(\frac{(\pi_k \triangleright \Gamma_k p \vdash^{\Theta}_{v} Z_k : G_k p)_{1 \le k \le m} \qquad p \in \mathbb{N}_{\ge 1}[i]}{\boxplus_k \Gamma_k p \vdash^{\Theta}_{c} f_p(Z_1, \ldots, Z_m) : Gp}\ \text{FUN}\right) := 1 + size_G(G) + \sum_{k=1}^{m} \underline{poly}(\pi_k)$$

(b) $\underline{poly}(\cdot)$ Cost Assignment for Computations.

Figure 8.8: $\underline{poly}(\cdot)$ Cost Assignment for each Typing Rule.

**Lemma 59** (Substitution)**.**

1. Given two derivations $\pi \triangleright \Gamma, x : P \vdash^{\Theta}_{c} M : A$ and $\rho \triangleright \vdash^{\Theta}_{v} V : P$, then there exists a derivation $\pi[\rho/x] \triangleright \Gamma \vdash^{\Theta}_{c} M[V/x] : A$ such that $\mathit{poly}(\pi[\rho/x]) \leq \mathit{poly}(\pi) + \underline{\mathit{poly}}(\rho)$.

2. Given two derivations $\varphi \triangleright \Gamma, x : P \vdash^{\Theta}_{v} U : A$ and $\rho \triangleright \vdash^{\Theta}_{v} V : P$, then there exists a derivation $\varphi[\rho/x] \triangleright \Gamma \vdash^{\Theta}_{c} U[V/x] : A$ such that $\mathit{poly}(\varphi[\rho/x]) \leq \mathit{poly}(\varphi) + \underline{\mathit{poly}}(\rho)$.

*Sketch Proof.* By mutual induction on $\pi$ and $\varphi$. See page 210 for more details. □

Another crucial property of the polynomial cost assignment is that it strictly decreases along term reduction. The following lemma formalizes this concept by showing that, for any well-typed term and store, each small-step reduction produces a term whose polynomial cost is strictly smaller than that of the original term.

**Lemma 60** (Measure Decreasing along Reduction)**.** Given $\pi \triangleright \vdash^{\Theta}_{c} M : A$ and store $e \in \mathrm{St}_{\Theta}$ such that $\mathbf{sm}^{\Theta;A}_{n}(M, e) = \sum_k a_k\ \delta(M_k, e_k)$, then for every $k$ there exists $\rho_k \triangleright \vdash^{\Theta}_{c} M_k : A$ such that $\mathit{poly}(\pi) > \mathit{poly}(\rho_k)$.

*Proof.* By induction on the definition of small step reduction (Figure 8.4). See page 216 for more details. □

Moreover, we prove the Subject Reduction Theorem (Theorem 18), which guarantees two critical properties for well-typed terms in λBLL. First, it ensures that every term obtained after a single small-step reduction remains well-typed with the same type, preserving type correctness. Second, the theorem guarantees that the assigned cost of the original term is always strictly higher than the costs of any terms it reduces to, reinforcing the idea that evaluation progresses by reducing complexity. Together, these properties are essential to maintain both type safety and the soundness of the polynomial cost analysis during execution.

**Theorem 18** (Subject Reduction)**.** For all $n \geq 1$, for every reference context $\Theta$, type $A$ and term $M \in \Lambda_n(\Theta, A)$, for all $e \in \mathrm{St}_{\Theta}$ such that $\mathbf{sm}^{\Theta;A}_{n}(M, e) = \sum_k a_k\ \delta(M_k, e_k)$ for some $a_k$, term $M_k$ and store $e_k$, then

1. $\forall k. \vdash^{\Theta n}_{c} M_k : An$ is derivable;

2. $\mathrm{cost}^{A,\Theta}_{n}(M, e) > \max_k \mathrm{cost}^{A,\Theta}_{n}(M_k, e_k)$

*Proof.*

1. The existence of a derivation $\vdash^{\Theta n}_{c} M_k : An$ has been already proved in Lemma 60.

2. By Definition 9 we have that $\mathrm{cost}^{A,\Theta}_{n}(M, e) = \inf_k \{\mathbf{sm}^{\Theta;A}_{n,k}(M, e) \neq \bot\}$, so we proceed by splitting in the following cases:
   - *Case* $\forall l \in \mathbb{N}.\mathbf{sm}^{\Theta;A}_{n,l}(M, e) = \bot$:
     In this case we have that $\mathrm{cost}^{A,\Theta}_{n}(M, e) = +\inf$ and so the inequality is trivial.

   - *Case* $\exists\ l \in \mathbb{N}.\mathbf{sm}^{\Theta;A}_{n,l}(M, e) \neq \bot$:
     By hypothesis we have that $\mathbf{sm}^{\Theta;A}_{n}(M, e) = \sum_k a_k\ \delta(M_k, e_k)$ and $\mathbf{sm}^{\Theta;A}_{n,l}(M, e) \neq \bot$, then $\mathbf{sm}^{\Theta;A}_{n,l}(M, e) = \sum_k a_k\ \mathscr{D}_k$ where $\mathbf{sm}^{\Theta;A}_{n,l_k}(M_k, e_k) = \mathscr{D}_k$ and $l = 1 + \max_k l_k$. Therefore, we have that $\mathrm{cost}^{A,\Theta}_{n}(M, e) \leq l = 1 + \max_k l_k$ and so we can conclude.

□

Finally, Polytime Soundness formalizes the key result that all well-typed term in λBLL evaluates to normal form within a number of steps bounded by a polynomial in the size of its input. The proof proceeds by leveraging the strictly decreasing nature of the polynomial cost assignment along reduction steps (Lemma 60) and the preservation of typing and cost bounds under reduction (Theorem 18).

**Theorem 19** (Polytime Soundness)**.** For every type derivation $\pi \triangleright \vdash_c^\Theta M : A$, then there exists a polynomial $\mathit{poly}(\pi)$ such that

$$\forall n \geq 1. \forall e \in \mathrm{St}_{\Theta n}.\ \mathrm{cost}_n^{A,\Theta}(M, e) \leq \mathit{poly}(\pi)(n).$$

*Proof.* By Theorem 17 we have two cases:

1. *Case* $M = \mathtt{return}\, V$:
   We have that $\mathrm{cost}_n^{A,\Theta}(M, e) = 0$ and so the inequality is trivial.

2. *Case* $\mathbf{sm}_n^{\Theta;A}(M, e) = \sum_k a_k\ \delta(M_k, e_k)$:
   By Lemma 60 we have that for every $k$ there exists a derivation $\rho_k \triangleright \vdash_c^\Theta M_k : A$ such that $\mathit{poly}(\pi) > \mathit{poly}(\rho_k)$. Therefore, we have that

$$1 + \max_k \mathit{poly}(\rho_k)(n) \leq \mathit{poly}(\pi)(n). \tag{8.1}$$

   By Theorem 18 case 2, we have that

$$\mathrm{cost}_n^{A,\Theta}(M, e) > \max_k\ \mathrm{cost}_n^{A,\Theta}(M_k, e_k). \tag{8.2}$$

   Summing up, we have

$$\begin{aligned} \mathrm{cost}_n^{A,\Theta}(M, e) &\overset{\text{Eq. } 8.2}{\leq} 1 + \max_k\ \mathrm{cost}_n^{A,\Theta}(M_k, e_k) \\ &\leq 1 + \max_k\ \mathit{poly}(\rho_k)(n) \\ &\overset{\text{Eq. } 8.1}{\leq} \mathit{poly}(\pi)(n) \end{aligned}$$

   and we can conclude.

□

### 8.4.2 Completeness for PPT

Theorem 19 implicitly tells us that algorithms formulated as typable λBLL terms are PPT, since the number of reduction steps performed is polynomially bounded and reduction can be simulated by a Turing machine [39, 7]. One can further prove that all PPT functions can be represented by λBLL terms. Given the freedom we have about picking more and more basic function symbols, this does not seem surprising: we are anyway allowed to throw in new basic function symbols whenever needed. However, one can prove that completeness for PPT can be achieved with a very minimal set of basic functions symbols only including cyclic shift functions on strings and functions testing the value of the first bit in a string.

**Remark 14.** Noticeably, soundness and completeness as presented above scale to *second-order* PPT, namely a notion of probabilistic polynomial-time function accessing an oracle. This is quite relevant in our setting, given our emphasis on cryptographic constructions, and the fact that adversaries for some of those have oracle access to the underlying primitive. A formal treatment of soundness and completeness for second-order PPT is left for future work.

# Detailed Proofs

## Proof Progress

**Theorem 17** (Progress)**.** For every $M \in \Lambda_o(\varnothing;\Theta;A)$ (i.e., $\vdash_c^{\Theta} M : A$ is derivable), then either

1. $M = \texttt{return}\, V$ for some value $V \in \mathcal{V}_o(\varnothing;\Theta;A)$ or

2. $\forall n \geq 1$, $\forall e \in \mathrm{St}_{\Theta n}$, $\exists D \in \mathbf{D}(\Lambda_n(\Theta;A) \times \mathrm{St}_{\Theta n})$ such that $\mathbf{sm}_n^{\Theta;A}(M) = D$.

*Case 1.* We have that $M = \texttt{return}\, V$, by ETA-typing rule we have that $V \in \mathcal{V}_o(\varnothing;\Theta;A)$ and we can conclude. □

*Case 2.* By induction on the derivation of the typing judgment $\vdash_c^{\Theta n} M : An$.

- *Rule FUN*:

$$\frac{\mathrm{typeof}(f) = G_1 \times \cdots \times G_m \to G \qquad (\vdash_v^{\Theta n} Z_k : G_k pn)_{1 \leq k \leq m} \qquad p \in \mathbb{N}_{\geq 1}[i]}{\vdash_c^{\Theta n} f_p(Z_1, \ldots, Z_m) : Gpn}$$

If $[\![f]\!]_{p(n)} : [\![G_1]\!]_{p(n)} \times \cdots \times [\![G_m]\!]_{p(n)} \to D([\![G]\!]_{p(n)})$, we have that $[\![f]\!]_{p(n)}(Z_1, \ldots, Z_m) = \sum_{j=1}^m a_j\, \delta_{M_j}$.
By operational semantics we have $\forall e.\mathbf{sm}_n^{\Theta;A}(f_p(Z_1, \ldots, Z_m), e) = \sum_{j=1}^m a_j\, \delta(M_j, e')$ and we can conclude.

- *Rule LET_PAIR*:

$$\frac{\vdash_v^{\Theta n} Z : Pn \otimes Qn \qquad x : Pn, y : Qn \vdash_c^{\Theta n} M : An}{\vdash_c^{\Theta n} \texttt{let}\, \langle x, y\rangle = Z\, \texttt{in}\, M : An}$$

We must have $Z = \langle Z_1, Z_2\rangle$ such that $\vdash_v^{\Theta n} Z_1 : Pn$ and $\vdash_v^{\Theta n} Z_2 : Qn$ are derivable, then by small step semantics we have that $\forall e.\mathbf{sm}_n^{\Theta;A}(\texttt{let}\, \langle x, y\rangle = \langle Z_1, Z_2\rangle\, \texttt{in}\, M, e) = \delta(N[Z_1/x, Z_2/y], e)$ and we can conclude.

- *Rule DER*:

$$\frac{\vdash_v^{\Theta n} Z :!_1 An}{\vdash_c^{\Theta n} \texttt{der}(Z) : An}$$

We must have $Z = !M$ such that $\vdash_c^{\Theta n} N : An$ is derivable, then by small step semantics we have that $\forall e.\mathbf{sm}_n^{\Theta;A}(\texttt{der}(!N), e) = \delta(N, e)$ and we can conclude.

- *Rule LET*:

$$\frac{\vdash_c^{\Theta n} N_1 : Pn \qquad x : Pn \vdash_c^{\Theta n} N_2 : An}{\vdash_c^{\Theta n} \texttt{let}\, x = N_1\, \texttt{in}\, N_2 : An}$$

By induction hypothesis on $N_1$ we have that either $N_1 = \texttt{return}\, V$ or $\forall e \in \mathrm{St}_{\Theta n}.\exists D \in \mathbf{D}(\Lambda_n(\Theta;A) \times \mathrm{St}_{\Theta n})$ such that $\mathbf{sm}_n^{\Theta;A}(N_1) = D$. We proceed by analyzing these two cases:

– Sub-case $N_1 = \texttt{return}\, V$:
We have that $M = \texttt{let}\, x = N_1 \,\texttt{in}\, N_2 = \texttt{let}\, x = \texttt{return}\, V \,\texttt{in}\, N_2$. By hypothesis of LET-typing rule we have that $\vdash_c^{\Theta n} N_1 : Pn$ is derivable, then in this case $\vdash_c^{\Theta n}$ $\texttt{return}\, V : Pn$ is derivable. By small step semantics we have that $\forall e.\mathbf{sm}_n^{\Theta;A}(\texttt{let}\, x = \texttt{return}\, V \,\texttt{in}\, N_2, e) = \delta(N_2[V/x], e)$ and we can conclude.

– Sub-case $D = \sum_k a_k \delta(M_k, e_k)$ such that $\mathbf{sm}_n^{\Theta;A}(N_1) = D$:
By small step semantics we have that $\forall e.\mathbf{sm}_n^{\Theta;A}(\texttt{let}\, x = N_1 \,\texttt{in}\, N_2, e) = \sum_k a_k\ \delta(\texttt{let}\, x = M_k \,\texttt{in}\, N_2, e_k)$ and we can conclude.

- *Rule APP*:

$$\frac{\vdash_c^{\Theta n} N : Pn \multimap An \qquad \vdash_v^{\Theta n} Z : Pn}{\vdash_c^{\Theta n} NZ : An}$$

By induction hypothesis on $N$ we have that either $N = \texttt{return}\, V$ or $\forall e \in \mathrm{St}_{\Theta n}.\exists D \in \mathbf{D}(\Lambda_n(\Theta; A) \times \mathrm{St}_{\Theta n})$ such that $\mathbf{sm}_n^{\Theta;A}(N) = D$. We proceed by analyzing these two cases:

– Sub-case $N_1 = \texttt{return}\, V$:
We have that $M = NZ = (\texttt{return}\, V)Z$. By hypothesis of APP-typing rule we have that $\vdash_c^{\Theta n} N : Pn \multimap An$ is derivable, then in this case $\vdash_c^{\Theta n} \texttt{return}\, V : Pn \multimap An$ is derivable, so $V$ is in the form $V = \lambda x.N'$ for some $N'$ such that $x : Pn \vdash_c^{\Theta n} N' : An$. By small step semantics we have that $\forall e.\mathbf{sm}_n^{\Theta;A}((\texttt{return}\, \lambda x.N')Z, e) = \delta(N'[Z/x], e)$ and we can conclude.

– Sub-case $D = \sum_k a_k \delta(M_k, e_k)$ such that $\mathbf{sm}_n^{\Theta;A}(N) = D$:
By small step semantics we have that $\forall e.\mathbf{sm}_n^{\Theta;A}(NZ, e) = \sum_k a_k\ \delta(M_k Z, e_k)$ and we can conclude.

- *Rule LOOP*:

$$\frac{\vdash_v^{\Theta n} V : Pn \multimap Pn \qquad \vdash_c^{\Theta n} N : Pn}{\vdash_c^{\Theta n} \texttt{loop}\, V\, p\, \texttt{times}\, \texttt{from}\, N : Pn}$$

By hypothesis of LOOP-typing rule we have that $\vdash_v^{\Theta n} V : Pn \multimap Pn$ is derivable, s $V$ is in the form $V = \lambda x.N'$ for some $N'$ such that $x : Pn \vdash_c^{\Theta n} N' : Pn$ derivable. By small step semantics we have to analyze the following two sub-cases:

– Sub-case $p = 1$:
By small step semantics we have that $\forall e.\mathbf{sm}_n^{\Theta;A}(\texttt{loop}\, \lambda x.N'\, 1\, \texttt{times}\, \texttt{from}\, N, e) = \delta(\texttt{let}\, x = N' \,\texttt{in}\, N, e)$ and we can conclude.

– Sub-case $p = 1 + q$ for some $q \in \mathbb{N}_{\geq 1}[i]$:
By small step semantics we have $\forall e.\mathbf{sm}_n^{\Theta;A}(\texttt{loop}\, \lambda x.N'\, 1 + qn\, \texttt{times}\, \texttt{from}\, N, e) = \delta(\texttt{loop}\, \lambda x.N'\, qn\, \texttt{times}\, \texttt{from}\, N, e)$

- *Rule CASE*:

$$\frac{\vdash_v^{\Theta n} Z : \mathbb{B} \qquad \vdash_c^{\Theta n} N_1 : An \qquad \vdash_c^{\Theta n} N_2 : An}{\vdash_c^{\Theta n} \texttt{if}\, Z\, \texttt{then}\, N_1\, \texttt{else}\, N_2 : An}$$

By hypothesis of CASE-typing rule we have that $\vdash_v^{\Theta n} Z : \mathbb{B}$ is derivable, so we have to analyze the following two sub-cases:

– Sub-case $Z = \mathbf{t}$:
By small step semantics we have $\forall e.\mathbf{sm}_n^{\Theta;A}(\texttt{if}\,\mathbf{t}\,\texttt{then}\,N_1\,\texttt{else}\,N_2, e) = \delta(N_1, e)$ and we can conclude.

– Sub-case $Z = \mathbf{f}$:
By small step semantics we have $\forall e.\mathbf{sm}_n^{\Theta;A}(\texttt{if}\,\mathbf{f}\,\texttt{then}\,N_1\,\texttt{else}\,N_2, e) = \delta(N_2, e)$ and we can conclude.

- *Rule SET*:

$$\frac{\vdash_v^{\Theta n} Z : G \qquad r : G \in \Theta n}{\vdash_c^{\Theta n} \texttt{set}\,r\,Z : \mathbb{U}}$$

By small step semantics we have $\forall e.\mathbf{sm}_n^{\Theta;A}(\texttt{set}\,r\,Z, e) = \delta(\texttt{return}\,*, e[Z/r])$ and we can conclude.

- *Rule GET*:

$$\frac{r : G \in \Theta n}{\vdash_c^{\Theta n} \texttt{get}\,r : G}$$

By small step semantics we have $\forall e.\mathbf{sm}_n^{\Theta;A}(\texttt{get}\,r, e) = \delta(\texttt{return}\,e(r), e)$ and we can conclude.

□

## Proofs Polytime Soundness

**Lemma 55.** For any ground type $G$, $\forall n \geq 1.\forall V \in [\![G]\!]_n$ then $\vdash_v^{\varnothing} V : Gn$.

*Proof.* By cases inspection on $G$.

- *Case* $G = \mathbb{U}$: by definition $[\![\mathbb{U}]\!] = \{\star\}$ and by UNIT-typing rule we have that $\vdash_v^{\varnothing} \star : \mathbb{U}$ is derivable.
- *Case* $G = \mathbb{S}[p]$: by definition $[\![\mathbb{S}[p]]\!] = \{0,1\}^{p(n)}$. Let us consider a string $s \in \{0,1\}^{p(n)}$ and by STRING-typing rule we have that $\vdash_v^{\varnothing} s : \mathbb{S}[p]n$ is derivable.
- *Case* $G = \mathbb{B}$: by definition $[\![\mathbb{B}]\!] = \{\mathbf{t}, \mathbf{f}\}$ and by TRUE-typing rule (resp. FALSE-typing rule) we have that $\vdash_v^{\varnothing} \mathbf{t} : \mathbb{B}$ (resp. $\vdash_v^{\varnothing} \mathbf{f} : \mathbb{B}$) is derivable.

□

**Lemma 56.** For any ground type $G$ and value $V$ with derivation $\pi \rhd \vdash_v^{\Theta} V : G$, we have that $poly(\pi) = size_G(G)$.

*Proof.* We proceed by induction on the last rule of $\pi$. For a value of ground type, the last rule can only be:

- *Case UNIT*: we have that $\pi \rhd \vdash_v^{\Theta} \star : \mathbb{U}$ and by definition we have that $poly(\pi) = size_G(\mathbb{U})$.
- *Case TRUE/FALSE*: we have that $\pi \rhd \vdash_v^{\Theta} \mathbf{t} : \mathbb{B}$ or $\pi \rhd \vdash_v^{\Theta} \mathbf{f} : \mathbb{B}$ and by definition in both cases we have that $poly(\pi) = size_G(\mathbb{B})$.

- *Case STRING*: let us consider a string $s \in \{0,1\}^{p(n)}$, we have that $\pi \triangleright \vdash_v^{\Theta} s : \mathbb{S}[p]$ and by definition we have that $poly(\pi) = size_G(\mathbb{S}[p])$.

$\square$

**Lemma 57.** For any ground type $G$ and value $V$ with derivation $\pi \triangleright \Gamma \vdash_v^{\Theta} V : G$, we have that $\underline{poly}(\pi) = 0$.

*Proof.* We proceed by induction on the last rule of $\pi$. For a value of ground type, the last rule can only be:

- *Case UNIT*: we have that $\pi \triangleright \vdash_v^{\Theta} \star : \mathbb{U}$ and by definition we have that $\underline{poly}(\pi) = 1$.
- *Case TRUE/FALSE*: we have that $\pi \triangleright \vdash_v^{\Theta} \mathbf{t} : \mathbb{B}$ or $\pi \triangleright \vdash_v^{\Theta} \mathbf{f} : \mathbb{B}$ and by definition in both cases we have that $\underline{poly}(\pi) = 0$.
- *Case STRING*: let us consider a string $s \in \{0,1\}^{p(n)}$, we have that $\pi \triangleright \vdash_v^{\Theta} s : \mathbb{S}[p]$ and by definition we have that $\underline{poly}(\pi) = 0$.

$\square$

**Lemma 58.**

A. For a value $V$, positive types $P$ and $Q$, if $\pi \triangleright \vdash_v^{\Theta} V : P \boxplus Q$, then there exist derivations $\pi_1 \triangleright \vdash_v^{\Theta} V : P$ and $\pi_2 \triangleright \vdash_v^{\Theta} V : Q$ such that $\underline{poly}(\pi) = \underline{poly}(\pi_1) + \underline{poly}(\pi_2)$.

B. For a positive value $Z$, positive type $P$ and polynomial $p$ in $\mathbb{N}_{\geq 1}[i]$, if $\pi \triangleright \vdash_v^{\Theta} Z : p * P$, then there exists a derivation $\pi' \triangleright \vdash_v^{\Theta} Z : P$ such that $\underline{poly}(\pi) = p \times \underline{poly}(\pi')$.

*Case A.* By induction on the definition of $P \boxplus Q$.

- *Case $P = Q = G$ for some ground type $G$*:
  By definition of $\boxplus$ we have that $G \boxplus G = G$, therefore we have that $\pi \triangleright \vdash_v^{\Theta} V : G$. By Lemma 57 we have that $\underline{poly}(\pi) = 0$ and $\underline{poly}(\pi_1) = \underline{poly}(\pi_2) = 0$.

- *Case $P = P_1 \otimes P_2$ and $Q = Q_1 \otimes Q_2$*:
  By definition of $\boxplus$ we have that $(P_1 \otimes P_2) \boxplus (Q_1 \otimes Q_2) = (P_1 \boxplus Q_1) \otimes (P_2 \boxplus Q_2)$. By hypothesis we have that $\pi \triangleright \vdash_v^{\Theta} V : (P_1 \boxplus Q_1) \otimes (P_2 \boxplus Q_2)$, so $V$ must be in the form $Z = \langle Z_1, Z_2 \rangle$ and by TENSOR-typing rule we have $\pi_1 \triangleright \vdash_v^{\Theta} Z_1 : P_1 \boxplus Q_1$ and $\pi_2 \triangleright \vdash_v^{\Theta} Z_2 : P_2 \boxplus Q_2$. Moreover, by definition $\underline{poly}(\pi) = \underline{poly}(\pi_1) + \underline{poly}(\pi_2)$.
  By IH on $\pi_1$ we have $\pi_{1,1} \triangleright \vdash_v^{\Theta} Z_1 : P_1$ and $\pi_{1,2} \triangleright \vdash_v^{\Theta} Z_1 : Q_1$ such that $\underline{poly}(\pi_1) = \underline{poly}(\pi_{1,1}) + \underline{poly}(\pi_{1,2})$.
  By IH on $\pi_2$ we have $\pi_{2,1} \triangleright \vdash_v^{\Theta} Z_2 : P_2$ and $\pi_{2,2} \triangleright \vdash_v^{\Theta} Z_2 : Q_2$ such that $\underline{poly}(\pi_2) = \underline{poly}(\pi_{2,1}) + \underline{poly}(\pi_{2,2})$.
  Therefore, we have that

$$\begin{aligned}\underline{poly}(\pi) &= \underline{poly}(\pi_{1,1}) + \underline{poly}(\pi_{1,2}) + \underline{poly}(\pi_{2,1}) + \underline{poly}(\pi_{2,2}) \\ &\overset{\text{IH on } \pi_1}{=} \underline{poly}(\pi_1) + \underline{poly}(\pi_{2,1}) + \underline{poly}(\pi_{2,2}) \\ &\overset{\text{IH on } \pi_2}{=} \underline{poly}(\pi_1) + \underline{poly}(\pi_2)\end{aligned}$$

  and we can conclude.

- *Case $P = !_p A$ and $Q = !_q A$:*
  By definition of $\boxplus$ we have that $!_p A \boxplus !_q A = !_{p+q} A$. By hypothesis we have that $\pi \triangleright \vdash^{\Theta}_{v} V : !_{p+q} A$, so $V$ must be in the form $Z = !M$ for some term $M$ and by BANG-typing rule we have $\pi' \triangleright \vdash^{\Theta}_{v} M : A$. Moreover, by definition $\underline{poly}(\pi) = (p+q) \times \underline{poly}(\pi')$.
  Let $\pi_1 \triangleright \vdash^{\Theta}_{v} !M : !_p A$ which is derivable by BANG-typing rule and $\pi'$, by definition we have that $\underline{poly}(\pi_1) = p \times \underline{poly}(\pi')$.
  Let $\pi_2 \triangleright \vdash^{\Theta}_{v} !M : !_q A$ which is derivable by BANG-typing rule and $\pi'$, by definition we have that $\underline{poly}(\pi_2) = q \times \underline{poly}(\pi')$.
  Therefore, we have that

$$\begin{aligned} \underline{poly}(\pi) &= (p+q) \times \underline{poly}(\pi') \\ &= (p \times \underline{poly}(\pi')) + (q \times \underline{poly}(\pi')) \\ &= \underline{poly}(\pi_1) + \underline{poly}(\pi_2) \end{aligned}$$

  and we can conclude.

□

*Case B.* By induction on the definition of $p * P$.

- *Case $P = G$ for some ground type $G$:*
  By definition $p * G = G$ and by hypothesis we have $\pi \triangleright \vdash^{\Theta}_{v} Z : G$, then we take $\pi' = \pi$. By Lemma 57 we have that $\underline{poly}(\pi) = 0 = p \times 0 = p \times \underline{poly}(\pi')$ and we can conclude.

- *Case $P = Q_1 \otimes Q_2$:*
  By definition $p * (Q_1 \otimes Q_2) = (p * Q_1) \otimes (p * Q_2)$. By hypothesis we have that $\pi \triangleright \vdash^{\Theta}_{v} Z : (p * Q_1) \otimes (p * Q_2)$, so $Z$ must be in the form $Z = \langle Z_1, Z_2 \rangle$ and by TENSOR-typing rule we have $\pi_1 \triangleright \vdash^{\Theta}_{v} Z_1 : p * Q_1$ and $\pi_2 \triangleright \vdash^{\Theta}_{v} Z_2 : p * Q_2$. Moreover, by definition $\underline{poly}(\pi) = \underline{poly}(\pi_1) + \underline{poly}(\pi_2)$.
  By IH on $\pi_1$ we have that $\pi'_1 \triangleright \vdash^{\Theta}_{v} Z_1 : Q_1$ such that $\underline{poly}(\pi_1) = p \times \underline{poly}(\pi'_1)$.
  By IH on $\pi_2$ we have that $\pi'_2 \triangleright \vdash^{\Theta}_{v} Z_2 : Q_2$ such that $\underline{poly}(\pi_2) = p \times \underline{poly}(\pi'_2)$.
  Let $\pi' \triangleright \vdash^{\Theta}_{v} \langle Z_1, Z_2 \rangle : Q_1 \otimes Q_2$, by definition $\underline{poly}(\pi') = \underline{poly}(\pi'_1) + \underline{poly}(\pi'_2)$. Therefore, we have that

$$\begin{aligned} \underline{poly}(\pi) &= \underline{poly}(\pi_1) + poly(\pi_2) \\ &\overset{IHs}{=} (p \times \underline{poly}(\pi'_1)) + (p \times \underline{poly}(\pi'_2)) \\ &= p \times (\underline{poly}(\pi'_1) + \underline{poly}(\pi'_2)) \\ &= p \times \underline{poly}(\pi') \end{aligned}$$

  and we can conclude.

- *Case $P = !_q A$:*
  By definition $p * (!_q A) = !_{p \times q} A$. By hypothesis we have that $\pi \triangleright \vdash^{\Theta}_{v} Z : !_{p \times q} A$, so $Z$ must be in the form $Z = !M$ for some term $M$ and by BANG-typing rule we have $\pi_1 \triangleright \vdash^{\Theta}_{v} M : A$. By definition $\underline{poly}(\pi) = (p \times q) \times \underline{poly}(\pi_1)$.
  Let $\pi' \triangleright \vdash^{\Theta}_{v} !M : !_q A$, by definition $poly(\pi') = q \times \underline{poly}(\pi_1)$. Therefore, we have that

$$poly(\pi) = (p \times q) \times \underline{poly}(\pi_1)$$

$$
\begin{aligned}
&= p \times (q \times \underline{poly}(\pi_1)) \\
&= p \times \underline{poly}(\pi')
\end{aligned}
$$

and we can conclude.

□

**Lemma 59** (Substitution)**.**

1. Given two derivations $\pi \triangleright \Gamma, x : P \vdash_c^\Theta M : A$ and $\rho \triangleright \vdash_v^\Theta V : P$, then there exists a derivation $\pi[\rho/x] \triangleright \Gamma \vdash_c^\Theta M[V/x] : A$ such that $poly(\pi[\rho/x]) \leq poly(\pi) + \underline{poly}(\rho)$.

2. Given two derivations $\varphi \triangleright \Gamma, x : P \vdash_v^\Theta U : A$ and $\rho \triangleright \vdash_v^\Theta V : P$, then there exists a derivation $\varphi[\rho/x] \triangleright \Gamma \vdash_c^\Theta U[V/x] : A$ such that $poly(\varphi[\rho/x]) \leq poly(\varphi) + \underline{poly}(\rho)$.

*Proof.* By mutual induction on $\pi$ and $\varphi$.

- *Case* $\varphi \triangleright x : P \vdash_v^\Theta x : P$:
  By hypothesis we have $\rho \triangleright \vdash_v^\Theta V : P$, then we have a derivation $\varphi[\rho/x] \triangleright \vdash_v^\Theta V : P$. Moreover, by definitions of *poly* and $\underline{poly}$ we have $poly(\varphi[\rho/x]) = size_P(P) = \underline{poly}(\rho)$. Therefore, we can conclude as we have that $poly(\varphi[\rho/x]) = size_P(P) = \underline{poly}(\rho) \leq poly(\varphi) + \underline{poly}(\rho)$.

- *Case* $\varphi \triangleright p * \Gamma \vdash_v^\Theta !N :!_p A$:
  We have that $p * \Gamma = p * \Delta, x : P$, so we have that $P = p * Q$ for some positive type $Q$. By BANG-typing rule we have $\varphi_1 \triangleright \Delta, x : Q \vdash_c^\Theta N : A$ and $poly(\varphi) = p \times poly(\varphi_1)$. By hypothesis we have $\rho \triangleright \vdash_v^\Theta V : p * Q$ and by Lemma 58 case B, we have that there exists $\rho' \triangleright \vdash_v^\Theta V : Q$ such that
  $$\underline{poly}(\rho) = p \times \underline{poly}(\rho') \tag{8.3}$$
  By IH on $\varphi_1$ we have that there exists $\varphi[\rho'/x] \triangleright \Delta \vdash_c^\Theta N[V/x] : A$ such that $poly(\varphi[\rho'/x]) \leq poly(\varphi_1) + \underline{poly}(\rho')$.
  Therefore, let $\varphi[\rho/x] \triangleright p * \Delta \vdash_v^\Theta !N[V/x] :!_p A$, we have that
  $$
  \begin{aligned}
  poly(\varphi[\rho/x]) &= p \times poly(\varphi[\rho'/x]) \\
  &\overset{IH}{\leq} p \times (poly(\varphi_1) + \underline{poly}(\rho')) \\
  &= p \times poly(\varphi_1) + p \times \underline{poly}(\rho') \\
  &\overset{\text{Eq.8.3}}{=} p \times \underline{poly}(\varphi_1) + poly(\rho) \\
  &= poly(\varphi) + \underline{poly}(\rho)
  \end{aligned}
  $$
  and we can conclude.

- *Case* $\varphi \triangleright \Gamma_1 \boxplus \Gamma_2 \vdash_v^\Theta \langle Z_1, Z_2 \rangle : Q_1 \otimes Q_2$:
  We have that $\Gamma_1 \boxplus \Gamma_2 = \Delta_1 \boxplus \Delta_2, x : P$ for some $\Delta_1, \Delta_2$ and by hypothesis we have $\rho \triangleright \vdash_v^\Theta V : P$. We proceed by analyzing the following sub-cases:

  - Sub-case $x \in \Gamma_1$ and $x \notin \Gamma_2$:
    We have that $\Gamma_1 = \Delta_1, x : P$ and $\Gamma_2 = \Delta_2$. By TENSOR-typing rule we have that $\varphi_1 \triangleright \Delta_1, x : P \vdash_v^\Theta Z_1 : Q_1$ and $\varphi_2 \triangleright \Delta_2 \vdash_v^\Theta Z_2 : Q_2$. Moreover, by definition and $poly(\varphi) = 1 + poly(\varphi_1) + poly(\varphi_2)$.

By IH on $\varphi_1$ we have that there exists $\varphi_1[\rho/x] \triangleright \Delta_1 \vdash_v^\Theta Z_1[V/x] : Q_1$ such that $poly(\varphi_1[\rho/x]) \leq poly(\varphi_1) + \underline{poly}(\rho)$.
Therefore, let $\varphi[\rho/x] \triangleright \Delta_1 \boxplus \Delta_2 \vdash_v^\Theta \langle Z_1[V/x], Z_2\rangle : Q_1 \otimes Q_2$, we have that

$$\begin{aligned} poly(\varphi[\rho/x]) &= 1 + poly(\varphi_1[\rho/x]) + poly(\varphi_2) \\ &\overset{IH}{\leq} 1 + poly(\varphi_1) + \underline{poly}(\rho) + poly(\varphi_2) \\ &= poly(\varphi) + \underline{poly}(\rho) \end{aligned}$$

and we can conclude.

– Sub-case $x \notin \Gamma_1$ and $x \in \Gamma_2$: similar to the previous sub-case.

– Sub-case $P = P_1 \boxplus P_2$:
We have that $\Gamma_1 = \Delta_1, x : P_1$ and $\Gamma_2 = \Delta_2, x : P_2$. By TENSOR-typing rule we have that $\varphi_1 \triangleright \Delta_1, x : P_1 \vdash_v^\Theta Z_1 : Q_1$ and $\varphi_2 \triangleright \Delta_2, x : P_2 \vdash_v^\Theta Z_2 : Q_2$. Moreover, by definition and $poly(\varphi) = 1 + poly(\varphi_1) + poly(\varphi_2)$.
By hypothesis we have $\rho \triangleright \vdash_v^\Theta V : P$ and by Lemma 58 case A, we have that there exists $\rho_1 \triangleright \vdash_v^\Theta V : P_1$ and $\rho_2 \triangleright \vdash_v^\Theta V : P_2$ such that

$$\underline{poly}(\rho) = \underline{poly}(\rho_1) + \underline{poly}(\rho_2) \tag{8.4}$$

By IH on $\varphi_1$ we have that there exists $\varphi_1[\rho_1/x] \triangleright \Delta_1 \vdash_v^\Theta Z_1[V/x] : Q_1$ such that $poly(\varphi_1[\rho_1/x]) \leq poly(\varphi_1) + \underline{poly}(\rho_1)$.
By IH on $\varphi_2$ we have that there exists $\varphi_2[\rho_2/x] \triangleright \Delta_2 \vdash_v^\Theta Z_2[V/x] : Q_2$ such that $poly(\varphi_2[\rho_2/x]) \leq poly(\varphi_2) + \underline{poly}(\rho_2)$.
Therefore, let $\varphi[\rho/x] \triangleright \Delta_1 \boxplus \Delta_2 \vdash_v^\Theta \langle Z_1[V/x], Z_2[V/x]\rangle : Q_1 \otimes Q_2$, we have that

$$\begin{aligned} poly(\varphi[\rho/x]) &= 1 + poly(\varphi_1[\rho_1/x]) + poly(\varphi_2[\rho_2/x]) \\ &\overset{IHs}{\leq} 1 + poly(\varphi_1) + \underline{poly}(\rho_1) + poly(\varphi_2) + \underline{poly}(\rho_2) \\ &\overset{\text{Eq.8.4}}{=} 1 + poly(\varphi_1) + poly(\varphi_2) + \underline{poly}(\rho) \\ &= poly(\varphi) + \underline{poly}(\rho) \end{aligned}$$

and we can conclude.

- *Case* $\varphi \triangleright \Gamma, x : P \vdash_v^\Theta \lambda y.N : P' \multimap A'$:
By LAM-typing rule we have $\varphi_1 \triangleright \Gamma, x : P, y : P' \vdash_c^\Theta N : A'$ and $poly(\varphi) = 1 + poly(\varphi_1)$.
Moreover, by hypothesis we have $\rho \triangleright \vdash_v^\Theta V : P$.
By IH on $\varphi_1$ we have that there exists $\varphi_1[\rho/x] \triangleright \Gamma, y : P' \vdash_c^\Theta N[V/x] : A'$ such that $poly(\varphi_1[\rho/x]) \leq poly(\varphi_1) + \underline{poly}(\rho)$.
Therefore, let $\varphi[\rho/x] \triangleright \Gamma \vdash_v^\Theta \lambda y.N[V/x] : P' \multimap A'$, we have that

$$\begin{aligned} poly(\varphi[\rho/x]) &= 1 + poly(\varphi_1[\rho/x]) \\ &\overset{IH}{\leq} 1 + poly(\varphi_1) + \underline{poly}(\rho) \\ &= poly(\varphi) + \underline{poly}(\rho) \end{aligned}$$

and we can conclude.

- *Case* $\pi \triangleright \Gamma, x : P \vdash_c^\Theta \mathtt{return}\, V' : A$:
  By ETA-typing rule we have $\pi_1 \triangleright \Gamma, x : P \vdash_v^\Theta V' : A$ and $poly(\pi) = 1 + poly(\pi_1)$. Moreover, by hypothesis we have $\rho \triangleright \vdash_v^\Theta V : P$.
  By IH on $\pi_1$ we have that there exists $\pi_1[\rho/x] \triangleright \Gamma \vdash_v^\Theta V'[V/x] : A$ such that $poly(\pi_1[\rho/x]) \leq poly(\pi_1) + \underline{poly}(\rho)$.
  Therefore, let $\pi[\rho/x] \triangleright \Gamma \vdash_c^\Theta \mathtt{return}\, V'[V/x] : A$, we have that

$$\begin{aligned} poly(\pi[\rho/x]) &= 1 + poly(\pi_1[\rho/x]) \\ &\overset{IH}{\leq} 1 + poly(\pi_1) + \underline{poly}(\rho) \\ &= poly(\pi) + \underline{poly}(\rho) \end{aligned}$$

  and we can conclude.

- *Case* $\pi \triangleright \Gamma, x : P \vdash_c^\Theta \mathtt{der}(Z) : A$:
  By DER-typing rule we have $\pi_1 \triangleright \Gamma, x : P \vdash_v^\Theta Z :!_1 A$ and $poly(\pi) = 1 + poly(\pi_1)$. Moreover, by hypothesis we have $\rho \triangleright \vdash_v^\Theta V : P$.
  By IH on $\pi_1$ we have that there exists $\pi_1[\rho/x] \triangleright \Gamma \vdash_v^\Theta Z[V/x] :!_1 A$ such that $poly(\pi_1[\rho/x]) \leq poly(\pi_1) + \underline{poly}(\rho)$.
  Therefore, let $\pi[\rho/x] \triangleright \Gamma \vdash_c^\Theta \mathtt{der}(Z[V/x]) : A$, we have that

$$\begin{aligned} poly(\pi[\rho/x]) &= 1 + poly(\pi_1[\rho/x]) \\ &\overset{IH}{\leq} 1 + poly(\pi_1) + \underline{poly}(\rho) \\ &= poly(\pi) + \underline{poly}(\rho) \end{aligned}$$

  and we can conclude.

- *Case* $\pi \triangleright \Gamma, x : P \vdash_c^\Theta \mathtt{set}\, r\, Z : \mathbb{U}$:
  By SET-typing rule we have $\pi_1 \triangleright \Gamma, x : P \vdash_v^\Theta Z : G$ where $r : G \in \Theta$ and $poly(\pi) = 2 + poly(\pi_1)$. Moreover, by hypothesis we have $\rho \triangleright \vdash_v^\Theta V : P$.
  By IH on $\pi_1$ we have that there exists $\pi_1[\rho/x] \triangleright \Gamma \vdash_v^\Theta Z[V/x] : G$ such that $poly(\pi_1[\rho/x]) \leq poly(\pi_1) + \underline{poly}(\rho)$.
  Therefore, let $\pi[\rho/x] \triangleright \Gamma \vdash_c^\Theta \mathtt{set}\, r\, Z[V/x] : \mathbb{U}$, we have that

$$\begin{aligned} poly(\pi[\rho/x]) &= 2 + poly(\pi_1[\rho/x]) \\ &\overset{IH}{\leq} 2 + poly(\pi) + \underline{poly}(\rho) \\ &= 2 + poly(\pi_1) + \underline{poly}(\rho) \end{aligned}$$

  and we can conclude.

- *Case* $\pi \triangleright \Gamma_1 p \boxplus \Gamma_2 p \vdash_c^\Theta f_p(Z_1, Z_2) : Gp$:
  We have that $\Gamma_1 p \boxplus \Gamma_2 p = \Delta_1 p \boxplus \Delta_2 p, x : P$ for some $\Delta_1, \Delta_2$ and by hypothesis we have $\rho \triangleright \vdash_v^\Theta V : P$. We proceed by analyzing the following sub-cases:

  – Sub-case $x \in \Gamma_1 p$ and $x \notin \Gamma_2 p$:
    We have that $\Gamma_1 p = \Delta_1 p, x : P$ and $\Gamma_2 p = \Delta_2 p$. By FUN-typing rule $\mathsf{typeof}(f) : G_1 \times G_2 \to G$ and we have $\pi_1 \triangleright \Delta_1 p, x : P \vdash_v^\Theta Z_1 : G_1 p$ and $\pi_2 \triangleright \Delta_2 p \vdash_v^\Theta Z_2 : G_2 p$.

Moreover, by definition we have $\mathit{poly}(\pi) = 1 + \mathit{poly}(\pi_1) + \mathit{poly}(\pi_2) + \mathit{size}_G(G)$.
By IH on $\pi_1$ we have that there exists $\pi_1[\rho/x] \triangleright \Delta_1 p \vdash_v^\Theta Z_1[V/x] : G_1 p$ such that $\mathit{poly}(\pi_1[\rho/x]) \leq \mathit{poly}(\pi_1) + \underline{\mathit{poly}}(\rho)$.
Therefore, let $\pi[\rho/x] \triangleright \Delta_1 p \boxplus \Delta_2 p \vdash_c^\Theta f_p(Z_1[V/x], Z_2) : Gp$, we have that

$$\begin{aligned}\mathit{poly}(\pi[\rho/x]) &= 1 + \mathit{poly}(\pi_1[\rho/x]) + \mathit{poly}(\pi_2) + \mathit{size}_G(G)\\ &\overset{\text{IH}}{\leq} 1 + \mathit{poly}(\pi_1) + \underline{\mathit{poly}}(\rho) + \mathit{poly}(\pi_2) + \mathit{size}_G(G)\\ &= \mathit{poly}(\pi) + \underline{\mathit{poly}}(\rho)\end{aligned}$$

and we can conclude.

- Sub-case $x \notin \Gamma_1 p$ and $x \in \Gamma_2 p$: similar to the previous sub-case.
- Sub-case $P = P_1 \boxplus P_2$:
  We have that $\Gamma_1 p = \Delta_1 p, x : P_1$ and $\Gamma_2 p = \Delta_2 p, x : P_2$. By FUN-typing rule $\mathsf{typeof}(f) : G_1 \times G_2 \to G$ and we have $\pi_1 \triangleright \Delta_1 p, x : P \vdash_v^\Theta Z_1 : G_1 p$ and $\pi_2 \triangleright \Delta_2 p, x : P_2 \vdash_v^\Theta Z_2 : G_2 p$. Moreover, by definition we have $\mathit{poly}(\pi) = 1 + \mathit{poly}(\pi_1) + \mathit{poly}(\pi_2) + \mathit{size}_G(G)$.
  By hypothesis we have $\rho \triangleright \vdash_v^\Theta V : P$ and by Lemma 58 case A we have that there exists $\rho_1 \triangleright \vdash_v^\Theta V : P_1$ and $\rho_2 \triangleright \vdash_v^\Theta V : P_2$ such that $\underline{\mathit{poly}}(\rho) = \underline{\mathit{poly}}(\rho_1) + \underline{\mathit{poly}}(\rho_2)$.
  By IH on $\pi_1$ we have that there exists $\pi_1[\rho_1/x] \triangleright \Delta_1 p \vdash_v^\Theta Z_1[V/x] : G_1 p$ such that $\mathit{poly}(\pi_1[\rho_1/x]) \leq \mathit{poly}(\pi_1) + \underline{\mathit{poly}}(\rho_1)$.
  By IH on $\pi_2$ we have that there exists $\pi_2[\rho_2/x] \triangleright \Delta_2 p \vdash_v^\Theta Z_2[V/x] : G_2 p$ such that $\mathit{poly}(\pi_2[\rho_2/x]) \leq \mathit{poly}(\pi_2) + \underline{\mathit{poly}}(\rho_2)$.
  Therefore, let $\pi[\rho/x] \triangleright \Delta_1 p \boxplus \Delta_2 p \vdash_c^\Theta f_p(Z_1[V/x], Z_2[V/x]) : Gp$, we have that

$$\begin{aligned}\mathit{poly}(\pi[\rho/x]) &= 1 + \mathit{poly}(\pi_1[\rho_1/x]) + \mathit{poly}(\pi_2[\rho_2/x]) + \mathit{size}_G(G)\\ &\overset{\text{IHs}}{\leq} 1 + \mathit{poly}(\pi_1) + \underline{\mathit{poly}}(\rho_1) + \mathit{poly}(\pi_2) + \underline{\mathit{poly}}(\rho_2) + \mathit{size}_G(G)\\ &= 1 + \mathit{poly}(\pi_1) + \mathit{poly}(\pi_2) + \mathit{size}_G(G) + \underline{\mathit{poly}}(\rho)\\ &= \mathit{poly}(\pi) + \underline{\mathit{poly}}(\rho)\end{aligned}$$

  and we can conclude.

- *Case* $\pi \triangleright \Gamma_1 \boxplus \Gamma_2 \vdash_c^\Theta \mathtt{let}\,\langle y_1, y_2\rangle = Z\,\mathtt{in}\,N : A$:
  We have that $\Gamma_1 \boxplus \Gamma_2 = \Delta_1 \boxplus \Delta_2, x : P$ for some $\Delta_1, \Delta_2$ and by hypothesis we have $\rho \triangleright \vdash_v^\Theta V : P$. We proceed by analyzing the following sub-cases:

  - Sub-case $x \in \Gamma_1$ and $x \notin \Gamma_2$:
    We have that $\Gamma_1 = \Delta_1, x : P$ and $\Gamma_2 = \Delta_2$. By LET_PAIR-typing rule we have $\pi_1 \triangleright \Delta_1, x : P \vdash_v^\Theta Z : Q_1 \otimes Q_2$ and $\pi_2 \triangleright \Delta_2 \vdash_c^\Theta N : A$. Moreover, by definition we have $\mathit{poly}(\pi) = 1 + \mathit{poly}(\pi_1) + \mathit{poly}(\pi_2)$.
    By IH on $\pi_1$ we have that there exists $\pi_1[\rho/x] \triangleright \Delta_1 \vdash_v^\Theta Z[V/x] : Q_1 \otimes Q_2$ such that $\mathit{poly}(\pi_1[\rho/x]) \leq \mathit{poly}(\pi_1) + \underline{\mathit{poly}}(\rho)$.
    Therefore, let $\pi[\rho/x] \triangleright \Delta_1 \boxplus \Delta_2 \vdash_c^\Theta \mathtt{let}\,\langle y_1, y_2\rangle = Z[V/x]\,\mathtt{in}\,N : A$, we have that

$$\begin{aligned}\mathit{poly}(\pi[\rho/x]) &= 1 + \mathit{poly}(\pi_1[\rho/x]) + \mathit{poly}(\pi_2)\\ &\overset{IH}{\leq} 1 + \mathit{poly}(\pi_1) + \underline{\mathit{poly}}(\rho) + \mathit{poly}(\pi_2)\\ &= \mathit{poly}(\pi) + \underline{\mathit{poly}}(\rho)\end{aligned}$$

    and we can conclude.

- Sub-case $x \notin \Gamma_1$ and $x \in \Gamma_2$:
  We have that $\Gamma_2 = \Delta_2, x : P$ and $\Gamma_1 = \Delta_1$. By LET_PAIR-typing rule we have $\pi_1 \triangleright \Delta_1, x : P \vdash_v^\Theta Z : Q_1 \otimes Q_2$ and $\pi_2 \triangleright \Delta_2 \vdash_c^\Theta N : A$. Moreover, by definition we have $poly(\pi) = 1 + poly(\pi_1) + poly(\pi_2)$.
  By IH on $\pi_2$ we have that there exists $\pi_2[\rho/x] \triangleright \Delta_2 \vdash_v^\Theta N[V/x] : A$ such that $poly(\pi_2[\rho/x]) \leq poly(\pi_2) + \underline{poly}(\rho)$.
  Therefore, let $\pi[\rho/x] \triangleright \Delta_1 \boxplus \Delta_2 \vdash_c^\Theta \texttt{let}\, \langle y_1, y_2 \rangle = Z\, \texttt{in}\, N[V/x] : A$, we have that

$$\begin{aligned} poly(\pi[\rho/x]) &= 1 + poly(\pi_1) + poly(\pi_2[\rho/x]) \\ &\overset{IH}{\leq} 1 + poly(\pi_1) + poly(\pi_2) + \underline{poly}(\rho) \\ &= poly(\pi) + \underline{poly}(\rho) \end{aligned}$$

  and we can conclude.
- Sub-case $P = P_1 \boxplus P_2$:
  We have that $\Gamma_1 = \Delta_1, x : P_1$ and $\Gamma_2 = \Delta_2, x : P_2$. By LET_PAIR-typing rule we have $\pi_1 \triangleright \Delta_1, x : P_1 \vdash_v^\Theta Z : Q_1 \otimes Q_2$ and $\pi_2 \triangleright \Delta_2, x : P_2 \vdash_c^\Theta N : A$. Moreover, by definition we have $poly(\pi) = 1 + poly(\pi_1) + poly(\pi_2)$.
  By hypothesis we have $\rho \triangleright \vdash_v^\Theta V : P$ and by Lemma 58 case A we have that there exists $\rho_1 \triangleright \vdash_v^\Theta V : P_1$ and $\rho_2 \triangleright \vdash_v^\Theta V : P_2$ such that

$$\underline{poly}(\rho) = \underline{poly}(\rho_1) + \underline{poly}(\rho_2). \tag{8.5}$$

  By IH on $\pi_1$ we have that there exists $\pi_1[\rho_1/x] \triangleright \Delta_1 \vdash_v^\Theta Z[V/x] : Q_1 \otimes Q_2$ such that $poly(\pi_1[\rho_1/x]) \leq poly(\pi_1) + \underline{poly}(\rho_1)$.
  By IH on $\pi_2$ we have that there exists $\pi_2[\rho_2/x] \triangleright \Delta_2 \vdash_c^\Theta N[V/x] : A$ such that $poly(\pi_1[\rho_2/x]) \leq poly(\pi_2) + \underline{poly}(\rho_2)$.
  Therefore, let $\pi[\rho/x] \triangleright \Delta_1 \boxplus \Delta_2 \vdash_c^\Theta \texttt{let}\, \langle y_1, y_2 \rangle = Z[V/x]\, \texttt{in}\, N[V/x] : A$, we have that

$$\begin{aligned} poly(\pi[\rho/x]) &= 1 + poly(\pi_1[\rho_1/x]) + poly(\pi_2[\rho_2/x]) \\ &\overset{IHs}{\leq} 1 + poly(\pi_1) + \underline{poly}(\rho_1) + poly(\pi_2) + \underline{poly}(\rho_2) \\ &\overset{\text{Eq.8.5}}{=} 1 + poly(\pi_1) + poly(\pi_2) + \underline{poly}(\rho) \\ &= poly(\pi) + \underline{poly}(\rho) \end{aligned}$$

  and we can conclude.

- *Case* $\pi \triangleright \Gamma_1 \boxplus \Gamma_2 \vdash_c^\Theta \texttt{let}\, y = Z\, \texttt{in}\, N : A$ *and case* $\pi \triangleright \Gamma_1 \boxplus \Gamma_2 \vdash_c^\Theta NZ : A$: similar to the previous case.

- *Case* $\pi \triangleright (p * \Gamma_1) \boxplus \Gamma_2 \vdash_c^\Theta \texttt{loop}\, V'\, p\, \texttt{times}\, \texttt{from}\, N : Q$:
  We have that $(p * \Gamma_1) \boxplus \Gamma_2 = (p * \Delta_1) \boxplus \Delta_2, x : P$ for some $\Delta_1, \Delta_2$ and by hypothesis we have $\rho \triangleright \vdash_v^\Theta V : P$. We proceed by analyzing the following sub-cases:

  - Sub-case $x \in \Gamma_1$ and $x \notin \Gamma_2$:
    Since $x \in \Gamma_1$, we have that $P = p * P'$ for some positive type $P'$. Therefore, we have $\Gamma_1 = \Delta_1, x : P'$ and $\Gamma_1 = \Delta_2$. By LOOP-typing rule we have $\pi_1 \triangleright \Delta_1, x : P' \vdash_v^\Theta V' : Q \multimap Q$ and $\pi_2 \triangleright \Delta_2 \vdash_c^\Theta N : Q$. Moreover, by definition we have that $poly(\pi) = 1 + p + p \times poly(\pi_1) + poly(\pi_2)$.

By hypothesis we have $\rho \triangleright \vdash_v^\Theta V : p * P'$ and by Lemma 58 case B we have that there exists $\rho' \triangleright \vdash_v^\Theta V : P'$ such that

$$\underline{poly}(\rho) = p \times \underline{poly}(\rho'). \tag{8.6}$$

By IH on $\pi_1$ we have that there exists $\pi_1[\rho'/x] \triangleright \Delta_1 \vdash_v^\Theta V'[V/x] : Q \multimap Q$ such that $poly(\pi_1[\rho'/x]) \leq poly(\pi_1) + \underline{poly}(\rho')$.
Therefore, let $\pi[\rho/x] \triangleright (p * \Delta_1) \boxplus \Delta_2 \vdash_c^\Theta \texttt{loop}\,(V'[V/x])\,p\,\texttt{times}\,\texttt{from}\,N : Q$, we have that

$$\begin{aligned}
poly(\pi[\rho/x]) &= 1 + p + p \times poly(\pi_1[\rho'/x]) + poly(\pi_2) \\
&\overset{IH}{\leq} 1 + p + p \times (poly(\pi_1) + \underline{poly}(\rho')) + poly(\pi_2) \\
&= 1 + p + p \times poly(\pi_1) + p \times \underline{poly}(\rho') + poly(\pi_2) \\
&\overset{\text{Eq.8.6}}{=} 1 + p + p \times poly(\pi_1) + \underline{poly}(\rho) + poly(\pi_2) \\
&= poly(\pi) + \underline{poly}(\rho)
\end{aligned}$$

and we can conclude.

– Sub-case $x \notin \Gamma_1$ and $x \in \Gamma_2$:
We have that $\Gamma_1 = \Delta_1$ and $\Gamma_2 = \Delta_2, x : P$. By LOOP-typing rule we have $\pi_1 \triangleright \Delta_1 \vdash_v^\Theta V' : Q \multimap Q$ and $\pi_2 \triangleright \Delta_2, x : P' \vdash_c^\Theta N : Q$. Moreover, by definition we have that $poly(\pi) = 1 + p + p \times poly(\pi_1) + poly(\pi_2)$.
By IH on $\pi_2$ we have that there exists $\pi_2[\rho/x] \triangleright \Delta_2 \vdash_c^\Theta N[V/x] : Q$ such that $poly(\pi_2[\rho/x]) \leq poly(\pi_2) + \underline{poly}(\rho)$.
Therefore, let $\pi[\rho/x] \triangleright (p * \Delta_1) \boxplus \Delta_2 \vdash_c^\Theta \texttt{loop}\,V'\,p\,\texttt{times}\,\texttt{from}\,(N[V/x]) : Q$, we have that

$$\begin{aligned}
poly(\pi[\rho/x]) &= 1 + p + p \times poly(\pi_1) + poly(\pi_2[\rho/x]) \\
&\overset{IH}{\leq} 1 + p + p \times poly(\pi_1) + poly(\pi_2) + \underline{poly}(\rho) \\
&= poly(\pi) + \underline{poly}(\rho)
\end{aligned}$$

and we can conclude.

– Sub-case $x \in \Gamma_1$ and $x \in \Gamma_2$:
We have that $\Gamma_1 = \Delta_1, x : P_1$ and $\Gamma_2 = \Delta_2, x : P_2$, and $P = (p * P_1) \boxplus P_2$. By LOOP-typing rule we have $\pi_1 \triangleright \Delta_1, x : P_1 \vdash_v^\Theta V' : Q \multimap Q$ and $\pi_2 \triangleright \Delta_2, x : P_2 \vdash_c^\Theta N : Q$. Moreover, by definition we have that $poly(\pi) = 1 + p + p \times poly(\pi_1) + poly(\pi_2)$.
In this case case by hypothesis we have $\rho \triangleright \vdash_v^\Theta V : (p * P_1) \boxplus P_2$. By Lemma 58 case A on $\rho$ we have that there exists $\rho_1 \triangleright \vdash_v^\Theta V : p * P_1$ and $\rho_2 \triangleright \vdash_v^\Theta V : P_2$ such that $\underline{poly}(\rho) = \underline{poly}(\rho_1) + \underline{poly}(\rho_2)$. Moreover, by Lemma 58 case B on $\rho_1$ we have that there exists $\rho_1' \triangleright \vdash_v^\Theta V : P_1$ such that $\underline{poly}(\rho_1) = p \times \underline{poly}(\rho_1')$. Summing up, we have that

$$\underline{poly}(\rho) = p \times \underline{poly}(\rho_1') + \underline{poly}(\rho_2). \tag{8.7}$$

By IH on $\pi_1$ we have that there exists $\pi_1[\rho_1'/x] \triangleright \Delta_1 \vdash_v^\Theta V'[V/x] : Q \multimap Q$ such that $poly(\pi_1[\rho_1'/x]) \leq poly(\pi_1) + \underline{poly}(\rho_1')$.
By IH on $\pi_2$ we have that there exists $\pi_2[\rho_2/x] \triangleright \Delta_2 \vdash_c^\Theta N[V/x] : Q$ such that $poly(\pi_2[\rho_2/x]) \leq poly(\pi_2) + \underline{poly}(\rho_2)$.
Therefore, let $\pi[\rho/x] \triangleright (p * \Delta_1) \boxplus \Delta_2 \vdash_c^\Theta \texttt{loop}\,(V'[V/x])\,p\,\texttt{times}\,\texttt{from}\,(N[V/x]) : Q$, we have that

$$poly(\pi[\rho/x]) = 1 + p + p \times poly(\pi_2[\rho_1'/x]) + \underline{poly}(\pi_2[\rho_2/x])$$

$$\begin{aligned}
&\overset{IHs}{\leq} 1 + p + p \times (poly(\pi_1) + poly(\rho'_1)) + poly(\pi_2) + \underline{poly}(\rho_2) \\
&= 1 + p + p \times poly(\pi_1) + p \times \underline{poly}(\rho'_1) + poly(\pi_2) + \underline{poly}(\rho_2) \\
&= poly(\pi) + p \times \underline{poly}(\rho'_1) + \underline{poly}(\rho_2) \\
&\overset{\text{Eq.8.7}}{=} poly(\pi) + \underline{poly}(\rho)
\end{aligned}$$

and we can conclude.

- *Case* $\pi \triangleright \Gamma_1 \boxplus \Gamma_2 \vdash_c^\Theta \texttt{if}\, Z\, \texttt{then}\, N_1\, \texttt{else}\, N_2 : A$:
Recall that in Subsection 8.4.1, we restrict attention to `if`-constructs where both branches are closed terms, ensuring that their evaluation cost does not depend on the surrounding context and can be analyzed independently. Therefore we have $\Gamma_2$ is the empty environment and $\Gamma_1 \boxplus \Gamma_2 = \Delta_1, x : P$ for some $\Delta_1$. Moreover, by hypothesis we have $\rho \triangleright \vdash_v^\Theta V : P$. By CASE-typing rule we have $\pi_g \triangleright \Delta_1, x : P \vdash_v^\Theta Z : \mathbb{B}$, $\pi_1 \triangleright \Delta_2 \vdash_c^\Theta N_1 : A$ and $\pi_2 \triangleright \Delta_2 \vdash_c^\Theta N_2 : A$. Moreover, by definition we have $poly(\pi) = 1 + poly(\pi_g) + poly(\pi_1) + poly(\pi_2)$
By IH on $\pi_g$ we have that there exists $\pi_g[\rho/x] \triangleright \Delta_1 \vdash_v^\Theta Z[V/x] : \mathbb{B}$ such that $poly(\pi_g[\rho/x]) \leq poly(\pi_g) + \underline{poly}(\rho)$.
Therefore, let $\pi[\rho/x] \triangleright \Delta_1 \boxplus \Delta_2 \vdash_c^\Theta \texttt{if}\, Z[V/x]\, \texttt{then}\, N_1\, \texttt{else}\, N_2 : A$, we have that

$$\begin{aligned}
poly(\pi[\rho/x]) &= 1 + poly(\pi_g[\rho/x]) + poly(\pi_1) + poly(\pi_2) \\
&\overset{IH}{\leq} 1 + poly(\pi_g) + \underline{poly}(\rho) + poly(\pi_1) + poly(\pi_2) \\
&= poly(\pi) + poly(\rho)
\end{aligned}$$

and we can conclude.

□

**Lemma 60** (Measure Decreasing along Reduction)**.** Given $\pi \triangleright \vdash_c^\Theta M : A$ and store $e \in \mathrm{St}_\Theta$ such that $\mathbf{sm}_n^{\Theta;A}(M, e) = \sum_k a_k\ \delta(M_k, e_k)$, then for every $k$ there exists $\rho_k \triangleright \vdash_c^\Theta M_k : A$ such that $poly(\pi) > poly(\rho_k)$.

*Proof.* By induction on the definition of small step reduction (Figure 8.4).

- *Case* $\pi \triangleright \vdash_c^\Theta \texttt{let}\, x = \texttt{return}\, V\, \texttt{in}\, N : A$ *and* $\mathbf{sm}_n^{\Theta;A}(M, e) = \delta(N[V/x], e)$:
By LET-typing rule we have $\pi_1 \triangleright \vdash_c^\Theta \texttt{return}\, V : P$ and $\pi_2 \triangleright x : P \vdash_c^\Theta N : A$ and by definition
$$poly(\pi) = 1 + poly(\pi_1) + poly(\pi_2). \tag{8.8}$$
Moreover, by ETA-typing rule applied to $\pi_1$ we have $\pi'_1 \triangleright \vdash_v^\Theta V : P$ and by definition
$$poly(\pi_1) = 1 + poly(\pi'_1). \tag{8.9}$$
By Lemma 59 applied to $\pi_2$ and $\pi'_1$, we have that there exists $\pi_2[\pi'_1/x] \triangleright \vdash_c^\Theta N[V/x] : A$ such that $poly(\pi_2[\pi'_1/x]) \leq poly(\pi_2) + \underline{poly}(\pi'_1)$. Moreover, by using Lemma 54 we have that $\underline{poly}(\cdot)$ is always lower or equal to $poly(\cdot)$, then we have:
$$poly(\pi_2[\pi'_1/x]) \leq poly(\pi_2) + poly(\pi'_1). \tag{8.10}$$
Therefore, we have that
$$poly(\pi) \overset{\text{Eq. 8.8}}{=} 1 + poly(\pi_1) + poly(\pi_2)$$

$$\overset{\text{Eq. 8.9}}{=} 2 + poly(\pi_1') + poly(\pi_2)$$
$$> poly(\pi_2[\pi_1'/x])$$
$$\overset{\text{Eq. 8.10}}{\leq} poly(\pi_1') + poly(\pi_2)$$

and we can conclude.

- *Case* $\pi \triangleright \vdash_c^{\Theta} NZ : A$ *and* $\mathbf{sm}_n^{\Theta;A}(M, e) = \delta(N_k Z, e_k)$:
  By APP-typing rule we have $\pi_1 \triangleright \vdash_c^{\Theta} N : P \multimap A$ and $\pi_2 \triangleright \vdash_v^{\Theta} Z : P$ and by definition

  $$poly(\pi) = 1 + poly(\pi_1) + poly(\pi_2). \tag{8.11}$$

  Moreover, by small step semantics $\mathbf{sm}_n^{\Theta;A}(N, e) = \sum_k a_k\ \delta(N_k, e_k)$.
  By IH on $\pi_1$ we have that for every $k$ there exists $\rho_k' \triangleright \vdash_c^{\Theta} N_k : P \multimap A$ such that $poly(\pi_1) > poly(\rho_k')$.
  Let $\rho_k \triangleright \vdash_c^{\Theta} N_k Z : A$, then by definition we have

  $$poly(\rho_k) = 1 + poly(\rho_k') + poly(\pi_2). \tag{8.12}$$

  Therefore, we have that

  $$poly(\pi) \overset{\text{Eq. 8.11}}{=} 1 + poly(\pi_1) + poly(\pi_2) \overset{IH}{>} 1 + poly(\rho_k') + poly(\pi_2) \overset{\text{Eq. 8.12}}{=} poly(\rho_k)$$

  this holds for every $k$ and so we can conclude.

- *Case* $\pi \triangleright \vdash_c^{\Theta} \texttt{let}\, x = N \,\texttt{in}\, N' : A$ *and* $\mathbf{sm}_n^{\Theta;A}(M, e) = \sum_k a_k\ \delta(\texttt{let}\, x = N_k \,\texttt{in}\, N', e_k)$:
  By LET-typing rule we have $\pi_1 \triangleright \vdash_c^{\Theta} N : P$ and $\pi_2 \triangleright x : P \vdash_c^{\Theta} N' : A$ and by definition

  $$poly(\pi) = 1 + poly(\pi_1) + poly(\pi_2). \tag{8.13}$$

  Moreover, by small step semantics $\mathbf{sm}_n^{\Theta;A}(N, e) = \sum_k a_k\ \delta(N_k, e_k)$.
  By IH on $\pi_1$ we have that for every $k$ there exists $\rho_k' \triangleright \vdash_c^{\Theta} N_k : P$ such that $poly(\pi_1) > poly(\rho_k')$.
  Let $\rho_k \triangleright \vdash_c^{\Theta} \texttt{let}\, x = N_k \,\texttt{in}\, N' : A$, then by definition we have

  $$poly(\rho_k) = 1 + poly(\rho_k') + poly(\pi_2). \tag{8.14}$$

  Therefore, we have that

  $$poly(\pi) \overset{\text{Eq. 8.13}}{=} 1 + poly(\pi_1) + poly(\pi_2) \overset{IH}{>} 1 + poly(\rho_k') + poly(\pi_2) \overset{\text{Eq. 8.14}}{=} poly(\rho_k)$$

  this holds for every $k$ and so we can conclude.

- *Case* $\pi \triangleright \vdash_c^{\Theta} \texttt{let}\, \langle x_1, x_2 \rangle = \langle Z_1, Z_2 \rangle \,\texttt{in}\, N : A$ *and* $\mathbf{sm}_n^{\Theta;A}(M, e) = \delta(N[Z_1/x_1, Z_2/x_2], e)$:
  By LET_PAIR-typing rule we have $\pi_1 \triangleright \vdash_v^{\Theta} \langle Z_1, Z_2 \rangle : P_1 \otimes P_2$ and $\pi_2 \triangleright x_1 : P_1, x_2 : P_2 \vdash_c^{\Theta} N : A$ and by definition

  $$poly(\pi) = 1 + poly(\pi_1) + poly(\pi_2). \tag{8.15}$$

  Moreover, by TENSOR-typing rule applied to $\pi_1$ we have $\pi_1' \triangleright \vdash_v^{\Theta} Z_1 : P_1$ and $\pi_1'' \triangleright \vdash_v^{\Theta} Z_2 : P_2$, and by definition

  $$poly(\pi_1) = 1 + poly(\pi_1') + poly(\pi_1''). \tag{8.16}$$

By Lemma 59 applied twice, we have that there exists $\pi_2[\pi_1'/x_1, \pi_1''/x_2] \triangleright \vdash_c^{\Theta} N[Z_1/x_1, Z_2/x_2] : A$ such that $poly(\pi_2[\pi_1'/x_1, \pi_1''/x_2]) \leq poly(\pi_2) + \underline{poly}(\pi_1') + \underline{poly}(\pi_1'')$. Moreover, by using Lemma 54 we have that $\underline{poly}(\cdot)$ is always lower or equal to $poly(\cdot)$, then we have:

$$poly(\pi_2[\pi_1'/x_1, \pi_1''/x_2]) \leq poly(\pi_2) + poly(\pi_1') + poly(\pi_1''). \tag{8.17}$$

Therefore, we have that

$$\begin{aligned} poly(\pi) &\overset{\text{Eq. 8.15}}{=} 1 + poly(\pi_1) + poly(\pi_2) \\ &> poly(\pi_2[\pi_1'/x_1, \pi_1''/x_2]) \\ &\overset{\text{Eq. 8.17}}{\leq} 1 + poly(\pi_2) + poly(\pi_1') + poly(\pi_1'') \\ &\overset{\text{Eq. 8.16}}{=} 1 + poly(\pi_1) + poly(\pi_2) \end{aligned}$$

and we can conclude.

- *Case* $\pi \triangleright \vdash_c^{\Theta} \texttt{der}(!N) : A$ *and* $\mathbf{sm}_n^{\Theta;A}(M, e) = \delta(N, e)$:
  By DER-typing rule we have $\pi_1 \triangleright \vdash_v^{\Theta} !N :!_1 A$ and by definition

$$poly(\pi) = 1 + poly(\pi_1). \tag{8.18}$$

Moreover, by BANG-typing rule applied to $\pi_1$ we have $\pi_1' \triangleright \vdash_c^{\Theta} N : A$ and by definition

$$poly(\pi_1) = 1 \times poly(\pi_1') = poly(\pi_1'). \tag{8.19}$$

Therefore, we have that

$$poly(\pi) \overset{\text{Eq. 8.18}}{=} 1 + poly(\pi_1) > poly(\pi_1') \overset{\text{Eq. 8.19}}{=} poly(\pi_1)$$

and we can conclude.

- *Case* $\pi \triangleright \vdash_c^{\Theta} (\texttt{return}\,(\lambda x.N))Z : A$ *and* $\mathbf{sm}_n^{\Theta;A}(M, e) = \delta(N[Z/x], e)$:
  Since by hypothesis we have $\pi \triangleright \vdash_c^{\Theta} (\texttt{return}\,(\lambda x.N))Z : A$, we proceed by analyzing such derivation as follows:
  - By APP-typing rule we have $\pi_1 \triangleright \vdash_c^{\Theta} \texttt{return}\,(\lambda x.N) : P \multimap A$ and $\pi_2 \triangleright \vdash_v^{\Theta} Z : P$, and by definition we have $poly(\pi) = 1 + poly(\pi_1) + poly(\pi_2)$.
  - By ETA-typing rule applied to $\pi_1$ we have $\pi_1' \triangleright \vdash_v^{\Theta} \lambda x.N : P \multimap A$ and by definition $poly(\pi_1) = poly(\pi_1')$.
  - By LAM-typing rule applied to $\pi_1'$ we have $\pi_1'' \triangleright x : P \vdash_c^{\Theta} N : A$ and by definition $poly(\pi_1') = 1 + poly(\pi_1'')$.

Summing up, we gave that

$$poly(\pi) = 2 + poly(\pi_1'') + poly(\pi_2). \tag{8.20}$$

By Lemma 59 applied to $\pi_1''$ and $\pi_2$, we have that there exists $\pi_1''[\pi_2/x] \triangleright \vdash_c^{\Theta} N[Z/x] : A$ such that $poly(\pi_1''[\pi_2/x]) \leq poly(\pi_1'') + \underline{poly}(\pi_2)$. Moreover, by using Lemma 54 we have that $\underline{poly}(\cdot)$ is always lower or equal to $poly(\cdot)$, then we have:

$$poly(\pi_1''[\pi_2/x]) \leq poly(\pi_1'') + poly(\pi_2). \tag{8.21}$$

Therefore, we have that

$$poly(\pi) \overset{\text{Eq. 8.20}}{=} 2 + poly(\pi_1'') + poly(\pi_2) > poly(\pi_1''[\pi_2/x]) \overset{\text{Eq. 8.21}}{\leq} poly(\pi_1'') + poly(\pi_2)$$

and we can conclude.

- *Case* $\pi \triangleright \vdash_c^{\Theta} \texttt{if}\, \mathbf{t}\, \texttt{then}\, N_1\, \texttt{else}\, N_2 : A$ *and* $\mathbf{sm}_n^{\Theta;A}(M, e) = \delta(N_1, e)$:
By CASE-typing rule we have $\pi_g \triangleright \vdash_v^{\Theta} \mathbf{t} : \mathbb{B}$, $\pi_1 \triangleright \vdash_c^{\Theta} N_1 : A$ and $\pi_2 \triangleright \vdash_c^{\Theta} N_2 : A$ and by definition
$$poly(\pi) = 1 + poly(\pi_g) + poly(\pi_1) + poly(\pi_2). \tag{8.22}$$
Therefore, we have that
$$poly(\pi) \overset{\text{Eq. 8.22}}{=} 1 + poly(\pi_g) + poly(\pi_1) + poly(\pi_2) > poly(\pi_1)$$
and we can conclude.

- *Case* $\pi \triangleright \vdash_c^{\Theta} \texttt{if}\, \mathbf{f}\, \texttt{then}\, N_1\, \texttt{else}\, N_2 : A$ *and* $\mathbf{sm}_n^{\Theta;A}(M, e) = \delta(N_2, e)$:
By CASE-typing rule we have $\pi_g \triangleright \vdash_v^{\Theta} \mathbf{t} : \mathbb{B}$, $\pi_1 \triangleright \vdash_c^{\Theta} N_1 : A$ and $\pi_2 \triangleright \vdash_c^{\Theta} N_2 : A$ and by definition
$$poly(\pi) = 1 + poly(\pi_g) + poly(\pi_1) + poly(\pi_2). \tag{8.23}$$
Therefore, we have that
$$poly(\pi) \overset{\text{Eq. 8.23}}{=} 1 + poly(\pi_g) + poly(\pi_1) + poly(\pi_2) > poly(\pi_2)$$
and we can conclude.

- *Case* $\pi \triangleright \vdash_c^{\Theta} \texttt{set}\, r\, V : \mathbb{U}$ *and* $\mathbf{sm}_n^{\Theta;A}(M, e) = \delta(\texttt{return}\, \star, e[V/r])$:
By SET-typing rule we have $\pi_1 \triangleright \vdash_v^{\Theta} V : G$ and $r : G \in \Theta$, by definition
$$poly(\pi) = 2 + poly(\pi_1). \tag{8.24}$$
Let $\rho \triangleright \vdash_c^{\Theta} \texttt{return}\, \star : \mathbb{U}$, then by ETA-typing rule we have $\rho' \triangleright \vdash_v^{\Theta} \star : \mathbb{U}$ and by definition
$$poly(\rho) = 1 + poly(\rho') = 1 + 1 = 2. \tag{8.25}$$
Therefore, we have that $poly(\pi) \overset{\text{Eq. 8.24}}{=} 2 + poly(\pi_1) > poly(\rho) \overset{\text{Eq. 8.25}}{=} 2$ and we can conclude since it is easy to check that $\forall n \geq 1.\ poly(\pi_1)n \geq 1$.

- *Case* $\pi \triangleright \vdash_c^{\Theta} \texttt{get}\, r : G$ *and* $\mathbf{sm}_n^{\Theta;A}(M, e) = \delta(\texttt{return}\, e(r), e)$:
By GET-typing rule we have $r : G \in \Theta$ and by definition
$$poly(\pi) = 2 + size_G(G). \tag{8.26}$$
Let $\rho \triangleright \vdash_c^{\Theta} \texttt{return}\, e(r) : G$, then by ETA-typing rule and since $e(r)$ is of ground type $G$ and by Lemma 55 we have $\rho' \triangleright \vdash_v^{\varnothing} e(r) : G$. By definition $poly(\rho) = 1 + poly(\rho')$. Moreover, by Lemma 56 applied to $\rho'$ we have that $poly(\rho') = size_G(G)$. Summing up, we have
$$poly(\rho) = 1 + size_G(G). \tag{8.27}$$
Therefore, we have that $poly(\pi) \overset{\text{Eq. 8.26}}{=} 2 + size_G(G) > poly(\rho) \overset{\text{Eq. 8.27}}{=} 1 + size_G(G)$ and we can conclude.

- *Case* $\pi \triangleright \vdash_c^{\Theta} f_p(Z_1, \ldots, Z_m) : Gp$ *and* $\mathbf{sm}_n^{\Theta;A}(M, e) = \sum_k a_k\ \delta(U_k, e_k) = [\![f]\!]_p(n)(Z_1, \ldots, Z_m)$:
By FUN-typing rule we have $\mathsf{typeof}(f) = G_1 \times \ldots \times G_m \to G$ and $\pi_{1j} \triangleright \vdash_v^{\Theta} Z_j : G_j$ for $1 \le j \le m$. Moreover, by definition we have

$$poly(\pi) = 1 + size_G(G) + \sum_{j=1}^{m} poly(\pi_{1j}). \tag{8.28}$$

Since $U_k \in [\![Gp]\!]_n$ for every $k$, by Lemma 55 we have $\rho_k \triangleright \vdash_v^{\varnothing} U_k : Gp(n)$. Moreover, by Lemma 56 applied to $\rho_k$ for every $k$, we have that

$$poly(\rho_k) = size_G(G). \tag{8.29}$$

Therefore, we have that

$$poly(\pi) \overset{\text{Eq. } 8.28}{=} 1 + size_G(G) + \sum_{j=1}^{m} poly(\pi_{1j}) > poly(\rho_k) \overset{\text{Eq. } 8.29}{=} size_G(G)$$

this holds for every $k$ and so we can conclude.

- *Case* $\pi \triangleright \vdash_c^{\Theta} \mathtt{loop}\,(\lambda x.N_2)\,1\,\mathtt{times\,from}\,N_1 : P$ *and* $\mathbf{sm}_n^{\Theta;A}(M, e) = \sum_k a_k\ \delta(\mathtt{let}\,x = N_1\,\mathtt{in}\,N_2, e)$:
Since by hypothesis we have $\pi \triangleright \vdash_c^{\Theta} \mathtt{loop}\,(\lambda x.N_2)\,1\,\mathtt{times\,from}\,N_1 : P$, we proceed by analyzing such derivation as follows:

  - By LOOP-typing rule we have $\pi_1 \triangleright \vdash_v^{\Theta} \lambda x.N_2 : P \multimap P$ and $\pi_2 \triangleright \vdash_c^{\Theta} N_1 : P$. Moreover, by definition $poly(\pi) = 1 + 1 + 1 \times poly(\pi_1) + poly(\pi_2) = 2 + poly(\pi_1) + poly(\pi_2)$.
  - By LAM-typing rule applied to $\pi_1$ we have $\pi_1' \triangleright x : P \vdash_c^{\Theta} N_2 : P$. Moreover, by definition we have that $poly(\pi_1) = 1 + poly(\pi_1')$.

  Summing up, we have that

$$poly(\pi) = 3 + poly(\pi_1') + poly(\pi_2). \tag{8.30}$$

  Let $\rho \triangleright \vdash_c^{\Theta} \mathtt{let}\,x = N_1\,\mathtt{in}\,N_2 : P$, then by definition

$$poly(\rho) = 1 + poly(\pi_1') + poly(\pi_2). \tag{8.31}$$

  Therefore, we have that

$$poly(\pi) \overset{\text{Eq. } 8.30}{=} 3 + poly(\pi_1') + poly(\pi_2) > poly(\rho) \overset{\text{Eq. } 8.31}{=} 1 + poly(\pi_1') + poly(\pi_2)$$

  and we can conclude.

- *Case* $\pi \triangleright \vdash_c^{\Theta} \mathtt{loop}\,(\lambda x.N_2)\,k+1\,\mathtt{times\,from}\,N_1 : P$ *and* $\mathbf{sm}_n^{\Theta;A}(M, e) = \sum_k a_k\ \delta(\mathtt{let}\,x = (\mathtt{loop}\,(\lambda x.N_2)\,k\,\mathtt{times\,from}\,N_1)\,\mathtt{in}\,N_2, e)$:
Since by hypothesis we have $\pi \triangleright \vdash_c^{\Theta} \mathtt{loop}\,(\lambda x.N_2)\,k+1\,\mathtt{times\,from}\,N_1 : P$, we proceed by analyzing such derivation as follows:

  - By LOOP-typing rule we have $\pi_1 \triangleright \vdash_v^{\Theta} \lambda x.N_2 : P \multimap P$ and $\pi_2 \triangleright \vdash_c^{\Theta} N_1 : P$. Moreover, by definition $poly(\pi) = 1 + k + 1 + (k+1) \times poly(\pi_1) + poly(\pi_2) = 2 + k + (k+1) \times poly(\pi_1) + poly(\pi_2)$.

- By LAM-typing rule applied to $\pi_1$ we have $\pi'_1 \triangleright x : P \vdash^{\Theta}_{c} N_2 : P$. Moreover, by definition we have that $\mathit{poly}(\pi_1) = 1 + \mathit{poly}(\pi'_1)$.

Summing up, we have that

$$\begin{aligned}\mathit{poly}(\pi) &= 2 + k + (k+1) \times \mathit{poly}(\pi_1) + \mathit{poly}(\pi_2)\\ &= 2 + k + (k+1) \times (1 + \mathit{poly}(\pi'_1)) + \mathit{poly}(\pi_2)\\ &= 2 + k + (k+1) \times 1 + (k+1) \times \mathit{poly}(\pi'_1) + \mathit{poly}(\pi_2)\\ &= 2 + k + k + 1 + (k+1) \times \mathit{poly}(\pi'_1) + \mathit{poly}(\pi_2)\\ &= 3 + 2k + (k+1) \times \mathit{poly}(\pi'_1) + \mathit{poly}(\pi_2)\end{aligned}$$

Thus, we have

$$\mathit{poly}(\pi) = 3 + 2k + (k+1) \times \mathit{poly}(\pi'_1) + \mathit{poly}(\pi_2). \tag{8.32}$$

Let $\rho \triangleright \vdash^{\Theta}_{c} \mathtt{let}\, x = (\mathtt{loop}\,(\lambda x.N_2)\, k\, \mathtt{times}\, \mathtt{from}\, N_1)\, \mathtt{in}\, N_2 : P$ , we proceed by analyzing such derivation as follows:

- By LET-typing rule we have $\rho_1 \vdash^{\Theta}_{c} \mathtt{loop}\,(\lambda x.N_2)\, k\, \mathtt{times}\, \mathtt{from}\, N_1 : P$ and $\pi'_1 \triangleright x : P \vdash^{\Theta}_{c} N_2 : P$. By definition we have $\mathit{poly}(\rho) = 1 + \mathit{poly}(\rho_1) + \mathit{poly}(\pi'_1)$.
- By LOOP-typing rule applied to $\rho_1$ we have $\pi_1 \triangleright \vdash^{\Theta}_{v} \lambda x.N_2 : P \multimap P$ and $\pi_2 \triangleright \vdash^{\Theta}_{c} N_1 : P$. Moreover, by definition $\mathit{poly}(\rho_1) = 1 + k + k \times \mathit{poly}(\pi_1) + \mathit{poly}(\pi_2)$.
- By LAM-typing rule applied to $\pi_1$ we have $\pi'_1 \triangleright x : P \vdash^{\Theta}_{c} N_2 : P$. Moreover, by definition we have that $\mathit{poly}(\pi_1) = 1 + \mathit{poly}(\pi'_1)$.

Summing up, we have that

$$\begin{aligned}\mathit{poly}(\rho) &= 1 + \mathit{poly}(\rho_1) + \mathit{poly}(\pi'_1)\\ &= 1 + (1 + k + k \times \mathit{poly}(\pi_1) + \mathit{poly}(\pi_2)) + \mathit{poly}(\pi'_1)\\ &= 1 + (1 + k + k \times (1 + \mathit{poly}(\pi'_1)) + \mathit{poly}(\pi_2)) + \mathit{poly}(\pi'_1)\\ &= 2 + k + k \times (1 + \mathit{poly}(\pi'_1)) + \mathit{poly}(\pi_2) + \mathit{poly}(\pi'_1)\\ &= 2 + k + k + k \times \mathit{poly}(\pi'_1) + \mathit{poly}(\pi_2) + \mathit{poly}(\pi'_1)\\ &= 2 + 2k + k \times \mathit{poly}(\pi'_1) + \mathit{poly}(\pi'_1) + \mathit{poly}(\pi_2)\\ &= 2 + 2k + (k+1) \times \mathit{poly}(\pi'_1) + \mathit{poly}(\pi_2)\end{aligned}$$

Thus, we have

$$\mathit{poly}(\rho) = 2 + 2k + (k+1) \times \mathit{poly}(\pi'_1) + \mathit{poly}(\pi_2) \tag{8.33}$$

Therefore, we have that

$$\begin{aligned}\mathit{poly}(\pi) &\overset{\text{Eq. 8.32}}{=} 3 + 2k + (k+1) \times \mathit{poly}(\pi'_1) + \mathit{poly}(\pi_2)\\ &> \mathit{poly}(\rho) \overset{\text{Eq. 8.33}}{=} 2 + 2k + (k+1) \times \mathit{poly}(\pi'_1) + \mathit{poly}(\pi_2)\end{aligned}$$

and we can conclude.

□

Chapter 9

# Computational Indistinguishability

This chapter introduces a logical relation that we prove sound with respect to computational indistinguishability, a core concept in cryptographic security. We begin in Section 9.1 by shifting from reasoning about terms which are closed with respect to the security parameter to families of open terms indexed over all security parameters, an essential formulation for modelling computational indistinguishability. We then define $\Theta$-contextual indistinguishability in Section 9.2, which accurately formalizes the notion of computational indistinguishability within the $\lambda$BLL framework. Building on the metric foundations from Section 9.3, Section 9.4 introduces a logical metric that measures the behavioural differences between $\lambda$BLL terms. From this, we derive the indistinguishability logical relation, which connects terms whose behavioural distance is negligible, guaranteeing that they cannot be efficiently distinguished. This relation is proved to be sound with respect to $\Theta$-contextual indistinguishability, providing the semantic foundation for our security analysis in Chapter 10, where we establish the CPA security of an encryption scheme induced by a pseudorandom function within the $\lambda$BLL framework.

## 9.1 Term Relations

In Section 8.3, we introduced indexed families $\Lambda(\Gamma;\Theta;A)$ and $\mathcal{V}(\Gamma;\Theta;A)$ of sets of terms that are closed with respect to the security parameter variable $i$. These sets serve as the semantic universe in which we interpret the types and terms of our calculus, assuming the security parameter has already been fixed. However, in this chapter, our attention shifts to the study of computational indistinguishability in which the security parameter is treated as a free variable ranging over all positive natural numbers. Consequently, indistinguishability concerns not closed terms at a fixed security parameter, but rather families of open terms whose behaviour may vary with each concrete instantiation of the security parameter. This formulation is essential for capturing the cryptographic requirement that two programs remain indistinguishable by any efficient adversary across all sufficiently large values of the security parameter. This shift in perspective from fixed-index semantics to parameterized relational reasoning plays a central role in formulating and proving cryptographic security within the type-theoretic and semantic framework of $\lambda$BLL.

Formally, for a variable context $\Gamma$, a reference context $\Theta$ and a type $A$, let $\Lambda_o(\Gamma;\Theta;A)$ and $\mathcal{V}_o(\Gamma;\Theta;A)$ be the sets of derivable computation terms and values respectively which are open for the security parameter variable $i$:

$$\Lambda_o(\Gamma;\Theta;A) := \{M \in \Lambda \mid \Gamma \vdash_c^{\Theta} M : A\} \quad \text{and} \quad \mathcal{V}_o(\Gamma;\Theta;A) := \{V \in \mathcal{V} \mid \Gamma \vdash_v^{\Theta} V : A\}.$$

Lemma 61 provides a crucial bridge between symbolic reasoning and concrete instantiation in the semantics of $\lambda$BLL. It ensures that well-typed computations and values, when parameterized by the security parameter, remain well-typed under any substitution with a concrete $n \geq 1$. This result justifies the use of open terms with a free security parameter in our logical framework.

**Lemma 61.** The following rules are derivable for all $n \geq 1$:

$$\frac{\Gamma \vdash_c^{\Theta} M : A}{\Gamma n \vdash_c^{\Theta n} Mn : An} \qquad\qquad \frac{\Gamma \vdash_v^{\Theta} V : A}{\Gamma n \vdash_v^{\Theta n} Vn : An}$$

It implies that if $M$ is in $\Lambda_o(\Gamma; \Theta; A)$, then $Mn$ is in $\Lambda_n(\Gamma; \Theta; A)$ for all $n \geq 1$ and a similar statement holds for values in $\mathcal{V}_o(\Gamma; \Theta; A)$.

We now formalize the notion of a term relations used to reason about computational indistinguishability. Since computational indistinguishability must hold across all instantiations of the security parameter and may involve terms with free variables, we define relations not only on closed terms but also on open ones. These relations are parameterized by the reference context $\Theta$, which governs the global store.

**Definition 10.** For a fixed reference context $\Theta$, a $\Theta$-*open term relation* $\mathcal{R}$ is an indexed family of pairs of relations

$$\{(\mathcal{RC}(\Gamma; \Theta; A), \mathcal{RV}(\Gamma; \Theta; A))\}_{\Gamma, A}$$

where for all variable contexts $\Gamma$,and types $A$,

$$\mathcal{RC}(\Gamma; \Theta; A) \subseteq \Lambda_o(\Gamma; \Theta; A) \times \Lambda_o(\Gamma; \Theta; A) \qquad \mathcal{RV}(\Gamma; \Theta; A) \subseteq \mathcal{V}_o(\Gamma; \Theta; A) \times \mathcal{V}_o(\Gamma; \Theta; A).$$

A *closed* (for term variables) $\Theta$-*term relation* $\mathcal{R}$ is an indexed family of pairs of relations

$$\{(\mathcal{RC}(\Theta; A), \mathcal{RV}(\Theta; A))\}_{\Gamma, A}$$

with $\mathcal{RC}(\Theta; A) \subseteq \Lambda_o(\Theta; A) \times \Lambda_o(\Theta; A)$ and $\mathcal{RV}(\Theta; A) \subseteq \mathcal{V}_o(\Theta; A) \times \mathcal{V}_o(\Theta; A)$.

Moreover, every open term relation induces a closed term relation by restricting to empty term variable contexts. For the other direction, we use the standard notion of *open extension* of a closed relation via substitutions with positive values.

Let us consider fixed reference contexts $\Theta \subseteq \Xi$ and an open $\Xi$-term relation $\mathcal{R}$. The term relation $\mathcal{R}$ is said to be $\Theta$-compatible if it includes its own compatible closure $\mathbf{Cl}(\Theta)(\mathcal{R})$, as defined in Figure 9.1. This means that for any terms and types, if two terms are related under the compatible closure, they must also be related under the original relation. Formally, this ensures that the relation is closed under the syntactic constructs with respects to contexts acting only on $\Theta$. This restriction to contexts acting only on $\Theta$ is evident in the last two rules of Figure 9.1, where the reference $r$ must come from $\Theta$ rather than the full context $\Xi$.

**Definition 11.** For fixed reference contexts $\Theta \subseteq \Xi$, an open $\Xi$-term relation $\mathcal{R} = (\mathcal{RV}, \mathcal{RC})$ is $\Theta$-*compatible* if it contains its compatible closure $\mathbf{Cl}(\Theta)(\mathcal{R})$ with respect to contexts acting only on $\Theta$ (given in Figure 9.1). Formally, for every variable context $\Gamma$, and type $A$, we have:

$$\begin{aligned} (M, N) \in \mathbf{Cl}(\Theta)(\mathcal{RC})(\Gamma; \Xi; A) \quad &\Rightarrow \quad (M, N) \in \mathcal{RC}(\Gamma; \Xi; A) \\ (U, V) \in \mathbf{Cl}(\Theta)(\mathcal{RV})(\Gamma; \Xi; A) \quad &\Rightarrow \quad (U, V) \in \mathcal{RV}(\Gamma; \Xi; A). \end{aligned}$$

$$\frac{}{(x,x) \in \mathbf{Cl}(\Theta)(\mathcal{RV})(x : P; \Xi; P)} \qquad \frac{}{(\mathbf{t},\mathbf{t}) \in \mathbf{Cl}(\Theta)(\mathcal{RV})(\varnothing; \Xi; \mathbb{B})}$$

$$\frac{}{(\mathbf{f},\mathbf{f}) \in \mathbf{Cl}(\Theta)(\mathcal{RV})(\varnothing; \Xi; \mathbb{B})} \qquad \frac{\text{length}(s) = n_0 \qquad p : i \mapsto n_0 \text{ constant polynomial}}{(s,s) \in \mathbf{Cl}(\Theta)(\mathcal{RV})(\varnothing; \Xi; \mathbb{S}[p])}$$

$$\frac{}{(\star,\star) \in \mathbf{Cl}(\Theta)(\mathcal{RV})(\varnothing; \Xi; \mathbb{U})}$$

$$\frac{\text{typeof}(f) = G_1 \times \cdots \times G_n \to G \qquad \{(V_k, W_k) \in \mathcal{RV}(\Gamma_k p; \Xi; G_k p)\}_{1 \leq k \leq n} \qquad p \in \mathbb{N}_{\geq 1}[i]}{(f_p(V_1,\dots,V_n), f_p(W_1,\dots,V_n)) \in \mathbf{Cl}(\Theta)(\mathcal{RC})(\boxplus \Gamma_k p; \Xi; Gp)}$$

$$\frac{(U,U') \in \mathcal{RV}(\Gamma; \Xi; P) \qquad (V,V') \in \mathcal{RV}(\Delta; \Xi; Q)}{(\langle U, V\rangle, \langle U', V'\rangle) \in \mathbf{Cl}(\Theta)(\mathcal{RV})(\Gamma \boxplus \Delta; \Xi; P \otimes Q)} \qquad \frac{(M,N) \in \mathcal{RC}(\Gamma; \Xi; A)}{(!M, !N) \in \mathbf{Cl}(\Theta)(\mathcal{RV})(p * \Gamma; \Xi; !_p A)}$$

$$\frac{(Z,Z') \in \mathcal{RV}(\Gamma; \Xi; P \otimes P') \qquad (M,M') \in \mathcal{RC}(x : P, y : P', \Delta; \Xi; A)}{(\texttt{let}\,\langle x,y\rangle = Z \,\texttt{in}\, M, \texttt{let}\,\langle x,y\rangle = Z' \,\texttt{in}\, M') \in \mathbf{Cl}(\Theta)(\mathcal{RC})(\Gamma \boxplus \Delta; \Xi; A)}$$

$$\frac{(Z,Z') \in \mathcal{RV}(\Gamma; \Xi; !_1 A)}{(\texttt{der}(Z), \texttt{der}(Z')) \in \mathbf{Cl}(\Theta)(\mathcal{RC})(\Gamma; \Xi; A)} \qquad \frac{(M,N) \in \mathcal{RC}(\Theta)(\Gamma, x : P; \Xi; A)}{(\lambda x.M, \lambda x.N) \in \mathbf{Cl}(\Theta)(\mathcal{RV})(\Gamma; \Xi; P \multimap A)}$$

$$\frac{(M,M') \in \mathcal{RC}(\Gamma; \Xi; P \multimap A) \qquad (Z,Z') \in \mathcal{RV}(\Delta; \Xi; P)}{(MZ, M'Z') \in \mathbf{Cl}(\Theta)(\mathcal{RC})(\Gamma \boxplus \Delta; \Xi; A)}$$

$$\frac{(N,N') \in \mathcal{RC}(\Gamma; \Xi; P) \qquad (M,M') \in \mathcal{RC}(x : P, \Delta; \Xi; A)}{(\texttt{let}\, x = N \,\texttt{in}\, M, \texttt{let}\, x = N' \,\texttt{in}\, M') \in \mathbf{Cl}(\Theta)(\mathcal{RC})(\Gamma \boxplus \Delta; \Xi A)}$$

$$\frac{(V,V') \in \mathcal{RV}(\Gamma; \Xi; P \multimap P) \qquad (M,M') \in \mathcal{RC}(\Delta; \Xi; P)}{(\texttt{loop}\, V\, p\, \texttt{times}\,\texttt{from}\, M, \texttt{loop}\, V'\, p\, \texttt{times}\,\texttt{from}\, M') \in \mathbf{Cl}(\Theta)(\mathcal{RC})(p * \Gamma \boxplus \Delta; \Xi; P)}$$

$$\frac{(Z,Z') \in \mathcal{RV}(\Gamma; \Xi; \mathbb{B}) \qquad (M,M') \in \mathcal{RC}(\Delta; \Xi; A) \qquad (N,N') \in \mathcal{RC}(\Delta; \Xi; A)}{(\texttt{if}\, Z \,\texttt{then}\, M \,\texttt{else}\, N, \texttt{if}\, Z' \,\texttt{then}\, M' \,\texttt{else}\, N') \in \mathbf{Cl}(\Theta)(\mathcal{RC})(\Gamma \boxplus \Delta; \Xi; A)}$$

$$\frac{(r : G) \in \Theta \qquad (Z,Z') \in \mathcal{RV}(\Gamma; \Xi; G)}{(\texttt{set}\, r\, Z, \texttt{set}\, r\, Z') \in \mathbf{Cl}(\Theta)(\mathcal{RC})(\Gamma; \Xi; \mathbb{U})} \qquad \frac{(r : G) \in \Theta}{(\texttt{get}\, r, \texttt{get}\, r) \in \mathbf{Cl}(\Theta)(\mathcal{RC})(\varnothing; \Xi; G)}$$

Figure 9.1: $\Theta$-compatible closure $\mathbf{Cl}(\Theta)(\mathcal{R})$ of a $\Xi$-term relation $\mathcal{R}$.

## 9.2 Contextual Indistinguishability

The notion of behavioural equivalence used in this chapter is computational indistinguishability, a standard semantic criterion in cryptography. It states that two programs are equivalent if no adversary running in probabilistic polynomial time can distinguish their outputs with more than negligible advantage. In this setting, a function is negligible if it grows asymptotically slower than the inverse of any polynomial [21]:

**Definition 12.** A function $\varepsilon : \mathbb{N} \to \mathbb{R}_+$ is *negligible* if for all $k \in \mathbb{N}$, there exists $N \in \mathbb{N}$ such that for all $n \geq N$, $\varepsilon(n) < \frac{1}{n^k}$.

In our framework, adversaries are represented by well-typed contexts in the language $\lambda$BLL, where both the expressiveness and the cost model are constrained to ensure polynomial-time soundness (see Section 8.4). We consider single-hole contexts generated by the following grammars:

$$E ::= [\cdot]_{\mathcal{V}} \mid \langle E, Z\rangle \mid \langle Z, E\rangle \mid !C \mid \lambda x.C \qquad \text{(Value Contexts)}$$

$$\begin{aligned} C ::=\ & [\cdot]_\Lambda \mid \texttt{return}\, E \mid M E \mid C Z \mid \texttt{let}\, \langle x, y\rangle = E\, \texttt{in}\, M \\ & \mid \texttt{let}\, \langle x, y\rangle = Z\, \texttt{in}\, C \mid \texttt{der}(E) \mid \texttt{let}\, x = N\, \texttt{in}\, C \\ & \mid \texttt{if}\, Z\, \texttt{then}\, C\, \texttt{else}\, D \mid \texttt{loop}\, V\, p\, \texttt{times}\, \texttt{from}\, C \\ & \mid \texttt{let}\, x = C\, \texttt{in}\, M \mid \texttt{loop}\, E\, p\, \texttt{times}\, \texttt{from}\, N \\ & \mid \texttt{set}\, r\, E \mid f_p(Z_1, \ldots, E, \ldots, Z_m) \end{aligned} \qquad \text{(Computations Contexts)}$$

The following definition introduces the notion of indexed families of contexts, which are central to defining the logical relation for computational indistinguishability. They enable a precise modelling of adversaries through contextual interactions with open terms. More precisely, we consider typed single-hole contexts that govern how terms of a certain type can be inserted into larger program structures while preserving type correctness.

**Definition 13.** For fixed reference contexts $\Theta \subseteq \Xi$, we define indexed families of contexts:

$$\{\mathrm{Con}(\Theta, \Xi)(\Gamma;^v A \mid \Delta;^v B)\}_{\Gamma,A,\Delta,B} \qquad \{\mathrm{Con}(\Theta, \Xi)(\Gamma;^v A \mid \Delta;^c B)\}_{\Gamma,A,\Delta,B}$$
$$\{\mathrm{Con}(\Theta, \Xi)(\Gamma;^c A \mid \Delta;^v B)\}_{\Gamma,A,\Delta,B} \qquad \{\mathrm{Con}(\Theta, \Xi)(\Gamma;^c A \mid \Delta;^c B)\}_{\Gamma,A,\Delta,B}$$

inductively (main cases are given in Figure 9.2) that verify the property that for every derivable computation term $\Gamma \vdash_c^\Theta M : A$ and $E$ in $\mathrm{Con}(\Theta, \Xi)(\Gamma;^c A \mid \Delta;^v B)$, we have

$$\Delta \vdash_v^\Xi E[M] : B$$

and for every $C$ in $\mathrm{Con}(\Theta, \Xi)(\Gamma;^c A \mid \Delta;^c B)$, we have

$$\Delta \vdash_c^\Xi C[M] : B.$$

Similarly, for every derivable value term $\Gamma \vdash_v^\Theta V : A$ and $E$ in $\mathrm{Con}(\Theta, \Xi)(\Gamma;^v A \mid \Delta;^v B)$, we have

$$\Delta \vdash_v^\Xi E[V] : B$$

and for every $C$ in $\mathrm{Con}(\Theta, \Xi)(\Gamma;^v A \mid \Delta;^c B)$, we have

$$\Delta \vdash_c^\Xi C[V] : B.$$

$$\frac{}{[\cdot]_{\mathcal{V}} \in \mathrm{Con}(\Theta,\Xi)(\Gamma;^{v} A \mid \Gamma;^{v} A)} \qquad \frac{}{[\cdot]_{\Lambda} \in \mathrm{Con}(\Theta,\Xi)(\Gamma;^{c} A \mid \Gamma;^{c} A)}$$

$$\frac{r : G \in \Theta \qquad E \in \mathrm{Con}(\Theta,\Xi)(\Gamma;^{c} A \mid \Delta;^{v} G)}{\mathtt{set}\, r\, E \in \mathrm{Con}(\Theta,\Xi)(\Gamma;^{c} A \mid \Delta;^{c} \mathbb{U})} \qquad \frac{r : G \in \Theta \qquad E \in \mathrm{Con}(\Theta,\Xi)(\Gamma;^{v} A \mid \Delta;^{v} G)}{\mathtt{set}\, r\, E \in \mathrm{Con}(\Theta,\Xi)(\Gamma;^{v} A \mid \Delta;^{c} \mathbb{U})}$$

$$\frac{C \in \mathrm{Con}(\Theta,\Xi)(\Gamma;^{c} A \mid \Delta, x : P;^{c} B)}{\lambda x.C \in \mathrm{Con}(\Theta,\Xi)(\Gamma;^{c} A \mid \Delta;^{v} P \multimap B)} \qquad \frac{C \in \mathrm{Con}(\Theta,\Xi)(\Gamma;^{v} A \mid \Delta, x : P;^{c} B)}{\lambda x.C \in \mathrm{Con}(\Theta,\Xi)(\Gamma;^{v} A \mid \Delta;^{v} P \multimap B)}$$

$$\frac{C \in \mathrm{Con}(\Theta,\Xi)(\Gamma;^{c} A \mid \Delta;^{c} P \multimap B) \qquad Z \in \mathcal{V}_o(\Sigma;\Theta;P)}{CZ \in \mathrm{Con}(\Theta,\Xi)(\Gamma;^{c} A \mid \Delta \boxplus \Sigma;^{c} B)}$$

$$\frac{C \in \mathrm{Con}(\Theta,\Xi)(\Gamma;^{v} A \mid \Delta;^{c} P \multimap B) \qquad Z \in \mathcal{V}_o(\Sigma;\Theta;P)}{CZ \in \mathrm{Con}(\Theta,\Xi)(\Gamma;^{v} A \mid \Delta \boxplus \Sigma;^{c} B)}$$

$$\frac{E \in \mathrm{Con}(\Theta,\Xi)(\Gamma;^{c} A \mid \Delta;^{v} P) \qquad M \in \Lambda_o(\Sigma;\Theta;P \multimap B)}{ME \in \mathrm{Con}(\Theta,\Xi)(\Gamma;^{c} A \mid \Delta \boxplus \Sigma;^{c} B)}$$

$$\frac{E \in \mathrm{Con}(\Theta,\Xi)(\Gamma;^{v} A \mid \Delta;^{v} P) \qquad M \in \Lambda_o(\Sigma;\Theta;P \multimap B)}{ME \in \mathrm{Con}(\Theta,\Xi)(\Gamma;^{v} A \mid \Delta \boxplus \Sigma;^{c} B)}$$

$$\frac{E \in \mathrm{Con}(\Theta,\Xi)(\Gamma;^{v} A \mid \Delta;^{v} P) \qquad Z \in \mathcal{V}(\Sigma;\Theta;Q)}{\langle E, Z\rangle \in \mathrm{Con}(\Theta,\Xi)(\Gamma;^{v} A \mid \Delta \boxplus \Sigma;^{v} P \otimes Q)}$$

$$\frac{E \in \mathrm{Con}(\Theta,\Xi)(\Gamma;^{v} A \mid \Sigma;^{v} Q) \qquad Z \in \mathcal{V}(\Delta;\Theta;P)}{\langle Z, E\rangle \in \mathrm{Con}(\Theta,\Xi)(\Gamma;^{v} A \mid \Delta \boxplus \Sigma;^{v} P \otimes Q)}$$

$$\frac{E \in \mathrm{Con}(\Theta,\Xi)(\Gamma;^{v} A \mid \Delta;^{v} P \otimes Q) \qquad M \in \Lambda_o(x : P, y : Q, \Sigma;\Theta;B)}{\mathtt{let}\, \langle x, y\rangle = E\, \mathtt{in}\, M \in \mathrm{Con}(\Theta,\Xi)(\Gamma;^{v} A \mid \Delta \boxplus \Sigma;^{c} B)}$$

$$\frac{E \in \mathrm{Con}(\Theta,\Xi)(\Gamma;^{c} A \mid \Delta;^{v} P \otimes Q) \qquad M \in \Lambda_o(x : P, y : Q, \Sigma;\Theta;B)}{\mathtt{let}\, \langle x, y\rangle = E\, \mathtt{in}\, M \in \mathrm{Con}(\Theta,\Xi)(\Gamma;^{c} A \mid \Delta \boxplus \Sigma;^{c} B)}$$

Figure 9.2: $\lambda$BLL Context Families (a few cases)

**Remark 15.** Recall that we assume $\Theta \subseteq \Xi$, it is worth noting that in the rules of Figure 9.2, all the subterms $M$ and positive values $Z$ are typed under the reference context $\Theta$. Similarly, in the `set` rules (second line), the references $r$ must come from $\Theta$, not from the full context $\Xi$. This reflects the restriction that only references explicitly tracked in $\Theta$, i.e., those under adversarial control, can be manipulated or assigned within these contexts.

In order to reason about program behaviour in the presence of references, particularly when distinguishing between observable and internal state, we introduce an equivalence relation on stores that reflects agreement over a designated set of observable locations.

**Definition 14.** For fixed reference contexts $\Theta \subseteq \Xi$, we define an equivalence relation $\equiv_\Theta$ on $\mathrm{St}_\Xi$ as follows:

$$\forall e, e' \in \mathrm{St}_\Xi, e \equiv_\Theta e' \quad :\Leftrightarrow \quad \forall r \in \Theta, e(r) = e'(r).$$

Finally, in order to capture the idea that two open computations or values are computationally indistinguishable from the perspective of an observer with limited access to the store, we define $\Theta$-contextual indistinguishability. Given a set of observable references $\Theta \subseteq \Xi$, two computations (resp. values) are $\Theta$-contextually indistinguishable if no context, restricted to observing and manipulating only $\Theta$ (see Remark 15), can distinguish between them with more than negligible probability. This contextual definition generalizes the standard cryptographic notion of computational indistinguishability to higher-order, stateful programs by quantifying over all adversarial contexts compatible with $\Theta$.

**Definition 15.** For fixed reference contexts $\Theta \subseteq \Xi$ and terms $M, N$ in $\Lambda_o(\Gamma; \Xi; A)$, we say that $M$ and $N$ are $\Theta$-*contextually indistinguishable* if for every closing context $C$ in $\mathrm{Con}(\Theta, \Xi)(\Gamma;^c A \mid \varnothing;^c \mathbb{B})$ (so that $C[M]$ and $C[N]$ are in $\Lambda_o(\Xi; \mathbb{B})$), there exists a negligible function $\varepsilon : \mathbb{N} \to \mathbb{R}_+$ such that for every $n \geq 1$, $e, e' \in \mathrm{St}_{\Xi n}$ and subset $X \subseteq \mathrm{St}_{\Xi n}/\equiv_{\Theta n}$,

$$e \equiv_{\Theta n} e' \quad \Rightarrow \quad |(\!|C[M]n|\!)_n^{\Xi,\mathbb{B}}(e)(\mathbf{t}, X) - (\!|C[N]n|\!)_n^{\Xi,\mathbb{B}}(e')(\mathbf{t}, X)| \leq \varepsilon(n).$$

Similarly, for values $U, V$ in $\mathcal{V}_o(\Gamma; \Xi; A)$, we say that $U$ and $V$ are $\Theta$-*contextually indistinguishable* if for every closing context $C$ in $\mathrm{Con}(\Theta, \Xi)(\Gamma;^v A \mid \varnothing;^c \mathbb{B})$ (so that $C[U]$ and $C[V]$ are in $\Lambda_o(\Xi; \mathbb{B})$), there exists a negligible function $\varepsilon : \mathbb{N} \to \mathbb{R}_+$ such that for every $n \geq 1$, $e, e' \in \mathrm{St}_{\Xi n}$ and subset $X \subseteq \mathrm{St}_{\Xi n}/\equiv_{\Theta n}$,

$$e \equiv_{\Theta n} e' \quad \Rightarrow \quad |(\!|C[U]n|\!)_n^{\Xi,\mathbb{B}}(e)(\mathbf{t}, X) - (\!|C[V]n|\!)_n^{\Xi,\mathbb{B}}(e')(\mathbf{t}, X)| \leq \varepsilon(n).$$

We adopt a coinductive characterization of contextual indistinguishability, following the approach in [58, 76] used for contextual equivalence in applicative bisimilarity. The $\Theta$-contextual indistinguishability relation can alternatively be defined as the largest open $\lambda$BLL-term relation that is both $\Theta$-*compatible* and $\Theta$-*adequate*. $\Theta$-Compatibility ensures that the relation is closed under contexts that use only references in $\Theta$: if $(M, N)$ belongs to the relation and $C$ is a context, then $(C[M], C[N])$ must also be in the relation. Adequacy, on the other hand, depends on the specific observational behaviour under consideration. For contextual equivalence, this typically refers to termination, or in the case of non-deterministic calculi, the probability of convergence. In our setting, the relevant notion is $\Theta$-computational indistinguishability:

**Definition 16.** For fixed reference contexts $\Theta \subseteq \Xi$, an open $\Xi$-$\lambda$BLL-term relation $\mathcal{R} = (\mathcal{RV}, \mathcal{RC})$ is $\Theta$-*adequate* if for every pair of terms $(M, N)$ in $\mathcal{RC}(\varnothing; \Xi; \mathbb{B})$, there exists a negligible function $\varepsilon : \mathbb{N} \to \mathbb{R}_+$ such that for every $n \geq 1$, $e, e' \in \mathrm{St}_{\Xi n}$ and subset $X \subseteq \mathrm{St}_{\Xi n}/\equiv_{\Theta n}$,

$$e \equiv_{\Theta n} e' \quad \Rightarrow \quad |(\!|Mn|\!)_n^{\Xi,\mathbb{B}}(e)(\mathbf{t}, X) - (\!|Nn|\!)_n^{\Xi,\mathbb{B}}(e')(\mathbf{t}, X)| \leq \varepsilon(n).$$

We obtain that the predicate of Θ-adequacy on open λBLL-term relations is closed under countable unions and relational composition.

**Lemma 62.** For fixed reference contexts $\Theta \subseteq \Xi$, Θ-contextual indistinguishability is the largest Θ-adequate and Θ-compatible open Ξ-λBLL-relation and we denote it by $\sim_{\Theta \subseteq \Xi}$.

This coinductive characterization provides a useful proof principle to show soundness: any open Ξ-relation $\mathcal{R}$ that is both Θ-adequate and Θ-compatible must be included in $\sim_{\Theta \subseteq \Xi}$ and is therefore *sound* for Θ-contextual indistinguishability (i.e. any pair of terms $(M, N)$ in $\mathcal{R}$ are Θ-contextually indistinguishable). We also obtain as a corollary that if $\Theta_2 \subseteq \Theta_1$, then two terms that are indistinguishable by contexts using references in $\Theta_1$ should also indistinguishable by contexts using references from a smaller context $\Theta_2$:

**Corollary 8.** For fixed reference contexts $\Theta_2 \subseteq \Theta_1 \subseteq \Xi$, we have $\sim_{\Theta_1 \subseteq \Xi} \quad \subseteq \quad \sim_{\Theta_2 \subseteq \Xi}$ .

## 9.3 Background on Metrics

In order to reason about program behaviour quantitatively in λBLL, we extend beyond traditional binary equivalence and introduce a metric framework that assigns real-valued distances between programs. This approach allows for a fine-grained characterization of behavioural differences, particularly in the presence of probabilistic effects and mutable state, thereby supporting more precise analyses of program similarity.

We begin by introducing weighted relations, which generalize traditional relations by assigning to each pair of elements a real value in the interval $[0, 1]$, interpreted as a distance. These values are combined using truncated addition, a natural operation that models the accumulation of effects in the language. The resulting structure is grounded in the Łukasiewicz quantale, which provides an algebraic framework for compositional reasoning about distances. Next, we consider pseudo-metric spaces as a specific instance of weighted relations. A pseudo-metric space equips each pair of elements with a distance satisfying reflexivity, symmetry, and the triangle inequality. These spaces serve as the semantic domain in which program terms and their behaviours are interpreted. Finally, we extend the treatment of computational effects from the category of sets to that of metric spaces. Through the Kantorovich lifting, we define a metric on probabilistic computations by evaluating the minimal cost required to couple their outcomes. This construction enables a principled comparison of computations that accounts for both probabilistic behaviour and interaction with mutable state, thus providing the foundation for the logical metric developed in the subsequent section.

**Weighted Relations.** For metric reasoning on λBLL, we consider distances valued in the unit real interval $[0, 1]$ equipped with the operation of *truncated addition* $x \oplus y := \min\{1, x + y\}$ for $x, y \in [0, 1]$. We have in particular that $1 = 1 \oplus 1$ which has a direct correspondence with the fact that values of ground type are arbitrarily duplicable in λBLL (see Remark 16). We also make use of *truncated subtraction*, defined as: $x \ominus y := \max\{0, x - y\}$, for $x, y \in [0, 1]$.

Recall that a (unital) *quantale* is a tuple $(\mathcal{Q}, \leq, \otimes, 1)$ where $(\mathcal{Q}, \leq)$ is a complete lattice, $(\mathcal{Q}, \otimes, 1)$ is a monoid and $\otimes$ distributes over arbitrary joins [99]. The unit interval with the opposite of the natural order (the natural order is defined as: $x \leq y$ if and only if there exists $z$ such that $x \oplus z = y$) can be equipped with a quantale structure $\mathcal{L} = ([0, 1], \geq, \oplus, 0)$, called the *Łukasiewicz quantale*.

For sets $X$ and $Y$, an *$\mathcal{L}$-weighted relation* $R : X \twoheadrightarrow Y$ from $X$ to $Y$ consists of a function $X \times Y \to [0, 1]$. They form a category, which we denote by $\mathbf{Rel}_{\mathcal{L}}$, where the identity $\mathrm{id}_X : X \twoheadrightarrow X$ maps a pair $(x, y)$ to 0 if $x = y$ and 1 otherwise. The composite of two relations $R : X \twoheadrightarrow Y$ and $S : Y \nrightarrow Z$ is the relation $S \circ R : X \nrightarrow Z$ mapping a pair $(x, z)$ to $\inf_{y \in Y} R(x, y) \oplus S(y, z)$. The category $\mathbf{Rel}_{\mathcal{L}}$ can be equipped with a *dual* (transpose) operation mapping a relation $R : X \nrightarrow Y$ to the relation $R^{\text{op}} : Y \nrightarrow X$ which simply maps $(y, x)$ to $R(x, y)$. Any function $f : X \to Y$, induces a $\mathcal{L}$-relation via its graph $\mathbf{gr}(f) : X \nrightarrow Y$ mapping a pair $(x, y)$ to 0 if $f(x) = y$ and 1 otherwise.

**Pseudo-metric Spaces.** If we restrict to the special case of weighted endo-relations $R : X \nrightarrow X$ that are reflexive ($R \leq \text{id}_X$), symmetric ($R = R^{\text{op}}$) and transitive ($R \leq R \circ R$), we obtain the notion of pseudo-metric space:

**Definition 17.** A *pseudo-metric space* consists of a pair $(X, d_X)$ where $X$ is a set and $d_X$ is a function from $X \times X \to [0, 1]$ satisfying the following axioms:

- reflexivity: for all $x$ in $X$, $d_X(x, x) = 0$;
- symmetry: for all $x, y$ in $X$, $d_X(x, y) = d(y, x)$;
- triangular inequality: for all $x, y, z$ in $X$, $d_X(x, z) \leq d_X(x, y) \oplus d_X(y, z)$

If $d_X$ further satisfies the separation axiom (for all $x, y$, $d_X(x, y) = 0$ implies $x = y$), then $(X, d_X)$ is a *metric space.*

For the rest of the paper, even if we do not assume that the separation axiom holds, we will just say metric space instead of pseudo-metric space. Note that metric space with the *discrete metric* $\text{disc} : X \times X \to [0, 1]$ mapping a pair $(x, y)$ to 0 if $x = y$ and 1 otherwise corresponds exactly to the identity weighted relation defined above.

**Definition 18.** For two metric spaces $(X, d_X)$ and $(Y, d_Y)$, a function $f : X \to Y$ is said to be *non-expansive* if for all $x, x'$ in $X$, $d_Y(f(x), f(x')) \leq d_X(x, x')$. We denote by **PMet** the category of pseudo-metric spaces and non-expansive maps.

The category **PMet** is equivalent to the category of $\mathcal{L}$-enriched categories and $\mathcal{L}$-enriched functors between them [62]. We recall below some properties of **PMet** which we will use to define the logical metric in the following section, they are all instances of more general statements on quantale-enriched categories and we refer the reader to [62] for a complete account. **PMet** is symmetric monoidal closed with tensor product $(X, d_X) \otimes (Y d_Y)$ given by $(X \times Y, d_{X \otimes Y})$ where for all $x, x' \in X$ and $y, y' \in Y$,

$$d_{X \otimes Y}((x, y), (x', y')) := d_X(x, x') \oplus d_Y(y, y').$$

The unit is given by $\mathbf{1} = (\{\star\}, \text{disc})$ and the linear hom $X \multimap Y$ has underlying set $\mathbf{PMet}(X, Y)$ (the set of non-expansive maps from $X$ to $Y$) and distance $d_{X \multimap Y}(f, g) := \sup_{x \in X} d_Y(f(x), g(x))$. For any $k \geq 1$ and $x \in [0, 1]$, we define inductively $k \cdot x$ as $1 \cdot x := x$ and $(k+1) \cdot x := (k \cdot x) \oplus x$. This operation induces a *scaling* operation $k \cdot (X, d_X) := (X, k \cdot d_X)$ on **PMet** which we use to model the graded bang of λBLL. Note that if $f : (X, d_X) \to (Y, d_Y)$ is non-expansive, then $f : (X, k \cdot d_X) \to (Y, k \cdot d_Y)$ is also non-expansive for all $k \in \mathbb{N}$.

**Extending Monadic Effects from Sets to Metric Spaces.** In order to define the logical metric on computation terms, we need to extend the effect monad $\mathbf{T}_\Theta$ defined in Section 8.2 from sets to metric spaces. To do so, we follow the standard approach of monad extensions from sets to quantale weighted relations [62, 12]. It is well-known that the distribution monad $\mathbf{D}$ (and therefore the monads $\mathbf{T}_\Theta$ as well) on sets only *laxly* extends to weighted relations via *Kantorovich lifting* [69, 25, 13]. The Kantorovich lifting for distributions fits into the more general framework of *Barr extensions* for monads from sets to quantale relations [62].

In this section, we only give the explicit definition of how the Barr lax extension $\overline{\mathbf{T}}_\Theta$ of the effect monad acts on metric spaces and we refer the reader to [49, 114] for more background on lax extensions for weighted relations. The Kantorovich lifting can be formulated in terms of couplings for probability distributions:

**Definition 19.** For sets $X, Y$ and distributions $\mu \in \mathbf{D}(X)$, $\psi \in \mathbf{D}(Y)$, a *coupling over* $\mu$ *and* $\nu$ is a distribution $\gamma \in \mathbf{D}(X \times Y)$ such that

$$\forall x \in X, \mu(x) = \sum_{y \in Y} \gamma(x,y) \text{ and } \forall y \in Y, \nu(y) = \sum_{x \in X} \gamma(x,y).$$

We denote by $\Omega(\mu,\nu)$ the set of all couplings over $\mu$ and $\nu$.

For a metric space $(X, d_X)$, the Kantorovich lifting of the distance $d_X$ is the distance $\mathbf{K}(d_X)$ on $\mathbf{D}(X)$ mapping distributions $\mu, \nu \in \mathbf{D}(X)$ to

$$\mathbf{K}(d_X)(\mu,\nu) := \inf_{\gamma \in \Omega(\mu,\nu)} \sum_{x_1, x_2 \in X} \gamma(x_1, x_2) \cdot d_X(\mu(x_1), \nu(x_2)).$$

While there are many other possible choices of metrics on distribution spaces besides the Kantorovich distance $\mathbf{K}(d_X)$ [50], it is the smallest among the ones which laxly extends to weighted relations and it also coincides with the *statistical distance* (or *total variation distance*) when $d_X$ is the discrete metric.

**Definition 20.** For a set $X$, the *statistical distance* $d_{\text{stat}} : \mathbf{D}(X) \times \mathbf{D}(X) \to [0,1]$ maps two distributions $\mu, \nu \in \mathbf{D}(X)$ to

$$d_{\text{stat}}(\mu,\nu) := \frac{1}{2} \cdot \sum_{x \in X} |\mu(x) - \nu(x)| = \sup_{A \subseteq X} |\mu(A) - \nu(A)|.$$

For fixed reference contexts $\Theta \subseteq \Xi$ that is closed for the security parameter and a metric space $(X, d_X)$, we define $\overline{\mathbf{T}}_{\Theta \subseteq \Xi}(X, d_X)$ to be the metric space with underlying set $\mathbf{T}_{\Xi}(X)$ and distance $\overline{\mathbf{T}}_{\Theta \subseteq \Xi}(d_X)$ mapping functions $\varphi, \psi : \mathrm{St}_{\Xi} \to \mathbf{D}(X \times \mathrm{St}_{\Xi})$ to

$$\overline{\mathbf{T}}_{\Theta \subseteq \Xi}(d_X)(\varphi,\psi) := \sup_{e \equiv_\Theta e'} \mathbf{K}(d_X \otimes d_{\Theta \subseteq \Xi})(\varphi(e), \psi(e')). \tag{9.1}$$

where $d_{\Theta \subseteq \Xi}$ is the metric on $\mathrm{St}_{\Xi}$ mapping a pair $(e, e')$ to 0 if $e \equiv_\Theta e'$ and to 1 otherwise.

## 9.4 Logical Relation for Computational Indistinguishability

The notions introduced in the previous section form the groundwork for defining a logical metric for $\lambda$BLL terms, as detailed in Subsection 9.4.1. By leveraging the framework of metric spaces, along with truncated addition and the Kantorovich lifting, we can lift the monadic semantics of the language into a quantitative setting where distances between terms can be meaningfully measured. This logical metric facilitates precise reasoning about approximate equivalence and program indistinguishability by assigning real-valued distances to pairs of terms. This construction also supports the definition of a logical relation in Subsection 9.4.2, which is shown to be sound with respect to $\Theta$-contextual indistinguishability, capturing when two programs cannot be distinguished by any context that observes only the references in $\Theta$ with more than negligible probability.

### 9.4.1 Logical Metric for $\lambda$BLL terms

A logical metric generalizes logical relations by assigning a numerical distance that reflects how different two terms are, rather than simply stating whether they are equivalent. It replaces the binary structure of traditional logical relations with a real-valued measure between 0 and 1, capturing the degree of behavioural difference between terms.

For fixed reference contexts $\Theta \subseteq \Xi$, we now have all the ingredients to define a logical metric for $\lambda$BLL terms using the lax extension of the monad $\mathbf{T}_\Xi$ to metric spaces.

**Definition 21.** We define a family of metrics on closed computations and values indexed by the security parameter:

$$\mathbf{dV}_n(\Theta \subseteq \Xi; A) : \mathcal{V}_n(\Xi; A) \times \mathcal{V}_n(\Xi; A) \to [0,1] \text{ and } \mathbf{dC}_n(\Theta \subseteq \Xi; A) : \Lambda_n(\Xi; A) \times \Lambda_n(\Xi; A) \to [0,1]$$

by mutual induction on the type $A$:

$$\begin{aligned}
\mathbf{dV}_n(\Theta \subseteq \Xi; \mathbb{S}[p])(s, s') &:= \mathrm{disc}_{\mathbb{S}[p(n)]}(s, s') \\
\mathbf{dV}_n(\Theta \subseteq \Xi; \mathbb{B})(W, W') &:= \mathrm{disc}_{\mathbb{B}}(W, W') \\
\mathbf{dV}_n(\Theta \subseteq \Xi; \mathbb{U})(\star, \star) &:= \mathrm{disc}_{\mathbb{U}}(\star, \star) = 0 \\
\mathbf{dV}_n(\Theta \subseteq \Xi; !_p A)(!M, !N) &:= \oplus_{p(n)} \mathbf{dC}_n(\Theta \subseteq \Xi; A)(M, N) \\
\mathbf{dV}_n(\Theta \subseteq \Xi; P \otimes Q)(\langle U, V\rangle, \langle U', V'\rangle) &:= \mathbf{dV}_n(\Theta \subseteq \Xi; P)(U, U') \oplus \mathbf{dV}_n(\Theta \subseteq \Xi; Q)(V, V') \\
\mathbf{dV}_n(\Theta \subseteq \Xi; P \multimap A)(\lambda x.M, \lambda y.N) &:= \sup_{U,V \in \mathcal{V}_n(\Theta; P)} \begin{pmatrix} \mathbf{dC}_n(\Theta \subseteq \Xi; A)(M[V/x], N[U/y]) \\ \ominus \\ \mathbf{dV}_n(\Theta \subseteq \Xi; P)(U, V) \end{pmatrix} \\
\mathbf{dC}_n(\Theta \subseteq \Xi; A)(M, N) &:= \overline{\mathbf{T}}_{\Theta \subseteq \Xi}(\mathbf{dV}_n(\Theta \subseteq \Xi; A))(\llparenthesis M \rrparenthesis_n^{\Xi,A}, \llparenthesis N \rrparenthesis_n^{\Xi,A})
\end{aligned}$$

where disc denotes the discrete metric and $\overline{\mathbf{T}}_{\Theta \subseteq \Xi}$ is the lax extension of the functor $\mathbf{T}_\Xi : \mathbf{Set} \to \mathbf{Set}$ defined in Equation 9.1. In the definition of $\mathbf{dV}_n(\Theta \subseteq \Xi; P \multimap A)$, we are using the fact that $\Theta \subseteq \Xi$ implies that $\mathcal{V}_n(\Theta; P) \subseteq \mathcal{V}_n(\Xi; P)$ and therefore $\mathbf{dV}_n(\Theta \subseteq \Xi; P)(U, V)$ is well-defined.

In our setting, the metric version of the fundamental lemma states that substitution by positive value terms which only have access to the smaller set of locations is a non-expansive operation:

**Lemma 63** (Fundamental Lemma for Logical Metrics)**.** For a context $\Gamma = x_1 : P_1, \ldots, x_m : P_m$ and a term $M$ in $\Lambda_n(\Xi; \Gamma; A)$ with $n \geq 1$, for every closed positive values $Z_j, Z'_j \in \mathcal{V}_n(\Theta; P_j)$ with $1 \leq j \leq m$, we have

$$\mathbf{dC}_n(\Theta \subseteq \Xi; A)(M\rho, M\rho') \leq \bigoplus_{1 \leq j \leq m} \mathbf{dV}_n(\Theta \subseteq \Xi; P_j)(Z_j, Z'_j)$$

where $\rho := [Z_1/x_1, \ldots, Z_m/x_m]$ and $\rho' := [Z'_1/x_1, \ldots, Z'_m/x_m]$. A similar statement holds for open value terms in $\mathcal{V}_n(\Xi; \Gamma; A)$.

**Remark 16.** A key ingredient in the proof of the fundamental lemma is the equality

$$\mathbf{dV}_n(\Theta \subseteq \Xi; P \boxplus Q)(V, W) = \mathbf{dV}_n(\Theta \subseteq \Xi; P)(V, W) \oplus \mathbf{dV}_n(\Theta \subseteq \Xi; Q)(V, W)$$

for closed values $V, W$, it allows to keep a precise track of how distances are amplified when contexts are added $\Gamma \boxplus \Delta$ in rules such as `let` or $\otimes$ for example. We can see here that the main motivation behind using truncated addition $\oplus$ in our setting is that it allows for the additional flexibility of having ground types being duplicable without loosing the ability to measure distances for higher types: if $P$ and $Q$ are equal to some ground type $G$, and $V \neq W$, then the equality above indeed rewrites to $1 = 1 \oplus 1$ which would not be possible if we had considered for example the Lawvere quantale with regular addition instead of the Łukasiewicz quantale with truncated addition.

### 9.4.2 Indistinguishability Logical Relation

We now define the indistinguishability logical relation, a qualitative notion derived from the logical metric. It characterizes when two $\lambda$BLL terms are computationally indistinguishable in the cryptographic sense, meaning that no context, restricted to observing a subset of references $\Theta$, can distinguish them with more than negligible probability. In fact, this relation is proven to be sound with respect to $\Theta$-contextual indistinguishability, defined in Section 9.2, thereby establishing a connection between metric-based reasoning and cryptographic security guarantees.

Formally, we define a closed (for term variables) $\lambda$BLL-relation $\mathbf{Ind}(\Theta \subseteq \Xi)$ with

$$\mathbf{IndC}(\Theta \subseteq \Xi; A) \subseteq \Lambda_o(\Xi; A) \times \Lambda_o(\Xi; A) \quad \text{and} \quad \mathbf{IndV}(\Theta \subseteq \Xi; A) \subseteq \mathcal{V}_o(\Xi; A) \times \mathcal{V}_o(\Xi; A).$$

For computations $M, N$ in $\Lambda_o(\Xi; A)$, the pair $(M, N)$ is in $\mathbf{IndC}(\Theta \subseteq \Xi; A)$ if there exists a negligible function $\varepsilon : \mathbb{N} \to \mathbb{R}_+$ such that for all $n \geq 1$,

$$\mathbf{dC}_n(\Theta \subseteq \Xi; A)(Mn, Nn) \leq \varepsilon(n).$$

The relation on values $\mathbf{IndV}(\Theta \subseteq \Xi; A)$ is defined similarly via the logical metric on values $\mathbf{dV}(\Theta \subseteq \Xi; A)$.

The fundamental lemma for the indistinguishability logical relation can now be directly derivable from the non-expansiveness of the logical metric (Lemma 63) and basic closure properties of negligible functions:

**Lemma 64.** For a variable context $\Gamma = x_1 : P_1, \ldots, x_m : P_m$ and closed positive values $(Z_k, Z'_k)$ in $\mathbf{IndV}(\Theta \subseteq \Xi; P_k)$ with $1 \leq k \leq m$, we have for all $M \in \Lambda_o(\Gamma; \Xi; A)$ and $U \in \mathcal{V}_o(\Gamma; \Xi; A)$,

$$(M\rho, M\rho') \in \mathbf{IndC}(\Theta \subseteq \Xi; A) \text{ and } (U\rho, U\rho') \in \mathbf{IndV}(\Theta \subseteq \Xi; A)$$

where $\rho := [Z_1/x_1, \ldots, Z_m/x_m]$ and $\rho' := [Z'_1/x_1, \ldots, Z'_m/x_m]$. In particular, for a closed term $M \in \Lambda_o(\Xi; A)$, we have $(M, M) \in \mathbf{IndC}(\Theta \subseteq \Xi; A)$.

**Theorem 20.** The $\Theta$-open extension of $\mathbf{Ind}(\Theta \subseteq \Xi)$ is $\Theta$-adequate and $\Theta$-compatible.

Since the $\Theta$-contextual indistinguishability relation $\sim_{\Theta \subseteq \Xi}$ is the largest $\Theta$-compatible $\Theta$-adequate relation, it contains $\mathbf{Ind}(\Theta \subseteq \Xi)$ by Lemma 62, which implies that $\mathbf{Ind}(\Theta \subseteq \Xi)$ is *sound* for $\Theta$-contextual indistinguishability. Full abstraction on the other hand is not possible within our framework: since base types are equipped with the discrete metric whose Kantorovich lifting coincides with statistical distance, we cannot hope to capture the whole contextual indistinguishability relation as it is well-known that statistical closeness is strictly included in computational indistinguishability (e.g. Proposition 3.2.3 in [55]).

**Remark 17** (The Statistical Nature of Logical Indistinguishability)**.** Our indistinguishability logical relation is inherently statistical, as it is defined through an approximate logical relation that depends on a logical metric quantifying the behavioural distance between terms. This relation identifies two terms as indistinguishable when the behavioural distance between them is bounded by a negligible quantity, reflecting the approximate and statistical nature of equivalence that underpins our framework. This concept aligns with the relations commonly used in cryptographic proofs, where indistinguishability is defined in terms of the probabilistic indistinguishability of distributions or the advantage of an adversary in distinguishing between two scenarios. In cryptographic settings, this often translates to showing that no efficient adversary can distinguish between two distributions or systems with more than negligible probability. Our approximate logical relation directly aligns with cryptographic security proofs, as it facilitates the statistical measurement of equivalence between terms, enabling the demonstration of security in probabilistic terms that are consistent with the established cryptographic approach of proving computational indistinguishability.

There is a notable conceptual alignment between our approach and the method introduced by Shoup in [104], known as *sequences of games*. Both approaches aim to establish security by making stepwise modifications that preserve computational indistinguishability, relying on probabilistic rather than exact equivalence and thereby reflecting their inherently statistical nature. In Shoup's method, a series of gradually modified games is constructed, each differing from the previous by a small, controlled change. The goal is to argue that these changes do not significantly affect the adversary's ability to distinguish the games. At each step, the adversary's distinguishing advantage is carefully bounded, and through a chain of such transitions, it is shown that the initial and final games are computationally indistinguishable. This step-by-step reasoning enables modular security proofs, where complex systems can be analyzed by reducing them to simpler, incrementally different components. Similarly, our approach relies on an approximate logical relation based on a logical metric, which quantifies the behavioural distance between terms. Cryptographic proofs are expressed as structured sequences of equational steps, each justified by the logical relation to guarantee that the overall transformation preserves computational indistinguishability. While both approaches pursue the same foundational goal, they diverge significantly in structure and scope. Shoup's method is explicitly game-based and adversarial, relying on the presence of an external adversary whose advantage is bounded at each step. In contrast, our approach is semantic and compositional, based on the internal behaviour of programs as captured by the logical metric.

Chapter 10

# Proving the Security of an Encryption Scheme Equationally

This chapter is devoted to the formal modelling and verification of cryptographic constructions within the $\lambda$BLL framework. Our objective is to show how equational reasoning based on the $\Theta$-contextual indistinguishability can be effectively used to capture and verify security properties in higher order cryptographic settings.

We begin in Section 10.1 by introducing a set of function symbols and associated equations that enrich the expressive power of $\lambda$BLL. These function symbols abstract key computational patterns, such as randomness generation, bitstring manipulation, and memory operations, while the equations serve as semantic constraints that describe their behaviour under evaluation. Together, these constructs provide a general and flexible foundation for reasoning about computational effects. In Section 10.2, we apply this extended framework to the formalization of a standard cryptographic proof: the CPA security of an encryption scheme induced by a pseudorandom function $F$. We encode the core components of the reduction-based proof as $\lambda$BLL terms, setting the stage for a formal analysis. Section 10.3 completes the formalization by establishing the CPA security of $\Pi_F$ through equational reasoning. Relying on the equations introduced in Section 10.1, we show how the indistinguishability properties required for the security proof can be derived within the $\lambda$BLL framework. This case study illustrates the applicability of our framework to cryptographic reasoning and highlights the compositional nature of our framework in a higher order setting.

## 10.1 Useful Functional Symbols and Equations in $\lambda$BLL

In this section, we introduce a collection of useful function symbols in $\lambda$BLL, along with a set of equations, some of them involving the introduced function symbols. These function symbols capture key computational patterns such as randomness, bitstring manipulation, and memory operations. The accompanying equations serve as semantic constraints or rewriting rules that specify how the function symbols behave under evaluation, enabling precise and structured reasoning within the $\lambda$BLL framework. Beyond their use in the specific setting addressed in this work, these constructs are designed to be general and versatile, supporting formal modelling in a wide range of contexts where reasoning about effects are required.

| **Function Symbol** | **Type** | **Interpretation** |
|---|---|---|
| **Functional Symbols Generating Random Values** | | |
| `flipcoin` | $\mathbb{B}$ | Function generating a random boolean value. |
| `random` | $\mathbb{S}[i]$ | Function generating a random bitstring in $\{0,1\}^i$. |
| **Function Symbols for Manipulating Bitstrings and Boolean Values** | | |
| `not` | $\mathbb{B} \to \mathbb{B}$ | Function computing the negation of a boolean value. |
| `xor` | $\mathbb{S}[i] \times \mathbb{S}[i] \to \mathbb{S}[i]$ | Function computing the bitwise exclusive-or of two binary strings with the same length. |
| `zero` | $\mathbb{S}[i]$ | Function that returns a bitstring consisting solely of zeros. |
| `concat` | $\mathbb{S}[i] \times \mathbb{S}[i] \to \mathbb{S}[2i]$ | Function computing the concatenation of two binary strings with the same length. |
| **Function Symbols for Ledger Manipulation** | | |
| `initTable`$_p$ | $\mathbb{S}[p \times (2i+1)]$ | Function that initializes a ledger by generating a string of length $p \times (2i+1)$ filled entirely with zeros. |
| `isdefined`$_p$ | $\mathbb{S}[p \times (2i+1)] \times \mathbb{S}[i] \to \mathbb{B}$ | Function that takes as input a ledger and a string of length $i$, and returns true if the row, whose first $i$ bits match the input string, is active (i.e., the last bit of the row is set to 1), and false otherwise. |
| `getValue`$_p$ | $\mathbb{S}[p \times (2i+1)] \times \mathbb{S}[i] \to \mathbb{S}[i]$ | Function that, given a ledger and an input string of length $i$, searches for a row in the ledger whose first $i$ bits are equal to the input string and whose final bit is set to 1 (indicating that the row is active), and returns the substring spanning from position $i+1$ to position $2i$ within the selected row. If no matching active row is found, it returns a dummy value. |
| `modify`$_p$ | $\mathbb{S}[p \times (2i+1)] \times \mathbb{S}[i] \times \mathbb{S}[i] \to \mathbb{S}[p \times (2i+1)]$ | Function that, given a ledger and two input strings of length $i$, identifies the first inactive row in the ledger, inserts into that row the concatenation of the two strings followed by a bit set to 1 (indicating activation), and returns the resulting updated ledger. |

Table 10.1: Table of function symbols and their interpretations. In the last section of the table we consider a ledger modelled by a string of length $p \times (2i+1)$, where $p$ is a polynomial in $\mathbb{N}_{\geq 1}[i]$. This string represents a table with $p$ rows, where each row consists of two strings of length $i$ followed by a single bit, which is set to 1 if the corresponding row is active, and 0 otherwise.

### 10.1.1 Functional Symbols

Table 10.1 defines a collection of function symbols that operate over boolean values and bitstrings, and are particularly designed to support random generation, basic data manipulation, and operations on a specialized structure called *ledger*. These function symbols, while simple in isolation, collectively form a powerful framework capable of modelling both computation and stateful interaction. The symbols in Table 10.1 are grouped into three categories based on their purpose.

**Remark 18** (Interpretation of Function Symbols). In our framework, each function symbol in $\mathcal{F}$ denotes a mapping that yields a probability distribution over values of its output type, as defined in Section 8.2. When we say, then, that one of these function symbols is interpreted as a mapping that returns a specific value $v$, we mean that it yields the Dirac distribution centred at $v$, denoted $\delta_v$, which is the distribution that assigns probability 1 to $v$ and 0 to all other outcomes.

The first group of function symbols encapsulates operations related to randomness, modelling nondeterministic or probabilistic behaviour within the λBLL framework. The symbol `flipcoin` denotes a function that returns a randomly selected boolean value, effectively simulating the outcome of a fair coin toss, yielding either true or false with equal probability. The second symbol, `random`, generates a bitstring of length $i$, where each bit is chosen independently and uniformly at random.

The second group of function symbols provides primitives for direct manipulation of boolean values and bitstrings, enabling fine-grained reasoning over low-level data representations. The function symbol `not` corresponds to standard boolean negation. The `xor` symbol represents the bitwise exclusive-or operation, $\oplus$, applied to two bitstrings of equal length. The `zero` function is a utility that produces a bitstring of a specified length composed entirely of zero bits, often used for initialization or padding purposes. Finally, `concat` combines two bitstrings of equal length into one longer bitstring by appending one to the other, resulting in a string twice as long. These symbols collectively support structured and compositional modelling of bit-level computations within the λBLL framework.

The last and most complex set of symbols deals with a ledger, a structure that behaves like a table. The ledger is represented by a flat binary string segmented into rows of length $2i + 1$. Each row is made up of two bitstrings of length $i$ and an additional bit that indicates whether the row is active (set to 1) or inactive (set to 0). The total size of the ledger is defined to accommodate $p$ such rows, where $p$ is a polynomial in $\mathbb{N}_{\geq 1}[i]$. It is crucial to ensure that the number of rows in the ledger is polynomially bounded. Otherwise, the resulting string could exceed the length limits imposed by λBLL, which allows only strings of polynomial length as defined by the grammar of ground types in Figure 8.1.

The function symbol `initTable` constructs an empty ledger, returning a bitstring initialized entirely with zeros, representing a table with no defined entries. The `isdefined` symbol allows us to check whether a given input string already exists in the ledger. It scan the ledger to determine whether there exists a row whose first $i$ bits match the input string and whose last bit is set to 1. If such a row is found, the input string is considered to be defined within the ledger and the function symbol returns true. The function `getValue` retrieves the value associated with a given input string in the ledger. It searches for an active row whose first $i$ bits match the input string. If such a row exists, the function symbol returns the associated value, i.e., the substring spanning from position $i + 1$ to position $2i$ within the selected row. The function is thus partial, and it returns a dummy value when no matching active row is found. Lastly, `modify` enables the insertion of a new pair into the ledger. It searches for the first inactive row, writes the input key and value into the respective segments, and sets the final bit to 1, marking the row as active. Through repeated applications of `modify`, the ledger can evolve dynamically, supporting incremental updates in a formally tractable manner.

In order to illustrate how these functions operate, we explain their behaviour through the following example.

**Example 4.** Consider a ledger of type $\mathbb{S}[p \times (2i+1)]$, where the polynomial $p$ is defined as $3i$ and we set $i$ to 4. This gives us a bitstring of length $3i \times (2i+1) = 12 \times 9 = 108$. As described earlier, this ledger can be seen as a table consisting of 12 rows, where each row contains two binary strings of length 4, followed by a single bit indicating whether the row is active (1) or inactive (0).

From a data structure perspective, the ledger corresponds to a dictionary or associative array structure. In this context, the first two columns form a key-value pair: the first binary string serves as the key, and the second as its associated value. The final bit in each row functions as a tag or status indicator, marking whether the key-value pair is currently active. This analogy highlights how the formal ledger representation aligns with common data structures used in computer science to store collections of key-value pairs along with additional metadata.

We begin by initializing the ledger using the function `initTable`. This produces a table, denoted as $tab_0$, where all entries are zero, as shown in Figure 10.1a.

Next, we populate the ledger step by step using the function symbol `modify`. We first insert the pair $(0100, 1000)$ into $tab_0$, producing $tab_1$. Then, we insert the pair $(0010, 1100)$ into $tab_1$ to get $tab_2$, and finally we add $(0000, 0000)$ to obtain $tab_3$. The resulting table, $tab_3$, is illustrated in Figure 10.1b. Observe that the third row of $tab_3$, corresponding to the last inserted pair, has its tag set to 1, indicating that this entry is active. In contrast, the subsequent row has its tag set to 0, marking it as inactive. This tagging mechanism clearly distinguishes active entries from inactive ones within the ledger.

| | **Key** | **Value** | **Tag** |
|---|---|---|---|
| **1** | 0000 | 0000 | 0 |
| **2** | 0000 | 0000 | 0 |
| ... | | ... | |
| **12** | 0000 | 0000 | 0 |

(a) Initial empty ledger $tab_0$ returned by `initTable`.

| | **Key** | **Value** | **Tag** |
|---|---|---|---|
| **1** | 0100 | 1000 | 1 |
| **2** | 0010 | 1100 | 1 |
| **3** | 0000 | 0000 | 1 |
| **4** | 0000 | 0000 | 0 |
| ... | | ... | |
| **12** | 0000 | 0000 | 0 |

(b) Ledger $tab_3$ after applying three successive `modify` operations.

Figure 10.1: Evolution of a ledger from initialization to populated state.

In order to determine whether the key 0000 is present in the ledger $tab_3$, we evaluate the function symbol `isdefined`$(tab_3, 0000)$, which returns true, since an active row with key 0000 exists in the table. Specifically, the function scans the table in Figure 10.1b from top to bottom, examining each row. It first encounters the key 0100 in the first row and 0010 in the second, both active but not matching 0000. Upon reaching the third row, it finds the key 0000 with the active flag set to 1. At this point, the function stops and returns true, confirming that the key is defined in the ledger. Similarly, we can retrieve the value associated with the key 0010 by evaluating `getValue`$(tab_3, 0010)$. The function performs a linear scan of the table and checks each active row. It skips the first row (0100) and then matches the key 0010 in the second row, where the value is 1100 and the tag is 1. Since the row is active and the key matches, the function returns 1100 as result.

Functions like `flipcoin` and `random` are particularly useful in modelling randomized algorithms or systems with nondeterministic behaviour. This allows formal reasoning tools to explore all possible branches or equivalence classes that arise due to randomness, which is essential when verifying the security properties of cryptographic protocols or the fairness of probabilistic algorithms.

Bitstring manipulation is likewise essential for formal models of computation, particularly in domains where low-level data representation matters, such as cryptographic algorithms and protocol analysis. In computational cryptography, bitstrings serve as the fundamental representation for keys, ciphertexts, nonces, and other core elements. Therefore, operations on bitstrings are crucial for modelling encryption schemes, hash functions, and digital signatures, as well as for analyzing their security properties within formal frameworks.

The ledger introduces a formalized notion of state and memory within the system, providing a structured representation of dynamic data. The representation of the ledger as a flat sequence of rows provides a compact and analyzable representation of memory. The structure of the ledger can be exploited to model a random function – a theoretical construct in which each unique input is deterministically assigned to a random output upon first use. Specifically, the `isdefined` function symbol can be used to check if a value has already been associated with a given input, and `modify` to record new input-output pairs. In this way, the ledger can emulate the behaviour of such a function in a consistent and analyzable manner. This is exactly how the ledger will be employed in our model.

### 10.1.2 Equations

The equations in Figure 10.2 define a set of equivalences between terms in the language λBLL, which are grounded in the notion of Θ-contextual indistinguishability, formally introduced earlier in Section 9.2 and denoted by $\sim_{\Theta\subseteq\Xi}$. Intuitively, these equations specify when two terms are considered behaviourally equivalent in any context that cannot distinguish between them based on a specified set of observable references. In the following, we describe some key equations from Figure 10.2 to illustrate their role and interpretation, as well as their versatility in a wide range of contexts. Please note that, since Θ-contextual indistinguishability is an open Ξ-term relation, the equations presented concern both the terms and their associated typing judgments. For the sake of simplicity Figure 10.2 shows only the terms.

**Safe Discarding.** The safe discarding equations in Figure 10.2, namely `safeDisc1`, `safeDisc2` and `safeDisc3`, describe how terms in the language λBLL can be reduced to simpler terms under certain conditions. These equations ensure that reductions preserve the context and behaviour of the terms, while adhering to satisfying safety conditions related to the references accessed by the involved terms.

Before delving into the specifics of each safe discarding equation, we first outline the additional conditions they impose on the references in Θ. These conditions play a key role in ensuring that the involved terms satisfy Θ-contextual indistinguishability, as defined in Definition 13.

The condition $\mathcal{R}(M) \cap \Theta = \emptyset$ ensures that the term $M$, which will be removed by the equations, does not access any references from the context Θ. Since Θ represents the set of references accessible to the observer (see Remark 15), allowing $M$ to access references in Θ and applying an equation that eliminates $M$ could modify the state observable by the context, thus violating equivalence. This condition prevents such discrepancies and maintains equivalence. For instance, by omitting this condition `safeDisc1` would become $\texttt{let}\, x = \texttt{set}\, r\, V\, \texttt{in}\, N \sim_{\Theta\subseteq\Xi} N$ with $r \in \Theta$. However, this violates Θ-contextual indistinguishability because the term $M$ (i.e., $\texttt{set}\, r\, V$) modifies the state of Θ by assigning $V$ to the reference $r$. This produces observable

$$\texttt{let}\,x = M\,\texttt{in}\,N \sim_{\Theta\subseteq\Xi} N$$
$$\text{if } x \notin FV(N) \text{ and } \mathcal{R}(M) \cap \mathcal{R}(N) = \emptyset \text{ and} \qquad (\texttt{safeDisc1})$$
$$\mathcal{R}(M) \cap \Theta = \emptyset \text{ and } HOFV(M) = \emptyset$$

$$\texttt{let}\,\langle x, y\rangle = M\,\texttt{in}\,N \sim_{\Theta\subseteq\Xi} N$$
$$\text{if } x, y \notin FV(N) \text{ and } \mathcal{R}(M) \cap \mathcal{R}(N) = \emptyset \text{ and} \qquad (\texttt{safeDisc2})$$
$$\mathcal{R}(M) \cap \Theta = \emptyset \text{ and } HOFV(M) = \emptyset$$

$$\begin{pmatrix}\texttt{let}\,b = \texttt{flipcoin}\,\texttt{in} \\ \texttt{if}\,b\,\texttt{then}\,M\,\texttt{else}\,(\texttt{let}\,z = M\,\texttt{in}\,\texttt{not}(z))\end{pmatrix} \sim_{\Theta\subseteq\Xi} \texttt{flipcoin} \qquad (\texttt{safeDisc3})$$
$$\text{if } \mathcal{R}(M) \cap \Theta = \emptyset \text{ and } HOFV(M) = \emptyset$$

$$\begin{pmatrix}\texttt{if}\ b\ \texttt{then}\ \texttt{let}\,x = M\,\texttt{in}\,N \\ \texttt{else}\ \texttt{let}\,x = L\,\texttt{in}\,N\end{pmatrix} \sim_{\Theta\subseteq\Xi} \texttt{let}\,x = (\texttt{if}\,b\,\texttt{then}\,M\,\texttt{else}\ L)\,\texttt{in}\,N \quad (\texttt{ifLet})$$

$$\texttt{let}\,x_1 = M_1\,\texttt{in}\,\texttt{let}\,x_2 = M_2\,\texttt{in}\,N \sim_{\Theta\subseteq\Xi} \texttt{let}\,x_2 = M_2\,\texttt{in}\,\texttt{let}\,x_1 = M_1\,\texttt{in}\,N$$
$$\text{if } x_i \notin FV(M_{3-i}) \text{ for } i \in \{1,2\},\ \ \mathcal{R}(M_1) \cap \mathcal{R}(M_2) = \emptyset \qquad (\texttt{letCom})$$
$$\text{and } HOFV(M_1) \cup HOFV(M_2) = \emptyset$$

$$\texttt{let}\,x = M_1\,\texttt{in}\,\texttt{let}\,y = M_2\,\texttt{in}\,N \sim_{\Theta\subseteq\Xi} \texttt{let}\,y = (\texttt{let}\,x = M_1\,\texttt{in}\,M_2)\,\texttt{in}\,N \quad (\texttt{letAss})$$
$$\text{if } x \notin FV(N)$$

$$\texttt{let}\,x = M\,\texttt{in}\,\texttt{return}\,x \sim_{\Theta\subseteq\Xi} M \qquad (\texttt{reduc1})$$

$$(\texttt{return}\,\lambda x.M)Z \sim_{\Theta\subseteq\Xi} M[Z/x] \qquad (\texttt{reduc2})$$

$$\texttt{der}(!M) \sim_{\Theta\subseteq\Xi} M \qquad (\texttt{reduc3})$$

$$\texttt{let}\,x = \texttt{return}\,V\,\texttt{in}\,M \sim_{\Theta\subseteq\Xi} M[V/x] \qquad (\texttt{reduc4})$$

$$\texttt{let}\,\langle x, y\rangle = \langle Z, Z'\rangle\,\texttt{in}\,M \sim_{\Theta\subseteq\Xi} M[Z/x, Z'/y] \qquad (\texttt{reduc5})$$

$$\texttt{let}\,x = \texttt{random}\,\texttt{in}\,\texttt{xor}(x, y) \sim_{\Theta\subseteq\Xi} \texttt{random} \qquad (\texttt{randXor})$$

$$\begin{pmatrix}\texttt{let}\,x = \texttt{random}\,\texttt{in} \\ \texttt{let}\,y = \texttt{random}\,\texttt{in} \\ \texttt{concat}(x, y)\end{pmatrix} \sim_{\Theta\subseteq\Xi} \texttt{random} \qquad (\texttt{randConcat})$$

$$\begin{pmatrix}\texttt{let}\,r = \texttt{random}\,\texttt{in} \\ \texttt{let}\,z = \texttt{isdefined}(y, r)\,\texttt{in} \\ \texttt{if}\,z\,\texttt{then}\,M\,\texttt{else}\ N\end{pmatrix} \sim_{\Theta\subseteq\Xi} \texttt{let}\,r = \texttt{random}\,\texttt{in}\,N \qquad (\texttt{randT})$$

Figure 10.2: Equations in $\lambda$BLL. Let $\mathcal{R}(M)$ be the set of references accessed by the term $M$. The equations in this figure hold for all reference contexts $\Theta, \Xi$ such that $\Theta \subseteq \Xi$, except where stated otherwise.

effects depending on the state of $\Theta$, allowing an observer which has access to $\Theta$ to distinguish between the terms, thus violating contextual equivalence.

The condition that the free variables in $M$, which will be removed by the equations, do not include higher-order variables ($HOFV(M) = \emptyset$), is crucial for preserving contextual equivalence, as these higher-order variables might modify references and potentially altering the behaviour of the term from an observer's perspective. For instance, a higher-order variable could modify references in $\Theta$, making the term's behaviour dependent from the context $\Theta$, since it contains observable references (see Remark 15). By excluding higher-order free variables, we ensure that the term remains independent from the context, thereby preserving $\Theta$-contextual indistinguishability and avoiding violations of consistency.

Collectively, these conditions constrain the behaviour of the term $M$ to ensure that its elimination during the equational step does not introduce observable effects involving references in $\Theta$. By requiring that $M$ neither accesses references in $\Theta$ nor depends on higher-order variables that might do so, these conditions help preserve $\Theta$-contextual indistinguishability and uphold consistency across all contexts.

Consider Equation `safeDisc1`, the Equation `safeDisc2` being an extension to pairs of variables. Equation `safeDisc1` deals with a term where we bind a variable $x$ to a term $M$. Specifically, the equation tells us that we can eliminate the binding and simply reduce the term to $N$, but only if the conditions are satisfied. First, $x$ should not appear as a free variable in $N$, meaning that removing the binding won't change how $N$ behaves. Additionally, the second condition is $\mathcal{R}(M) \cap \mathcal{R}(N) = \emptyset$, meaning that the term $M$ and the term $N$ must not share any references. This is important because if they did share references, reducing the term could lead to unexpected side effect. In our framework, shared references could lead to observable behaviour that may be distinguishable within a context, potentially breaking the contextual equivalence $\sim_{\Theta \subseteq \Xi}$. These conditions along with the additional conditions described above ensure the term remains safe to reduce, making the simplification process both efficient and reliable. This is especially beneficial in scenarios such as memory management and compiler optimizations, where removing unnecessary bindings or computations can significantly improve efficiency without compromising the correctness of the program.

The Equation `safeDisc3` captures a decision-making process driven by a random event, such as the outcome of a fair coin toss. It asserts that any conditional, governed by a boolean generated by flipping a fair coin, where the branches return a boolean value computed by a term $M$ or its negation, can be replaced by the coin flip itself, without needing to explicitly handle the decision branches. In more detail, the expression on the left-hand side of Equation `safeDisc3` performs a random choice by evaluating $b = \texttt{flipcoin}$, and then returns either the boolean result of evaluating $M$ or its negation, depending on the outcome of the coin toss. Specifically, if $b$ is true, it performs $M$; otherwise, it performs $\texttt{let}\, z = M\, \texttt{in}\ \not{M}$. Since `flipcoin` produces a fair and unbiased outcome, both branches are selected with equal probability. As a result, regardless of the actual value of $M$, the overall expression always yields a uniformly random boolean. This is because one branch returns the output of $M$, and the other returns its logical opposite, so exactly one of the two possible boolean values will be produced, each with probability $\frac{1}{2}$. Therefore, the entire construction behaves exactly like as `flipcoin`, making the two terms indistinguishable. More specifically, this holds in our framework only if the conditions restricting the behaviour of $M$, described earlier, are satisfied.

**Equation randT.** The equation `randT` captures a common pattern, such as testing whether a freshly generated random value already exists in a data structure. In this equation, a fresh random value $r$ is generated and immediately checked against a ledger $y$ using the `isdefined` function symbol. Since $r$ is sampled uniformly at random and independently of the contents of the ledger using `random`, the likelihood that it has already been defined is negligible in practice. Therefore, the test $\texttt{isdefined}(y, r)$ evaluates to false, due to the independence and uniqueness of freshly generated randomness. As a result, the conditional selects the else-branch with high probability, and the entire expression simplifies to directly executing $N$ after generating $r$.

The versatility of this equation stems from its ability to model situations where random values need to be generated and then checked for uniqueness within a given context. More precisely, terms like the left-hand side of Equation randT are common in a variety of computational settings, from cryptographic protocols to gaming systems, databases, and beyond. For instance, imagine a system that generates a unique session token for each new user session. The system would first generate a random token and then check if this token has already been used. If the token is already in use, the system would retry with a new one; otherwise, it would proceed with creating the session. In cryptographic protocols, nonces (numbers used once) must be generated randomly and checked to ensure they haven't been used previously. Similarly, in a game system, a random number might be drawn, and if that number has been drawn before, the system retries until it finds a unique one.

**Equations for Random Bitstring Manipulation.** The $\lambda$BLL language includes equations that simplify operations involving random bitstrings, capturing key principles of uniform randomness under transformation.

The randXor equation in the $\lambda$BLL language simplifies terms involving the bitwise exclusive-or operation between a random bitstring value and another bitstring value of the same length. Specifically, it states that the term $\texttt{let}\, x = \texttt{random}\, \texttt{in}\, \texttt{xor}(x, y)$ is contextually indistinguishable from just `random`. This is because performing bitwise exclusive-or between a random value $x$ and any other value $y$ does not change the randomness of $x$; the result remains a random value. This equation is particularly useful in cryptographic and randomized algorithms.

Equation randConcat formalizes the intuition that concatenating two independent random bitstrings produces a longer bitstring that remains uniformly random. As a result, the combined operation can be abstracted by a single invocation of the `random` function symbol, adjusted to reflect the new length. Note that Equation randConcat is well-defined because, in our calculus, function symbols are associated with a term constructors indexed by a polynomial $p$ in $\mathbb{N}_{\geq 1}[i]$. On the left-hand side of the equation, the constructor for `random` is parameterized by $i$, while on the right-hand side, it is parameterized by $2i$, reflecting the length of the concatenated bitstrings.

## 10.2 Modelling the CPA Security Components of $\Pi_F$

For the sake of readability, in the last two sections of this chapter we introduce the following syntactic sugar:

$$VZ \triangleq (\texttt{return}\, V)Z \tag{10.1}$$

In $\lambda$BLL, application is written as $MZ$, and we use `return` to lift values into computations. This notation allows us to simplify expressions by conceptually promoting values to computations, thereby reducing the need for deeply nested `return` construct.

Consider the standard proof by reduction that establishes CPA security for the encryption scheme $\Pi_F$, induced by a pseudorandom function $F$, as outlined in Theorem 16. In this section we first introduce the $\lambda$BLL model for the CPA-experiment, as discussed in Subsection 7.1. Next, we present a model for the components of the proof related to the CPA security of $\Pi_F$. Specifically, we instantiate the $\texttt{PrivK}^{CPA}$ experiment, defined earlier, for the schemes $\Pi_F$ and $\widehat{\Pi}$, and we give the term for the distinguisher $D_A$ using an unspecified adversary $A$ as a subroutine.

### 10.2.1 Model for CPA-Experiment

The first step is to formalize the desired security property, in our case the CPA security, and to do so we model in $\lambda$BLL the experiment discussed in Subsection 7.1, that we report again in Figure 10.3. Let us consider the experiment $\mathtt{PrivK}^{CPA}_{A,\Pi}$ for a generic encryption scheme $\Pi$. First,

$$
\begin{array}{l}
\mathtt{PrivK}^{CPA}_{A,\Pi}(n): \\
\quad k \leftarrow Gen(1^n) \\
\quad m_0, m_1 \leftarrow A^{Enc_k(\cdot)}(1^n) \\
\quad b \leftarrow \{0,1\} \\
\quad c \leftarrow Enc(k, m_b) \\
\quad g \leftarrow A^{Enc_k(\cdot)}(c) \\
\quad \mathtt{return}\ (b = g)
\end{array}
$$

Figure 10.3: Pseudocode for $\mathtt{PrivK}^{CPA}$ Experiment.

we define the types for all the components of the experiment (encryption scheme, adversary and oracle) that appear as subterms. The implementation of the terms modelling the algorithms *Enc* and *Gen* and the oracle $Enc_k(\cdot)$ is closely tied to the encryption scheme being analyzed. Therefore, the instantiation of these terms will be made explicit only after the encryption scheme $\Pi$ has been fixed. In the next paragraph, we will instantiate these terms for the schemes $\Pi_F$ and $\widehat{\Pi}$, which are involved in the proof of Theorem 16. Recall that in our calculus the security parameter is modelled by a polynomial variable $i$. The terms *Enc* and *Gen* are typed as follows

$$\vdash^{\Theta}_{v} Enc : (\mathbb{S}[i] \otimes \mathbb{S}[i]) \multimap \mathbb{S}[2i] \qquad\qquad \vdash^{\Theta}_{v} Gen : \mathbb{U} \multimap \mathbb{S}[i]$$

Please observe that the output of *Enc* is a string of length $2i$ because we are modelling $\Pi_F$, where the ciphertext consists of two strings, each of type $\mathbb{S}[i]$: a random value and the encrypted message. Since both parts have length $i$, the ciphertext is twice as long as the original message.

The oracle $Enc_k(\cdot)$ of Figure 10.3 is modelled by a term that takes a key and a message as input and returns the ciphertext obtained by calling the term which models the encryption algorithm of $\Pi$. Formally, the term *Oracle* is typed as $\vdash^{\Theta}_{c} Oracle : \mathbb{S}[i] \multimap\ !_{q+1}(\mathbb{S}[i] \multimap \mathbb{S}[2i])$ where $q$ is a polynomial in $\mathbb{N}_{\geq 1}[i]$ that limits the numbers of interactions between the adversary and the oracle. The bang in the *Oracle* type is indexed by the polynomial $q+1$, this is closely tied to the way we will model the experiment later (see Figure 10.4 and Remark 20).

It is worth noting that the type derivations of the terms introduced so far are parameterized by a reference context $\Theta$, which varies depending on the encryption scheme under consideration. Anticipating what will be discussed later, the reference context $\Theta$ will have different content in the model for the encryption scheme $\Pi_F$ compared to the one used in the model for $\widehat{\Pi}$. More precisely, in the former case it will be empty, whereas in the latter it will contain a reference used to model the random function $f$.

In the experiment for CPA security shown in Figure 10.3, the adversary's behaviour is divided into two main phases. These phases are modelled by two distinct $\lambda$BLL terms with the following types:

$$
\begin{array}{l}
\vdash^{\varnothing}_{v} Adv_1 :!_{q_1}(\mathbb{S}[i] \multimap \mathbb{S}[2i]) \multimap ((\mathbb{S}[i] \otimes \mathbb{S}[i]) \otimes \mathbb{S}[r]) \\
\vdash^{\varnothing}_{v} Adv_2 : (!_{q_2}(\mathbb{S}[i] \multimap \mathbb{S}[2i]) \otimes \mathbb{S}[r]) \otimes \mathbb{S}[2i] \multimap \mathbb{B}
\end{array}
$$

where $q_1$, $q_2$ and $r$ are polynomials in $\mathbb{N}_{\geq 1}[i]$.

**Remark 19.** Since we have modelled the two phases of the adversary with distinct terms, we assume that the polynomial $q$ indexing the type of the oracle is equal to $q_1 + q_2$, as it bounds the total number of interactions between the oracle and the adversary. This reflects the fact that, for a given security parameter $n$, the adversary may perform at most $q_1[n/i]$ queries in the first phase and at most $q_2[n/i]$ queries in the second phase resulting in a total of at most $q[n/i]$ queries to the oracle.

The first phase is modelled by the term $Adv_1$, where the adversary is given access to an encryption oracle of type $!_{q_1}(\mathbb{S}[i] \multimap \mathbb{S}[2i])$. Based on its interactions with the oracle, the adversary derives relevant information, such as patterns in the ciphertexts or other characteristics of the encryption process, and generates two messages, $m_0$ and $m_1$. This term returns the two messages along with a third string of type $\mathbb{S}[r]$ that encodes the internal state of the adversary, including the informations it has gathered during its interaction with the oracle. It is important to observe that the string encoding the adversary's internal state has length bounded by the polynomial $r$, as the adversary operates within probabilistic polynomial time (see Section 8.4 and Remark 14) and, consequently, cannot generate strings whose length exceeds a polynomial bound.

The term $Adv_2$ models the second phase by taking as input the encryption oracle, the ciphertext, and the string produced in the first phase that encodes the adversary's internal state. The goal of this phase is to analyze the ciphertext and determine which of the two previously generated messages it corresponds to. To do this, the adversary continues to interact with the oracle, using the information encoded in the state string to support its analysis of the ciphertext. The term returns $\mathbf{t}$ (resp. $\mathbf{f}$) if it guesses that the message encrypted by the experiment is $m_1$ (resp. $m_0$).

The precise implementations of $Adv_1$ and $Adv_2$ are not known, however, the calculus is complete and so it can express all polytime functions.

Finally, we define the term for the experiment as in Figure 10.4. After generating a fresh key $k$ and building the encryption oracle $o$ with it, the experiment proceeds by first allowing $Adv_1$ the possibility of generating two distinct messages $m_0$ and $m_1$ (see lines 3-4 of Figure 10.4). Then, the uniform sampling from the set $\{0, 1\}$ in Figure 10.3 is replaced in our model by the function symbol `flipcoin`, with $\mathsf{typeof}(\texttt{flipcoin}) = \mathbb{B}$, interpreted as a map computing a random boolean $b$. At the end, the last three lines of Figure 10.3 are modelled using an if-then-else construct based on the value of $b$ and using the term $Adv_2$, which models the second phase of the adversary.

In order to emphasize the correspondence with the pseudocode given in Figure 10.3, we apply the syntactic sugar from Equation 10.1 in the first line of the term shown in Figure 10.4, since *Gen* is a value. In the third line, we additionally use the following syntactic sugar:

$$\texttt{let}\,\langle x, y\rangle = M\,\texttt{in}\,N \triangleq \texttt{let}\,w = M\,\texttt{in}\,\texttt{let}\,\langle x, y\rangle = w\,\texttt{in}\,N \tag{10.2}$$

Furthermore, the syntactic sugar from Equation 10.1 is also applied in the if-then-else construct (last two lines of Figure 10.4), where the `return` is omitted when applying values to messages in both branches.

**Remark 20** (Typing of *PrivKCPA*)**.** Recall that the term *Oracle* is of type $\mathbb{S}[i] \multimap\ !_{q+1}(\mathbb{S}[i] \multimap \mathbb{S}[2i])$, so the variable $o$ in *PrivKCPA* is of type $!_{q+1}(\mathbb{S}[i] \multimap \mathbb{S}[2i])$. It is important to observe that the polynomial on which the type of $o$ is indexed must be $q + 1$ where $q = q_1 + q_2$, as described in Remark 19. This arises from the fact that the oracle is invoked at most a polynomial number $q_1$ of times by the fist phase of the adversary ($Adv_1$ in Line 3 of Figure 10.4), a polynomial

$$
\begin{aligned}
&\mathit{PrivKCPA} \triangleq\\
&\quad \texttt{let}\, k = \mathit{Gen} * \,\texttt{in}\\
&\quad \texttt{let}\, o = \mathit{Oracle}\ k\,\texttt{in}\\
&\quad \texttt{let}\, \langle \mathit{msgs}, s\rangle = \mathit{Adv}_1\ o\,\texttt{in}\\
&\quad \texttt{let}\, \langle m_0, m_1\rangle = \mathit{msgs}\,\texttt{in}\\
&\quad \texttt{let}\, b = \texttt{flipcoin}\,\texttt{in}\\
&\quad \texttt{if}\ b\ \texttt{then}\ (\lambda m.\texttt{let}\, c = (\texttt{der}(o)\ m)\,\texttt{in}\, \mathit{Adv}_2\langle\langle o, s\rangle, c\rangle)m_1\\
&\quad \texttt{else let}\, z = (\lambda m.\texttt{let}\, c = (\texttt{der}(o)\ m)\,\texttt{in}\, \mathit{Adv}_2\langle\langle o, s\rangle, c\rangle)m_0\,\texttt{in}\,\texttt{not}(z)
\end{aligned}
$$

Figure 10.4: $\lambda$BLL Term for $\texttt{PrivK}^{CPA}$ Experiment.

number $q_2$ of times by the second phase of the adversary ($\mathit{Adv}_2$ in Figure 10.4) and once by the experiment to create the ciphertext $c$ (see the last two lines of Figure 10.4).

### 10.2.2 Model for the Components of the Proof

Before formalizing the proof of security, we have to model its components. More precisely, we have to define a $\lambda$BLL term for the experiment using the encryption scheme $\Pi_F$ (called $\texttt{PrivK}^{CPA}_{A,\Pi_F}$) and one for the experiment using the encryption scheme $\widehat{\Pi}$ (called $\texttt{PrivK}^{CPA}_{A,\widehat{\Pi}}$). Furthermore, it is essential to define a $\lambda$BLL term able to model the distinguisher $D_A$ used in the proof by reduction.

In both the cryptographic experiment and the distinguisher, we need to construct two instances of the $\lambda$BLL term that models these components, namely one that interacts with the pseudorandom function $F$ and one that interacts with a truly random function $f$. Both $F$ and $f$ can be modelled in $\lambda$BLL, but while the former can be taken as a term which does not use any reference, the latter can only be captured by a stateful computation — one cannot hope to pick uniformly at random a function on $n$-bit strings in polynomial time in $n$ without the help of some bookkeeping mechanism. The latter will actually be implemented by a *ledger*, as described in Section 10.1, whose purpose is to keep track of the previous strings on which the function has been queried, so that randomness can be generated *only when needed*. Since the type of $f$ reflects the presence of ledger, the type of the components interacting with $f$ must be updated accordingly, as we are going to describe in the following.

The model for $\texttt{PrivK}^{CPA}_{A,\Pi_F}$, which we call $\mathit{PrivKCPA}^F$, is obtained instantiating the term $\mathit{PrivKCPA}$ in Figure 10.4 with $\mathit{Enc}^F$, $\mathit{Gen}^F$ and $\mathit{Oracle}^F$ as given in Figure 10.5. Specifically, the key generation algorithm of $\Pi_F$ is modelled by the term $\mathit{Gen}^F$, which generates the key by using the function symbol $\texttt{random}$, with $\texttt{typeof}(\texttt{random}) = \mathbb{S}[i]$. A pseudorandom function can be instantiated in various ways, including constructions based cryptographic hash functions [22], block ciphers such as AES [80] or pseudorandom generators as in [56]. In the following, we model the pseudorandom function by introducing a generic value $F$ typed as $\vdash^{\varnothing}_{v} F : \mathbb{S}[i] \otimes \mathbb{S}[i] \multimap \mathbb{S}[i]$, reflecting the fact that the encryption scheme $\Pi_F$ is parameterized by an arbitrary pseudorandom function $F$. The encryption algorithm of $\Pi_F$ is modelled by the term $\mathit{Enc}^F$. It generates the random value $r$ using the function symbol $\texttt{random}$ and computes the pseudorandom value $x$ by querying the term $F$ on the key $k$ and the random string $r$. Then, it proceeds by combining the message $m$ and the pseudorandom value $x$ using the function symbol $\texttt{xor}$. This step creates a transformed version of the message, denoted by $y$. Finally, the term concatenates the random

$$
\begin{array}{ll}
 & Enc^F \triangleq \\
 & \quad \lambda e. \\
 & \quad \texttt{let}\, \langle k, m\rangle = e\, \texttt{in} \\
 & \quad \texttt{let}\, r = \texttt{random}\, \texttt{in} \\
Gen^F \triangleq & \quad \texttt{let}\, x = F\langle k, r\rangle\, \texttt{in} \\
\quad \lambda u. & \quad \texttt{let}\, y = \texttt{xor}(x, m)\, \texttt{in} \\
\quad \texttt{let}\, k = \texttt{random}\, \texttt{in} & \quad \texttt{let}\, c = \texttt{concat}(r, y)\, \texttt{in} \\
\quad \texttt{return}\, k & \quad \texttt{return}\, c
\end{array}
$$

$$Oracle^F \triangleq \texttt{return}\,(\lambda k.\texttt{return}\,!(\texttt{return}\,\lambda m.Enc^F\langle k, m\rangle))$$

Figure 10.5: Models for subterms of $PrivKCPA^F$

value $r$ and the transformed message $y$ by using the function symbol `concat`, and returns the resulting concatenated output.

As previously mentioned, it is worth noting that the terms modelling the algorithms of the encryption scheme $\Pi_F$, formally defined in Figure 10.5, do not use any references, and therefore the reference context $\Theta$ under which they are typed is empty.

Note also that in the term $Oracle^F$ in Figure 10.5, we use the syntactic sugar from Equation 10.1, as $Enc^F$ is a value.

The model for $\texttt{PrivK}^{CPA}_{\mathcal{A},\widehat{\Pi}}$, which we call $\widehat{PrivKCPA}$, is obtained from the term $PrivKCPA$ in Figure 10.4 instantiating the terms $Enc$, $Gen$ and $Oracle$ according to $\widehat{\Pi}$, see Figure 10.6. The models for these terms are more intricate, as the encryption scheme $\widehat{\Pi}$ uses a random function $f$ as key. This random function is characterized by the property that each distinct input is deterministically associated with a uniformly random output upon its first use, and that this output remains fixed for all subsequent evaluations of $f$ on the same input. In order to model this behaviour within our framework, we rely on the notion of ledger which can be manipulated using several function symbols, as described in Section 10.1. In Figure 10.6 the ledger is stored in the reference $rand$ of type $\mathbb{S}[(q+1)\times(2i+1)]$ to keep track of the random string associated to each possible value of $r$. This reference can be viewed as a table with $q+1$ rows, where each row contains two strings of length $i$ and a final bit indicating whether the row is active (set to 1) or inactive (set to 0). Let us observe that the function $f$, which is the oracle in the encryption scheme $\widehat{\Pi}$, can be queried at most $q+1$ times — once by the experiment and up to a polynomial $q$ number of times by the adversary. The size of the ledger reflects this constraint, as it contains exactly $q+1$ rows, corresponding to the maximum number of distinct values of $r$ on which $f$ can be queried.

It is important to observe that, in order to model the random function $f$, the terms representing the algorithms of the encryption scheme $\widehat{\Pi}$ operate on the ledger stored in the reference $rand$. Consequently, the types of the terms in Figure 10.6 must be adapted by defining a reference context $\Upsilon = \{rand : \mathbb{S}[(q+1)\times(2i+1)]\}$ and setting the context $\Theta$ in the type derivation of these terms to this reference context.

The key generation algorithm of $\widehat{\Pi}$ is modelled by the term $\widehat{Gen}$. It initializes the table representing the ledger by setting the value of the reference $rand$ to the output of the function symbol `initTable`, which generates a string of length $(q+1)\times(2i+1)$ filled entirely with zeros. Finally, it returns a dummy value generated using the function symbol `zero`, as the key will not

$$
\begin{array}{l}
\widehat{Gen} \triangleq \\
\quad \lambda u. \\
\quad \texttt{let}\, tab = \texttt{initTable}\,\texttt{in} \\
\quad \texttt{let}\, - = \texttt{set}\, rand\, tab\, \texttt{in} \\
\quad \texttt{let}\, k = \texttt{zero}\,\texttt{in} \\
\quad \texttt{return}\, k
\end{array}
\qquad
\begin{array}{l}
\widehat{Enc} \triangleq \\
\quad \lambda e. \\
\quad \texttt{let}\, \langle k, m\rangle = e\, \texttt{in} \\
\quad \texttt{let}\, tab = \texttt{get}\, rand\, \texttt{in} \\
\quad \texttt{let}\, r = \texttt{random}\, \texttt{in} \\
\quad \texttt{let}\, test = \texttt{isdefined}(tab, r)\, \texttt{in} \\
\quad \texttt{if}\ test\ \texttt{then} \\
\quad\ \ \texttt{let}\, x = \texttt{getValue}(tab, r)\, \texttt{in} \\
\quad\ \ \texttt{let}\, y = \texttt{xor}(x, m)\, \texttt{in} \\
\quad\ \ \texttt{let}\, c = \texttt{concat}(r, y)\, \texttt{in} \\
\quad\ \ \texttt{return}\, c \\
\quad \texttt{else} \\
\quad\ \ \texttt{let}\, x = \texttt{random}\, \texttt{in} \\
\quad\ \ \texttt{let}\, y = \texttt{xor}(x, m)\, \texttt{in} \\
\quad\ \ \texttt{let}\, newtab = \texttt{modify}(tab, r, x)\, \texttt{in} \\
\quad\ \ \texttt{let}\, - = \texttt{set}\, rand\, newtab\, \texttt{in} \\
\quad\ \ \texttt{let}\, c = \texttt{concat}(r, y)\, \texttt{in} \\
\quad\ \ \texttt{return}\, c
\end{array}
$$

$$\widehat{Oracle} \triangleq \texttt{return}\,(\lambda k.\texttt{return}\,!(\texttt{return}\,\lambda m.\widehat{Enc}\langle k, m\rangle))$$

Figure 10.6: Models for subterms of $\widehat{PrivKCPA}$.

be used in the context of $\widehat{PrivKCPA}$.

The term $\widehat{Enc}$ models the encryption procedure of $\widehat{\Pi}$ built from a random function using a ledger. It takes as input a key-message pair $\langle k, m\rangle$, generates a fresh random value $r$, and checks, using `isdefined`, whether $r$ has already been used to query the random function. If so, it retrieves the corresponding output $x$ using `getValue`, computes the ciphertext as $\texttt{concat}(r, \texttt{xor}(x, m))$, and returns it. Otherwise, if $r$ is a new input value for the random function, the term generates a fresh output $x$ using `random`, records this new association in the ledger using `modify`, and then returns the ciphertext constructed in the same way.

Please observe that in the term $\widehat{Oracle}$ in Figure 10.6, we use the syntactic sugar from Equation 10.1, as $\widehat{Enc}$ is a value.

The distinguisher $D_A$ used in the first step of the proof of Theorem 16 emulates the experiment $\texttt{PrivK}^{CPA}$. The model for the distinguisher is very similar to the one for the experiment but this time the oracle $o$ is taken as input, in this way it is possible to use the same term to model both $D^{F_k(\cdot)}$ and $D^{f(\cdot)}$. Formally, the term is well-typed by the following type judgement $\vdash_c^{\Theta} D :!_{q+1}(\mathbb{S}[i] \multimap \mathbb{S}[2i]) \multimap \mathbb{B}$ and it is defined as follows

$$\begin{aligned}
&D \triangleq \\
&\quad \texttt{return}\,(\lambda o. \\
&\quad \texttt{let}\,\langle msgs, s\rangle = Adv_1\ o\,\texttt{in} \\
&\quad \texttt{let}\,\langle m_0, m_1\rangle = msgs\,\texttt{in} \\
&\quad \texttt{let}\,b = \texttt{flipcoin}\,\texttt{in} \\
&\quad \texttt{if}\ b\ \texttt{then}\ (\lambda m.\texttt{let}\,c = (\texttt{der}(o)\ m)\,\texttt{in}\,Adv_2\langle\langle o, s\rangle, c\rangle)m_1 \\
&\quad \texttt{else}\ \texttt{let}\,z = (\lambda m.\texttt{let}\,c = (\texttt{der}(o)\ m)\,\texttt{in}\,Adv_2\langle\langle o, s\rangle, c\rangle)m_0\,\texttt{in}\,\texttt{not}(z))
\end{aligned} \tag{10.3}$$

Observe that in the second line of the term $D$, we use the syntactic sugar from Equation 10.2, as we also did in Figure 10.4. Likewise, the syntactic sugar from Equation 10.1 is applied in the if-then-else construct, where the `return` is omitted when applying values to messages in both branches, following the same convention used in Figure 10.4. Therefore, we can model the distinguishers $D^{F_k(\cdot)}$ and $D^{f(\cdot)}$ as follows

$$\begin{aligned}
D^F &\triangleq \texttt{let}\,k = Gen^F * \ \texttt{in}\,\texttt{let}\,o = M^F\ k\,\texttt{in}\,\texttt{let}\,e = EncOracle\ o\,\texttt{in}\,D\ e \\
\widehat{D} &\triangleq \texttt{let}\,k = \widehat{Gen} * \ \texttt{in}\,\texttt{let}\,o = \widehat{M}\ k\,\texttt{in}\,\texttt{let}\,e = EncOracle\ o\,\texttt{in}\,D\ e
\end{aligned} \tag{10.4}$$

where the subterm $M^F$ (resp. $\widehat{M}$) models the pseudorandom (resp. random) function $F$ (resp. $f$). Furthermore, as in Figure 10.4, we apply the syntactic sugar from Equation 10.1 also in the terms in Equation 10.4, as $Gen^F$ and $\widehat{Gen}$ are values.

In contrast to earlier constructions, the function oracle and the encryption oracle are defined as separate components: the function oracle is first constructed independently — based on either $F$ or $f$ — and is subsequently provided as input to the encryption oracle. This separation allows the encryption oracle to be defined generically over an abstract oracle, thereby isolating the influence of the underlying function's nature (pseudorandom or random) on the behaviour of the resulting ciphertexts.

The encryption oracle *EncOracle*, defined in Equation (10.5), returns an encryption function parameterized by a function oracle $o$. Upon receiving a message $m$, this function samples a fresh random value $r$, queries the function oracle on $r$, and computes the ciphertext as the concatenation of $r$ with the exclusive-or of $m$ and the output of the function oracle. This term is well-typed under the following type derivation $\vdash_c^{\varnothing} EncOracle :!_{q+1}(\mathbb{S}[i] \multimap \mathbb{S}[i]) \multimap !_{q+1}(\mathbb{S}[i] \multimap \mathbb{S}[2i])$.

$$\begin{aligned}
EncOracle \triangleq\ &\texttt{return}\,(\lambda o.\texttt{return}\,(!(\texttt{return}\,(\lambda m. \\
&\quad \texttt{let}\,r = \texttt{random}\,\texttt{in} \\
&\quad \texttt{let}\,x = (\texttt{der}(o)r)\,\texttt{in} \\
&\quad \texttt{let}\,y = \texttt{xor}(x, m)\,\texttt{in} \\
&\quad \texttt{let}\,c = \texttt{concat}(r, y)\,\texttt{in} \\
&\quad \texttt{return}\,c))))
\end{aligned} \tag{10.5}$$

The precise definitions of the function oracles $M^F$ and $\widehat{M}$ are provided in Figure 10.7, while their respective types are as follows:

$$\begin{aligned}
&\vdash_c^{\varnothing} M^F : \mathbb{S}[i] \multimap !_{q+1}(\mathbb{S}[i] \multimap \mathbb{S}[i]) \\
&\vdash_c^{\Upsilon} \widehat{M} : \mathbb{S}[i] \multimap !_{q+1}(\mathbb{S}[i] \multimap \mathbb{S}[i])
\end{aligned}$$

$M^F$ :

$\texttt{return}\,(\lambda k.\texttt{return}\,(!(\texttt{return}\,(\lambda r.F\langle k, r\rangle)))))$

(a) subterm with pseudorandom function.

$\widehat{M}$ :

$\texttt{return}\,(\lambda k.\texttt{return}\,(!(\texttt{return}\,(\lambda r.$

$\texttt{let}\, tab = \texttt{get}\, rand\, \texttt{in}$

$\texttt{let}\, test = \texttt{isdefined}(tab, r)\, \texttt{in}$

$\texttt{if}\; test\; \texttt{then}$

$\quad \texttt{let}\, x = \texttt{getValue}(tab, r)\, \texttt{in}$

$\quad \texttt{return}\, x$

$\texttt{else}$

$\quad \texttt{let}\, x = \texttt{random}\, \texttt{in}$

$\quad \texttt{let}\, newtab = \texttt{modify}(tab, r, x)\, \texttt{in}$

$\quad \texttt{let}\, - = \texttt{set}\, rand\, newtab\, \texttt{in}$

$\quad \texttt{return}\, x))))$

(b) subterm with random function.

Figure 10.7: Models for function oracles used by the distinguisher.

## 10.3 Establishing CPA Security of $\Pi_F$ Equationally

As discussed in Section 7.1, the security of $\Pi_F$ is established via a proof by reduction (see Theorem 16). Specifically, it proceeds contrapositively, turning any hypothetical adversary *Adv* for $\Pi_F$ into a distinguisher $D$ for $F$, namely an algorithm designed to distinguish $F$ from a truly random function. If $D$ can be proved successful whenever *Adv* is successful, we can conclude that $\Pi_F$ is secure whenever $F$ is pseudorandom, both notions being spelled out as the *non-existence* of adversaries of the appropriate kind.

In the previous subsection, we have defined how to construct the distinguisher $D$ using the adversary *Adv* as a subroutine, where *Adv* is modelled through its two phases, $Adv_1$ and $Adv_2$. The idea was to design $D$ in such a way as to create the right environment around *Adv*, letting it believe that it is interacting with the experiment $\texttt{PrivK}^{CPA}$, and exploiting its capabilities for the sake of distinguishing $F$ from a random function. In the context of $\lambda$BLL, the distinguisher $D$ becomes the term defined in Equation 10.3 of type $!_{q+1}(\mathbb{S}[i] \multimap \mathbb{S}[2i]) \multimap \mathbb{B}$. Then, we gave two instances on $D$ namely that interacting with the pseudorandom function $F$, which we indicate as $D^F$, and that interacting with a genuinely random function $f$, indicated as $\widehat{D}$. The term modelling $F$ is instantiated as a term which does not use any reference as described in Figure 10.7a. In contrast, the model for $f$ can only be captured by a stateful computation using a ledger. In fact, the latter is actually implemented by using the reference *rand*, as shown in Figure 10.7b. We have also defined an encryption scheme $\widehat{\Pi}$ that is structurally the same as $\Pi_F$, but operates using a truly random functions instead of pseudorandom ones. The components of $\widehat{\Pi}$ are specified in Figure 10.6, while those of $\Pi_F$ are given in Figure 10.5. Accordingly, we have constructed a variation $\widehat{PrivKCPA}$ on $PrivKCPA^F$.

Let us consider the reference context $\Upsilon = \{rand : \mathbb{S}[(q+1)\times(2i+1)]\}$ under which the terms defined in the previous subsection can be typed. The security of $\Pi_F$ becomes the equation $PrivKCPA^F \sim_{\varnothing\subseteq\Upsilon} \texttt{flipcoin}$, whereas $\texttt{flipcoin}$ is a function symbol of boolean type, returning each possible result with probability $\frac{1}{2}$, while $\sim_{\varnothing\subseteq\Upsilon}$ is an instance of our $\Theta$-contextual indistinguishability relation defined in Section 9.2. The aforementioned equation can be proved under the hypothesis that $F$ is pseudorandom, and this last condition also becomes an equation. This time, however, the terms to be compared are $D^F$ and $\widehat{D}$, which are defined in Equation 10.4. The security proof then proceeds by contraposition, as explained schematically in Figure 10.8: from the negation of the thesis, the negation of the hypothesis is derived and this is done by proving that both on the right and on the left sides of the diagram it is possible to link the terms through the relation $\sim_{\varnothing\subseteq\Upsilon}$. In this context, it is clear that the use of Θ-contextual indistinguishability and logical relations becomes useful. In particular, the equations listed in Figure 10.2 can be used, as discussed in the following. Noticeably, all of them can be proved sound for Θ-contextual indistinguishability through logical relations and in particular through the Equations introduced in Section 10.1.

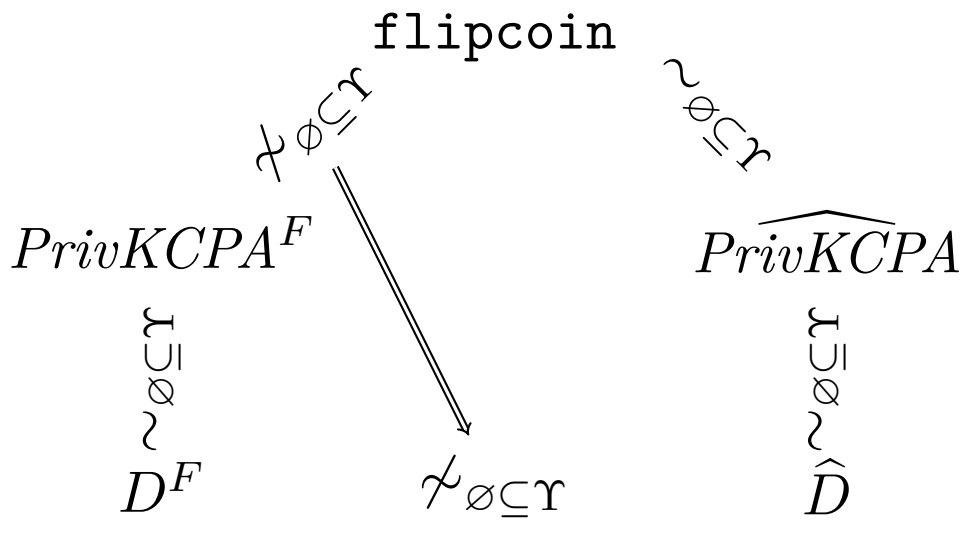

Figure 10.8: Outline Proof of Security

Consider the diagram in Figure 10.8. The goal is to prove that, on both the left and right sides of the diagram, we can use the relation $\sim_{\varnothing\subseteq\Upsilon}$ to link the respective terms. On the left side, we have to show that the equation $D^F \sim_{\varnothing\subseteq\Upsilon} \mathit{PrivKCPA}^F$ holds. Similarly, on the right side, we have to show that $\widehat{D} \sim_{\varnothing\subseteq\Upsilon} \widehat{\mathit{PrivKCPA}}$. Observe that these two equations corresponds respectively to Equation 7.2 and Equation 7.3 in Theorem 16. In our setting, demonstrating that both of these relations hold is relatively straightforward, due to the way the terms involved have been defined.

We begin by focusing on the bottom-right side of the diagram. The first step is to establish the relation $\mathtt{let}\, o = \widehat{M}\ k\, \mathtt{in}\, \mathit{EncOracle}\ o \sim_{\varnothing\subseteq\Upsilon} \widehat{\mathit{Oracle}}\ k$, which is proved in Lemma 65. Once this equivalence is established, the desired relation $\widehat{D} \sim_{\varnothing\subseteq\Upsilon} \widehat{\mathit{PrivKCPA}}$ follows by successive application of Equation `letAss` and Lemma 65, concluding with a straightforward reduction via Equation `reduc2`, as detailed in the proof of Theorem 21.

**Lemma 65.**

$$\mathtt{let}\, o = \widehat{M}\ k\, \mathtt{in}\, \mathit{EncOracle}\ o \sim_{\varnothing\subseteq\Upsilon} \widehat{\mathit{Oracle}}\ k$$

*Proof.* We begin by considering the term $\widehat{\mathit{Oracle}}\ k$. First, we unfold the definition of $\widehat{\mathit{Oracle}}$ as defined in Figure 10.6, which returns a curried function ultimately applying $\widehat{\mathit{Enc}}$ to the pair $\langle k, m\rangle$. We apply `reduc2`, which simplifies applying a returned lambda to an argument, we eliminate the lambda and substitute $k$ directly. Next, we unfold the definition of $\widehat{\mathit{Enc}}$ using its definition in Figure 10.6. We simplify the resulting term using Equation `reduc2` to reduce an application of returned lambda abstractions to a value and Equation `reduc5` to eliminate the pattern match and directly substituting $k$ and $m$ in the body. Observe that in these steps $\Theta = \Xi = \Upsilon = \{\mathit{rand} : \mathbb{S}[(q+1) \times (2i+1)]\}$, since the closing context involved in this step uses the reference *rand* by performing a `get` operation on it.

$$\widehat{\mathit{Oracle}}\ k \triangleq \mathtt{return}\,(\lambda k.\mathtt{return}\,!(\mathtt{return}\,\lambda m.\widehat{\mathit{Enc}}\langle k, m\rangle))\ k$$
$$\overset{\mathtt{reduc2}}{\sim_{\varnothing\subseteq\Upsilon}} \quad \mathtt{return}\,!(\mathtt{return}\,\lambda m.\widehat{\mathit{Enc}}\langle k, m\rangle)$$

$$
\triangleq
\begin{array}{l}
\texttt{return}\,(!(\texttt{return}\,(\lambda m.\\
\quad(\lambda e.\\
\qquad \texttt{let}\,\langle k,m\rangle = e\,\texttt{in}\\
\qquad \texttt{let}\,tab = \texttt{get}\,rand\,\texttt{in}\\
\qquad \texttt{let}\,r = \texttt{random}\,\texttt{in}\\
\qquad \texttt{let}\,test = \texttt{isdefined}(tab,r)\,\texttt{in}\\
\qquad \texttt{if}\ test\ \texttt{then}\\
\qquad\ \ \texttt{let}\,x = \texttt{getValue}(tab,r)\,\texttt{in}\\
\qquad\ \ \texttt{let}\,y = \texttt{xor}(x,m)\,\texttt{in}\\
\qquad\ \ \texttt{let}\,c = \texttt{concat}(r,y)\,\texttt{in}\\
\qquad\ \ \texttt{return}\,c\\
\qquad \texttt{else}\\
\qquad\ \ \texttt{let}\,x = \texttt{random}\,\texttt{in}\\
\qquad\ \ \texttt{let}\,y = \texttt{xor}(x,m)\,\texttt{in}\\
\qquad\ \ \texttt{let}\,newtab = \texttt{modify}(tab,r,x)\,\texttt{in}\\
\qquad\ \ \texttt{let}\,- = \texttt{set}\,rand\,newtab\,\texttt{in}\\
\qquad\ \ \texttt{let}\,c = \texttt{concat}(r,y)\,\texttt{in}\\
\qquad\ \ \texttt{return}\,c\\
\quad)\langle k,m\rangle)))
\end{array}
\overset{\texttt{reduc2}}{\sim}_{\varnothing\subseteq\Upsilon}
\begin{array}{l}
\texttt{return}\,(!(\texttt{return}\,(\lambda m.\\
\quad \texttt{let}\,\langle k,m\rangle = \langle k,m\rangle\,\texttt{in}\\
\quad \texttt{let}\,tab = \texttt{get}\,rand\,\texttt{in}\\
\quad \texttt{let}\,r = \texttt{random}\,\texttt{in}\\
\quad \texttt{let}\,test = \texttt{isdefined}(tab,r)\,\texttt{in}\\
\quad \texttt{if}\ test\ \texttt{then}\\
\qquad \texttt{let}\,x = \texttt{getValue}(tab,r)\,\texttt{in}\\
\qquad \texttt{let}\,y = \texttt{xor}(x,m)\,\texttt{in}\\
\qquad \texttt{let}\,c = \texttt{concat}(r,y)\,\texttt{in}\\
\qquad \texttt{return}\,c\\
\quad \texttt{else}\\
\qquad \texttt{let}\,x = \texttt{random}\,\texttt{in}\\
\qquad \texttt{let}\,y = \texttt{xor}(x,m)\,\texttt{in}\\
\qquad \texttt{let}\,newtab = \texttt{modify}(tab,r,x)\,\texttt{in}\\
\qquad \texttt{let}\,- = \texttt{set}\,rand\,newtab\,\texttt{in}\\
\qquad \texttt{let}\,c = \texttt{concat}(r,y)\,\texttt{in}\\
\qquad \texttt{return}\,c\\
)))
\end{array}
$$

$$
\overset{\texttt{reduc5}}{\sim}_{\varnothing\subseteq\Upsilon}
\begin{array}{l}
\texttt{return}\,(!(\texttt{return}\,(\lambda m.\\
\quad \texttt{let}\,tab = \texttt{get}\,rand\,\texttt{in}\\
\quad \texttt{let}\,r = \texttt{random}\,\texttt{in}\\
\quad \texttt{let}\,test = \texttt{isdefined}(tab,r)\,\texttt{in}\\
\quad \texttt{if}\ test\ \texttt{then}\\
\qquad \texttt{let}\,x = \texttt{getValue}(tab,r)\,\texttt{in}\\
\qquad \texttt{let}\,y = \texttt{xor}(x,m)\,\texttt{in}\\
\qquad \texttt{let}\,c = \texttt{concat}(r,y)\,\texttt{in}\\
\qquad \texttt{return}\,c\\
\quad \texttt{else}\\
\qquad \texttt{let}\,x = \texttt{random}\,\texttt{in}\\
\qquad \texttt{let}\,y = \texttt{xor}(x,m)\,\texttt{in}\\
\qquad \texttt{let}\,newtab = \texttt{modify}(tab,r,x)\,\texttt{in}\\
\qquad \texttt{let}\,- = \texttt{set}\,rand\,newtab\,\texttt{in}\\
\qquad \texttt{let}\,c = \texttt{concat}(r,y)\,\texttt{in}\\
\qquad \texttt{return}\,c\\
)))
\end{array}
\tag{10.6}
$$

Now, the goal is to show that the term $\texttt{let}\,o = \widehat{M}\ k\,\texttt{in}\,EncOracle\ o$ is $\Theta$-contextual indistinguishable from the term above in Equation 10.6, we proceed by gradually unfolding and simplifying it using the equations listed in Figure 10.2. We begin by unfolding the definition of *EncOracle* as defined in Equation 10.5, which is a curried function returning an encryption behavior when applied to an oracle $o$. We apply Equation reduc2 to eliminate the lambda abstraction and simplify the application of *EncOracle* to $o$. Next, we turn our attention to $\widehat{M}$ and we unfold its definition from Figure 10.7b. Applying $\widehat{M}$ to a key $k$ yields a function that, on input $r$, checks the ledger to determine whether a value has already been assigned to $r$. If so, it returns that value; otherwise, it samples a fresh random value, updates the table accordingly, and then returns the new value. We again apply reduc2 to evaluate this application. At this point, we continue simplifying by using the rules reduc4, reduc3, and reduc2.

$$\texttt{let}\ o = \widehat{M}\ k\ \texttt{in}\ \mathit{EncOracle}\ o$$

$$\triangleq \begin{array}{l}\texttt{let}\ o = \widehat{M}\ k\ \texttt{in}\\ \texttt{return}\ (\lambda o.\texttt{return}\ (!(\texttt{return}\ (\lambda m.\\ \quad \texttt{let}\ r = \texttt{random}\ \texttt{in}\\ \quad \texttt{let}\ x = (\texttt{der}(o)r)\ \texttt{in}\\ \quad \texttt{let}\ y = \texttt{xor}(x, m)\ \texttt{in}\\ \quad \texttt{let}\ c = \texttt{concat}(r, y)\ \texttt{in}\\ \quad \texttt{return}\ c\\ ))))\ o\end{array} \overset{\texttt{reduc2}}{\sim_{\varnothing \subseteq \Upsilon}} \begin{array}{l}\texttt{let}\ o = \widehat{M}\ k\ \texttt{in}\\ \texttt{return}\ (!(\texttt{return}\ (\lambda m.\\ \quad \texttt{let}\ r = \texttt{random}\ \texttt{in}\\ \quad \texttt{let}\ x = (\texttt{der}(o)r)\ \texttt{in}\\ \quad \texttt{let}\ y = \texttt{xor}(x, m)\ \texttt{in}\\ \quad \texttt{let}\ c = \texttt{concat}(r, y)\ \texttt{in}\\ \quad \texttt{return}\ c\\ )))\end{array}$$

$$\triangleq \begin{array}{l}\texttt{let}\ o = \left(\begin{array}{l}\texttt{return}\ (\lambda k.\texttt{return}\ (!(\texttt{return}\ (\lambda r.\\ \quad \texttt{let}\ \mathit{tab} = \texttt{get}\ \mathit{rand}\ \texttt{in}\\ \quad \texttt{let}\ \mathit{test} = \texttt{isdefined}(\mathit{tab}, r)\ \texttt{in}\\ \quad \texttt{if}\ \mathit{test}\ \texttt{then}\\ \quad\ \ \texttt{let}\ x = \texttt{getValue}(\mathit{tab}, r)\ \texttt{in}\\ \quad\ \ \texttt{return}\ x\\ \quad \texttt{else}\\ \quad\ \ \texttt{let}\ x = \texttt{random}\ \texttt{in}\\ \quad\ \ \texttt{let}\ \mathit{newtab} = \texttt{modify}(\mathit{tab}, r, x)\ \texttt{in}\\ \quad\ \ \texttt{let}\ - = \texttt{set}\ \mathit{rand}\ \mathit{newtab}\ \texttt{in}\\ \quad\ \ \texttt{return}\ x))))\end{array}\right) k\ \texttt{in}\\ \texttt{return}\ (!(\texttt{return}\ (\lambda m.\\ \quad \texttt{let}\ r = \texttt{random}\ \texttt{in}\\ \quad \texttt{let}\ x = (\texttt{der}(o)r)\ \texttt{in}\\ \quad \texttt{let}\ y = \texttt{xor}(x, m)\ \texttt{in}\\ \quad \texttt{let}\ c = \texttt{concat}(r, y)\ \texttt{in}\\ \quad \texttt{return}\ c\\ )))\end{array}$$

$$
\begin{array}{l}
\texttt{let}\ o = \begin{pmatrix}
\texttt{return}\,(!(\texttt{return}\,(\lambda r. \\
\quad \texttt{let}\ tab = \texttt{get}\ rand\ \texttt{in} \\
\quad \texttt{let}\ test = \texttt{isdefined}(tab, r)\ \texttt{in} \\
\quad \texttt{if}\ test\ \texttt{then} \\
\quad\ \ \texttt{let}\ x = \texttt{getValue}(tab, r)\ \texttt{in} \\
\quad\ \ \texttt{return}\ x \\
\quad \texttt{else} \\
\quad\ \ \texttt{let}\ x = \texttt{random}\ \texttt{in} \\
\quad\ \ \texttt{let}\ newtab = \texttt{modify}(tab, r, x)\ \texttt{in} \\
\quad\ \ \texttt{let}\ - = \texttt{set}\ rand\ newtab\ \texttt{in} \\
\quad\ \ \texttt{return}\ x)))
\end{pmatrix} \texttt{in}
\end{array}
$$

$$\overset{\texttt{reduc2}}{\sim_{\varnothing\subseteq\Upsilon}}$$

$$
\begin{array}{l}
\texttt{return}\,(!(\texttt{return}\,(\lambda m. \\
\quad \texttt{let}\ r = \texttt{random}\ \texttt{in} \\
\quad \texttt{let}\ x = (\texttt{der}(o) r)\ \texttt{in} \\
\quad \texttt{let}\ y = \texttt{xor}(x, m)\ \texttt{in} \\
\quad \texttt{let}\ c = \texttt{concat}(r, y)\ \texttt{in} \\
\quad \texttt{return}\ c \\
)))
\end{array}
$$

$$\overset{\texttt{reduc4}}{\sim_{\varnothing\subseteq\Upsilon}}$$

$$
\begin{array}{l}
\texttt{return}\,(!(\texttt{return}\,(\lambda m. \\
\quad \texttt{let}\ r = \texttt{random}\ \texttt{in} \\
\quad \texttt{let}\ x = \texttt{der}\left( ! \begin{pmatrix}
\texttt{return}\,(\lambda r. \\
\quad \texttt{let}\ tab = \texttt{get}\ rand\ \texttt{in} \\
\quad \texttt{let}\ test = \texttt{isdefined}(tab, r)\ \texttt{in} \\
\quad \texttt{if}\ test\ \texttt{then} \\
\quad\ \ \texttt{let}\ x = \texttt{getValue}(tab, r)\ \texttt{in} \\
\quad\ \ \texttt{return}\ x \\
\quad \texttt{else} \\
\quad\ \ \texttt{let}\ x = \texttt{random}\ \texttt{in} \\
\quad\ \ \texttt{let}\ newtab = \texttt{modify}(tab, r, x)\ \texttt{in} \\
\quad\ \ \texttt{let}\ - = \texttt{set}\ rand\ newtab\ \texttt{in} \\
\quad\ \ \texttt{return}\ x)
\end{pmatrix} \right) r\ \texttt{in} \\
\quad \texttt{let}\ y = \texttt{xor}(x, m)\ \texttt{in} \\
\quad \texttt{let}\ c = \texttt{concat}(r, y)\ \texttt{in} \\
\quad \texttt{return}\ c \\
)))
\end{array}
$$

$$
\overset{\texttt{reduc3}}{\sim}_{\varnothing\subseteq\Upsilon}
\begin{array}{l}
\texttt{return}\,(!(\texttt{return}\,(\lambda m.\\
\quad \texttt{let}\,r = \texttt{random}\,\texttt{in}\\
\quad \texttt{let}\,x = \left(\begin{array}{l}
\texttt{return}\,(\lambda r.\\
\quad \texttt{let}\,tab = \texttt{get}\,rand\,\texttt{in}\\
\quad \texttt{let}\,test = \texttt{isdefined}(tab, r)\,\texttt{in}\\
\quad \texttt{if}\,test\,\texttt{then}\\
\quad\; \texttt{let}\,x = \texttt{getValue}(tab, r)\,\texttt{in}\\
\quad\; \texttt{return}\,x\\
\quad \texttt{else}\\
\quad\; \texttt{let}\,x = \texttt{random}\,\texttt{in}\\
\quad\; \texttt{let}\,newtab = \texttt{modify}(tab, r, x)\,\texttt{in}\\
\quad\; \texttt{let}\,- = \texttt{set}\,rand\,newtab\,\texttt{in}\\
\quad\; \texttt{return}\,x)
\end{array}\right) r\,\texttt{in}\\
\quad \texttt{let}\,y = \texttt{xor}(x, m)\,\texttt{in}\\
\quad \texttt{let}\,c = \texttt{concat}(r, y)\,\texttt{in}\\
\quad \texttt{return}\,c\\
)))
\end{array}
$$

$$
\overset{\texttt{reduc2}}{\sim}_{\varnothing\subseteq\Upsilon}
\begin{array}{l}
\texttt{return}\,(!(\texttt{return}\,(\lambda m.\\
\quad \texttt{let}\,r = \texttt{random}\,\texttt{in}\\
\quad \texttt{let}\,x = \left(\begin{array}{l}
\quad \texttt{let}\,tab = \texttt{get}\,rand\,\texttt{in}\\
\quad \texttt{let}\,test = \texttt{isdefined}(tab, r)\,\texttt{in}\\
\quad \texttt{if}\,test\,\texttt{then}\\
\quad\; \texttt{let}\,x = \texttt{getValue}(tab, r)\,\texttt{in}\\
\quad\; \texttt{return}\,x\\
\quad \texttt{else}\\
\quad\; \texttt{let}\,x = \texttt{random}\,\texttt{in}\\
\quad\; \texttt{let}\,newtab = \texttt{modify}(tab, r, x)\,\texttt{in}\\
\quad\; \texttt{let}\,- = \texttt{set}\,rand\,newtab\,\texttt{in}\\
\quad\; \texttt{return}\,x
\end{array}\right) \texttt{in}\\
\quad \texttt{let}\,y = \texttt{xor}(x, m)\,\texttt{in}\\
\quad \texttt{let}\,c = \texttt{concat}(r, y)\,\texttt{in}\\
\quad \texttt{return}\,c\\
)))
\end{array}
$$

At this point of the proof, we want to simplify the internal structure of the encryption function produced when applying *EncOracle* to the oracle derived from $\widehat{M}$. We begin by applying Equation `letAss` to restructure nested let-bindings.

$$
\stackrel{\texttt{letAss}}{\sim_{\varnothing\subseteq\Upsilon}}
\begin{array}{l}
\texttt{return}\,(!(\texttt{return}\,(\lambda m.\\
\quad \texttt{let}\,r = \texttt{random}\,\texttt{in}\\
\quad \texttt{let}\,tab = \texttt{get}\,rand\,\texttt{in}\\
\quad \texttt{let}\,test = \texttt{isdefined}(tab, r)\,\texttt{in}\\
\quad \texttt{let}\,x = \left(\begin{array}{l}
\texttt{if}\ test\ \texttt{then}\\
\ \texttt{let}\,x = \texttt{getValue}(tab, r)\,\texttt{in}\\
\ \texttt{return}\,x\\
\texttt{else}\\
\ \texttt{let}\,x = \texttt{random}\,\texttt{in}\\
\ \texttt{let}\,newtab = \texttt{modify}(tab, r, x)\,\texttt{in}\\
\ \texttt{let}\,- = \texttt{set}\,rand\,newtab\,\texttt{in}\\
\ \texttt{return}\,x
\end{array}\right)\ \texttt{in}\\
\quad \texttt{let}\,y = \texttt{xor}(x, m)\,\texttt{in}\\
\quad \texttt{let}\,c = \texttt{concat}(r, y)\,\texttt{in}\\
\quad \texttt{return}\,c\\
)))
\end{array}
$$

Next, we apply Equation **ifLet**, which enables factoring a let binding out of a conditional expression by lifting the shared continuation, thereby restructuring the code to isolate common parts and simplify control flow. Formally, it is defined as follows

$$
\begin{pmatrix}\texttt{if}\ b\ \texttt{then}\ \texttt{let}\,x = M\,\texttt{in}\,N\\ \texttt{else}\ \texttt{let}\,x = L\,\texttt{in}\,N\end{pmatrix} \sim_{\Theta\subseteq\Xi} \texttt{let}\,x = (\texttt{if}\,b\,\texttt{then}\,M\,\texttt{else}\ L)\,\texttt{in}\,N
$$

In our particular instance, we instantiate this rule with the following terms:

$$
N = \begin{array}{l}\texttt{let}\,y = \texttt{xor}(x, m)\,\texttt{in}\\ \texttt{let}\,c = \texttt{concat}(r, y)\,\texttt{in}\\ \texttt{return}\,c\end{array}
$$

$$
M = \begin{array}{l}\texttt{let}\,x = \texttt{getValue}(tab, r)\,\texttt{in}\\ \texttt{return}\,x\end{array}
$$

$$
L = \begin{array}{l}\texttt{let}\,x = \texttt{random}\,\texttt{in}\\ \texttt{let}\,newtab = \texttt{modify}(tab, r, x)\,\texttt{in}\\ \texttt{let}\,- = \texttt{set}\,rand\,newtab\,\texttt{in}\\ \texttt{return}\,x\end{array}
$$

This step is key because it removes the nested conditional expression and exposes the conditional structure at the top level of the program. It separates the two branches clearly, allowing us to apply simple reductions and commutations within each branch independently and conclude.

```
return (!(return (λm.
  let r = random in
  let tab = get rand in
  let test = isdefined(tab, r) in
  if test then
      let x = (let x = getValue(tab, r) in
               return x) in
      let y = xor(x, m) in
      let c = concat(r, y) in
      return c
  else
      let x = (let x = random in
               let newtab = modify(tab, r, x) in
               let _ = set rand newtab in
               return x) in
      let y = xor(x, m) in
      let c = concat(r, y) in
      return c
)))
```

$$\overset{\texttt{ifLet}}{\sim}_{\varnothing\subseteq\Upsilon}$$

```
return (!(return (λm.
  let r = random in
  let tab = get rand in
  let test = isdefined(tab, r) in
  if test then
      let x = getValue(tab, r) in
      let y = xor(x, m) in
      let c = concat(r, y) in
      return c
  else
      let x = (let x = random in
               let newtab = modify(tab, r, x) in
               let _ = set rand newtab in
               return x) in
      let y = xor(x, m) in
      let c = concat(r, y) in
      return c
)))
```

$$\overset{\texttt{reduc1}}{\sim}_{\varnothing\subseteq\Upsilon}$$

$$
\begin{array}{ll}
& \texttt{return}\,(!(\texttt{return}\,(\lambda m. \\
& \quad \texttt{let}\, tab = \texttt{get}\, rand\, \texttt{in} \\
& \quad \texttt{let}\, r = \texttt{random}\, \texttt{in} \\
& \quad \texttt{let}\, test = \texttt{isdefined}(tab, r)\, \texttt{in} \\
& \quad \texttt{if}\; test\; \texttt{then} \\
& \qquad \texttt{let}\, x = \texttt{getValue}(tab, r)\, \texttt{in} \\
& \qquad \texttt{let}\, y = \texttt{xor}(x, m)\, \texttt{in} \\
& \qquad \texttt{let}\, c = \texttt{concat}(r, y)\, \texttt{in} \\
\overset{\texttt{letCom}}{\sim_{\varnothing \subseteq \Upsilon}} & \qquad \texttt{return}\, c \\
& \quad \texttt{else} \\
& \qquad \texttt{let}\, x = \begin{pmatrix} \texttt{let}\, x = \texttt{random}\, \texttt{in} \\ \texttt{let}\, newtab = \texttt{modify}(tab, r, x)\, \texttt{in} \\ \texttt{let}\, - = \texttt{set}\, rand\, newtab\, \texttt{in} \\ \texttt{return}\, x \end{pmatrix} \texttt{in} \\
& \qquad \texttt{let}\, y = \texttt{xor}(x, m)\, \texttt{in} \\
& \qquad \texttt{let}\, c = \texttt{concat}(r, y)\, \texttt{in} \\
& \qquad \texttt{return}\, c \\
& )))
\end{array}
$$

$$
\begin{array}{ll}
& \texttt{return}\,(!(\texttt{return}\,(\lambda m. \\
& \quad \texttt{let}\, tab = \texttt{get}\, rand\, \texttt{in} \\
& \quad \texttt{let}\, r = \texttt{random}\, \texttt{in} \\
& \quad \texttt{let}\, test = \texttt{isdefined}(tab, r)\, \texttt{in} \\
& \quad \texttt{if}\; test\; \texttt{then} \\
& \qquad \texttt{let}\, x = \texttt{getValue}(tab, r)\, \texttt{in} \\
& \qquad \texttt{let}\, y = \texttt{xor}(x, m)\, \texttt{in} \\
& \qquad \texttt{let}\, c = \texttt{concat}(r, y)\, \texttt{in} \\
& \qquad \texttt{return}\, c \\
=_\alpha & \quad \texttt{else} \\
& \qquad \texttt{let}\, x' = \begin{pmatrix} \texttt{let}\, x = \texttt{random}\, \texttt{in} \\ \texttt{let}\, newtab = \texttt{modify}(tab, r, x)\, \texttt{in} \\ \texttt{let}\, - = \texttt{set}\, rand\, newtab\, \texttt{in} \\ \texttt{return}\, x \end{pmatrix} \texttt{in} \\
& \qquad \texttt{let}\, y = \texttt{xor}(x', m)\, \texttt{in} \\
& \qquad \texttt{let}\, c = \texttt{concat}(r, y)\, \texttt{in} \\
& \qquad \texttt{return}\, c \\
& )))
\end{array}
$$

$$
\overset{\texttt{letAss}}{\sim}_{\varnothing\subseteq\Upsilon}
\begin{array}{l}
\texttt{return}\,(!(\texttt{return}\,(\lambda m.\\
\quad \texttt{let}\,tab = \texttt{get}\,rand\,\texttt{in}\\
\quad \texttt{let}\,r = \texttt{random}\,\texttt{in}\\
\quad \texttt{let}\,test = \texttt{isdefined}(tab, r)\,\texttt{in}\\
\quad \texttt{if}\ test\ \texttt{then}\\
\qquad \texttt{let}\,x = \texttt{getValue}(tab, r)\,\texttt{in}\\
\qquad \texttt{let}\,y = \texttt{xor}(x, m)\,\texttt{in}\\
\qquad \texttt{let}\,c = \texttt{concat}(r, y)\,\texttt{in}\\
\qquad \texttt{return}\,c\\
\quad \texttt{else}\\
\qquad \texttt{let}\,x = \texttt{random}\,\texttt{in}\\
\qquad \texttt{let}\,newtab = \texttt{modify}(tab, r, x)\,\texttt{in}\\
\qquad \texttt{let}\,- = \texttt{set}\,rand\,newtab\,\texttt{in}\\
\qquad \texttt{let}\,x' = \texttt{return}\,x\,\texttt{in}\\
\qquad \texttt{let}\,y = \texttt{xor}(x', m)\,\texttt{in}\\
\qquad \texttt{let}\,c = \texttt{concat}(r, y)\,\texttt{in}\\
\qquad \texttt{return}\,c\\
)))
\end{array}
\overset{\texttt{reduc4}}{\sim}_{\varnothing\subseteq\Upsilon}
\begin{array}{l}
\texttt{return}\,(!(\texttt{return}\,(\lambda m.\\
\quad \texttt{let}\,tab = \texttt{get}\,rand\,\texttt{in}\\
\quad \texttt{let}\,r = \texttt{random}\,\texttt{in}\\
\quad \texttt{let}\,test = \texttt{isdefined}(tab, r)\,\texttt{in}\\
\quad \texttt{if}\ test\ \texttt{then}\\
\qquad \texttt{let}\,x = \texttt{getValue}(tab, r)\,\texttt{in}\\
\qquad \texttt{let}\,y = \texttt{xor}(x, m)\,\texttt{in}\\
\qquad \texttt{let}\,c = \texttt{concat}(r, y)\,\texttt{in}\\
\qquad \texttt{return}\,c\\
\quad \texttt{else}\\
\qquad \texttt{let}\,x = \texttt{random}\,\texttt{in}\\
\qquad \texttt{let}\,newtab = \texttt{modify}(tab, r, x)\,\texttt{in}\\
\qquad \texttt{let}\,- = \texttt{set}\,rand\,newtab\,\texttt{in}\\
\qquad \texttt{let}\,y = \texttt{xor}(x, m)\,\texttt{in}\\
\qquad \texttt{let}\,c = \texttt{concat}(r, y)\,\texttt{in}\\
\qquad \texttt{return}\,c\\
)))
\end{array}
$$

$$
\overset{\texttt{letCom}}{\sim}_{\varnothing\subseteq\Upsilon}
\begin{array}{l}
\texttt{return}\,(!(\texttt{return}\,(\lambda m.\\
\quad \texttt{let}\,tab = \texttt{get}\,rand\,\texttt{in}\\
\quad \texttt{let}\,r = \texttt{random}\,\texttt{in}\\
\quad \texttt{let}\,test = \texttt{isdefined}(tab, r)\,\texttt{in}\\
\quad \texttt{if}\ test\ \texttt{then}\\
\qquad \texttt{let}\,x = \texttt{getValue}(tab, r)\,\texttt{in}\\
\qquad \texttt{let}\,y = \texttt{xor}(x, m)\,\texttt{in}\\
\qquad \texttt{let}\,c = \texttt{concat}(r, y)\,\texttt{in}\\
\qquad \texttt{return}\,c\\
\quad \texttt{else}\\
\qquad \texttt{let}\,x = \texttt{random}\,\texttt{in}\\
\qquad \texttt{let}\,y = \texttt{xor}(x, m)\,\texttt{in}\\
\qquad \texttt{let}\,newtab = \texttt{modify}(tab, r, x)\,\texttt{in}\\
\qquad \texttt{let}\,- = \texttt{set}\,rand\,newtab\,\texttt{in}\\
\qquad \texttt{let}\,c = \texttt{concat}(r, y)\,\texttt{in}\\
\qquad \texttt{return}\,c\\
)))
\end{array}
\quad = \text{Term in Equation 10.6}
$$

□

**Theorem 21.**

$$\widehat{D} \sim_{\varnothing\subseteq\Upsilon} \widehat{PrivKCPA}$$

*Proof.* We begin with the definition of $\widehat{D}$ given in the second line of Equation 10.4. The latter describes a term where a key $k$ is generated by $\widehat{Gen}$, and then the function oracle $\widehat{M}$ is applied to $k$ to produce an output $o$ and then it is passed to the term $EncOracle$ to produce $e$. This encryption oracle $e$ is then passed to the distinguisher $D$, which uses it to make a decision. We first apply Equation `letAss` to restructure the expression by flattening the nested let-bindings, allowing us to isolate the encryption oracle term. Next, we replace the term $\widehat{M}$ with $\widehat{Oracle}$, as they are Θ-contextually indistinguishable by Lemma 65. After this substitution, we use $\alpha$-equivalence to rename variables and better align the resulting structure with the definition of the distinguisher. We then expand $D$ using its definition from Equation 10.3, along with the syntactic sugar from Equation 10.1, to reveal its internal structure. Next, we apply Equation `reduc2` to simplify the structure by removing unnecessary bindings. This step eliminates some of the intermediate layers, yielding a simplified version of the distinguisher's interaction with the oracle. This leads to a term that is precisely $\widehat{PrivKCPA}$, completing the proof.

$$\widehat{D} \triangleq \texttt{let}\, k = \widehat{Gen} * \,\texttt{in}\,\texttt{let}\, o = \widehat{M}\ k\,\texttt{in}\,\texttt{let}\, e = EncOracle\ o\,\texttt{in}\, D\ e$$

$$\overset{\texttt{letAss}}{\sim_{\varnothing\subseteq\Upsilon}} \texttt{let}\, k = \widehat{Gen} * \,\texttt{in}\,\texttt{let}\, e = (\texttt{let}\, o = \widehat{M}\ k\,\texttt{in}\, EncOracle\ o)\,\texttt{in}\, D\ e$$

$$\overset{\text{Lemma 65}}{\sim_{\varnothing\subseteq\Upsilon}} \begin{array}{l} \texttt{let}\, k = \widehat{Gen} * \,\texttt{in} \\ \texttt{let}\, e = \widehat{Oracle}\ k\,\texttt{in} \\ D\ e \end{array}$$

$$=_\alpha \begin{array}{l} \texttt{let}\, k = \widehat{Gen} * \,\texttt{in} \\ \texttt{let}\, o = \widehat{Oracle}\ k\,\texttt{in} \\ D\ o \end{array}$$

$$\triangleq \begin{array}{l} \texttt{let}\, k = \widehat{Gen} * \,\texttt{in}\,\texttt{let}\, o = \widehat{Oracle}\ k\,\texttt{in} \\ \begin{pmatrix} \lambda o. \\ \texttt{let}\,\langle msgs, s\rangle = Adv_1\ o\,\texttt{in} \\ \texttt{let}\,\langle m_0, m_1\rangle = msgs\,\texttt{in} \\ \texttt{let}\, b = \texttt{flipcoin}\,\texttt{in} \\ \texttt{if}\ b\ \texttt{then}\ (\lambda m.\texttt{let}\, c = (\texttt{der}(o)\ m)\,\texttt{in}\, Adv_2\langle\langle o, s\rangle, c\rangle)m_1 \\ \texttt{else}\ \ \texttt{let}\, z = (\lambda m.\texttt{let}\, c = (\texttt{der}(o)\ m)\,\texttt{in}\, Adv_2\langle\langle o, s\rangle, c\rangle)m_0\,\texttt{in}\,\texttt{not}(z) \end{pmatrix} o \end{array}$$

$$\overset{\texttt{reduc2}}{\sim_{\varnothing\subseteq\Upsilon}} \begin{array}{l} \texttt{let}\, k = \widehat{Gen} * \,\texttt{in} \\ \texttt{let}\, o = \widehat{Oracle}\ k\,\texttt{in} \\ \texttt{let}\,\langle msgs, s\rangle = Adv_1\ o\,\texttt{in} \\ \texttt{let}\,\langle m_0, m_1\rangle = msgs\,\texttt{in} \\ \texttt{let}\, b = \texttt{flipcoin}\,\texttt{in} \\ \texttt{if}\ b\ \texttt{then}\ (\lambda m.\texttt{let}\, c = (\texttt{der}(o)\ m)\,\texttt{in}\, Adv_2\langle\langle o, s\rangle, c\rangle)m_1 \\ \texttt{else}\ \ \texttt{let}\, z = (\lambda m.\texttt{let}\, c = (\texttt{der}(o)\ m)\,\texttt{in}\, Adv_2\langle\langle o, s\rangle, c\rangle)m_0\,\texttt{in}\,\texttt{not}(z) \end{array}$$

$$\triangleq \widehat{PrivKCPA}$$

□

Dually, for the left side of the diagram, we first show that $M^F \sim_{\varnothing\subseteq\Upsilon} Oracle^F$ in Lemma 66 and then we prove the desired relation in Theorem 22. The proofs of these statements follow the same approach already described for the right side of the diagram.

**Lemma 66.**

$$\mathtt{let}\, o = M^F\ k\, \mathtt{in}\, EncOracle\ o \sim_{\varnothing\subseteq\Upsilon} Oracle^F\ k$$

*Proof.* We follow a similar approach to that in Lemma 66, starting by examining the term $Oracle^F\ k$ and simplifying it as follows.

$$\begin{aligned}
\widehat{Oracle}\ k &\triangleq \mathtt{return}\,(\lambda k.\mathtt{return}\,!(\mathtt{return}\,\lambda m.Enc^F\langle k,m\rangle))\ k\\
&\overset{\mathtt{reduc2}}{\sim_{\varnothing\subseteq\Upsilon}}\ \mathtt{return}\,!(\mathtt{return}\,\lambda m.Enc^F\langle k,m\rangle)
\end{aligned}$$

$$\triangleq \begin{array}{l}
\mathtt{return}\,(!(\mathtt{return}\,(\lambda m.\\
\quad(\lambda e.\\
\quad\quad \mathtt{let}\,\langle k,m\rangle = e\,\mathtt{in}\\
\quad\quad \mathtt{let}\,r = \mathtt{random}\,\mathtt{in}\\
\quad\quad \mathtt{let}\,x = F\langle k,r\rangle\,\mathtt{in}\\
\quad\quad \mathtt{let}\,y = \mathtt{xor}(x,m)\,\mathtt{in}\\
\quad\quad \mathtt{let}\,c = \mathtt{concat}(r,y)\,\mathtt{in}\\
\quad\quad \mathtt{return}\,c\\
\quad)\langle k,m\rangle\\
)))
\end{array}
\overset{\mathtt{reduc2}}{\sim_{\varnothing\subseteq\Upsilon}}
\begin{array}{l}
\mathtt{return}\,(!(\mathtt{return}\,(\lambda m.\\
\quad \mathtt{let}\,\langle k,m\rangle = \langle k,m\rangle\,\mathtt{in}\\
\quad \mathtt{let}\,r = \mathtt{random}\,\mathtt{in}\\
\quad \mathtt{let}\,x = F\langle k,r\rangle\,\mathtt{in}\\
\quad \mathtt{let}\,y = \mathtt{xor}(x,m)\,\mathtt{in}\\
\quad \mathtt{let}\,c = \mathtt{concat}(r,y)\,\mathtt{in}\\
\quad \mathtt{return}\,c\\
)))
\end{array}$$

$$\overset{\mathtt{reduc5}}{\sim_{\varnothing\subseteq\Upsilon}}
\begin{array}{l}
\mathtt{return}\,(!(\mathtt{return}\,(\lambda m.\\
\quad \mathtt{let}\,r = \mathtt{random}\,\mathtt{in}\\
\quad \mathtt{let}\,x = F\langle k,r\rangle\,\mathtt{in}\\
\quad \mathtt{let}\,y = \mathtt{xor}(x,m)\,\mathtt{in}\\
\quad \mathtt{let}\,c = \mathtt{concat}(r,y)\,\mathtt{in}\\
\quad \mathtt{return}\,c\\
)))
\end{array} \tag{10.7}$$

Now, the goal is to show that the term $\mathtt{let}\, o = M^F\ k\,\mathtt{in}\, EncOracle\ o$ is Θ-contextual indistinguishable from the term above in Equation 10.7, we proceed by gradually unfolding and simplifying it using the equations listed in Figure 10.2.

$$\mathtt{let}\, o = M^F\ k\,\mathtt{in}\, EncOracle\ o$$

$$\triangleq \begin{array}{l}
\mathtt{let}\,o = M^F\ k\,\mathtt{in}\\
\mathtt{return}\,(\lambda o.\mathtt{return}\,(!(\mathtt{return}\,(\lambda m.\\
\quad \mathtt{let}\,r = \mathtt{random}\,\mathtt{in}\\
\quad \mathtt{let}\,x = (\mathtt{der}(o)r)\,\mathtt{in}\\
\quad \mathtt{let}\,y = \mathtt{xor}(x,m)\,\mathtt{in}\\
\quad \mathtt{let}\,c = \mathtt{concat}(r,y)\,\mathtt{in}\\
\quad \mathtt{return}\,c\\
))))\ o
\end{array}
\overset{\mathtt{reduc2}}{\sim_{\varnothing\subseteq\Upsilon}}
\begin{array}{l}
\mathtt{let}\,o = \widehat{M}\ k\,\mathtt{in}\\
\mathtt{return}\,(!(\mathtt{return}\,(\lambda m.\\
\quad \mathtt{let}\,r = \mathtt{random}\,\mathtt{in}\\
\quad \mathtt{let}\,x = (\mathtt{der}(o)r)\,\mathtt{in}\\
\quad \mathtt{let}\,y = \mathtt{xor}(x,m)\,\mathtt{in}\\
\quad \mathtt{let}\,c = \mathtt{concat}(r,y)\,\mathtt{in}\\
\quad \mathtt{return}\,c\\
)))
\end{array}$$

$$\begin{aligned}
&\texttt{let}\, o = \Big(\texttt{return}\,(\lambda k.\texttt{return}\,(!(\texttt{return}\,(\lambda r.F\langle k,r\rangle)))))\Big)\, k\,\texttt{in}\\
&\texttt{return}\,(!(\texttt{return}\,(\lambda m.\\
&\quad \texttt{let}\, r = \texttt{random}\,\texttt{in}\\
\triangleq\quad &\quad \texttt{let}\, x = (\texttt{der}(o)r)\,\texttt{in}\\
&\quad \texttt{let}\, y = \texttt{xor}(x,m)\,\texttt{in}\\
&\quad \texttt{let}\, c = \texttt{concat}(r,y)\,\texttt{in}\\
&\quad \texttt{return}\, c\\
&)))
\end{aligned}$$

$$\begin{aligned}
&\texttt{let}\, o = \Big(\texttt{return}\,(!(\texttt{return}\,(\lambda r.F\langle k,r\rangle))))\Big)\;\texttt{in}\\
&\texttt{return}\,(!(\texttt{return}\,(\lambda m.\\
&\quad \texttt{let}\, r = \texttt{random}\,\texttt{in}\\
\overset{\texttt{reduc2}}{\sim_{\varnothing\subseteq\Upsilon}}\quad &\quad \texttt{let}\, x = (\texttt{der}(o)r)\,\texttt{in}\\
&\quad \texttt{let}\, y = \texttt{xor}(x,m)\,\texttt{in}\\
&\quad \texttt{let}\, c = \texttt{concat}(r,y)\,\texttt{in}\\
&\quad \texttt{return}\, c\\
&)))
\end{aligned}$$

$$\begin{aligned}
&\texttt{return}\,(!(\texttt{return}\,(\lambda m.\\
&\quad \texttt{let}\, r = \texttt{random}\,\texttt{in}\\
&\quad \texttt{let}\, x = (\texttt{der}\,\Big(!\,\Big(\texttt{return}\,(\lambda r.F\langle k,r\rangle)\Big)\Big)\, r)\,\texttt{in}\\
\overset{\texttt{reduc4}}{\sim_{\varnothing\subseteq\Upsilon}}\quad &\quad \texttt{let}\, y = \texttt{xor}(x,m)\,\texttt{in}\\
&\quad \texttt{let}\, c = \texttt{concat}(r,y)\,\texttt{in}\\
&\quad \texttt{return}\, c\\
&)))
\end{aligned}$$

$$\begin{aligned}
&\texttt{return}\,(!(\texttt{return}\,(\lambda m.\\
&\quad \texttt{let}\, r = \texttt{random}\,\texttt{in}\\
&\quad \texttt{let}\, x = \Big(\texttt{return}\,(\lambda r.F\langle k,r\rangle)\Big)\, r\,\texttt{in}\\
\overset{\texttt{reduc3}}{\sim_{\varnothing\subseteq\Upsilon}}\quad &\quad \texttt{let}\, y = \texttt{xor}(x,m)\,\texttt{in}\\
&\quad \texttt{let}\, c = \texttt{concat}(r,y)\,\texttt{in}\\
&\quad \texttt{return}\, c\\
&)))
\end{aligned}$$

$$\overset{\texttt{reduc2}}{\sim_{\varnothing\subseteq\Upsilon}} \quad \begin{array}{l} \texttt{return}\,(!(\texttt{return}\,(\lambda m. \\ \quad \texttt{let}\, r = \texttt{random}\,\texttt{in} \\ \quad \texttt{let}\, x = F\langle k, r\rangle\,\texttt{in} \\ \quad \texttt{let}\, y = \texttt{xor}(x, m)\,\texttt{in} \\ \quad \texttt{let}\, c = \texttt{concat}(r, y)\,\texttt{in} \\ \quad \texttt{return}\, c \\ ))) \end{array} \quad = \text{Term in Equation 10.7}$$

□

**Theorem 22.**

$$D^F \sim_{\varnothing\subseteq\Upsilon} PrivKCPA^F$$

*Proof.*

$$D^F \triangleq \texttt{let}\, k = Gen^F * \,\texttt{in}\,\texttt{let}\, o = M^F\ k\,\texttt{in}\,\texttt{let}\, e = EncOracle\ o\,\texttt{in}\, D\ e$$

$$\overset{\texttt{letAss}}{\sim_{\varnothing\subseteq\Upsilon}} \texttt{let}\, k = Gen^F * \,\texttt{in}\,\texttt{let}\, e = (\texttt{let}\, o = M^F\ k\,\texttt{in}\, EncOracle\ o)\,\texttt{in}\, D\ e$$

$$\overset{\text{Lemma 66}}{\sim_{\varnothing\subseteq\Upsilon}} \texttt{let}\, k = Gen^F * \,\texttt{in}\,\texttt{let}\, e = Oracle^F\ k\,\texttt{in}\, D\ e$$

$$=_\alpha \begin{array}{l} \texttt{let}\, k = Gen^F * \,\texttt{in} \\ \texttt{let}\, o = Oracle^F\ k\,\texttt{in} \\ D\ o \end{array}$$

$$\triangleq \begin{array}{l} \texttt{let}\, k = Gen^F * \,\texttt{in} \\ \texttt{let}\, o = Oracle^F\ k\,\texttt{in} \\ \left( \begin{array}{l} \lambda o. \\ \texttt{let}\, \langle msgs, s\rangle = Adv_1\ o\,\texttt{in} \\ \texttt{let}\, \langle m_0, m_1\rangle = msgs\,\texttt{in} \\ \texttt{let}\, b = \texttt{flipcoin}\,\texttt{in} \\ \texttt{if}\ b\ \texttt{then}\ (\lambda m.\texttt{let}\, c = (\texttt{der}(o)\ m)\,\texttt{in}\, Adv_2\langle\langle o, s\rangle, c\rangle)m_1 \\ \texttt{else}\ \texttt{let}\, z = (\lambda m.\texttt{let}\, c = (\texttt{der}(o)\ m)\,\texttt{in}\, Adv_2\langle\langle o, s\rangle, c\rangle)m_0\,\texttt{in}\,\texttt{not}(z) \end{array} \right) o \end{array}$$

$$\overset{\texttt{reduc2}}{\sim_{\varnothing\subseteq\Upsilon}} \begin{array}{l} \texttt{let}\, k = Gen^F * \,\texttt{in} \\ \texttt{let}\, o = Oracle^F\ k\,\texttt{in} \\ \texttt{let}\, \langle msgs, s\rangle = Adv_1\ o\,\texttt{in} \\ \texttt{let}\, \langle m_0, m_1\rangle = msgs\,\texttt{in} \\ \texttt{let}\, b = \texttt{flipcoin}\,\texttt{in} \\ \texttt{if}\ b\ \texttt{then}\ (\lambda m.\texttt{let}\, c = (\texttt{der}(o)\ m)\,\texttt{in}\, Adv_2\langle\langle o, s\rangle, c\rangle)m_1 \\ \texttt{else}\ \texttt{let}\, z = (\lambda m.\texttt{let}\, c = (\texttt{der}(o)\ m)\,\texttt{in}\, Adv_2\langle\langle o, s\rangle, c\rangle)m_0\,\texttt{in}\,\texttt{not}(z) \end{array}$$

$$\triangleq \quad PrivKCPA^F$$

□

Lastly, in order to complete the overall security proof, it remains to demonstrate the equation $\widehat{PrivKCPA} \sim_{\varnothing\subseteq\Upsilon}$ `flipcoin`, which corresponds to the top-right part of the diagram in Figure 10.8. In order to do so, we fist show that $\widehat{Enc} \sim_{\varnothing\subseteq\Upsilon} \lambda e.\mathtt{return\,random}$ as in Lemma 67 and we proceed by rewriting the oracle of $\widehat{PrivKCPA}$ as in Lemma 68. In the end, Theorem 23 conclude the proof.

**Lemma 67.**

$$\widehat{Enc} \sim_{\varnothing\subseteq\Upsilon} \lambda e.\mathtt{random}$$

*Proof.* We want to show that the encryption algorithm of $\widehat{\Pi}$, denoted by $\widehat{Enc}$, is $\Theta$-contextual indistinguishable from the term $\lambda e.\mathtt{return\,random}$. In other words, the behaviour of $\widehat{Enc}$ is essentially equivalent to a function that, given an input $e$, simply returns a random value, denoted by the function symbol `random`. We begin by unfolding $\widehat{Enc}$ as defined in Figure 10.6. Then, we simplify it by applying the Equation `randT` where $\Theta = \Xi = \Upsilon = \{rand : \mathbb{S}[(q+1) \times (2i+1)]\}$, since the closing context involved in this step uses the reference $rand$ by performing a `get` operation on it. Concretely, we obtain the following

$$\widehat{Enc} \triangleq \begin{array}{l}
\lambda e.\\
\mathtt{let}\,\langle k,m\rangle = e\,\mathtt{in}\\
\mathtt{let}\,tab = \mathtt{get}\,rand\,\mathtt{in}\\
\mathtt{let}\,r = \mathtt{random}\,\mathtt{in}\\
\mathtt{let}\,test = \mathtt{isdefined}(tab,r)\,\mathtt{in}\\
\mathtt{if}\ test\ \mathtt{then}\\
\quad \mathtt{let}\,x = \mathtt{getValue}(tab,r)\,\mathtt{in}\\
\quad \mathtt{let}\,y = \mathtt{xor}(x,m)\,\mathtt{in}\\
\quad \mathtt{let}\,c = \mathtt{concat}(r,y)\,\mathtt{in}\\
\quad \mathtt{return}\,c\\
\mathtt{else}\\
\quad \mathtt{let}\,x = \mathtt{random}\,\mathtt{in}\\
\quad \mathtt{let}\,y = \mathtt{xor}(x,m)\,\mathtt{in}\\
\quad \mathtt{let}\,newtab = \mathtt{modify}(tab,r,x)\,\mathtt{in}\\
\quad \mathtt{let}\,- = \mathtt{set}\,rand\,newtab\,\mathtt{in}\\
\quad \mathtt{let}\,c = \mathtt{concat}(r,y)\,\mathtt{in}\\
\quad \mathtt{return}\,c
\end{array}
\overset{\mathtt{randT}}{\sim_{\varnothing\subseteq\Upsilon}}
\begin{array}{l}
\lambda e.\\
\mathtt{let}\,\langle k,m\rangle = e\,\mathtt{in}\\
\mathtt{let}\,tab = \mathtt{get}\,rand\,\mathtt{in}\\
\mathtt{let}\,r = \mathtt{random}\,\mathtt{in}\\
\mathtt{let}\,x = \mathtt{random}\,\mathtt{in}\\
\mathtt{let}\,y = \mathtt{xor}(x,m)\,\mathtt{in}\\
\mathtt{let}\,newtab = \mathtt{modify}(tab,r,x)\,\mathtt{in}\\
\mathtt{let}\,- = \mathtt{set}\,rand\,newtab\,\mathtt{in}\\
\mathtt{let}\,c = \mathtt{concat}(r,y)\,\mathtt{in}\\
\mathtt{return}\,c
\end{array}$$

Next, we use the `letCom` rule several times to reorder the let bindings in the term, ensuring the structure is more streamlined. This is followed by a double application of the `letAss` rule, which groups related operations together. This reorganization of the term allows us to apply the equation `safeDisc1`, as all its conditions are satisfied, simplifying the term by eliminating unnecessary let bindings. Specifically, it allows the removal of a let binding if the bounded variable does not appear in the body of the let construct, and if the term being bound has no side effects that interact with the surrounding context. This reduction effectively streamlines the term by removing redundant assignments, resulting in a more concise term without changing its behaviour. Observe that the Equation `letCom` and the first application of Equation `letAss` are applied with $\Theta = \Xi = \Upsilon = \{rand : \mathbb{S}[(q+1) \times (2i+1)]\}$ since the closing context involved uses the reference $rand$. More precisely, in the steps where we apply `letCom`, the reference $rand$ is used by the closing context via a `set` operation. While in the first application of Equation

`letAss` the closing context uses the reference *rand* by performing a `get` operation on it.

$$
\overset{\texttt{letCom}}{\sim_{\varnothing\subseteq\Upsilon}}
\begin{array}{l}
\lambda e.\\
\texttt{let}\,\langle k,m\rangle = e\,\texttt{in}\\
\texttt{let}\,r = \texttt{random}\,\texttt{in}\\
\texttt{let}\,x = \texttt{random}\,\texttt{in}\\
\texttt{let}\,y = \texttt{xor}(x,m)\,\texttt{in}\\
\texttt{let}\,tab = \texttt{get}\,rand\,\texttt{in}\\
\texttt{let}\,newtab = \texttt{modify}(tab,r,x)\,\texttt{in}\\
\texttt{let}\,- = \texttt{set}\,rand\,newtab\,\texttt{in}\\
\texttt{let}\,c = \texttt{concat}(r,y)\,\texttt{in}\\
\texttt{return}\,c
\end{array}
$$

$$
\overset{\texttt{letAss}}{\sim_{\varnothing\subseteq\Upsilon}}
\begin{array}{l}
\lambda e.\\
\texttt{let}\,\langle k,m\rangle = e\,\texttt{in}\\
\texttt{let}\,r = \texttt{random}\,\texttt{in}\\
\texttt{let}\,x = \texttt{random}\,\texttt{in}\\
\texttt{let}\,y = \texttt{xor}(x,m)\,\texttt{in}\\
\texttt{let}\,- = \begin{pmatrix}\texttt{let}\,tab = \texttt{get}\,rand\,\texttt{in}\\ \texttt{let}\,newtab = \texttt{modify}(tab,r,x)\,\texttt{in}\\ \texttt{set}\,rand\,newtab\end{pmatrix}\,\texttt{in}\\
\texttt{let}\,c = \texttt{concat}(r,y)\,\texttt{in}\\
\texttt{return}\,c
\end{array}
\overset{\texttt{safeDisc1}}{\sim_{\varnothing\subseteq\Upsilon}}
\begin{array}{l}
\lambda e.\\
\texttt{let}\,\langle k,m\rangle = e\,\texttt{in}\\
\texttt{let}\,r = \texttt{random}\,\texttt{in}\\
\texttt{let}\,x = \texttt{random}\,\texttt{in}\\
\texttt{let}\,y = \texttt{xor}(x,m)\,\texttt{in}\\
\texttt{let}\,c = \texttt{concat}(r,y)\,\texttt{in}\\
\texttt{return}\,c
\end{array}
$$

We proceed by applying the associativity of let construct (Equation `letAss`) and then applying the `randXor` rule. More precisely, the Equation `randXor` states that performing a bitwise exclusive-or between a random generated string and another string of the same length is observationally equivalent to just generating a random string.

$$
\overset{\texttt{letAss}}{\sim_{\varnothing\subseteq\Upsilon}}
\begin{array}{l}
\lambda e.\\
\texttt{let}\,\langle k,m\rangle = e\,\texttt{in}\\
\texttt{let}\,r = \texttt{random}\,\texttt{in}\\
\texttt{let}\,y = (\texttt{let}\,x = \texttt{random}\,\texttt{in}\,\texttt{xor}(x,m))\,\texttt{in}\\
\texttt{let}\,c = \texttt{concat}(r,y)\,\texttt{in}\\
\texttt{return}\,c
\end{array}
\overset{\texttt{randXor}}{\sim_{\varnothing\subseteq\Upsilon}}
\begin{array}{l}
\lambda e.\\
\texttt{let}\,\langle k,m\rangle = e\,\texttt{in}\\
\texttt{let}\,r = \texttt{random}\,\texttt{in}\\
\texttt{let}\,y = \texttt{random}\,\texttt{in}\\
\texttt{let}\,c = \texttt{concat}(r,y)\,\texttt{in}\\
\texttt{return}\,c
\end{array}
$$

After performing two steps of `letAss`, we are able to apply `randConcat`, which asserts that two independently random strings are indistinguishable from any random string of the total length. Then, Equation `reduc1` simplifies the let binding by removing the intermediate variable $c$. Finally, Equation `safeDisc2` is applied since its conditions are satisfied and it simplifies the term by removing unnecessary let bindings. This sequence of reductions ultimately simplifies the original term to just returning a random value.

$$\overset{\texttt{letAss}}{\sim_{\varnothing\subseteq\Upsilon}} \begin{array}{l} \lambda e. \\ \texttt{let}\,\langle k,m\rangle = e\,\texttt{in} \\ \texttt{let}\,c = \begin{pmatrix} \texttt{let}\,r = \texttt{random}\,\texttt{in} \\ \texttt{let}\,y = \texttt{random}\,\texttt{in} \\ \texttt{concat}(r,y) \end{pmatrix} \texttt{in} \\ \texttt{return}\,c \end{array} \quad \overset{\texttt{randConcat}}{\sim_{\varnothing\subseteq\Upsilon}} \begin{array}{l} \lambda e. \\ \texttt{let}\,\langle k,m\rangle = e\,\texttt{in} \\ \texttt{let}\,c = \texttt{random}\,\texttt{in} \\ \texttt{return}\,c \end{array}$$

$$\overset{\texttt{reduc1}}{\sim_{\varnothing\subseteq\Upsilon}} \begin{array}{l} \lambda e. \\ \texttt{let}\,\langle k,m\rangle = e\,\texttt{in} \\ \texttt{random} \end{array} \quad \overset{\texttt{safeDisc2}}{\sim_{\varnothing\subseteq\Upsilon}} \begin{array}{l} \lambda e. \\ \texttt{random} \end{array}$$

□

**Lemma 68.**

$$\widehat{Oracle} \sim_{\varnothing\subseteq\Upsilon} \texttt{return}\,(\lambda k.\texttt{return}\,!(\texttt{return}\,\lambda m.\texttt{random}))$$

*Proof.*

$$\widehat{Oracle} \triangleq \texttt{return}\,(\lambda k.\texttt{return}\,!(\texttt{return}\,\lambda m.\widehat{Enc}\langle k,m\rangle))$$

$$\overset{\text{Lemma 67}}{\sim_{\varnothing\subseteq\Upsilon}} \texttt{return}\,(\lambda k.\texttt{return}\,!(\texttt{return}\,\lambda m.(\lambda e.\texttt{random})\langle k,m\rangle))$$

$$\overset{\texttt{reduc2}}{\sim_{\varnothing\subseteq\Upsilon}} \texttt{return}\,(\lambda k.\texttt{return}\,!(\texttt{return}\,\lambda m.\texttt{random}))$$

□

**Theorem 23.**

$$\widehat{PrivKCPA} \sim_{\varnothing\subseteq\Upsilon} \texttt{flipcoin}$$

*Proof.* In the following, we show that the term $\widehat{PrivKCPA}$ behaves in a way that is indistinguishable from a fair random binary choice, where the outcome is either 0 or 1 with equal probability. We begin by unfolding $\widehat{PrivKCPA}$ which is the variation of the term in Figure 10.4 for the scheme $\widehat{\Pi}$.

$$\widehat{PrivKCPA} \triangleq \begin{array}{l} \texttt{let}\,k = \widehat{Gen}\,*\,\texttt{in} \\ \texttt{let}\,o = \widehat{Oracle}\;k\,\texttt{in} \\ \texttt{let}\,\langle msgs, s\rangle = Adv_1\;o\,\texttt{in} \\ \texttt{let}\,\langle m_0, m_1\rangle = msgs\,\texttt{in} \\ \texttt{let}\,b = \texttt{flipcoin}\,\texttt{in} \\ \texttt{if}\;b\;\texttt{then} \\ \quad (\lambda m.\texttt{let}\,c = (\texttt{der}(o)\;m)\,\texttt{in}\,Adv_2\langle\langle o,s\rangle,c\rangle)m_1 \\ \texttt{else} \\ \quad \texttt{let}\,z = (\lambda m.\texttt{let}\,c = (\texttt{der}(o)\;m)\,\texttt{in}\,Adv_2\langle\langle o,s\rangle,c\rangle)m_0\,\texttt{in}\,\texttt{not}(z) \end{array}$$

In the second step, we apply Lemma 68 which states that $\widehat{Oracle}$ is observationally equivalent to a term that, when given a key $k$, produces a value by returning a function. This function, when invoked with a message $m$, generates a random value. We then simplify the term by applying simple reduction steps, as detailed below. Please note that these steps are performed in parallel across both branches of the if-then-else construct.

$$\begin{array}{ll}
 & \texttt{let}\, k = \widehat{Gen} * \texttt{ in} \\
 & \texttt{let}\, o = (\lambda k.\texttt{return}\,!(\texttt{return}\,\lambda m.\texttt{random}))\; k\,\texttt{in} \\
 & \texttt{let}\, \langle msgs, s\rangle = Adv_1\; o\,\texttt{in} \\
 & \texttt{let}\, \langle m_0, m_1\rangle = msgs\,\texttt{in} \\
\overset{\text{Lemma 68}}{\sim_{\varnothing\subseteq\Upsilon}} & \texttt{let}\, b = \texttt{flipcoin}\,\texttt{in} \\
 & \texttt{if}\; b\; \texttt{then} \\
 & \quad (\lambda m.\texttt{let}\, c = (\texttt{der}(o)\;\, m)\,\texttt{in}\, Adv_2\langle\langle o, s\rangle, c\rangle)m_1 \\
 & \texttt{else} \\
 & \quad \texttt{let}\, z = (\lambda m.\texttt{let}\, c = (\texttt{der}(o)\;\, m)\,\texttt{in}\, Adv_2\langle\langle o, s\rangle, c\rangle)m_0\,\texttt{in}\,\texttt{not}(z)
\end{array}$$

$$\begin{array}{ll}
 & \texttt{let}\, k = \widehat{Gen} * \texttt{ in} \\
 & \texttt{let}\, o = \texttt{return}\,!(\texttt{return}\,\lambda m.\texttt{random})\,\texttt{in} \\
 & \texttt{let}\, \langle msgs, s\rangle = Adv_1\; o\,\texttt{in} \\
 & \texttt{let}\, \langle m_0, m_1\rangle = msgs\,\texttt{in} \\
\overset{\texttt{reduc2}}{\sim_{\varnothing\subseteq\Upsilon}} & \texttt{let}\, b = \texttt{flipcoin}\,\texttt{in} \\
 & \texttt{if}\; b\; \texttt{then} \\
 & \quad (\lambda m.\texttt{let}\, c = (\texttt{der}(o)\;\, m)\,\texttt{in}\, Adv_2\langle\langle o, s\rangle, c\rangle)m_1 \\
 & \texttt{else} \\
 & \quad \texttt{let}\, z = (\lambda m.\texttt{let}\, c = (\texttt{der}(o)\;\, m)\,\texttt{in}\, Adv_2\langle\langle o, s\rangle, c\rangle)m_0\,\texttt{in}\,\texttt{not}(z)
\end{array}$$

$$\begin{array}{ll}
 & \texttt{let}\, k = \widehat{Gen} * \texttt{ in} \\
 & \texttt{let}\, \langle msgs, s\rangle = Adv_1\; !(\texttt{return}\,\lambda m.\texttt{random})\,\texttt{in} \\
 & \texttt{let}\, \langle m_0, m_1\rangle = msgs\,\texttt{in} \\
 & \texttt{let}\, b = \texttt{flipcoin}\,\texttt{in} \\
\overset{\texttt{reduc4}}{\sim_{\varnothing\subseteq\Upsilon}} & \texttt{if}\; b\; \texttt{then} \\
 & \quad \left(\lambda m.\left(\begin{array}{l}\texttt{let}\, c = \texttt{der}(!(\texttt{return}\,\lambda m.\texttt{random}))\;\, m\,\texttt{in} \\ Adv_2\langle\langle !(\texttt{return}\,\lambda m.\texttt{random}), s\rangle, c\rangle\end{array}\right)\right) m_1 \\
 & \texttt{else} \\
 & \quad \texttt{let}\, z = \left(\lambda m.\left(\begin{array}{l}\texttt{let}\, c = \texttt{der}(!(\texttt{return}\,\lambda m.\texttt{random}))\;\, m\,\texttt{in} \\ Adv_2\langle\langle !(\texttt{return}\,\lambda m.\texttt{random}), s\rangle, c\rangle\end{array}\right)\right) m_0\,\texttt{in}\,\texttt{not}(z)
\end{array}$$

$$\begin{array}{l}
\texttt{let}\, k = \widehat{Gen} * \texttt{ in} \\
\texttt{let}\, \langle msgs, s\rangle = Adv_1 \; !(\texttt{return}\, \lambda m.\texttt{random})\,\texttt{in} \\
\texttt{let}\, \langle m_0, m_1\rangle = msgs\,\texttt{in} \\
\texttt{let}\, b = \texttt{flipcoin}\,\texttt{in} \\
\texttt{if}\ b\ \texttt{then} \\
\quad \left(\lambda m. \begin{pmatrix} \texttt{let}\, c = (\texttt{return}\, \lambda m.\texttt{random})\ m\,\texttt{in} \\ Adv_2\langle\langle !(\texttt{return}\, \lambda m.\texttt{random}), s\rangle, c\rangle \end{pmatrix}\right) m_1 \\
\texttt{else} \\
\quad \texttt{let}\, z = \left(\lambda m. \begin{pmatrix} \texttt{let}\, c = (\texttt{return}\, \lambda m.\texttt{random})\ m\,\texttt{in} \\ Adv_2\langle\langle !(\texttt{return}\, \lambda m.\texttt{random}), s\rangle, c\rangle \end{pmatrix}\right) m_0\,\texttt{in}\,\texttt{not}(z)
\end{array}$$

$\overset{\texttt{reduc3}}{\sim_{\varnothing \subseteq \Upsilon}}$

$$\begin{array}{l}
\texttt{let}\, k = \widehat{Gen} * \texttt{ in} \\
\texttt{let}\, \langle msgs, s\rangle = Adv_1 \; !(\texttt{return}\, \lambda m.\texttt{random})\,\texttt{in} \\
\texttt{let}\, \langle m_0, m_1\rangle = msgs\,\texttt{in} \\
\texttt{let}\, b = \texttt{flipcoin}\,\texttt{in} \\
\texttt{if}\ b\ \texttt{then} \\
\quad \left(\lambda m. \begin{pmatrix} \texttt{let}\, c = \texttt{random}\,\texttt{in} \\ Adv_2\langle\langle !(\texttt{return}\, \lambda m.\texttt{random}), s\rangle, c\rangle \end{pmatrix}\right) m_1 \\
\texttt{else} \\
\quad \texttt{let}\, z = \left(\lambda m. \begin{pmatrix} \texttt{let}\, c = \texttt{random}\,\texttt{in} \\ Adv_2\langle\langle !(\texttt{return}\, \lambda m.\texttt{random}), s\rangle, c\rangle \end{pmatrix}\right) m_0\,\texttt{in}\,\texttt{not}(z)
\end{array}$$

$\overset{\texttt{reduc2}}{\sim_{\varnothing \subseteq \Upsilon}}$

$$\begin{array}{l}
\texttt{let}\, k = \widehat{Gen} * \texttt{ in} \\
\texttt{let}\, \langle msgs, s\rangle = Adv_1 \; (\texttt{return}\,!(\texttt{return}\, \lambda m.\texttt{random}))\,\texttt{in} \\
\texttt{let}\, \langle m_0, m_1\rangle = msgs\,\texttt{in} \\
\texttt{let}\, b = \texttt{flipcoin}\,\texttt{in} \\
\texttt{if}\ b\ \texttt{then} \\
\quad \texttt{let}\, c = \texttt{random}\,\texttt{in}\, Adv_2\langle\langle !(\texttt{return}\, \lambda m.\texttt{random}), s\rangle, c\rangle \\
\texttt{else} \\
\quad \texttt{let}\, z = (\texttt{let}\, c = \texttt{random}\,\texttt{in}\, Adv_2\langle\langle !(\texttt{return}\, \lambda m.\texttt{random}), s\rangle, c\rangle)\,\texttt{in}\,\texttt{not}(z)
\end{array}$$

$\overset{\texttt{reduc2}}{\sim_{\varnothing \subseteq \Upsilon}}$

Observe that in the last equation the if-then-else subterm is in the form $\texttt{if}\, b\, \texttt{then}\, M\, \texttt{else}\, (\texttt{let}\, z = M\, \texttt{in}\, \texttt{not}(z))$, where $b$ is generated by the function symbol `flipcoin`. Therefore, Equation `safeDisc3` applies, as all its conditions are satisfied, and we can conclude by performing some simple steps as follows

$$\begin{aligned}
&\texttt{let}\, k = \widehat{Gen} * \,\texttt{in} \\
\overset{\texttt{safeDisc3}}{\sim_{\varnothing\subseteq\Upsilon}}\quad &\texttt{let}\, \langle msgs, s\rangle = Adv_1\, !(\texttt{return}\, \lambda m.\texttt{random})\,\texttt{in} \\
&\texttt{let}\, \langle m_0, m_1\rangle = msgs\,\texttt{in} \\
&\texttt{flipcoin}
\end{aligned}$$

$$\begin{aligned}
&\texttt{let}\, \langle msgs, s\rangle = Adv_1\, !(\texttt{return}\, \lambda m.\texttt{random})\,\texttt{in} \\
\overset{\texttt{safeDisc1}}{\sim_{\varnothing\subseteq\Upsilon}}\quad &\texttt{let}\, \langle m_0, m_1\rangle = msgs\,\texttt{in} \\
&\texttt{flipcoin}
\end{aligned}$$

$$\begin{aligned}
\overset{\texttt{letAss}}{\sim_{\varnothing\subseteq\Upsilon}}\quad &\texttt{let}\, \langle m_0, m_1\rangle = (\texttt{let}\, \langle msgs, s\rangle = Adv_1\, !(\texttt{return}\, \lambda m.\texttt{random})\,\texttt{in}\, msgs)\ \texttt{in} \\
&\texttt{flipcoin}
\end{aligned}$$

$$\overset{\texttt{safeDisc2}}{\sim_{\varnothing\subseteq\Upsilon}}\quad \texttt{flipcoin}$$

□

Summing up, let us turn our attention back to the diagram in Figure 10.8, which offers a structured representation of the overall security proof within the λBLL framework. First, we establish the key equivalences involving the distinguishers: specifically, Theorem 21 shows that $D^F \sim_{\varnothing\subseteq\Upsilon} PrivKCPA^F$, while Theorem 22 establishes that $\widehat{D} \sim_{\varnothing\subseteq\Upsilon} \widehat{PrivKCPA}$. The final step in the proof involves showing that $\widehat{PrivKCPA} \sim_{\varnothing\subseteq\Upsilon}$ `flipcoin`, corresponding to the top-right part of the diagram. This crucial equivalence is addressed by first proving, in Lemma 67, that the encryption function used in $\widehat{PrivKCPA}$ satisfies $\widehat{Enc} \sim_{\varnothing\subseteq\Upsilon} \lambda e.\texttt{return}\,\texttt{random}$. Subsequently, Lemma 68 formalizes the rewriting of the oracle in $\widehat{PrivKCPA}$ based on this result. These components collectively allow us to prove Theorem 23, which concludes the argument by establishing that the behaviour of $\widehat{PrivKCPA}$ is indistinguishable from fair coin toss.

This sequence of results is derived through a series of equational steps, each of which can be rigorously justified by the logical relation defined in Section 9.4. These steps collectively form a formal proof of the CPA security for the encryption scheme $\Pi_F$ within the λBLL framework. The argument leverages the expressive power of equational reasoning, applied in a precise and compositional manner, to capture the essential structure of the proof by reduction. By systematically mapping the behaviour of each component in the proof to an idealized counterpart through the logical relation, we establish that the encryption scheme satisfies the necessary indistinguishability guarantees. This approach not only evidences the efficacy of the λBLL framework in formalizing cryptographic security properties but also highlights its capacity to rigorously verify these properties through an equational and structured methodology within a higher-order setting. The ability of λBLL to model both computational and probabilistic behaviours in a unified framework further strengthens its potential to reason about complex cryptographic constructs, ensuring the soundness and robustness of security proofs.

# Conclusion and Future Perspectives of CryptoBLL

In this part of the thesis, we have developed $\lambda$BLL, a higher-order, probabilistic, linear $\lambda$-calculus, as a semantically robust and computationally faithful framework for reasoning about computational indistinguishability. This work lies at the intersection of programming language theory and computational cryptography and addresses the longstanding challenge of designing formal, language-based techniques that are expressive enough to model cryptographic protocols, capture probabilistic behaviour accurately, and respect polytime constraints. At the same time, it aims to keep the framework simple by abstracting away underlying complexity and masking low-level details of probability and computation, thereby supporting clear and tractable reasoning.

The $\lambda$BLL calculus has a Curry-Howard correspondence with Bounded Linear Logic and extends existing frameworks by incorporating graded modalities, probabilistic effects, and mutable state in a disciplined and compositional way. Through the design of a sound operational semantics and a type system that tracks complexity via indexed types and effect annotations, we have established that all typable terms correspond to probabilistic polytime computations, including second-order constructions, an essential feature for modelling adversarial behaviour with oracle access in cryptographic settings.

Central to our framework is the definition of an approximate logical relation formulated as a metric, which generalizes traditional logical relations by incorporating a quantitative notion of indistinguishability. This metric relation is shown to be sound with respect to $\Theta$-contextual indistinguishability, an approximate observational equivalence formalized in terms of adversarial advantage. This connection bridges the gap between abstract formalism and practical cryptographic security, allowing for the translation of security guarantees into mathematically verifiable properties. Crucially, the metric relation supports equational reasoning by reducing the task of proving program indistinguishability to showing that their distance in the logical metric is negligible. This approach eliminates the need to consider all possible contexts explicitly and instead provides a compositional and scalable framework for establishing security properties.

We have shown the expressiveness of the framework by formalizing the CPA security proof of an encryption scheme induced by a pseudorandom function. We began by introducing a set of function symbols and equations that enrich the expressive power of $\lambda$BLL, enabling us to model computational patterns such as randomness generation, bitstring manipulation, and memory operations. We then formalized a standard cryptographic proof by encoding the components of a reduction-based argument as $\lambda$BLL terms. The proof was concluded by establishing the CPA security of the encryption scheme through equational reasoning, with each step justified by the logical relation. This example highlights the capability of $\lambda$BLL to effectively model crypto-graphic constructs within higher-order settings, facilitating a rigorous and structured approach to security proofs. By providing a formal framework that captures both the probabilistic behaviour and computational complexity of cryptographic protocols, $\lambda$BLL ensures the soundness and robustness of these proofs. Its integration of equational reasoning and effects enables precise analysis of cryptographic systems, offering a powerful tool for modelling adversarial behaviour and verifying security properties in a compositional and scalable manner. This demonstrates the framework's potential to support secure, high-level cryptographic reasoning while maintaining rigorous formal guarantees.

In conclusion, this part of thesis advances the theoretical and practical foundations of cryptographic verification by introducing a language-based framework in which security properties can be proved in an equational way. Specifically, computational indistinguishability can be reasoned about equationally, compositionally, and with semantic rigour even in higher order settings.

Potential directions for future work include transitioning to a logic based on higher-order logic, which would facilitate the integration of both relational and logical reasoning. This would enable a more sophisticated and flexible framework for reasoning about cryptographic protocols, combining the advantages of relational methods with the expressive capabilities of higher-order logic. By incorporating quantification over functions and predicates, such a framework could support more intricate cryptographic constructs and protocols, requiring higher levels of abstraction. Additionally, this approach could enable the formal verification of advanced cryptographic properties, such as compositional security guarantees and higher-order security reductions. A relevant reference in this domain is the work of Aguirre and co-authors [9], which explores the interplay between relational and logical techniques in the context of cryptography. Extending this approach within the $\lambda$BLL framework could significantly enhance the scalability, modularity, and rigour of cryptographic reasoning, particularly in analyzing and proving security properties of complex cryptographic primitives.

# Conclusion

This thesis explores the role of linearity as a unifying principle across two distinct domains: automatic differentiation and computational cryptography. Through the lens of PL theory and formal systems, we have shown how linear logic and linear $\lambda$-calculi offer robust foundations for reasoning about resource-sensitive computation in both numerical and security-critical contexts.

In the first part (ADLL), we focused on the formalization of AD transformations using a linear $\lambda$-calculus, $\lambda$LL, grounded in the Curry-Howard correspondence with a fragment of linear logic. We presented a type-directed encoding of the Linear A system and its core transformations (forward, unzipping and transpose) as defined in the JAX Autodiff framework. Our formalization captures both the qualitative correctness and quantitative efficiency of these transformations, establishing extensional equivalence modulo resource cost. Notably, our approach improves modularity by demonstrating that the unzipping step, central to reverse-mode AD in JAX, can be safely skipped in $\lambda$LL, thereby offering benefits in terms of program structure and potential for parallelism. This shows how linear logic provides not only a formal foundation for AD, but also practical insights into optimizing its implementation in frameworks like JAX.

In the second part (CryptoBLL), we introduced the calculus $\lambda$BLL, a linear $\lambda$-calculus designed for reasoning about computational indistinguishability in cryptographic settings. Building on Bounded Linear Logic, $\lambda$BLL integrates probabilistic effects, mutable references and linear types with complexity annotations to ensure that all typable programs correspond to probabilistic polynomial-time computations. We defined an approximate logical relation that enables equational reasoning about computational indistinguishability, abstracting away from low-level adversarial simulations. Through a detailed case study formalizing a CPA security proof, we demonstrated the expressiveness of the calculus and the compositionality of its reasoning principles in a higher-order setting. The use of a metric logical relation bridges semantic rigour with cryptographic intuition, enabling precise and scalable formal verification of cryptographic protocols.

Figure 10.9 graphically describes how core concepts from PL theory and linear logic recur across the two contributions of this thesis: ADLL and CryptoBLL. Although these systems are developed for different domains, automatic differentiation and cryptography, they draw on a shared theoretical foundation. The diagram visually organizes these parallels, emphasizing how analogous tools (such as evaluation strategies, logical systems, cost models, and logical relations) are instantiated differently but serve similar foundational roles. In the following discussion, we unpack each of these correspondences, moving across the central band of linear logic to explore how both ADLL and CryptoBLL apply and adapt the same conceptual tools to address domain-specific goals.

What unifies the two contributions from a logical perspective is the Curry-Howard correspondence between the type systems of $\lambda$LL and $\lambda$BLL with Linear Logic. More precisely, $\lambda$LL has a correspondence with the fragment of Linear Logic that accurately captures the specific role played by linearity in Autodiff. Instead, $\lambda$BLL has a correspondence with Bounded Linear Logic, a refinement of Linear Logic able to model the complexity constraints of attackers in the realm of computational cryptography. Furthermore, both parts embrace the modularity offered by linear logic. In ADLL, modularity enables reuse and parallelization of derivative computations; in CryptoBLL, it supports compositional proofs of security where subcomponents can be analyzed independently and securely combined. In both cases, linearity avoids unwanted duplication or erasure of data, maintaining structural invariants that are essential for correctness or security.

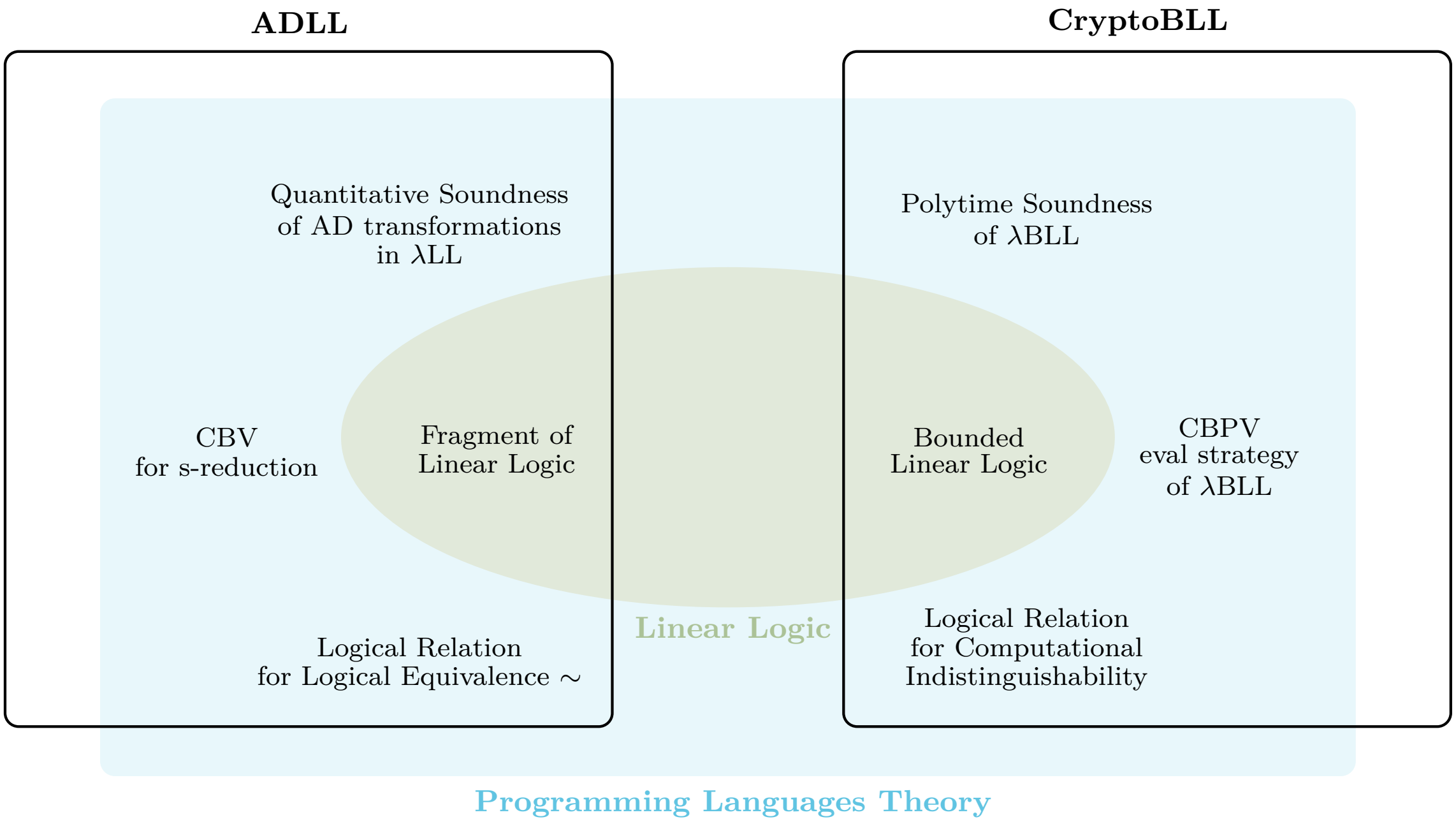

Figure 10.9: Diagram illustrating the shared theoretical foundations between ADLL and CryptoBLL, highlighting the use of linear logic, evaluation strategies, cost models, and logical relations in both domains.

From the syntactic perspective both $\lambda$LL and $\lambda$BLL extend linear $\lambda$-calculi with domain-specific primitives. Specifically, $\lambda$LL extends linear $\lambda$-calculi to the ground type of the real numbers $\mathbb{R}$ with a set of functional symbols representing differentiable functions, which are the core elements of AD. In contrast, $\lambda$BLL enriches linear $\lambda$-calculi with function symbols representing probabilistic polynomial-time computations, namely the basic building blocks of any cryptographic protocol, and global references, which enable the representation of shared mutable state.

Both systems leverage their evaluation strategies and cost models to formalize and guarantee soundness within their respective domains, highlighting a unified approach to managing both evaluation and resource consumption. In the ADLL setting, $\lambda$LL adopts a CBV-style reduction strategy, restricted to a class of terms called *safe*, with a cost model designed to bound the number of numeric steps. This provides the basis for establishing the quantitative soundness of the AD system. In contrast, in the CryptoBLL part, $\lambda$BLL adopts a CBPV evaluation strategy, and its cost model is tailored to establish the probabilistic polynomial-time soundness of the calculus in the context of cryptographic computations.

Another shared foundational element of the two contributions is the use of logical relations, a semantic technique originally developed for reasoning about program behaviour and equivalence in typed $\lambda$-calculi. In ADLL, a binary logical relation $\sim$ is used to relate programs that are extensionally equivalent on the ground type of real numbers, which supports the formulation and proof of soundness theorems for AD transformations. This relation is qualitative in nature, capturing when two programs compute the same result, possibly with different internal structure. In contrast, CryptoBLL employs a quantitative logical relation, defined via a logical metric that measures behavioural distance between terms. This approximation is crucial for proving soundness with respect to computational indistinguishability, a central notion in cryptography that allows small behavioural differences between terms, as long as no efficient adversary can distinguish them with more than negligible advantage. In both cases, logical relations are parameterized by the type structure of a linear $\lambda$-calculus. This shared methodology exemplifies the unifying strength of type-indexed, compositional semantics, especially when combined with linearity. It enables reasoning about both what a program computes and how it computes it, including the resources it consumes, the flow of data within the program, and how its behaviour is perceived by an observer or adversary.

In conclusion, this thesis has explored the use of linear logic and PL theory to address key challenges in two distinct domains: automatic differentiation and computational cryptography. By formalizing and analyzing the evaluation strategies, cost models, and logical relations that underpin the two systems, we have shown how the same foundational tools from PL theory can be adapted to solve different problems effectively. The ADLL part of the thesis demonstrated how linear logic can be used to ensure the soundness of AD transformations, with an emphasis on efficiency and resource management. The CryptoBLL part applied similar principles to cryptographic protocols, using Bounded Linear Logic to establish the probabilistic polynomial-time soundness necessary for cryptographic applications in the computational model. These correspondences reveal the versatility of linear logic as a unifying theoretical framework for diverse applications in programming and security. By highlighting these parallels, this work contributes to the understanding of how foundational concepts can be adapted across different domains, fostering more efficient, secure, and sound computational systems. Ultimately, this thesis demonstrates the power of formal tools to bridge the gap between fields as seemingly disparate as machine learning and cryptography, laying the groundwork for further exploration and innovation in both areas.

# Bibliography

## Bibliography ADLL

# Bibliography CryptoBLL